%% file: motion.tex
\pdfoutput=1
\documentclass{article}
\usepackage{fancyhdr}
\usepackage{amsmath}
\usepackage{amssymb}
\usepackage{amsfonts}
\usepackage{bm} 
\usepackage{cite}
\usepackage{graphicx}
\usepackage{textcomp}
\usepackage{calc}
\usepackage{mathrsfs}
\usepackage{color}
\usepackage{tensor}

\makeatletter
  \@addtoreset{equation}{section}
  
 \renewcommand\section{\@startsection {section}{1}{\z@}%
       {-3.5ex \@plus -1ex \@minus -.2ex}%
       {2.3ex \@plus.2ex}%
       {\normalfont\Large\bfseries\centering}}
  \renewcommand\subsection{\@startsection{subsection}{2}{\z@}%
       {-3.25ex\@plus -1ex \@minus -.2ex}%
       {1.5ex \@plus .2ex}%
       {\normalfont\large\bfseries\centering}}
  \renewcommand{\sectionmark}[1]%
    {\markboth{\thesection\hspace*{10pt} #1}{}} 
  \renewcommand{\subsectionmark}[1]%
    {\markright{\thesubsection\hspace*{10pt} #1}} 
\renewcommand{\contentsname}{Contents} 
\renewcommand{\tableofcontents}{%
    \section*{\contentsname
        \@mkboth{\contentsname}{\contentsname}}%
    \@starttoc{toc}%
}
\renewcommand{\listfigurename}{List of figures} 
\renewcommand{\listoffigures}{%
    \section*{\listfigurename
      \@mkboth{\listfigurename}{\listfigurename}}%
    \@starttoc{lof}%
}
\makeatother 

\newcommand{\base}[2]{e^{#1}_{#2}}

\newcommand{\e}{\varepsilon}
\newcommand{\A}[2]{\hat A_{#1}^{(#2)}}
\newcommand{\B}[2]{\hat B_{#1}^{(#2)}}
\newcommand{\C}[2]{\hat C_{#1}^{(#2)}}
\newcommand{\D}[2]{\hat D_{#1}^{(#2)}}
\newcommand{\E}[2]{\hat E_{#1}^{(#2)}}
\newcommand{\F}[2]{\hat F_{#1}^{(#2)}}
\newcommand{\G}[2]{\hat G_{#1}^{(#2)}}
\renewcommand{\H}[2]{\hat H_{#1}^{(#2)}}
\newcommand{\I}[2]{\hat I_{#1}^{(#2)}}
\newcommand{\K}[2]{\hat K_{#1}^{(#2)}}
\newcommand{\rad}{\mathcal{R}}
\newcommand{\advr}{r_{\rm adv}}

\newcommand{\tail}{h^{\rm tail}}
\newcommand{\etide}{\mathcal{E}}
\newcommand{\btide}{\mathcal{B}}
\newcommand{\hmn}[2]{h^{(#2)}_{#1}}

\newcommand{\del}[1]{\nabla_{\!\!#1}}
\newcommand{\Lie}[1]{\pounds_{\!#1}}
\DeclareMathOperator{\STF}{STF}
\newcommand{\man}{\mathcal{M}}
\newcommand{\exact}[1]{\mathsf{#1}}
\newcommand{\ddR}[3]{\delta^2 R^{(#2)}_{#1}\!\left[#3\right]}

\DeclareFontFamily{OT1}{rsfs}{} 
\DeclareFontShape{OT1}{rsfs}{m}{n}{<-7> rsfs5 
    <7-10> rsfs7 <10-> rsfs10}{}   
\DeclareMathAlphabet{\scr}{OT1}{rsfs}{m}{n}

\begin{document}
\thispagestyle{empty} 

\begin{center} 
\LARGE \sc 
The motion of point particles in curved spacetime 
\end{center} 

\vspace*{.1in} 
\begin{center} 
\large
Eric Poisson, Adam Pound, and Ian Vega \\ 
\small 
Department of Physics, University of Guelph, 
Guelph, Ontario, Canada N1G 2W1 
\end{center} 

\begin{center} 
Major update of Living Reviews article, with final revisions
(September 25, 2011)  
\end{center} 

\vspace*{.1in} 
\begin{abstract} 
This review is concerned with the motion of a point scalar charge, a
point electric charge, and a point mass in a specified background
spacetime. In each of the three cases the particle produces a field 
that behaves as outgoing radiation in the wave zone, and therefore
removes energy from the particle. In the near zone the field acts on
the particle and gives rise to a self-force that prevents the
particle from moving on a geodesic of the background spacetime. The
self-force contains both conservative and dissipative terms, and the 
latter are responsible for the radiation reaction. The work done by
the self-force matches the energy radiated away by the particle. 

The field's action on the particle is difficult to calculate because
of its singular nature: the field diverges at the position of the
particle. But it is possible to isolate the field's singular part and
show that it exerts no force on the particle --- its only effect is to
contribute to the particle's inertia. What remains after subtraction
is a regular field that is fully responsible for the
self-force. Because this field satisfies a homogeneous wave equation,  
it can be thought of as a free field that interacts with the particle;
it is this interaction that gives rise to the self-force.  

The mathematical tools required to derive the equations of motion of a
point scalar charge, a point electric charge, and a point mass in a
specified background spacetime are developed here from scratch. The
review begins with a discussion of the basic theory of bitensors 
(part \ref{part1}). It then applies the theory to the construction of
convenient coordinate systems to chart a neighbourhood of the
particle's word line (part \ref{part2}). It continues with a thorough
discussion of Green's functions in curved spacetime (part
\ref{part3}). The review presents a detailed derivation of each
of the three equations of motion (part \ref{part4}).  Because the
notion of a point mass is problematic in general relativity, the
review concludes (part \ref{part5}) with an alternative derivation of
the equations of motion that applies to a small body of arbitrary
internal structure.  
\end{abstract} 

\vspace*{.1in}
\tableofcontents

\newpage
\input{introduction}

\input{literature}

\newpage
\hrule
\hrule
\part{General theory of bitensors}
\label{part1}
\hrule
\hrule 
\vspace*{.25in} 
\input{part1}

\newpage
\hrule
\hrule
\part{Coordinate systems} 
\label{part2}
\hrule
\hrule
\vspace*{.25in} 
\input{part2} 

\newpage
\hrule
\hrule
\part{Green's functions} 
\label{part3}
\hrule
\hrule
\vspace*{.25in} 
\input{part3}

\newpage
\hrule
\hrule
\part{Motion of point particles} 
\label{part4}
\hrule
\hrule
\vspace*{.25in} 
\input{part4}

\newpage
\hrule
\hrule
\part{Motion of a small body} 
\label{part5}
\hrule
\hrule
\vspace*{.25in} 
\input{part5}
\input{conclusion}

\newpage
\hrule
\hrule
\part*{Appendices} 
\label{appendices}
\hrule
\hrule
\vspace*{.25in} 
\appendix
\input{appendices}

\newpage
\bibliography{motion}

\end{document}

%% file: introduction.tex
%
\section{Introduction and summary} 
\label{1}

\subsection{Invitation}
\label{1.1} 

The motion of a point electric charge in flat spacetime was the 
subject of active investigation since the early work of Lorentz, 
Abrahams, Poincar\'e, and Dirac \cite{dirac:38}, until Gralla, Harte,
and Wald produced a definitive derivation of the equations of motion   
\cite{gralla-etal:09} with all the rigour that one should demand,
without recourse to postulates and renormalization procedures.  
(The field's early history is well related in
Ref.~\cite{rohrlich:90}.) In 1960 DeWitt and Brehme  
\cite{dewitt-brehme:60} generalized Dirac's result to curved
spacetimes, and their calculation was corrected by Hobbs
\cite{hobbs:68} several years later. In 1997 the  
motion of a point mass in a curved background spacetime was
investigated by Mino, Sasaki, and Tanaka \cite{mino-etal:97a}, who
derived an expression for the particle's acceleration (which is not
zero unless the particle is a test mass); the same equations of motion
were later obtained by Quinn and Wald \cite{quinn-wald:97} using an
axiomatic approach. The case of a point scalar charge was finally
considered by Quinn in 2000 \cite{quinn:00}, and this led to the
realization that the mass of a scalar particle is not necessarily a
constant of the motion.   

This article reviews the achievements described in the preceding
paragraph; it is concerned with the motion of a point scalar charge 
$q$, a point electric charge $e$, and a point mass $m$ in a specified
background spacetime with metric $g_{\alpha\beta}$. These particles
carry with them fields that behave as outgoing radiation in the wave 
zone. The radiation removes energy and angular momentum from the
particle, which then undergoes a radiation reaction --- its world line
cannot be simply a geodesic of the background spacetime. The
particle's motion is affected by the near-zone field which acts
directly on the particle and produces a {\it self-force}. In curved
spacetime the self-force contains a radiation-reaction component that
is directly associated with dissipative effects, but it contains also
a conservative component that is not associated with energy or
angular-momentum transport. The self-force is proportional to $q^2$ in
the case of a scalar charge, proportional to $e^2$ in the case of an
electric charge, and proportional to $m^2$ in the case of a point mass.  

In this review we derive the equations that govern the motion of 
a point particle in a curved background spacetime. The presentation is 
entirely self-contained, and all relevant materials are developed  
{\it ab initio}. The reader, however, is assumed to have a solid grasp  
of differential geometry and a deep understanding of general
relativity. The reader is also assumed to have unlimited stamina, 
for the road to the equations of motion is a long one. One must first
assimilate the basic theory of bitensors (part \ref{part1}), then
apply the theory to construct convenient coordinate systems to chart a 
neighbourhood of the particle's world line (part \ref{part2}). One
must next formulate a theory of Green's functions in curved spacetimes
(part \ref{part3}), and finally calculate the scalar, electromagnetic,
and gravitational fields near the world line and figure out how
they should act on the particle (part \ref{part4}). A dedicated
reader, correctly skeptical that sense can be made of a point mass in
general relativity, will also want to work through the last portion of
the review (part \ref{part5}), which provides a derivation of the
equations of motion for a small, but physically extended, body; this 
reader will be reassured to find that the extended body follows the
same motion as the point mass. The review is very long, but the
satisfaction derived, we hope, will be commensurate.     

In this introductory section we set the stage and present an
impressionistic survey of what the review contains. This should help
the reader get oriented and acquainted with some of the ideas and some
of the notation. Enjoy! 

\subsection{Radiation reaction in flat spacetime}  
\label{1.2}

Let us first consider the relatively simple and well-understood case
of a point electric charge $e$ moving in flat spacetime
\cite{rohrlich:90, jackson:99, teitelboim-etal:80}. The charge
produces an electromagnetic vector potential $A^\alpha$ that satisfies
the wave equation    
\begin{equation} 
\Box A^\alpha = - 4\pi j^\alpha
\label{1.2.1}
\end{equation} 
together with the Lorenz gauge condition $\partial_\alpha A^\alpha 
= 0$. (On page 294 Jackson \cite{jackson:99} explains why the term
``Lorenz gauge'' is preferable to ``Lorentz gauge''.) 
The vector $j^\alpha$ is the charge's current density, which is 
formally written in terms of a four-dimensional Dirac functional
supported on the charge's world line: the density is zero everywhere, 
except at the particle's position where it is infinite. For
concreteness we will imagine that the particle moves around a
centre (perhaps another charge, which is taken to be fixed) and that
it emits outgoing radiation. We expect that the charge will undergo a
radiation reaction and that it will spiral down toward the
centre. This effect must be accounted for by the equations of motion,
and these must therefore include the action of the charge's own field,
which is the only available agent that could be responsible for the
radiation reaction. We seek to determine this self-force acting on the 
particle.   

An immediate difficulty presents itself: the vector potential, and 
also the electromagnetic field tensor, diverge on the particle's world
line, because the field of a point charge is necessarily infinite at
the charge's position. This behaviour makes it most difficult to
decide how the field is supposed to act on the particle.  

Difficult but not impossible. To find a way around this problem we 
note first that {\it the situation considered here, in which the
radiation is propagating outward and the charge is spiraling inward,
breaks the time-reversal invariance of Maxwell's theory}. A specific 
time direction was adopted when, among all possible solutions to the
wave equation, we chose $A^\alpha_{\rm ret}$, the {\it retarded
solution}, as the physically relevant solution. Choosing instead 
the {\it advanced solution} $A^\alpha_{\rm adv}$ would produce a 
time-reversed picture in which the radiation is propagating inward 
and the charge is spiraling outward. Alternatively, choosing the
linear superposition  
\begin{equation}
A^\alpha_{\rm S} = 
\frac{1}{2} \bigl( A^\alpha_{\rm ret} + A^\alpha_{\rm adv} \bigr)
\label{1.2.2}
\end{equation}
would restore time-reversal invariance: outgoing and incoming
radiation would be present in equal amounts, there would be no net
loss nor gain of energy by the system, and the charge would 
undergo no radiation reaction. In Eq.~(\ref{1.2.2}) the subscript `S'    
stands for `symmetric', as the vector potential depends symmetrically
upon future and past.  

Our second key observation is that while the potential of
Eq.~(\ref{1.2.2}) does not exert a force on the charged
particle, {\it it is just as singular as the retarded potential in the
vicinity of the world line}. This follows from the fact that 
$A^\alpha_{\rm ret}$, $A^\alpha_{\rm adv}$, and $A^\alpha_{\rm S}$ all
satisfy Eq.~(\ref{1.2.1}), whose source term is infinite on the world
line. So while the wave-zone behaviours of these solutions are very
different (with the retarded solution describing outgoing waves, the
advanced solution describing incoming waves, and the symmetric
solution describing standing waves), the three vector potentials share
the same singular behaviour near the world line --- all three
electromagnetic fields are dominated by the particle's Coulomb field
and the different asymptotic conditions make no difference close to
the particle. This observation gives us an alternative interpretation 
for the subscript `S': it stands for `singular' as well as
`symmetric'. 

Because $A^\alpha_{\rm S}$ is just as singular as 
$A^\alpha_{\rm ret}$, removing it from the retarded solution gives
rise to a potential that is well behaved in a neighbourhood of the
world line. And because $A^\alpha_{\rm S}$ is known not to affect the
motion of the charged particle, {\it this new potential must be
entirely responsible for the radiation reaction}. We therefore
introduce the new potential  
\begin{equation} 
A^\alpha_{\rm R} = A^\alpha_{\rm ret} - A^\alpha_{\rm S} = 
\frac{1}{2} \bigl( A^\alpha_{\rm ret} - A^\alpha_{\rm adv} \bigr)
\label{1.2.3}
\end{equation}
and postulate that it, and it alone, exerts a force on the
particle. The subscript `R' stands for `regular', because 
$A^\alpha_{\rm R}$ is nonsingular on the world line. This property can
be directly inferred from the fact that the regular potential
satisfies the homogeneous version of Eq.~(\ref{1.2.1}), 
$\Box A^\alpha_{\rm R} = 0$; there is no singular source to produce a
singular behaviour on the world line. Since $A^\alpha_{\rm R}$
satisfies the homogeneous wave equation, it can be thought of as a
free radiation field, and the subscript `R' could also stand for
`radiative'. 

The self-action of the charge's own field is now clarified: a singular 
potential $A^\alpha_{\rm S}$ can be removed from the retarded
potential and shown not to affect the motion of the particle.  
What remains is a well-behaved potential $A^\alpha_{\rm R}$ 
that must be solely responsible for the radiation reaction. From the
regular potential we form an electromagnetic field tensor 
$F^{\rm R}_{\alpha\beta} = \partial_\alpha A^{\rm R}_\beta 
- \partial_\beta A^{\rm R}_\alpha$ and we take the particle's
equations of motion to be 
\begin{equation} 
m a_\mu = f_\mu^{\rm ext} + e F^{\rm R}_{\mu\nu} u^\nu, 
\label{1.2.4}
\end{equation} 
where $u^\mu = d z^\mu/d\tau$ is the charge's four-velocity 
[$z^\mu(\tau)$ gives the description of the world line and $\tau$ is
proper time], $a^\mu = du^\mu/d\tau$ its acceleration, $m$ its
(renormalized) mass, and $f^\mu_{\rm ext}$ an external force also
acting on the particle. Calculation of the regular field yields the
more concrete expression   
\begin{equation} 
m a^\mu = f^\mu_{\rm ext} + \frac{2e^2}{3m} \bigl( \delta^\mu_{\ \nu}
+ u^\mu u_\nu \bigr) \frac{d f^\nu_{\rm ext}}{d\tau},  
\label{1.2.5}
\end{equation} 
in which the second term is the self-force that is responsible for the
radiation reaction. We observe that the self-force is proportional to
$e^2$, it is orthogonal to the four-velocity, and it depends on the
rate of change of the external force. This is the result that was
first derived by Dirac \cite{dirac:38}. (Dirac's original expression
actually involved the rate of change of the acceleration vector on the
right-hand side. The resulting equation gives rise to the well-known
problem of runaway solutions. To avoid such unphysical behaviour we
have submitted Dirac's equation to a reduction-of-order procedure
whereby $d a^\nu/d\tau$ is replaced with 
$m^{-1} d f^\nu_{\rm ext}/d\tau$. This procedure is explained and
justified, for example, in Refs.~\cite{landau-lifshitz:b2,
  flanagan-wald:96}, and further discussed in Sec.~\ref{conclusion}
below.)       

To establish that the singular field exerts no force on the particle
requires a careful analysis that is presented in the bulk of the
paper. What really happens is that, because the particle is a monopole
source for the electromagnetic field, the singular field is locally
isotropic around the particle; it therefore exerts no force, but
contributes to the particle's inertia and renormalizes its mass.  In
fact, one could do without a decomposition of the field into singular
and regular solutions, and instead construct the force by using the
retarded field and averaging it over a small sphere around the 
particle, as was done by Quinn and Wald \cite{quinn-wald:97}. In the
body of this review we will use both methods and emphasize the
equivalence of the results. We will, however, give some emphasis to
the decomposition because it provides a compelling physical
interpretation of the self-force as an interaction with a free
electromagnetic field. 

\subsection{Green's functions in flat spacetime} 
\label{1.3} 

To see how Eq.~(\ref{1.2.5}) can eventually be generalized to curved 
spacetimes, we introduce a new layer of mathematical formalism and
show that the decomposition of the retarded potential into
singular and regular pieces can be performed at
the level of the Green's functions associated with
Eq.~(\ref{1.2.1}). The retarded solution to the wave equation can be
expressed as  
\begin{equation} 
A^\alpha_{\rm ret}(x) = \int G^{\ \alpha}_{+\beta'}(x,x')   
  j^{\beta'}(x')\, dV', 
\label{1.3.1}
\end{equation}
in terms of the retarded Green's function 
$G^{\ \alpha}_{+\beta'}(x,x') = \delta^\alpha_{\beta'}
\delta(t-t'-|\bm{x}-\bm{x'}|)/|\bm{x}-\bm{x'}|$. Here $x = (t,\bm{x})$
is an arbitrary field point, $x' = (t',\bm{x'})$ is a source point,
and $dV' := d^4 x'$; tensors at $x$ are identified with unprimed
indices, while primed indices refer to tensors at $x'$. Similarly, the
advanced solution can be expressed as 
\begin{equation} 
A^\alpha_{\rm adv}(x) = \int G^{\ \alpha}_{-\beta'}(x,x')   
  j^{\beta'}(x')\, dV',  
\label{1.3.2}
\end{equation}
in terms of the advanced Green's function 
$G^{\ \alpha}_{-\beta'}(x,x') = \delta^\alpha_{\beta'}
\delta(t-t'+|\bm{x}-\bm{x'}|)/|\bm{x}-\bm{x'}|$. The retarded Green's
function is zero whenever $x$ lies outside of the future light cone of  
$x'$, and $G^{\ \alpha}_{+\beta'}(x,x')$ is infinite at these
points. On the other hand, the advanced Green's function is zero 
whenever $x$ lies outside of the past light cone of $x'$, and  
$G^{\ \alpha}_{-\beta'}(x,x')$ is infinite at these points. The
retarded and advanced Green's functions satisfy the reciprocity
relation  
\begin{equation} 
G^-_{\beta'\alpha}(x',x) = G^+_{\alpha\beta'}(x,x'); 
\label{1.3.3}
\end{equation} 
this states that the retarded Green's function becomes the advanced
Green's function (and vice versa) when $x$ and $x'$ are interchanged.  

From the retarded and advanced Green's functions we can define a 
singular Green's function by 
\begin{equation} 
G^{\ \alpha}_{{\rm S}\,\beta'}(x,x') = \frac{1}{2} \Bigl[ 
G^{\ \alpha}_{+\beta'}(x,x') 
+ G^{\ \alpha}_{-\beta'}(x,x') \Bigr] 
\label{1.3.4}
\end{equation}
and a regular two-point function by 
\begin{equation} 
G^{\ \alpha}_{{\rm R}\,\beta'}(x,x') = 
G^{\ \alpha}_{+\beta'}(x,x') - G^{\ \alpha}_{{\rm S}\,\beta'}(x,x') 
= \frac{1}{2} \Bigl[ 
G^{\ \alpha}_{+\beta'}(x,x') 
- G^{\ \alpha}_{-\beta'}(x,x') \Bigr]. 
\label{1.3.5}
\end{equation}
By virtue of Eq.~(\ref{1.3.3}) the singular Green's function is
symmetric in its indices and arguments: 
$G^{\rm S}_{\beta'\alpha}(x',x) 
= G^{\rm S}_{\alpha\beta'}(x,x')$. The regular two-point function,  
on the other hand, is antisymmetric. The potential 
\begin{equation} 
A^\alpha_{\rm S}(x) = \int G^{\ \alpha}_{{\rm S}\,\beta'}(x,x')   
  j^{\beta'}(x')\, dV' 
\label{1.3.6}
\end{equation}
satisfies the wave equation of Eq.~(\ref{1.2.1}) and is singular on
the world line, while 
\begin{equation} 
A^\alpha_{\rm R}(x) = \int G^{\ \alpha}_{{\rm R}\,\beta'}(x,x')   
  j^{\beta'}(x')\, dV'  
\label{1.3.7}
\end{equation}
satisfies the homogeneous equation $\Box A^\alpha = 0$ and is 
well behaved on the world line.  

Equation (\ref{1.3.1}) implies that the retarded potential at $x$ is 
generated by a single event in spacetime: the intersection of the
world line and $x$'s past light cone (see Fig.~1). We shall call this
the {\it retarded point} associated with $x$ and denote it $z(u)$; $u$
is the {\it retarded time}, the value of the proper-time parameter at 
the retarded point. Similarly we find that the advanced potential of
Eq.~(\ref{1.3.2}) is generated by the intersection of the world 
line and the future light cone of the field point $x$. We shall call
this the {\it advanced point} associated with $x$ and denote it
$z(v)$; $v$ is the {\it advanced time}, the value of the proper-time
parameter at the advanced point. 

\begin{figure}
\begin{center}
\vspace*{-20pt} 
\includegraphics[width=0.6\linewidth]{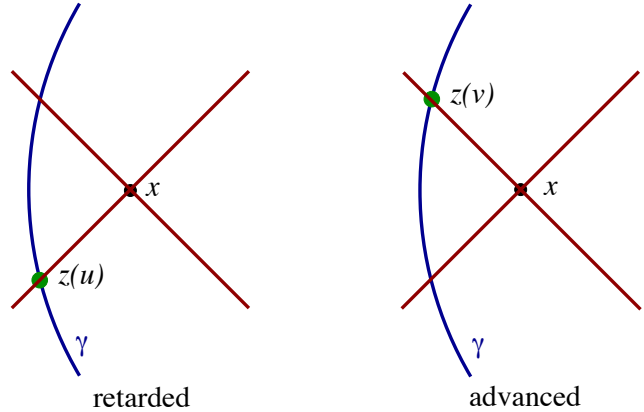}
\vspace*{-20pt}
\end{center} 
\caption{In flat spacetime, the retarded potential at $x$ depends on
the particle's state of motion at the retarded point $z(u)$ on the
world line; the advanced potential depends on the state of motion at
the advanced point $z(v)$.} 
\end{figure} 

\subsection{Green's functions in curved spacetime} 
\label{1.4}

In a curved spacetime with metric $g_{\alpha\beta}$ the wave equation
for the vector potential becomes 
\begin{equation} 
\Box A^\alpha - R^\alpha_{\ \beta} A^\beta = -4\pi j^\alpha, 
\label{1.4.1}
\end{equation}
where $\Box = g^{\alpha\beta} \nabla_\alpha \nabla_\beta$ is the
covariant wave operator and $R_{\alpha\beta}$ is the spacetime's
Ricci tensor; the Lorenz gauge conditions becomes $\nabla_\alpha
A^\alpha = 0$, and $\nabla_\alpha$ denotes covariant
differentiation. Retarded and advanced Green's functions can be 
defined for this equation, and solutions to Eq.~(\ref{1.4.1}) take
the same form as in Eqs.~(\ref{1.3.1}) and (\ref{1.3.2}), except that
$dV'$ now stands for $\sqrt{-g(x')}\, d^4 x'$.  

The causal structure of the Green's functions is richer in curved 
spacetime: While in flat spacetime the retarded Green's function has
support only on the future light cone of $x'$, in curved spacetime its 
support extends {\it inside} the light cone as well; 
$G^{\ \alpha}_{+\beta'}(x,x')$ is therefore nonzero when $x \in
I^+(x')$, which denotes the chronological future of $x'$. This
property reflects the fact that in curved spacetime, electromagnetic
waves propagate not just at the speed of light, but at {\it all speeds
smaller than or equal to the speed of light}; the delay is caused by
an interaction between the radiation and the spacetime curvature. A
direct implication of this property is that the retarded potential at
$x$ is now generated by the point charge during its entire history
prior to the retarded time $u$ associated with $x$: the potential
depends on the particle's state of motion for all times $\tau \leq u$
(see Fig.~2).   

\begin{figure}
\begin{center}
\vspace*{-20pt} 
\includegraphics[width=0.6\linewidth]{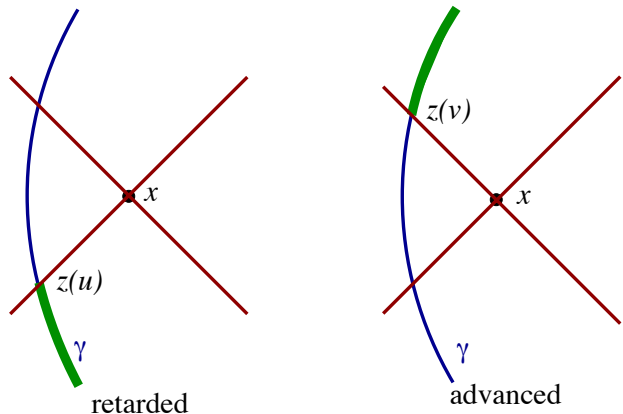}
\vspace*{-20pt} 
\end{center}
\caption{In curved spacetime, the retarded potential at $x$ depends on 
the particle's history before the retarded time $u$; the advanced
potential depends on the particle's history after the advanced time
$v$.}  
\end{figure} 

Similar statements can be made about the advanced Green's function and
the advanced solution to the wave equation. While in flat spacetime
the advanced Green's function has support only on the past light cone
of $x'$, in curved spacetime its support extends inside the light
cone, and $G^{\ \alpha}_{-\beta'}(x,x')$ is nonzero when $x
\in I^-(x')$, which denotes the chronological past of $x'$. This
implies that the advanced potential at $x$ is generated by the point 
charge during its entire {\it future} history following the advanced 
time $v$ associated with $x$: the potential depends on the
particle's state of motion for all times $\tau \geq v$.  

The physically relevant solution to Eq.~(\ref{1.4.1}) is obviously the
retarded potential $A^\alpha_{\rm ret}(x)$, and as in flat spacetime,
this diverges on the world line. The cause of this singular
behaviour is still the pointlike nature of the source, and the
presence of spacetime curvature does not change the fact that the
potential diverges at the position of the particle. Once more this
behaviour makes it difficult to figure out how the retarded field is
supposed to act on the particle and determine its motion. As in flat
spacetime we shall attempt to decompose the retarded solution into a
singular part that exerts no force, and a regular part that
produces the entire self-force.  

To decompose the retarded Green's function into singular and regular
parts is not a straightforward task in curved spacetime. The
flat-spacetime definition for the singular Green's function,
Eq.~(\ref{1.3.4}), cannot be adopted without modification: While the  
combination half-retarded plus half-advanced Green's functions does
have the property of being symmetric, and while the resulting vector
potential would be a solution to Eq.~(\ref{1.4.1}), this candidate for
the singular Green's function would produce a self-force with an
unacceptable dependence on the particle's future history. For suppose
that we made this choice. Then the regular two-point function would be
given by the combination half-retarded minus half-advanced Green's
functions, just as in flat spacetime. The resulting potential would
satisfy the homogeneous wave equation, and it would be regular on the
world line, but it would also depend on the particle's 
entire history, both past (through the retarded Green's function) and
future (through the advanced Green's function). More precisely stated,
we would find that the regular potential at $x$ depends on the
particle's state of motion at all times $\tau$ outside the interval 
$u < \tau < v$; in the limit where $x$ approaches the world line, this
interval shrinks to nothing, and we would find that the regular
potential is generated by the complete history of the particle. A
self-force constructed from this potential would be highly noncausal,
and we are compelled to reject these definitions for the singular and
regular Green's functions. 

The proper definitions were identified by Detweiler and Whiting
\cite{detweiler-whiting:03}, who proposed the following generalization
to Eq.~(\ref{1.3.4}):  
\begin{equation} 
G^{\ \alpha}_{{\rm S}\,\beta'}(x,x') = \frac{1}{2} \Bigl[  
G^{\ \alpha}_{+\beta'}(x,x') 
+ G^{\ \alpha}_{-\beta'}(x,x')
- H^\alpha_{\ \beta'}(x,x') \Bigr]. 
\label{1.4.2} 
\end{equation}  
The two-point function $H^\alpha_{\ \beta'}(x,x')$ is introduced
specifically to cure the pathology described in the preceding
paragraph. It is symmetric in its indices and arguments, so that 
$G^{\rm S}_{\alpha\beta'}(x,x')$ will be also (since the retarded 
and advanced Green's functions are still linked by a reciprocity
relation); and it is a solution to the homogeneous wave equation,
$\Box H^\alpha_{\ \beta'}(x,x') 
- R^\alpha_{\ \gamma}(x) H^\gamma_{\ \beta'}(x,x') = 0$, so that the 
singular, retarded, and advanced Green's functions will all satisfy
the same wave equation. Furthermore, and this is its key property, the
two-point function is defined to agree with the advanced Green's
function when $x$ is in the chronological past of $x'$:  
$H^\alpha_{\ \beta'}(x,x') = G^{\ \alpha}_{-\beta'}(x,x')$ when 
$x \in I^-(x')$. This ensures that 
$G^{\ \alpha}_{{\rm S}\,\beta'}(x,x')$ vanishes when $x$ is in the 
chronological past of $x'$. In fact, reciprocity implies that 
$H^\alpha_{\ \beta'}(x,x')$ will also agree with the {\it retarded}
Green's function when $x$ is in the chronological future of $x'$, and
it follows that the symmetric Green's function vanishes also when $x$
is in the chronological future of $x'$.  

The potential $A^\alpha_{\rm S}(x)$ constructed from the singular 
Green's function can now be seen to depend on the particle's state of 
motion at times $\tau$ restricted to the interval $u \leq \tau \leq
v$ (see Fig.~3). Because this potential satisfies Eq.~(\ref{1.4.1}),
it is just as singular as the retarded potential in the vicinity of
the world line. And because the singular Green's function is symmetric
in its arguments, the singular potential can be shown to exert no
force on the charged particle. (This requires a lengthy analysis that
will be presented in the bulk of the paper.) 

\begin{figure}
\begin{center}
\vspace*{-20pt} 
\includegraphics[width=0.6\linewidth]{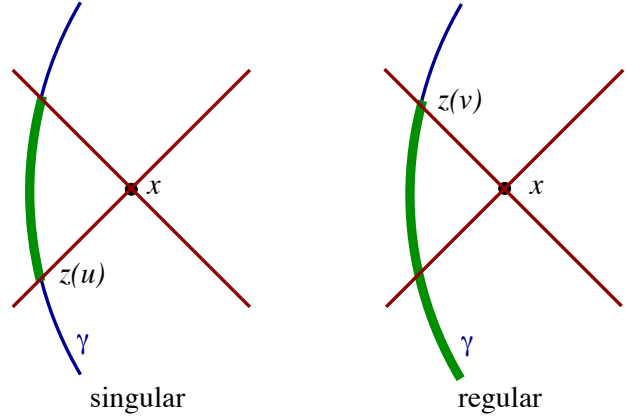}
\vspace*{-20pt}
\end{center} 
\caption{In curved spacetime, the singular potential at $x$ depends on  
the particle's history during the interval $u \leq \tau \leq v$; for
the regular potential the relevant interval is $-\infty < \tau 
\leq v$.}  
\end{figure} 

The Detweiler-Whiting \cite{detweiler-whiting:03} definition for the 
regular two-point function is then  
\begin{equation} 
G^{\ \alpha}_{{\rm R}\,\beta'}(x,x') = 
G^{\ \alpha}_{+\beta'}(x,x') - G^{\ \alpha}_{{\rm S}\,\beta'}(x,x') 
= \frac{1}{2} \Bigl[ 
G^{\ \alpha}_{+\beta'}(x,x') 
- G^{\ \alpha}_{-\beta'}(x,x') + H^\alpha_{\ \beta'}(x,x') \Bigr].  
\label{1.4.3}
\end{equation}
The potential $A^\alpha_{\rm R}(x)$ constructed from this depends on
the particle's state of motion at all times $\tau$ prior to the
advanced time $v$: $\tau \leq v$. Because this potential satisfies the 
homogeneous wave equation, it is well behaved on the world line and
its action on the point charge is well defined. And because the 
singular potential $A^\alpha_{\rm S}(x)$ can be shown to exert no
force on the particle, we conclude that $A^\alpha_{\rm R}(x)$ alone is   
responsible for the self-force.  

From the regular potential we form an electromagnetic field tensor  
$F^{\rm R}_{\alpha\beta} = \nabla_\alpha A^{\rm R}_\beta 
- \nabla_\beta A^{\rm R}_\alpha$ and the curved-spacetime
generalization to Eq.~(\ref{1.2.4}) is 
\begin{equation} 
m a_\mu = f_\mu^{\rm ext} + e F^{\rm R}_{\mu\nu} u^\nu, 
\label{1.4.4}
\end{equation} 
where $u^\mu = d z^\mu/d\tau$ is again the charge's four-velocity, but    
$a^\mu = Du^\mu/d\tau$ is now its covariant acceleration. 

\subsection{World line and retarded coordinates} 
\label{1.5} 

To flesh out the ideas contained in the preceding subsection we 
add yet another layer of mathematical formalism and construct a
convenient coordinate system to chart a neighbourhood of 
the particle's world line. In the next subsection we will display
explicit expressions for the retarded, singular, and regular fields
of a point electric charge.  

Let $\gamma$ be the world line of a point particle in a curved
spacetime. It is described by parametric relations 
$z^\mu(\tau)$ in which $\tau$ is proper time. Its tangent vector is
$u^\mu = dz^\mu/d\tau$ and its acceleration is 
$a^\mu = D u^\mu/d\tau$; we shall also encounter 
$\dot{a}^\mu := D a^\mu/d\tau$. 

On $\gamma$ we erect an orthonormal basis that consists of the
four-velocity $u^\mu$ and three spatial vectors $\base{\mu}{a}$
labelled by a frame index $a = (1,2,3)$. These vectors satisfy the
relations $g_{\mu\nu} u^\mu u^\nu = -1$, $g_{\mu\nu} u^\mu
\base{\nu}{a} = 0$, and $g_{\mu\nu} \base{\mu}{a} \base{\nu}{b} =
\delta_{ab}$. We take the spatial vectors to be Fermi-Walker 
transported on the world line: $D \base{\mu}{a} / d\tau = a_a u^\mu$,
where  
\begin{equation}
a_a(\tau) = a_\mu \base{\mu}{a} 
\label{1.5.1}
\end{equation}
are frame components of the acceleration vector; it is easy to show
that Fermi-Walker transport preserves the orthonormality of the basis
vectors. We shall use the tetrad to decompose various tensors
evaluated on the world line. An example was already given in
Eq.~(\ref{1.5.1}) but we shall also encounter frame components of the
Riemann tensor,   
\begin{equation}
R_{a0b0}(\tau) = R_{\mu\lambda\nu\rho} \base{\mu}{a} u^\lambda 
\base{\nu}{b} u^\rho, \qquad 
R_{a0bc}(\tau) = R_{\mu\lambda\nu\rho} \base{\mu}{a} u^\lambda 
\base{\nu}{b} \base{\rho}{c}, \qquad 
R_{abcd}(\tau) = R_{\mu\lambda\nu\rho} \base{\mu}{a} \base{\lambda}{b}   
\base{\nu}{c} \base{\rho}{d},
\label{1.5.2}
\end{equation}
as well as frame components of the Ricci tensor, 
\begin{equation} 
R_{00}(\tau) = R_{\mu\nu} u^\mu u^\nu, \qquad 
R_{a0}(\tau) = R_{\mu\nu} \base{\mu}{a} u^\nu, \qquad 
R_{ab}(\tau) = R_{\mu\nu} \base{\mu}{a} \base{\nu}{b}. 
\label{1.5.3}
\end{equation} 
We shall use $\delta_{ab} = \mbox{diag}(1,1,1)$ and its inverse
$\delta^{ab} = \mbox{diag}(1,1,1)$ to lower and raise frame indices,
respectively. 

Consider a point $x$ in a neighbourhood of the world line $\gamma$. We
assume that $x$ is sufficiently close to the world line that a unique
geodesic links $x$ to any neighbouring point $z$ on $\gamma$. The
two-point function $\sigma(x,z)$, known as {\it Synge's world
function} \cite{synge:60}, is numerically equal to half the squared
geodesic distance between $z$ and $x$; it is positive if $x$ and $z$
are spacelike related, negative if they are timelike related, and
$\sigma(x,z)$ is zero if $x$ and $z$ are linked by a null geodesic. We
denote its gradient $\partial \sigma / \partial z^\mu$ by
$\sigma_\mu(x,z)$, and $-\sigma^\mu$ gives a meaningful notion of a
separation vector (pointing from $z$ to $x$).       

To construct a coordinate system in this neighbourhood we locate the 
unique point $x' := z(u)$ on $\gamma$ which is linked to $x$ by a
future-directed null geodesic (this geodesic is directed from $x'$ to
$x$); we shall refer to $x'$ as the {\it retarded point} associated
with $x$, and $u$ will be called the {\it retarded time}. To tensors
at $x'$ we assign indices $\alpha'$, $\beta'$, \ldots; this will
distinguish them from tensors at a generic point $z(\tau)$ on the
world line, to which we have assigned indices $\mu$, $\nu$, \ldots. We
have $\sigma(x,x') = 0$ and $-\sigma^{\alpha'}(x,x')$ is a null vector
that can be interpreted as the separation between $x'$ and $x$.  

\begin{figure}
\begin{center}
\vspace*{-20pt} 
\includegraphics[width=0.5\linewidth]{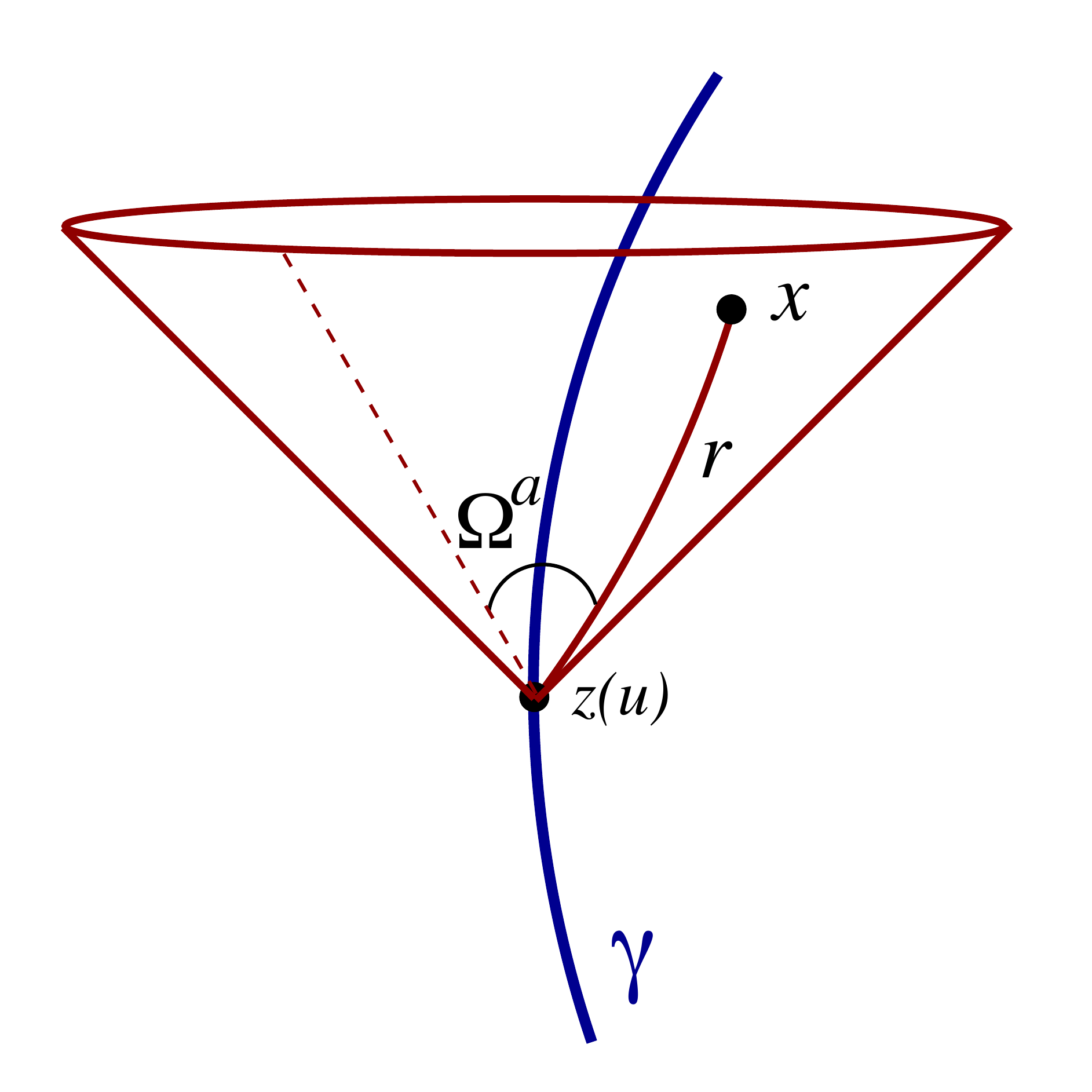}
\vspace*{-20pt}
\end{center} 
\caption{Retarded coordinates of a point $x$ relative to a world
line $\gamma$. The retarded time $u$ selects a particular null cone,
the unit vector $\Omega^a := \hat{x}^a/r$ selects a particular
generator of this null cone, and the retarded distance $r$ selects a
particular point on this generator.}  
\end{figure} 

The {\it retarded coordinates} of the point $x$ are $(u,\hat{x}^a)$,
where $\hat{x}^a = - \base{a}{\alpha'} \sigma^{\alpha'}$ are the frame
components of the separation vector. They come with a straightforward
interpretation (see Fig.~4). The invariant quantity    
\begin{equation}
r := \sqrt{ \delta_{ab} \hat{x}^a \hat{x}^b } = u_{\alpha'}
\sigma^{\alpha'}  
\label{1.5.4}
\end{equation} 
is an affine parameter on the null geodesic that links $x$ to $x'$; it
can be loosely interpreted as the time delay between $x$ and $x'$ as
measured by an observer moving with the particle. This therefore gives
a meaningful notion of distance between $x$ and the retarded point,
and we shall call $r$ the {\it retarded distance} between $x$ and the
world line. The unit vector  
\begin{equation}
\Omega^a = \hat{x}^a/r  
\label{1.5.5}
\end{equation} 
is constant on the null geodesic that links $x$ to $x'$. Because
$\Omega^a$ is a different constant on each null geodesic that emanates
from $x'$, keeping $u$ fixed and varying $\Omega^a$ produces a
congruence of null geodesics that generate the future light cone of
the point $x'$ (the congruence is hypersurface orthogonal). Each light
cone can thus be labelled by its retarded time $u$, each generator on
a given light cone can be labelled by its direction vector $\Omega^a$,
and each point on a given generator can be labelled by its retarded
distance $r$. We therefore have a good coordinate system in a
neighbourhood of $\gamma$. 

To tensors at $x$ we assign indices $\alpha$, $\beta$, \ldots. These
tensors will be decomposed in a tetrad 
$(\base{\alpha}{0},\base{\alpha}{a})$ that is constructed as
follows: Given $x$ we locate its associated retarded point $x'$ on the
world line, as well as the null geodesic that links these two
points; we then take the tetrad $(u^{\alpha'},\base{\alpha'}{a})$ at   
$x'$ and parallel transport it to $x$ along the null geodesic to
obtain $(\base{\alpha}{0},\base{\alpha}{a})$.      

\subsection{Retarded, singular, and regular electromagnetic fields
of a point electric charge} 
\label{1.6}
 
The retarded solution to Eq.~(\ref{1.4.1}) is 
\begin{equation}
A^\alpha(x) = e \int_\gamma G^{\ \alpha}_{+\mu}(x,z) u^\mu\, d\tau, 
\label{1.6.1}
\end{equation}
where the integration is over the world line of the point electric
charge. Because the retarded solution is the physically relevant
solution to the wave equation, it will not be necessary to put a
label `ret' on the vector potential. 

From the vector potential we form the electromagnetic field tensor
$F_{\alpha\beta}$, which we decompose in the tetrad 
$(\base{\alpha}{0},\base{\alpha}{a})$ introduced at the end of
Sec.~\ref{1.5}. We then express the frame components of the field  
tensor in retarded coordinates, in the form of an expansion in powers
of $r$. This gives 
\begin{eqnarray} 
F_{a0}(u,r,\Omega^a) &:=& F_{\alpha\beta}(x) 
\base{\alpha}{a}(x) \base{\beta}{0}(x) 
\nonumber \\  
&=& \frac{e}{r^2} \Omega_a 
- \frac{e}{r} \bigl( a_a - a_b \Omega^b \Omega_a \bigr) 
+ \frac{1}{3} e R_{b0c0} \Omega^b \Omega^c \Omega_a 
- \frac{1}{6} e \bigl( 5R_{a0b0} \Omega^b + R_{ab0c} \Omega^b \Omega^c
\bigr) 
\nonumber \\ & & \mbox{}
+ \frac{1}{12} e \bigl( 5 R_{00} + R_{bc} \Omega^b\Omega^c + R \bigr)
\Omega_a 
+ \frac{1}{3} e R_{a0} - \frac{1}{6} e R_{ab} \Omega^b 
+ F_{a0}^{\rm tail} + O(r), 
\label{1.6.2} \\ 
F_{ab}(u,r,\Omega^a) &:=& F_{\alpha\beta}(x) 
\base{\alpha}{a}(x) \base{\beta}{b}(x) 
\nonumber \\  
&=& \frac{e}{r} \bigl( a_a \Omega_b - \Omega_a a_b \bigr) 
+ \frac{1}{2} e \bigl( R_{a0bc} - R_{b0ac} + R_{a0c0} \Omega_b 
- \Omega_a R_{b0c0} \bigr) \Omega^c 
\nonumber \\ & & \mbox{}
- \frac{1}{2} e \bigl( R_{a0} \Omega_b - \Omega_a R_{b0} \bigr)
+ F_{ab}^{\rm tail} + O(r),  
\label{1.6.3}
\end{eqnarray} 
where 
\begin{equation} 
F_{a0}^{\rm tail} = F_{\alpha'\beta'}^{\rm tail}(x') \base{\alpha'}{a}
u^{\beta'}, \qquad 
F_{ab}^{\rm tail} = F_{\alpha'\beta'}^{\rm tail}(x') \base{\alpha'}{a}
\base{\beta'}{b} 
\label{1.6.4}
\end{equation}
are the frame components of the ``tail part'' of the field, which is
given by 
\begin{equation}
F_{\alpha'\beta'}^{\rm tail}(x') = 2 e \int_{-\infty}^{u^-}
\nabla_{[\alpha'} G_{+\beta']\mu}(x',z) u^\mu\, d\tau.  
\label{1.6.5}
\end{equation} 
In these expressions, all tensors (or their frame components) are 
evaluated at the retarded point $x' := z(u)$ associated with $x$; 
for example, $a_a := a_a(u) := a_{\alpha'}
\base{\alpha'}{a}$. The tail part of the electromagnetic field tensor
is written as an integral over the portion of the world line that
corresponds to the interval $-\infty < \tau \leq u^- := u - 0^+$;
this represents the past history of the particle. The integral is cut
short at $u^-$ to avoid the singular behaviour of the retarded Green's 
function when $z(\tau)$ coincides with $x'$; the portion of the
Green's function involved in the tail integral is smooth, and the
singularity at coincidence is completely accounted for by the other
terms in Eqs.~(\ref{1.6.2}) and (\ref{1.6.3}).  

The expansion of $F_{\alpha\beta}(x)$ near the world line does indeed
reveal many singular terms. We first recognize terms that diverge when 
$r \to 0$; for example the Coulomb field $F_{a0}$ diverges as $r^{-2}$ 
when we approach the world line. But there are also terms that, though  
they stay bounded in the limit, possess a directional ambiguity at
$r=0$; for example $F_{ab}$ contains a term proportional to
$R_{a0bc}\Omega^c$ whose limit depends on the direction of approach.  

This singularity structure is perfectly reproduced by the singular
field $F^{\rm S}_{\alpha\beta}$ obtained from the potential 
\begin{equation} 
A^\alpha_{\rm S}(x) = e \int_\gamma 
G^{\ \alpha}_{{\rm S}\,\mu}(x,z) u^\mu\, d\tau,  
\label{1.6.6}
\end{equation} 
where $G^{\ \alpha}_{{\rm S}\,\mu}(x,z)$ is the singular Green's
function of Eq.~(\ref{1.4.2}). Near the world line the singular field
is given by 
\begin{eqnarray} 
F^{\rm S}_{a0}(u,r,\Omega^a) &:=& F^{\rm S}_{\alpha\beta}(x)  
\base{\alpha}{a}(x) \base{\beta}{0}(x) 
\nonumber \\  
&=& \frac{e}{r^2} \Omega_a 
- \frac{e}{r} \bigl( a_a - a_b \Omega^b \Omega_a \bigr) 
- \frac{2}{3} e \dot{a}_a 
+ \frac{1}{3} e R_{b0c0} \Omega^b \Omega^c \Omega_a 
- \frac{1}{6} e \bigl( 5R_{a0b0} \Omega^b + R_{ab0c} \Omega^b \Omega^c
\bigr) 
\nonumber \\ & & \mbox{}
+ \frac{1}{12} e \bigl( 5 R_{00} + R_{bc} \Omega^b\Omega^c + R \bigr)
\Omega_a 
- \frac{1}{6} e R_{ab} \Omega^b 
+ O(r), 
\label{1.6.7} \\ 
F^{\rm S}_{ab}(u,r,\Omega^a) &:=& F^{\rm S}_{\alpha\beta}(x)  
\base{\alpha}{a}(x) \base{\beta}{b}(x) 
\nonumber \\  
&=& \frac{e}{r} \bigl( a_a \Omega_b - \Omega_a a_b \bigr) 
+ \frac{1}{2} e \bigl( R_{a0bc} - R_{b0ac} + R_{a0c0} \Omega_b 
- \Omega_a R_{b0c0} \bigr) \Omega^c 
\nonumber \\ & & \mbox{}
- \frac{1}{2} e \bigl( R_{a0} \Omega_b - \Omega_a R_{b0} \bigr)
+ O(r).   
\label{1.6.8}
\end{eqnarray} 
Comparison of these expressions with Eqs.~(\ref{1.6.2}) and
(\ref{1.6.3}) does indeed reveal that all singular terms are shared by
both fields. 

The difference between the retarded and singular fields defines the
regular field $F^{\rm R}_{\alpha\beta}(x)$. Its frame components are  
\begin{eqnarray} 
F^{\rm R}_{a0} &=& \frac{2}{3} e \dot{a}_a + \frac{1}{3} e R_{a0}  
+ F_{a0}^{\rm tail} + O(r), 
\label{1.6.9} \\ 
F^{\rm R}_{ab} &=& F_{ab}^{\rm tail} + O(r), 
\label{1.6.10}
\end{eqnarray} 
and at $x'$ the regular field becomes 
\begin{equation} 
F^{\rm R}_{\alpha'\beta'} = 
2e u_{[\alpha'} \bigl( g_{\beta']\gamma'}  
+ u_{\beta']} u_{\gamma'} \bigr) 
\biggl( \frac{2}{3} \dot{a}^{\gamma'} 
+ \frac{1}{3} R^{\gamma'}_{\ \delta'} u^{\delta'}
\biggr) + F_{\alpha'\beta'}^{\rm tail}, 
\label{1.6.11}
\end{equation}
where $\dot{a}^{\gamma'} = D a^{\gamma'}/d\tau$ is the rate of change
of the acceleration vector, and where the tail term was given by
Eq.~(\ref{1.6.5}). We see that $F^{\rm R}_{\alpha\beta}(x)$ is a
regular tensor field, even on the world line.   

\subsection{Motion of an electric charge in curved spacetime} 
\label{1.7} 

We have argued in Sec.~\ref{1.4} that the self-force acting on a point
electric charge is produced by the regular field, and that the
charge's equations of motion should take the form of $m a_\mu =
f_\mu^{\rm ext} + e F^{\rm R}_{\mu\nu} u^\nu$, where $f_\mu^{\rm ext}$
is an external force also acting on the particle. Substituting
Eq.~(\ref{1.6.11}) gives 
\begin{equation} 
m a^\mu = f_{\rm ext}^\mu 
+ e^2 \bigl( \delta^\mu_{\ \nu} + u^\mu u_\nu \bigr) 
\biggl( \frac{2}{3m} \frac{D f_{\rm ext}^\nu}{d \tau}   
+ \frac{1}{3} R^{\nu}_{\ \lambda} u^{\lambda} \biggr)  
+ 2 e^2 u_\nu \int_{-\infty}^{\tau^-}     
\nabla^{[\mu} G^{\ \nu]}_{+\,\lambda'}\bigl(z(\tau),z(\tau')\bigr)   
u^{\lambda'}\, d\tau',  
\label{1.7.1}  
\end{equation}   
in which all tensors are evaluated at $z(\tau)$, the current position
of the particle on the world line. The primed indices in the tail  
integral refer to a point $z(\tau')$ which represents a prior
position; the integration is cut short at $\tau' = \tau^- := \tau
- 0^+$ to avoid the singular behaviour of the retarded Green's
function at coincidence. To get Eq.~(\ref{1.7.1}) we have reduced the
order of the differential equation by replacing $\dot{a}^{\nu}$
with $m^{-1} \dot{f}^{\nu}_{\rm ext}$ on the right-hand side; this
procedure was explained at the end of Sec.~\ref{1.2}. 

Equation (\ref{1.7.1}) is the result that was first derived by DeWitt
and Brehme \cite{dewitt-brehme:60} and later corrected by Hobbs
\cite{hobbs:68}. (The original version of the equation did not include the
Ricci-tensor term.) In flat spacetime the Ricci tensor is zero, the
tail integral disappears (because the Green's function vanishes
everywhere within the domain of integration), and Eq.~(\ref{1.7.1})
reduces to Dirac's result of Eq.~(\ref{1.2.5}). In curved spacetime
the self-force does not vanish even when the electric charge is moving
freely, in the absence of an external force: it is then given by the
tail integral, which represents radiation emitted earlier and coming
back to the particle after interacting with the spacetime
curvature. This delayed action implies that in general, the self-force
is nonlocal in time: it depends not only on the current state of
motion of the particle, but also on its past history. Lest this
behaviour should seem mysterious, it may help to keep in mind
that the physical process that leads to Eq.~(\ref{1.7.1}) is simply an
interaction between the charge and a free electromagnetic field
$F^{\rm R}_{\alpha\beta}$; it is this field that carries the
information about the charge's past.   

\subsection{Motion of a scalar charge in curved spacetime} 
\label{1.8}

The dynamics of a point scalar charge can be formulated in a way that 
stays fairly close to the electromagnetic theory. The particle's 
charge $q$ produces a scalar field $\Phi(x)$ which satisfies a wave 
equation 
\begin{equation} 
\bigl( \Box - \xi R \bigr) \Phi = -4\pi \mu 
\label{1.8.1}
\end{equation} 
that is very similar to Eq.~(\ref{1.4.1}). Here, $R$ is the
spacetime's Ricci scalar, and $\xi$ is an arbitrary coupling constant;
the scalar charge density $\mu(x)$ is given by a four-dimensional
Dirac functional supported on the particle's world line $\gamma$. The
retarded solution to the wave equation is 
\begin{equation}
\Phi(x) = q \int_\gamma G_+(x,z)\, d\tau, 
\label{1.8.2}
\end{equation}
where $G_+(x,z)$ is the retarded Green's function associated with
Eq.~(\ref{1.8.1}). The field exerts a force on the particle, whose
equations of motion are 
\begin{equation} 
m a^\mu = q \bigl( g^{\mu\nu} + u^\mu u^\nu \bigr) \nabla_\nu
\Phi,    
\label{1.8.3}
\end{equation} 
where $m$ is the particle's mass; this equation is very similar to the
Lorentz-force law. But the dynamics of a scalar charge comes with a
twist: If Eqs.~(\ref{1.8.1}) and (\ref{1.8.3}) are to follow from a 
variational principle, {\it the particle's mass should not be expected
to be a constant of the motion}. It is found instead to satisfy the
differential equation 
\begin{equation} 
\frac{d m}{d\tau} = -q u^\mu \nabla_\mu \Phi,  
\label{1.8.4}
\end{equation} 
and in general $m$ will vary with proper time. This phenomenon is
linked to the fact that a scalar field has zero spin: the particle can
radiate monopole waves and the radiated energy can come at the
expense of the rest mass. 

The scalar field of Eq.~(\ref{1.8.2}) diverges on the world line and
its singular part $\Phi_{\rm S}(x)$ must be removed before
Eqs.~(\ref{1.8.3}) and (\ref{1.8.4}) can be evaluated. This procedure
produces the regular field $\Phi_{\rm R}(x)$, and it is this field
(which satisfies the homogeneous wave equation) that gives rise to a
self-force. The gradient of the regular field takes the form of 
\begin{equation} 
\nabla_\mu \Phi_{\rm R} = - \frac{1}{12} (1-6\xi) q R u_{\mu}  
+ q \bigl( g_{\mu\nu} + u_{\mu} u_{\nu} \bigr) 
\biggl( \frac{1}{3} \dot{a}^{\nu} 
+ \frac{1}{6} R^{\nu}_{\ \lambda} u^{\lambda}
\biggr) + \Phi_{\mu}^{\rm tail}
\label{1.8.5}
\end{equation}
when it is evaluated on the world line. The last term is the tail
integral 
\begin{equation} 
\Phi_\mu^{\rm tail} = q \int_{-\infty}^{\tau^-}  
\nabla_{\mu} G_+\bigl(z(\tau),z(\tau')\bigr)\, d\tau',    
\label{1.8.6}
\end{equation}
and this brings the dependence on the particle's past.  

Substitution of Eq.~(\ref{1.8.5}) into Eqs.~(\ref{1.8.3}) and
(\ref{1.8.4}) gives the equations of motion of a point scalar
charge. (At this stage we introduce an external force 
$f_{\rm ext}^\mu$ and reduce the order of the differential equation.)
The acceleration is given by   
\begin{equation} 
m a^\mu = f_{\rm ext}^\mu 
+ q^2 \bigl( \delta^\mu_{\ \nu} + u^\mu u_\nu \bigr) 
\Biggl[ \frac{1}{3m} \frac{D f_{\rm ext}^\nu}{d \tau}   
+ \frac{1}{6} R^{\nu}_{\ \lambda} u^{\lambda} 
+ \int_{-\infty}^{\tau^-}   
\nabla^{\nu} G_+\bigl(z(\tau),z(\tau')\bigr)\, d\tau' \Biggr] 
\label{1.8.7} 
\end{equation}   
and the mass changes according to 
\begin{equation}
\frac{d m}{d\tau} = - \frac{1}{12} (1-6\xi) q^2 R 
- q^2 u^\mu \int_{-\infty}^{\tau^-} \nabla_{\mu}
G_+\bigl(z(\tau),z(\tau')\bigr)\, d\tau'. 
\label{1.8.8}
\end{equation} 
These equations were first derived by Quinn \cite{quinn:00}. (His
analysis was restricted to a minimally coupled scalar field, so that
$\xi = 0$ in his expressions. We extended Quinn's results to an
arbitrary coupling counstant for this review.)  

In flat spacetime the Ricci-tensor term and the tail integral
disappear and Eq.~(\ref{1.8.7}) takes the form of Eq.~(\ref{1.2.5})
with $q^2/(3m)$ replacing the factor of $2e^2/(3m)$. In this simple
case Eq.~(\ref{1.8.8}) reduces to $dm/d\tau = 0$ and the mass is in
fact a constant. This property remains true in a conformally flat 
spacetime when the wave equation is conformally invariant 
($\xi = 1/6$): in this case the Green's function possesses only a
light-cone part and the right-hand side of Eq.~(\ref{1.8.8})
vanishes. In generic situations the mass of a point scalar charge will
vary with proper time.   
         
\subsection{Motion of a point mass, or a small body, in a background
spacetime} 
\label{1.9}

The case of a point mass moving in a specified background spacetime
presents itself with a serious conceptual challenge, as the
fundamental equations of the theory are nonlinear and the very notion
of a ``point mass'' is somewhat misguided. Nevertheless, to the extent
that the perturbation $h_{\alpha\beta}(x)$ created by the point mass
can be considered to be ``small'', the problem can be formulated in
close analogy with what was presented before. 

We take the metric $g_{\alpha\beta}$ of the background spacetime to be
a solution of the Einstein field equations in vacuum. (We impose this
condition globally.) We describe the gravitational perturbation
produced by a point particle of mass $m$ in terms of
trace-reversed potentials $\gamma_{\alpha\beta}$ defined by
\begin{equation}
\gamma_{\alpha\beta} = h_{\alpha\beta} - \frac{1}{2} \bigl(
g^{\gamma\delta} h_{\gamma\delta} \bigr) g_{\alpha\beta}, 
\label{1.9.1}
\end{equation} 
where $h_{\alpha\beta}$ is the difference between 
${\sf g}_{\alpha\beta}$, the actual metric of the perturbed spacetime,
and $g_{\alpha\beta}$. The potentials satisfy the wave equation 
\begin{equation} 
\Box \gamma^{\alpha\beta} + 2 R_{\gamma\ \delta}^{\ \alpha\ \beta} 
\gamma^{\gamma\delta} = -16\pi T^{\alpha\beta} + O(m^2) 
\label{1.9.2}
\end{equation}
together with the Lorenz gauge condition 
$\gamma^{\alpha\beta}_{\ \ \ ;\beta} = 0$. Here and below,
covariant differentiation refers to a connection that is compatible
with the background metric,   
$\Box = g^{\alpha\beta} \nabla_\alpha \nabla_\beta$ is the wave 
operator for the background spacetime, and $T^{\alpha\beta}$ is the  
energy-momentum tensor of the point mass; this is given by a Dirac
distribution supported on the particle's world line $\gamma$. The
retarded solution is  
\begin{equation} 
\gamma^{\alpha\beta}(x) = 4 m \int_\gamma 
G^{\ \alpha\beta}_{+\ \mu\nu}(x,z) u^\mu u^\nu\, d\tau
+ O(m^2), 
\label{1.9.3}
\end{equation}
where $G^{\ \alpha\beta}_{+\ \mu\nu}(x,z)$ is the retarded Green's
function associated with Eq.~(\ref{1.9.2}). The perturbation
$h_{\alpha\beta}(x)$ can be recovered by inverting
Eq.~(\ref{1.9.1}).   

Equations of motion for the point mass can be obtained by formally 
demanding that the motion be geodesic in the perturbed spacetime with
metric ${\sf g}_{\alpha\beta} = g_{\alpha\beta} 
+ h_{\alpha\beta}$. After a mapping to the background
spacetime, the equations of motion take the form of 
\begin{equation}
a^\mu = -\frac{1}{2} \bigl( g^{\mu\nu} + u^\mu
u^\nu \bigr) \bigl( 2 h_{\nu\lambda;\rho} - h_{\lambda\rho;\nu} \bigr)
u^\lambda u^\rho + O(m^2).  
\label{1.9.4} 
\end{equation}   
The acceleration is thus proportional to $m$; in the test-mass limit 
the world line of the particle is a geodesic of the background
spacetime.   

We now remove $h^{\rm S}_{\alpha\beta}(x)$ from the
retarded perturbation and postulate that it is the regular field 
$h^{\rm R}_{\alpha\beta}(x)$ that should act on the particle. (Note
that $\gamma^{\rm S}_{\alpha\beta}$ satisfies the same wave equation
as the retarded potentials, but that $\gamma^{\rm R}_{\alpha\beta}$ is
a free gravitational field that satisfies the homogeneous wave
equation.) On the world line we have 
\begin{equation} 
h^{\rm R}_{\mu\nu;\lambda} = -4 m \Bigl( u_{(\mu}
R_{\nu)\rho\lambda\xi} + R_{\mu\rho\nu\xi} u_\lambda \Bigr) u^\rho
u^\xi + h^{\rm tail}_{\mu\nu\lambda}, 
\label{1.9.5}
\end{equation}
where the tail term is given by 
\begin{equation}
h^{\rm tail}_{\mu\nu\lambda} = 4 m \int_{-\infty}^{\tau^-}
\nabla_\lambda \biggl( G_{+\mu\nu\mu'\nu'}
- \frac{1}{2} g_{\mu\nu} G^{\ \ \rho}_{+\ \rho\mu'\nu'}
\biggr) \bigl( z(\tau), z(\tau')\bigr) u^{\mu'} u^{\nu'}\, d\tau'. 
\label{1.9.6} 
\end{equation}     
When Eq.~(\ref{1.9.5}) is substituted into Eq.~(\ref{1.9.4}) we find
that the terms that involve the Riemann tensor cancel out, and we
are left with  
\begin{equation}
a^\mu = -\frac{1}{2} \bigl( g^{\mu\nu} + u^\mu
u^\nu \bigr) \bigl( 2 h^{\rm tail}_{\nu\lambda\rho} 
- h^{\rm tail}_{\lambda\rho\nu} \bigr) u^\lambda u^\rho
+ O(m^2). 
\label{1.9.7} 
\end{equation}   
Only the tail integral appears in the final form of the equations
of motion. It involves the current position $z(\tau)$ of the particle,
at which all tensors with unprimed indices are evaluated, as well as
all prior positions $z(\tau')$, at which tensors with primed indices
are evaluated. As before the integral is cut short at $\tau' = \tau^-
:= \tau - 0^+$ to avoid the singular behaviour of the retarded
Green's function at coincidence.

The equations of motion of Eq.~(\ref{1.9.7}) were first derived by
Mino, Sasaki, and Tanaka \cite{mino-etal:97a}, and then reproduced with
a different analysis by Quinn and Wald \cite{quinn-wald:97}. They are
now known as the MiSaTaQuWa equations of motion. As noted by these
authors, the MiSaTaQuWa equation has the appearance of the geodesic
equation in a metric 
$g_{\alpha\beta}+h^{\rm tail}_{\alpha\beta}$. Detweiler and 
Whiting \cite{detweiler-whiting:03} have contributed the more
compelling interpretation that the motion is actually geodesic in a
spacetime with metric $g_{\alpha\beta}+h^{\rm R}_{\alpha\beta}$. The
distinction is important: Unlike the first version of the metric, the
Detweiler-Whiting metric is regular on the world line and 
satisfies the Einstein field equations in vacuum; and because it is 
a solution to the field equations,  it can be viewed as a physical
metric --- specifically, the metric of the background spacetime
perturbed by a free field produced by the particle at an
earlier stage of its history. 

While Eq.~(\ref{1.9.7}) does indeed give the correct equations of
motion for a small mass $m$ moving in a background spacetime with
metric $g_{\alpha\beta}$, the derivation outlined here leaves much to
be desired --- to what extent should we trust an analysis based on the  
existence of a point mass? As a partial answer to this question, Mino,
Sasaki, and Tanaka \cite{mino-etal:97a} produced an alternative
derivation of their result, which involved a small nonrotating black
hole instead of a point mass. In this alternative derivation, the
metric of the black hole perturbed by the tidal gravitational field of
the external universe is matched to the metric of the background
spacetime perturbed by the moving black hole. Demanding that this
metric be a solution to the vacuum field equations determines the
motion of the black hole: it must move according to
Eq.~(\ref{1.9.7}). This alternative derivation (which was given a
different implementation in Ref.~\cite{poisson:04a}) is entirely free
of singularities (except deep within the black hole), and it suggests
that the MiSaTaQuWa equations can be trusted to describe the
motion of any gravitating body in a curved background spacetime (so
long as the body's internal structure can be ignored). This
derivation, however, was limited to the case of a non-rotating black
hole, and it relied on a number of unjustified and sometimes unstated
assumptions \cite{gralla-wald:08, pound:10a, pound:10b}. The conclusion
was made firm by the more rigorous analysis of Gralla and Wald
\cite{gralla-wald:08}  (as extended by Pound \cite{pound:10a}), who
showed that the MiSaTaQuWa equations apply to any sufficiently compact 
body of arbitrary internal structure.  

It is important to understand that unlike Eqs.~(\ref{1.7.1}) and
(\ref{1.8.7}), which are true tensorial equations, Eq.~(\ref{1.9.7}) 
reflects a specific choice of coordinate system and its form would not
be preserved under a coordinate transformation. In other words, 
{\it the MiSaTaQuWa equations are not gauge invariant}, and they
depend upon the Lorenz gauge condition 
$\gamma^{\alpha\beta}_{\ \ \ ;\beta} = O(m^2)$. Barack and Ori
\cite{barack-ori:01} have shown that under a coordinate transformation
of the form $x^\alpha \to x^\alpha + \xi^\alpha$, where $x^\alpha$ are
the coordinates of the background spacetime and $\xi^\alpha$ is a
smooth vector field of order $m$, the particle's acceleration changes
according to $a^\mu \to a^\mu + a[\xi]^\mu$, where 
\begin{equation}
a[\xi]^\mu = \bigl( \delta^\mu_{\ \nu} + u^\mu u_\nu \bigr) \biggl( 
\frac{D^2 \xi^\nu}{d\tau^2} + R^\nu_{\ \rho\omega\lambda} u^\rho
\xi^\omega u^\lambda \biggr) 
\label{1.9.8}
\end{equation} 
is the ``gauge acceleration''; $D^2 \xi^\nu/d\tau^2 
= (\xi^\nu_{\ ;\mu} u^\mu)_{;\rho} u^\rho$ is the second covariant
derivative of $\xi^\nu$ in the direction of the world line. This 
implies that the particle's acceleration can be altered at will by a
gauge transformation; $\xi^\alpha$ could even be chosen so as to
produce $a^\mu = 0$, making the motion geodesic after all. This
observation provides a dramatic illustration of the following point: 
{\it The MiSaTaQuWa equations of motion are not gauge invariant and
they cannot by themselves produce a meaningful answer to a well-posed
physical question; to obtain such answers it is necessary
to combine the equations of motion with the metric perturbation 
$h_{\alpha\beta}$ so as to form gauge-invariant quantities that will
correspond to direct observables.} This point is very important and
cannot be over-emphasized.      

The gravitational self-force possesses a physical significance that is
not shared by its scalar and electromagnetic analogues, because the
motion of a small body in the strong gravitational field of a much
larger body is a problem of direct relevance to gravitational-wave
astronomy. Indeed, {\it extreme-mass-ratio inspirals}, involving
solar-mass compact objects moving around massive black holes of the  
sort found in galactic cores, have been identified as promising
sources of low-frequency gravitational waves for space-based
interferometric detectors such as the proposed Laser Interferometer
Space Antenna (LISA \cite{LISA}). These systems involve highly
eccentric, nonequatorial, and relativistic orbits around rapidly
rotating black holes, and the waves produced by such orbital motions   
are rich in information concerning the strongest gravitational
fields in the Universe. This information will be extractable from the
LISA data stream, but the extraction depends on sophisticated
data-analysis strategies that require a detailed and accurate modeling 
of the source. This modeling involves formulating the equations of
motion for the small body in the field of the rotating black hole, as
well as a consistent incorporation of the motion into a
wave-generation formalism. In short, the extraction of this wealth of
information relies on a successful evaluation of the gravitational
self-force.     

The finite-mass corrections to the orbital motion are
important. For concreteness, let us assume that the orbiting body is a
black hole of mass $m = 10\ M_\odot$ and that the central black hole
has a mass $M = 10^6\ M_\odot$. Let us also assume that the small
black hole is in the deep field of the large hole, near the innermost
stable circular orbit, so that its orbital period $P$ is of the order
of minutes. The gravitational waves produced by the orbital motion
have frequencies $f$ of the order of the mHz, which is well within
LISA's frequency band. The radiative losses drive the orbital motion   
toward a final plunge into the large black hole; this occurs over a
radiation-reaction timescale $(M/m) P$ of the order of a year, during
which the system will go through a number of wave cycles of  
the order of $M/m = 10^5$. {\it The role of the gravitational
self-force is precisely to describe this orbital evolution toward the
final plunge.} While at any given time the self-force provides 
fractional corrections of order $m/M = 10^{-5}$ to the motion of
the small black hole, these build up over a number of orbital cycles
of order $M/m = 10^5$ to produce a large cumulative effect. As will be
discussed in some detail in Sec.~\ref{sec:physical}, the gravitational
self-force is important, because it drives large secular changes in
the orbital motion of an extreme-mass-ratio binary.       

\subsection{Case study: static electric charge in Schwarzschild
  spacetime}  
\label{1.10} 

One of the first self-force calculations ever performed for a curved
spacetime was presented by Smith and Will
\cite{smith-will:80}. They considered an electric charge $e$ held in
place at position $r = r_0$ outside a Schwarzschild black hole of mass 
$M$. Such a static particle must be maintained in position with an
external force that compensates for the black hole's attraction. For a
particle without electric charge this force is directed outward, and
its radial component in Schwarzschild coordinates is given by 
$f_{\rm ext}^r = \frac{1}{2} m f'$, where $m$ is the particle's mass,
$f := 1-2M/r_0$ is the usual metric factor, and a prime indicates
differentiation with respect to $r_0$, so that $f' = 2M/r_0^2$. Smith
and Will found that for a particle of charge $e$, the external force
is given instead by $f_{\rm ext}^r = \frac{1}{2} m f'
- e^2 M f^{1/2}/r_0^3$. The second term is contributed by the  
electromagnetic self-force, and implies that the external force is 
{\it smaller} for a charged particle. This means that the
electromagnetic self-force acting on the particle is 
{\it directed outward} and given by 
\begin{equation} 
f^r_{\rm self} =  \frac{e^2 M}{r_0^3} f^{1/2}. 
\label{1.10.1} 
\end{equation}
This is a {\it repulsive force}. It was shown by Zel'nikov and Frolov
\cite{zelnikov-frolov:82} that the same expression applies to a static
charge outside a Reissner-Nordstr\"om black hole of mass $M$ and
charge $Q$, provided that $f$ is replaced by the more general 
expression $f = 1 - 2M/r_0 + Q^2/r_0^2$.   

The repulsive nature of the electromagnetic self-force acting on a
static charge outside a black hole is unexpected. In an attempt
to gain some intuition about this result, it is useful to recall that
a black-hole horizon always acts as perfect conductor, because the  
electrostatic potential $\varphi := -A_t$ is necessarily uniform across   
its surface. It is then tempting to imagine that the self-force should
result from a fictitious distribution of induced charge on the horizon,
and that it could be estimated on the basis of an elementary model
involving a spherical conductor. Let us, therefore, calculate the
electric field produced by a point charge $e$ situated outside a
spherical conductor of radius $R$. The charge is placed at a distance 
$r_0$ from the centre of the conductor, which is taken at first to 
be grounded. The electrostatic potential produced by the charge
can easily be obtained with the method of images. It is found that an
image charge $e' = -eR/r_0$ is situated at a distance $r'_0 = R^2/r_0$
from the centre of the conductor, and the potential is given by $\varphi
= e/s + e'/s'$, where $s$ is the distance to the charge, while $s'$ is
the distance to the image charge. The first term can be identified
with the singular potential $\varphi_{\rm S}$, and the associated
electric field exerts no force on the point charge. The second term is
the regular potential $\varphi_{\rm R}$, and the associated field is
entirely responsible for the self-force. The regular electric field is
$E^r_{\rm R} = -\partial_r \varphi_{\rm R}$, and the self-force is 
$f^r_{\rm self} = e E^r_{\rm R}$. A simple computation returns 
\begin{equation} 
f^r_{\rm self}  = -\frac{e^2 R}{r_0^3 (1-R^2/r_0^2)}.  
\label{1.10.2a} 
\end{equation} 
This is an attractive self-force, because the total induced charge on
the conducting surface is equal to $e'$, which is opposite in sign
to $e$. With $R$ identified with $M$ up to a numerical factor, we find
that our intuition has produced the expected factor of $e^2 M/r_0^3$,
but that it gives rise to the {\it wrong sign} for the self-force. An
attempt to refine this computation by removing the net charge $e'$ on 
the conductor (to mimic more closely the black-hole horizon, which
cannot support a net charge) produces a wrong dependence on $r_0$ in
addition to the same wrong sign. In this case the conductor is
maintained at a constant potential $\phi_0 = -e'/R$, and the situation
involves a second image charge $-e'$ situated at $r=0$. It is easy to
see that in this case,  
\begin{equation} 
f^r_{\rm self}  = -\frac{e^2 R^3}{r_0^5 (1-R^2/r_0^2)}.    
\label{1.10.2b} 
\end{equation} 
This is still an attractive force, which is weaker than the force of
Eq.~(\ref{1.10.2a}) by a factor of $(R/r_0)^2$; the force is now
exerted by an image dipole instead of a single image charge.  

The computation of the self-force in the black-hole case is almost as
straightforward. The exact solution to Maxwell's equations that
describes a point charge $e$ situated $r=r_0$ and $\theta=0$ in the 
Schwarzschild spacetime is given by   
\begin{equation} 
\varphi = \varphi^{\rm S} + \varphi^{\rm R}, 
\label{1.10.3} 
\end{equation} 
where  
\begin{equation} 
\varphi^{\rm S} = \frac{e}{r_0 r} 
\frac{ (r-M)(r_0-M) - M^2\cos\theta }
{ \bigl[ (r-M)^2 - 2(r-M)(r_0-M)\cos\theta + (r_0-M)^2
  - M^2\sin^2\theta \bigr]^{1/2}}, 
\label{1.10.4} 
\end{equation} 
is the solution first discovered by Copson in 1928 \cite{copson:28},
while    
\begin{equation} 
\varphi^{\rm R} = \frac{eM/r_0}{r} 
\label{1.10.5} 
\end{equation} 
is the monopole field that was added by Linet \cite{linet:76} to 
obtain the correct asymptotic behaviour $\varphi \sim e/r$ when $r$ is
much larger than $r_0$. It is easy to see that Copson's potential
behaves as $e(1-M/r_0)/r$ at large distances, which reveals that in 
addition to $e$, $\varphi^{\rm S}$ comes with an additional (and
unphysical) charge $-eM/r_0$ situated at $r=0$. This charge
must be removed by adding to $\varphi^{\rm S}$ a potential that (i) is a 
solution to the vacuum Maxwell equations, (ii) is regular everywhere
except at $r=0$, and (iii) carries the opposite charge $+eM/r_0$;
this potential must be a pure monopole, because higher multipoles
would produce a singularity on the horizon, and it is given uniquely
by $\varphi^{\rm R}$. The Copson solution was generalized to
Reissner-Nordstr\"om spacetime by L\'eaut\'e and Linet
\cite{leaute-linet:76}, who also showed that the regular potential of 
Eq.~(\ref{1.10.5}) requires no modification.  
   
The identification of Copson's potential with the singular potential  
$\varphi^{\rm S}$ is motivated by the fact that its associated electric
field $F_{tr}^{\rm S} = \partial_r \varphi^{\rm S}$ is isotropic around
the charge $e$ and therefore exerts no force. The self-force comes
entirely from the monopole potential, which describes a (fictitious)   
charge $+eM/r_0$ situated at $r=0$. Because this charge is of the same 
sign as the original charge $e$, the self-force is repulsive. More
precisely stated, we find that the regular piece of the electric field
is given by  
\begin{equation} 
F^{\rm R}_{tr} = -\frac{eM/r_0}{r^2}, 
\label{1.10.6}
\end{equation} 
and that it produces the self-force of Eq.~(\ref{1.10.1}). The simple 
picture described here, in which the electromagnetic self-force is
produced by a fictitious charge $eM/r_0$ situated at the centre of the
black hole, is not easily extracted from the derivation presented
originally by Smith and Will \cite{smith-will:80}. To the best of our
knowledge, the monopolar origin of the self-force was first noticed by
Alan Wiseman \cite{wiseman:00}. (In his paper, Wiseman computed the
{\it scalar self-force} acting on a static particle in Schwarzschild
spacetime, and found a zero answer. In this case, the analogue of the
Copson solution for the scalar potential happens to satisfy the
correct asymptotic conditions, and there is no need to add another
solution to it. Because the scalar potential is precisely equal to the 
singular potential, the self-force vanishes.)    

We should remark that the identification of $\varphi^S$ and
$\varphi^R$ with the Detweiler-Whiting singular and regular fields,
respectively, is a matter of conjecture. Although $\varphi^S$ and
$\varphi^R$ satisfy the essential properties of the Detweiler-Whiting
decomposition --- being, respectively, a regular homogenous solution
and a singular solution sourced by the particle --- one should accept 
the possibility that they may not be the actual Detweiler-Whiting
fields. It is a topic for future research to investigate the precise
relation between the Copson field and the Detweiler-Whiting singular
field.  

It is instructive to compare the electromagnetic self-force produced
by the presence of a grounded conductor to the self-force produced by
the presence of a black hole. In the case of a conductor, the total
induced charge on the conducting surface is $e' = -eR/r_0$, and it is
this charge that is responsible for the attractive self-force; the
induced charge is supplied by the electrodes that keep the conductor
grounded. In the case of a black hole, there is no external apparatus
that can supply such a charge, and the total induced charge on the
horizon necessarily vanishes. The origin of the self-force is
therefore very different in this case. As we have seen, the self-force
is produced by a fictitious charge $eM/r_0$ situated at the centre of
black hole; and because this charge is positive, the self-force is
repulsive.    

\subsection{Organization of this review} 
\label{1.11} 

After a detailed review of the literature in Sec.~\ref{lit_surv}, 
the main body of the review begins in Part \ref{part1} (Secs.~\ref{2}
to \ref{6}) with a description of the general theory of bitensors,
the name designating tensorial functions of two points in
spacetime. We introduce Synge's world function $\sigma(x,x')$ and
its derivatives in Sec.~\ref{2}, the parallel propagator 
$g^\alpha_{\ \alpha'}(x,x')$ in Sec.~\ref{4}, and the van Vleck
determinant $\Delta(x,x')$ in Sec.~\ref{6}. An important portion of
the theory (covered in Secs.~\ref{3} and \ref{5}) is concerned with
the expansion of bitensors when $x$ is very close to $x'$; expansions
such as those displayed in Eqs.~(\ref{1.6.2}) and (\ref{1.6.3}) are
based on these techniques. The presentation in Part \ref{part1}
borrows heavily from Synge's book \cite{synge:60} and the article by
DeWitt and Brehme \cite{dewitt-brehme:60}. These two sources use
different conventions for the Riemann tensor, and we have adopted
Synge's conventions (which agree with those of Misner, Thorne, and
Wheeler \cite{MTW:73}). The reader is therefore warned that formulae
derived in Part \ref{part1} may look superficially different from
those found in DeWitt and Brehme.  

In Part \ref{part2} (Secs.~\ref{7} to \ref{10}) we introduce a number
of coordinate systems that play an important role in later parts of
the review. As a warmup exercise we first construct (in Sec.~\ref{7}) 
Riemann normal coordinates in a neighbourhood of a reference point
$x'$. We then move on (in Sec.~\ref{8}) to Fermi normal coordinates
\cite{manasse-misner:63}, which are defined in a neighbourhood of a
world line $\gamma$. The retarded coordinates, which are also based at
a world line and which were briefly introduced in Sec.~\ref{1.5}, are
covered systematically in Sec.~\ref{9}. The relationship between 
Fermi and retarded coordinates is worked out in Sec.~\ref{10}, which
also locates the advanced point $z(v)$ associated with a field point
$x$. The presentation in Part \ref{part2} borrows heavily from Synge's
book \cite{synge:60}. In fact, we are much indebted to Synge for initiating
the construction of retarded coordinates in a neighbourhood of a world 
line. We have implemented his program quite differently (Synge was
interested in a large neighbourhood of the world line in a weakly
curved spacetime, while we are interested in a small neighbourhood in a  
strongly curved spacetime), but the idea is originally his.    

In Part \ref{part3} (Secs.~\ref{11} to \ref{15}) we review the theory
of Green's functions for (scalar, vectorial, and tensorial) wave
equations in curved spacetime. We begin in Sec.~\ref{11} with a
pedagogical introduction to the retarded and advanced Green's
functions for a massive scalar field in flat spacetime; in this simple
context the all-important Hadamard decomposition \cite{hadamard:23} of 
the Green's function into ``light-cone'' and ``tail'' parts can be
displayed explicitly. The invariant Dirac functional is defined in
Sec.~\ref{12} along with its restrictions on the past and future null
cones of a reference point $x'$. The retarded, advanced, singular, and
regular Green's functions for the scalar wave equation are
introduced in Sec.~\ref{13}. In Secs.~\ref{14} and \ref{15} we cover
the vectorial and tensorial wave equations, respectively. The
presentation in Part \ref{part3} is based partly on the paper by
DeWitt and Brehme \cite{dewitt-brehme:60}, but it is inspired mostly
by Friedlander's book \cite{friedlander:75}. The reader should be
warned that in one important aspect, our notation differs from the
notation of DeWitt and Brehme: While they denote the tail part of the
Green's function by $-v(x,x')$, we have taken the liberty of
eliminating the silly minus sign and call it instead $+V(x,x')$. The
reader should also note that all our Green's functions are normalized
in the same way, with a factor of $-4\pi$ multiplying a
four-dimensional Dirac functional of the right-hand side of the wave
equation. (The gravitational Green's function is sometimes normalized
with a $-16\pi$ on the right-hand side.)  

In Part \ref{part4} (Secs.~\ref{16} to \ref{18}) we compute the
retarded, singular, and regular fields associated with a point
scalar charge (Sec.~\ref{16}), a point electric charge
(Sec.~\ref{17}), and a point mass (Sec.~\ref{18}). We provide two
different derivations for each of the equations of motion. The first
type of derivation was outlined previously: We follow Detweiler and
Whiting \cite{detweiler-whiting:03} and postulate that only the regular
field exerts a force on the particle. In the second type of derivation
we take guidance from Quinn and Wald \cite{quinn-wald:97} and
postulate that the net force exerted on a point particle is given by
an average of the retarded field over a surface of constant proper
distance orthogonal to the world line --- this rest-frame average is
easily carried out in Fermi normal coordinates. The averaged field is
still infinite on the world line, but the divergence points in the
direction of the acceleration vector and it can thus be removed by
mass renormalization. Such calculations show that while the singular
field does not affect the motion of the particle, it nonetheless
contributes to its inertia. 

In Part~\ref{part5} (Secs.~\ref{20} to \ref{global_solution}), we show
that at linear order in the body's mass $m$, an extended body behaves
just as a point mass, and except for the effects of the body's spin,
the world line representing its mean motion is governed by the
MiSaTaQuWa equation. At this order, therefore, the picture of a point
particle interacting with its own field, and the results obtained from
this picture, is justified. Our derivation utilizes the method of
matched asymptotic expansions, with an inner expansion accurate near
the body and an outer expansion accurate everywhere else. The equation
of motion of the body's world line, suitably defined, is calculated by
solving the Einstein equation in a buffer region around the body,
where both expansions are accurate.       

Concluding remarks are presented in Sec.~\ref{conclusion}, and
technical developments that are required in Part~\ref{part5} are  
relegated to Appendices. Throughout this review we use geometrized
units and adopt the notations and conventions of Misner, Thorne, and
Wheeler \cite{MTW:73}.      

\subsection{Changes relative to the 2004 edition} 
\label{1.12} 

This 2010 version of the review is a major update of the original
article published in 2004. Two additional authors, Adam Pound and Ian
Vega, have joined the article's original author, and each one has
contributed a major piece of the update. The literature survey   
presented in Sec.~\ref{lit_surv} was contributed by Ian Vega, and
Part~\ref{part5} (Secs.~\ref{20} to \ref{global_solution})
was contributed by Adam Pound. Part~\ref{part5} replaces a section of
the 2004 article in which the motion of a small black hole was derived
by the method of matched asymptotic expansions; this material can
still be found in Ref.~\cite{poisson:04a}, but Pound's work provides a
much more satisfactory foundation for the gravitational self-force.
The case study of Sec.~\ref{1.10} is new, and the ``exact''
formulation of the dynamics of a point mass in Sec.~\ref{18.1} is a
major improvement from the original article. The concluding remarks of
Sec.~\ref{conclusion}, contributed mostly by Adam Pound, are also
updated from the 2004 article. 

\subsection*{Acknowledgments} 

Our understanding of the work presented in this review was shaped by a 
series of annual meetings named after the movie director Frank
Capra. The first of these meetings took place in 1998 and was held at
Capra's ranch in Southern California; the ranch now belongs to
Caltech, Capra's alma mater. Subsequent meetings were held in Dublin
(1999), Pasadena (2000), Potsdam (2001), State College PA (2002),
Kyoto (2003), Brownsville (2004), Oxford (2005), Milwaukee (2006),
Hunstville (2007), Orl\'eans (2008), Bloomington (2009), and Waterloo
(2010).  At these meetings and elsewhere we have enjoyed many
instructive conversations with Paul Anderson, Warren Anderson, Patrick
Brady, Claude Barrab\`es, Leor Barack, Luc Blanchet, Lior Burko,
Manuella Campanelli, Marc Casals, Peter Diener, Steve Detweiler, Sam
Dolan, Steve Drasco, Scott Field, Eanna Flanagan, John Friedman,
Ryuichi Fujita, Chad Galley, Darren Golbourn, Sam Gralla, Roland Haas,
Abraham Harte, Wataru Hikida, Tanja Hinderer, Scott Hughes, Werner
Israel, Toby Keidl, Dong-hoon Kim, Alexandre Le Tiec, Carlos Lousto,
Eirini Messaritaki, Yasushi Mino, Hiroyuki Nakano, Amos Ori, Larry
Price, Eran Rosenthal, Norichika Sago, Misao Sasaki, Abhay Shah,
Carlos Sopuerta, Alessandro Spallicci, Hideyuki Tagoshi, Takahiro
Tanaka, Jonathan Thornburg, Bill Unruh, Bob Wald, Niels Warburton,
Barry Wardell, Alan Wiseman, and Bernard Whiting. This work was
supported by the Natural Sciences and Engineering Research Council of
Canada.

%% file: literature.tex
%
\section{Computing the self-force: a 2010 
literature survey} 
\label{lit_surv}

Much progress has been achieved in the development of practical
methods for computing the self-force.  We briefly summarize
these efforts in this section, with the goal of introducing the main
ideas and some key issues. A more detailed coverage of the various
implementations can be found in Barack's excellent review
\cite{barack:09}. The 2005 collection of reviews published in 
\emph{Classical and Quantum Gravity} \cite{cqg-collection} is also
recommended for an introduction to the various aspects of self-force
theory and numerics. Among our favourites in this collection are the
reviews by Detweiler \cite{detweiler:05} and Whiting
\cite{whiting:05}.   
  
An important point to bear in mind is that all the methods covered
here mainly compute the self-force on a particle moving on a
\emph{fixed world line} of the background spacetime. A few numerical
codes based on the radiative approximation have allowed orbits to
evolve according to energy and angular-momentum balance. As
will be emphasized below, however, these calculations miss out on
important conservative effects that are only accounted for by the full
self-force. Work is currently underway to develop methods to let the
self-force alter the motion of the particle in a self-consistent
manner.  

\subsection{Early work: DeWitt and DeWitt; Smith and Will} 

The first evaluation of the electromagnetic self-force in curved
spacetime was carried out by DeWitt and DeWitt \cite{dewitt-dewitt:64}
for a charge moving freely in a weakly curved spacetime characterized
by a Newtonian potential $\Phi \ll 1$. In this context the right-hand
side of Eq.~(\ref{1.7.1}) reduces to the tail integral, because the
particle moves in a vacuum region of the spacetime, and there is no  
external force acting on the charge. They found that the spatial
components of the self-force are given by 
\begin{equation} 
\bm{f}_{\rm em} = e^2 \frac{M}{r^3}\, \bm{\hat{r}} + \frac{2}{3} e^2
\frac{d \bm{g}}{dt}, 
\label{1.11.1}
\end{equation} 
where $M$ is the total mass contained in the spacetime, $r =
|\bm{x}|$ is the distance from the centre of mass, $\bm{\hat{r}} =
\bm{x}/r$, and $\bm{g} = -\bm{\nabla} \Phi$ is the Newtonian
gravitational field. (In these expressions the bold-faced symbols
represent vectors in three-dimensional flat space.) 
The first term on the right-hand side of Eq.~(\ref{1.11.1}) is a
conservative correction to the Newtonian force $m \bm{g}$. The
second term is the standard radiation-reaction force; although it 
comes from the tail integral, this is the same result that would be  
obtained in flat spacetime if an external force $m \bm{g}$ were acting  
on the particle. This agreement is necessary, but remarkable!  

A similar expression was obtained by Pfenning and Poisson
\cite{pfenning-poisson:02} for the case of a scalar charge. Here  
\begin{equation} 
\bm{f}_{\rm scalar} = 2 \xi q^2 \frac{M}{r^3}\, \bm{\hat{r}} 
+ \frac{1}{3} q^2 \frac{d \bm{g}}{dt}, 
\label{1.11.2}
\end{equation} 
where $\xi$ is the coupling of the scalar field to the spacetime
curvature; the conservative term disappears when the field is
minimally coupled. Pfenning and Poisson also computed the
gravitational self-force acting on a point mass moving in a
weakly curved spacetime. The expression they obtained is in complete 
agreement (within its domain of validity) with the standard
post-Newtonian equations of motion.   

The force required to hold an electric charge in place in a
Schwarzschild spacetime was computed, without approximations, by 
Smith and Will \cite{smith-will:80}. As we reviewed previously in
Sec.~\ref{1.10}, the self-force contribution to the total force is
given by 
\begin{equation} 
f^r_{\rm self} =  e^2\frac{M}{r^3} f^{1/2}, 
\end{equation}
where $M$ is the black-hole mass, $r$ the position of the charge (in
Schwarzschild coordinates), and $f := 1-2M/r$. When $r \gg M$, this
expression agrees with the conservative term in Eq.~(\ref{1.11.1}). 
This result was generalized to Reissner-Nordstr\"om spacetime by 
Zel'nikov and Frolov \cite{zelnikov-frolov:82}. Wiseman
\cite{wiseman:00} calculated the self-force acting on a static scalar
charge in Schwarzschild spacetime. He found that in this case the
self-force vanishes. This result is not incompatible with
Eq.~(\ref{1.11.2}), even for nonminimal coupling, because the
computation of the weak-field self-force requires the presence of
matter, while Wiseman's scalar charge lives in a purely vacuum
spacetime.   

\subsection{Mode-sum method}

Self-force calculations involving a sum over modes were pioneered by
Barack and Ori \cite{barack-ori:00, barack:00}, and the method was  
further developed by Barack, Ori, Mino, Nakano, and Sasaki
\cite{barack-etal:02, barack:01, barack-ori:02, barack-ori:03a,
  barack-ori:03b, mino-etal:03}; a somewhat related approach was also
considered by Lousto \cite{lousto:00}. It has now emerged as
the method of choice for self-force calculations in spacetimes such as
Schwarzschild and Kerr. Our understanding of the method was greatly
improved by the Detweiler-Whiting decomposition
\cite{detweiler-whiting:03} of the retarded field into singular and
regular pieces, as outlined in Secs.~\ref{1.4} and \ref{1.8}, and
subsequent work by Detweiler, Whiting, and their collaborators
\cite{detweiler-etal:03}.  

\subsubsection*{Detweiler-Whiting decomposition; mode decomposition;
  regularization parameters} 

For simplicity we consider the problem of computing the self-force
acting on a particle with a scalar charge $q$ moving on a world line
$\gamma$. (The electromagnetic and gravitational problems are
conceptually similar, and they will be discussed below.) The potential
$\Phi$ produced by the particle satisfies Eq.~(\ref{1.8.1}), which we
rewrite schematically as 
\begin{equation}
\Box \Phi = q \delta(x,z),
\label{schem} 
\end{equation}
where $\Box$ is the wave operator in curved spacetime, and 
$\delta(x,z)$ represents a distributional source that depends 
on the world line $\gamma$ through its coordinate representation
$z(\tau)$. From the perspective of the Detweiler-Whiting
decomposition, the scalar self-force is given by 
\begin{equation}
  F_\alpha = q\nabla_\alpha \Phi_{\rm R} := 
q\bigl( \nabla_\alpha\Phi -\nabla_\alpha\Phi_{\rm S} \bigr),  
  \label{eqn:sfdw}
\end{equation}
where $\Phi$, $\Phi_{\rm S}$, and $\Phi_{\rm R}$ are the retarded,
singular, and regular potentials, respectively. To evaluate the
self-force, then, is to compute the gradient of the regular potential.   

From the point of view of Eq~(\ref{eqn:sfdw}), the task of computing
the self-force appears conceptually straightforward: Either (i)
compute the retarded and singular potentials, subtract them, and take
a gradient of the difference; or (ii) compute the gradients of the
retarded and singular potentials, and then subtract the
gradients. Indeed, this is the basic idea for most methods of
self-force computations. However, the apparent simplicity of this
sequence of steps is complicated by the following facts: (i) except
for a very limited number of cases, the retarded potential of a point
particle cannot be computed analytically and must therefore be
obtained by numerical means; and (ii) both the retarded and singular
potential diverge at the particle's position. Thus, any sort of
subtraction will generally have to be performed numerically, and for
this to be possible, one requires representations of the retarded and
singular potentials (and/or their gradients) in terms of finite
quantities.

In a mode-sum method, these difficulties are overcome with a
decomposition of the potential in spherical-harmonic functions: 
\begin{equation}
  \Phi = \sum_{lm} \Phi^{lm}(t,r) Y^{lm}(\theta,\phi). 
  \label{eqn:shd.scalar}
\end{equation}
When the background spacetime is spherically symmetric,
Eq.~(\ref{schem}) gives rise to a fully decoupled set of reduced wave
equations for the mode coefficients $\Phi^{lm}(t,r)$,
and these are easily integrated with simple numerical methods. 
The dimensional reduction of the wave equation implies that each
$\Phi^{lm}(t,r)$ is finite and continuous (though nondifferentiable)
at the position of the particle. There is, therefore, no obstacle to
evaluating each $l$-mode of the field, defined by 
\begin{equation}
  (\nabla_\alpha \Phi)_l:= \lim_{x\rightarrow z}\sum^l_{m=-l} 
\nabla_\alpha [\Phi^{lm}(t,r)Y^{lm}(\theta,\phi)]. 
\end{equation}
The sum over modes, however, must reproduce the singular field
evaluated at the particle's position, and this is infinite; the mode
sum, therefore, does not converge. 

Fortunately, there is a piece of each $l$-mode that does not
contribute to the self-force, and that can be subtracted out; this
piece is the corresponding $l$-mode of the
singular field $\nabla_\alpha \Phi_{\rm S}$. Because the retarded and 
singular fields share the same singularity structure near the
particle's world line (as described in Sec.~\ref{1.6}), the subtraction
produces a mode decomposition of the regular field $\nabla_\alpha
\Phi_{\rm R}$. And because this field is regular at the particle's
position, the sum over all modes $q(\nabla_\alpha \Phi_{\rm R})_l$
is guaranteed to converge to the correct value for the self-force. The
key to the mode-sum method, therefore, is the ability to express the
singular field as a mode decomposition. 

This can be done because the singular field, unlike the retarded
field, can always be expressed as a local expansion in powers of the
distance to the particle; such an expansion was displayed in
Eqs.~(\ref{1.6.7}) and (\ref{1.6.8}). (In a few special cases the
singular field is actually known exactly \cite{copson:28, linet:76, 
  burko-etal:02, haas-poisson:05, shankar-whiting:07}.) This local
expansion can then be turned into a multipole decomposition. Barack
and Ori \cite{barack-ori:02, barack-etal:02, barack-ori:03a,
  barack-ori:03b, barack:09}, and then Mino, Nakano, and Sasaki
\cite{mino-etal:03}, were the first to show that this produces the
following generic structure:    
\begin{equation}
(\nabla_\alpha \Phi_{\textrm{S}})_l = 
(l+ {\textstyle \frac{1}{2}}) A_\alpha + 
B_\alpha + \frac{C_\alpha}{l+\frac{1}{2}} 
+ \frac{D_\alpha}{(l-\frac{1}{2})(l+\frac{3}{2})} 
+ \frac{E_\alpha}{(l-\frac{3}{2})(l-\frac{1}{2})(l+\frac{3}{2})(l+\frac{5}{2})}
+ \cdots,  
\label{eqn:asymp}
\end{equation}
where $A_\alpha$, $B_\alpha$, $C_\alpha$, and so on are
$l$-independent functions that depend on the choice of field
(i.e. scalar, electromagnetic, or gravitational), the choice of
spacetime, and the particle's state of motion. These so-called
\emph{regularization parameters} are now ubiquitous in the self-force
literature, and they can all be determined from the local expansion
for the singular field. The number of regularization parameters that
can be obtained depends on the accuracy of the expansion. For example,
expansions accurate through order $r^0$ such as Eqs.~(\ref{1.6.7}) and 
(\ref{1.6.8}) permit the determination of $A_\alpha$, $B_\alpha$, and
$C_\alpha$; to obtain $D_\alpha$ one requires the terms of order
$r$, and to get $E_\alpha$ the expansion must be carried out through
order $r^2$. The particular polynomials in $l$ that accompany the
regularization parameters were first identified by Detweiler and his 
collaborators \cite{detweiler-etal:03}. Because the $D_\alpha$ term is
generated by terms of order $r$ in the local expansion of the singular
field, the sum of $[(l-\frac{1}{2})(l+\frac{3}{2})]^{-1}$ from $l=0$
to $l=\infty$ evaluates to zero. The sum of the polynomial in front of
$E_\alpha$ also evaluates to zero, and this property is shared by all
remaining terms in Eq.~(\ref{eqn:asymp}). 

\subsubsection*{Mode sum} 

With these elements in place, the self-force is finally 
computed by implementing the mode-sum formula
\begin{eqnarray}
  F_\alpha &=& q\sum_{l=0}^L \biggl[ (\nabla_\alpha\Phi)_l 
- (l+ {\textstyle \frac{1}{2}}) A_\alpha 
- B_\alpha - \frac{C_\alpha}{l+\frac{1}{2}} 
- \frac{D_\alpha}{(l-\frac{1}{2})(l+\frac{3}{2})} 
\nonumber \\ & & \mbox{} 
- \frac{E_\alpha}{(l-\frac{3}{2})(l-\frac{1}{2})(l+\frac{3}{2})(l+\frac{5}{2})}
- \cdots\biggr] + \mbox{remainder},
\label{eqn:modesumformula}
\end{eqnarray}
where the infinite sum over $l$ is truncated to a maximum mode number
$L$. (This truncation is necessary in practice, because in general the
modes must be determined numerically.) The remainder consists of the
remaining terms in the sum, from $l=L+1$ to $l=\infty$; it is easy to
see that since the next regularization term would scale as $l^{-6}$
for large $l$, the remainder scales as $L^{-5}$, and can be made
negligible by summing to a suitably large value of $l$. This
observation motivates the inclusion of the $D_\alpha$ and $E_\alpha$
terms within the mode sum, even though their complete sums evaluate to
zero. These terms are useful because the sum must necessarily be
truncated, and they permit a more rapid convergence of the mode
sum. For example, exclusion of the $D_\alpha$ and $E_\alpha$ terms in
Eq.~(\ref{eqn:modesumformula}) would produce a remainder that scales
as $L^{-1}$ instead of $L^{-5}$; while this is sufficient for
convergence, the rate of convergence is too slow to permit
high-accuracy computations. Rapid convergence therefore relies on a
knowledge of as many regularization parameters as possible, but
unfortunately these parameters are not easy to calculate. To date,
only $A_\alpha$, $B_\alpha$, $C_\alpha$, and $D_\alpha$ have been
calculated for general orbits in Schwarzschild spacetime
\cite{detweiler-etal:03, haas-poisson:06}, and only  
$A_\alpha$, $B_\alpha$, $C_\alpha$ have been calculated for orbits in
Kerr spacetime \cite{barack-ori:03b}. It is possible,
however, to estimate a few additional regularization parameters by
fitting numerical results to the structure of Eq.~(\ref{eqn:asymp});
this clever trick was first exploited by Detweiler and his
collaborators \cite{detweiler-etal:03} to achieve extremely high
numerical accuracies. This trick is now applied routinely in mode-sum
computations of the self-force. 

\subsubsection*{Case study: static electric charge in extreme
  Reissner-Nordstr\"om spacetime}
 
The practical use of the mode-sum method can be illustrated with the
help of a specific example that can be worked out fully and
exactly. We consider, as in Sec.~\ref{1.10}, an electric charge $e$
held in place at position $r=r_0$ in the spacetime of an extreme
Reissner-Nordstr\"om black hole of mass $M$ and charge $Q=M$. The
reason for selecting this spacetime resides in the resulting
simplicity of the spherical-harmonic modes for the electromagnetic
field.      

The metric of the extreme Reissner-Nordstr\"om spacetime is given by 
\begin{equation}
ds^2 = -f\, dt^2 + f^{-1} dr^2 + r^2 d\Omega^2,
\end{equation}
where $f = (1 - M/r)^2$. The only nonzero component of the
electromagnetic field tensor is $F_{tr} = -E_r$, and this is
decomposed as  
\begin{equation}
F_{tr} = \sum_{lm} F^{lm}_{tr}(r) Y^{lm}(\theta,\phi). 
\end{equation}
This field diverges at $r=r_0$, but the modes $F^{lm}_{tr}(r)$ are
finite, though discontinuous. The multipole coefficients of the field
are defined to be 
\begin{equation}
(F_{tr})_l = \lim \sum^{l}_{m=-l} F^{lm}_{tr} Y^{lm}, 
\label{mult_def1} 
\end{equation}
where the limit is taken in the direction of the particle's
position. The charge can be placed on the axis $\theta = 0$,
and this choice produces an axisymmetric field with contributions from
$m=0$ only. Because $Y^{l0} = [(2l+1)/4\pi]^{1/2}P_l(\cos\theta)$
and $P_l(1) =1$, we have 
\begin{equation}
(F_{tr})_l= \sqrt{\frac{2l+1}{4\pi}}\lim_{\Delta \rightarrow 0}
F_{tr}^{l0}(r_0 +\Delta).
\label{mult_def2} 
\end{equation}
The sign of $\Delta$ is arbitrary, and $(F_{tr})_l$ depends on the
direction in which $r_0$ is approached. 

The charge density of a static particle can also be decomposed in  
spherical harmonics, and the mode coefficients are given by  
\begin{equation}
r^2j^{l0}_t = e\sqrt{\frac{2l+1}{4\pi}} f_0 \delta(r-r_0),
\end{equation}
where $f_0 = (1-M/r_0)^2$. If we let
\begin{equation}
\Phi^l := -r^2 F^{l0}_{tr}, 
\end{equation}
then Gauss's law in the extreme Reissner-Nordstr\"om spacetime can be
shown to reduce to 
\begin{equation}
(f\Phi')' - \frac{l(l+1)}{r^2}\Phi = 
4\pi e \sqrt{\frac{2l+1}{4\pi}} f_0 \delta'(r-r_0), 
\label{gauss} 
\end{equation}
in which a prime indicates differentiation with respect to $r$, and
the index $l$ on $\Phi$ is omitted to simplify the expressions. The
solution to Eq.~(\ref{gauss}) can be expressed as $\Phi(r) = \Phi_>(r) 
\Theta(r-r_0) + \Phi_<(r) \Theta(r_0-r)$, where $\Phi_>$ and $\Phi_<$
are each required to satisfy the homogeneous equation 
$(f\Phi')' - l(l+1)\Phi/r^2 = 0$, as well as the junction conditions   
\begin{equation}
[\Phi] = 4\pi e\sqrt{\frac{2l+1}{4\pi}}, \qquad 
[\Phi'] =0,
\end{equation}
with $[\Phi] := \Phi_>(r_0) - \Phi_<(r_0)$ denoting the jump across
$r=r_0$. 

For $l = 0$ the general solution to the homogeneous equation is
$c_1 r^* + c_2$, where $c_1$ and $c_2$ are constants and 
$r^* = \int f^{-1}\, dr$. The solution for $r<r_0$ must be regular at
$r=M$, and we select $\Phi_< = \mbox{constant}$. The solution for
$r>r_0$ must produce a field that decays as $r^{-2}$ at large $r$, and
we again select $\Phi_> = \mbox{constant}$. Since each constant is
proportional to the total charge enclosed within a sphere of radius
$r$, we arrive at 
\begin{equation} 
\Phi_< = 0, \qquad 
\Phi_> = \sqrt{4\pi} e, \qquad (l = 0). 
\end{equation} 
For $l \neq 0$ the solutions to the homogeneous equation are  
\begin{equation} 
\Phi_< = c_1 e \biggl( \frac{r-M}{r_0-M} \biggr)^l 
\bigl( l r + M \bigr) 
\end{equation} 
and 
\begin{equation} 
\Phi_> = c_2 e \biggl( \frac{r_0-M}{r-M} \biggr)^{l+1}  
\bigl[ (l +1) r - M \bigr].
\end{equation} 
The constants $c_1$ and $c_2$ are determined by the junction
conditions, and we get 
\begin{equation} 
c_1 = -\sqrt{\frac{4\pi}{2l+1}} \frac{1}{r_0}, 
\qquad 
c_2 = \sqrt{\frac{4\pi}{2l+1}} \frac{1}{r_0}. 
\end{equation} 
The modes of the electromagnetic field are now completely determined. 

According to the foregoing results, and recalling the definition of
Eq.~(\ref{mult_def2}), the multipole coefficients of the
electromagnetic field at $r=r_0 + 0^+$ are given by     
\begin{equation}
\bigl( F^>_{tr} \bigr)_0 = -\frac{e}{r_0^2}, \qquad 
\bigl( F^>_{tr} \bigr)_l = 
e \bigl( l + {\textstyle \frac{1}{2}} \bigr) 
\biggl( -\frac{1}{r_0^2} \biggr) 
- \frac{e}{2r_0^3} (r_0 - 2M).   
\label{eqn:mult_in}
\end{equation}  
For $r=r_0 + 0^-$ we have instead 
\begin{equation}
\bigl( F^<_{tr} \bigr)_0 = 0, \qquad 
\bigl( F^<_{tr} \bigr)_l = 
e \bigl( l + {\textstyle \frac{1}{2}} \bigr) 
\biggl( +\frac{1}{r_0^2} \biggr) 
- \frac{e}{2r_0^3} (r_0 - 2M).   
\label{eqn:mult_out}
\end{equation}  
We observe that the multipole coefficients lead to a diverging mode
sum. We also observe, however, that the multipole structure is
identical to the decomposition of the singular field displayed in
Eq.~(\ref{eqn:asymp}). Comparison of the two expressions allows 
us to determine the regularization parameters for the given situation,
and we obtain 
\begin{equation} 
A = \mp \frac{e}{r_0^2}, \qquad
B = -\frac{e}{2r_0^3} (r_0 - 2M), \qquad
C = D = E = \cdots = 0. 
\end{equation} 
Regularization of the mode sum via Eq.~(\ref{eqn:modesumformula})
reveals that the modes $l \neq 0$ give rise to the singular field,
while the regular field comes entirely from the mode $l = 0$. In
this case, therefore, we can state that the {\it exact expression} for
the regular field evaluated at the position of the particle is 
$F^{\sf R}_{tr} = (F_{tr})_0 - \frac{1}{2} A - B$, or 
$F^{\sf R}_{tr}(r_0) = -eM/r_0^3$. Because the regular field
must be a solution to the vacuum Maxwell equations, its monopole
structure guarantees that its value at any position is given by 
\begin{equation} 
F^{\rm R}_{tr}(r) = -\frac{eM/r_0}{r^2}.   
\end{equation}   
This is the field of an image charge $e' = +eM/r_0$ situated at the
centre of the black hole.  

The self-force acting on the static charge is then 
\begin{equation} 
f^r = -e \sqrt{f_0} F^{\rm R}_{tr}(r_0) =  
\frac{e^2 M}{r_0^3} \sqrt{f_0} = \frac{e^2 M}{r_0^3} (1-M/r_0).  
\end{equation} 
This expression agrees with the Smith-Will force of
Eq.~(\ref{1.10.1}). The interpretation of the result in terms of an
interaction between $e$ and the image charge $e'$ was elaborated in
Sec.~\ref{1.10}.   

\subsubsection*{Computations in Schwarzschild spacetime} 

The mode-sum method was successfully implemented in Schwarzschild
spacetime to compute the scalar and electromagnetic self-forces on a
static particle \cite{burko:00b, burko-etal:01} . It was used to
calculate the scalar self-force on a particle moving on a
radial trajectory \cite{barack-burko:00}, circular orbit 
\cite{burko:00c,detweiler-etal:03,haas-poisson:06,canizares-sopuerta:09},
and a generic bound orbit \cite{haas:07}. It was also developed to 
compute the electromagnetic self-force on a particle moving on a
generic bound orbit \cite{haasPhD}, as well as the gravitational
self-force on a point mass moving on circular \cite{barack-sago:07,
  akcay:10} and eccentric orbits \cite{barack-sago:09}. The mode-sum
method was also used to compute unambiguous physical effects
associated with the gravitational self-force
\cite{detweiler:08,sago-etal:08,barack-etal:10}, and these results
were involved in detailed comparisons with post-Newtonian theory 
\cite{detweiler:08,blanchet-etal:10a,
  blanchet-etal:10b,damour:10,barack-etal:10}. These
achievements will be described in more detail in
Sec.~\ref{sec:physical}.    

An issue that arises in computations of the electromagnetic and
gravitational self-forces is the choice of gauge. While the self-force
formalism is solidly grounded in the Lorenz gauge (which allows the 
formulation of a wave equation for the potentials, the decomposition
of the retarded field into singular and regular pieces, and the
computation of regularization parameters), it is often convenient to
carry out the numerical computations in other gauges, such as the
popular Regge-Wheeler gauge and the Chrzanowski radiation gauge
described below. Compatibility of calculations carried out in
different gauges has been debated in the literature. It is clear that
the singular field is gauge invariant when the transformation between
the Lorenz gauge and the adopted gauge is smooth on the particle's
world line; in such cases the regularization parameters also are gauge
invariant \cite{barack-ori:01}, the transformation affects the regular
field only, and the self-force changes according to
Eq.~(\ref{1.9.8}). The transformations between 
the Lorenz gauge and the Regge-Wheeler and radiation gauges are not
regular on the world line, however, and in such cases the
regularization of the retarded field must be handled with extreme
care.    

\subsubsection*{Computations in Kerr spacetime; metric reconstruction} 
 
The reliance of the mode-sum method on a spherical-harmonic
decomposition makes it generally impractical to apply to self-force
computations in Kerr spacetime. Wave equations in this spacetime
are better analyzed in terms of a \emph{spheroidal}-harmonic
decomposition, which simultaneously requires a Fourier decomposition
of the field's time dependence. (The eigenvalue equation for the
angular functions depends on the mode's frequency.) For a static
particle, however, the situation simplifies, and Burko and Liu
\cite{burko-liu:01} were able to apply the method to calculate the 
self-force on a static scalar charge in Kerr spacetime. 

More recently, Warburton and Barack \cite{warburton-barack:10} 
carried out mode-sum calculations of the scalar self-force on a
particle moving on equatorial orbits of a Kerr black hole. They first
solve for the spheroidal multipoles of the retarded potential, and
then re-express them in terms of spherical-harmonic
multipoles. Fortunately, they find that a spheroidal multipole is well 
represented by summing over a limited number of spherical
multipoles. The Warburton-Barack work represents the first successful
computations of the self-force in Kerr spacetime, and it reveals the
interesting effect of the black hole's spin on the behaviour of the
self-force. 

The analysis of the scalar wave equation in terms of spheroidal
functions and a Fourier decomposition permits a complete separation of 
the variables. For decoupling and separation to occur in the case of a
gravitational perturbation, it is necessary to formulate the
perturbation equations in terms of Newman-Penrose (NP) quantities   
\cite{teukolsky:73}, and to work with the Teukolsky equation that
governs their behaviour. Several computer codes are now available that
are capable of integrating the Teukolsky equation when the source is a
point mass moving on an arbitrary geodesic of the Kerr spacetime. (A
survey of these codes is given below.) Once a solution to the
Teukolsky equation is at hand, however, there still remains the
additional task of recovering the metric perturbation from this
solution, a problem referred to as \emph{metric reconstruction}.     

Reconstruction of the metric perturbation from solutions to the
Teukolsky equation was tackled in the past in the pioneering
efforts of Chrzanowski \cite{chrzanowski:75}, Cohen and Kegeles
\cite{cohen-kegeles:74,kegeles-cohen:79}, Stewart \cite{stewart:79},
and Wald \cite{wald:78}. These works have established a procedure,
typically attributed to Chrzanowski, that returns the metric
perturbation in a so-called radiation gauge. An important limitation
of this method, however, is that it applies only to vacuum solutions
to the Teukolsky equation. This makes the standard Chrzanowski
procedure inapplicable in the self-force context, because a point
particle must necessarily act as a source of the perturbation. Some
methods were devised to extend the Chrzanowski procedure to   
accomodate point sources in specific circumstances
\cite{lousto-whiting:02,ori:03}, but these were not developed 
sufficiently to permit the computation of a self-force. See
Ref.~\cite{whiting-price:05} for a review of metric   
reconstruction from the perspective of self-force calculations.  

A remarkable breakthrough in the application of metric-reconstruction 
methods in self-force calculations was achieved by Keidl, Wiseman, and
Friedman \cite{keidl-etal:06,keidlPhD,keidl-etal:10}, who were able to
compute a self-force starting from a Teukolsky equation 
sourced by a point particle. They did it first for the case of an
electric charge and a point mass held at a fixed position in a
Schwarzschild spacetime \cite{keidl-etal:06}, and then for the case of
a point mass moving on a circular orbit around a Schwarzschild black
hole \cite{keidl-etal:10}. The key conceptual advance is the
realization that, according to the Detweiler-Whiting perspective, the
self-force is produced by a regularized field that satisfies vacuum
field equations in a neighbourhood of the particle. The regular field
can therefore be submitted to the Chrzanowski procedure and
reconstructed from a source-free solution to the Teukolsky equation.    

More concretely, suppose that we have access to the Weyl scalar
$\psi_0$ produced by a point mass moving on a geodesic of a Kerr
spacetime. To compute the self-force from this, one first calculates
the singular Weyl scalar $\psi^{\rm S}_0$ from the Detweiler-Whiting
singular field $h^{\rm S}_{\alpha\beta}$, and subtracts it from
$\psi_0$. The result is a regularized Weyl scalar $\psi^{\rm R}_0$,
which is a solution to the homogeneous Teukolsky equation. This 
sets the stage for the metric-reconstruction procedure, which
returns (a piece of) the regular field $h^{\rm R}_{\alpha\beta}$ in
the radiation gauge selected by Chrzanowski. The computation must be
completed by adding the pieces of the metric perturbation that are not
contained in $\psi_0$; these are referred to either as the
nonradiative degrees of freedom (since $\psi_0$ is purely radiative),
or as the $l=0$ and $l=1$ field multipoles (because the sum over
multipoles that make up $\psi_0$ begins at $l=2$). A method to
complete the Chrzanowski reconstruction of $h^{\rm R}_{\alpha\beta}$
was devised by Keidl {\it et al.}~\cite{keidl-etal:06,keidl-etal:10},
and the end result leads directly to the gravitational self-force. The
relevance of the $l=0$ and $l=1$ modes to the gravitational self-force
was emphasized by Detweiler and Poisson \cite{detweiler-poisson:03}.  

\subsubsection*{Time-domain versus frequency-domain methods}  

When calculating the spherical-harmonic components $\Phi^{lm}(t,r)$ of
the retarded potential $\Phi$ --- refer back to
Eq.~(\ref{eqn:shd.scalar}) --- one can choose to work either directly
in the time domain, or perform a Fourier decomposition of the time 
dependence and work instead in the frequency domain. While the
time-domain method requires the integration of a partial differential
equation in $t$ and $r$, the frequency-domain method gives rise to set
of ordinary differential equations in $r$, one for each frequency
$\omega$. For particles moving on circular or slightly eccentric
orbits in Schwarzschild spacetime, the frequency spectrum is
limited to a small number of discrete frequencies, and a
frequency-domain method is easy to implement and yields highly
accurate results. As the orbital eccentricity increases, however, the
frequency spectrum broadens, and the computational burden of summing
over all frequency components becomes more
significant. Frequency-domain methods are less efficient for large
eccentricities, the case of most relevance for extreme-mass-ratio
inspirals, and it becomes advantageous to replace them with
time-domain methods. (See Ref.~\cite{barton-etal:08} for a
quantitative study of this claim.) This observation has motivated the
development of accurate evolution codes for wave equations in $1+1$
dimensions.  

Such codes must be able to accomodate point-particle sources, and 
various strategies have been pursued to represent a Dirac distribution
on a numerical grid, including the use of very narrow Gaussian pulses 
\cite{lopez-aleman-etal:03,khanna:04,burko-khanna:07} and of ``finite 
impulse representations'' \cite{sundararajan-etal:07}. These methods
do a good job with waveform and radiative flux calculations far away
from the particle, but are of very limited accuracy when computing the
potential in a neighborhood of the particle. A numerical method
designed to provide an {\it exact representation} of a Dirac
distribution in a time-domain computation was devised by Lousto and
Price \cite{lousto-price:97} (see also
Ref.~\cite{martel-poisson:02}). It was implemented by Haas
\cite{haas:07,haasPhD} for the specific purpose  of evaluating
$\Phi^{lm}(t,r)$ at the position of particle and computing the
self-force. Similar codes were developed by other workers for scalar
\cite{vega-detweiler:08} and gravitational \cite{barack-sago:07,
  barack-sago:09} self-force calculations.

Most extant time-domain codes are based on finite-difference
techniques, but codes based on pseudo-spectral methods have also been
developed \cite{field-etal:09, field-etal:10, canizares-sopuerta:09,
canizares-etal:10}. Spectral codes are a powerful alternative to
finite-difference codes, especially when dealing with smooth
functions, because they produce much faster convergence. The  
fact that self-force calculations deal with point sources and field
modes that are not differentiable might suggest that spectral
convergence should not be expected in this case. This objection can be
countered, however, by placing the particle at the boundary between
two spectral domains. Functions are then smooth in each domain, and 
discontinuities are handled by formulating appropriate boundary
conditions; spectral convergence is thereby achieved. 

\subsection{Effective-source method}

The mode-sum methods reviewed in the preceding subsection have been
developed and applied extensively, but they do not exhaust the range
of approaches that may be exploited to compute a self-force. Another 
set of methods, devised by Barack and his collaborators
\cite{barack-golbourn:07, barack-etal:07, dolan-barack:11} as well as
Vega and his collaborators
\cite{vega-detweiler:08,vega-etal:09,vegaPhD}, begin by  
recognizing that an approximation to the exact singular potential can
be used to regularize the delta-function source term of the original
field equation. We shall explain this idea in the simple context of a
scalar potential $\Phi$.  

We continue to write the wave equation for the retarded potential
$\Phi$ in the schematic form  
\begin{equation}
\Box \Phi = q \delta(x,z),
  \label{eqn:schematic}
\end{equation}
where $\Box$ is the wave operator in curved spacetime, and
$\delta(x,z)$ is a distributional source term that depends on the
particle's world line $\gamma$ through its coordinate representation
$z(\tau)$. By construction, the exact singular potential $\Phi_{\rm S}$
satisfies the same equation, and an approximation to the singular
potential, denoted $\tilde{\Phi}_{\rm S}$, will generally satisfy an
equation of the form 
\begin{equation}
\Box \tilde{\Phi}_{\rm S} = q \delta(x,x_0) + O(r^n) 
\end{equation}
for some integer $n > 0$, where $r$ is a measure of distance to the
world line. A ``better'' approximation to the singular potential is
one with a higher value of $n$.  From the approximated singular
potential we form an approximation to the regular potential by writing  
\begin{equation} 
\tilde{\Phi}_{\rm R} := \Phi - W \tilde{\Phi}_{\rm S}, 
\end{equation} 
where $W$ is a window function whose properties will be specified
below. The approximated regular potential is governed by the wave 
equation 
\begin{equation}
\Box \tilde{\Phi}_{\rm R} = q \delta(x,z) 
- \Box \bigl( W \tilde{\Phi}_{\rm S} \bigr) := S(x,z),  
\end{equation}
and the right-hand side of this equation defines the effective source 
term $S(x,z)$. This equation is much less singular than
Eq.~(\ref{eqn:schematic}), and it can be integrated using 
numerical methods designed to handle smooth functions.  

To see this, we write the effective source more specifically as 
\begin{equation}
S(x,z) = -\tilde{\Phi}_{\rm S} \Box W 
- 2\nabla_\alpha W\nabla^\alpha \tilde{\Phi}_{\rm S}  
- W \Box\tilde{\Phi}_{\rm S} + q\delta(x,z).
\end{equation}
With the window function $W$ designed to approach unity as $x \to z$,
we find that the delta function that arises from the third term on the
right-hand side precisely cancels out the fourth term. To keep the
other terms in $S$ well behaved on the world line, we further 
restrict the window function to satisfy $\nabla_\alpha W =
O(r^p)$ with $p \geq 2$; this ensures that multiplication by 
$\nabla_\alpha \tilde{\Phi}_{\rm S} = O(r^{-2})$ leaves behind a
bounded quantity. In addition, we demand that $\Box W = O(r^q)$ with     
$q \geq 1$, so that multiplication by $\tilde{\Phi}_{\rm S} 
= O(r^{-1})$ again produces a bounded quantity. It is also useful to
require that $W(x)$ have compact (spatial) support, to ensure that the
effective source term $S(x,z)$ does not extend beyond a reasonably
small neighbourhood of the world line; this property also has the
virtue of making $\tilde{\Phi}_{\rm R}$ precisely equal to the
retarded potential $\Phi$ outside the support of the window
function. This implies, in particular, that $\tilde{\Phi}_{\rm R}$ can
be used directly to compute radiative fluxes at infinity. Another
considerable virtue of these specifications for the window function is
that they guarantee that the gradient of $\tilde{\Phi}_{\rm R}$ is
directly tied to the self-force. We indeed see that   
\begin{align}
  \lim_{x \rightarrow z}\nabla_\alpha \tilde{\Phi}_{\rm R} 
&= \lim_{x\rightarrow z} \bigl( \nabla_\alpha \Phi 
- W \nabla_\alpha \tilde{\Phi}_{\rm S} \bigr)  
  - \lim_{x\rightarrow z}\tilde{\Phi}_{\rm S} \nabla_\alpha W 
\nonumber \\ &= 
\lim_{x\rightarrow z} \bigl( \nabla_\alpha \Phi 
- \nabla_\alpha \tilde{\Phi}_{\rm S} \bigr) 
\nonumber \\ &= 
q^{-1} F_\alpha, 
\label{<++>}
\end{align}
with the second line following by virtue of the imposed conditions on
$W$, and the third line from the properties of the approximated
singular field.  

The effective-source method therefore consists of integrating the wave
equation 
\begin{equation}
\Box \tilde{\Phi}_{\rm R} = S(x,z),
\label{eqn:effwave}
\end{equation}
for the approximated regular potential $\tilde{\Phi}_{\rm R}$, with
a source term $S(x,z)$ that has become a regular function (of limited
differentiability) of the spacetime coordinates $x$. The method is
also known as a ``puncture approach,'' in reference to a similar
regularization strategy employed in numerical relativity. It is well
suited to a $3+1$ integration of the wave equation, which can be
implemented on mature codes already in   
circulation within the numerical-relativity community. An important
advantage of a $3+1$ implementation is that it is largely indifferent
to the choice of background spacetime, and largely insensitive to the
symmetries possessed by this spacetime; a self-force in Kerr spacetime
is in principle just as easy to obtain as a self-force in Schwarzschild
spacetime. 

The method is also well suited to a self-consistent implementation of
the self-force, in which the motion of the particle is not fixed in
advance, but determined by the action of the computed self-force. 
This amounts to combining Eq.~(\ref{eqn:effwave}) with the self-force 
equation  
\begin{equation}
m\frac{Du^\mu}{d\tau} = q \bigl( g^{\mu\nu}
+u^\mu u^\nu \bigr)\nabla_\nu \tilde{\Phi}_{\rm R}, 
\end{equation}
in which the field is evaluated on the dynamically determined world
line. The system of equations is integrated jointly, and
self-consistently. The $3+1$ version of the effective-source approach
presents a unique opportunity for the numerical-relativity community
to get involved in self-force computations, with only a minimal amount
of infrastructure development. This was advocated by Vega and
Detweiler \cite{vega-detweiler:08}, who first demonstrated the
viability of the approach with a $1+1$ time-domain code for a scalar
charge on a circular orbit around a Schwarzschild black hole. An
implementation with two separate $3+1$ codes imported from numerical
relativity was also accomplished \cite{vega-etal:09}.  

The work of Barack and collaborators
\cite{barack-golbourn:07, barack-etal:07} is a particular
implementation of the effective-source approach in a $2+1$ numerical
calculation of the scalar self-force in Kerr spacetime. (See also the
independent implementation by Lousto and Nakano
\cite{lousto-nakano:08}.) Instead of starting with
Eq.~(\ref{eqn:schematic}), they first decompose $\Phi$ according to 
\begin{equation}
\Phi(x) = \sum_m \Phi^m(t,r,\theta) \exp(im\phi) 
\end{equation}
and formulate reduced wave equations for the Fourier coefficients
$\Phi^m$. Each coefficient is then regularized with an 
appropriate singular field $\tilde{\Phi}^m_{\rm S}$, which eliminates
the delta-funtion from Eq.~(\ref{eqn:schematic}). This gives rise to
regularized source terms for the reduced wave equations,
which can then be integrated with a $2+1$ evolution code. In the final
stage of the computation, the self-force is recovered by summing over
the regularized Fourier coefficients. This strategy, known as the 
\emph{$m$-mode regularization scheme}, is currently under active 
development. Recently it was successfully applied by Dolan and Barack
\cite{dolan-barack:11} to compute the self-force on a scalar charge in
circular orbit around a Schwarzschild black hole.

\subsection{Quasilocal approach with ``matched expansions''} 

As was seen in Eqs.~(\ref{1.7.1}), (\ref{1.8.7}), and (\ref{1.9.6}),
the self-force can be expressed as an integral over the past world
line of the particle, the integrand involving the Green's function for
the appropriate wave equation. Attempts have been made to compute the
Green's function directly \cite{dewitt-dewitt:64, pfenning-poisson:02,
  burko-etal:02, haas-poisson:05}, and to evaluate the world-line
integral. The quasilocal approach, first introduced by Anderson and
his collaborators \cite{anderson-hu:04,anderson-etal:06,
  anderson-wiseman:05, anderson-etal:05}, is based on the expectation
that the world-line integral might be dominated by the particle's recent
past, so that the Green's function can be represented by its Hadamard 
expansion, which is restricted to the normal convex neighbourhood of
the particle's current position. To help with this enterprise,
Ottewill and his collaborators \cite{ottewill-wardell:08, wardellPhD, 
ottewill-wardell:09, casals-etal:09b} have pushed the Hadamard
expansion to a very high order of accuracy, building on earlier work
by D\'ecanini and Folacci \cite{decanini-folacci:06}. 

The weak-field calculations performed by DeWitt and DeWitt
\cite{dewitt-dewitt:64} and Pfenning and Poisson
\cite{pfenning-poisson:02} suggest that the world-line integral is
not, in fact, dominated by the recent past. Instead, most of the
self-force is produced by signals that leave the particle at some
time in the past, scatter off the central mass, and reconnect with the
particle at the current time; such signals mark the boundary of the
normal convex neighbourhood. The quasilocal evaluation of the
world-line integral must therefore be supplemented with contributions
from the distant past, and this requires a representation of the
Green's function that is not limited to the normal convex
neighbourhood. In some spacetimes it is possible to express the
Green's function as an expansion in quasi-normal modes, as was  
demonstrated by Casals and his collaborators for a static scalar
charge in the Nariai spacetime \cite{casals-etal:09a}.  Their study
provided significant insights into the geometrical structure of
Green's functions in curved spacetime, and increased our understanding
of the non-local character of the self-force.    

\subsection{Adiabatic approximations}
\label{subsec:adiabatic} 

The accurate computation of long-term waveforms from
extreme-mass-ratio inspirals (EMRIs) involves a lengthy sequence of 
calculations that include the calculation of the self-force. One can
already imagine the difficulty of numerically integrating the coupled
linearized Einstein equation for durations much longer than has ever
been attempted by existing numerical codes. While doing so, the code
would also have to evaluate the self-force reasonably often (if not at
each time step) in order to remain close to the true dynamics of the
point mass. Moreover, gravitational-wave data analysis via matched
filtering require full evolutions of the sort just described for a
large sample of systems parameters. All these considerations underlie
the desire for simplified approximations to fully self-consistent
self-force EMRI models. 
 
The \emph{adiabatic approximation} refers to one such class of
potentially useful approximations.  The basic assumption is that the
secular effects of the self-force occur on a timescale that is
much longer than the orbital period. In an extreme-mass-ratio
binary, this assumption is valid during the early stage of inspiral;
it breaks down in the final moments, when the orbit transitions to
a quasi-radial infall called the plunge. From the adiabaticity
assumption, numerous approximations have been formulated: For example, 
(i) since the particle's orbit deviates only slowly from geodesic
motion, the self-force can be calculated from a field sourced by a
geodesic; (ii) since the radiation-reaction timescale $t_{rr}$, over
which a significant shrinking of the orbit occurs due to the
self-force, is much longer than the orbital period, periodic effects
of the self-force can be neglected; and (iii) conservative effects of
the self-force can be neglected (the \emph{radiative approximation}). 

A seminal example of an adiabatic approximation is the Peters-Mathews 
formalism \cite{peters-mathews:63, peters:64}, which determines the
long-term evolution of a binary orbit by equating the time-averaged
rate of change of the orbital energy $E$ and angular momentum $L$ to,
respectively, the flux of gravitational-wave energy and angular
momentum at infinity.  This formalism was used to successfully predict
the decreasing orbital period of the Hulse-Taylor pulsar, before more
sophisticated methods, based on post-Newtonian equations of motion
expanded to 2.5{\sc pn} order, were incorporated in times-of-arrival
formulae.   

In the hope of achieving similar success in the context of the
self-force, considerable work has been done to formulate a similar 
approximation for the case of an extreme-mass-ratio inspiral 
\cite{mino:03, mino:05, mino:06,
  hughes:00, drasco-etal:05,
  drasco-hughes:06,sago-etal:05,sago-etal:06, ganz-etal:07,
  mino-price:08, hinderer-flanagan:08}. Bound geodesics in Kerr
spacetime are specified by three constants of motion --- the energy
$E$, angular momentum $L$, and Carter constant $C$. If one
could easily calculate the rates of change of these quantities, using
a method analogous to the Peters-Mathews formalism, then one could
determine an approximation to the long-term orbital evolution of the
small body in an EMRI, avoiding the lengthy process of regularization
involved in the direct integration of the self-forced equation of
motion. In the early 1980s, Gal'tsov \cite{galtsov:82}  
showed that the average rates of change of $E$ and $L$, as calculated 
from balance equations that assume geodesic source motion, agree with 
the averaged rates of change induced by a self-force constructed from
a radiative Green's function defined as
$G_{\text{rad}} := \tfrac{1}{2}(G_- - G_+)$. As
discussed in Sec.~\ref{1.4}, this is equal to the regular two-point
function $G_{\rm R}$ in flat spacetime, but 
$G_{\rm rad} \neq G_{\rm R}$ in curved spacetime; because of its
time-asymmetry, it is purely dissipative. Mino \cite{mino:03}, 
building on the work of Gal'tsov, was able to show that the true
self-force and the dissipative force constructed from $G_{\rm rad}$
cause the same averaged rates of change of all three constants of
motion, lending credence to the radiative approximation. Since then,
the radiative Green's function was used to derive explicit expressions
for the rates of change of $E$, $L$, and $C$ in terms of the
particle's orbit and wave amplitudes at infinity \cite{sago-etal:05,
  sago-etal:06, ganz-etal:07}, and radiative approximations based on
such expressions have been concretely implemented by Drasco, Hughes
and their collaborators
\cite{hughes-etal:05,drasco-etal:05,drasco-hughes:06}. 

The relevance of the conservative part of the self-force --- the part  
left out when using $G_{\rm rad}$ --- was analyzed in numerous recent
publications 
\cite{burko:01, pound-etal:05, pound-poisson:08a, pound-poisson:08b, 
  hinderer-flanagan:08, huerta-gair:09}. As was shown by Pound
{\it et al.}~\cite{pound-etal:05, pound-poisson:08a,
pound-poisson:08b}, neglect of the conservative effects of the
self-force generically leads to long-term errors in the phase of an
orbit and the gravitational wave it produces. These phasing errors
are due to orbital precession and a direct shift in orbital
frequency.  This shift can be understood by considering a conservative
force acting on a circular orbit: the force is radial, it alters the
centripetal acceleration, and the frequency associated with a given
orbital radius is affected. Despite these errors, a radiative
approximation may still suffice for gravitational-wave detection 
\cite{hinderer-flanagan:08}; for circular orbits, which have minimal 
conservative effects, radiative approximations may suffice even for
parameter-estimation \cite{huerta-gair:09}.  However, at this point in 
time, these analyses remain inconclusive because they
all rely on extrapolations from post-Newtonian results for the
conservative part of the self-force.  For a more comprehensive
discussion of these issues, the reader is referred to
Ref.~\cite{hinderer-flanagan:08, pound:10c}.

Hinderer and Flanagan performed the most comprehensive study of these
issues \cite{flanagan-hinderer:10}, utilizing a two-timescale expansion
\cite{kevorkian-cole:96,pound:10b} of the field equations and
self-forced equations of motion in an EMRI. In this method, all
dynamical variables are written in terms of two time coordinates: a
fast time $t$ and a slow time $\tilde t:=(m/M) t$. In the case of
an EMRI, the dynamical variables are the metric and the phase-space
variables of the world line. The fast-time dependence captures
evolution on the orbital timescale $\sim M$, while the
slow-time dependence captures evolution on the radiation-reaction
timescale $\sim M^2/m$. One obtains a sequence of
fast-time and slow-time equations by expanding the full equations in
the limit of small $m$ while treating the two time coordinates as
independent. Solving the leading-order fast-time 
equation, in which $\tilde t$ is held fixed, yields a metric
perturbation sourced by a geodesic, as one would expect from the
linearized field equations for a point particle. The leading-order
effects of the self-force are incorporated by solving the slow-time
equation and letting $\tilde t$ vary. (Solving the next-higher-order
slow-time equation determines similar effects, but also the
backreaction that causes the parameters of the large black hole to
change slowly.)

Using this method, Hinderer and Flanagan identified the effects of the
various pieces of the self-force. To describe this we write the
self-force as 
\begin{equation}
f^\mu=\frac{m}{M} \Bigl( f^\mu_{(1){\rm rr}} 
+ f^\mu_{(1){\rm c}} \Bigr) +\frac{m^2}{M^2} \Bigl(f^\mu_{(2){\rm rr}}
+ f^\mu_{(2){\rm c}} \Bigr)+\cdots,
\end{equation}
where `rr' denotes a radiation-reaction, or dissipative, piece of the 
force, and `c' denotes a conservative piece. Hinderer and Flanagan's
principal result is a formula for the orbital phase (which directly
determines the phase of the emitted gravitational waves) in terms of
these quantities:
\begin{equation}
\phi = \frac{M^2}{m}\left(\phi^{(0)}(\tilde
t)+\frac{m}{M}\phi^{(1)}(\tilde t)+\cdots\right),
\label{eq:phase} 
\end{equation}
where $\phi^{(0)}$ depends on an averaged piece of 
$f^\mu_{(1){\rm rr}}$, while $\phi^{(1)}$ depends on 
$f^\mu_{(1){\rm c}}$, the oscillatory piece of $f^\mu_{(1){\rm rr}}$,
and the averaged piece of $f^\mu_{(2){\rm rr}}$. From this result, we
see that the radiative approximation yields the leading-order phase,
but fails to determine the first subleading correction. We also see
that the approximations (i)--(iii) mentioned above are consistent (so
long as the parameters of the `geodesic' source are allowed to vary
slowly) at leading order in the two-timescale expansion, but diverge
from one another beyond that order. Hence, we see
that an adiabatic approximation is generically insufficient to extract
parameters from a waveform, since doing so requires a description of
the inspiral accurate up to small (i.e., smaller than order-1)
errors. But we also see that an adiabatic approximation based on the
radiative Green's function may be an excellent approximation for other
purposes, such as detection.  

To understand this result, consider the following naive analysis of a
quasicircular EMRI --- that is, an orbit that would be circular were it
not for the action of the self-force, and which is slowly spiraling
into the large central body. We write the orbital frequency as
$\omega^{(0)}(E)+(m/M)\omega^{(1)}_1(E)+\cdots$, where
$\omega^{(0)}(E)$ is the frequency as a function of energy  
on a circular geodesic, and
$(m/M)\omega^{(1)}_1(E)$ is the correction to this due to the
conservative part of the first-order self-force (part of the
correction also arises due to oscillatory zeroth-order effects
combining with oscillatory first-order effects, but for simplicity we
ignore this contribution). Neglecting oscillatory effects, we write
the energy in terms only of its slow-time dependence: 
$E=E^{(0)}(\tilde t) + (m/M) E^{(1)}(\tilde t)+\cdots$. The
leading-order term $E^{(0)}$ is determined by the dissipative part of
first-order self-force, while $E^{(1)}$ is determined by both the
dissipative part of the second-order force and a combination of
conservative and dissipative parts of the first-order
force. Substituting this into the frequency, we arrive at
\begin{equation}
\omega = \omega^{(0)}(E^{(0)})
+ \frac{m}{M} \left[\omega^{(1)}_1(E^{(0)})
+ \omega^{(1)}_2(E^{(0)},E^{(1)})\right]+\cdots,
\end{equation}
where $\omega^{(1)}_2=E^{(1)} \partial \omega^{(0)}/\partial E$, in 
which the partial derivative is evaluated at 
$E = E^{(0)}$. Integrating this over a radiation-reaction time, we
arrive at the orbital phase of Eq.~(\ref{eq:phase}). (In a complete
description, $E(t)$ will have oscillatory pieces, which are functions
of $t$ rather than $\tilde t$, and one must know these in order to
correctly determine $\phi^{(1)}$.) Such a result remains valid even
for generic orbits, where, for example, orbital precession due to the
conservative force contributes to the analogue of $\omega^{(1)}_1$.

\subsection{Physical consequences of the self-force}
\label{sec:physical}

To be of relevance to gravitational-wave astronomy, the paramount goal 
of the self-force community remains the computation of waveforms
that properly encode the long-term dynamical evolution of an
extreme-mass-ratio binary. This requires a fully consistent
orbital evolution fed to a wave-generation formalism, and to this day
the completion of this program remains as a future challenge. In the
meantime, a somewhat less ambitious, though no less compelling,
undertaking is that of probing the physical consequences of the
self-force on the motion of point particles. 

\subsubsection*{Scalar charge in cosmological spacetimes} 

The intriguing phenomenon of a scalar charge changing its rest mass
because of an interaction with its self-field was studied by Burko,
Harte, and Poisson \cite{burko-etal:02} and Haas and Poisson
\cite{haas-poisson:05} in the simple context of a particle at rest in
an expanding universe. The scalar Green's function could be computed 
explicitly for a wide class of cosmological spacetimes, and the action
of the field on the particle determined without approximations. It is
found that for certain cosmological models, the mass decreases and
then increases back to its original value. For other models, the mass
is restored only to a fraction of its original value. For de Sitter
spacetime, the particle radiates all of its rest mass into monopole
scalar waves. 

\subsubsection*{Physical consequences of the gravitational self-force}  

The earliest calculation of a gravitational self-force was performed
by Barack and Lousto for the case of a point mass plunging radially
into a Schwarzschild black hole \cite{barack-lousto:02}. The
calculation, however, depended on a specific choice of gauge and did
not identify unambiguous physical consequences of the self-force. To
obtain such consequences, it is necessary to combine the self-force
(computed in whatever gauge) with the metric perturbation
(computed in the same gauge) in the calculation of a 
well-defined observable that could in principle be measured. For
example, the conservative pieces of the self-force and metric
perturbation can be combined to calculate the shifts in orbital 
frequencies that originate from the gravitational effects of the small
body; an application of such a calculation would be to determine the
shift (as measured by frequency) in the innermost stable circular
orbit of an extreme-mass-ratio binary, or the shift in the rate of
periastron advance for eccentric orbits. Such calculations, however,
must exclude all dissipative aspects of the self-force, because these
introduce an inherent ambiguity in the determination of orbital
frequencies.    

A calculation of this kind was recently achieved by Barack and Sago  
\cite{barack-sago:09, barack-sago:10}, who computed the shift in the   
innermost stable circular orbit of a Schwarzschild black hole caused
by the conservative piece of the gravitational self-force. The shift in
orbital radius is gauge dependent (and was reported in the Lorenz
gauge by Barack and Sago), but the shift in orbital frequency is
measurable and therefore gauge invariant. Their main result --- a
genuine milestone in self-force computations --- is that the
fractional shift in frequency is equal to $0.4870 m/M$; the frequency
is driven upward by the gravitational self-force. Barack and Sago
compare this shift to the ambiguity created by the dissipative piece
of the self-force, which was previously investigated by 
Ori and Thorne \cite{ori-thorne:00} and Sundararajan
\cite{sundararajan:08}; they find that the conservative shift is very
small compared with the dissipative ambiguity. In a follow-up
analysis, Barack, Damour, and Sago \cite{barack-etal:10} computed the  
conservative shift in the rate of periastron advance of slightly
eccentric orbits in Schwarzschild spacetime. 

Conservative shifts in the innermost stable circular orbit of a
Schwarzschild black hole were first obtained in the context of the
scalar self-force by Diaz-Rivera 
{\it et al.}~\cite{diazrivera-etal:04}; in this case they obtain a 
fractional shift of $0.0291657q^2/(mM)$, and here also the frequency
is driven upward. 

\subsubsection*{Detweiler's redshift factor} 

In another effort to extract physical consequences from the
gravitational self-force on a particle in circular motion in
Schwarzschild spacetime, Detweiler discovered \cite{detweiler:08} that
$u^t$, the time component of the velocity vector in Schwarzschild
coordinates, is invariant with respect to a class of gauge
transformations that preserve the helical symmetry of the perturbed
spacetime. Detweiler further showed that $1/u^t$ is an observable: it is
the redshift that a photon suffers when it propagates from the
orbiting body to an observer situated at a large distance on the
orbital axis. This gauge-invariant quantity can be calculated together
with the orbital frequency $\Omega$, which is a second gauge-invariant
quantity that can be constructed for circular orbits in Schwarzschild
spacetime. Both $u^t$ and $\Omega$ acquire corrections of fractional
order $m/M$ from the self-force and the metric perturbation. While the
functions $u^t(r)$ and $\Omega(r)$ are still gauge dependent, because
of the dependence on the radial coordinate $r$, elimination of $r$
from these relations permits the construction of $u^t(\Omega)$, which
is gauge invariant. A plot of $u^t$ as a function of $\Omega$
therefore contains physically unambiguous information about the
gravitational self-force.  

The computation of the gauge-invariant relation $u^t(\Omega)$ opened
the door to a detailed comparison between the predictions of the
self-force formalism to those of post-Newtonian theory. This was first
pursued by Detweiler \cite{detweiler:08}, who compared $u^t(\Omega)$
as determined accurately through second post-Newtonian order, to
self-force results obtained numerically; he reported full consistency
at the expected level of accuracy. This comparison was pushed to the
third post-Newtonian order \cite{blanchet-etal:10a, blanchet-etal:10b,
  damour:10, barack-etal:10}. Agreement is remarkable, and it conveys 
a rather deep point about the methods of calculation. The computation
of $u^t(\Omega)$, in the context of both the self-force and
post-Newtonian theory, requires regularization of the metric
perturbation created by the point mass. In the self-force calculation,
removal of the singular field is achieved with the Detweiler-Whiting
prescription, while in post-Newtonian theory it is performed with a
very different prescription based on dimensional regularization. Each
prescription could have returned a different regularized field, and
therefore a different expression for $u^t(\Omega)$. This, remarkably,
does not happen; the singular fields are ``physically the same'' in
the self-force and post-Newtonian calculations.      

A generalization of Detweiler's redshift invariant to eccentric orbits
was recently proposed and computed by Barack and Sago
\cite{barack-sago:11}, who report consistency with corresponding
post-Newtonian results in the weak-field regime. They also computed
the influence of the conservative gravitational self-force on the
periastron advance of slightly eccentric orbits, and compared their
results with full numerical relativity simulations for modest
mass-ratio binaries. Thus, in spite of the unavailability of
self-consistent waveforms, it is becoming clear that self-force
calculations are already proving to be of value: they inform
post-Newtonian calculations and serve as benchmarks for numerical 
relativity.

%% file: part1.tex
%
\section{Synge's world function}
\label{2}

\subsection{Definition} 
\label{2.1}

In this and the following sections we will construct a number of 
{\it bitensors}, tensorial functions of two points in spacetime. The
first is $x'$, which we call the ``base point'', and to which
we assign indices $\alpha'$, $\beta'$, etc. The second is $x$, which
we call the ``field point'', and to which we assign indices
$\alpha$, $\beta$, etc. We assume that $x$ belongs to ${\cal N}(x')$,
the {\it normal convex neighbourhood} of $x'$; this is the set of
points that are linked to $x'$ by a {\it unique} geodesic. The
geodesic segment $\beta$ that links $x$ to $x'$ is described by relations
$z^\mu(\lambda)$ in which $\lambda$ is an affine parameter that ranges 
from $\lambda_0$ to $\lambda_1$; we have $z(\lambda_0) := x'$ and
$z(\lambda_1) := x$. To an arbitrary point $z$ on the geodesic we
assign indices $\mu$, $\nu$, etc. The vector $t^\mu = dz^\mu/d\lambda$
is tangent to the geodesic, and it obeys the geodesic equation 
$D t^\mu/d\lambda = 0$. The situation is illustrated in Fig.~5. 

\begin{figure}[b]
\begin{center}
\vspace*{-20pt} 
\includegraphics[width=0.5\linewidth]{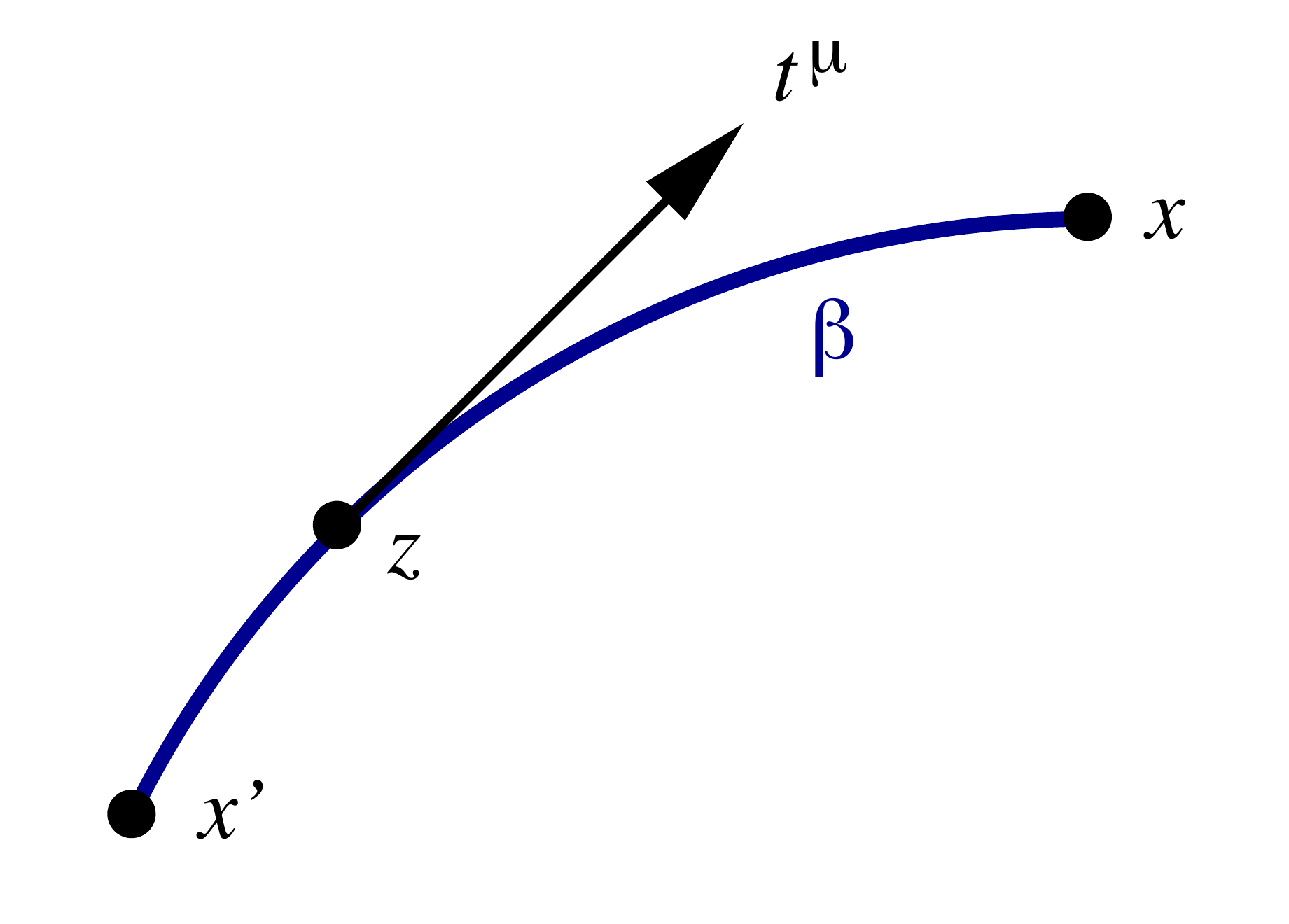}
\vspace*{-20pt}
\end{center} 
\caption{The base point $x'$, the field point $x$, and the geodesic
segment $\beta$ that links them. The geodesic is described by parametric
relations $z^\mu(\lambda)$ and $t^\mu = dz^\mu/d\lambda$ is its
tangent vector.}
\end{figure} 
 
Synge's world function is a scalar function of the base point $x'$ and
the field point $x$. It is defined by 
\begin{equation}
\sigma(x,x') = \frac{1}{2} (\lambda_1 - \lambda_0)
\int_{\lambda_0}^{\lambda_1} g_{\mu\nu}(z) 
t^\mu t^\nu\, d\lambda, 
\label{2.1.1}
\end{equation}       
and the integral is evaluated on the geodesic $\beta$ that links $x$
to $x'$. You may notice that $\sigma$ is invariant under a constant
rescaling of the affine parameter, $\lambda \to \bar{\lambda} = a
\lambda + b$, where $a$ and $b$ are constants. 

By virtue of the geodesic equation, the quantity $\varepsilon :=
g_{\mu\nu} t^\mu t^\nu$ is constant on the geodesic. The world
function is therefore numerically equal to $\frac{1}{2} \varepsilon
(\lambda_1-\lambda_0)^2$. If the geodesic is timelike, then $\lambda$
can be set equal to the proper time $\tau$, which implies that
$\varepsilon = -1$ and $\sigma = -\frac{1}{2} (\Delta \tau)^2$. If the
geodesic is spacelike, then $\lambda$ can be set equal to the proper
distance $s$, which implies that $\varepsilon = 1$ and $\sigma =
\frac{1}{2} (\Delta s)^2$. If the geodesic is null, then $\sigma =
0$. Quite generally, therefore, the world function is half the squared
geodesic distance between the points $x'$ and $x$.  

In flat spacetime, the geodesic linking $x$ to $x'$ is a straight
line, and $\sigma = \frac{1}{2} \eta_{\alpha\beta} (x-x')^\alpha
(x-x')^\beta$ in Lorentzian coordinates.  

\subsection{Differentiation of the world function} 
\label{2.2}

The world function $\sigma(x,x')$ can be differentiated with respect
to either argument. We let $\sigma_\alpha = \partial \sigma / \partial
x^\alpha$ be its partial derivative with respect to $x$, and
$\sigma_{\alpha'} = \partial \sigma / \partial x^{\alpha'}$ its
partial derivative with respect to $x'$. It is clear that
$\sigma_\alpha$ behaves as a dual vector with respect to tensorial 
operations carried out at $x$, but as a scalar with respect to
operations carried out $x'$. Similarly, $\sigma_{\alpha'}$ is a
scalar at $x$ but a dual vector at $x'$. 

We let $\sigma_{\alpha\beta} := \nabla_\beta \sigma_{\alpha}$ be
the covariant derivative of $\sigma_{\alpha}$ with respect to $x$;
this is a rank-2 tensor at $x$ and a scalar at $x'$. Because $\sigma$
is a scalar at $x$, we have that this tensor is symmetric:
$\sigma_{\beta\alpha} = \sigma_{\alpha\beta}$. Similarly, we let
$\sigma_{\alpha\beta'} := \partial_{\beta'} \sigma_{\alpha} = 
\partial^2 \sigma / \partial x^{\beta'} \partial x^\alpha$ be the
partial derivative of $\sigma_{\alpha}$ with respect to $x'$; this
is a dual vector both at $x$ and $x'$. We can also define 
$\sigma_{\alpha'\beta} := \partial_\beta \sigma_{\alpha'} =
\partial^2 \sigma / \partial x^{\beta} \partial x^{\alpha'}$ to be the
partial derivative of $\sigma_{\alpha'}$ with respect to $x$. Because
partial derivatives commute, these bitensors are equal:
$\sigma_{\beta'\alpha} = \sigma_{\alpha\beta'}$. Finally, we let
$\sigma_{\alpha'\beta'} := \nabla_{\beta'} \sigma_{\alpha'}$ be
the covariant derivative of $\sigma_{\alpha'}$ with respect to $x'$;
this is a symmetric rank-2 tensor at $x'$ and a scalar at $x$. 

The notation is easily extended to any number of derivatives. For
example, we let $\sigma_{\alpha\beta\gamma\delta'} := 
\nabla_{\delta'} \nabla_\gamma \nabla_\beta \nabla_\alpha \sigma$, 
which is a rank-3 tensor at $x$ and a dual vector at $x'$. This
bitensor is symmetric in the pair of indices $\alpha$ and $\beta$, but
not in the pairs $\alpha$ and $\gamma$, nor $\beta$ and
$\gamma$. Because $\nabla_{\delta'}$ is here an ordinary partial
derivative with respect to $x'$, the bitensor is symmetric in any pair
of indices involving $\delta'$. The ordering of the primed index
relative to the unprimed indices is therefore irrelevant: the same
bitensor can be written as $\sigma_{\delta'\alpha\beta\gamma}$ or
$\sigma_{\alpha\delta'\beta\gamma}$ or
$\sigma_{\alpha\beta\delta'\gamma}$, making sure that the ordering of
the unprimed indices is not altered.    
       
More generally, we can show that derivatives of any bitensor
$\Omega_{\cdots}(x,x')$ satisfy the property  
\begin{equation}
\Omega_{\cdots ; \beta \alpha' \cdots} 
= \Omega_{\cdots ; \alpha' \beta \cdots}, 
\label{2.2.1} 
\end{equation}
in which ``$\cdots$'' stands for any combination of primed and
unprimed indices. We start by establishing the symmetry of
$\Omega_{\cdots ;\alpha \beta'}$ with respect to the pair $\alpha$ and  
$\beta'$. This is most easily done by adopting Fermi normal
coordinates (see Sec.~\ref{8}) adapted to the geodesic $\beta$ and 
setting the connection to zero both at $x$ and $x'$. In these
coordinates, the bitensor $\Omega_{\cdots ;\alpha}$ is the partial
derivative of $\Omega_{\cdots}$ with respect to $x^\alpha$, and 
$\Omega_{\cdots ;\alpha\beta'}$ is obtained by taking an additional
partial derivative with respect to $x^{\beta'}$. These two operations
commute, and $\Omega_{\cdots ;\beta' \alpha} = 
\Omega_{\cdots ;\alpha \beta'}$ follows as a bitensorial
identity. Equation (\ref{2.2.1}) then follows by further
differentiation with respect to either $x$ or $x'$.     

The message of Eq.~(\ref{2.2.1}), when applied to derivatives of the 
world function, is that while the ordering of the primed and
unprimed indices relative to themselves is important, their 
ordering with respect to each other is arbitrary. For example, 
$\sigma_{\alpha'\beta'\gamma\delta'\epsilon} =
\sigma_{\alpha'\beta'\delta'\gamma\epsilon} =
\sigma_{\gamma\epsilon\alpha'\beta'\delta'}$. 
 
\subsection{Evaluation of first derivatives} 
\label{2.3}

We can compute $\sigma_\alpha$ by examining how $\sigma$ varies when    
the field point $x$ moves. We let the new field point be $x +
\delta x$, and $\delta \sigma := \sigma(x+\delta x,x') -
\sigma(x,x')$ is the corresponding variation of the world function. We
let $\beta + \delta \beta$ be the unique geodesic segment that links $x + 
\delta x$ to $x'$; it is described by relations $z^\mu(\lambda) 
+ \delta z^\mu(\lambda)$, in which the affine parameter is scaled in
such a way that it runs from $\lambda_0$ to $\lambda_1$ also on the
new geodesic. We note that $\delta z (\lambda_0) = \delta x' = 0$
and $\delta z (\lambda_1) = \delta x$.  

Working to first order in the variations, Eq.~(\ref{2.1.1}) implies   
\[ 
\delta \sigma = \Delta \lambda \int_{\lambda_0}^{\lambda_1} 
\biggl( g_{\mu\nu} \dot{z}^\mu\, \delta\dot{z}^\nu + \frac{1}{2}\, 
g_{\mu\nu,\lambda} \dot{z}^\mu \dot{z}^\nu\, \delta z^\lambda
\biggr)\, d\lambda, 
\]
where $\Delta \lambda = \lambda_1 - \lambda_0$, an overdot
indicates differentiation with respect to $\lambda$, and the metric
and its derivatives are evaluated on $\beta$. Integrating the 
first term by parts gives 
\[ 
\delta \sigma = \Delta \lambda \Bigl[ g_{\mu\nu}
\dot{z}^\mu\, \delta z^\nu \Bigr]^{\lambda_1}_{\lambda_0} \\
- \Delta \lambda \int_{\lambda_0}^{\lambda_1} 
\Bigl( g_{\mu\nu} \ddot{z}^\nu + \Gamma_{\mu\nu\lambda}  
\dot{z}^\nu \dot{z}^\lambda \Bigr)\, \delta z^\mu\, d\lambda. 
\] 
The integral vanishes because $z^\mu(\lambda)$ satisfies the geodesic 
equation. The boundary term at $\lambda_0$ is zero because the 
variation $\delta z^\mu$ vanishes there. We are left with 
$\delta \sigma = \Delta \lambda g_{\alpha\beta} t^\alpha 
\delta x^\beta$, or
\begin{equation} 
\sigma_\alpha(x,x') = (\lambda_1 - \lambda_0)\, 
g_{\alpha\beta} t^\beta,  
\label{2.3.1}
\end{equation} 
in which the metric and the tangent vector are both evaluated at
$x$. Apart from a factor $\Delta \lambda$, we see that
$\sigma^\alpha(x,x')$ is equal to the geodesic's tangent vector at
$x$. If in Eq.~(\ref{2.3.1}) we replace $x$ by a generic point
$z(\lambda)$ on $\beta$, and if we correspondingly replace
$\lambda_1$ by $\lambda$, we obtain $\sigma^\mu(z,x') = (\lambda -
\lambda_0) t^\mu$; we therefore see that $\sigma^\mu(z,x')$ is a
rescaled tangent vector on the geodesic.    

A virtually identical calculation reveals how $\sigma$ varies under a 
change of base point $x'$. Here the variation of the geodesic is such
that $\delta z (\lambda_0) = \delta x'$ and $\delta z (\lambda_1) =
\delta x = 0$, and we obtain $\delta \sigma = - \Delta \lambda 
g_{\alpha'\beta'} t^{\alpha'} \delta x^{\beta'}$. This shows that 
\begin{equation}  
\sigma_{\alpha'}(x,x') = -(\lambda_1 - \lambda_0)\,
g_{\alpha'\beta'} t^{\beta'},   
\label{2.3.2}
\end{equation} 
in which the metric and the tangent vector are both evaluated at
$x'$. Apart from a factor $\Delta \lambda$, we see that
$\sigma^{\alpha'}(x,x')$ is minus the geodesic's tangent 
vector at $x'$. 

It is interesting to compute the norm of $\sigma_{\alpha}$. According
to Eq.~(\ref{2.3.1}) we have $g_{\alpha\beta} \sigma^{\alpha}
\sigma^{\beta} = (\Delta \lambda)^2 g_{\alpha\beta} t^\alpha t^\beta =  
(\Delta \lambda)^2 \varepsilon$. According to Eq.~(\ref{2.1.1}),
this is equal to $2 \sigma$. We have obtained 
\begin{equation}
g^{\alpha\beta} \sigma_\alpha \sigma_\beta = 2 \sigma,  
\label{2.3.3}
\end{equation} 
and similarly,  
\begin{equation}
g^{\alpha'\beta'} \sigma_{\alpha'} \sigma_{\beta'} = 2 \sigma.  
\label{2.3.4} 
\end{equation} 
These important relations will be the starting point of many
computations to be described below.  

We note that in flat spacetime, $\sigma_\alpha = \eta_{\alpha\beta}
(x-x')^\beta$ and $\sigma_{\alpha'} = - \eta_{\alpha\beta} 
(x-x')^\beta$ in Lorentzian coordinates. From this it follows that
$\sigma_{\alpha\beta} = \sigma_{\alpha'\beta'} =
-\sigma_{\alpha\beta'} = -\sigma_{\alpha'\beta} = \eta_{\alpha\beta}$,
and finally, $g^{\alpha\beta} \sigma_{\alpha\beta} = 4 =
g^{\alpha'\beta'} \sigma_{\alpha'\beta'}$.   

\subsection{Congruence of geodesics emanating from $x'$} 
\label{2.4}

If the base point $x'$ is kept fixed, $\sigma$ can be considered to be
an ordinary scalar function of $x$. According to Eq.~(\ref{2.3.3}),
this function is a solution to the nonlinear
differential equation $\frac{1}{2} g^{\alpha\beta} \sigma_\alpha 
\sigma_\beta = \sigma$. Suppose that we are presented with such a
scalar field. What can we say about it? 

An additional differentiation of the defining equation reveals that
the vector $\sigma^\alpha := \sigma^{;\alpha}$ satisfies  
\begin{equation}
\sigma^\alpha_{\ ;\beta} \sigma^\beta = \sigma^\alpha, 
\label{2.4.1}
\end{equation}
which is the geodesic equation in a non-affine parameterization. The  
vector field is therefore tangent to a congruence of geodesics. The
geodesics are timelike where $\sigma < 0$, they are spacelike where
$\sigma > 0$, and they are null where $\sigma = 0$. Here, for
concreteness, we shall consider only the timelike subset of the
congruence.    

The vector 
\begin{equation}
u^\alpha = \frac{\sigma^\alpha}{|2\sigma|^{1/2}} 
\label{2.4.2}
\end{equation}
is a normalized tangent vector that satisfies the geodesic equation 
in affine-parameter form: $u^\alpha_{\ ;\beta} u^\beta = 0$. The 
parameter $\lambda$ is then proper time $\tau$. If $\lambda^*$
denotes the original parameterization of the geodesics, we have that
$d\lambda^*/d\tau = |2\sigma|^{-1/2}$, and we see that the original 
parameterization is singular at $\sigma = 0$. 

In the affine parameterization, the expansion of the congruence is
calculated to be 
\begin{equation}
\theta = \frac{\theta^*}{|2\sigma|^{1/2}}, \qquad
\theta^* := \sigma^{\alpha}_{\ ;\alpha} - 1, 
\label{2.4.3}
\end{equation} 
where $\theta^* = (\delta V)^{-1} (d/d\lambda^*) (\delta V)$ is the
expansion in the original parameterization ($\delta V$ is the
congruence's cross-sectional volume). While $\theta^*$ is well 
behaved in the limit $\sigma \to 0$ (we shall see below that $\theta^* 
\to 3$), we have that $\theta \to \infty$. This means that the point
$x'$ at which $\sigma = 0$ is a caustic of the congruence: all
geodesics emanate from this point.  

These considerations, which all follow from a postulated relation
$\frac{1}{2} g^{\alpha\beta} \sigma_\alpha \sigma_\beta = \sigma$,
are clearly compatible with our preceding explicit construction of the  
world function.            

\section{Coincidence limits} 
\label{3}

It is useful to determine the limiting behaviour of the bitensors 
$\sigma_{\cdots}$ as $x$ approaches $x'$. We introduce the notation 
\[
\bigl[ \Omega_{\cdots} \bigr] 
= \lim_{x \to x'} \Omega_{\cdots}(x,x') 
= \mbox{a tensor at $x'$}   
\]
to designate the limit of any bitensor $\Omega_{\cdots}(x,x')$ as $x$
approaches $x'$; this is called the {\it coincidence limit} of the
bitensor. We assume that the coincidence limit is a unique tensorial
function of the base point $x'$, independent of the direction in which
the limit is taken. In other words, if the limit is computed by
letting $\lambda \to \lambda_0$ after evaluating
$\Omega_{\cdots}(z,x')$ as a function of $\lambda$ on a specified
geodesic $\beta$, it is assumed that the answer does not depend on the
choice of geodesic.    

\subsection{Computation of coincidence limits} 
\label{3.1}

From Eqs.~(\ref{2.1.1}), (\ref{2.3.1}), and (\ref{2.3.2}) we already
have    
\begin{equation} 
\bigl[ \sigma \bigr] = 0, \qquad
\bigl[ \sigma_\alpha \bigr] = \bigl[ \sigma_{\alpha'} \bigr] = 0. 
\label{3.1.1}
\end{equation} 
Additional results are obtained by repeated differentiation of the 
relations (\ref{2.3.3}) and (\ref{2.3.4}). For example,
Eq.~(\ref{2.3.3}) implies $\sigma_\gamma = g^{\alpha\beta}
\sigma_\alpha \sigma_{\beta\gamma} = \sigma^\beta
\sigma_{\beta\gamma}$, or $(g_{\beta\gamma} - \sigma_{\beta\gamma})
t^\beta = 0$ after using Eq.~(\ref{2.3.1}). From the assumption stated
in the preceding paragraph, $\sigma_{\beta\gamma}$ becomes independent
of $t^\beta$ in the limit $x \to x'$, and we arrive at
$[\sigma_{\alpha\beta}] = g_{\alpha'\beta'}$. By very similar
calculations we obtain all other coincidence limits for the second
derivatives of the world function. The results are   
\begin{equation} 
\bigl[ \sigma_{\alpha\beta} \bigr] =  
\bigl[ \sigma_{\alpha'\beta'} \bigr] = 
g_{\alpha'\beta'}, \qquad
\bigl[ \sigma_{\alpha\beta'} \bigr] = 
\bigl[ \sigma_{\alpha'\beta} \bigr] = - g_{\alpha'\beta'}. 
\label{3.1.2}
\end{equation}
From these relations we infer that
$[\sigma^\alpha_{\ \alpha}] = 4$, so that $[\theta^*] = 3$, where
$\theta^*$ was defined in Eq.~(\ref{2.4.3}). 

To generate coincidence limits of bitensors involving primed 
indices, it is efficient to invoke Synge's rule, 
\begin{equation} 
\bigl[ \sigma_{\cdots \alpha'} \bigr] = \bigl[ \sigma_{\cdots}
\bigr]_{;\alpha'} - \bigl[ \sigma_{\cdots \alpha} \bigr], 
\label{3.1.3}
\end{equation} 
in which ``$\cdots$'' designates any combination of primed and
unprimed indices; this rule will be established below. For example,
according to Synge's rule we have $[\sigma_{\alpha\beta'}] =
[\sigma_\alpha]_{;\beta'} - [\sigma_{\alpha\beta}]$, and since the 
coincidence limit of $\sigma_\alpha$ is zero, this gives us  
$[\sigma_{\alpha\beta'}] = - [\sigma_{\alpha\beta}] =
-g_{\alpha'\beta'}$, as was stated in Eq.~(\ref{3.1.2}). Similarly,  
$[\sigma_{\alpha'\beta'}] = [\sigma_{\alpha'}]_{;\beta'} -
[\sigma_{\alpha'\beta}] = - [\sigma_{\beta\alpha'}] =
g_{\alpha'\beta'}$. The results of Eq.~(\ref{3.1.2}) can thus all be 
generated from the known result for $[\sigma_{\alpha\beta}]$. 

The coincidence limits of Eq.~(\ref{3.1.2}) were derived from the
relation $\sigma_\alpha = \sigma^\delta_{\ \alpha} \sigma_\delta$. We
now differentiate this twice more and obtain
$\sigma_{\alpha\beta\gamma} = 
\sigma^\delta_{\ \alpha\beta\gamma} \sigma_\delta 
+ \sigma^\delta_{\ \alpha\beta} \sigma_{\delta \gamma}
+ \sigma^\delta_{\ \alpha\gamma} \sigma_{\delta \beta} 
+ \sigma^\delta_{\ \alpha} \sigma_{\delta\beta\gamma}$. 
At coincidence we have  
\[
\bigl[ \sigma_{\alpha\beta\gamma} \bigr] = \bigl[ \sigma^\delta_{\
\alpha\beta} \bigr] g_{\delta'\gamma'} + \bigl[ \sigma^\delta_{\
\alpha\gamma} \bigr] g_{\delta'\beta'} + \delta^{\delta'}_{\ \alpha'}
\bigl[ \sigma_{\delta\beta\gamma} \bigr], 
\]
or $[\sigma_{\gamma\alpha\beta}] + [\sigma_{\beta\alpha\gamma}] = 0$
if we recognize that the operations of raising or lowering indices and 
taking the limit $x \to x'$ commute. Noting the symmetries of
$\sigma_{\alpha\beta}$, this gives us
$[\sigma_{\alpha\gamma\beta}] + [\sigma_{\alpha\beta\gamma}] = 0$, 
or $2[\sigma_{\alpha\beta\gamma}] - [R^\delta_{\ \alpha\beta\gamma}
\sigma_\delta] = 0$, or $2[\sigma_{\alpha\beta\gamma}] =
R^{\delta'}_{\ \alpha'\beta'\gamma'}[\sigma_{\delta'}]$. Since the
last factor is zero, we arrive at  
\begin{equation}
\bigl[\sigma_{\alpha\beta\gamma} \bigr] = 
\bigl[\sigma_{\alpha\beta\gamma'} \bigr] = 
\bigl[\sigma_{\alpha\beta'\gamma'} \bigr] = 
\bigl[\sigma_{\alpha'\beta'\gamma'} \bigr] = 0. 
\label{3.1.4}
\end{equation}
The last three results were derived from $[\sigma_{\alpha\beta\gamma}]  
= 0$ by employing Synge's rule. 

We now differentiate the relation $\sigma_\alpha = \sigma^\delta_{\
\alpha} \sigma_\delta$ three times and obtain 
\[ 
\sigma_{\alpha\beta\gamma\delta} = 
\sigma^\epsilon_{\ \alpha\beta\gamma\delta} \sigma_\epsilon 
+ \sigma^\epsilon_{\ \alpha\beta\gamma} \sigma_{\epsilon\delta} 
+ \sigma^\epsilon_{\ \alpha\beta\delta} \sigma_{\epsilon\gamma}  
+ \sigma^\epsilon_{\ \alpha\gamma\delta} \sigma_{\epsilon\beta}  
+ \sigma^\epsilon_{\ \alpha\beta} \sigma_{\epsilon\gamma\delta} 
+ \sigma^\epsilon_{\ \alpha\gamma} \sigma_{\epsilon\beta\delta} 
+ \sigma^\epsilon_{\ \alpha\delta} \sigma_{\epsilon\beta\gamma}
+ \sigma^\epsilon_{\ \alpha} \sigma_{\epsilon\beta\gamma\delta}. 
\]
At coincidence this reduces to $[\sigma_{\alpha\beta\gamma\delta}] 
+ [\sigma_{\alpha\delta\beta\gamma}] 
+ [\sigma_{\alpha\gamma\beta\delta}] = 0$. To simplify the third term
we differentiate Ricci's identity $\sigma_{\alpha\gamma\beta} =
\sigma_{\alpha\beta\gamma} - R^\epsilon_{\ \alpha\beta\gamma}
\sigma_\epsilon$ with respect to $x^\delta$ and then take the
coincidence limit. This gives us $[\sigma_{\alpha\gamma\beta\delta}] 
= [\sigma_{\alpha\beta\gamma\delta}] +
R_{\alpha'\delta'\beta'\gamma'}$. The same manipulations on the second
term give $[\sigma_{\alpha\delta\beta\gamma}] =
[\sigma_{\alpha\beta\delta\gamma}] +
R_{\alpha'\gamma'\beta'\delta'}$. Using the identity
$\sigma_{\alpha\beta\delta\gamma} = \sigma_{\alpha\beta\gamma\delta} 
- R^\epsilon_{\ \alpha\gamma\delta} \sigma_{\epsilon\beta}   
- R^\epsilon_{\ \beta\gamma\delta} \sigma_{\alpha\epsilon}$ and the
symmetries of the Riemann tensor, it is then easy to show that
$[\sigma_{\alpha\beta\delta\gamma}] =
[\sigma_{\alpha\beta\gamma\delta}]$. Gathering the results, we obtain 
$3 [\sigma_{\alpha\beta\gamma\delta}] +
R_{\alpha'\gamma'\beta'\delta'} + R_{\alpha'\delta'\beta'\gamma'} = 
0$, and Synge's rule allows us to generalize this to any combination
of primed and unprimed indices. Our final results are 
\begin{eqnarray} 
\bigl[ \sigma_{\alpha\beta\gamma\delta} \bigr] &=& 
- \frac{1}{3} \bigl( R_{\alpha'\gamma'\beta'\delta'} 
+ R_{\alpha'\delta'\beta'\gamma'} \bigr), 
\qquad 
\bigl[ \sigma_{\alpha\beta\gamma\delta'} \bigr] = 
\frac{1}{3} \bigl( R_{\alpha'\gamma'\beta'\delta'} 
+ R_{\alpha'\delta'\beta'\gamma'} \bigr),
\nonumber \\
\bigl[ \sigma_{\alpha\beta\gamma'\delta'} \bigr] &=&  
- \frac{1}{3} \bigl( R_{\alpha'\gamma'\beta'\delta'} 
+ R_{\alpha'\delta'\beta'\gamma'} \bigr), 
\qquad 
\bigl[ \sigma_{\alpha\beta'\gamma'\delta'} \bigr] =  
- \frac{1}{3} \bigl( R_{\alpha'\beta'\gamma'\delta'} 
+ R_{\alpha'\gamma'\beta'\delta'} \bigr),
\nonumber \\ 
\bigl[ \sigma_{\alpha'\beta'\gamma'\delta'} \bigr] &=&  
- \frac{1}{3} \bigl( R_{\alpha'\gamma'\beta'\delta'} 
+ R_{\alpha'\delta'\beta'\gamma'} \bigr). 
\label{3.1.5}
\end{eqnarray} 

\subsection{Derivation of Synge's rule} 
\label{3.2}

We begin with {\it any} bitensor $\Omega_{AB'}(x,x')$ in which $A =
\alpha \cdots \beta$ is a multi-index that represents any number of
unprimed indices, and $B' = \gamma' \cdots \delta'$ a multi-index
that represents any number of primed indices. (It does not matter
whether the primed and unprimed indices are segregated or mixed.) On
the geodesic $\beta$ that links $x$ to $x'$ we introduce an ordinary
tensor $P^M(z)$ where $M$ is a multi-index that contains the same
number of indices as $A$. This tensor is arbitrary, but we assume that
it is parallel transported on $\beta$; this means that it satisfies 
$P^{A}_{\ \ ;\alpha} t^\alpha = 0$ at $x$. Similarly, we
introduce an ordinary tensor $Q^{N}(z)$ in which $N$ contains the same
number of indices as $B'$. This tensor is arbitrary, but we assume
that it is parallel transported on $\beta$; at $x'$ it satisfies 
$Q^{B'}_{\ \ ;\alpha'} t^{\alpha'} = 0$. With $\Omega$, 
$P$, and $Q$ we form a biscalar $H(x,x')$ defined by   
\[
H(x,x') = \Omega_{AB'}(x,x') P^A(x) Q^{B'}(x').   
\]
Having specified the geodesic that links $x$ to $x'$, we can consider  
$H$ to be a function of $\lambda_0$ and $\lambda_1$. If $\lambda_1$ is 
not much larger than $\lambda_0$ (so that $x$ is not far from $x'$),
we can express $H(\lambda_1,\lambda_0)$ as 
\[
H(\lambda_1,\lambda_0) = H(\lambda_0,\lambda_0) + (\lambda_1 -
\lambda_0)\, \frac{\partial H}{\partial \lambda_1} \biggr|_{\lambda_1
= \lambda_0} + \cdots.
\]
Alternatively, 
\[
H(\lambda_1,\lambda_0) = H(\lambda_1,\lambda_1) - (\lambda_1 -
\lambda_0)\, \frac{\partial H}{\partial \lambda_0} \biggr|_{\lambda_0
= \lambda_1} + \cdots, 
\]
and these two expressions give 
\[
\frac{d}{d\lambda_0} H(\lambda_0,\lambda_0) = 
\frac{\partial H}{\partial \lambda_0} \biggr|_{\lambda_0 =
\lambda_1} + \frac{\partial H}{\partial \lambda_1} \biggr|_{\lambda_1
= \lambda_0},
\]
because the left-hand side is the limit of $[H(\lambda_1,\lambda_1) - 
H(\lambda_0,\lambda_0)]/(\lambda_1-\lambda_0)$ when $\lambda_1 \to
\lambda_0$. The partial derivative of $H$ with respect to $\lambda_0$
is equal to $\Omega_{AB';\alpha'} t^{\alpha'} P^A Q^{B'}$, and in
the limit this becomes
$[\Omega_{AB';\alpha'}] t^{\alpha'} P^{A'} Q^{B'}$. Similarly, the 
partial derivative of $H$ with respect to $\lambda_1$ is  
$\Omega_{AB';\alpha} t^{\alpha} P^A Q^{B'}$, and in the limit
$\lambda_1 \to \lambda_0$ this becomes $[\Omega_{AB';\alpha}]
t^{\alpha'} P^{A'} Q^{B'}$. Finally, $H(\lambda_0,\lambda_0) =
[\Omega_{AB'}] P^{A'} Q^{B'}$, and its derivative with respect to 
$\lambda_0$ is $[\Omega_{AB'}]_{;\alpha'} t^{\alpha'} P^{A'} Q^{B'}$. 
Gathering the results we find that  
\[
\Bigl\{ \bigl[ \Omega_{AB'} \bigr]_{;\alpha'} 
- \bigl[ \Omega_{AB';\alpha'} \bigr]  
- \bigl[ \Omega_{AB';\alpha} \bigr] \Bigr\} 
t^{\alpha'} P^{A'} Q^{B'} = 0,  
\]
and the final statement of Synge's rule, 
\begin{equation} 
\bigl[ \Omega_{AB'} \bigr]_{;\alpha'} = 
\bigl[ \Omega_{AB';\alpha'} \bigr] 
+ \bigl[ \Omega_{AB';\alpha} \bigr],  
\label{3.2.1}
\end{equation}
follows from the fact that the tensors $P^M$ and $Q^N$, and the
direction of the selected geodesic $\beta$, are all
arbitrary. Equation (\ref{3.2.1}) reduces to Eq.~(\ref{3.1.3}) when
$\sigma_{\cdots}$ is substituted in place of $\Omega_{AB'}$.   

\section{Parallel propagator} 
\label{4}

\subsection{Tetrad on $\beta$} 
\label{4.1}

On the geodesic segment $\beta$ that links $x$ to $x'$ we introduce an 
orthonormal basis $\base{\mu}{\sf a}(z)$ that is parallel transported
on the geodesic. The frame indices $\sf a$, $\sf b$, \ldots, run from
0 to 3 and the basis vectors satisfy 
\begin{equation}
g_{\mu\nu}\, \base{\mu}{\sf a} \base{\nu}{\sf b} = \eta_{\sf ab},
\qquad   
\frac{D \base{\mu}{\sf a}}{d\lambda} = 0,  
\label{4.1.1}
\end{equation}    
where $\eta_{\sf ab} = \mbox{diag}(-1,1,1,1)$ is the Minkowski metric 
(which we shall use to raise and lower frame indices). We have the  
completeness relations 
\begin{equation}
g^{\mu\nu} = \eta^{\sf ab}\, \base{\mu}{\sf a} \base{\nu}{\sf b}, 
\label{4.1.2}
\end{equation} 
and we define a dual tetrad $\base{\sf a}{\mu}(z)$ by  
\begin{equation}           
\base{\sf a}{\mu} := \eta^{\sf ab} g_{\mu\nu}\, \base{\nu}{\sf b};  
\label{4.1.3}
\end{equation}
this is also parallel transported on $\beta$. In terms of the dual
tetrad the completeness relations take the form 
\begin{equation}
g_{\mu\nu} = \eta_{\sf ab}\, \base{\sf a}{\mu} \base{\sf b}{\nu},  
\label{4.1.4}
\end{equation} 
and it is easy to show that the tetrad and its dual satisfy
$\base{\sf a}{\mu} \base{\mu}{\sf b} = \delta^{\sf a}_{\ \sf b}$ and
$\base{\sf a}{\nu} \base{\mu}{\sf a} = \delta^\mu_{\ \nu}$. Equations 
(\ref{4.1.1})--(\ref{4.1.4}) hold everywhere on $\beta$. In 
particular, with an appropriate change of notation they hold at  
$x'$ and $x$; for example, $g_{\alpha\beta} = \eta_{\sf ab}\,
\base{\sf a}{\alpha} \base{\sf b}{\beta}$ is the metric at $x$. 

(You will have noticed that we use {\sf sans-serif} symbols for the
frame indices. This is to distinguish them from another set of frame
indices that will appear below. The frame indices introduced here run
from 0 to 3; those to be introduced later will run from 1 to 3.)      

\subsection{Definition and properties of the parallel propagator}  
\label{4.2}

Any vector field $A^\mu(z)$ on $\beta$ can be decomposed in the basis 
$\base{\mu}{\sf a}$: $A^\mu = A^{\sf a}\, \base{\mu}{\sf a}$, and
the vector's frame components are given by $A^{\sf a} 
= A^\mu\, \base{\sf a}{\mu}$. If $A^\mu$ is parallel transported on
the geodesic, then the coefficients $A^{\sf a}$ are constants. The
vector at $x$ can then be expressed as $A^\alpha 
= (A^{\alpha'}\, \base{\sf a}{\alpha'}) \base{\alpha}{\sf a}$, or   
\begin{equation}
A^{\alpha}(x) = g^\alpha_{\ \alpha'}(x,x')\, A^{\alpha'}(x'), 
\qquad
g^\alpha_{\ \alpha'}(x,x') := \base{\alpha}{\sf a}(x)\, 
\base{\sf a}{\alpha'}(x').
\label{4.2.1}
\end{equation} 
The object $g^\alpha_{\ \alpha'} = \base{\alpha}{\sf a} 
\base{\sf a}{\alpha'}$ is the {\it parallel propagator}: it takes a
vector at $x'$ and parallel-transports it to $x$ along the unique
geodesic that links these points.  

Similarly, we find that  
\begin{equation}
A^{\alpha'}(x') = g^{\alpha'}_{\ \alpha}(x',x)\, A^{\alpha}(x),
\qquad 
g^{\alpha'}_{\ \alpha}(x',x) := \base{\alpha'}{\sf a}(x')\,
\base{\sf a}{\alpha}(x),
\label{4.2.2} 
\end{equation} 
and we see that $g^{\alpha'}_{\ \alpha} = \base{\alpha'}{\sf a} 
\base{\sf a}{\alpha}$ performs the inverse operation: it takes a
vector at $x$ and parallel-transports it back to $x'$. Clearly, 
\begin{equation} 
g^\alpha_{\ \alpha'} g^{\alpha'}_{\ \beta} = \delta^\alpha_{\ \beta},
\qquad 
g^{\alpha'}_{\ \alpha} g^{\alpha}_{\ \beta'} = 
\delta^{\alpha'}_{\ \beta'},
\label{4.2.3}
\end{equation} 
and these relations formally express the fact that 
$g^{\alpha'}_{\ \alpha}$ is the inverse of 
$g^\alpha_{\ \alpha'}$.   

The relation $g^\alpha_{\ \alpha'} = \base{\alpha}{\sf a} 
\base{\sf a}{\alpha'}$ can also be expressed as 
$g_\alpha^{\ \alpha'} = \base{\sf a}{\alpha} \base{\alpha'}{\sf a}$,
and this reveals that 
\begin{equation}
g_\alpha^{\ \alpha'}(x,x') = g^{\alpha'}_{\ \alpha}(x',x), 
\qquad
g_{\alpha'}^{\ \alpha}(x',x) = g^\alpha_{\ \alpha'}(x,x').   
\label{4.2.4}
\end{equation}
The ordering of the indices, and the ordering of the arguments, are
arbitrary.  

The action of the parallel propagator on tensors of arbitrary rank is 
easy to figure out. For example, suppose that the dual vector $p_\mu =
p_a\, \base{a}{\mu}$ is parallel transported on $\beta$. Then the
frame components $p_{\sf a} = p_\mu\, \base{\mu}{\sf a}$ are
constants, and the dual vector at $x$ can be expressed as $p_\alpha =
(p_{\alpha'} \base{\alpha'}{\sf a}) \base{\alpha}{\sf a}$, or 
\begin{equation}
p_\alpha(x) = g^{\alpha'}_{\ \alpha}(x',x)\, p_{\alpha'}(x').
\label{4.2.5}
\end{equation}
It is therefore the inverse propagator $g^{\alpha'}_{\ \alpha}$ that
takes a dual vector at $x'$ and parallel-transports it to $x$. As
another example, it is easy to show that a tensor $A^{\alpha\beta}$ at
$x$ obtained by parallel transport from $x'$ must be given by 
\begin{equation}
A^{\alpha\beta}(x) = g^\alpha_{\ \alpha'}(x,x') 
g^\beta_{\ \beta'}(x,x')\, A^{\alpha'\beta'}(x'). 
\label{4.2.6}
\end{equation}       
Here we need two occurrences of the parallel propagator, one for each 
tensorial index. Because the metric tensor is covariantly constant, it
is automatically parallel transported on $\beta$, and a special case
of Eq.~(\ref{4.2.6}) is $g_{\alpha\beta} = 
g^{\alpha'}_{\ \alpha} g^{\beta'}_{\ \beta}\, g_{\alpha'\beta'}$.   

Because the basis vectors are parallel transported on $\beta$,
they satisfy $e^\alpha_{{\sf a};\beta} \sigma^\beta = 0$ at $x$ and 
$e^{\alpha'}_{{\sf a};\beta'} \sigma^{\beta'} = 0$ at $x'$. This
immediately implies that the parallel propagators must satisfy 
\begin{equation}
g^{\alpha}_{\ \alpha';\beta} \sigma^{\beta} =  
g^{\alpha}_{\ \alpha';\beta'} \sigma^{\beta'} = 0,
\qquad
g^{\alpha'}_{\ \alpha;\beta} \sigma^{\beta} = 
g^{\alpha'}_{\ \alpha;\beta'} \sigma^{\beta'} = 0. 
\label{4.2.7}
\end{equation} 
Another useful property of the parallel propagator follows from the
fact that if $t^\mu = dz^\mu/d\lambda$ is tangent to the geodesic
connecting $x$ to $x'$, then $t^\alpha = g^\alpha_{\ \alpha'}
t^{\alpha'}$. Using Eqs.~(\ref{2.3.1}) and (\ref{2.3.2}), this
observation gives us the relations 
\begin{equation}
\sigma_{\alpha} = -g^{\alpha'}_{\ \alpha} \sigma_{\alpha'}, \qquad 
\sigma_{\alpha'} = -g^{\alpha}_{\ \alpha'} \sigma_{\alpha}. 
\label{4.2.8}
\end{equation} 
 
\subsection{Coincidence limits} 
\label{4.3}

Equation (\ref{4.2.1}) and the completeness relations of
Eqs.~(\ref{4.1.2}) or (\ref{4.1.4}) imply that 
\begin{equation} 
\bigl[ g^\alpha_{\ \beta'} \bigr] = \delta^{\alpha'}_{\ \beta'}. 
\label{4.3.1}
\end{equation} 
Other coincidence limits are obtained by differentiation of 
Eqs.~(\ref{4.2.7}). For example, the relation 
$g^{\alpha}_{\ \beta';\gamma} \sigma^\gamma = 0$ implies 
$g^{\alpha}_{\ \beta';\gamma\delta} \sigma^\gamma + 
g^{\alpha}_{\ \beta';\gamma} \sigma^\gamma_{\ \delta} = 0$,  
and at coincidence we have 
\begin{equation} 
\bigl[ g^\alpha_{\ \beta';\gamma} \bigr] = 
\bigl[ g^\alpha_{\ \beta';\gamma'} \bigr] = 0; 
\label{4.3.2}
\end{equation}
the second result was obtained by applying Synge's rule on the first
result. Further differentiation gives 
\[
g^{\alpha}_{\ \beta';\gamma\delta\epsilon} \sigma^\gamma +  
g^{\alpha}_{\ \beta';\gamma\delta} \sigma^\gamma_{\ \epsilon} +  
g^{\alpha}_{\ \beta';\gamma\epsilon} \sigma^\gamma_{\ \delta} + 
g^{\alpha}_{\ \beta';\gamma} \sigma^\gamma_{\ \delta\epsilon} = 0, 
\]
and at coincidence we have $[g^{\alpha}_{\ \beta';\gamma\delta}] +
[g^{\alpha}_{\ \beta';\delta\gamma}] = 0$, or 
$2[g^{\alpha}_{\ \beta';\gamma\delta}] 
+ R^{\alpha'}_{\ \beta'\gamma'\delta'} = 0$. The coincidence limit for 
$g^{\alpha}_{\ \beta';\gamma\delta'} 
= g^{\alpha}_{\ \beta';\delta'\gamma}$ can then be obtained from
Synge's rule, and an additional application of the rule gives
$[g^{\alpha}_{\ \beta';\gamma'\delta'}]$. Our results are 
\begin{eqnarray} 
\bigl[ g^{\alpha}_{\ \beta';\gamma\delta} \bigr] &=& 
-\frac{1}{2}\, R^{\alpha'}_{\ \beta'\gamma'\delta'}, 
\qquad
\bigl[ g^{\alpha}_{\ \beta';\gamma\delta'} \bigr] = 
\frac{1}{2}\, R^{\alpha'}_{\ \beta'\gamma'\delta'}, 
\nonumber \\ 
& & \label{4.3.3} \\
\bigl[ g^{\alpha}_{\ \beta';\gamma'\delta} \bigr] &=& 
-\frac{1}{2}\, R^{\alpha'}_{\ \beta'\gamma'\delta'}, 
\qquad
\bigl[ g^{\alpha}_{\ \beta';\gamma'\delta'} \bigr] = 
\frac{1}{2}\, R^{\alpha'}_{\ \beta'\gamma'\delta'}. 
\nonumber
\end{eqnarray} 

\section{Expansion of bitensors near coincidence} 
\label{5}

\subsection{General method} 
\label{5.1}

We would like to express a bitensor $\Omega_{\alpha'\beta'}(x,x')$
near coincidence as an expansion in powers of
$-\sigma^{\alpha'}(x,x')$, the closest analogue in curved spacetime to
the flat-spacetime quantity $(x-x')^\alpha$. For concreteness we shall
consider the case of rank-2 bitensor, and for the moment we will
assume that the tensorial indices all refer to the base point $x'$.   

The expansion we seek is of the form 
\begin{equation}
\Omega_{\alpha'\beta'}(x,x') = A_{\alpha'\beta'}
+ A_{\alpha'\beta'\gamma'}\, \sigma^{\gamma'}  
+ \frac{1}{2}\, A_{\alpha'\beta'\gamma'\delta'}\, 
  \sigma^{\gamma'} \sigma^{\delta'} 
+ O(\epsilon^3), 
\label{5.1.1}
\end{equation} 
in which the ``expansion coefficients'' $A_{\alpha'\beta'}$,
$A_{\alpha'\beta'\gamma'}$, and $A_{\alpha'\beta'\gamma'\delta'}$ are
all ordinary tensors at $x'$; this last tensor is symmetric in the
pair of indices $\gamma'$ and $\delta'$, and $\epsilon$ measures the 
size of a typical component of $\sigma^{\alpha'}$. 

To find the expansion coefficients we differentiate Eq.~(\ref{5.1.1}) 
repeatedly and take coincidence limits. Equation (\ref{5.1.1})   
immediately implies $[\Omega_{\rm \alpha'\beta'}] =
A_{\alpha'\beta'}$. After one differentiation we obtain
$\Omega_{\alpha'\beta';\gamma'} = A_{\alpha'\beta';\gamma'} +
A_{\alpha'\beta'\epsilon';\gamma'} \sigma^{\epsilon'} +
A_{\alpha'\beta'\epsilon'} \sigma^{\epsilon'}_{\ \gamma'} +
\frac{1}{2}\, A_{\alpha'\beta'\epsilon'\iota';\gamma'}
\sigma^{\epsilon'} \sigma^{\iota'} + A_{\alpha'\beta'\epsilon'\iota'}
\sigma^{\epsilon'} \sigma^{\iota'}_{\ \gamma'} + O(\epsilon^2)$, and
at coincidence this reduces to $[\Omega_{\alpha'\beta';\gamma'}] = 
A_{\alpha'\beta';\gamma'} + A_{\alpha'\beta'\gamma'}$. Taking the
coincidence limit after two differentiations yields
$[\Omega_{\alpha'\beta';\gamma'\delta'}] =
A_{\alpha'\beta';\gamma'\delta'} + A_{\alpha'\beta'\gamma';\delta'} +
A_{\alpha'\beta'\delta';\gamma'} + A_{\alpha'\beta'\gamma'\delta'}$. 
The expansion coefficients are therefore  
\begin{eqnarray}
A_{\alpha'\beta'} &=& \bigl[ \Omega_{\alpha'\beta'} \bigr], 
\nonumber \\
A_{\alpha'\beta'\gamma'} &=& \bigl[ \Omega_{\alpha'\beta';\gamma'}
\bigr] - A_{\alpha'\beta';\gamma'}, 
\nonumber \\
A_{\alpha'\beta'\gamma'\delta'} &=& 
\bigl[ \Omega_{\alpha'\beta';\gamma'\delta'} \bigr] 
- A_{\alpha'\beta';\gamma'\delta'}
- A_{\alpha'\beta'\gamma';\delta'} 
- A_{\alpha'\beta'\delta';\gamma'}. 
\label{5.1.2}
\end{eqnarray}  
These results are to be substituted into Eq.~(\ref{5.1.1}), and this
gives us $\Omega_{\alpha'\beta'}(x,x')$ to second order in $\epsilon$.    

Suppose now that the bitensor is $\Omega_{\alpha'\beta}$, with one
index referring to $x'$ and the other to $x$. The previous procedure 
can be applied directly if we introduce an auxiliary bitensor 
$\tilde{\Omega}_{\alpha'\beta'} := g^\beta_{\ \beta'}
\Omega_{\alpha'\beta}$ whose indices all refer to the point $x'$. Then
$\tilde{\Omega}_{\alpha'\beta'}$ can be expanded as in
Eq.~(\ref{5.1.1}), and the original bitensor is reconstructed as 
$\Omega_{\alpha'\beta} = g^{\beta'}_{\ \beta} 
\tilde{\Omega}_{\alpha'\beta'}$, or 
\begin{equation}
\Omega_{\alpha'\beta}(x,x') = g^{\beta'}_{\ \beta}
\biggl( B_{\alpha'\beta'}
+ B_{\alpha'\beta'\gamma'}\, \sigma^{\gamma'}  
+ \frac{1}{2}\, B_{\alpha'\beta'\gamma'\delta'}\, 
  \sigma^{\gamma'} \sigma^{\delta'} \biggr) 
+ O(\epsilon^3). 
\label{5.1.3}
\end{equation} 
The expansion coefficients can be obtained from the coincidence limits
of $\tilde{\Omega}_{\alpha'\beta'}$ and its derivatives. It is 
convenient, however, to express them directly in terms of the original
bitensor $\Omega_{\alpha'\beta}$ by substituting the relation
$\tilde{\Omega}_{\alpha'\beta'} = g^\beta_{\ \beta'}
\Omega_{\alpha'\beta}$ and its derivatives. After using the results of 
Eq.~(\ref{4.3.1})--(\ref{4.3.3}) we find 
\begin{eqnarray}
B_{\alpha'\beta'} &=& \bigl[ \Omega_{\alpha'\beta} \bigr], 
\nonumber \\
B_{\alpha'\beta'\gamma'} &=& \bigl[ \Omega_{\alpha'\beta;\gamma'}
\bigr] - B_{\alpha'\beta';\gamma'}, 
\nonumber \\
B_{\alpha'\beta'\gamma'\delta'} &=& 
\bigl[ \Omega_{\alpha'\beta;\gamma'\delta'} \bigr] 
+ \frac{1}{2}\, B_{\alpha'\epsilon'} 
  R^{\epsilon'}_{\ \beta'\gamma'\delta'}
- B_{\alpha'\beta';\gamma'\delta'} 
- B_{\alpha'\beta'\gamma';\delta'} 
- B_{\alpha'\beta'\delta';\gamma'}. 
\label{5.1.4}
\end{eqnarray}  
The only difference with respect to Eq.~(\ref{5.1.3}) is the presence
of a Riemann-tensor term in $B_{\alpha'\beta'\gamma'\delta'}$.  

Suppose finally that the bitensor to be expanded is
$\Omega_{\alpha\beta}$, whose indices all refer to $x$. Much as we did 
before, we introduce an auxiliary bitensor
$\tilde{\Omega}_{\alpha'\beta'} =  
g^\alpha_{\ \alpha'} g^\beta_{\ \beta'} \Omega_{\alpha\beta}$ whose
indices all refer to $x'$, we expand
$\tilde{\Omega}_{\alpha'\beta'}$ as in Eq.~(\ref{5.1.1}), and we then 
reconstruct the original bitensor. This gives us 
\begin{equation}
\Omega_{\alpha\beta}(x,x') = g^{\alpha'}_{\ \alpha}
g^{\beta'}_{\ \beta} \biggl( C_{\alpha'\beta'}    
+ C_{\alpha'\beta'\gamma'}\, \sigma^{\gamma'}  
+ \frac{1}{2}\, C_{\alpha'\beta'\gamma'\delta'}\, 
  \sigma^{\gamma'} \sigma^{\delta'} \biggr) 
+ O(\epsilon^3), 
\label{5.1.5}
\end{equation} 
and the expansion coefficients are now 
\begin{eqnarray}
\hspace*{-10pt} 
C_{\alpha'\beta'} &=& \bigl[ \Omega_{\alpha\beta} \bigr], 
\nonumber \\
\hspace*{-10pt} 
C_{\alpha'\beta'\gamma'} &=& \bigl[ \Omega_{\alpha\beta;\gamma'} 
\bigr] - C_{\alpha'\beta';\gamma'}, 
\nonumber \\
\hspace*{-10pt} 
C_{\alpha'\beta'\gamma'\delta'} &=& 
\bigl[ \Omega_{\alpha\beta;\gamma'\delta'} \bigr] 
+ \frac{1}{2}\, C_{\alpha'\epsilon'} 
  R^{\epsilon'}_{\ \beta'\gamma'\delta'}
+ \frac{1}{2}\, C_{\epsilon'\beta'} 
  R^{\epsilon'}_{\ \alpha'\gamma'\delta'}
- C_{\alpha'\beta';\gamma'\delta'} 
- C_{\alpha'\beta'\gamma';\delta'} 
- C_{\alpha'\beta'\delta';\gamma'}. 
\label{5.1.6}
\end{eqnarray}  
This differs from Eq.~(\ref{5.1.4}) by the presence of an additional  
Riemann-tensor term in $C_{\alpha'\beta'\gamma'\delta'}$. 

\subsection{Special cases} 
\label{5.2}

We now apply the general expansion method developed in the preceding
subsection to the bitensors $\sigma_{\alpha'\beta'}$,
$\sigma_{\alpha'\beta}$, and $\sigma_{\alpha\beta}$. In the first
instance we have $A_{\alpha'\beta'} = g_{\alpha'\beta'}$,
$A_{\alpha'\beta'\gamma'} = 0$, and $A_{\alpha'\beta'\gamma'\delta'} =
-\frac{1}{3}(R_{\alpha'\gamma'\beta'\delta'} 
+ R_{\alpha'\delta'\beta'\gamma'})$. In the second instance we have   
$B_{\alpha'\beta'} = -g_{\alpha'\beta'}$,
$B_{\alpha'\beta'\gamma'} = 0$, and $B_{\alpha'\beta'\gamma'\delta'} = 
-\frac{1}{3}(R_{\beta'\alpha'\gamma'\delta'} 
+ R_{\beta'\gamma'\alpha'\delta'}) 
- \frac{1}{2} R_{\alpha'\beta'\gamma'\delta'} = 
-\frac{1}{3} R_{\alpha'\delta'\beta'\gamma'} 
-\frac{1}{6} R_{\alpha'\beta'\gamma'\delta'}$. In the third instance
we have $C_{\alpha'\beta'} = g_{\alpha'\beta'}$,
$C_{\alpha'\beta'\gamma'} = 0$, and $C_{\alpha'\beta'\gamma'\delta'} =  
-\frac{1}{3} (R_{\alpha'\gamma'\beta'\delta'} 
+ R_{\alpha'\delta'\beta'\gamma'})$. This gives us the expansions 
\begin{eqnarray} 
\sigma_{\alpha'\beta'} &=& g_{\alpha'\beta'} - \frac{1}{3}\,
R_{\alpha'\gamma'\beta'\delta'}\, \sigma^{\gamma'} \sigma^{\delta'}
+ O(\epsilon^3), 
\label{5.2.1} \\ 
\sigma_{\alpha'\beta} &=& -g^{\beta'}_{\ \beta} 
\Bigl( g_{\alpha'\beta'} 
+ \frac{1}{6}\, R_{\alpha'\gamma'\beta'\delta'}\, \sigma^{\gamma'}
\sigma^{\delta'} \Bigr) + O(\epsilon^3),
\label{5.2.2} \\ 
\sigma_{\alpha\beta} &=& g^{\alpha'}_{\ \alpha} g^{\beta'}_{\ \beta'} 
\Bigl( g_{\alpha'\beta'} - \frac{1}{3}\,
R_{\alpha'\gamma'\beta'\delta'}\, \sigma^{\gamma'} 
\sigma^{\delta'} \Bigr) + O(\epsilon^3).
\label{5.2.3} 
\end{eqnarray}
Taking the trace of the last equation returns 
$\sigma^\alpha_{\ \alpha} = 4 
- \frac{1}{3} R_{\gamma'\delta'}\, \sigma^{\gamma'}
\sigma^{\delta'} + O(\epsilon^3)$, or 
\begin{equation}
\theta^* = 3 - \frac{1}{3}\, R_{\alpha'\beta'}\, \sigma^{\alpha'} 
\sigma^{\beta'} + O(\epsilon^3), 
\label{5.2.4}
\end{equation}
where $\theta^* := \sigma^{\alpha}_{\ \alpha} - 1$ was shown in
Sec.~\ref{2.4} to describe the expansion of the congruence of
geodesics that emanate from $x'$. Equation (\ref{5.2.4}) reveals that
timelike geodesics are focused if the Ricci tensor is nonzero and
the strong energy condition holds: when $R_{\alpha'\beta'}\,
\sigma^{\alpha'} \sigma^{\beta'} > 0$ we see that $\theta^*$ is
smaller than 3, the value it would take in flat spacetime. 

The expansion method can easily be extended to bitensors of other
tensorial ranks. In particular, it can be adapted to give expansions
of the first derivatives of the parallel propagator. The expansions
\begin{equation}
g^\alpha_{\ \beta';\gamma'} = \frac{1}{2}\, g^\alpha_{\ \alpha'}
R^{\alpha'}_{\ \beta'\gamma'\delta'}\, \sigma^{\delta'} 
+ O(\epsilon^2), \qquad
g^\alpha_{\ \beta';\gamma} = \frac{1}{2}\, 
g^\alpha_{\ \alpha'} g^{\gamma'}_{\ \gamma} 
R^{\alpha'}_{\ \beta'\gamma'\delta'}\, \sigma^{\delta'} 
+ O(\epsilon^2) 
\label{5.2.5}
\end{equation}
and thus easy to establish, and they will be needed in part
\ref{part3} of this review.   

\subsection{Expansion of tensors} 
\label{5.3}

The expansion method can also be applied to ordinary tensor
fields. For concreteness, suppose that we wish to express a rank-2
tensor $A_{\alpha\beta}$ at a point $x$ in terms of its values (and
that of its covariant derivatives) at a neighbouring point $x'$. The
tensor can be written as an expansion in powers
of $-\sigma^{\alpha'}(x,x')$ and in this case we have 
\begin{equation}
A_{\alpha\beta}(x) = g^{\alpha'}_{\ \alpha} g^{\beta'}_{\ \beta} 
\biggl( A_{\alpha'\beta'} 
- A_{\alpha'\beta';\gamma'}\, \sigma^{\gamma'}   
+ \frac{1}{2}\, A_{\alpha'\beta';\gamma'\delta'}\, 
  \sigma^{\gamma'} \sigma^{\delta'} \biggr) 
+ O(\epsilon^3).  
\label{5.3.1}
\end{equation} 
If the tensor field is parallel transported on the geodesic $\beta$
that links $x$ to $x'$, then Eq.~(\ref{5.3.1}) reduces to
Eq.~(\ref{4.2.6}). The extension of this formula to tensors of other
ranks is obvious.   

To derive this result we express $A_{\mu\nu}(z)$, the restriction of 
the tensor field on $\beta$, in terms of its tetrad components
$A_{\sf ab}(\lambda) = A_{\mu\nu} \base{\mu}{\sf a} 
\base{\nu}{\sf b}$. Recall from Sec.~\ref{4.1} that 
$\base{\mu}{\sf a}$ is an orthonormal basis that
is parallel transported on $\beta$; recall also that the affine
parameter $\lambda$ ranges from $\lambda_0$ (its value at $x'$) to
$\lambda_1$ (its value at $x$). We have $A_{\alpha'\beta'}(x') = 
A_{\sf ab}(\lambda_0) \base{\sf a}{\alpha'} \base{\sf b}{\beta'}$,  
$A_{\alpha\beta}(x) = A_{\sf ab}(\lambda_1)
\base{\sf a}{\alpha} \base{\sf b}{\beta}$, and $A_{\sf ab}(\lambda_1)$
can be expressed in terms of quantities at $\lambda = \lambda_0$ by
straightforward Taylor expansion. Since, for example, 
\[
(\lambda_1 - \lambda_0) \frac{d A_{\sf ab}}{d\lambda}
\biggr|_{\lambda_0} = (\lambda_1 - \lambda_0) \bigl(A_{\mu\nu}
\base{\mu}{\sf a} \base{\nu}{\sf b} \bigr)_{;\lambda} t^\lambda
\Bigr|_{\lambda_0} = (\lambda_1 - \lambda_0) A_{\mu\nu;\lambda}
\base{\mu}{\sf a} \base{\nu}{\sf b} t^\lambda \Bigr|_{\lambda_0} 
= - A_{\alpha'\beta';\gamma'} \base{\alpha'}{\sf a} 
\base{\beta'}{\sf b} \sigma^{\gamma'}, 
\]
where we have used Eq.~(\ref{2.3.2}), we arrive at Eq.~(\ref{5.3.1}) 
after involving Eq.~(\ref{4.2.2}). 

\section{van Vleck determinant} 
\label{6}

\subsection{Definition and properties} 
\label{6.1}

The van Vleck biscalar $\Delta(x,x')$ is defined by 
\begin{equation}
\Delta(x,x') := \mbox{det}\bigl[ \Delta^{\alpha'}_{\ \beta'}(x,x')
\bigr], \qquad 
\Delta^{\alpha'}_{\ \beta'}(x,x') := -g^{\alpha'}_{\ \alpha}(x',x) 
\sigma^\alpha_{\ \beta'}(x,x').    
\label{6.1.1}
\end{equation}
As we shall show below, it can also be expressed as  
\begin{equation} 
\Delta(x,x') = - \frac{ \mbox{det}\bigl[ -\sigma_{\alpha\beta'}(x,x')
\bigr]} {\sqrt{-g}\sqrt{-g'}}, 
\label{6.1.2}
\end{equation}
where $g$ is the metric determinant at $x$ and $g'$ the metric
determinant at $x'$. 

Equations (\ref{3.1.2}) and (\ref{4.3.1}) imply that at coincidence, 
$[\Delta^{\alpha'}_{\ \beta'}] = \delta^{\alpha'}_{\ \beta'}$ and
$[\Delta] = 1$. Equation (\ref{5.2.2}), on the other hand, implies
that near coincidence, 
\begin{equation} 
\Delta^{\alpha'}_{\ \beta'} = \delta^{\alpha'}_{\ \beta'} +
\frac{1}{6}\, R^{\alpha'}_{\ \gamma'\beta'\delta'}\, \sigma^{\gamma'}
\sigma^{\delta'} + O(\epsilon^3), 
\label{6.1.3}
\end{equation}
so that 
\begin{equation}
\Delta = 1 + \frac{1}{6}\, R_{\alpha'\beta'}\, \sigma^{\alpha'}
\sigma^{\beta'} + O(\epsilon^3). 
\label{6.1.4}
\end{equation}  
This last result follows from the fact that for a ``small'' matrix
$\bm{a}$, $\mbox{det}(\bm{1} + \bm{a}) = 1 + \mbox{tr}(\bm{a}) +
O(\bm{a}^2)$.      

We shall prove below that the van Vleck determinant satisfies the
differential equation 
\begin{equation}
\frac{1}{\Delta} \bigl( \Delta \sigma^\alpha \bigr)_{;\alpha} = 4, 
\label{6.1.5}
\end{equation}
which can also be written as $(\ln \Delta)_{,\alpha} \sigma^\alpha = 4
- \sigma^\alpha_{\ \alpha}$, or 
\begin{equation}
\frac{d}{d\lambda^*}(\ln \Delta) = 3 - \theta^*  
\label{6.1.6}
\end{equation}
in the notation introduced in Sec.~\ref{2.4}. Equation (\ref{6.1.6})
reveals that the behaviour of the van Vleck determinant is
governed by the expansion of the congruence of geodesics that emanate
from $x'$. If $\theta^* < 3$, then the congruence expands less rapidly
than it would in flat spacetime, and $\Delta$ {\it increases} along
the geodesics. If, on the other hand, $\theta^* > 3$, then the
congruence expands more rapidly than it would in flat spacetime, and
$\Delta$ {\it decreases} along the geodesics. Thus, $\Delta > 1$
indicates that the geodesics are undergoing focusing, while $\Delta <
1$ indicates that the geodesics are undergoing defocusing. The
connection between the van Vleck determinant and the strong energy
condition is well illustrated by Eq.~(\ref{6.1.4}): the sign of
$\Delta - 1$ near $x'$ is determined by the sign of
$R_{\alpha'\beta'}\, \sigma^{\alpha'} \sigma^{\beta'}$.   

\subsection{Derivations} 
\label{6.2}

To show that Eq.~(\ref{6.1.2}) follows from Eq.~(\ref{6.1.1}) we
rewrite the completeness relations at $x$, $g^{\alpha\beta} =
\eta^{\sf ab} \base{\alpha}{\sf a} \base{\beta}{\sf b}$, in the matrix
form $\bm{g}^{-1} = \bm{E} \bm{\eta} \bm{E}^T$, where $\bm{E}$ denotes
the $4 \times 4$ matrix whose entries correspond to
$\base{\alpha}{\sf a}$. (In this translation we put tensor and frame
indices on an equal footing.) With $e$ denoting the determinant of this
matrix, we have $1/g = -e^2$, or $e = 1/\sqrt{-g}$. Similarly, we 
rewrite the completeness relations at $x'$, $g^{\alpha'\beta'} =
\eta^{\sf ab} \base{\alpha'}{\sf a} \base{\beta'}{\sf b}$, in the
matrix form $\bm{g'}^{-1} = \bm{E'} \bm{\eta} \bm{E'}^T$, where
$\bm{E'}$ is the matrix corresponding to $\base{\alpha'}{\sf a}$. With
$e'$ denoting its determinant, we have $1/g' = -e^{\prime 2}$, or $e'
= 1/\sqrt{-g'}$. Now, the parallel propagator is defined by
$g^{\alpha}_{\ \alpha'} = \eta^{\sf ab} g_{\alpha'\beta'}
\base{\alpha}{\sf a} \base{\beta'}{\sf b}$, and the matrix form of
this equation is $\bm{\hat{g}} = \bm{E} \bm{\eta} \bm{E'}^T
\bm{g'}^T$. The determinant of the parallel propagator is therefore
$\hat{g} = -ee'g' = \sqrt{-g'}/\sqrt{-g}$. So we have 
\begin{equation}
\mbox{det}\bigl[ g^{\alpha}_{\ \alpha'} \bigr] =
\frac{\sqrt{-g'}}{\sqrt{-g}}, \qquad 
\mbox{det}\bigl[ g^{\alpha'}_{\ \alpha} \bigr] =
\frac{\sqrt{-g}}{\sqrt{-g'}},
\label{6.2.1}
\end{equation}
and Eq.~(\ref{6.1.2}) follows from the fact that the matrix form of 
Eq.~(\ref{6.1.1}) is $\bm{\Delta} = - \bm{\hat{g}}^{-1} \bm{g}^{-1}  
\bm{\sigma}$, where $\bm{\sigma}$ is the matrix corresponding to
$\sigma_{\alpha\beta'}$. 

To establish Eq.~(\ref{6.1.5}) we differentiate the relation 
$\sigma = \frac{1}{2} \sigma^\gamma \sigma_\gamma$ twice and obtain 
$\sigma_{\alpha\beta'} = \sigma^\gamma_{\ \alpha}
\sigma_{\gamma\beta'} + \sigma^\gamma \sigma_{\gamma\alpha\beta'}$. If 
we replace the last factor by $\sigma_{\alpha\beta'\gamma}$ and
multiply both sides by $-g^{\alpha'\alpha}$ we find
\[
\Delta^{\alpha'}_{\ \beta'} = -g^{\alpha'\alpha} \bigl(   
\sigma^\gamma_{\ \alpha} \sigma_{\gamma\beta'} 
+ \sigma^\gamma \sigma_{\alpha\beta'\gamma} \bigr).  
\]
In this expression we make the substitution $\sigma_{\alpha\beta'} =
-g_{\alpha\alpha'} \Delta^{\alpha'}_{\ \beta'}$, which follows
directly from Eq.~(\ref{6.1.1}). This gives us  
\begin{equation} 
\Delta^{\alpha'}_{\ \beta'} = g^{\alpha'}_{\ \alpha} 
g^{\gamma}_{\ \gamma'} \sigma^\alpha_{\ \gamma} 
\Delta^{\gamma'}_{\ \beta'} 
+ \Delta^{\alpha'}_{\ \beta';\gamma} \sigma^\gamma, 
\label{6.2.2}
\end{equation}  
where we have used Eq.~(\ref{4.2.7}). At this stage we introduce
an inverse $(\Delta^{-1})^{\alpha'}_{\ \beta'}$ to the van Vleck
bitensor, defined by $\Delta^{\alpha'}_{\ \beta'} 
(\Delta^{-1})^{\beta'}_{\ \gamma'} = 
\delta^{\alpha'}_{\ \gamma'}$. After multiplying both sides of
Eq.~(\ref{6.2.2}) by $(\Delta^{-1})^{\beta'}_{\ \gamma'}$ we find  
\[
\delta^{\alpha'}_{\ \beta'} = g^{\alpha'}_{\ \alpha} 
g^{\beta}_{\ \beta'} \sigma^\alpha_{\ \beta} 
+ (\Delta^{-1})^{\gamma'}_{\ \beta'}
\Delta^{\alpha'}_{\ \gamma';\gamma} \sigma^\gamma, 
\]
and taking the trace of this equation yields 
\[
4 = \sigma^{\alpha}_{\ \alpha} + (\Delta^{-1})^{\beta'}_{\ \alpha'} 
\Delta^{\alpha'}_{\ \beta';\gamma} \sigma^\gamma.  
\]
We now recall the identity $\delta \ln \mbox{det}\bm{M} =
\mbox{Tr}(\bm{M}^{-1} \delta \bm{M})$, which relates the variation of
a determinant to the variation of the matrix elements. It implies, in
particular, that $(\Delta^{-1})^{\beta'}_{\ \alpha'}  
\Delta^{\alpha'}_{\ \beta';\gamma} = (\ln \Delta)_{,\gamma}$, and we
finally obtain  
\begin{equation}
4 = \sigma^\alpha_{\ \alpha} + (\ln \Delta)_{,\alpha} \sigma^\alpha, 
\label{6.2.3}
\end{equation} 
which is equivalent to Eq.~(\ref{6.1.5}) or Eq.~(\ref{6.1.6}).

%% file: part2.tex
%
\section{Riemann normal coordinates} 
\label{7}

\subsection{Definition and coordinate transformation} 
\label{7.1}

Given a fixed base point $x'$ and a tetrad 
$\base{\alpha'}{\sf a}(x')$, we assign to a neighbouring point $x$ the
four coordinates  
\begin{equation}
\hat{x}^{\sf a} = - \base{\sf a}{\alpha'}(x')\,
\sigma^{\alpha'}(x,x'),   
\label{7.1.1}
\end{equation}
where $\base{\sf a}{\alpha'} = \eta^{\sf ab} g_{\alpha'\beta'}
\base{\beta'}{\sf b}$ is the dual tetrad attached to $x'$. The new  
coordinates $\hat{x}^{\sf a}$ are called {\it Riemann normal
coordinates} (RNC), and they are such that $\eta_{\sf ab} 
\hat{x}^{\sf a} \hat{x}^{\sf b} = \eta_{\sf ab} 
\base{\sf a}{\alpha'} \base{\sf b}{\beta'} \sigma^{\alpha'}
\sigma^{\beta'} = g_{\alpha'\beta'} \sigma^{\alpha'} \sigma^{\beta'}$, 
or  
\begin{equation}
\eta_{\sf ab} \hat{x}^{\sf a} \hat{x}^{\sf b} = 2\sigma(x,x'). 
\label{7.1.2}
\end{equation}
Thus, $\eta_{\sf ab} \hat{x}^{\sf a} \hat{x}^{\sf b}$ is the squared
geodesic distance between $x$ and the base point $x'$. It is obvious
that $x'$ is at the origin of the RNC, where $\hat{x}^{\sf a} = 0$.       

If we move the point $x$ to $x + \delta x$, the new coordinates change
to $\hat{x}^{\sf a} + \delta \hat{x}^{\sf a} = - \base{\sf a}{\alpha'} 
\sigma^{\alpha'}(x+\delta x,x') = \hat{x}^{\sf a} 
- \base{\sf a}{\alpha'} \sigma^{\alpha'}_{\ \beta}\, \delta x^\beta$,
so that   
\begin{equation}
d \hat{x}^{\sf a} = - \base{\sf a}{\alpha'} 
\sigma^{\alpha'}_{\ \beta}\, d x^\beta. 
\label{7.1.3}
\end{equation}
The coordinate transformation is therefore determined by $\partial 
\hat{x}^{\sf a}/\partial x^\beta = -\base{\sf a}{\alpha'} 
\sigma^{\alpha'}_{\ \beta}$, and at coincidence we have 
\begin{equation} 
\biggl[ \frac{\partial \hat{x}^{\sf a}}{\partial x^\alpha} \biggr] = 
\base{\sf a}{\alpha'}, 
\qquad 
\biggl[ \frac{\partial x^\alpha}{\partial \hat{x}^{\sf a}} \biggr] = 
\base{\alpha'}{\sf a}; 
\label{7.1.4}
\end{equation} 
the second result follows from the identities
$\base{\sf a}{\alpha'} \base{\alpha'}{\sf b} = 
\delta^{\sf a}_{\ \sf b}$ and $\base{\alpha'}{\sf a} 
\base{\sf a}{\beta'}  = \delta^{\alpha'}_{\ \beta'}$.  

It is interesting to note that the Jacobian of the
transformation of Eq.~(\ref{7.1.3}), $J := 
\mbox{det}(\partial \hat{x}^{\sf a}/\partial x^\beta)$, is given by $J
= \sqrt{-g} \Delta(x,x')$, where $g$ is the determinant of the metric
in the original coordinates, and $\Delta(x,x')$ is the Van Vleck
determinant of Eq.~(\ref{6.1.2}). This result follows simply by
writing the coordinate transformation in the form $\partial
\hat{x}^{\sf a}/\partial x^\beta = -\eta^{\sf ab} 
\base{\alpha'}{\sf b} \sigma_{\alpha'\beta}$ and computing the product
of the determinants. It allows us to deduce that in RNC, the
determinant of the metric is given by
\begin{equation} 
\sqrt{-g(\mbox{RNC})} = \frac{1}{\Delta(x,x')}.  
\label{7.1.5}
\end{equation} 
It is easy to show that the geodesics emanating from $x'$ are straight
coordinate lines in RNC. The proper volume of a small comoving region
is then equal to $d V = \Delta^{-1}\, d^4 \hat{x}$, and this is
smaller than the flat-spacetime value of $d^4 \hat{x}$ if $\Delta >
1$, that is, if the geodesics are focused by the spacetime curvature.        

\subsection{Metric near $x'$} 
\label{7.2}

We now would like to invert Eq.~(\ref{7.1.3}) in order to express the 
line element $ds^2 = g_{\alpha\beta}\, dx^\alpha dx^\beta$ in terms of
the displacements $d\hat{x}^{\sf a}$. We shall do this approximately,
by working in a small neighbourhood of $x'$. We recall the expansion
of Eq.~(\ref{5.2.2}),    
\[
\sigma^{\alpha'}_{\ \beta} = - g^{\beta'}_{\ \beta} 
\biggl( \delta^{\alpha'}_{\ \beta'} + \frac{1}{6}\, 
R^{\alpha'}_{\ \gamma'\beta'\delta'} \sigma^{\gamma'} \sigma^{\delta'}
\biggr) + O(\epsilon^3),
\]
and in this we substitute the frame decomposition of the Riemann
tensor,  
$R^{\alpha'}_{\ \gamma'\beta'\delta'} = R^{\sf a}_{\ \sf cbd}\, 
\base{\alpha'}{\sf a} \base{\sf c}{\gamma'} \base{\sf b}{\beta'}
\base{\sf d}{\delta'}$, and the tetrad decomposition of the parallel  
propagator, $g^{\beta'}_{\ \beta} = \base{\beta'}{\sf b} 
\base{\sf b}{\beta}$, where $\base{\sf b}{\beta}(x)$ is the dual
tetrad at $x$ obtained by parallel transport of 
$\base{\sf b}{\beta'}(x')$. After some algebra we obtain   
\[
\sigma^{\alpha'}_{\ \beta} = 
- \base{\alpha'}{\sf a} \base{\sf a}{\beta} 
- \frac{1}{6}\, R^{\sf a}_{\ \sf cbd}\, \base{\alpha'}{\sf a} 
\base{\sf b}{\beta} \hat{x}^{\sf c} \hat{x}^{\sf d} 
+ O(\epsilon^3),  
\]
where we have used Eq.~(\ref{7.1.1}). Substituting this into
Eq.~(\ref{7.1.3}) yields
\begin{equation}
d \hat{x}^{\sf a} = \biggl[ \delta^{\sf a}_{\ \sf b} 
+ \frac{1}{6}\, R^{\sf a}_{\ \sf cbd} \hat{x}^{\sf c} \hat{x}^{\sf d} 
+ O(x^3)\biggr] \base{\sf b}{\beta}\, d x^\beta,  
\label{7.2.1}
\end{equation} 
and this is easily inverted to give 
\begin{equation}
\base{\sf a}{\alpha}\, d x^\alpha = \biggl[ \delta^{\sf a}_{\ \sf b} - 
\frac{1}{6}\, R^{\sf a}_{\ \sf cbd} \hat{x}^{\sf c} \hat{x}^{\sf d} 
+ O(x^3) \biggr] d \hat{x}^{\sf b}.   
\label{7.2.2}
\end{equation} 
This is the desired approximate inversion of
Eq.~(\ref{7.1.3}). It is useful to note that Eq.~(\ref{7.2.2}), when
specialized from the arbitrary coordinates $x^\alpha$ to 
$\hat{x}^{\sf a}$, gives us the components of the dual tetrad at $x$
in RNC. And since $\base{\alpha'}{\sf a} = \delta^{\alpha'}_{\ \sf a}$ in
RNC, we immediately obtain the components of the parallel propagator: 
$g^{\sf a'}_{\ \sf b} = \delta^{\sf a}_{\ \sf b} - \frac{1}{6} 
R^{\sf a}_{\ \sf cbd} \hat{x}^{\sf c} \hat{x}^{\sf d} + O(x^3)$. 

We are now in a position to calculate the metric in the new
coordinates. We have $ds^2 = g_{\alpha\beta}\, dx^\alpha dx^\beta =
(\eta_{\sf ab} \base{\sf a}{\alpha} \base{\sf b}{\beta}) dx^\alpha
dx^\beta = \eta_{\sf ab} (\base{\sf a}{\alpha}\, dx^\alpha) 
(\base{\sf b}{\beta}\, dx^\beta)$, and in this we substitute
Eq.~(\ref{7.2.2}). The final result is $ds^2 = g_{\sf ab}\, 
d\hat{x}^{\sf a} d\hat{x}^{\sf b}$, with  
\begin{equation}
g_{\sf ab} = \eta_{\sf ab} - \frac{1}{3} R_{\sf acbd} 
\hat{x}^{\sf c} \hat{x}^{\sf d} + O(x^3).  
\label{7.2.3}
\end{equation} 
The quantities $R_{\sf acbd}$ appearing in Eq.~(\ref{7.2.3}) are the 
frame components of the Riemann tensor evaluated at the base point
$x'$,  
\begin{equation}
R_{\sf acbd} := R_{\alpha'\gamma'\beta'\delta'}\, \base{\alpha'}{\sf a} 
\base{\gamma'}{\sf c} \base{\beta'}{\sf b} \base{\delta'}{\sf d}, 
\label{7.2.4}
\end{equation}
and these are independent of $\hat{x}^{\sf a}$. They are also, by
virtue of Eq.~(\ref{7.1.4}), the components of the (base-point)
Riemann tensor in RNC, because Eq.~(\ref{7.2.4}) can also
be expressed as 
\[
R_{\sf acdb} = R_{\alpha'\gamma'\beta'\delta'} 
\biggl[ \frac{\partial x^\alpha}{\partial \hat{x}^{\sf a}} \biggr] 
\biggl[ \frac{\partial x^\gamma}{\partial \hat{x}^{\sf c}} \biggr] 
\biggl[ \frac{\partial x^\beta}{\partial \hat{x}^{\sf b}} \biggr] 
\biggl[ \frac{\partial x^\delta}{\partial \hat{x}^{\sf d}} \biggr],  
\]
which is the standard transformation law for tensor components.    

It is obvious from Eq.~(\ref{7.2.3}) that $g_{\sf ab}(x') = 
\eta_{\sf ab}$ and $\Gamma^{\sf a}_{\ \sf bc}(x') = 0$, where
$\Gamma^{\sf a}_{\ \sf bc} = -\frac{1}{3} 
(R^{\sf a}_{\ \sf bcd} + R^{\sf a}_{\ \sf cbd}) \hat{x}^{\sf d} 
+ O(x^2)$ is the connection compatible with the metric 
$g_{\sf ab}$. The Riemann normal coordinates therefore provide a
constructive proof of the local flatness theorem.   

\section{Fermi normal coordinates} 
\label{8}

\subsection{Fermi-Walker transport} 
\label{8.1}

Let $\gamma$ be a timelike curve described by parametric relations
$z^\mu(\tau)$ in which $\tau$ is proper time. Let $u^\mu =
dz^\mu/d\tau$ be the curve's normalized tangent vector, and let $a^\mu
= D u^\mu/d\tau$ be its acceleration vector.  

A vector field $v^{\mu}$ is said to be {\it Fermi-Walker transported}
on $\gamma$ if it is a solution to the differential equation 
\begin{equation}
\frac{D v^{\mu}}{d\tau} = \bigl( v_{\nu} a^{\nu} \bigr)
u^{\mu} - \bigl( v_{\nu} u^{\nu} \bigr) a^{\mu}. 
\label{8.1.1}
\end{equation} 
Notice that this reduces to parallel transport when $a^{\mu} = 0$
and $\gamma$ is a geodesic. 

The operation of Fermi-Walker (FW) transport satisfies two important  
properties. The first is that $u^{\mu}$ is automatically FW
transported along $\gamma$; this follows at once from
Eq.~(\ref{8.1.1}) and the fact that $u^{\mu}$ is orthogonal to
$a^{\mu}$. The second is that if the vectors $v^{\mu}$ and
$w^{\mu}$ are both FW transported along $\gamma$, then their inner
product $v_{\mu} w^{\mu}$ is constant on $\gamma$:
$D(v_{\mu} w^{\mu})/d\tau = 0$; this also follows immediately
from Eq.~(\ref{8.1.1}).  

\subsection{Tetrad and dual tetrad on $\gamma$} 
\label{8.2}

Let $\bar{z}$ be an arbitrary reference point on $\gamma$. At this
point we erect an orthonormal tetrad $(u^{\bar{\mu}},
\base{\bar{\mu}}{a})$ where, as a modification to former usage, the
frame index $a$ runs from 1 to 3. We then propagate each frame vector
on $\gamma$ by FW transport; this guarantees that the tetrad remains 
orthonormal everywhere on $\gamma$. At a generic point $z(\tau)$ we
have  
\begin{equation}
\frac{D \base{\mu}{a}}{d \tau} = \bigl( a_{\nu} \base{\nu}{a}
\bigr) u^{\mu}, \qquad
g_{\mu\nu} u^{\mu} u^{\nu} = -1, \qquad
g_{\mu\nu} \base{\mu}{a} u^{\nu} = 0, \qquad
g_{\mu\nu} \base{\mu}{a} \base{\nu}{b} = \delta_{ab}. 
\label{8.2.1}
\end{equation} 
From the tetrad on $\gamma$ we define a dual tetrad $(\base{0}{\mu}, 
\base{a}{\mu})$ by the relations 
\begin{equation} 
\base{0}{\mu} = - u_{\mu}, \qquad
\base{a}{\mu} = \delta^{ab} g_{\mu\nu} \base{\nu}{b}; 
\label{8.2.2}
\end{equation}
this also is FW transported on $\gamma$. The tetrad and its dual give 
rise to the completeness relations 
\begin{equation}
g^{\mu\nu} = -u^{\mu} u^{\nu} 
+ \delta^{ab} \base{\mu}{a} \base{\nu}{b}, \qquad
g_{\mu\nu} = -\base{0}{\mu} \base{0}{\nu} 
+ \delta_{ab}\, \base{a}{\mu} \base{b}{\nu}.
\label{8.2.3}
\end{equation} 

\subsection{Fermi normal coordinates} 
\label{8.3}

To construct the Fermi normal coordinates (FNC) of a point $x$ in 
the normal convex neighbourhood of $\gamma$ we locate the unique
{\it spacelike geodesic} $\beta$ that passes through $x$ and
intersects $\gamma$ {\it orthogonally}. We denote the intersection
point by $\bar{x} := z(t)$, with $t$ denoting the value of the 
proper-time parameter at this point. To tensors at $\bar{x}$ we
assign indices $\bar{\alpha}$, $\bar{\beta}$, and so on. The FNC of
$x$ are defined by  
\begin{equation}
\hat{x}^0 = t, \qquad
\hat{x}^a = -\base{a}{\bar{\alpha}}(\bar{x}) 
\sigma^{\bar{\alpha}}(x,\bar{x}), \qquad
\sigma_{\bar{\alpha}}(x,\bar{x}) u^{\bar{\alpha}}(\bar{x}) = 0; 
\label{8.3.1}
\end{equation}
the last statement determines $\bar{x}$ from the requirement that 
$-\sigma^{\bar{\alpha}}$, the vector tangent to $\beta$ at $\bar{x}$,
be orthogonal to $u^{\bar{\alpha}}$, the vector tangent to
$\gamma$. From the definition of the FNC and the completeness
relations of Eq.~(\ref{8.2.3}) it follows that 
\begin{equation}
s^2 := \delta_{a b} \hat{x}^a \hat{x}^b = 2 \sigma(x,\bar{x}),  
\label{8.3.2}
\end{equation} 
so that $s$ is the spatial distance between $\bar{x}$ and $x$ along
the geodesic $\beta$. This statement gives an immediate meaning to
$\hat{x}^a$, the spatial Fermi normal coordinates, and the time
coordinate $\hat{x}^0$ is simply proper time at the intersection point
$\bar{x}$. The situation is illustrated in Fig.~6.  

\begin{figure}
\begin{center}
\vspace*{-20pt} 
\includegraphics[width=0.5\linewidth]{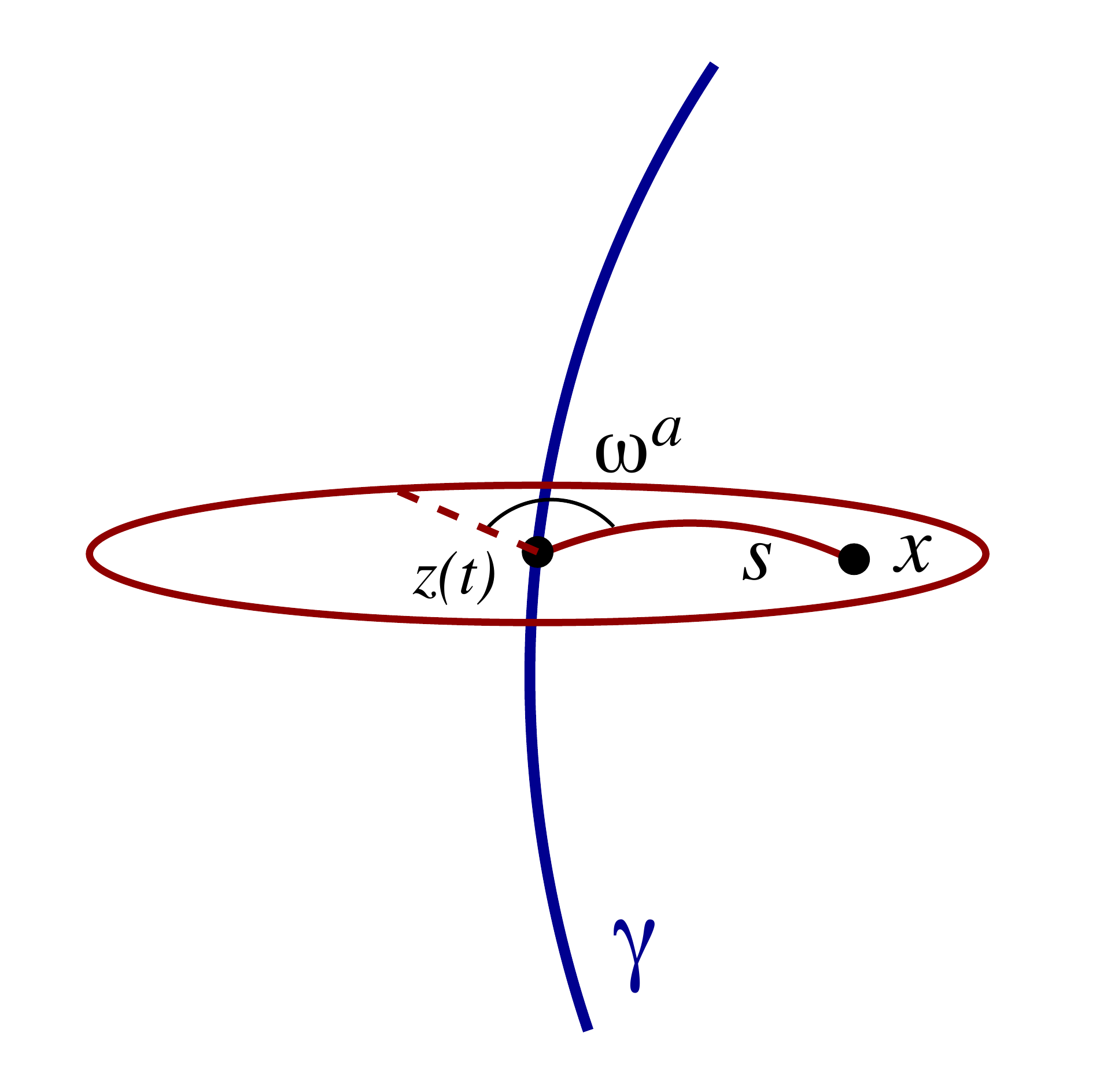}
\vspace*{-20pt}
\end{center} 
\caption{Fermi normal coordinates of a point $x$ relative to a world 
line $\gamma$. The time coordinate $t$ selects a particular point on
the word line, and the disk represents the set of spacelike
geodesics that intersect $\gamma$ orthogonally at $z(t)$. The   
unit vector $\omega^a := \hat{x}^a/s$ selects a particular
geodesic among this set, and the spatial distance $s$ selects a
particular point on this geodesic.}    
\end{figure} 

Suppose that $x$ is moved to $x + \delta x$. This typically induces a 
change in the spacelike geodesic $\beta$, which moves to $\beta +
\delta \beta$, and a corresponding change in the intersection point
$\bar{x}$, which moves to $x'' := \bar{x} + \delta \bar{x}$, with
$\delta x^{\bar{\alpha}} = u^{\bar{\alpha}} \delta t$. The FNC of the
new point are then $\hat{x}^0(x+\delta x) = t + \delta t$ and 
$\hat{x}^a(x + \delta x) = -\base{a}{\alpha''}(x'') 
\sigma^{\alpha''}(x + \delta x, x'')$, with $x''$ determined by 
$\sigma_{\alpha''}(x + \delta x,x'') u^{\alpha''}(x'') = 0$. Expanding
these relations to first order in the displacements, and simplifying
using Eqs.~(\ref{8.2.1}), yields      
\begin{equation}
dt = \mu\, \sigma_{\bar{\alpha}\beta} u^{\bar{\alpha}}\, 
d x^{\beta}, \qquad 
d \hat{x}^a = -\base{a}{\bar{\alpha}} \bigl( 
\sigma^{\bar{\alpha}}_{\ \beta} 
+ \mu\, \sigma^{\bar{\alpha}}_{\ \bar{\beta}} u^{\bar{\beta}} 
\sigma_{\beta \bar{\gamma}} u^{\bar{\gamma}} \bigr) d x^{\beta},  
\label{8.3.3}
\end{equation} 
where $\mu$ is determined by $\mu^{-1} = - 
( \sigma_{\bar{\alpha}\bar{\beta}} u^{\bar{\alpha}} u^{\bar{\beta}} 
+ \sigma_{\bar{\alpha}} a^{\bar{\alpha}})$.    

\subsection{Coordinate displacements near $\gamma$}  
\label{8.4}

The relations of Eq.~(\ref{8.3.3}) can be expressed as expansions in
powers of $s$, the spatial distance from $\bar{x}$ to $x$. For this we
use the expansions of Eqs.~(\ref{5.2.1}) and (\ref{5.2.2}), in which
we substitute $\sigma^{\bar{\alpha}} = -\base{\bar{\alpha}}{a}
\hat{x}^a$ and $g^{\bar{\alpha}}_{\ \alpha} = u^{\bar{\alpha}}
\bar{e}^{0}_{\alpha} + \base{\bar{\alpha}}{a} \bar{e}^{a}_{\alpha}$, 
where $(\bar{e}^{0}_{\alpha},\bar{e}^{a}_{\alpha})$ is a dual tetrad
at $x$ obtained by parallel transport of
$(-u_{\bar{\alpha}},\base{a}{\bar{\alpha}})$ on the 
spacelike geodesic $\beta$. After some algebra we obtain  
\[
\mu^{-1} = 1 + a_a \hat{x}^a + \frac{1}{3} R_{0c0d} \hat{x}^c
\hat{x}^d + O(s^3), 
\]
where $a_a(t) := a_{\bar{\alpha}} \base{\bar{\alpha}}{a}$ are
frame components of the acceleration vector, and $R_{0c0d}(t) := 
R_{\bar{\alpha}\bar{\gamma}\bar{\beta}\bar{\delta}} u^{\bar{\alpha}}
\base{\bar{\gamma}}{c} u^{\bar{\beta}} \base{\bar{\delta}}{d}$ are
frame components of the Riemann tensor evaluated on $\gamma$. This
last result is easily inverted to give  
\[
\mu = 1 - a_a \hat{x}^a + \bigl( a_a \hat{x}^a \bigr)^2      
- \frac{1}{3} R_{0c0d} \hat{x}^c \hat{x}^d + O(s^3).
\]
Proceeding similarly for the other relations of Eq.~(\ref{8.3.3}), we 
obtain 
\begin{equation}
dt = \biggl[ 1 - a_a \hat{x}^a 
+ \bigl( a_a \hat{x}^a \bigr)^2       
- \frac{1}{2} R_{0c0d} \hat{x}^c \hat{x}^d + O(s^3) \biggr] 
\bigl( \bar{e}^{0}_{\beta} d x^\beta \bigr) 
+ \biggl[ -\frac{1}{6} R_{0cbd} \hat{x}^c \hat{x}^d + O(s^3) \biggr]  
\bigl( \bar{e}^{b}_{\beta} d x^\beta \bigr)
\label{8.4.1}
\end{equation} 
and
\begin{equation}
d \hat{x}^a = \biggl[ \frac{1}{2} R^a_{\ c0d} \hat{x}^c \hat{x}^d  
+ O(s^3) \biggr] 
\bigl( \bar{e}^{0}_{\beta} d x^\beta \bigr) 
+ \biggl[ \delta^a_{\ b} +\frac{1}{6} R^a_{\ cbd} \hat{x}^c \hat{x}^d  
+ O(s^3) \biggr] 
\bigl( \bar{e}^{b}_{\beta} d x^\beta \bigr),
\label{8.4.2}
\end{equation} 
where $R_{ac0d}(t) :=
R_{\bar{\alpha}\bar{\gamma}\bar{\beta}\bar{\delta}}  
\base{\bar{\alpha}}{a} \base{\bar{\gamma}}{c} u^{\bar{\beta}}
\base{\bar{\delta}}{d}$ and $R_{acbd}(t) := 
R_{\bar{\alpha}\bar{\gamma}\bar{\beta}\bar{\delta}}  
\base{\bar{\alpha}}{a} \base{\bar{\gamma}}{c} \base{\bar{\beta}}{b} 
\base{\bar{\delta}}{d}$ are additional frame components of the Riemann 
tensor evaluated on $\gamma$. (Note that frame indices are raised with 
$\delta^{ab}$.) 

As a special case of Eqs.~(\ref{8.4.1}) and (\ref{8.4.2}) we find that  
\begin{equation}
\frac{\partial t}{\partial x^\alpha} \biggr|_\gamma =
-u_{\bar{\alpha}}, \qquad 
\frac{\partial \hat{x}^a}{\partial x^\alpha} \biggr|_\gamma =
\base{a}{\bar{\alpha}},
\label{8.4.3}
\end{equation}
because in the limit $x \to \bar{x}$, the dual tetrad
$(\bar{e}^{0}_{\alpha},\bar{e}^{a}_{\alpha})$ at $x$ coincides with
the dual tetrad $(-u_{\bar{\alpha}},\base{a}{\bar{\alpha}})$ at
$\bar{x}$. It follows that on $\gamma$, the transformation matrix
between the original coordinates $x^\alpha$ and the FNC
$(t,\hat{x}^a)$ is formed by the Fermi-Walker transported tetrad:  
\begin{equation} 
\frac{\partial x^\alpha}{\partial t} \biggr|_\gamma =
u^{\bar{\alpha}}, \qquad
\frac{\partial x^\alpha}{\partial \hat{x}^a} \biggr|_\gamma =
\base{\bar{\alpha}}{a}. 
\label{8.4.4}
\end{equation} 
This implies that the frame components of the acceleration vector,
$a_a(t)$, are also the {\it components} of the acceleration
vector in FNC; orthogonality between $u^{\bar{\alpha}}$ and
$a^{\bar{\alpha}}$ means that $a_0 = 0$. Similarly, $R_{0c0d}(t)$, 
$R_{0cbd}(t)$, and $R_{acbd}(t)$ are the {\it components} of the
Riemann tensor (evaluated on $\gamma$) in Fermi normal
coordinates.   

\subsection{Metric near $\gamma$} 
\label{8.5}

Inversion of Eqs.~(\ref{8.4.1}) and (\ref{8.4.2}) gives 
\begin{equation} 
\bar{e}^{0}_{\alpha} dx^\alpha = \biggl[ 1 + a_a \hat{x}^a 
+ \frac{1}{2} R_{0c0d} \hat{x}^c \hat{x}^d 
+ O(s^3) \biggr]\, dt  
+ \biggl[ \frac{1}{6} R_{0cbd} \hat{x}^c \hat{x}^d 
+ O(s^3) \biggr]\, d\hat{x}^b 
\label{8.5.1}
\end{equation}
and 
\begin{equation} 
\bar{e}^{a}_{\alpha} dx^\alpha = \biggl[ \delta^a_{\ b} 
- \frac{1}{6} R^a_{\ cbd} \hat{x}^c \hat{x}^d  
+ O(s^3) \biggr]\, d\hat{x}^b 
+ \biggl[ - \frac{1}{2} R^a_{\ c0d} \hat{x}^c \hat{x}^d  
+ O(s^3) \biggr]\, dt. 
\label{8.5.2}
\end{equation} 
These relations, when specialized to the FNC, give the components of
the dual tetrad at $x$. They can also be used to compute the metric at
$x$, after invoking the completeness relations $g_{\alpha\beta} =
-\bar{e}^{0}_{\alpha} \bar{e}^{0}_{\beta} + \delta_{ab}
\bar{e}^{a}_{\alpha} \bar{e}^{b}_{\beta}$. This gives 
\begin{eqnarray}
g_{tt} &=& - \Bigl[ 1 + 2 a_a \hat{x}^a 
+ \bigl( a_a \hat{x}^a \bigr)^2 
+ R_{0c0d} \hat{x}^c \hat{x}^d + O(s^3) \Bigr], 
\label{8.5.3} \\ 
g_{ta} &=& - \frac{2}{3} R_{0cad} \hat{x}^c \hat{x}^d + O(s^3),  
\label{8.5.4} \\ 
g_{ab} &=& \delta_{ab} - \frac{1}{3} R_{acbd} \hat{x}^c \hat{x}^d  
+ O(s^3). 
\label{8.5.5}
\end{eqnarray} 
This is the metric near $\gamma$ in the Fermi normal
coordinates. Recall that $a_a(t)$ are the components of the
acceleration vector of $\gamma$ --- the timelike curve described by 
$\hat{x}^a = 0$ --- while $R_{0c0d}(t)$, $R_{0cbd}(t)$,
and $R_{acbd}(t)$ are the components of the Riemann tensor evaluated
on $\gamma$.  

Notice that on $\gamma$, the metric of
Eqs.~(\ref{8.5.3})--(\ref{8.5.5}) reduces to $g_{tt} = -1$ and $g_{ab}
= \delta_{ab}$. On the other hand, the nonvanishing Christoffel
symbols (on $\gamma$) are $\Gamma^t_{\ ta} = \Gamma^a_{\ tt} = a_a$;
these are zero (and the FNC enforce local flatness on the entire
curve) when $\gamma$ is a geodesic.     

\subsection{Thorne-Hartle-Zhang coordinates}
\label{8.6}

The form of the metric can be simplified when the Ricci tensor
vanishes on the world line:  
\begin{equation}
R_{\mu\nu}(z) = 0. 
\label{8.6.1}
\end{equation}
In such circumstances, a transformation from the Fermi normal
coordinates $(t,\hat{x}^a)$ to the {\it Thorne-Hartle-Zhang (THZ)
coordinates} $(t,\hat{y}^a)$ brings the metric to the form 
\begin{eqnarray}
g_{tt} &=& - \Bigl[ 1 + 2 a_a \hat{y}^a 
+ \bigl( a_a \hat{y}^a \bigr)^2 
+ R_{0c0d} \hat{y}^c \hat{y}^d + O(s^3) \Bigr], 
\label{8.6.2} \\ 
g_{ta} &=& -\frac{2}{3} R_{0cad} \hat{y}^c \hat{y}^d + O(s^3),  
\label{8.6.3} \\ 
g_{ab} &=& \delta_{ab} \bigl( 1 
- R_{0c0d} \hat{y}^c \hat{y}^d  \bigr) + O(s^3). 
\label{8.6.4}
\end{eqnarray} 
We see that the transformation leaves $g_{tt}$ and $g_{ta}$ unchanged,
but that it diagonalizes $g_{ab}$. This metric was first displayed in 
Ref.~\cite{thorne-hartle:85} and the coordinate transformation was
later produced by Zhang \cite{zhang:86}. 

The key to the simplification comes from Eq.~(\ref{8.6.1}), which 
dramatically reduces the number of independent components of the 
Riemann tensor. In particular, Eq.~(\ref{8.6.1}) implies that the
frame components $R_{acbd}$ of the Riemann tensor are completely
determined by ${\cal E}_{ab} := R_{0a0b}$, which in this special
case is a symmetric-tracefree tensor. To prove this we invoke the
completeness relations of Eq.~(\ref{8.2.3}) and take frame components
of Eq.~(\ref{8.6.1}). This produces the three independent equations 
\[
\delta^{cd} R_{acbd} = {\cal E}_{ab}, \qquad
\delta^{cd} R_{0cad} = 0, \qquad
\delta^{cd} {\cal E}_{cd} = 0, 
\]
the last of which stating that ${\cal E}_{ab}$ has a vanishing
trace. Taking the trace of the first equation gives 
$\delta^{ab} \delta^{cd} R_{acbd} = 0$, and this implies that
$R_{acbd}$ has five independent components. Since this is also the
number of independent components of ${\cal E}_{ab}$, we see that the
first equation can be inverted --- $R_{acbd}$ can be expressed in
terms of ${\cal E}_{ab}$. A complete listing of the relevant relations
is $R_{1212} = {\cal E}_{11} + {\cal E}_{22} = -{\cal E}_{33}$,
$R_{1213} = {\cal E}_{23}$, $R_{1223} = -{\cal E}_{13}$, $R_{1313} =  
{\cal E}_{11} + {\cal E}_{33} = -{\cal E}_{22}$, $R_{1323} = 
{\cal E}_{12}$,  and $R_{2323} = {\cal E}_{22} + {\cal E}_{33} = -
{\cal E}_{11}$. These are summarized by 
\begin{equation} 
R_{acbd} = \delta_{ab} {\cal E}_{cd} + \delta_{cd} {\cal E}_{ab} -
\delta_{ad} {\cal E}_{bc} - \delta_{bc} {\cal E}_{ad}, 
\label{8.6.5}
\end{equation}
and ${\cal E}_{ab} := R_{0a0b}$ satisfies
$\delta^{ab} {\cal E}_{ab} = 0$. 

We may also note that the relation $\delta^{cd} R_{0cad} = 0$,
together with the usual symmetries of the Riemann tensor, imply that
$R_{0cad}$ too possesses five independent components. These may thus
be related to another symmetric-tracefree tensor ${\cal B}_{ab}$. We 
take the independent components to be $R_{0112} := 
-{\cal B}_{13}$, $R_{0113} := {\cal B}_{12}$, $R_{0123} :=  
-{\cal B}_{11}$, $R_{0212} := -{\cal B}_{23}$, and $R_{0213}
:= {\cal B}_{22}$, and it is easy to see that all other
components can be expressed in terms of these. For example, $R_{0223}
= -R_{0113} = -{\cal B}_{12}$, $R_{0312} = -R_{0123} + R_{0213} =
{\cal B}_{11} + {\cal B}_{22} = -{\cal B}_{33}$, $R_{0313}
= -R_{0212} = {\cal B}_{23}$, and $R_{0323} = R_{0112} = 
-{\cal B}_{13}$. These relations are summarized by 
\begin{equation}
R_{0abc} = -\varepsilon_{bcd} {\cal B}^d_{\ a}, 
\label{8.6.6}
\end{equation}
where $\varepsilon_{abc}$ is the three-dimensional permutation
symbol. The inverse relation is ${\cal B}^a_{\ b} = -\frac{1}{2}
\varepsilon^{acd} R_{0bcd}$.   

Substitution of Eq.~(\ref{8.6.5}) into Eq.~(\ref{8.5.5}) gives  
\[
g_{ab} = \delta_{ab} \Bigl( 1 - \frac{1}{3} {\cal E}_{cd} \hat{x}^c
\hat{x}^d \Bigr) - \frac{1}{3} \bigl( \hat{x}_c \hat{x}^c \bigr)
{\cal E}_{ab} + \frac{1}{3} \hat{x}_a {\cal E}_{bc} \hat{x}^c +  
\frac{1}{3} \hat{x}_b {\cal E}_{ac} \hat{x}^c + O(s^3), 
\]
and we have not yet achieved the simple form of Eq.~(\ref{8.6.4}). The  
missing step is the transformation from the FNC $\hat{x}^a$ to the
THZ coordinates $\hat{y}^a$. This is given by 
\begin{equation}
\hat{y}^a = \hat{x}^a + \xi^a, \qquad
\xi^a = - \frac{1}{6} \bigl( \hat{x}_c \hat{x}^c \bigr)  
{\cal E}_{ab} \hat{x}^b + \frac{1}{3} \hat{x}_a {\cal E}_{bc}
\hat{x}^b \hat{x}^c + O(s^4). 
\label{8.6.7}
\end{equation} 
It is easy to see that this transformation does not affect $g_{tt}$
nor $g_{ta}$ at orders $s$ and $s^2$. The remaining components of the
metric, however, transform according to $g_{ab}(\mbox{THZ}) =
g_{ab}(\mbox{FNC}) - \xi_{a;b} - \xi_{b;a}$, where  
\[
\xi_{a;b} = \frac{1}{3} \delta_{ab} {\cal E}_{cd} \hat{x}^c \hat{x}^d
- \frac{1}{6} \bigl( \hat{x}_c \hat{x}^c \bigr) {\cal E}_{ab}  
- \frac{1}{3} {\cal E}_{ac} \hat{x}^c \hat{x}_b 
+ \frac{2}{3} \hat{x}_a {\cal E}_{bc} \hat{x}^c + O(s^3).
\]
It follows that $g_{ab}^{\rm THZ} = \delta_{ab}(1 - {\cal E}_{cd}
\hat{y}^c \hat{y}^d) + O(\hat{y}^3)$, which is just the same statement
as in Eq.~(\ref{8.6.4}).   

Alternative expressions for the components of the THZ metric are  
\begin{eqnarray}
g_{tt} &=& - \Bigl[ 1 + 2 a_a \hat{y}^a 
+ \bigl( a_a \hat{y}^a \bigr)^2 
+ {\cal E}_{ab} \hat{y}^a \hat{y}^b + O(s^3) \Bigr],  
\label{8.6.8} \\ 
g_{ta} &=& - \frac{2}{3} \varepsilon_{abc} {\cal B}^b_{\ d} \hat{y}^c
\hat{y}^d + O(s^3), 
\label{8.6.9} \\ 
g_{ab} &=& \delta_{ab} \bigl( 1 
- {\cal E}_{cd} \hat{y}^c \hat{y}^d  \bigr)  + O(s^3). 
\label{8.6.10}
\end{eqnarray} 

\section{Retarded coordinates}
\label{9}

\subsection{Geometrical elements}
\label{9.1}

We introduce the same geometrical elements as in Sec.~\ref{8}: we have
a timelike curve $\gamma$ described by relations $z^\mu(\tau)$, its
normalized tangent vector $u^\mu = dz^\mu/d\tau$, and its acceleration
vector $a^\mu = D u^\mu/d\tau$. We also have an orthonormal triad 
$\base{\mu}{a}$ that is FW transported on the world line according to   
\begin{equation}
\frac{D \base{\mu}{a}}{d \tau} = a_a u^{\mu},  
\label{9.1.1}
\end{equation}  
where $a_a(\tau) = a_{\mu} \base{\mu}{a}$ are the frame components of
the acceleration vector. Finally, we have a dual tetrad
$(\base{0}{\mu},\base{a}{\mu})$, with $\base{0}{\mu} = -u_\mu$ and
$\base{a}{\mu} = \delta^{ab} g_{\mu\nu} \base{\nu}{b}$. The tetrad and
its dual give rise to the completeness relations  
\begin{equation}
g^{\mu\nu} = -u^{\mu} u^{\nu} 
+ \delta^{ab} \base{\mu}{a} \base{\nu}{b}, \qquad
g_{\mu\nu} = -\base{0}{\mu} \base{0}{\nu} 
+ \delta_{ab}\, \base{a}{\mu} \base{b}{\nu}, 
\label{9.1.2}
\end{equation} 
which are the same as in Eq.~(\ref{8.2.3}). 

The Fermi normal coordinates of Sec.~\ref{8} were constructed on the
basis of a spacelike geodesic connecting a field point $x$ to the
world line. The retarded coordinates are based
instead on a {\it null geodesic} going from the world line to the
field point. We thus let $x$ be within the normal convex neighbourhood
of $\gamma$, $\beta$ be the unique future-directed null geodesic that
goes from the world line to $x$, and $x' := z(u)$ be the point at
which $\beta$ intersects the world line, with $u$ denoting the value
of the proper-time parameter at this point.   

From the tetrad at $x'$ we obtain another tetrad
$(\base{\alpha}{0},\base{\alpha}{a})$ at $x$ by parallel transport 
on $\beta$. By raising the frame index and lowering the vectorial
index we also obtain a dual tetrad at $x$: $\base{0}{\alpha} =
-g_{\alpha\beta} \base{\beta}{0}$ and $\base{a}{\alpha} = \delta^{ab}
g_{\alpha\beta} \base{\beta}{b}$. The metric at $x$ can be then be
expressed as 
\begin{equation}
g_{\alpha\beta} = -\base{0}{\alpha} \base{0}{\beta} + \delta_{ab} 
\base{a}{\alpha} \base{b}{\beta},   
\label{9.1.3}
\end{equation} 
and the parallel propagator from $x'$ to $x$ is given by 
\begin{equation} 
g^{\alpha}_{\ \alpha'}(x,x') = -\base{\alpha}{0} u_{\alpha'} + 
\base{\alpha}{a} \base{a}{\alpha'}, \qquad 
g^{\alpha'}_{\ \alpha}(x',x) = u^{\alpha'} \base{0}{\alpha} + 
\base{\alpha'}{a} \base{a}{\alpha}. 
\label{9.1.4}  
\end{equation} 
 
\subsection{Definition of the retarded coordinates} 
\label{9.2}

The quasi-Cartesian version of the retarded coordinates are defined by  
\begin{equation}
\hat{x}^0 = u, \qquad
\hat{x}^a = -\base{a}{\alpha'}(x') \sigma^{\alpha'}(x,x'), \qquad 
\sigma(x,x') = 0; 
\label{9.2.1}
\end{equation}
the last statement indicates that $x'$ and $x$ are linked by a null
geodesic. From the fact that $\sigma^{\alpha'}$ is a null vector we
obtain  
\begin{equation} 
r := (\delta_{ab} \hat{x}^a \hat{x}^b)^{1/2} 
= u_{\alpha'} \sigma^{\alpha'}, 
\label{9.2.2}
\end{equation} 
and $r$ is a positive quantity by virtue of the fact that $\beta$ is a 
future-directed null geodesic --- this makes $\sigma^{\alpha'}$
past-directed. In flat spacetime, $\sigma^{\alpha'} =
-(x-x')^{\alpha}$, and in a Lorentz frame that is momentarily comoving
with the world line, $r = t-t' > 0$; with the speed of light set equal
to unity, $r$ is also the spatial distance between $x'$ and $x$ as
measured in this frame. In curved spacetime, the quantity $r =
u_{\alpha'} \sigma^{\alpha'}$ can still be called the {\it retarded 
distance} between the point $x$ and the world line. Another
consequence of Eq.~(\ref{9.2.1}) is that 
\begin{equation}
\sigma^{\alpha'} = -r \bigl( u^{\alpha'} + \Omega^a \base{\alpha'}{a}
\bigr), 
\label{9.2.3}
\end{equation}
where $\Omega^a := \hat{x}^a/r$ is a unit spatial vector that
satisfies $\delta_{ab} \Omega^a \Omega^b = 1$. 

A straightforward calculation reveals that under a displacement of the
point $x$, the retarded coordinates change according to  
\begin{equation} 
d u = -k_\alpha\, d x^\alpha, \qquad 
d \hat{x}^a = - \bigl( r a^a - \omega^a_{\ b} \hat{x}^b 
+ \base{a}{\alpha'} \sigma^{\alpha'}_{\ \beta'} u^{\beta'} 
\bigr)\, du 
- \base{a}{\alpha'} \sigma^{\alpha'}_{\ \beta}\, d x^\beta, 
\label{9.2.4}
\end{equation}
where $k_\alpha = \sigma_{\alpha}/r$ is a future-directed null vector
at $x$ that is tangent to the geodesic $\beta$. To obtain these
results we must keep in mind that a displacement of $x$ typically
induces a simultaneous displacement of $x'$ because the new points 
$x + \delta x$ and $x' + \delta x'$ must also be linked by a null 
geodesic. We therefore have $0 = \sigma(x+\delta x, x' + \delta x') 
= \sigma_{\alpha}\, \delta x^\alpha + \sigma_{\alpha'}\, 
\delta x^{\alpha'}$, and the first relation of Eq.~(\ref{9.2.4}) 
follows from the fact that a displacement along the world line is
described by $\delta x^{\alpha'} = u^{\alpha'}\, \delta u$.  
 
\subsection{The scalar field $r(x)$ and the vector field
$k^\alpha(x)$}  
\label{9.3}

If we keep $x'$ linked to $x$ by the relation $\sigma(x,x') = 0$, then 
the quantity 
\begin{equation}
r(x) = \sigma_{\alpha'}(x,x') u^{\alpha'}(x') 
\label{9.3.1}
\end{equation}
can be viewed as an ordinary scalar field defined in a neighbourhood
of $\gamma$. We can compute the gradient of $r$ by finding how $r$
changes under a displacement of $x$ (which again induces a
displacement of $x'$). The result is 
\begin{equation} 
\partial_\beta r = - \bigl( \sigma_{\alpha'} a^{\alpha'} +
\sigma_{\alpha'\beta'} u^{\alpha'} u^{\beta'} \bigr) k_\beta +
\sigma_{\alpha'\beta} u^{\alpha'}.
\label{9.3.2}
\end{equation} 

Similarly, we can view 
\begin{equation} 
k^{\alpha}(x) = \frac{\sigma^{\alpha}(x,x')}{r(x)} 
\label{9.3.3}
\end{equation}
as an ordinary vector field, which is tangent to the congruence of
null geodesics that emanate from $x'$. It is easy to check that this 
vector satisfies the identities 
\begin{equation} 
\sigma_{\alpha\beta} k^\beta = k_\alpha, \qquad 
\sigma_{\alpha'\beta} k^\beta = \frac{\sigma_{\alpha'}}{r}, 
\label{9.3.4}
\end{equation}
from which we also obtain $\sigma_{\alpha'\beta} u^{\alpha'} k^\beta 
= 1$. From this last result and Eq.~(\ref{9.3.2}) we deduce the
important relation  
\begin{equation}
k^\alpha \partial_\alpha r = 1. 
\label{9.3.5}
\end{equation} 
In addition, combining the general statement $\sigma^{\alpha} = 
-g^{\alpha}_{\ \alpha'} \sigma^{\alpha'}$ --- cf.\ Eq.~(\ref{4.2.8})
--- with Eq.~(\ref{9.2.3}) gives
\begin{equation} 
k^\alpha = g^{\alpha}_{\ \alpha'} \bigl( u^{\alpha'} + \Omega^a
\base{\alpha'}{a} \bigr); 
\label{9.3.6}
\end{equation} 
the vector at $x$ is therefore obtained by parallel transport of
$u^{\alpha'} + \Omega^a \base{\alpha'}{a}$ on $\beta$. From this and
Eq.~(\ref{9.1.4}) we get the alternative expression 
\begin{equation} 
k^\alpha = \base{\alpha}{0} + \Omega^a \base{\alpha}{a},  
\label{9.3.7}
\end{equation} 
which confirms that $k^\alpha$ is a future-directed null vector field
(recall that $\Omega^a = \hat{x}^a/r$ is a unit vector).    

The covariant derivative of $k_\alpha$ can be computed by finding 
how the vector changes under a displacement of $x$. (It is in fact
easier to calculate first how $r k_\alpha$ changes, and then
substitute our previous expression for $\partial_\beta r$.) The result
is  
\begin{equation} 
r k_{\alpha;\beta} = \sigma_{\alpha\beta} - k_{\alpha} \sigma_{\beta
\gamma'} u^{\gamma'} - k_{\beta} \sigma_{\alpha \gamma'} u^{\gamma'} +
\bigl( \sigma_{\alpha'} a^{\alpha'} + \sigma_{\alpha'\beta'}
u^{\alpha'} u^{\beta'} \bigr) k_{\alpha} k_{\beta}. 
\label{9.3.8}
\end{equation} 
From this we infer that $k^\alpha$ satisfies the geodesic equation in
affine-parameter form, $k^\alpha_{\ ;\beta} k^\beta = 0$, and
Eq.~(\ref{9.3.5}) informs us that the affine parameter is in fact
$r$. A displacement along a member of the congruence is therefore
given by $dx^\alpha = k^\alpha\, dr$. Specializing to retarded
coordinates, and using Eqs.~(\ref{9.2.4}) and (\ref{9.3.4}), we find
that this statement becomes $du = 0$ and $d\hat{x}^a = (\hat{x}^a/r)\,
dr$, which integrate to $u = \mbox{constant}$ and $\hat{x}^a = r
\Omega^a$, respectively, with $\Omega^a$ still denoting a constant
unit vector. We have found that the congruence of null geodesics
emanating from $x'$ is described by   
\begin{equation}
u = \mbox{constant}, \qquad
\hat{x}^a = r \Omega^a(\theta^A)
\label{9.3.9}
\end{equation} 
in the retarded coordinates. Here, the two angles $\theta^A$ ($A = 1,
2$) serve to parameterize the unit vector $\Omega^a$, which is
independent of $r$.       

Equation (\ref{9.3.8}) also implies that the expansion of the
congruence is given by 
\begin{equation} 
\theta = k^\alpha_{\ ;\alpha} = \frac{\sigma^\alpha_{\ \alpha} 
- 2}{r}.
\label{9.3.10}
\end{equation} 
Using Eq.~(\ref{5.2.4}), we find that this becomes $r\theta = 2 
- \frac{1}{3} R_{\alpha'\beta'} \sigma^{\alpha'} \sigma^{\beta'} 
+ O(r^3)$, or 
\begin{equation}
r \theta =  
2 - \frac{1}{3} r^2 \bigl( R_{00} + 2 R_{0a} \Omega^a + R_{ab}
\Omega^a \Omega^b \bigr) + O(r^3)
\label{9.3.11} 
\end{equation}
after using Eq.~(\ref{9.2.3}). Here, 
$R_{00} = R_{\alpha'\beta'} u^{\alpha'} u^{\beta'}$,
$R_{0a} = R_{\alpha'\beta'} u^{\alpha'} \base{\beta'}{a}$, and  
$R_{ab} = R_{\alpha'\beta'} \base{\alpha'}{a} \base{\beta'}{b}$
are the frame components of the Ricci tensor evaluated at $x'$. This
result confirms that the congruence is singular at $r=0$, because
$\theta$ diverges as $2/r$ in this limit; the caustic coincides with
the point $x'$.     

Finally, we infer from Eq.~(\ref{9.3.8}) that $k^{\alpha}$ is 
hypersurface orthogonal. This, together with the property that
$k^\alpha$ satisfies the geodesic equation in affine-parameter form, 
implies that there exists a scalar field $u(x)$ such that 
\begin{equation}
k_\alpha = -\partial_\alpha u. 
\label{9.3.12}
\end{equation} 
This scalar field was already identified in Eq.~(\ref{9.2.4}): it is 
numerically equal to the proper-time parameter of the world line at
$x'$. We conclude that the geodesics to which $k^\alpha$ is
tangent are the generators of the null cone $u = \mbox{constant}$. As 
Eq.~(\ref{9.3.9}) indicates, a specific generator is selected by
choosing a direction $\Omega^a$ (which can be parameterized by two
angles $\theta^A$), and $r$ is an affine parameter on each
generator. The geometrical meaning of the retarded coordinates is now
completely clear; it is illustrated in Fig.~7. 

\begin{figure}
\begin{center}
\vspace*{-20pt} 
\includegraphics[width=0.5\linewidth]{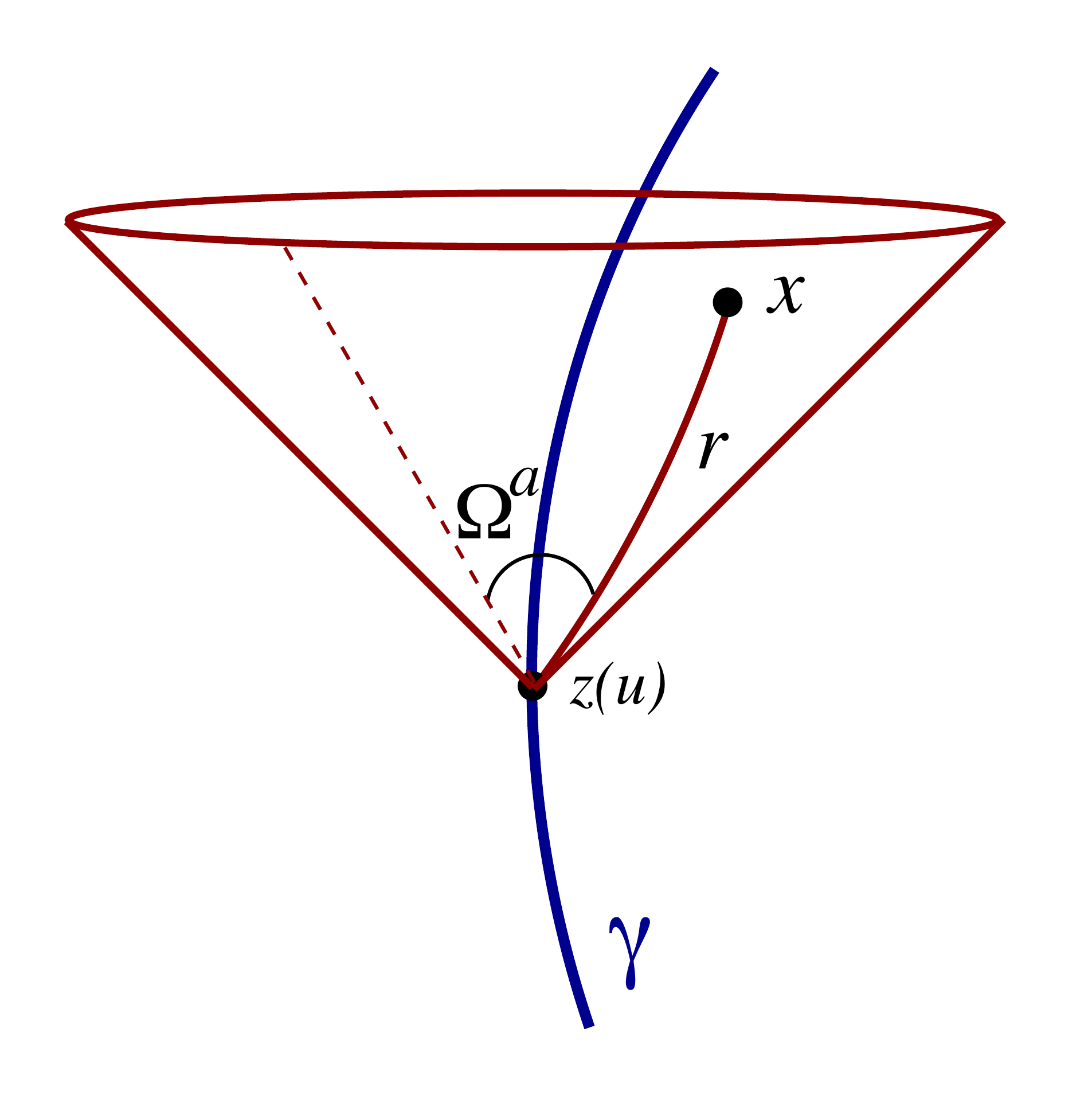}
\vspace*{-20pt}
\end{center} 
\caption{Retarded coordinates of a point $x$ relative to a world
line $\gamma$. The retarded time $u$ selects a particular null cone,
the unit vector $\Omega^a := \hat{x}^a/r$ selects a particular
generator of this null cone, and the retarded distance $r$ selects a
particular point on this generator. This figure is identical to
Fig.~4.} 
\end{figure} 

\subsection{Frame components of tensor fields on the world line}   
\label{9.4}

The metric at $x$ in the retarded coordinates will be expressed in
terms of frame components of vectors and tensors evaluated on the
world line $\gamma$. For example, if $a^{\alpha'}$ is the acceleration
vector at $x'$, then as we have seen,  
\begin{equation}
a_a(u) = a_{\alpha'}\, \base{\alpha'}{a}
\label{9.4.1}
\end{equation}
are the frame components of the acceleration at proper time $u$.  

Similarly,   
\begin{equation}
R_{a0b0}(u) = R_{\alpha'\gamma'\beta'\delta'}\, \base{\alpha'}{a} 
u^{\gamma'} \base{\beta'}{b} u^{\delta'}, \quad
R_{a0bd}(u) = R_{\alpha'\gamma'\beta'\delta'}\, \base{\alpha'}{a}
u^{\gamma'} \base{\beta'}{b} \base{\delta'}{d}, \quad
R_{acbd}(u) = R_{\alpha'\gamma'\beta'\delta'}\, \base{\alpha'}{a}
\base{\gamma'}{c} \base{\beta'}{b} \base{\delta'}{d} 
\label{9.4.2}
\end{equation}
are the frame components of the Riemann tensor evaluated on
$\gamma$. From these we form the useful combinations  
\begin{eqnarray}
S_{ab}(u,\theta^A) &=& R_{a0b0} + R_{a0bc} \Omega^c 
+ R_{b0ac} \Omega^c + R_{acbd} \Omega^c \Omega^d = S_{ba}, 
\label{9.4.3} \\
S_{a}(u,\theta^A) &=& S_{ab}\Omega^b = R_{a0b0} \Omega^b - R_{ab0c} 
\Omega^b \Omega^c, 
\label{9.4.4} \\
S(u,\theta^A) &=& S_{a} \Omega^a = R_{a0b0} \Omega^a \Omega^b, 
\label{9.4.5}
\end{eqnarray}
in which the quantities $\Omega^a := \hat{x}^a / r$ depend on the 
angles $\theta^A$ only --- they are independent of $u$ and $r$. 

We have previously introduced the frame components of the Ricci tensor
in Eq.~(\ref{9.3.11}). The identity  
\begin{equation}    
R_{00} + 2 R_{0a} \Omega^a + R_{ab} \Omega^a \Omega^b 
= \delta^{ab} S_{ab} - S
\label{9.4.6}
\end{equation}
follows easily from Eqs.~(\ref{9.4.3})--(\ref{9.4.5}) and the
definition of the Ricci tensor. 

In Sec.~\ref{8} we saw that the frame components of a given tensor
were also the components of this tensor (evaluated on the world line)
in the Fermi normal coordinates. We should not expect this
property to be true also in the case of the retarded coordinates: 
{\it the frame components  of a tensor are not to be
identified with the components of this tensor in the retarded
coordinates}. The reason is that the retarded coordinates are in fact 
{\it singular} on the world line. As we shall see, they give rise to a 
metric that possesses a directional ambiguity at $r = 0$. (This can 
easily be seen in Minkowski spacetime by performing the coordinate
transformation $u = t - \sqrt{x^2+y^2+z^2}$.) Components of tensors
are therefore not defined on the world line, although they are
perfectly well defined for $r \neq 0$. Frame components, on the other
hand, are well defined both off and on the world line, and working
with them will eliminate any difficulty associated with the singular
nature of the retarded coordinates.      

\subsection{Coordinate displacements near $\gamma$} 
\label{9.5}

The changes in the quasi-Cartesian retarded coordinates under a
displacement of $x$ are given by Eq.~(\ref{9.2.4}). In these we
substitute the standard expansions for $\sigma_{\alpha'\beta'}$ and
$\sigma_{\alpha'\beta}$, as given by Eqs.~(\ref{5.2.1}) and
(\ref{5.2.2}), as well as Eqs.~(\ref{9.2.3}) and
(\ref{9.3.6}). After a straightforward (but fairly lengthy)
calculation, we obtain the following expressions for the coordinate
displacements:  
\begin{eqnarray}
du &=& \bigl( \base{0}{\alpha}\, dx^\alpha \bigr) 
- \Omega_a \bigl( \base{b}{\alpha}\, dx^\alpha \bigr),  
\label{9.5.1} \\ 
d\hat{x}^a &=& - \Bigl[ r a^a + \frac{1}{2} r^2 S^a + O(r^3) \Bigr]
\bigl( \base{0}{\alpha}\, dx^\alpha \bigr) 
\nonumber \\ 
& & \mbox{} + \Bigl[ \delta^a_{\ b} + \Bigl( r a^a 
+ \frac{1}{3} r^2 S^a \Bigr) \Omega_b + \frac{1}{6} r^2
S^a_{\ b} + O(r^3) \Bigr] \bigl( \base{b}{\alpha}\, dx^\alpha \bigr).    
\label{9.5.2}
\end{eqnarray} 
Notice that the result for $du$ is exact, but that $d\hat{x}^a$ is
expressed as an expansion in powers of $r$. 

These results can also be expressed in the form of gradients of the
retarded coordinates: 
\begin{eqnarray} 
\partial_\alpha u &=& \base{0}{\alpha} - \Omega_a \base{a}{\alpha},   
\label{9.5.3} \\ 
\partial_\alpha \hat{x}^a &=& - \Bigl[ r a^a 
+ \frac{1}{2} r^2 S^a + O(r^3) \Bigr] \base{0}{\alpha} 
\nonumber \\ & & \mbox{} 
+ \Bigl[ \delta^a_{\ b} + \Bigl( r a^a 
+ \frac{1}{3} r^2 S^a \Bigr) \Omega_b 
+ \frac{1}{6} r^2 S^a_{\ b} + O(r^3) \Bigr] \base{b}{\alpha}. 
\label{9.5.4}
\end{eqnarray} 
Notice that Eq.~(\ref{9.5.3}) follows immediately from
Eqs.~(\ref{9.3.7}) and (\ref{9.3.12}). From Eq.~(\ref{9.5.4}) and the 
identity $\partial_\alpha r = \Omega_a \partial_\alpha \hat{x}^a$ we
also infer 
\begin{equation} 
\partial_\alpha r = - \Bigl[ r a_a \Omega^a + \frac{1}{2} r^2 S 
+ O(r^3) \Bigr] \base{0}{\alpha} + \Bigl[ \Bigl(1 + r a_b \Omega^b 
+ \frac{1}{3} r^2 S \Bigr) \Omega_a + \frac{1}{6} r^2 S_a + O(r^3)
\Bigr] \base{a}{\alpha}, 
\label{9.5.5}
\end{equation} 
where we have used the facts that $S_a = S_{ab} \Omega^b$ and $S = S_a 
\Omega^a$; these last results were derived in Eqs.~(\ref{9.4.4}) and
(\ref{9.4.5}). It may be checked that Eq.~(\ref{9.5.5}) agrees with
Eq.~(\ref{9.3.2}).  
    
\subsection{Metric near $\gamma$}   
\label{9.6}

It is straightforward (but fairly tedious) to invert the relations of
Eqs.~(\ref{9.5.1}) and (\ref{9.5.2}) and solve for $\base{0}{\alpha}\, 
dx^\alpha$ and $\base{a}{\alpha}\, dx^\alpha$. The results are  
\begin{eqnarray} 
\base{0}{\alpha}\, dx^\alpha &=& \Bigl[ 1 + r a_a \Omega^a + \frac{1}{2} 
r^2 S + O(r^3) \Bigr]\, du + \Bigl[ \Bigl( 1 + \frac{1}{6} r^2 S
\Bigr) \Omega_a - \frac{1}{6} r^2 S_a + O(r^3) \Bigr]\, d\hat{x}^a, 
\label{9.6.1} \\ 
\base{a}{\alpha}\, dx^\alpha &=& \Bigl[ r a^a 
+ \frac{1}{2} r^2 S^a + O(r^3) \Bigr]\, du 
+ \Bigl[ \delta^a_{\ b} - \frac{1}{6} r^2 S^a_{\ b} 
+ \frac{1}{6} r^2 S^a \Omega_b + O(r^3) \Bigr]\, d\hat{x}^b.
\label{9.6.2}
\end{eqnarray}  
These relations, when specialized to the retarded coordinates, give us 
the components of the dual tetrad $(\base{0}{\alpha},
\base{a}{\alpha})$ at $x$. The metric is then computed by using the 
completeness relations of Eq.~(\ref{9.1.3}). We find 
\begin{eqnarray}
g_{uu} &=& - \bigl( 1 + r a_a \Omega^a \bigr)^2 + r^2 a^2
- r^2 S + O(r^3),  
\label{9.6.3} \\ 
g_{ua} &=& -\Bigl( 1 + r a_b \Omega^b + \frac{2}{3} r^2 S \Bigr) 
\Omega_a + r a_a + \frac{2}{3} r^2 S_a + O(r^3), 
\label{9.6.4} \\
g_{ab} &=& \delta_{ab} - \Bigl( 1 + \frac{1}{3} r^2 S \Bigr) \Omega_a 
\Omega_b - \frac{1}{3} r^2 S_{ab} + \frac{1}{3} r^2 \bigl( S_a
\Omega_b + \Omega_a S_b \bigr) + O(r^3),  
\label{9.6.5}
\end{eqnarray} 
where $a^2 := \delta_{ab} a^a a^b$. 
We see (as was pointed out in Sec.~\ref{9.4}) that the metric
possesses a directional ambiguity on the world line: the metric at
$r=0$ still depends on the vector $\Omega^a = \hat{x}^a/r$ that
specifies the direction to the point $x$. The retarded coordinates are
therefore singular on the world line, and tensor components cannot be
defined on $\gamma$. 

By setting $S_{ab} = S_a = S = 0$ in Eqs.~(\ref{9.6.3})--(\ref{9.6.5})
we obtain the metric of flat spacetime in the retarded
coordinates. This we express as  
\begin{eqnarray} 
\eta_{uu} &=& - \bigl( 1 + r a_a \Omega^a \bigr)^2 
+ r^2 a^2, \nonumber \\   
\eta_{ua} &=& - \bigl( 1 + r a_b \Omega^b \bigr) \Omega_a 
+ r a_a, 
\label{9.6.6} \\ 
\eta_{ab} &=& \delta_{ab} - \Omega_a \Omega_b. \nonumber 
\end{eqnarray}
In spite of the directional ambiguity, the metric of flat spacetime
has a unit determinant everywhere, and it is easily inverted:   
\begin{equation}
\eta^{uu} = 0, \qquad
\eta^{ua} = -\Omega^a, \qquad
\eta^{ab} = \delta^{ab} + r \bigl(a^a \Omega^b 
+ \Omega^a a^b \bigr).  
\label{9.6.7}
\end{equation}
The inverse metric also is ambiguous on the world line.  

To invert the curved-spacetime metric of
Eqs.~(\ref{9.6.3})--(\ref{9.6.5}) we express it as $g_{\alpha\beta} =
\eta_{\alpha\beta} + h_{\alpha\beta} + O(r^3)$ and treat
$h_{\alpha\beta} = O(r^2)$ as a perturbation. The inverse metric is
then $g^{\alpha\beta} = \eta^{\alpha\beta} - \eta^{\alpha\gamma}
\eta^{\beta\delta} h_{\gamma \delta} + O(r^3)$, or  
\begin{eqnarray} 
\hspace*{-15pt} g^{uu} &=& 0, 
\label{9.6.8} \\
\hspace*{-15pt} g^{ua} &=& -\Omega^a, 
\label{9.6.9} \\
\hspace*{-15pt} g^{ab} &=& \delta^{ab} + r \bigl( a^a \Omega^b 
+ \Omega^a a^b \bigr) + \frac{1}{3} r^2 S^{ab} 
+ \frac{1}{3} r^2 \bigl( S^a \Omega^b + \Omega^a S^b \bigr) + O(r^3).  
\label{9.6.10}
\end{eqnarray} 
The results for $g^{uu}$ and $g^{ua}$ are exact, and they follow from 
the general relations $g^{\alpha\beta} (\partial_\alpha u)
(\partial_\beta u) = 0$ and $g^{\alpha\beta} (\partial_\alpha u)
(\partial_\beta r) = -1$ that are derived from Eqs.~(\ref{9.3.5}) and 
(\ref{9.3.12}).  

The metric determinant is computed from $\sqrt{-g} = 1 + \frac{1}{2} 
\eta^{\alpha\beta} h_{\alpha\beta} + O(r^3)$, which gives 
\begin{equation}
\sqrt{-g} = 1 - \frac{1}{6} r^2 \bigl( \delta^{ab} S_{ab} - S \bigr) +
O(r^3) = 1 - \frac{1}{6} r^2 \bigl( R_{00} + 2 R_{0a} \Omega^a +
R_{ab} \Omega^a \Omega^b \bigr) + O(r^3), 
\label{9.6.11}
\end{equation}
where we have substituted the identity of
Eq.~(\ref{9.4.6}). Comparison with Eq.~(\ref{9.3.11}) gives us
the interesting relation $\sqrt{-g} = \frac{1}{2} r \theta + O(r^3)$,
where $\theta$ is the expansion of the generators of the null cones 
$u = \mbox{constant}$.  

\subsection{Transformation to angular coordinates} 
\label{9.7}

Because the vector $\Omega^a = \hat{x}^a/r$ satisfies $\delta_{ab}
\Omega^a \Omega^b = 1$, it can be parameterized by two angles
$\theta^A$. A canonical choice for the parameterization is $\Omega^a =
(\sin\theta \cos\phi, \sin\theta \sin\phi, \cos\theta)$. It is then
convenient to perform a coordinate transformation from $\hat{x}^a$ to 
$(r,\theta^A)$, using the relations $\hat{x}^a = r
\Omega^a(\theta^A)$. (Recall from Sec.~\ref{9.3} that the angles  
$\theta^A$ are constant on the generators of the null cones $u =
\mbox{constant}$, and that $r$ is an affine parameter on these
generators. The relations $\hat{x}^a = r \Omega^a$ therefore
describe the behaviour of the generators.) The differential form of
the coordinate transformation is     
\begin{equation}
d\hat{x}^a = \Omega^a\, dr + r \Omega^a_A\, d\theta^A, 
\label{9.7.1}
\end{equation}
where the transformation matrix 
\begin{equation} 
\Omega^a_A := \frac{\partial \Omega^a}{\partial \theta^A}
\label{9.7.2}
\end{equation} 
satisfies the identity $\Omega_a \Omega^a_A = 0$. 

We introduce the quantities 
\begin{equation}
\Omega_{AB} := \delta_{ab} \Omega^a_A \Omega^b_B, 
\label{9.7.3}
\end{equation}
which act as a (nonphysical) metric in the subspace spanned by the
angular coordinates. In the canonical parameterization, $\Omega_{AB} = 
\mbox{diag}(1,\sin^2\theta)$. We use the inverse of $\Omega_{AB}$, 
denoted $\Omega^{AB}$, to raise upper-case latin indices. We then
define the new object 
\begin{equation}
\Omega^A_a := \delta_{ab} \Omega^{AB} \Omega^b_B 
\label{9.7.4}
\end{equation} 
which satisfies the identities 
\begin{equation}
\Omega^A_a \Omega^a_B = \delta^A_B, \qquad
\Omega^a_A \Omega^A_b = \delta^a_{\ b} - \Omega^a \Omega_b. 
\label{9.7.5}
\end{equation} 
The second result follows from the fact that both sides are
simultaneously symmetric in $a$ and $b$, orthogonal to $\Omega_a$ and 
$\Omega^b$, and have the same trace.  

From the preceding results we establish that the transformation from
$\hat{x}^a$ to $(r,\theta^A)$ is accomplished by 
\begin{equation}
\frac{\partial \hat{x}^a}{\partial r} = \Omega^a, \qquad
\frac{\partial \hat{x}^a}{\partial \theta^A} = r \Omega^a_A, 
\label{9.7.6}
\end{equation}
while the transformation from $(r,\theta^A)$ to $\hat{x}^a$ is
accomplished by 
\begin{equation}
\frac{\partial r}{\partial \hat{x}^a} = \Omega_a, \qquad 
\frac{\partial \theta^A}{\partial \hat{x}^a} = \frac{1}{r}\Omega^A_a.  
\label{9.7.7}
\end{equation}
With these transformation rules it is easy to show that in the
angular coordinates, the metric is given by  
\begin{eqnarray}
g_{uu} &=& - \bigl( 1 + r a_a \Omega^a \bigr)^2 
+ r^2 a^2 - r^2 S + O(r^3),  
\label{9.7.8} \\
g_{ur} &=& -1, 
\label{9.7.9} \\ 
g_{uA} &=& r \Bigl[ r a_a  
+ \frac{2}{3} r^2 S_a + O(r^3) \Bigr] \Omega^a_A, 
\label{9.7.10} \\ 
g_{AB} &=& r^2 \Bigl[ \Omega_{AB} - \frac{1}{3} r^2 S_{ab} \Omega^a_A
\Omega^b_B + O(r^3) \Bigr]. 
\label{9.7.11}
\end{eqnarray} 
The results $g_{ru} = -1$, $g_{rr} = 0$, and $g_{rA} = 0$ are exact,
and they follow from the fact that in the retarded coordinates,
$k_\alpha\, dx^\alpha = - du$ and $k^\alpha \partial_\alpha =
\partial_r$.   

The nonvanishing components of the inverse metric are  
\begin{eqnarray}
g^{ur} &=& -1, 
\label{9.7.12} \\ 
g^{rr} &=& 1 + 2r a_a \Omega^a + r^2 S + O(r^3),   
\label{9.7.13} \\ 
g^{rA} &=& \frac{1}{r} \Bigl[ r a^a  + \frac{2}{3}r^2 S^a 
+ O(r^3) \Bigr] \Omega^A_a, 
\label{9.7.14} \\
g^{AB} &=& \frac{1}{r^2} \Bigl[ \Omega^{AB} + \frac{1}{3} r^2 S^{ab}
\Omega^A_a \Omega^B_b + O(r^3) \Bigr].  
\label{9.7.15}
\end{eqnarray} 
The results $g^{uu} = 0$, $g^{ur} = -1$, and $g^{uA} = 0$ are exact,
and they follow from the same reasoning as before.  

Finally, we note that in the angular coordinates, the metric
determinant is given by  
\begin{equation}
\sqrt{-g} = r^2 \sqrt{\Omega} \Bigl[ 1 - \frac{1}{6} r^2 \bigl( R_{00}
+ 2 R_{0a} \Omega^a + R_{ab} \Omega^a \Omega^b \bigr) + O(r^3) \Bigr], 
\label{9.7.16}
\end{equation}
where $\Omega$ is the determinant of $\Omega_{AB}$; in the canonical 
parameterization, $\sqrt{\Omega} = \sin\theta$. 
 
\subsection{Specialization to $a^\mu = 0 = R_{\mu\nu}$}  
\label{9.8} 

In this subsection we specialize our previous results to a situation 
where $\gamma$ is a geodesic on which the Ricci tensor vanishes. We
therefore set $a^\mu = 0 = R_{\mu\nu}$ everywhere on $\gamma$.        

We have seen in Sec.~\ref{8.6} that when the Ricci tensor vanishes on
$\gamma$, all frame components of the Riemann tensor can be expressed
in terms of the symmetric-tracefree tensors 
${\cal E}_{ab}(u)$ and ${\cal B}_{ab}(u)$. The relations are 
$R_{a0b0} = {\cal E}_{ab}$, $R_{a0bc} = \varepsilon_{bcd} 
{\cal B}^d_{\ a}$, and $R_{acbd} = \delta_{ab} {\cal E}_{cd} 
+ \delta_{cd} {\cal E}_{ab} - \delta_{ad} {\cal E}_{bc} 
- \delta_{bc} {\cal E}_{ad}$. These can be substituted into 
Eqs.~(\ref{9.4.3})--(\ref{9.4.5}) to give 
\begin{eqnarray}
S_{ab}(u,\theta^A) &=& 2 {\cal E}_{ab} 
- \Omega_a {\cal E}_{bc} \Omega^c 
- \Omega_b {\cal E}_{ac} \Omega^c 
+ \delta_{ab} {\cal E}_{bc} \Omega^c \Omega^d 
+ \varepsilon_{acd} \Omega^c {\cal B}^d_{\ b} 
+ \varepsilon_{bcd} \Omega^c {\cal B}^d_{\ a}, 
\label{9.8.1} \\ 
S_{a}(u,\theta^A) &=& {\cal E}_{ab} \Omega^b 
+ \varepsilon_{abc} \Omega^b {\cal B}^c_{\ d} \Omega^d, 
\label{9.8.2} \\ 
S(u,\theta^A) &=& {\cal E}_{ab} \Omega^a \Omega^b.  
\label{9.8.3}
\end{eqnarray} 
In these expressions the dependence on retarded time $u$ is contained
in ${\cal E}_{ab}$ and ${\cal B}_{ab}$, while the angular dependence
is encoded in the unit vector $\Omega^a$. 

It is convenient to introduce the irreducible quantities 
\begin{eqnarray}  
{\cal E}^* &:=& {\cal E}_{ab} \Omega^a \Omega^b, 
\label{9.8.4} \\ 
{\cal E}^*_a &:=& \bigl(\delta_a^{\ b} - \Omega_a \Omega^b \bigr) 
{\cal E}_{bc} \Omega^c,
\label{9.8.5} \\ 
{\cal E}^*_{ab} &:=& 2 {\cal E}_{ab} 
- 2\Omega_a {\cal E}_{bc} \Omega^c  
- 2\Omega_b {\cal E}_{ac} \Omega^c
+ (\delta_{ab} + \Omega_a \Omega_b) {\cal E}^*, 
\label{9.8.6} \\ 
{\cal B}^*_a &:=& \varepsilon_{abc} \Omega^b {\cal B}^c_{\ d} \Omega^d, 
\label{9.8.7} \\ 
{\cal B}^*_{ab} &:=& \varepsilon_{acd} \Omega^c {\cal B}^d_{\ e} 
\bigl(\delta^e_{\ b} - \Omega^e \Omega_b \bigr) 
+ \varepsilon_{bcd} \Omega^c {\cal B}^d_{\ e} 
\bigl(\delta^e_{\ a} - \Omega^e \Omega_a \bigr).  
\label{9.8.8}
\end{eqnarray}    
These are all orthogonal to $\Omega^a$: ${\cal E}^*_a \Omega^a = 
{\cal B}^*_a \Omega^a = 0$ and ${\cal E}^*_{ab} \Omega^b = 
{\cal B}^*_{ab} \Omega^b = 0$. In terms of these
Eqs.~(\ref{9.8.1})--(\ref{9.8.3}) become 
\begin{eqnarray} 
S_{ab} &=& {\cal E}_{ab}^* + \Omega_a {\cal E}^*_b + {\cal E}^*_a 
\Omega_b + \Omega_a \Omega_b {\cal E}^* + {\cal B}^*_{ab} 
+ \Omega_a {\cal B}^*_b + {\cal B}^*_a \Omega_b, 
\label{9.8.9} \\ 
S_{a} &=& {\cal E}^*_a + \Omega_a {\cal E}^* + {\cal B}^*_a, 
\label{9.8.10} \\ 
S &=& {\cal E}^*. 
\label{9.8.11}
\end{eqnarray} 

When Eqs.~(\ref{9.8.9})--(\ref{9.8.11}) are substituted into the
metric tensor of Eqs.~(\ref{9.6.3})--(\ref{9.6.5}) --- in which $a_a$
is set equal to zero --- we obtain the compact expressions  
\begin{eqnarray}  
g_{uu} &=& - 1 - r^2 {\cal E}^* + O(r^3),
\label{9.8.12} \\ 
g_{ua} &=& -\Omega_a + \frac{2}{3} r^2 \bigl( {\cal E}^*_a 
+ {\cal B}^*_a \bigr) + O(r^3), 
\label{9.8.13} \\ 
g_{ab} &=& \delta_{ab} - \Omega_a \Omega_b - \frac{1}{3} r^2  
\bigl( {\cal E}^*_{ab} + {\cal B}^*_{ab} \bigr) + O(r^3). 
\label{9.8.14}
\end{eqnarray}   
The metric becomes 
\begin{eqnarray} 
g_{uu} &=& - 1 - r^2 {\cal E}^* + O(r^3),
\label{9.8.15} \\
g_{ur} &=& - 1, 
\label{9.8.16} \\
g_{uA} &=& \frac{2}{3} r^3 \bigl( {\cal E}^*_A 
+ {\cal B}^*_A \bigr) + O(r^4), 
\label{9.8.17} \\ 
g_{AB} &=& r^2 \Omega_{AB} - \frac{1}{3} r^4  
\bigl( {\cal E}^*_{AB} + {\cal B}^*_{AB} \bigr) + O(r^5)  
\label{9.8.18}
\end{eqnarray}   
after transforming to angular coordinates using the rules of 
Eq.~(\ref{9.7.6}). Here we have introduced the projections 
\begin{eqnarray} 
{\cal E}^*_A &:=& {\cal E}^*_{a} \Omega^a_A 
= {\cal E}_{ab} \Omega^a_A \Omega^b, 
\label{9.8.19} \\
{\cal E}^*_{AB} &:=& {\cal E}^*_{ab} \Omega^a_A \Omega^b_B 
= 2 {\cal E}_{ab} \Omega^a_A \Omega^b_B + {\cal E}^* \Omega_{AB}, 
\label{9.8.20} \\
{\cal B}^*_A &:=& {\cal B}^*_{a} \Omega^a_A 
= \varepsilon_{abc} \Omega^a_A \Omega^b {\cal B}^c_{\ d} \Omega^d, 
\label{9.8.21} \\
{\cal B}^*_{AB} &:=& {\cal B}^*_{ab} \Omega^a_A \Omega^b_B 
= 2 \varepsilon_{acd} \Omega^c {\cal B}^d_{\ b} \Omega^{a}_{(A}  
\Omega^{b}_{B)}.  
\label{9.8.22}
\end{eqnarray} 
It may be noted that the inverse relations are ${\cal E}^*_a    
= {\cal E}^*_{A} \Omega^A_a$, ${\cal B}^*_a 
= {\cal B}^*_{A} \Omega^A_a$, ${\cal E}^*_{ab} 
= {\cal E}^*_{AB} \Omega^A_a \Omega^B_b$, and ${\cal B}^*_{ab}  
= {\cal B}^*_{AB} \Omega^A_a \Omega^B_b$, where $\Omega^A_a$ was
introduced in Eq.~(\ref{9.7.4}). 
  
\section{Transformation between Fermi and retarded coordinates;
advanced point}  
\label{10}

A point $x$ in the normal convex neighbourhood of a world line
$\gamma$ can be assigned a set of Fermi normal coordinates
(as in Sec.~\ref{8}), or it can be assigned a set of retarded
coordinates (Sec.~\ref{9}). These coordinate systems can obviously be 
related to one another, and our first task in this section (which will
occupy us in Secs.~\ref{10.1}--\ref{10.3}) will be to derive the
transformation rules. We begin by refining our notation so as to
eliminate any danger of ambiguity.    

\begin{figure}
\begin{center}
\vspace*{-20pt} 
\includegraphics[width=0.5\linewidth]{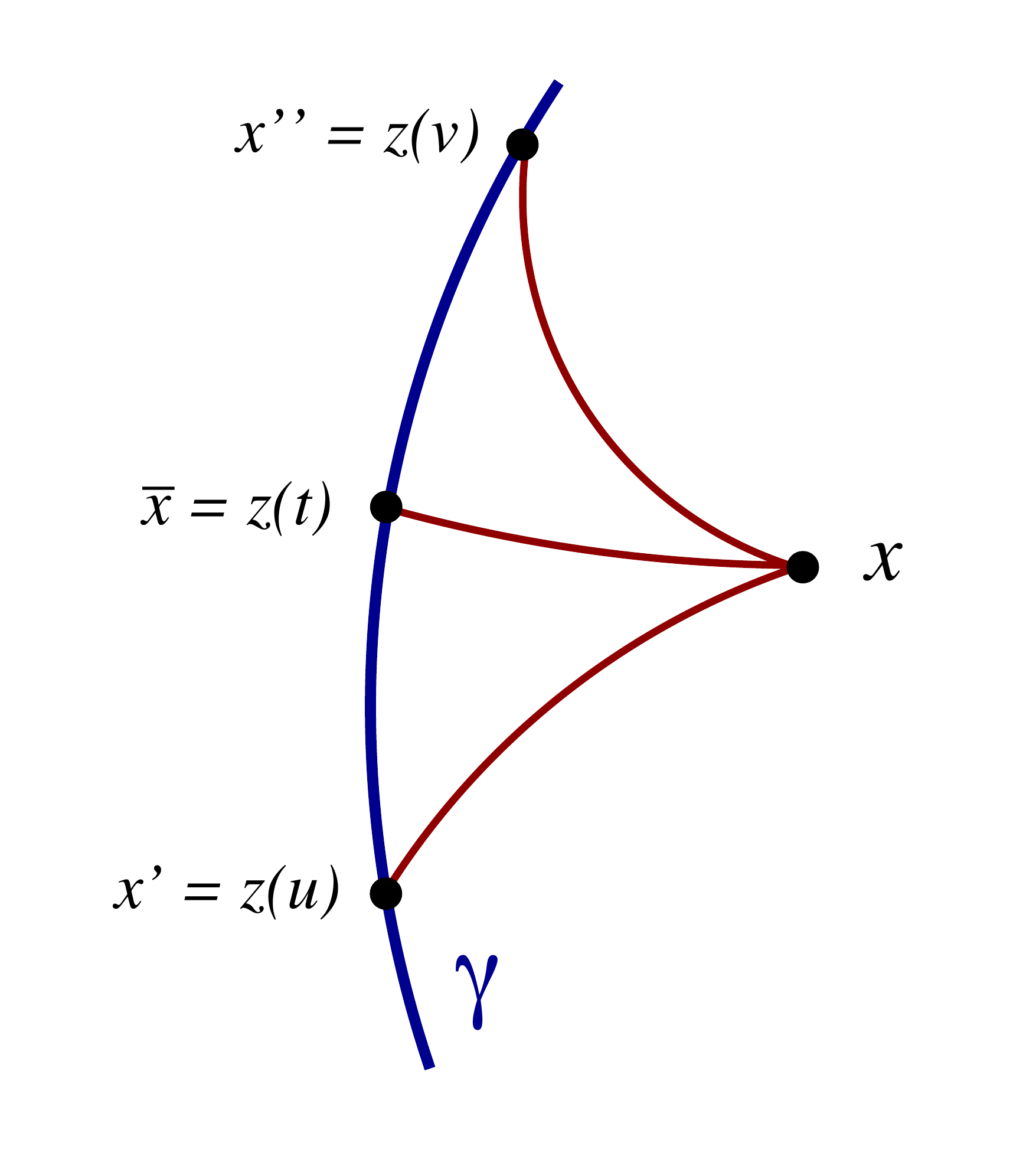}
\vspace*{-20pt}
\end{center} 
\caption{The retarded, simultaneous, and advanced points on a world
line $\gamma$. The retarded point $x' := z(u)$ is linked to $x$ by
a future-directed null geodesic. The simultaneous point $\bar{x}
:= z(t)$ is linked to $x$ by a spacelike geodesic that intersects
$\gamma$ orthogonally. The advanced point $x'' := z(v)$ is linked
to $x$ by a past-directed null geodesic.}  
\end{figure} 

The Fermi normal coordinates of $x$ refer to a point $\bar{x} :=
z(t)$ on $\gamma$ that is related to $x$ by a spacelike geodesic that
intersects $\gamma$ orthogonally; see Fig.~8. We refer to this point
as $x$'s {\it simultaneous point}, and to tensors at $\bar{x}$ we
assign indices $\bar{\alpha}$, $\bar{\beta}$, etc. We let
$(t,s\omega^a)$ be the Fermi normal coordinates of $x$, with $t$
denoting the value of $\gamma$'s proper-time parameter at $\bar{x}$,
$s = \sqrt{2\sigma(x,\bar{x})}$ representing the proper distance from 
$\bar{x}$ to $x$ along the spacelike geodesic, and $\omega^a$ denoting
a unit vector ($\delta_{a b} \omega^a \omega^b = 1$) that determines
the direction of the geodesic. The Fermi normal coordinates are
defined by $s \omega^a = -\base{a}{\bar{\alpha}}
\sigma^{\bar{\alpha}}$ and $\sigma_{\bar{\alpha}} u^{\bar{\alpha}} =
0$. Finally, we denote by $(\bar{e}^{\alpha}_0, \bar{e}^\alpha_a)$ the
tetrad at $x$ that is obtained by parallel transport of
$(u^{\bar{\alpha}}, \base{\bar{\alpha}}{a})$ on the spacelike
geodesic.  

The retarded coordinates of $x$ refer to a point $x' := z(u)$ on
$\gamma$ that is linked to $x$ by a future-directed null geodesic; see
Fig.~8. We refer to this point as $x$'s {\it retarded point}, and to
tensors at $x'$ we assign indices $\alpha'$, $\beta'$, etc. We let 
$(u,r\Omega^a)$ be the retarded coordinates of $x$, with $u$ denoting
the value of $\gamma$'s proper-time parameter at $x'$, $r =
\sigma_{\alpha'} u^{\alpha'}$ representing the affine-parameter
distance from $x'$ to $x$ along the null geodesic, and $\Omega^a$
denoting a unit vector ($\delta_{a b} \Omega^a \Omega^b = 1$) that
determines the direction of the geodesic. The retarded coordinates are
defined by $r \Omega^a = -\base{a}{\alpha'} \sigma^{\alpha'}$ and
$\sigma(x,x') = 0$. Finally, we denote by
$(\base{\alpha}{0},\base{\alpha}{a})$ the tetrad at $x$ that is
obtained by parallel transport of $(u^{\alpha'}, \base{\alpha'}{a})$
on the null geodesic.     

The reader who does not wish to follow the details of this discussion
can be informed that: (i) our results concerning the transformation 
from the retarded coordinates $(u,r,\Omega^a)$ to the Fermi normal
coordinates $(t,s,\omega^a)$ are contained in
Eqs.~(\ref{10.1.1})--(\ref{10.1.3}) below; (ii) our results  
concerning the transformation from the Fermi normal coordinates
$(t,s,\omega^a)$ to the retarded coordinates $(u,r,\Omega^a)$ are
contained in Eqs.~(\ref{10.2.1})--(\ref{10.2.3}); (iii) the
decomposition of each member of $(\bar{e}^{\alpha}_0,
\bar{e}^\alpha_a)$ in the tetrad
$(\base{\alpha}{0},\base{\alpha}{a})$ is given in retarded coordinates
by Eqs.~(\ref{10.3.1}) and (\ref{10.3.2}); and (iv) the
decomposition of each member of $(\base{\alpha}{0},\base{\alpha}{a})$
in the tetrad $(\bar{e}^{\alpha}_0, \bar{e}^\alpha_a)$ is given in
Fermi normal coordinates by Eqs.~(\ref{10.3.3}) and (\ref{10.3.4}). 

Our final task will be to define, along with the retarded and
simultaneous points, an {\it advanced point} $x''$ on the world line 
$\gamma$; see Fig.~8. This is taken on in Sec.~\ref{10.4}. 
  
\subsection{From retarded to Fermi coordinates} 
\label{10.1}

Quantities at $\bar{x} := z(t)$ can be related to quantities at
$x' := z(u)$ by Taylor expansion along the world line $\gamma$. To
implement this strategy we must first find an expression for $\Delta
:= t - u$. (Although we use the same notation, this should not be
confused with the van Vleck determinant introduced in Sec.~\ref{6}.) 

Consider the function $p(\tau)$ of the proper-time parameter $\tau$
defined by 
\[
p(\tau) = \sigma_\mu\bigl(x,z(\tau)\bigr) u^\mu(\tau),  
\]
in which $x$ is kept fixed and in which $z(\tau)$ is an arbitrary
point on the world line. We have that $p(u) = r$ and $p(t) = 0$, and 
$\Delta$ can ultimately be obtained by expressing $p(t)$ as $p(u +
\Delta)$ and expanding in powers of $\Delta$. Formally, 
\[
p(t) = p(u) + \dot{p}(u) \Delta + \frac{1}{2} \ddot{p}(u) \Delta^2 +
\frac{1}{6} p^{(3)}(u) \Delta^3 + O(\Delta^4), 
\]
where overdots (or a number within brackets) indicate repeated 
differentiation with respect to $\tau$. We have  
\begin{eqnarray*} 
\dot{p}(u) &=& \sigma_{\alpha'\beta'} u^{\alpha'} u^{\beta'} +
\sigma_{\alpha'} a^{\alpha'}, \\
\ddot{p}(u) &=& \sigma_{\alpha'\beta'\gamma'} u^{\alpha'} u^{\beta'} 
u^{\gamma'} + 3 \sigma_{\alpha'\beta'} u^{\alpha'} a^{\beta'} 
+ \sigma_{\alpha'} \dot{a}^{\alpha'}, \\ 
p^{(3)}(u) &=& \sigma_{\alpha'\beta'\gamma'\delta'} u^{\alpha'}
u^{\beta'} u^{\gamma'} u^{\delta'} + \sigma_{\alpha'\beta'\gamma'}
\bigl( 5 a^{\alpha'} u^{\beta'} u^{\gamma'} + u^{\alpha'} u^{\beta'}
a^{\gamma'} \bigr) + \sigma_{\alpha'\beta'} \bigl( 3 a^{\alpha'}
a^{\beta'} + 4 u^{\alpha'} \dot{a}^{\beta'} \bigr) 
+ \sigma_{\alpha'} \ddot{a}^{\alpha'},
\end{eqnarray*}
where $a^{\mu} = D u^{\mu}/d\tau$, $\dot{a}^{\mu} = D a^{\mu}/d\tau$,
and $\ddot{a}^{\mu} = D \dot{a}^{\mu}/d\tau$. 

We now express all of this in retarded coordinates by invoking the 
expansion of Eq.~(\ref{5.2.1}) for $\sigma_{\alpha'\beta'}$ (as well
as additional expansions for the higher derivatives of the world 
function, obtained by further differentiation of this result) and the
relation $\sigma^{\alpha'} = - r(u^{\alpha'} + \Omega^a
\base{\alpha'}{a})$ first derived in Eq.~(\ref{9.2.3}). With a degree
of accuracy sufficient for our purposes we obtain   
\begin{eqnarray*} 
\dot{p}(u) &=& -\Bigl[ 1 + r a_a \Omega^a + \frac{1}{3} r^2 S + O(r^3)
\Bigr], \\ 
\ddot{p}(u) &=& - r\bigl( \dot{a}_0 + \dot{a}_a \Omega^a \bigr) +
O(r^2), \\ 
p^{(3)}(u) &=& \dot{a}_0 + O(r), 
\end{eqnarray*}
where $S = R_{a0b0} \Omega^a \Omega^b$ was first introduced in
Eq.~(\ref{9.4.5}), and where $\dot{a}_0 := \dot{a}_{\alpha'} 
u^{\alpha'}$, $\dot{a}_a := \dot{a}_{\alpha'} \base{\alpha'}{a}$
are the frame components of the covariant derivative of the
acceleration vector. To arrive at these results we made use of the 
identity $a_{\alpha'} a^{\alpha'} + \dot{a}_{\alpha'} u^{\alpha'} = 0$
that follows from the fact that $a^\mu$ is orthogonal to
$u^\mu$. Notice that there is no distinction between the two possible 
interpretations $\dot{a}_a := d a_a / d\tau$ and $\dot{a}_a :=
\dot{a}_\mu \base{\mu}{a}$ for the quantity $\dot{a}_a(\tau)$; their
equality follows at once from the substitution of 
$D \base{\mu}{a}/d\tau = a_a u^\mu$ (which states that the basis
vectors are Fermi-Walker transported on the world line) into the
identity $d a_a / d\tau = D(a_\nu \base{\nu}{a})/d\tau$.       

Collecting our results we obtain
\[
r = \Bigl[ 1 + r a_a \Omega^a + \frac{1}{3} r^2 S + O(r^3)
\Bigr] \Delta + \frac{1}{2} r \Bigl[ \dot{a}_0 + \dot{a}_a \Omega^a +
O(r) \Bigr] \Delta^2 - \frac{1}{6} \Bigl[ \dot{a}_0 + O(r) \Bigr]
\Delta^3 + O(\Delta^4), 
\] 
which can readily be solved for $\Delta := t - u$ expressed as an
expansion in powers of $r$. The final result is  
\begin{equation} 
t = u + r \biggl\{ 1 - r a_a(u) \Omega^a + r^2 \bigl[ a_a(u)
\Omega^a \bigr]^2 - \frac{1}{3} r^2 \dot{a}_0(u) - \frac{1}{2} r^2 
\dot{a}_a(u) \Omega^a - \frac{1}{3} r^2 R_{a0b0}(u) \Omega^a \Omega^b
+ O(r^3) \biggr\},  
\label{10.1.1}
\end{equation}
where we show explicitly that all frame components are evaluated at
the retarded point $z(u)$.  

To obtain relations between the spatial coordinates we consider the
functions 
\[
p_a(\tau) = -\sigma_\mu\bigl(x,z(\tau)\bigr) \base{\mu}{a}(\tau),  
\]
in which $x$ is fixed and $z(\tau)$ is an arbitrary point on
$\gamma$. We have that the retarded coordinates are given by $r
\Omega^a = p^a(u)$, while the Fermi coordinates are given instead by
$s \omega^a = p^a(t) = p^a(u + \Delta)$. This last expression can be
expanded in powers of $\Delta$, producing 
\[
s\omega^a = p^a(u) + \dot{p}^a(u) \Delta + \frac{1}{2} \ddot{p}^a(u)
\Delta^2 + \frac{1}{6} p^{a(3)}(u) \Delta^3 + O(\Delta^4) 
\]
with 
\begin{eqnarray*}
\dot{p}_a(u) &=& -\sigma_{\alpha'\beta'} \base{\alpha'}{a} u^{\beta'}
- \bigl(\sigma_{\alpha'} u^{\alpha'}\bigr) \bigl( a_{\beta'}
  \base{\beta'}{a} \bigr) \\ 
&=& -r a_a - \frac{1}{3} r^2 S_a + O(r^3), \\ 
\ddot{p}_a(u) &=& -\sigma_{\alpha'\beta'\gamma'} \base{\alpha'}{a}
  u^{\beta'} u^{\gamma'} - \bigl( 2 \sigma_{\alpha'\beta'} u^{\alpha'}
u^{\beta'} + \sigma_{\alpha'} a^{\alpha'} \bigr) \bigl(a_{\gamma'}
\base{\gamma'}{a} \bigr) - \sigma_{\alpha'\beta'} \base{\alpha'}{a}
a^{\beta'} - \bigl(\sigma_{\alpha'} u^{\alpha'}\bigr)
\bigl(\dot{a}_{\beta'} \base{\beta'}{a} \bigr) \\ 
&=& \bigl(1 + r a_b \Omega^b \bigr) a_a - r \dot{a}_a 
+ \frac{1}{3} r R_{a0b0}\Omega^b + O(r^2), \\ 
p^{(3)}_a(u) &=& -\sigma_{\alpha'\beta'\gamma'\delta'}
\base{\alpha'}{a} u^{\beta'} u^{\gamma'} u^{\delta'} 
- \bigl( 3 \sigma_{\alpha'\beta'\gamma'} u^{\alpha'}
u^{\beta'} u^{\gamma'} + 6 \sigma_{\alpha'\beta'} u^{\alpha'}
a^{\beta'} + \sigma_{\alpha'} \dot{a}^{\alpha'} + \sigma_{\alpha'}
u^{\alpha'} \dot{a}_{\beta'} u^{\beta'} \bigr) \bigl(a_{\delta'}
\base{\delta'}{a} \bigr) \\ 
& & \mbox{} - \sigma_{\alpha'\beta'\gamma'}
\base{\alpha'}{a} \bigl( 2 a^{\beta'} u^{\gamma'} + u^{\beta'}
a^{\gamma'} \bigr) - \bigl(3 \sigma_{\alpha'\beta'} u^{\alpha'}
u^{\beta'} + 2 \sigma_{\alpha'} a^{\alpha'} \bigr)
\bigl(\dot{a}_{\gamma'} \base{\gamma'}{a} \bigr) -
\sigma_{\alpha'\beta'} \base{\alpha'}{a} \dot{a}^{\beta'} \\ 
& & \mbox{} - \bigl(\sigma_{\alpha'} u^{\alpha'} \bigr) \bigl(
\ddot{a}_{\beta'} \base{\beta'}{a} \bigr) \\ 
&=& 2 \dot{a}_a + O(r).  
\end{eqnarray*} 
To arrive at these results we have used the same expansions as before
and re-introduced $S_a = R_{a0b0} \Omega^b - R_{ab0c} \Omega^b
\Omega^c$, as it was first defined in Eq.~(\ref{9.4.4}).  

Collecting our results we obtain 
\begin{eqnarray*} 
s\omega^a &=& r\Omega^a - r \Bigl[ a^a + \frac{1}{3} r S^a + O(r^2) 
\Bigr] \Delta + \frac{1}{2} \Bigl[ \bigl(1 + r a_b \Omega^b\bigr) a^a
- r \dot{a}^a + \frac{1}{3} r R^a_{\ 0b0}\Omega^b + O(r^2) \Bigr]
\Delta^2 \\ & & \mbox{} 
+ \frac{1}{3} \Bigl[ \dot{a}^a + O(r) \Bigr] \Delta^3 
+ O(\Delta^4),   
\end{eqnarray*} 
which becomes 
\begin{equation}
s\omega^a = r \biggl\{ \Omega^a - \frac{1}{2} r \bigl[ 1 - r a_b(u)
\Omega^b \bigr] a^a(u) - \frac{1}{6} r^2 \dot{a}^a(u) - \frac{1}{6}
r^2 R^a_{\ 0b0}(u)\Omega^b + \frac{1}{3} r^2 R^a_{\ b0c}(u)
\Omega^b\Omega^c + O(r^3) \biggr\} 
\label{10.1.2}
\end{equation} 
after substituting Eq.~(\ref{10.1.1}) for $\Delta := t-u$. From
squaring Eq.~(\ref{10.1.2}) and using the identity $\delta_{ab} 
\omega^a \omega^b = 1$ we can also deduce 
\begin{equation} 
s = r \biggl\{ 1 - \frac{1}{2} r a_a(u) \Omega^a + \frac{3}{8} r^2
\bigl[ a_a(u) \Omega^a \bigr]^2 - \frac{1}{8} r^2 \dot{a}_0(u) -
\frac{1}{6} r^2 \dot{a}_a(u) \Omega^a - \frac{1}{6} r^2 R_{a0b0}(u)
\Omega^a \Omega^b + O(r^3) \biggr\}
\label{10.1.3} 
\end{equation}  
for the spatial distance between $x$ and $z(t)$. 

\subsection{From Fermi to retarded coordinates} 
\label{10.2}

The techniques developed in the preceding subsection can easily be
adapted to the task of relating the retarded coordinates of $x$ to its
Fermi normal coordinates. Here we use $\bar{x} := z(t)$ as
the reference point and express all quantities at $x' := z(u)$ as
Taylor expansions about $\tau = t$. 

We begin by considering the function  
\[
\sigma(\tau) = \sigma\bigl(x,z(\tau)\bigr) 
\]
of the proper-time parameter $\tau$ on $\gamma$. We have that
$\sigma(t) = \frac{1}{2} s^2$ and $\sigma(u) = 0$, and $\Delta :=
t-u$ is now obtained by expressing $\sigma(u)$ as $\sigma(t-\Delta)$
and expanding in powers of $\Delta$. Using the fact that
$\dot{\sigma}(\tau) = p(\tau)$, we have 
\[
\sigma(u) = \sigma(t) - p(t) \Delta + \frac{1}{2} \dot{p}(t) \Delta^2
- \frac{1}{6} \ddot{p}(t) \Delta^3 + \frac{1}{24} p^{(3)}(t)
  \Delta^4 + O(\Delta^5). 
\]
Expressions for the derivatives of $p(\tau)$ evaluated at $\tau = t$
can be constructed from results derived previously in Sec.~\ref{10.1}:
it suffices to replace all primed indices by barred indices and then
substitute the relation $\sigma^{\bar{\alpha}} = -s \omega^a
\base{\bar{\alpha}}{a}$ that follows immediately from
Eq.~(\ref{8.3.1}). This gives  
\begin{eqnarray*} 
\dot{p}(t) &=& -\Bigl[ 1 + s a_a \omega^a + \frac{1}{3} s^2 R_{a0b0} 
\omega^a \omega^b + O(s^3) \Bigr], \\ 
\ddot{p}(t) &=& - s \dot{a}_a \omega^a + O(s^2), \\ 
p^{(3)}(t) &=& \dot{a}_0 + O(s), 
\end{eqnarray*} 
and then 
\[
s^2 = \Bigl[ 1 + s a_a \omega^a + \frac{1}{3} s^2 R_{a0b0} 
\omega^a \omega^b + O(s^3) \Bigr] \Delta^2 - \frac{1}{3} s \Bigl[
\dot{a}_a \omega^a + O(s) \Bigr] \Delta^3 - \frac{1}{12} \Bigl[
\dot{a}_0 + O(s) \Bigr] \Delta^4 + O(\Delta^5) 
\]
after recalling that $p(t) = 0$. Solving for $\Delta$ as an expansion
in powers of $s$ returns  
\begin{equation} 
u = t - s \biggl\{ 1 - \frac{1}{2} s a_a(t) \omega^a + \frac{3}{8} s^2
\bigl[ a_a(t) \omega^a \bigr]^2 + \frac{1}{24} s^2 \dot{a}_0(t) 
+ \frac{1}{6} s^2 \dot{a}_a(t) \omega^a  
- \frac{1}{6} s^2 R_{a0b0}(t) \omega^a \omega^b + O(s^3) \biggr\}, 
\label{10.2.1} 
\end{equation}
in which we emphasize that all frame components are evaluated at the 
simultaneous point $z(t)$.  

An expression for $r = p(u)$ can be obtained by expanding
$p(t-\Delta)$ in powers of $\Delta$. We have  
\[
r = - \dot{p}(t) \Delta + \frac{1}{2} \ddot{p}(t) \Delta^2 -
\frac{1}{6} p^{(3)}(t) \Delta^3 + O(\Delta^4), 
\]
and substitution of our previous results gives 
\begin{equation} 
r = s \biggl\{ 1 + \frac{1}{2} s a_a(t) \omega^a - \frac{1}{8} s^2
\bigl[ a_a(t) \omega^a \bigr]^2 - \frac{1}{8} s^2 \dot{a}_0(t) 
- \frac{1}{3} s^2 \dot{a}_a(t) \omega^a 
+ \frac{1}{6} s^2 R_{a0b0}(t) \omega^a \omega^b + O(s^3) \biggr\}  
\label{10.2.2} 
\end{equation}     
for the retarded distance between $x$ and $z(u)$. 

Finally, the retarded coordinates $r\Omega^a = p^a(u)$ can be related
to the Fermi coordinates by expanding $p^a(t-\Delta)$ in powers of
$\Delta$, so that  
\[
r\Omega^a = p^a(t) - \dot{p}^a(t) \Delta + \frac{1}{2} \ddot{p}^a(t)
\Delta^2 - \frac{1}{6} p^{a(3)}(t) \Delta^3 + O(\Delta^4). 
\]
Results from the preceding subsection can again be imported with mild
alterations, and we find 
\begin{eqnarray*} 
\dot{p}_a(t) &=& \frac{1}{3} s^2 R_{ab0c} \omega^b \omega^c + O(s^3), 
\\ 
\ddot{p}_a(t) &=& \bigl( 1 + s a_b \omega^b \bigr) a_a + \frac{1}{3} s
R_{a0b0} \omega^b + O(s^2), \\ 
p^{(3)}_a(t) &=& 2 \dot{a}_a(t) + O(s).  
\end{eqnarray*} 
This, together with Eq.~(\ref{10.2.1}), gives 
\begin{equation}
r\Omega^a = s \biggl\{ \omega^a + \frac{1}{2} s a^a(t) - \frac{1}{3}
s^2 \dot{a}^a(t) - \frac{1}{3} s^2 R^a_{\ b0c}(t) \omega^b \omega^c +
\frac{1}{6} s^2 R^a_{\ 0b0}(t) \omega^b + O(s^3) \biggr\}.  
\label{10.2.3}
\end{equation}   
It may be checked that squaring this equation and using the identity
$\delta_{ab} \Omega^a \Omega^b = 1$ returns the same result as
Eq.~(\ref{10.2.2}).   

\subsection{Transformation of the tetrads at $x$} 
\label{10.3} 

Recall that we have constructed two sets of basis vectors at $x$. The
first set is the tetrad $(\bar{e}^{\alpha}_0, \bar{e}^\alpha_a)$ that
is obtained by parallel transport of $(u^{\bar{\alpha}},
\base{\bar{\alpha}}{a})$ on the spacelike geodesic that links $x$ to
the simultaneous point $\bar{x} := z(t)$. The second set is the
tetrad $(\base{\alpha}{0},\base{\alpha}{a})$ that is obtained by
parallel transport of $(u^{\alpha'}, \base{\alpha'}{a})$ on the null
geodesic that links $x$ to the retarded point $x' := z(u)$. Since
each tetrad forms a complete set of basis vectors, each member of
$(\bar{e}^{\alpha}_0, \bar{e}^\alpha_a)$ can be decomposed in the
tetrad $(\base{\alpha}{0},\base{\alpha}{a})$, and correspondingly,
each member of $(\base{\alpha}{0},\base{\alpha}{a})$ can be decomposed
in the tetrad $(\bar{e}^{\alpha}_0, \bar{e}^\alpha_a)$. These
decompositions are worked out in this subsection. For this purpose we 
shall consider the functions 
\[
p^\alpha(\tau) = g^\alpha_{\ \mu}\bigl(x,z(\tau)\bigr) u^\mu(\tau),
\qquad 
p^\alpha_a(\tau) = g^\alpha_{\ \mu}\bigl(x,z(\tau)\bigr)
\base{\mu}{a}(\tau), 
\]
in which $x$ is a fixed point in a neighbourhood of $\gamma$,
$z(\tau)$ is an arbitrary point on the world line, and 
$g^\alpha_{\ \mu}(x,z)$ is the parallel propagator on the unique
geodesic that links $x$ to $z$. We have $\bar{e}^\alpha_0 =
p^\alpha(t)$, $\bar{e}^\alpha_a = p^\alpha_a(t)$, $\base{\alpha}{0} =
p^\alpha(u)$, and $\base{\alpha}{a} = p^\alpha_a(u)$. 

We begin with the decomposition of $(\bar{e}^{\alpha}_0,
\bar{e}^\alpha_a)$ in the tetrad $(\base{\alpha}{0},\base{\alpha}{a})$
associated with the retarded point $z(u)$. This decomposition will be
expressed in the retarded coordinates as an expansion in powers of
$r$. As in Sec.~\ref{8.1} we express quantities at $z(t)$ in terms 
of quantities at $z(u)$ by expanding in powers of $\Delta := t-u$.  
We have
\[
\bar{e}^\alpha_0 = p^\alpha(u) + \dot{p}^\alpha(u) \Delta +
\frac{1}{2} \ddot{p}^\alpha(u) \Delta^2 + O(\Delta^3), 
\]
with 
\begin{eqnarray*}
\dot{p}^\alpha(u) &=& g^\alpha_{\ \alpha';\beta'} u^{\alpha'}
u^{\beta'} + g^{\alpha}_{\ \alpha'} a^{\alpha'} \\ 
&=& \Bigl[ a^a + \frac{1}{2} r R^a_{\ 0b0} \Omega^b + O(r^2) \Bigr]
\base{\alpha}{a}, \\ 
\ddot{p}^\alpha(u) &=& g^\alpha_{\ \alpha';\beta'\gamma'} u^{\alpha'} 
u^{\beta'} u^{\gamma'} + g^{\alpha}_{\ \alpha';\beta'} \bigl( 2
a^{\alpha'} u^{\beta'} + u^{\alpha'} a^{\beta'} \bigr) 
+ g^\alpha_{\ \alpha'} \dot{a}^{\alpha'} \\ 
&=& \Bigl[ -\dot{a}_0 + O(r) \Bigr] \base{\alpha}{0} + \Bigl[
\dot{a}^a + O(r) \Bigr] \base{\alpha}{a}, 
\end{eqnarray*} 
where we have used the expansions of Eq.~(\ref{5.2.5}) as well as the
decompositions of Eq.~(\ref{9.1.4}). Collecting these results and
substituting Eq.~(\ref{10.1.1}) for $\Delta$ yields  
\begin{equation} 
\bar{e}^\alpha_0 = \Bigl[ 1 - \frac{1}{2} r^2 \dot{a}_0(u) + O(r^3) 
\Bigr]\, \base{\alpha}{0} + \Bigl[ r \bigl( 1 - a_b \Omega^b \bigr)
a^a(u) + \frac{1}{2} r^2 \dot{a}^a(u) + \frac{1}{2} r^2 
R^a_{\ 0b0}(u) \Omega^b + O(r^3) \Bigr]\, \base{\alpha}{a}.  
\label{10.3.1}       
\end{equation} 
Similarly, we have  
\[
\bar{e}^\alpha_a = p^\alpha_a(u) + \dot{p}^\alpha_a(u) \Delta + 
\frac{1}{2} \ddot{p}^\alpha_a(u) \Delta^2 + O(\Delta^3), 
\]
with 
\begin{eqnarray*} 
\dot{p}^\alpha_a(u) &=& g^\alpha_{\ \alpha';\beta'} \base{\alpha'}{a}
u^{\beta'} + \bigl( g^\alpha_{\ \alpha'} u^{\alpha'} \bigr) \bigl(
a_{\beta'} \base{\beta'}{a} \bigr) \\ 
&=& \Bigl[a_a + \frac{1}{2} r R_{a0b0} \Omega^b + O(r^2) \Bigr]
\base{\alpha}{0} + \Bigl[ - \frac{1}{2} r R^b_{\ a0c} \Omega^c +
O(r^2) \Bigr] \base{\alpha}{b}, \\ 
\ddot{p}^\alpha_a(u) &=& g^\alpha_{\ \alpha';\beta'\gamma'}
\base{\alpha'}{a} u^{\beta'} u^{\gamma'} + g^\alpha_{\ \alpha';\beta'}
\bigl( 2 u^{\alpha'} u^{\beta'} a_{\gamma'} \base{\gamma'}{a} +
\base{\alpha'}{a} a^{\beta'} \bigr) + \bigl( g^\alpha_{\ \alpha'}
a^{\alpha'} \bigr) \bigl( a_{\beta'} \base{\beta'}{a} \bigr) + \bigl(
g^\alpha_{\ \alpha'} u^{\alpha'} \bigr) \bigl( \dot{a}_{\beta'}
\base{\beta'}{a} \bigr) \\ 
&=& \Bigl[ \dot{a}_a + O(r) \Bigr] \base{\alpha}{0} + \Bigl[ a_a a^b +
O(r) \Bigr] \base{\alpha}{b},
\end{eqnarray*} 
and all this gives 
\begin{eqnarray} 
\bar{e}^\alpha_a &=& \Bigl[ \delta^b_{\ a} + \frac{1}{2} r^2 a^b(u)
a_a(u) - \frac{1}{2} r^2 R^b_{\ a0c}(u) \Omega^c + O(r^3) \Bigr]\,
\base{\alpha}{b} 
\nonumber \\ & & \mbox{} 
+ \Bigl[ r \bigl( 1 - r a_b \Omega^b \bigr) a_a(u) +
\frac{1}{2} r^2 \dot{a}_a(u) + \frac{1}{2} r^2 R_{a0b0}(u) \Omega^b +
O(r^3) \Bigr]\, \base{\alpha}{0}. 
\label{10.3.2} 
\end{eqnarray}   

We now turn to the decomposition of $(\base{\alpha}{0},
\base{\alpha}{a})$ in the tetrad $(\bar{e}^{\alpha}_0,
\bar{e}^\alpha_a)$ associated with the simultaneous point $z(t)$. This
decomposition will be expressed in the Fermi normal coordinates as an
expansion in powers of $s$. Here, as in Sec.~\ref{8.2}, we shall
express quantities at $z(u)$ in terms of quantities at $z(t)$. We
begin with  
\[
\base{\alpha}{0} = p^\alpha(t) - \dot{p}^\alpha(t) \Delta +
\frac{1}{2} \ddot{p}^\alpha(t) \Delta^2 + O(\Delta^3) 
\]
and we evaluate the derivatives of $p^\alpha(\tau)$ at $\tau = t$. To
accomplish this we rely on our previous results (replacing primed
indices with barred indices), on the expansions of Eq.~(\ref{5.2.5}),
and on the decomposition of $g^\alpha_{\ \bar{\alpha}}(x,\bar{x})$ in
the tetrads at $x$ and $\bar{x}$. This gives 
\begin{eqnarray*} 
\dot{p}^\alpha(t) &=& \Bigl[ a^a + \frac{1}{2} s R^a_{\ 0b0} \omega^b +
O(s^2) \Bigr] \bar{e}^\alpha_a, \\ 
\ddot{p}^\alpha(t) &=& \Bigl[ -\dot{a}_0 + O(s) \Bigr]
\bar{e}^\alpha_0 + \Bigl[ \dot{a}^a + O(s) \Bigr] \bar{e}^\alpha_a,  
\end{eqnarray*} 
and we finally obtain 
\begin{equation} 
\base{\alpha}{0} = \Bigl[ 1 - \frac{1}{2} s^2 \dot{a}_0(t) + O(s^3) 
\Bigr]\, \bar{e}^\alpha_0 + \Bigl[ -s \Bigl( 1 - \frac{1}{2} s a_b
\omega^b \Bigr) a^a(t) + \frac{1}{2} s^2 \dot{a}^a(t) 
- \frac{1}{2} s^2 R^a_{\ 0b0}(t) \omega^b + O(s^3) \Bigr]\,
\bar{e}^\alpha_a.  
\label{10.3.3}
\end{equation}    
Similarly, we write 
\[
\base{\alpha}{a} = p^\alpha_a(t) - \dot{p}^\alpha_a(t) \Delta +
\frac{1}{2} \ddot{p}^\alpha_a(t) \Delta^2 + O(\Delta^3), 
\]
in which we substitute 
\begin{eqnarray*} 
\dot{p}^\alpha_a(t) &=& \Bigl[ a_a + \frac{1}{2} s R_{a0b0} \omega^b +
O(s^2) \Bigr] \bar{e}^\alpha_0 + \Bigl[ -\frac{1}{2} s R^b_{\ a0c}
\omega^c + O(s^2) \Bigr] \bar{e}^\alpha_b, \\ 
\ddot{p}^\alpha_a(t) &=& \Bigl[ \dot{a}_a + O(s) \Bigr]
\bar{e}^\alpha_0 + \Bigl[ a_a a^b + O(s) \Bigr] \bar{e}^\alpha_b, 
\end{eqnarray*} 
as well as Eq.~(\ref{10.2.1}) for $\Delta := t - u$. Our final
result is  
\begin{eqnarray} 
\base{\alpha}{a} &=& \Bigl[ \delta^b_{\ a} + \frac{1}{2} s^2 a^b(t) 
a_a(t) + \frac{1}{2} s^2 R^b_{\ a0c}(t) \omega^c + O(s^3) \Bigr]\,
\bar{e}^\alpha_b  
\nonumber \\ & & \mbox{} 
+ \Bigl[ -s \Bigl( 1 - \frac{1}{2} s a_b \omega^b \Bigr) a_a(t) +
\frac{1}{2} s^2 \dot{a}_a(t) - \frac{1}{2} s^2 R_{a0b0}(u) \omega^b + 
O(s^3) \Bigr]\, \bar{e}^\alpha_0.   
\label{10.3.4} 
\end{eqnarray}   

\subsection{Advanced point}  
\label{10.4}

It will prove convenient to introduce on the world line, along with
the retarded and simultaneous points, an {\it advanced point}
associated with the field point $x$. The advanced point will be
denoted $x'' := z(v)$, with $v$ denoting the value of the
proper-time parameter at $x''$; to tensors at this point we assign
indices $\alpha''$, $\beta''$, etc. The advanced point is linked to
$x$ by a {\it past-directed null geodesic} (refer back to Fig.~8), and
it can be located by solving $\sigma(x,x'') = 0$ together with the
requirement that $\sigma^{\alpha''}(x,x'')$ be a future-directed null
vector. The affine-parameter distance between $x$ and $x''$ along the
null geodesic is given by  
\begin{equation} 
r_{\rm adv} = - \sigma_{\alpha''} u^{\alpha''}, 
\label{10.4.1}
\end{equation} 
and we shall call this the {\it advanced distance} between $x$ and the
world line. Notice that $r_{\rm adv}$ is a positive quantity. 

We wish first to find an expression for $v$ in terms of the retarded
coordinates of $x$. For this purpose we define $\Delta^{\!\prime}
:= v - u$ and re-introduce the function $\sigma(\tau) :=
\sigma(x,z(\tau))$ first considered in Sec.~\ref{10.2}. We have that
$\sigma(v) = \sigma(u) = 0$, and $\Delta^{\!\prime}$ can ultimately be
obtained by expressing $\sigma(v)$ as $\sigma(u+\Delta^{\!\prime})$
and expanding in powers of $\Delta^{\!\prime}$. Recalling that
$\dot{\sigma}(\tau) = p(\tau)$, we have  
\[
\sigma(v) = \sigma(u) + p(u) \Delta^{\!\prime} + \frac{1}{2}
\dot{p}(u) \Delta^{\!\prime 2} + \frac{1}{6} \ddot{p}(u)
\Delta^{\!\prime 3} + \frac{1}{24} p^{(3)}(u) \Delta^{\!\prime 4}
+ O(\Delta^{\!\prime 5}).  
\]
Using the expressions for the derivatives of $p(\tau)$ that were first 
obtained in Sec.~\ref{10.1}, we write this as 
\[ 
r = \frac{1}{2} \Bigl[ 1 + r a_a \Omega^a + \frac{1}{3} r^2 S + O(r^3)  
\Bigr] \Delta^{\!\prime} + \frac{1}{6} r \Bigl[ \dot{a}_0 +
\dot{a}_a \Omega^a + O(r) \Bigr] \Delta^{\!\prime 2} - \frac{1}{24}
\Bigl[ \dot{a}_0 + O(r) \Bigr] \Delta^{\!\prime 3} +
O(\Delta^{\!\prime 4}).  
\]
Solving for $\Delta^{\!\prime}$ as an expansion in powers of $r$, we
obtain  
\begin{equation} 
v = u + 2 r \biggl\{ 1 - r a_a(u) \Omega^a + r^2 \bigl[ a_a(u)
\Omega^a \bigr]^2 - \frac{1}{3} r^2 \dot{a}_0(u) - \frac{2}{3} r^2  
\dot{a}_a(u) \Omega^a - \frac{1}{3} r^2 R_{a0b0}(u) \Omega^a \Omega^b 
+ O(r^3) \biggr\}, 
\label{10.4.2} 
\end{equation} 
in which all frame components are evaluated at the retarded point
$z(u)$. 

Our next task is to derive an expression for the advanced distance 
$r_{\rm adv}$. For this purpose we observe that $r_{\rm adv} = -p(v) = 
-p(u+\Delta^{\!\prime})$, which we can expand in powers of
$\Delta^{\!\prime} := v - u$. This gives 
\[
r_{\rm adv} = -p(u) - \dot{p}(u) \Delta^{\!\prime} - \frac{1}{2}
\ddot{p}(u) \Delta^{\!\prime 2} - \frac{1}{6} p^{(3)}(u)
\Delta^{\!\prime 3} + O(\Delta^{\!\prime 4}),  
\]
which then becomes 
\[
r_{\rm adv} = -r + \Bigl[ 1 + r a_a \Omega^a + \frac{1}{3} r^2 S + 
O(r^3) \Bigr] \Delta^{\!\prime} + \frac{1}{2} r \Bigl[ \dot{a}_0 +
\dot{a}_a \Omega^a + O(r) \Bigr] \Delta^{\!\prime 2} - \frac{1}{6}
\Bigl[ \dot{a}_0 + O(r) \Bigr] \Delta^{\!\prime 3} +
O(\Delta^{\!\prime 4}).  
\] 
After substituting Eq.~(\ref{10.4.2}) for $\Delta^{\!\prime}$ and 
witnessing a number of cancellations, we arrive at the simple
expression    
\begin{equation} 
r_{\rm adv} = r \biggl[ 1 + \frac{2}{3} r^2 \dot{a}_a(u) \Omega^a +
O(r^3) \biggr].  
\label{10.4.3} 
\end{equation} 

From Eqs.~(\ref{9.5.3}), (\ref{9.5.4}), and (\ref{10.4.2}) we deduce
that the gradient of the advanced time $v$ is given by   
\begin{equation} 
\partial_\alpha v = \Bigl[ 1 - 2 r a_a \Omega^a + O(r^2) \Bigr]\,
\base{0}{\alpha} + \Bigr[ \Omega_a - 2 r a_a + O(r^2) \Bigr]\,
\base{a}{\alpha}, 
\label{10.4.4} 
\end{equation}
where the expansion in powers of $r$ was truncated to a sufficient
number of terms. Similarly, Eqs.~(\ref{9.5.4}), (\ref{9.5.5}), and  
(\ref{10.4.3}) imply that the gradient of the advanced distance is
given by  
\begin{eqnarray} 
\partial_\alpha r_{\rm adv} &=& \Bigl[ \Bigl( 1 + r a_b \Omega^b +  
\frac{4}{3} r^2 \dot{a}_b \Omega^b + \frac{1}{3} r^2 S \Bigr) \Omega_a
+ \frac{2}{3} r^2 \dot{a}_a + \frac{1}{6} r^2 S_a + O(r^3) \Bigl]\,
\base{a}{\alpha} \nonumber \\ & & \mbox{} 
+ \Bigl[ -r a_a \Omega^a - \frac{1}{2} r^2 S +
O(r^3) \Bigr]\, \base{0}{\alpha}, 
\label{10.4.5} 
\end{eqnarray} 
where $S_a$ and $S$ were first introduced in Eqs.~(\ref{9.4.4}) and 
(\ref{9.4.5}), respectively. We emphasize that in Eqs.~(\ref{10.4.4})
and (\ref{10.4.5}), all frame components are evaluated at the retarded
point $z(u)$. 

%% file: part3.tex
%
\section{Scalar Green's functions in flat spacetime} 
\label{11}  

\subsection{Green's equation for a massive scalar field} 
\label{11.1} 

To prepare the way for our discussion of Green's functions in curved  
spacetime, we consider first the slightly nontrivial case of a massive
scalar field $\Phi(x)$ in flat spacetime. This field satisfies the
wave equation 
\begin{equation} 
(\Box - k^2) \Phi(x) = -4\pi \mu(x), 
\label{11.1.1}
\end{equation} 
where $\Box = \eta^{\alpha\beta} \partial_\alpha \partial_\beta$ is   
the wave operator, $\mu(x)$ a prescribed source, and where the
parameter $k$ has a dimension of inverse length. We seek a Green's
function $G(x,x')$ such that a solution to Eq.~(\ref{11.1.1}) can be
expressed as 
\begin{equation} 
\Phi(x) = \int G(x,x') \mu(x')\, d^4 x', 
\label{11.1.2} 
\end{equation} 
where the integration is over all of Minkowski spacetime. The relevant
wave equation for the Green's function is
\begin{equation} 
(\Box - k^2) G(x,x') = -4\pi \delta_4(x-x'), 
\label{11.1.3}
\end{equation}
where $\delta_4(x-x') = \delta(t-t') \delta(x-x') \delta(y-y')
\delta(z-z')$ is a four-dimensional Dirac distribution in flat
spacetime. Two types of Green's functions will be of particular 
interest: the retarded Green's function, a solution to 
Eq.~(\ref{11.1.3}) with the property that it vanishes when $x$ is in the 
past of $x'$, and the advanced Green's function, which vanishes when
$x$ is in the future of $x'$.     

To solve Eq.~(\ref{11.1.3}) we appeal to Lorentz invariance and the
fact that the spacetime is homogeneous to argue that the retarded and 
advanced Green's functions must be given by expressions of the form   
\begin{equation}
G_{\rm ret}(x,x') = \theta(t-t') g(\sigma), \qquad 
G_{\rm adv}(x,x') = \theta(t'-t) g(\sigma),
\label{11.1.4}
\end{equation}
where $\sigma = \frac{1}{2} \eta_{\alpha\beta} (x-x')^\alpha
(x-x')^\beta$ is Synge's world function in flat spacetime, and where
$g(\sigma)$ is a function to be determined. For the remainder of this
section we set $x'=0$ without loss of generality.   

\subsection{Integration over the source} 
\label{11.2} 

The Dirac functional on the right-hand side of Eq.~(\ref{11.1.3}) is a
highly singular quantity, and we can avoid dealing with it by
integrating the equation over a small four-volume $V$ that contains 
$x' \equiv 0$. This volume is bounded by a closed hypersurface
$\partial V$. After using Gauss' theorem on the first term of
Eq.~(\ref{11.1.3}), we obtain $\oint_{\partial V} G^{;\alpha}
d\Sigma_\alpha - k^2 \int_V G\, dV = -4\pi$, where $d\Sigma_{\alpha}$ 
is a surface element on $\partial V$. Assuming that the integral of 
$G$ over $V$ goes to zero in the limit $V \to 0$, we have    
\begin{equation}
\lim_{V \to 0} \oint_{\partial V} G^{;\alpha}
d\Sigma_{\alpha} = -4\pi. 
\label{11.2.1}
\end{equation} 
It should be emphasized that the four-volume $V$ must contain the
point $x'$.  

To examine Eq.~(\ref{11.2.1}) we introduce coordinates 
$(w,\chi,\theta,\phi)$ defined by 
\[
t = w \cos\chi, \qquad
x = w \sin\chi \sin\theta \cos\phi, \qquad
y = w \sin\chi \sin\theta \sin\phi, \qquad
z = w \sin\chi \cos\theta, 
\]
and we let $\partial V$ be a surface of constant $w$. The metric of
flat spacetime is given by 
\[
ds^2 = -\cos2\chi\, dw^2 + 2w \sin2\chi\, dwd\chi + w^2\cos2\chi\,
d\chi^2 + w^2 \sin^2\chi\, d\Omega^2
\]
in the new coordinates, where $d\Omega^2 = d\theta^2 + \sin^2\theta\,
d\phi^2$. Notice that $w$ is a timelike coordinate when $\cos 2\chi >
0$, and that $\chi$ is then a spacelike coordinate; the roles are
reversed when $\cos 2\chi < 0$. Straightforward computations reveal
that in these coordinates, 
$\sigma = -\frac{1}{2} w^2 \cos 2\chi$, $\sqrt{-g} = w^3 \sin^2\chi
\sin\theta$, $g^{ww} = -\cos 2\chi$, $g^{w\chi} = w^{-1} \sin 2\chi$,
$g^{\chi\chi} = w^{-2} \cos2\chi$, and the only nonvanishing component
of the surface element is $d\Sigma_w = w^3 \sin^2\chi\, d\chi
d\Omega$, where $d\Omega = \sin\theta\, d\theta d\phi$. To calculate
the gradient of the Green's function we express it as $G = \theta(\pm
t) g(\sigma) = \theta(\pm w\cos\chi) g(-\frac{1}{2} w^2 \cos 2\chi)$,
with the upper (lower) sign belonging to the retarded (advanced)
Green's function. Calculation gives $G^{;\alpha} d\Sigma_{\alpha} = 
\theta(\pm \cos\chi) w^4 \sin^2 \chi g'(\sigma)\, d\chi d\Omega$, with
a prime indicating differentiation with respect to $\sigma$; it should
be noted that derivatives of the step function do not appear in this
expression. 

Integration of $G^{;\alpha} d\Sigma_{\alpha}$ with respect to
$d\Omega$ is immediate, and we find that Eq.~(\ref{11.2.1}) reduces to   
\begin{equation} 
\lim_{w \to 0} \int_0^\pi \theta(\pm \cos\chi) w^4 \sin^2\chi
g'(\sigma)\, d\chi = -1. 
\label{11.2.2}
\end{equation} 
For the retarded Green's function, the step function restricts the
domain of integration to $0 < \chi < \pi/2$, in which $\sigma$
increases from $-\frac{1}{2} w^2$ to $\frac{1}{2} w^2$. Changing the  
variable of integration from $\chi$ to $\sigma$ transforms
Eq.~(\ref{11.2.2}) into 
\begin{equation}
\lim_{\epsilon \to 0} \epsilon \int_{-\epsilon}^\epsilon
w(\sigma/\epsilon)\, g'(\sigma)\, d\sigma
= - 1, \qquad
w(\xi) := \sqrt{ \frac{1+\xi}{1-\xi} }, 
\label{11.2.3}
\end{equation}
where $\epsilon := \frac{1}{2} w^2$. For the advanced Green's 
function, the domain of integration is $\pi/2 < \chi < \pi$, in which 
$\sigma$ decreases from $\frac{1}{2} w^2$ to $-\frac{1}{2}
w^2$. Changing the variable of integration from $\chi$ to $\sigma$
also produces Eq.~(\ref{11.2.3}).    

\subsection{Singular part of $g(\sigma)$} 
\label{11.3} 

We have seen that Eq.~(\ref{11.2.3}) properly encodes the influence of 
the singular source term on both the retarded and advanced Green's
function. The function $g(\sigma)$ that enters into the expressions of
Eq.~(\ref{11.1.4}) must therefore be such that Eq.~(\ref{11.2.3}) is 
satisfied. It follows immediately that $g(\sigma)$ must be a singular
function, because for a smooth function the integral of
Eq.~(\ref{11.2.3}) would be of order $\epsilon$ and the left-hand side
of Eq.~(\ref{11.2.3}) could never be made equal to $-1$. The
singularity, however, must be integrable, and this leads us to assume
that $g'(\sigma)$ must be made out of Dirac $\delta$-functions and  
derivatives.  

We make the ansatz
\begin{equation}
g(\sigma) = V(\sigma) \theta(-\sigma) + A \delta(\sigma) + B
\delta'(\sigma) + C \delta''(\sigma) + \cdots, 
\label{11.3.1}
\end{equation} 
where $V(\sigma)$ is a smooth function, and $A$, $B$, $C$, \ldots are 
constants. The first term represents a function supported within the
past and future light cones of $x' \equiv 0$; we exclude a term
proportional to $\theta(\sigma)$ for reasons of causality. The other
terms are supported on the past and future light cones. It is
sufficient to take the coefficients in front of the $\delta$-functions
to be constants. To see this we invoke the distributional identities   
\begin{equation} 
\sigma \delta(\sigma) = 0 
\quad \rightarrow \quad 
\sigma \delta'(\sigma) + \delta(\sigma) = 0
\quad \rightarrow \quad 
\sigma \delta''(\sigma) + 2 \delta'(\sigma) = 0 
\quad \rightarrow \quad \cdots
\label{11.3.2}
\end{equation}
from which it follows that $\sigma^2 \delta'(\sigma) = \sigma^3
\delta''(\sigma) = \cdots = 0$. A term like $f(\sigma)
\delta(\sigma)$ is then distributionally equal to $f(0)
\delta(\sigma)$, while a term like $f(\sigma)
\delta'(\sigma)$ is distributionally equal to $f(0)
\delta'(\sigma) - f'(0)\delta(\sigma)$, and a term like $f(\sigma) 
\delta''(\sigma)$ is distributionally equal to $f(0)
\delta''(\sigma) - 2f'(0)\delta'(\sigma) + 2f''(0)
\delta(\sigma)$; here $f(\sigma)$ is an arbitrary test
function. Summing over such terms, we recover an expression of
the form of Eq.~(\ref{11.3.2}), and there is no need to make $A$, $B$,
$C$, \ldots functions of $\sigma$.  

Differentiation of Eq.~(\ref{11.3.1}) and substitution into
Eq.~(\ref{11.2.3}) yields 
\[
\epsilon \int_{-\epsilon}^\epsilon
w(\sigma/\epsilon)\, g'(\sigma)\, d\sigma        
= \epsilon \biggl[ \int_{-\epsilon}^\epsilon V'(\sigma)
w(\sigma/\epsilon)\, d\sigma - V(0) w(0) - \frac{A}{\epsilon}
\dot{w}(0) + \frac{B}{\epsilon^2} \ddot{w}(0) - \frac{C}{\epsilon^3}
w^{(3)}(0) + \cdots \biggr],  
\]
where overdots (or a number within brackets) indicate repeated
differentiation with respect to $\xi := \sigma/\epsilon$. The
limit $\epsilon \to 0$ exists if and only if $B = C =\cdots = 0$. In
the limit we must then have $A \dot{w}(0) = 1$, which implies
$A=1$. We conclude that $g(\sigma)$ must have the form of 
\begin{equation}
g(\sigma) = \delta(\sigma) + V(\sigma) \theta(-\sigma), 
\label{11.3.3} 
\end{equation}
with $V(\sigma)$ a smooth function that cannot be determined from
Eq.~(\ref{11.2.3}) alone.  

\subsection{Smooth part of $g(\sigma)$}
\label{11.4}

To determine $V(\sigma)$ we must go back to the differential equation
of Eq.~(\ref{11.1.3}). Because the singular structure of the Green's 
function is now under control, we can safely set $x \neq x'\equiv 0$
in the forthcoming operations. This means that the equation to solve
is in fact $(\Box - k^2) g(\sigma) = 0$, the homogeneous version of
Eq.~(\ref{11.1.3}). We have $\nabla_\alpha g = g' \sigma_\alpha$,
$\nabla_\alpha \nabla_\beta g = g'' \sigma_\alpha \sigma_\beta + g'
\sigma_{\alpha\beta}$, $\Box g = 2\sigma g'' + 4 g'$, so that Green's
equation reduces to the ordinary differential equation  
\begin{equation}
2\sigma g'' + 4 g' - k^2 g = 0. 
\label{11.4.1}
\end{equation} 
If we substitute Eq.~(\ref{11.3.3}) into this we get 
\[
-(2V + k^2) \delta(\sigma) + (2\sigma V'' + 4 V' - k^2 V)
 \theta(-\sigma) = 0, 
\]
where we have used the identities of Eq.~(\ref{11.3.2}). The left-hand 
side will vanish as a distribution if we set 
\begin{equation}
2\sigma V'' + 4 V' - k^2 V = 0, \qquad
V(0) = -\frac{1}{2} k^2. 
\label{11.4.2}
\end{equation}
These equations determine $V(\sigma)$ uniquely, even in the absence of
a second boundary condition at $\sigma = 0$, because the differential
equation is singular at $\sigma = 0$ while $V$ is known to be smooth.        

To solve Eq.~(\ref{11.4.2}) we let $V = F(z)/z$, with $z :=  
k\sqrt{-2\sigma}$. This gives rise to Bessel's equation for the new 
function $F$:
\[
z^2 F_{zz} + z F_z + (z^2 - 1) F = 0. 
\]
The solution that is well behaved near $z=0$ is $F = aJ_1(z)$, where 
$a$ is a constant to be determined. We have that $J_1(z) \sim
\frac{1}{2} z$ for small values of $z$, and it follows that $V \sim
a/2$. From Eq.~(\ref{11.4.2}) we see that $a = -k^2$. So we have found 
that the only acceptable solution to Eq.~(\ref{11.4.2}) is  
\begin{equation} 
V(\sigma) = -\frac{k}{\sqrt{-2\sigma}}\, 
J_1\bigl( k\sqrt{-2\sigma} \bigr).   
\label{11.4.3} 
\end{equation}

To summarize, the retarded and advanced solutions to
Eq.~(\ref{11.1.3}) are given by Eq.~(\ref{11.1.4}) with $g(\sigma)$
given by Eq.~(\ref{11.3.3}) and $V(\sigma)$ given by
Eq.~(\ref{11.4.3}).  

\subsection{Advanced distributional methods}
\label{11.5}

The techniques developed previously to find Green's functions for the
scalar wave equation are limited to flat spacetime, and they would not
be very useful for curved spacetimes. To pursue this  
generalization we must introduce more powerful distributional
methods. We do so in this subsection, and in the next we shall use
them to recover our previous results.    

Let $\theta_+(x,\Sigma)$ be a generalized step function, defined to be 
one when $x$ is in the future of the spacelike hypersurface $\Sigma$
and zero otherwise. Similarly, define
$\theta_-(x,\Sigma) := 1 - \theta_+(x,\Sigma)$ to be one when $x$ is 
in the past of the spacelike hypersurface $\Sigma$ and zero
otherwise. Then define the light-cone step functions   
\begin{equation} 
\theta_\pm(-\sigma) = \theta_\pm(x,\Sigma) \theta(-\sigma), \qquad 
x' \in \Sigma, 
\label{11.5.1}
\end{equation}
so that $\theta_+(-\sigma)$ is one if $x$ is within $I^+(x')$,
the chronological future of $x'$, and zero otherwise, and
$\theta_-(-\sigma)$ is one if $x$ is within $I^-(x')$, the
chronological past of $x'$, and zero otherwise; the choice of
hypersurface is immaterial so long as $\Sigma$ is spacelike and
contains the reference point $x'$. Notice that $\theta_+(-\sigma) +
\theta_-(-\sigma) = \theta(-\sigma)$. Define also the light-cone Dirac
functionals     
\begin{equation} 
\delta_\pm(\sigma) = \theta_\pm(x,\Sigma) \delta(\sigma), \qquad 
x' \in \Sigma,  
\label{11.5.2}
\end{equation}
so that $\delta_+(\sigma)$, when viewed as a function of $x$, is 
supported on the future light cone of $x'$, while $\delta_-(\sigma)$
is supported on its past light cone. Notice that $\delta_+(\sigma) +
\delta_-(\sigma) = \delta(\sigma)$. In Eqs.~(\ref{11.5.1}) and
(\ref{11.5.2}), $\sigma$ is the world function for flat spacetime; it
is negative when $x$ and $x'$ are timelike related, and positive when
they are spacelike related.    

The distributions $\theta_\pm(-\sigma)$ and $\delta_\pm(\sigma)$ are
not defined at $x=x'$ and they cannot be differentiated there. This
pathology can be avoided if we shift $\sigma$ by a small positive
quantity $\epsilon$. We can therefore use the distributions 
$\theta_\pm(-\sigma - \epsilon)$ and $\delta_\pm(\sigma + \epsilon)$  
in some sensitive computations, and then take the limit $\epsilon \to 
0^+$. Notice that the equation $\sigma + \epsilon = 0$ describes a
two-branch hyperboloid that is located just {\it within} the light 
cone of the reference point $x'$. The hyperboloid does not include
$x'$, and $\theta_+(x,\Sigma)$ is one everywhere on its future branch,
while $\theta_-(x,\Sigma)$ is one everywhere on its past branch. These
factors, therefore, become invisible to differential operators. For
example, $\theta_+'(-\sigma-\epsilon) = \theta_+(x,\Sigma)
\theta'(-\sigma-\epsilon) = -\theta_+(x,\Sigma)
\delta(\sigma+\epsilon) = -\delta_+(\sigma + \epsilon)$. This
manipulation shows that after the shift from $\sigma$ to $\sigma +
\epsilon$, the distributions of Eqs.~(\ref{11.5.1}) and (\ref{11.5.2})
can be straightforwardly differentiated with respect to $\sigma$.  

In the next paragraphs we shall establish the distributional
identities  
\begin{eqnarray} 
\lim_{\epsilon \to 0^+} \epsilon \delta_\pm(\sigma + \epsilon) &=& 0, 
\label{11.5.3} \\
\lim_{\epsilon \to 0^+} \epsilon \delta'_\pm(\sigma + \epsilon) &=& 0, 
\label{11.5.4} \\
\lim_{\epsilon \to 0^+} \epsilon \delta''_\pm(\sigma + \epsilon) &=&
2\pi \delta_4(x - x')  
\label{11.5.5} 
\end{eqnarray}
in four-dimensional flat spacetime. These will be used in the next
subsection to recover the Green's functions for the scalar wave
equation, and they will be generalized to curved spacetime in 
Sec.~\ref{12}.   

The derivation of Eqs.~(\ref{11.5.3})--(\ref{11.5.5}) relies on
a ``master'' distributional identity, formulated in three-dimensional
flat space:   
\begin{equation}
\lim_{\epsilon \to 0^+} \frac{\epsilon}{R^5} = \frac{2\pi}{3}
\delta_3(\bm{x}), \qquad 
R := \sqrt{r^2 + 2\epsilon}, 
\label{11.5.6}
\end{equation}
with $r := |\bm{x}| := \sqrt{x^2+y^2+z^2}$. This follows from
yet another identity, $\nabla^2 r^{-1} = -4\pi \delta_3(\bm{x})$, in
which we write the left-hand side as $\lim_{\epsilon \to 0^+}\nabla^2
R^{-1}$; since $R^{-1}$ is nonsingular at $\bm{x} = 0$ it can be
straightforwardly differentiated, and the result is $\nabla^2 R^{-1} =
-6\epsilon/R^5$, from which Eq.~(\ref{11.5.6}) follows.  

To prove Eq.~(\ref{11.5.3}) we must show that $\epsilon 
\delta_\pm(\sigma+\epsilon)$ vanishes as a distribution in the limit
$\epsilon \to 0^+$. For this we must prove that a functional of the 
form  
\[
A_\pm[f] = \lim_{\epsilon \to 0^+}  
\int \epsilon \delta_\pm(\sigma + \epsilon) f(x)\, d^4 x, 
\]
where $f(x) = f(t,\bm{x})$ is a smooth test function, vanishes for all
such functions $f$. Our first task will be to find a more convenient
expression for $\delta_\pm(\sigma+\epsilon)$. Once more we set $x' =
0$ (without loss of generality) and we note that $2(\sigma+\epsilon) =
-t^2 + r^2 + 2 \epsilon = -(t-R)(t+R)$, where we have used
Eq.~(\ref{11.5.6}). It follows that      
\begin{equation}
\delta_\pm(\sigma + \epsilon) = \frac{\delta(t \mp R)}{R},         
\label{11.5.7} 
\end{equation} 
and from this we find 
\[
A_\pm[f] = \lim_{\epsilon \to 0^+} \int \epsilon 
\frac{f(\pm R, \bm{x})}{R}\, d^3 x 
= \lim_{\epsilon \to 0^+}  
\int \frac{\epsilon}{R^5} R^4 f(\pm R, \bm{x})\, d^3 x 
= \frac{2\pi}{3} \int \delta_3(\bm{x}) r^4 f(\pm r, \bm{x})\, d^3 x
= 0, 
\] 
which establishes Eq.~(\ref{11.5.3}). 

The validity of Eq.~(\ref{11.5.4}) is established by a similar
computation. Here we must show that a functional of the form   
\[
B_\pm[f] = \lim_{\epsilon \to 0^+}  
\int \epsilon \delta'_\pm(\sigma + \epsilon) f(x)\, d^4 x 
\]
vanishes for all test functions $f$. We have  
\begin{eqnarray*}
B_\pm[f] &=& \lim_{\epsilon \to 0^+} \epsilon \frac{d}{d\epsilon}   
\int \delta_\pm(\sigma+\epsilon) f(x)\, d^4 x 
= \lim_{\epsilon \to 0^+} \epsilon \frac{d}{d\epsilon}
  \int \frac{f(\pm R, \bm{x})}{R}\, d^3 x 
= \lim_{\epsilon \to 0^+} \epsilon 
  \int \biggl( \pm \frac{\dot{f}}{R^2} - \frac{f}{R^3} \biggr)\, d^3 x 
\\
&=& \lim_{\epsilon \to 0^+} 
  \int \frac{\epsilon}{R^5} \bigl( \pm R^3 \dot{f} 
  - R^2 f \bigr)\, d^3 x 
= \frac{2\pi}{3} \int \delta_3(\bm{x}) \bigl( \pm r^3 \dot{f} 
  - r^2 f \bigr)\, d^3 x
= 0,
\end{eqnarray*}
and the identity of Eq.~(\ref{11.5.4}) is proved. In these
manipulations we have let an overdot indicate partial differentiation
with respect to $t$, and we have used $\partial R/\partial \epsilon =
1/R$.         

To establish Eq.~(\ref{11.5.5}) we consider the functional
\[
C_\pm[f] = \lim_{\epsilon \to 0^+}  
\int \epsilon \delta''_\pm(\sigma + \epsilon) f(x)\, d^4 x 
\]
and show that it evaluates to $2\pi f(0,\bm{0})$. We have 
\begin{eqnarray*}
C_\pm[f] &=& \lim_{\epsilon \to 0^+} \epsilon \frac{d^2}{d\epsilon^2}    
\int \delta_\pm(\sigma+\epsilon) f(x)\, d^4 x 
= \lim_{\epsilon \to 0^+} \epsilon \frac{d^2}{d\epsilon^2} 
  \int \frac{f(\pm R, \bm{x})}{R}\, d^3 x \\ 
&=& \lim_{\epsilon \to 0^+} \epsilon \int \biggl( \frac{\ddot{f}}{R^3} 
\mp 3 \frac{\dot{f}}{R^4} + 3 \frac{f}{R^5} \biggr)\, d^3 x 
= 2\pi \int \delta_3(\bm{x}) \biggl( \frac{1}{3} r^2 \ddot{f}  
    \pm r \dot{f} + f \biggr)\, d^3 x \\ 
&=& 2\pi f(0,\bm{0}), 
\end{eqnarray*} 
as required. This proves that Eq.~(\ref{11.5.5}) holds as a 
distributional identity in four-dimensional flat spacetime. 

\subsection{Alternative computation of the Green's functions} 
\label{11.6}

The retarded and advanced Green's functions for the scalar wave
equation are now defined as the limit of the functions 
$G^\epsilon_\pm(x,x')$ when $\epsilon \to 0^+$. For these we make   
the ansatz 
\begin{equation}
G^\epsilon_\pm(x,x') = \delta_\pm(\sigma + \epsilon) + V(\sigma)
\theta_\pm(-\sigma -\epsilon),
\label{11.6.1} 
\end{equation}
and we shall prove that $G^\epsilon_\pm(x,x')$ satisfies
Eq.~(\ref{11.1.3}) in the limit. We recall that the distributions
$\theta_\pm$ and $\delta_\pm$ were defined in the preceding
subsection, and we assume that $V(\sigma)$ is a smooth function of
$\sigma(x,x') = \frac{1}{2} \eta_{\alpha\beta} (x-x')^\alpha
(x-x')^\beta$; because this function is smooth, it is not necessary to
evaluate $V$ at $\sigma + \epsilon$ in Eq.~(\ref{11.6.1}). We recall
also that $\theta_+$ and $\delta_+$ are nonzero when $x$ is in the
future of $x'$, while $\theta_-$ and $\delta_-$ are nonzero when $x$
is in the past of $x'$. We will therefore prove that the retarded and
advanced Green's functions are of the form   
\begin{equation}
G_{\rm ret}(x,x') = \lim_{\epsilon \to 0^+} G_+^\epsilon(x,x')
= \theta_+(x,\Sigma) \bigl[ \delta(\sigma) + V(\sigma) \theta(-\sigma) 
\bigr] 
\label{11.6.2}
\end{equation}
and 
\begin{equation}
G_{\rm adv}(x,x') = \lim_{\epsilon \to 0^+} G_-^\epsilon(x,x')
= \theta_-(x,\Sigma) \bigl[ \delta(\sigma) + V(\sigma) \theta(-\sigma) 
\bigr],    
\label{11.6.3}
\end{equation}
where $\Sigma$ is a spacelike hypersurface that contains $x'$. We will 
also determine the form of the function $V(\sigma)$.    

The functions that appear in Eq.~(\ref{11.6.1}) can be
straightforwardly differentiated. The manipulations are similar to
what was done in Sec.~\ref{11.4}, and dropping all labels, we obtain
$(\Box - k^2) G = 2\sigma G'' + 4 G' - k^2 G$, with a prime indicating
differentiation with respect to $\sigma$. From Eq.~(\ref{11.6.1}) we
obtain $G' = \delta' - V \delta + V' \theta$ and $G'' = \delta'' - V
\delta' - 2V' \delta + V'' \theta$. The identities of
Eq.~(\ref{11.3.2}) can be expressed as $(\sigma +
\epsilon)\delta'(\sigma + \epsilon) = - \delta(\sigma + \epsilon)$ and
$(\sigma + \epsilon)\delta''(\sigma + \epsilon) = - 2\delta'(\sigma +
\epsilon)$, and combining this with our previous results gives  
\begin{eqnarray*} 
(\Box - k^2) G^\epsilon_\pm(x,x') &=& 
(-2V - k^2) \delta_\pm(\sigma + \epsilon)   
+ (2\sigma V'' + 4 V' - k^2 V) \theta_\pm(-\sigma - \epsilon) 
\\ & & \mbox{} 
- 2 \epsilon \delta''_\pm(\sigma+\epsilon) 
+ 2 V \epsilon \delta'_\pm(\sigma + \epsilon) 
+ 4 V' \epsilon \delta_\pm(\sigma + \epsilon). 
\end{eqnarray*}
According to Eq.~(\ref{11.5.3})--(\ref{11.5.5}), the last two terms on
the right-hand side disappear in the limit $\epsilon \to 0^+$, and the 
third term becomes $-4\pi \delta_4(x-x')$. Provided that the first two
terms vanish also, we recover $(\Box - k^2)G(x,x') = -4\pi
\delta_4(x-x')$ in the limit, as required. Thus, the limit of
$G^\epsilon_\pm(x,x')$ when $\epsilon \to 0^+$ will indeed satisfy 
Green's equation provided that $V(\sigma)$ is a solution to  
\begin{equation}
2\sigma V'' + 4 V' - k^2 V = 0, \qquad
V(0) = - \frac{1}{2} k^2; 
\label{11.6.4}
\end{equation} 
these are the same statements as in Eq.~(\ref{11.4.2}). The solution
to these equations was produced in Eq.~(\ref{11.4.3}): 
\begin{equation} 
V(\sigma) = -\frac{k}{\sqrt{-2\sigma}}\, 
J_1 \bigl( k\sqrt{-2\sigma} \bigr), 
\label{11.6.5}
\end{equation}
and this completely determines the Green's functions of
Eqs.~(\ref{11.6.2}) and (\ref{11.6.3}). 
 
\section{Distributions in curved spacetime} 
\label{12} 

The distributions introduced in Sec.~\ref{11.5} can also be defined in
a four-dimensional spacetime with metric $g_{\alpha\beta}$. Here we 
produce the relevant generalizations of the results derived in that
section. 

\subsection{Invariant Dirac distribution} 
\label{12.1} 

We first introduce $\delta_4(x,x')$, an {\it invariant} Dirac
functional in a four-dimensional curved spacetime. This is
defined by the relations 
\begin{equation} 
\int_V f(x) \delta_4(x,x') \sqrt{-g}\, d^4 x = f(x'), \qquad 
\int_{V'} f(x') \delta_4(x,x') \sqrt{-g'}\, d^4 x' = f(x),
\label{12.1.1}
\end{equation}
where $f(x)$ is a smooth test function, $V$ any four-dimensional
region that contains $x'$, and $V'$ any four-dimensional region that
contains $x$. These relations imply that $\delta_4(x,x')$ is symmetric
in its arguments, and it is easy to see that  
\begin{equation}
\delta_4(x,x') = \frac{\delta_4(x-x')}{\sqrt{-g}} 
= \frac{\delta_4(x-x')}{\sqrt{-g'}}
= (gg')^{-1/4} \delta_4(x-x'), 
\label{12.1.2} 
\end{equation}
where $\delta_4(x-x') = \delta(x^0 - x^{\prime 0}) \delta(x^1 - 
x^{\prime 1}) \delta(x^2 - x^{\prime 2}) \delta(x^3 - x^{\prime 3})$
is the ordinary (coordinate) four-dimensional Dirac functional. The 
relations of Eq.~(\ref{12.1.2}) are all equivalent because $f(x)
\delta_4(x,x') = f(x') \delta_4(x,x')$ is a distributional identity;
the last form is manifestly symmetric in $x$ and $x'$. 

The invariant Dirac distribution satisfies the identities 
\begin{eqnarray}
\Omega_{\cdots}(x,x') \delta_4(x,x') &=& \bigl[ \Omega_{\cdots} 
\bigr] \delta_4(x,x'), 
\nonumber \\ 
& & \label{12.1.3} \\ 
\bigl( g^\alpha_{\ \alpha'}(x,x') \delta_4(x,x') \bigr)_{;\alpha} =
-\partial_{\alpha'} \delta_4(x,x'), &  &
\bigl( g^{\alpha'}_{\ \alpha}(x',x) \delta_4(x,x') \bigr)_{;\alpha'} = 
-\partial_{\alpha} \delta_4(x,x'), 
\nonumber
\end{eqnarray}
where $\Omega_{\cdots}(x,x')$ is any bitensor and 
$g^\alpha_{\ \alpha'}(x,x')$, $g^{\alpha'}_{\ \alpha}(x,x')$ are 
parallel propagators. The first identity follows immediately from the
definition of the $\delta$-function. The second and third identities
are established by showing that integration against a test function
$f(x)$ gives the same result from both sides. For example, the first
of the Eqs.~(\ref{12.1.1}) implies   
\[
\int_V f(x) \partial_{\alpha'}\delta_4(x,x') 
\sqrt{-g}\, d^4 x = \partial_{\alpha'} f(x'),  
\]
and on the other hand, 
\[
-\int_V f(x) \bigl( g^\alpha_{\ \alpha'} \delta_4(x,x')
\bigr)_{;\alpha} \sqrt{-g}\, d^4 x = - \oint_{\partial V} f(x)
g^\alpha_{\ \alpha'} \delta_4(x,x') d\Sigma_\alpha 
+ \bigl[ f_{,\alpha} g^\alpha_{\ \alpha'} \bigr] 
= \partial_{\alpha'} f(x'), 
\]
which establishes the second identity of Eq.~(\ref{12.1.3}). Notice
that in these manipulations, the integrations involve {\it scalar} 
functions of the coordinates $x$; the fact that these functions are
also vectors with respect to $x'$ does not invalidate the
procedure. The third identity of Eq.~(\ref{12.1.3}) is proved in a 
similar way.  

\subsection{Light-cone distributions} 
\label{12.2}

For the remainder of Sec.~\ref{12} we assume that $x \in 
{\cal N}(x')$, so that a unique geodesic $\beta$ links these two
points. We then let $\sigma(x,x')$ be the curved spacetime world
function, and we define light-cone step functions by   
\begin{equation} 
\theta_\pm(-\sigma) = \theta_\pm(x,\Sigma) \theta(-\sigma), \qquad  
x' \in \Sigma,
\label{12.2.1}
\end{equation}
where $\theta_+(x,\Sigma)$ is one when $x$ is in the future of the
spacelike hypersurface $\Sigma$ and zero otherwise, and
$\theta_-(x,\Sigma) = 1 - \theta_+(x,\Sigma)$. These are immediate
generalizations to curved spacetime of the objects defined in flat
spacetime by Eq.~(\ref{11.5.1}). We have that $\theta_+(-\sigma)$ is
one when $x$ is within $I^+(x')$, the chronological future of
$x'$, and zero otherwise, and $\theta_-(-\sigma)$ is one when $x$ is
within $I^-(x')$, the chronological past of $x'$, and zero
otherwise. We also have $\theta_+(-\sigma) + \theta_-(-\sigma) = 
\theta(-\sigma)$.  

We define the curved-spacetime version of the light-cone Dirac
functionals by     
\begin{equation} 
\delta_\pm(\sigma) = \theta_\pm(x,\Sigma) \delta(\sigma), \qquad
x' \in \Sigma,
\label{12.2.2}
\end{equation}
an immediate generalization of Eq.~(\ref{11.5.2}). We have that 
$\delta_+(\sigma)$, when viewed as a function of $x$, is supported on 
the future light cone of $x'$, while $\delta_-(\sigma)$ is supported
on its past light cone. We also have $\delta_+(\sigma) +
\delta_-(\sigma) = \delta(\sigma)$, and we recall that $\sigma$ is
negative when $x$ and $x'$ are timelike related, and positive when
they are spacelike related. 

For the same reasons as those mentioned in Sec.~\ref{11.5}, it is
sometimes convenient to shift the argument of the step and
$\delta$-functions from $\sigma$ to $\sigma + \epsilon$, where
$\epsilon$ is a small positive quantity. With this shift, the
light-cone distributions can be straightforwardly differentiated with
respect to $\sigma$. For example, $\delta_\pm(\sigma + \epsilon) = 
-\theta'_\pm(-\sigma-\epsilon)$, with a prime indicating 
differentiation with respect to $\sigma$.  

We now prove that the identities of Eq.~(\ref{11.5.3})--(\ref{11.5.5}) 
generalize to 
\begin{eqnarray} 
\lim_{\epsilon \to 0^+} \epsilon \delta_\pm(\sigma + \epsilon) &=& 0, 
\label{12.2.3} \\
\lim_{\epsilon \to 0^+} \epsilon \delta'_\pm(\sigma + \epsilon) &=& 0, 
\label{12.2.4} \\
\lim_{\epsilon \to 0^+} \epsilon \delta''_\pm(\sigma + \epsilon) &=&
2\pi \delta_4(x,x')  
\label{12.2.5} 
\end{eqnarray}
in a four-dimensional curved spacetime; the only differences lie with  
the definition of the world function and the fact that it is the
invariant Dirac functional that appears in Eq.~(\ref{12.2.5}). To 
establish these identities in curved spacetime we use the fact that
they hold in flat spacetime --- as was shown in Sec.~\ref{11.5} ---
and that they are scalar relations that must be valid in any
coordinate system if they are found to hold in one. Let us then
examine Eqs.~(\ref{12.2.3})--(\ref{12.2.4}) in the Riemann normal
coordinates of Sec.~\ref{7}; these are denoted
$\hat{x}^\alpha$ and are based at $x'$. We have that
$\sigma(x,x') = \frac{1}{2} \eta_{\alpha\beta} \hat{x}^\alpha
\hat{x}^\beta$ and $\delta_4(x,x') = \Delta(x,x') \delta_4(x-x') =
\delta_4(x-x')$, where $\Delta(x,x')$ is the van Vleck determinant,
whose coincidence limit is unity. In Riemann normal coordinates,
therefore, Eqs.~(\ref{12.2.3})--(\ref{12.2.5}) take exactly the same
form as Eqs.~(\ref{11.5.3})--(\ref{11.5.5}). Because the identities
are true in flat spacetime, they must be true also in curved spacetime
(in Riemann normal coordinates based at $x'$); and because these are
scalar relations, they must be valid in any coordinate system.     

\section{Scalar Green's functions in curved spacetime} 
\label{13}

\subsection{Green's equation for a massless scalar field in curved
spacetime} 
\label{13.1} 

We consider a massless scalar field $\Phi(x)$ in a curved spacetime 
with metric $g_{\alpha\beta}$. The field satisfies the wave equation 
\begin{equation}
( \Box - \xi R ) \Phi(x) = -4\pi \mu(x),   
\label{13.1.1}
\end{equation}
where $\Box = g^{\alpha\beta} \nabla_\alpha \nabla_\beta$ is the wave
operator, $R$ the Ricci scalar, $\xi$ an arbitrary coupling constant,
and $\mu(x)$ is a prescribed source. We seek a Green's function
$G(x,x')$ such that a solution to Eq.~(\ref{13.1.1}) can be expressed
as 
\begin{equation}
\Phi(x) = \int G(x,x') \mu(x') \sqrt{-g'}\, d^4 x', 
\label{13.1.2}
\end{equation} 
where the integration is over the entire spacetime. The wave equation
for the Green's function is 
\begin{equation}
( \Box - \xi R ) G(x,x') = -4\pi \delta_4(x,x'),  
\label{13.1.3}
\end{equation}
where $\delta_4(x,x')$ is the invariant Dirac functional introduced in
Sec.~\ref{12.1}. It is easy to verify that the field defined by
Eq.~(\ref{13.1.2}) is truly a solution to Eq.~(\ref{13.1.1}). 

We let $G_+(x,x')$ be the retarded solution to
Eq.~(\ref{13.1.3}), and $G_-(x,x')$ is the advanced solution;
when viewed as functions of $x$, $G_+(x,x')$ is nonzero in the causal 
future of $x'$, while $G_-(x,x')$ is nonzero in its causal past. We
assume that the retarded and advanced Green's functions exist as
distributions and can be defined globally in the entire spacetime.    

\subsection{Hadamard construction of the Green's functions} 
\label{13.2} 

Assuming throughout this subsection that $x$ is restricted to the
normal convex neighbourhood of $x'$, we make the ansatz     
\begin{equation}
G_\pm(x,x') = U(x,x') \delta_\pm(\sigma) 
+ V(x,x') \theta_\pm(-\sigma), 
\label{13.2.1}
\end{equation}
where $U(x,x')$ and $V(x,x')$ are smooth biscalars; the fact that the
spacetime is no longer homogeneous means that these functions cannot
depend on $\sigma$ alone.    

Before we substitute the Green's functions of Eq.~(\ref{13.2.1}) into
the differential equation of Eq.~(\ref{13.1.3}) we proceed as in
Sec.~\ref{11.6} and shift $\sigma$ by the small positive quantity 
$\epsilon$. We shall therefore consider the distributions
\[
G^\epsilon_\pm(x,x') = U(x,x') \delta_\pm(\sigma+\epsilon)  
+ V(x,x') \theta_\pm(-\theta-\epsilon),
\]
and later recover the Green's functions by taking the limit  
$\epsilon \to 0^+$. Differentiation of these objects is
straightforward, and in the following manipulations we will repeatedly 
use the relation $\sigma^\alpha \sigma_\alpha = 2 \sigma$
satisfied by the world function. We will also use the distributional
identities $\sigma \delta_\pm(\sigma + \epsilon) = -\epsilon
\delta_\pm(\sigma + \epsilon)$, $\sigma \delta'_\pm(\sigma + \epsilon)
= -\delta_\pm(\sigma + \epsilon) - \epsilon \delta'_\pm(\sigma +
\epsilon)$, and $\sigma \delta''_\pm(\sigma + \epsilon) = - 2
\delta'(\sigma+\epsilon) - \epsilon \delta''(\sigma +
\epsilon)$. After a routine calculation we obtain 
\begin{eqnarray*}
( \Box - \xi R ) G^\epsilon_\pm &=& 
- 2\epsilon \delta''_\pm(\sigma + \epsilon) U 
+ 2\epsilon \delta'_\pm(\sigma + \epsilon) V 
+ \delta'_\pm(\sigma+\epsilon) \Bigl\{ 2 U_{,\alpha} \sigma^\alpha 
  + (\sigma^\alpha_{\ \alpha} - 4) U \Bigr\} 
\\ & & \mbox{} 
+ \delta_\pm(\sigma+\epsilon) \Bigl\{ - 2 V_{,\alpha} \sigma^\alpha 
  + (2-\sigma^\alpha_{\ \alpha}) V + (\Box - \xi R) U \Bigr\} 
+ \theta_\pm(-\sigma-\epsilon) \Bigl\{ (\Box - \xi R) V \Bigr\}, 
\end{eqnarray*} 
which becomes 
\begin{eqnarray}
( \Box - \xi R ) G_\pm &=&  
- 4\pi \delta_4(x,x') U  
+ \delta'_\pm(\sigma) \Bigl\{ 2 U_{,\alpha} \sigma^\alpha  
  + (\sigma^\alpha_{\ \alpha} - 4) U \Bigr\} 
\nonumber \\ & & \mbox{} 
+ \delta_\pm(\sigma) \Bigl\{ - 2 V_{,\alpha} \sigma^\alpha 
  + (2-\sigma^\alpha_{\ \alpha}) V + (\Box - \xi R) U \Bigr\} 
+ \theta_\pm(-\sigma) \Bigl\{ (\Box - \xi R) V \Bigr\} \qquad
\label{13.2.2} 
\end{eqnarray} 
in the limit $\epsilon \to 0^+$,  after using the identities of
Eqs.~(\ref{12.2.3})--(\ref{12.2.5}).   
 
According to Eq.~(\ref{13.1.3}), the right-hand side of
Eq.~(\ref{13.2.2}) should be equal to $-4\pi \delta_4(x,x')$. This
immediately gives us the coincidence condition   
\begin{equation}
\bigl[ U \bigr] = 1
\label{13.2.3} 
\end{equation} 
for the biscalar $U(x,x')$. To eliminate the $\delta'_\pm$ term we 
make its coefficient vanish:
\begin{equation} 
2 U_{,\alpha} \sigma^\alpha + (\sigma^\alpha_{\ \alpha} - 4) U = 0. 
\label{13.2.4} 
\end{equation}
As we shall now prove, these two equations determine $U(x,x')$ 
uniquely. 

Recall from Sec.~\ref{2.3} that $\sigma^\alpha$ is a vector at $x$
that is tangent to the unique geodesic $\beta$ that connects $x$ to
$x'$. This geodesic is affinely parameterized by $\lambda$ and a
displacement along $\beta$ is described by $dx^\alpha =
(\sigma^\alpha/\lambda) d\lambda$. The first term of
Eq.~(\ref{13.2.4}) therefore represents the logarithmic rate of change
of $U(x,x')$ along $\beta$, and this can be expressed as $2 \lambda  
dU/d\lambda$. For the second term we recall from Sec.~\ref{6.1} the
differential equation $\Delta^{-1} (\Delta \sigma^\alpha)_{;\alpha} 
= 4$ satisfied by $\Delta(x,x')$, the van Vleck determinant. This
gives us $\sigma^\alpha_{\ \alpha} - 4 = -\Delta^{-1} \Delta_{,\alpha} 
\sigma^\alpha = -\Delta^{-1} \lambda d \Delta/d\lambda$, and  
Eq.~(\ref{13.2.4}) becomes   
\[
\lambda \frac{d}{d\lambda} \bigl( 2 \ln U - \ln \Delta \bigr) = 0.
\]
It follows that $U^2/\Delta$ is constant on $\beta$, and this must
therefore be equal to its value at the starting point $x'$:
$U^2/\Delta = [U^2/\Delta] = 1$, by virtue of Eq.~(\ref{13.2.3}) and
the property $[\Delta] = 1$ of the van Vleck determinant. Because this  
statement must be true for all geodesics $\beta$ that emanate from
$x'$, we have found that the unique solution to Eqs.~(\ref{13.2.3})
and (\ref{13.2.4}) is  
\begin{equation}
U(x,x') = \Delta^{1/2}(x,x').  
\label{13.2.5}
\end{equation}   

We must still consider the remaining terms in Eq.~(\ref{13.2.2}). The 
$\delta_\pm$ term can be eliminated by demanding that its coefficient
vanish when $\sigma = 0$. This, however, does not constrain its value
away from the light cone, and we thus obtain information about
$V|_{\sigma = 0}$ only. Denoting this by $\check{V}(x,x')$ --- the
restriction of $V(x,x')$ on the light cone $\sigma(x,x') = 0$ --- we
have  
\begin{equation} 
\check{V}_{,\alpha} \sigma^\alpha 
+ \frac{1}{2} \bigl( \sigma^\alpha_{\ \alpha} - 2 \bigr) \check{V} 
= \frac{1}{2} \bigl( \Box - \xi R \bigr) U \Bigr|_{\sigma=0}, 
\label{13.2.6}
\end{equation}
where we indicate that the right-hand side also must be
restricted to the light cone. The first term of Eq.~(\ref{13.2.6}) can 
be expressed as $\lambda d\check{V}/d\lambda$ and this equation can be  
integrated along any null geodesic that generates the null cone
$\sigma(x,x') = 0$. For these integrations to be well posed, however, 
we must provide initial values at $x = x'$. As we shall now see, these
can be inferred from Eq.~(\ref{13.2.6}) and the fact that $V(x,x')$
must be smooth at coincidence. 

Equations (\ref{6.1.4}) and (\ref{13.2.5}) imply that near
coincidence, $U(x,x')$ admits the expansion    
\begin{equation} 
U = 1 + \frac{1}{12} R_{\alpha'\beta'} \sigma^{\alpha'}
\sigma^{\beta'} + O(\lambda^3), 
\label{13.2.7}
\end{equation} 
where $R_{\alpha'\beta'}$ is the Ricci tensor at $x'$ and $\lambda$ 
is the affine-parameter distance to $x$ (which can be either on or off   
the light cone). Differentiation of this relation gives  
\begin{equation}
U_{,\alpha} = -\frac{1}{6} g^{\alpha'}_{\ \alpha} R_{\alpha'\beta'}
\sigma^{\beta'} + O(\lambda^2), \qquad 
U_{,\alpha'} = \frac{1}{6} R_{\alpha'\beta'} \sigma^{\beta'} +
O(\lambda^2), 
\label{13.2.8}
\end{equation} 
and eventually, 
\begin{equation}
\bigl[ \Box U \bigr] = \frac{1}{6} R(x'). 
\label{13.2.9} 
\end{equation} 
Using also $[\sigma^\alpha_{\ \alpha}] = 4$, we find that the
coincidence limit of Eq.~(\ref{13.2.6}) gives  
\begin{equation}
\bigl[ V \bigr] = \frac{1}{12} \Bigl( 1 - 6\xi \Bigr) R(x'), 
\label{13.2.10} 
\end{equation}
and this provides the initial values required for the integration of 
Eq.~(\ref{13.2.6}) on the null cone.   

Equations (\ref{13.2.6}) and (\ref{13.2.10}) give us a means to
construct $\check{V}(x,x')$, the restriction of $V(x,x')$ on the null
cone $\sigma(x,x') = 0$. These values can then be used as
characteristic data for the wave equation   
\begin{equation}
(\Box - \xi R) V(x,x') = 0, 
\label{13.2.11} 
\end{equation} 
which is obtained by elimination of the $\theta_\pm$ term in
Eq.~(\ref{13.2.2}). While this certainly does not constitute a
practical method to compute the biscalar $V(x,x')$, these
considerations show that $V(x,x')$ exists and is unique.   

To summarize: We have shown that with $U(x,x')$ given by 
Eq.~(\ref{13.2.5}) and $V(x,x')$ determined uniquely by the wave 
equation of Eq.~(\ref{13.2.11}) and the characteristic data
constructed with Eqs.~(\ref{13.2.6}) and (\ref{13.2.10}), the retarded
and advanced Green's functions of Eq.~(\ref{13.2.1}) do indeed satisfy 
Eq.~(\ref{13.1.3}). It should be emphasized that the construction
provided in this subsection is restricted to ${\cal N}(x')$, the
normal convex neighbourhood of the reference point $x'$.  

\subsection{Reciprocity} 
\label{13.3} 

We shall now establish the following reciprocity relation between the
(globally defined) retarded and advanced Green's functions:   
\begin{equation}
G_-(x',x) = G_+(x,x').
\label{13.3.1}   
\end{equation}
Before we get to the proof we observe that by virtue of
Eq.~(\ref{13.3.1}), the biscalar $V(x,x')$ must be symmetric in its 
arguments:   
\begin{equation}
V(x',x) = V(x,x'). 
\label{13.3.2}  
\end{equation}
To go from Eq.~(\ref{13.3.1}) to Eq.~(\ref{13.3.2}) we simply
note that when $x \in {\cal N}(x')$ and belongs to $I^+(x')$, then 
$G_+(x,x') = V(x,x')$ and $G_-(x',x) = V(x',x)$.  

To prove the reciprocity relation we invoke the identities 
\[
G_+(x,x') (\Box - \xi R) G_-(x,x'') = -4\pi G_+(x,x') \delta_4(x,x'') 
\]
and 
\[
G_-(x,x'') (\Box - \xi R) G_+(x,x') = -4\pi G_-(x,x'')
\delta_4(x,x')
\]
and take their difference. On the left-hand side we have
\[
G_+(x,x') \Box G_-(x,x'') - G_-(x,x'') \Box G_+(x,x') = \nabla_\alpha 
\Bigl( G_+(x,x') \nabla^\alpha G_-(x,x'') - G_-(x,x'') \nabla^\alpha
G_+(x,x') \Bigr), 
\]
while the right-hand side gives 
\[
-4\pi \Bigl( G_+(x,x') \delta_4(x,x'') - G_-(x,x'') \delta_4(x,x')
\Bigr). 
\]
Integrating both sides over a large four-dimensional region $V$ that
contains both $x'$ and $x''$, we obtain 
\[
\oint_{\partial V} \Bigl( G_+(x,x') \nabla^\alpha G_-(x,x'') -
G_-(x,x'') \nabla^\alpha G_+(x,x') \Bigr)\, d\Sigma_\alpha = 
-4\pi \Bigl( G_+(x'',x') - G_-(x',x'') \Bigr), 
\]
where $\partial V$ is the boundary of $V$. Assuming that the Green's
functions fall off sufficiently rapidly at infinity (in the limit
$\partial V \to \infty$; this statement imposes some
restriction on the spacetime's asymptotic structure), we have that the
left-hand side of the equation evaluates to zero in the limit. This
gives us the statement $G_+(x'',x') = G_-(x',x'')$, which is just
Eq.~(\ref{13.3.1}) with $x''$ replacing $x$.  

\subsection{Kirchhoff representation}  
\label{13.4} 

Suppose that the values for a scalar field $\Phi(x')$ and its normal
derivative $n^{\alpha'} \nabla_{\alpha'} \Phi(x')$ are known on a
spacelike hypersurface $\Sigma$. Suppose also that the scalar field 
satisfies the homogeneous wave equation  
\begin{equation}
(\Box - \xi R) \Phi(x) = 0. 
\label{13.4.1}
\end{equation} 
Then the value of the field at a point $x$ in the future of $\Sigma$ 
is given by Kirchhoff's formula,  
\begin{equation} 
\Phi(x) = -\frac{1}{4\pi} \int_\Sigma \Bigl( G_+(x,x')
\nabla^{\alpha'} \Phi(x') - \Phi(x') \nabla^{\alpha'} G_+(x,x')
\Bigr)\, d\Sigma_{\alpha'}, 
\label{13.4.2} 
\end{equation}
where $d\Sigma_{\alpha'}$ is the surface element on $\Sigma$. If
$n_{\alpha'}$ is the future-directed unit normal, then
$d\Sigma_{\alpha'} = - n_{\alpha'} dV$, with $dV$ denoting the
invariant volume element on $\Sigma$; notice that $d\Sigma_{\alpha'}$
is past directed.  

To establish this result we start with the equations 
\[
G_-(x',x) (\Box' - \xi R') \Phi(x') = 0, \qquad 
\Phi(x') (\Box' - \xi R') G_-(x',x) = -4\pi \delta_4(x',x) \Phi(x'),  
\]
in which $x$ and $x'$ refer to arbitrary points in spacetime. Taking
their difference gives 
\[
\nabla_{\alpha'} \Bigl( G_-(x',x) \nabla^{\alpha'} \Phi(x') - \Phi(x')
\nabla^{\alpha'} G_-(x',x) \Bigr) = 4\pi \delta_4(x',x) \Phi(x'), 
\]
and this we integrate over a four-dimensional region $V$ that is
bounded in the past by the hypersurface $\Sigma$. We suppose that $V$ 
contains $x$ and we obtain  
\[
\oint_{\partial V} \Bigl( G_-(x',x) \nabla^{\alpha'} \Phi(x') -
\Phi(x') \nabla^{\alpha'} G_-(x',x) \Bigr)\, d\Sigma_{\alpha'} 
= 4\pi \Phi(x), 
\]
where $d\Sigma_{\alpha'}$ is the outward-directed surface element on
the boundary $\partial V$. Assuming that the Green's function falls
off sufficiently rapidly into the future, we have that the only
contribution to the hypersurface integral is the one that comes from
$\Sigma$. Since the surface element on $\Sigma$ points in the
direction opposite to the outward-directed surface element on
$\partial V$, we must change the sign of the left-hand side to be
consistent with the convention adopted previously. With this change 
we have 
\[
\Phi(x) = -\frac{1}{4\pi} \oint_{\partial V} \Bigl( G_-(x',x)
\nabla^{\alpha'} \Phi(x') - \Phi(x') \nabla^{\alpha'} G_-(x',x)
\Bigr)\, d\Sigma_{\alpha'},
\]
which is the same statement as Eq.~(\ref{13.4.2}) if we take into
account the reciprocity relation of Eq.~(\ref{13.3.1}). 

\subsection{Singular and regular Green's functions}  
\label{13.5} 

In part \ref{part4} of this review we will compute the retarded field
of a moving scalar charge, and we will analyze its singularity
structure near the world line; this will be part of our effort to
understand the effect of the field on the particle's motion. The
retarded solution to the scalar wave equation is the physically
relevant solution because it properly incorporates outgoing-wave
boundary conditions at infinity --- the advanced solution would come
instead with incoming-wave boundary conditions. The retarded field is
singular on the world line because a point particle produces a Coulomb
field that diverges at the particle's position. In view of this
singular behaviour, it is a subtle matter to describe the field's
action on the particle, and to formulate meaningful equations of
motion.       

When facing this problem in flat spacetime (recall the discussion of
Sec.~\ref{1.3}) it is convenient to decompose the retarded Green's
function $G_+(x,x')$ into a {\it singular} Green's function 
$G_{\rm S}(x,x') := \frac{1}{2}
[G_+(x,x') + G_-(x,x')]$ and a {\it regular} two-point
function $G_{\rm R}(x,x') := \frac{1}{2} [G_+(x,x') 
- G_-(x,x')]$. The singular Green's function takes its name from
the fact that it produces a field with the same singularity
structure as the retarded solution: the diverging field near the
particle is insensitive to the boundary conditions imposed at 
infinity. We note also that $G_{\rm S}(x,x')$ satisfies the same wave 
equation as the retarded Green's function (with a Dirac functional as
a source), and that by virtue of the reciprocity relations, it is
symmetric in its arguments. The regular two-point function, on the
other hand, takes its name from the fact that it satisfies the 
{\it homogeneous} wave equation, without the Dirac functional on the
right-hand side; it produces a field that is regular on the world line
of the moving scalar charge. (We reserve the term ``Green's function''
to a two-point function that satisfies the wave equation with a Dirac
distribution on the right-hand side; when the source term is absent,
the object is called a ``two-point function''.)   

Because the singular Green's function is symmetric in its argument, it
does not distinguish between past and future, and it produces a field
that contains equal amounts of outgoing and incoming radiation --- the
singular solution describes a standing wave at infinity.  
Removing $G_{\rm S}(x,x')$ from the retarded Green's
function will have the effect of removing the singular
behaviour of the field {\it without affecting the motion of the
particle}. The motion is not affected because it is intimately
tied to the boundary conditions: If the waves are outgoing, the  
particle loses energy to the radiation and its motion is affected; if
the waves are incoming, the particle gains energy from the radiation
and its motion is affected differently. With equal amounts of outgoing
and incoming radiation, the particle neither loses nor gains energy
and its interaction with the scalar field cannot affect its
motion. Thus, subtracting $G_{\rm S}(x,x')$ from the retarded
Green's function eliminates the singular part of the field without
affecting the motion of the scalar charge. The subtraction leaves 
behind the regular two-point function, which produces a field that is
regular on the world line; it is this field that will govern the  
motion of the particle. The action of this field is well defined, 
and it properly encodes the outgoing-wave boundary conditions: the 
particle will lose energy to the radiation.       

In this subsection we attempt a decomposition of the curved-spacetime
retarded Green's function into singular and regular pieces. The
flat-spacetime relations will have to be amended, however, because of
the fact that in a curved spacetime, the advanced Green's function is
generally nonzero when $x'$ is in the chronological future of
$x$. This implies that the value of the advanced field at $x$ depends
on events $x'$ that will unfold {\it in the future}; this dependence
would be inherited by the regular field (which acts on the particle
and determines its motion) if the naive definition 
$G_{\rm R}(x,x') := \frac{1}{2} [G_{+}(x,x') - G_{-}(x,x')]$ were to
be adopted.    

We shall not adopt this definition. Instead, we shall follow Detweiler
and Whiting \cite{detweiler-whiting:03} and introduce a singular Green's
function with the properties    
\begin{description} 
\item[\qquad {\sf S1}:] $G_{\rm S}(x,x')$ satisfies the inhomogeneous
scalar wave equation,     
\begin{equation} 
(\Box - \xi R) G_{\rm S}(x,x') = -4\pi \delta_4(x,x'); 
\label{13.5.1} 
\end{equation} 
\item[\qquad {\sf S2}:] $G_{\rm S}(x,x')$ is symmetric in its
arguments,  
\begin{equation} 
G_{\rm S}(x',x) = G_{\rm S}(x,x');
\label{13.5.2}
\end{equation} 
\item[\qquad {\sf S3}:] $G_{\rm S}(x,x')$ vanishes if $x$ is in the 
chronological past or future of $x'$, 
\begin{equation}
G_{\rm S}(x,x') = 0 \qquad \mbox{when $x \in I^\pm(x')$}. 
\label{13.5.3}
\end{equation} 
\end{description} 
Properties {\sf S1} and {\sf S2} ensure that the singular Green's
function will properly reproduce the singular behaviour of the
retarded solution without distinguishing between past and future; and  
as we shall see, property {\sf S3} ensures that the support of the
regular two-point function will not include the chronological future
of $x$. 

The regular two-point function is then defined by 
\begin{equation}
G_{\rm R}(x,x') = G_+(x,x') - G_{\rm S}(x,x'), 
\label{13.5.4}
\end{equation}
where $G_+(x,x')$ is the retarded Green's function. This comes with
the properties 
\begin{description} 
\item[\qquad {\sf R1}:] $G_{\rm R}(x,x')$ satisfies the homogeneous wave
equation, 
\begin{equation}
(\Box - \xi R) G_{\rm R}(x,x') = 0;  
\label{13.5.5}
\end{equation} 
\item[\qquad {\sf R2}:] $G_{\rm R}(x,x')$ agrees with the retarded
Green's function if $x$ is in the chronological future of $x'$, 
\begin{equation}
G_{\rm R}(x,x') = G_+(x,x') \qquad \mbox{when $x \in I^+(x')$}; 
\label{13.5.6}
\end{equation} 
\item[\qquad {\sf R3}:] $G_{\rm R}(x,x')$ vanishes if $x$ is in the 
chronological past of $x'$,  
\begin{equation}
G_{\rm R}(x,x') = 0 \qquad \mbox{when $x \in I^-(x')$}.  
\label{13.5.7}
\end{equation} 
\end{description}
Property {\sf R1} follows directly from Eq.~(\ref{13.5.4}) and
property {\sf S1} of the singular Green's function. Properties 
{\sf R2} and {\sf R3} follow from {\sf S3} and the fact that the
retarded Green's function vanishes if $x$ is in past of $x'$. The
properties of the regular two-point function ensure that the
corresponding regular field will be nonsingular at the world line, and  
will depend only on the past history of the scalar charge.  

We must still show that such singular and regular Green's functions 
can be constructed. This relies on the existence of a two-point
function $H(x,x')$ that would possess the properties  
\begin{description} 
\item[\qquad {\sf H1}:] $H(x,x')$ satisfies the homogeneous wave
equation, 
\begin{equation}
(\Box - \xi R) H(x,x') = 0; 
\label{13.5.8}
\end{equation} 
\item[\qquad {\sf H2}:] $H(x,x')$ is symmetric in its arguments, 
\begin{equation}
H(x',x) = H(x,x'); 
\label{13.5.9}
\end{equation} 
\item[\qquad {\sf H3}:] $H(x,x')$ agrees with the retarded 
Green's function if $x$ is in the chronological future of $x'$,  
\begin{equation}
H(x,x') = G_+(x,x') \qquad \mbox{when $x \in I^+(x')$};  
\label{13.5.10}
\end{equation} 
\item[\qquad {\sf H4}:] $H(x,x')$ agrees with the advanced  
Green's function if $x$ is in the chronological past of $x'$,  
\begin{equation}
H(x,x') = G_-(x,x') \qquad \mbox{when $x \in I^-(x')$}.   
\label{13.5.11}
\end{equation} 
\end{description}
With a biscalar $H(x,x')$ satisfying these relations, a singular
Green's function defined by    
\begin{equation}
G_{\rm S}(x,x') = \frac{1}{2} \Bigl[ G_+(x,x') + G_-(x,x') - H(x,x')
\Bigr]
\label{13.5.12}
\end{equation}  
will satisfy all the properties listed previously: {\sf S1} comes as a
consequence of {\sf H1} and the fact that both the advanced and the
retarded Green's functions are solutions to the inhomogeneous wave
equation, {\sf S2} follows directly from {\sf H2} and the definition
of Eq.~(\ref{13.5.12}), and {\sf S3} comes as a consequence of 
{\sf H3}, {\sf H4} and the properties of the retarded and advanced
Green's functions. 

The question is now: does such a function $H(x,x')$ exist? We will 
present a plausibility argument for an affirmative answer. Later in
this section we will see that $H(x,x')$ is guaranteed to exist in the
local convex neighbourhood of $x'$, where it is equal to
$V(x,x')$. And in Sec.~\ref{13.6} we will see that there exist
particular spacetimes for which $H(x,x')$ can be defined globally.   

To satisfy all of {\sf H1}--{\sf H4} might seem a tall order, but it
should be possible. We first note that property {\sf H4} is not
independent from the rest: it follows from {\sf H2}, {\sf H3},
and the reciprocity relation (\ref{13.3.1}) satisfied by the retarded
and advanced Green's functions. Let $x \in I^-(x')$, so that $x' \in
I^+(x)$. Then $H(x,x') = H(x',x)$ by {\sf H2}, and by {\sf H3} this is
equal to $G_+(x',x)$. But by the reciprocity relation this is also
equal to $G_-(x,x')$, and we have obtained {\sf H4}. Alternatively,
and this shall be our point of view in the next paragraph, we can
think of {\sf H3} as following from {\sf H2} and {\sf H4}.  

Because $H(x,x')$ satisfies the homogeneous wave equation (property 
{\sf H1}), it can be given the Kirkhoff representation of
Eq.~(\ref{13.4.2}): if $\Sigma$ is a spacelike hypersurface in the
past of both $x$ and $x'$, then    
\[
H(x,x') = -\frac{1}{4\pi} \int_\Sigma \Bigl( G_+(x,x'')
\nabla^{\alpha''} H(x'',x') - H(x'',x') \nabla^{\alpha''} G_+(x,x'')  
\Bigr)\, d\Sigma_{\alpha''},
\]
where $d\Sigma_{\alpha''}$ is a surface element on $\Sigma$. The 
hypersurface can be partitioned into two segments, $\Sigma^-(x')$ and
$\Sigma - \Sigma^-(x')$, with $\Sigma^-(x')$ denoting the intersection
of $\Sigma$ with $I^-(x')$. To enforce {\sf H4} it suffices to choose
for $H(x,x')$ initial data on $\Sigma^-(x')$ that agree with the
initial data for the advanced Green's function; because both functions
satisfy the homogeneous wave equation in $I^-(x')$, the agreement will
be preserved in the entire domain of dependence of $\Sigma^-(x')$. The
data on $\Sigma - \Sigma^-(x')$ is still free, and it should be
possible to choose it so as to make $H(x,x')$ symmetric. Assuming that
this can be done, we see that {\sf H2} is enforced and we conclude 
that the properties {\sf H1}, {\sf H2}, {\sf H3}, and {\sf H4} can all
be satisfied. 

When $x$ is restricted to the normal convex neighbourhood of $x'$,
properties {\sf H1}--{\sf H4} imply that  
\begin{equation} 
H(x,x') = V(x,x'); 
\label{13.5.13}
\end{equation} 
it should be stressed here that while $H(x,x')$ is assumed to be
defined globally in the entire spacetime, the existence of $V(x,x')$
is limited to ${\cal N}(x')$. With Eqs.~(\ref{13.2.1}) and
(\ref{13.5.12}) we find that the singular Green's function is given
explicitly by   
\begin{equation} 
G_{\rm S}(x,x') = \frac{1}{2} U(x,x') \delta(\sigma) - \frac{1}{2}
V(x,x') \theta(\sigma)
\label{13.5.14}
\end{equation}
in the normal convex neighbourhood. Equation (\ref{13.5.14}) shows
very clearly that the singular Green's function does not distinguish 
between past and future (property {\sf S2}), and that its support
excludes $I^\pm(x')$, in which $\theta(\sigma) = 0$ (property 
{\sf S3}). From Eq.~(\ref{13.5.4}) we get an analogous expression for
the regular two-point function: 
\begin{equation} 
G_{\rm R}(x,x') = \frac{1}{2} U(x,x') \Bigl[ \delta_+(\sigma) -
\delta_-(\sigma) \Bigr] + V(x,x') \Bigl[ \theta_+(-\sigma) +
\frac{1}{2} \theta(\sigma) \Bigr]. 
\label{13.5.15}
\end{equation} 
This reveals directly that the regular two-point function coincides
with $G_+(x,x')$ in $I^+(x')$, in which $\theta(\sigma) = 0$ and
$\theta_+(-\sigma) = 1$ (property {\sf R2}), and that its support does  
not include $I^-(x')$, in which $\theta(\sigma) = \theta_+(-\sigma) =
0$ (property {\sf R3}).     

\subsection{Example: Cosmological Green's functions}  
\label{13.6} 

To illustrate the general theory outlined in the previous subsections
we consider here the specific case of a minimally coupled ($\xi=0$)
scalar field in a cosmological spacetime with metric  
\begin{equation}
ds^2 = a^2(\eta)(-d\eta^2 + dx^2 + dy^2 + dz^2), 
\label{13.6.1}
\end{equation}
where $a(\eta)$ is the scale factor expressed in terms of
conformal time. For concreteness we take the universe to be matter 
dominated, so that $a(\eta) = C \eta^2$, where $C$ is a constant. This
spacetime is one of the very few for which Green's functions can be
explicitly constructed. The calculation presented here was first
carried out by Burko, Harte, and Poisson \cite{burko-etal:02}; it can
be extended to other cosmologies \cite{haas-poisson:05}. 

To solve Green's equation $\Box G(x,x') = -4\pi \delta_4(x,x')$ we
first introduce a reduced Green's function $g(x,x')$ defined by  
\begin{equation}
G(x,x') = \frac{g(x,x')}{a(\eta) a(\eta')}. 
\label{13.6.2}
\end{equation} 
Substitution yields 
\begin{equation}
\biggl( - \frac{\partial^2}{\partial \eta^2} + \nabla^2 +
\frac{2}{\eta^2} \biggr) g(x,x') = 
-4\pi \delta(\eta - \eta') \delta_3(\bm{x} - \bm{x'}),    
\label{13.6.3}
\end{equation} 
where $\bm{x} = (x,y,z)$ is a vector in three-dimensional flat space,
and $\nabla^2$ is the Laplacian operator in this space. We next expand 
$g(x,x')$ in terms of plane-wave solutions to Laplace's equation,  
\begin{equation}
g(x,x') = \frac{1}{(2\pi)^3}\, \int \tilde{g}(\eta,\eta';\bm{k})\,
e^{i \bm{k} \cdot (\bm{x} - \bm{x'})}\, d^3 k,
\label{13.6.4} 
\end{equation}
and we substitute this back into Eq.~(\ref{13.6.3}). The result, after
also Fourier transforming $\delta_3(\bm{x}-\bm{x'})$, is an ordinary
differential equation for $\tilde{g}(\eta,\eta';\bm{k})$: 
\begin{equation}
\biggl( \frac{d^2}{d\eta^2} + k^2 - \frac{2}{\eta^2} \biggr)
\tilde{g} = 4\pi \delta(\eta - \eta'), 
\label{13.6.5}
\end{equation}
where $k^2 = \bm{k} \cdot \bm{k}$. To generate the retarded
Green's function we set  
\begin{equation}
\tilde{g}_+(\eta,\eta';\bm{k}) = \theta(\eta-\eta')\, 
\hat{g}(\eta,\eta';k),  
\label{13.6.6}
\end{equation}
in which we indicate that $\hat{g}$ depends only on the modulus of the 
vector $\bm{k}$. To generate the advanced Green's function we would
set instead $\tilde{g}_-(\eta,\eta';\bm{k}) = \theta(\eta'-\eta)\, 
\hat{g}(\eta,\eta';k)$. The following manipulations will refer
specifically to the retarded Green's function; they are easily 
adapted to the case of the advanced Green's function.  

Substitution of Eq.~(\ref{13.6.6}) into Eq.~(\ref{13.6.5}) reveals
that $\hat{g}$ must satisfy the homogeneous equation  
\begin{equation}
\biggl( \frac{d^2}{d\eta^2} + k^2 - \frac{2}{\eta^2} \biggr)
\hat{g} = 0, 
\label{13.6.7}
\end{equation}
together with the boundary conditions 
\begin{equation}
\hat{g}(\eta=\eta';k) = 0, \qquad
\frac{d\hat{g}}{d\eta}(\eta=\eta';k) = 4\pi. 
\label{13.6.8}
\end{equation} 
Inserting Eq.~(\ref{13.6.6}) into Eq.~(\ref{13.6.4}) and integrating
over the angular variables associated with the vector $\bm{k}$ yields   
\begin{equation}
g_+(x,x') = \frac{\theta(\Delta \eta)}{2\pi^2 R} \int_0^\infty  
\hat{g}(\eta,\eta';k)\, k \sin(kR)\, dk, 
\label{13.6.9}
\end{equation}
where $\Delta \eta := \eta - \eta'$ and 
$R := |\bm{x}-\bm{x'}|$.  

Equation (\ref{13.6.7}) has $\cos (k \Delta \eta) - (k \eta)^{-1} 
\sin (k \Delta \eta)$ and $\sin (k \Delta \eta) + (k \eta)^{-1} 
\cos (k \Delta \eta)$ as linearly independent solutions, and
$\hat{g}(\eta,\eta';k)$ must be given by a linear superposition. The
coefficients can be functions of $\eta'$, and after imposing  
Eqs.~(\ref{13.6.8}) we find that the appropriate combination is       
\begin{equation}
\hat{g}(\eta,\eta';k) = \frac{4\pi}{k}\, \biggl[ 
\biggl( 1 + \frac{1}{k^2 \eta \eta'} \biggr) \sin(k \Delta \eta) 
- \frac{\Delta \eta}{k \eta \eta'}\, \cos(k \Delta \eta) 
\biggr]. 
\label{13.6.10}
\end{equation}
Substituting this into Eq.~(\ref{13.6.9}) and using the identity 
$(2/\pi) \int_0^\infty \sin(\omega x) \sin(\omega x')\, d\omega
= \delta(x-x') - \delta(x+x')$ yields 
\[
g_+(x,x') = \frac{\delta(\Delta \eta - R)}{R} 
+ \frac{\theta(\Delta \eta)}{\eta \eta'}\, 
\frac{2}{\pi} \int_0^\infty \frac{1}{k}\,  
\sin(k \Delta \eta) \cos(k R)\, dk
\]
after integration by parts. The integral evaluates to 
$\theta(\Delta \eta - R)$. 

We have arrived at 
\begin{equation}
g_+(x,x') = \frac{\delta(\eta - \eta' 
- |\bm{x}-\bm{x'}|)}{|\bm{x}-\bm{x'}|}  
+ \frac{\theta(\eta - \eta' - |\bm{x}-\bm{x'}|)}{\eta \eta'}  
\label{13.6.11} 
\end{equation}
for our final expression for the retarded Green's function. The
advanced Green's function is given instead by 
\begin{equation}
g_-(x,x') = \frac{\delta(\eta - \eta' 
+ |\bm{x}-\bm{x'}|)}{|\bm{x}-\bm{x'}|}  
+ \frac{\theta(-\eta + \eta' - |\bm{x}-\bm{x'}|)}{\eta \eta'}.   
\label{13.6.12} 
\end{equation}
The distributions $g_\pm(x,x')$ are solutions to the reduced Green's
equation of Eq.~(\ref{13.6.3}). The actual Green's functions are
obtained by substituting Eqs.~(\ref{13.6.11}) and (\ref{13.6.12}) into 
Eq.~(\ref{13.6.2}). We note that the support of the retarded Green's
function is given by $\eta - \eta' \geq |\bm{x}-\bm{x'}|$, while the
support of the advanced Green's function is given by $\eta - \eta'
\leq -|\bm{x}-\bm{x'}|$.  

It may be verified that the symmetric two-point function 
\begin{equation}
h(x,x') = \frac{1}{\eta \eta'} 
\label{13.6.13}
\end{equation}
satisfies all of the properties {\sf H1}--{\sf H4} listed in
Sec.~\ref{13.5}; it may thus be used to define singular and regular 
Green's functions. According to
Eq.~(\ref{13.5.12}) the singular Green's function is given by 
\begin{eqnarray}
g_{\rm S}(x,x') &=& \frac{1}{2|\bm{x}-\bm{x'}|} \Bigl[ 
\delta(\eta - \eta' - |\bm{x}-\bm{x'}|)
+ \delta(\eta - \eta' + |\bm{x}-\bm{x'}|) \Bigr] 
\nonumber \\ & & \mbox{} 
+ \frac{1}{2\eta \eta'} \Bigr[ 
\theta(\eta - \eta' - |\bm{x}-\bm{x'}|)
- \theta(\eta - \eta' + |\bm{x}-\bm{x'}|) \Bigr]
\label{13.6.14}
\end{eqnarray} 
and its support is limited to the interval $-|\bm{x}-\bm{x'}| \leq
\eta-\eta' \leq |\bm{x}-\bm{x'}|$. According to Eq.~(\ref{13.5.4}) the
regular two-point function is given by 
\begin{eqnarray}
g_{\rm R}(x,x') &=& \frac{1}{2|\bm{x}-\bm{x'}|} \Bigl[ 
\delta(\eta - \eta' - |\bm{x}-\bm{x'}|)
- \delta(\eta - \eta' + |\bm{x}-\bm{x'}|) \Bigr] 
\nonumber \\ & & \mbox{} 
+ \frac{1}{2\eta \eta'} \Bigr[ 
\theta(\eta - \eta' - |\bm{x}-\bm{x'}|)
+ \theta(\eta - \eta' + |\bm{x}-\bm{x'}|) \Bigr]; 
\label{13.6.15}
\end{eqnarray} 
its support is given by $\eta-\eta' \geq - |\bm{x}-\bm{x'}|$ and for 
$\eta-\eta' \geq |\bm{x}-\bm{x'}|$ the regular two-point function
agrees with the retarded Green's function. 

As a final observation we note that for this cosmological spacetime,  
the normal convex neighbourhood of any point $x$ consists of the whole
spacetime manifold (which excludes the cosmological singularity at
$a = 0$). The Hadamard construction of the Green's functions is
therefore valid globally, a fact that is immediately revealed by
Eqs.~(\ref{13.6.11}) and (\ref{13.6.12}).   

\section{Electromagnetic Green's functions} 
\label{14} 

\subsection{Equations of electromagnetism} 
\label{14.1} 

The electromagnetic field tensor $F_{\alpha\beta} = \nabla_\alpha 
A_\beta - \nabla_\beta A_\alpha$ is expressed in terms of a vector
potential $A_{\alpha}$. In the Lorenz gauge $\nabla_\alpha A^\alpha =  
0$, the vector potential satisfies the wave equation  
\begin{equation} 
\Box A^\alpha - R^\alpha_{\ \beta} A^\beta = -4\pi j^\alpha, 
\label{14.1.1} 
\end{equation}
where $\Box = g^{\alpha\beta} \nabla_\alpha \nabla_\beta$ is the wave 
operator, $R^\alpha_{\ \beta}$ the Ricci tensor, and $j^\alpha$ a
prescribed current density. The wave equation
enforces the condition $\nabla_\alpha j^\alpha = 0$, which expresses
charge conservation.  

The solution to the wave equation is written as  
\begin{equation}
A^\alpha(x) = \int G^\alpha_{\ \beta'}(x,x') j^{\beta'}(x')
\sqrt{-g'}\, d^4 x', 
\label{14.1.2} 
\end{equation}
in terms of a Green's function $G^\alpha_{\ \beta'}(x,x')$ that
satisfies 
\begin{equation} 
\Box G^\alpha_{\ \beta'}(x,x') - R^\alpha_{\ \beta}(x) 
G^\beta_{\ \beta'}(x,x') = -4\pi g^\alpha_{\ \beta'}(x,x')
\delta_4(x,x'),  
\label{14.1.3} 
\end{equation}
where $g^\alpha_{\ \beta'}(x,x')$ is a parallel propagator and
$\delta_4(x,x')$ an invariant Dirac distribution. The parallel
propagator is inserted on the right-hand side of Eq.~(\ref{14.1.3}) to 
keep the index structure of the equation consistent from side to side;
because $g^\alpha_{\ \beta'}(x,x') \delta_4(x,x')$ is distributionally
equal to $[g^\alpha_{\ \beta'}] \delta_4(x,x') = 
\delta^{\alpha'}_{\ \beta'} \delta_4(x,x')$, it could have been
replaced by either $\delta^{\alpha'}_{\ \beta'}$ 
or $\delta^{\alpha}_{\ \beta}$. It is easy to check that by virtue of  
Eq.~(\ref{14.1.3}), the vector potential of Eq.~(\ref{14.1.2})
satisfies the wave equation of Eq.~(\ref{14.1.1}).   

We will assume that the retarded Green's function 
$G^{\ \alpha}_{+\beta'}(x,x')$, which is nonzero if $x$ is in the
causal future of $x'$, and the advanced Green's function 
$G^{\ \alpha}_{-\beta'}(x,x')$, which is nonzero if $x$ is in the
causal past of $x'$, exist as distributions and can be defined
globally in the entire spacetime.    

\subsection{Hadamard construction of the Green's functions} 
\label{14.2} 

Assuming throughout this subsection that $x$ is in the normal convex
neighbourhood of $x'$, we make the ansatz 
\begin{equation}
G^{\ \alpha}_{\pm\beta'}(x,x') = U^\alpha_{\ \beta'}(x,x')
\delta_\pm(\sigma) + V^\alpha_{\ \beta'}(x,x')
\theta_\pm(-\sigma), 
\label{14.2.1}
\end{equation}
where $\theta_\pm(-\sigma)$, $\delta_\pm(\sigma)$ are the
light-cone distributions introduced in Sec.~\ref{12.2}, and where   
$U^\alpha_{\ \beta'}(x,x')$, $V^\alpha_{\ \beta'}(x,x')$ are smooth 
bitensors. 

To conveniently manipulate the Green's functions we shift $\sigma$ by 
a small positive quantity $\epsilon$. The Green's functions are then 
recovered by the taking the limit of 
\[
G^{\epsilon\ \alpha}_{\pm \ \beta'}(x,x') := 
U^\alpha_{\ \beta'}(x,x') \delta_\pm(\sigma+\epsilon) 
+ V^{\alpha}_{\ \beta'}(x,x') \theta_\pm(-\sigma-\epsilon)
\]
as $\epsilon \to 0^+$. When we substitute this into the left-hand side
of Eq.~(\ref{14.1.3}) and then take the limit, we obtain  
\begin{eqnarray*}
\Box G^{\ \alpha}_{\pm \beta'} - R^\alpha_{\ \beta} 
G^{\ \beta}_{\pm \beta'} &=&  
- 4\pi \delta_4(x,x') U^\alpha_{\ \beta'}  
+ \delta'_\pm(\sigma) \Bigl\{ 2 U^\alpha_{\ \beta';\gamma}
  \sigma^\gamma 
  + (\sigma^\gamma_{\ \gamma} - 4) U^\alpha_{\ \beta'} \Bigr\} 
\\ & & \mbox{} 
+ \delta_\pm(\sigma) \Bigl\{ -2 V^\alpha_{\ \beta';\gamma}
  \sigma^\gamma
  + (2 - \sigma^\gamma_{\ \gamma}) V^\alpha_{\ \beta'}
  + \Box U^\alpha_{\ \beta'} 
  - R^\alpha_{\ \beta} U^\beta_{\ \beta'} \Bigr\}
\\ & & \mbox{} 
+ \theta_\pm(-\sigma) \Bigl\{ \Box V^\alpha_{\ \beta'} 
  - R^\alpha_{\ \beta} V^\beta_{\ \beta'} \Bigr\} 
\end{eqnarray*}
after a routine computation similar to the one presented at the
beginning of Sec.~\ref{13.2}. Comparison with Eq.~(\ref{14.1.3})
returns: (i) the equations   
\begin{equation}
\bigl[ U^\alpha_{\ \beta'} \bigr] = \bigl[g^\alpha_{\ \beta'} \bigr]
= \delta^{\alpha'}_{\ \beta'} 
\label{14.2.2}
\end{equation}
and 
\begin{equation} 
2 U^\alpha_{\ \beta';\gamma} \sigma^\gamma 
  + (\sigma^\gamma_{\ \gamma} - 4) U^\alpha_{\ \beta'} = 0
\label{14.2.3}
\end{equation}
that determine $U^\alpha_{\ \beta'}(x,x')$; (ii) the equation 
\begin{equation} 
\check{V}^\alpha_{\ \beta';\gamma} \sigma^\gamma 
+ \frac{1}{2} (\sigma^\gamma_{\ \gamma} - 2) \check{V}^\alpha_{\ \beta'}  
= \frac{1}{2} \bigl( \Box U^\alpha_{\ \beta'} 
  - R^\alpha_{\ \beta} U^\beta_{\ \beta'} \bigr) \Bigr|_{\sigma=0} 
\label{14.2.4}
\end{equation} 
that determines $\check{V}^\alpha_{\ \beta'}(x,x')$, the restriction
of $V^\alpha_{\ \beta'}(x,x')$ on the light cone $\sigma(x,x') = 0$; 
and (iii) the wave equation  
\begin{equation} 
\Box V^\alpha_{\ \beta'} - R^\alpha_{\ \beta} V^\beta_{\ \beta'} = 0
\label{14.2.5} 
\end{equation}
that determines $V^\alpha_{\ \beta'}(x,x')$ inside the light cone.  

Equation (\ref{14.2.3}) can be integrated along the unique geodesic 
$\beta$ that links $x'$ to $x$. The initial conditions are provided 
by Eq.~(\ref{14.2.2}), and if we set $U^\alpha_{\ \beta'}(x,x') = 
g^\alpha_{\ \beta'}(x,x') U(x,x')$, we find that these equations
reduce to Eqs.~(\ref{13.2.4}) and (\ref{13.2.3}),
respectively. According to Eq.~(\ref{13.2.5}), then, we have   
\begin{equation}
U^\alpha_{\ \beta'}(x,x') = g^\alpha_{\ \beta'}(x,x')
\Delta^{1/2}(x,x'),  
\label{14.2.6}
\end{equation} 
which reduces to 
\begin{equation} 
U^\alpha_{\ \beta'} = g^\alpha_{\ \beta'} \Bigl( 1 + \frac{1}{12}
R_{\gamma'\delta'} \sigma^{\gamma'} \sigma^{\delta'} + O(\lambda^3) 
\Bigr) 
\label{14.2.7}
\end{equation}
near coincidence, with $\lambda$ denoting the affine-parameter
distance between $x'$ and $x$. Differentiation of this relation gives  
\begin{eqnarray}
U^\alpha_{\ \beta';\gamma} &=& \frac{1}{2} g^{\gamma'}_{\ \gamma} 
\Bigl( g^\alpha_{\ \alpha'} R^{\alpha'}_{\ \beta'\gamma'\delta'} 
- \frac{1}{3} g^\alpha_{\ \beta'} R_{\gamma'\delta'} \Bigr) 
\sigma^{\delta'} + O(\lambda^2),
\label{14.2.8} \\  
U^\alpha_{\ \beta';\gamma'} &=& \frac{1}{2} 
\Bigl( g^\alpha_{\ \alpha'} R^{\alpha'}_{\ \beta'\gamma'\delta'} 
+ \frac{1}{3} g^\alpha_{\ \beta'} R_{\gamma'\delta'} \Bigr) 
\sigma^{\delta'} + O(\lambda^2),
\label{14.2.9}
\end{eqnarray}
and eventually, 
\begin{equation} 
\bigl[ \Box U^\alpha_{\ \beta'} \bigr] = \frac{1}{6}
\delta^{\alpha'}_{\ \beta'} R(x'). 
\label{14.2.10} 
\end{equation} 
 
Similarly, Eq.~(\ref{14.2.4}) can be integrated along each null
geodesic that generates the null cone $\sigma(x,x')=0$. The initial
values are obtained by taking the coincidence limit of this equation,
using Eqs.~(\ref{14.2.2}), (\ref{14.2.10}), and the additional
relation $[\sigma^\gamma_{\ \gamma}] = 4$. We arrive at   
\begin{equation} 
\bigl[ V^\alpha_{\ \beta'} \bigr] = -\frac{1}{2} \Bigl(
R^{\alpha'}_{\ \beta'} - \frac{1}{6} \delta^{\alpha'}_{\ \beta'} R'
\Bigr). 
\label{14.2.11}
\end{equation}
With the characteristic data obtained by integrating
Eq.~(\ref{14.2.4}), the wave equation of Eq.~(\ref{14.2.5}) admits a
unique solution.  

To summarize, the retarded and advanced electromagnetic Green's
functions are given by Eq.~(\ref{14.2.1}) with 
$U^\alpha_{\ \beta'}(x,x')$ given by Eq.~(\ref{14.2.6}) and     
$V^\alpha_{\ \beta'}(x,x')$ determined by Eq.~(\ref{14.2.5}) and the 
characteristic data constructed with Eqs.~(\ref{14.2.4}) and
(\ref{14.2.11}). It should be emphasized that the construction
provided in this subsection is restricted to ${\cal N}(x')$, the
normal convex neighbourhood of the reference point $x'$.   

\subsection{Reciprocity and Kirchhoff representation} 
\label{14.3} 

Like their scalar counterparts, the (globally defined) electromagnetic
Green's functions satisfy a reciprocity relation, the statement of
which is  
\begin{equation} 
G^-_{\beta'\alpha}(x',x) = G^+_{\alpha\beta'}(x,x'). 
\label{14.3.1} 
\end{equation} 
The derivation of Eq.~(\ref{14.3.1}) is virtually identical to what
was presented in Sec.~\ref{13.3}, and we shall not present the
details. It suffices to mention that it is based on the identities    
\[
G^+_{\alpha\beta'}(x,x') \Bigl( \Box G^{\ \alpha}_{-\gamma''}(x,x'') -  
R^\alpha_{\ \gamma} G^{\ \gamma}_{-\gamma''}(x,x'') \Bigl) = -4\pi
G^+_{\alpha\beta'}(x,x') g^\alpha_{\ \gamma''}(x,x'') \delta_4(x,x'') 
\]
and 
\[
G^-_{\alpha\gamma''}(x,x'') \Bigl( \Box G^{\ \alpha}_{+\beta'}(x,x') - 
R^\alpha_{\ \gamma} G^{\ \gamma}_{+\beta'}(x,x') \Bigl) = -4\pi
G^-_{\alpha\gamma''}(x,x'') g^\alpha_{\ \beta'}(x,x') \delta_4(x,x').   
\]
A direct consequence of the reciprocity relation is 
\begin{equation}
V_{\beta'\alpha}(x',x) = V_{\alpha\beta'}(x,x'), 
\label{14.3.2} 
\end{equation} 
the statement that the bitensor $V_{\alpha\beta'}(x,x')$ is symmetric
in its indices and arguments.  

The Kirchhoff representation for the electromagnetic vector potential
is formulated as follows. Suppose that $A^\alpha(x)$ satisfies the   
{\it homogeneous} version of Eq.~(\ref{14.1.1}) and that initial
values $A^{\alpha'}(x')$, $n^{\beta'} \nabla_{\beta'} A^{\alpha'}(x')$
are specified on a spacelike hypersurface $\Sigma$. Then the value of
the potential at a point $x$ in the future of $\Sigma$ is given by   
\begin{equation}
A^{\alpha}(x) = -\frac{1}{4\pi} \int_{\Sigma} \biggl(
G^{\ \alpha}_{+\beta'}(x,x') \nabla^{\gamma'} A^{\beta'}(x') -
A^{\beta'}(x') \nabla^{\gamma'} G^{\ \alpha}_{+\beta'}(x,x') \biggr)\, 
d\Sigma_{\gamma'}, 
\label{14.3.3} 
\end{equation}
where $d\Sigma_{\gamma'} = -n_{\gamma'} dV$ is a surface element on
$\Sigma$; $n_{\gamma'}$ is the future-directed unit normal and $dV$
is the invariant volume element on the hypersurface. The derivation of 
Eq.~(\ref{14.3.3}) is virtually identical to what was presented in
Sec.~\ref{13.4}.    

\subsection{Relation with scalar Green's functions} 
\label{14.3b} 

In a spacetime that satisfies the Einstein field equations in vacuum,
so that $R_{\alpha\beta} = 0$ everywhere in the spacetime, the
(retarded and advanced) electromagnetic Green's functions satisfy the
identities \cite{dewitt-brehme:60} 
\begin{equation} 
G^{\ \alpha}_{\pm \beta' ; \alpha} = -G_{\pm ;\beta'}, 
\label{14.3b.1} 
\end{equation} 
where $G_{\pm}$ are the corresponding scalar Green's functions.  

To prove this we differentiate Eq.~(\ref{14.1.3}) covariantly with
respect to $x^\alpha$ and use Eq.~(\ref{12.1.3}) to express the
right-hand side as $+4\pi \partial_{\beta'} \delta_4(x,x')$. After
repeated use of Ricci's identity to permute the ordering of the
covariant derivatives on the left-hand side, we arrive at the 
equation 
\begin{equation} 
\Box \bigl( -G^{\alpha}_{\ \beta';\alpha} \bigr) = -4\pi   
\partial_{\beta'} \delta_4(x,x');  
\label{14.3b.2}
\end{equation} 
all terms involving the Riemann tensor disappear by virtue of the fact
that the spacetime is Ricci-flat. Because Eq.~(\ref{14.3b.2}) is also
the differential equation satisfied by $G_{;\beta'}$, and because
the solutions are chosen to satisfy the same boundary conditions, we
have established the validity of Eq.~(\ref{14.3b.1}).   

\subsection{Singular and regular Green's functions} 
\label{14.4} 

We shall now construct singular and regular Green's functions for
the electromagnetic field. The treatment here parallels closely what
was presented in Sec.~\ref{13.5}, and the reader is referred to that
section for a more complete discussion.  

We begin by introducing the bitensor $H^\alpha_{\ \beta'}(x,x')$ with
properties  
\begin{description} 
\item[\qquad {\sf H1}:] $H^\alpha_{\ \beta'}(x,x')$ satisfies the
homogeneous wave equation, 
\begin{equation}
\Box H^\alpha_{\ \beta'}(x,x') 
- R^\alpha_{\ \beta}(x) H^\beta_{\ \beta'}(x,x') = 0;  
\label{14.4.1} 
\end{equation} 
\item[\qquad {\sf H2}:] $H^\alpha_{\ \beta'}(x,x')$ is symmetric in
its indices and arguments,   
\begin{equation}
H_{\beta'\alpha}(x',x) = H_{\alpha\beta'}(x,x'); 
\label{14.4.2}
\end{equation} 
\item[\qquad {\sf H3}:] $H^\alpha_{\ \beta'}(x,x')$ agrees with the
retarded Green's function if $x$ is in the chronological future of
$x'$,   
\begin{equation}
H^\alpha_{\ \beta'}(x,x') = G^{\ \alpha}_{+\beta'}(x,x') \qquad
\mbox{when $x \in I^+(x')$};   
\label{14.4.3} 
\end{equation} 
\item[\qquad {\sf H4}:] $H^\alpha_{\ \beta'}(x,x')$ agrees with the
advanced Green's function if $x$ is in the chronological past of $x'$,   
\begin{equation}
H^\alpha_{\ \beta'}(x,x') = G^{\ \alpha}_{-\beta'}(x,x') \qquad
\mbox{when $x \in I^-(x')$}.    
\label{14.4.4} 
\end{equation} 
\end{description}
It is easy to prove that property {\sf H4} follows from {\sf H2}, 
{\sf H3}, and the reciprocity relation (\ref{14.3.1}) satisfied by the 
retarded and advanced Green's functions. That such a bitensor exists
can be argued along the same lines as those presented in
Sec.~\ref{13.5}.  

Equipped with the bitensor $H^\alpha_{\ \beta'}(x,x')$ we define the
singular Green's function to be 
\begin{equation}
G^{\ \alpha}_{{\rm S}\,\beta'}(x,x') = \frac{1}{2} \Bigl[ 
G^{\ \alpha}_{+\beta'}(x,x') 
+ G^{\ \alpha}_{-\beta'}(x,x')
- H^\alpha_{\ \beta'}(x,x') \Bigr]. 
\label{14.4.5}
\end{equation} 
This comes with the properties  
\begin{description} 
\item[\qquad {\sf S1}:] $G^{\ \alpha}_{{\rm S}\,\beta'}(x,x')$
satisfies the inhomogeneous wave equation,     
\begin{equation} 
\Box G^{\ \alpha}_{{\rm S}\,\beta'}(x,x') - R^\alpha_{\ \beta}(x)  
G^{\ \beta}_{{\rm S}\,\beta'}(x,x') = -4\pi g^\alpha_{\ \beta'}(x,x') 
\delta_4(x,x'); 
\label{14.4.6}  
\end{equation} 
\item[\qquad {\sf S2}:] $G^{\ \alpha}_{{\rm S}\,\beta'}(x,x')$ is
symmetric in its indices and arguments,  
\begin{equation} 
G^{\rm S}_{\beta'\alpha}(x',x) = G^{\rm S}_{\alpha\beta'}(x,x'); 
\label{14.4.7}
\end{equation} 
\item[\qquad {\sf S3}:] $G^{\ \alpha}_{{\rm S}\,\beta'}(x,x')$
vanishes if $x$ is in the chronological past or future of $x'$,  
\begin{equation}
G^{\ \alpha}_{{\rm S}\,\beta'}(x,x') = 0 \qquad 
\mbox{when $x \in I^\pm(x')$}.  
\label{14.4.8} 
\end{equation} 
\end{description}  
These can be established as consequences of {\sf H1}--{\sf H4} and the
properties of the retarded and advanced Green's functions.  

The regular two-point function is then defined by 
\begin{equation} 
G^{\ \,\alpha}_{{\rm R}\,\beta'}(x,x') = 
G^{\ \alpha}_{+\beta'}(x,x') 
- G^{\ \alpha}_{{\rm S}\,\beta'}(x,x'), 
\label{14.4.9}
\end{equation}
and it comes with the properties 
\begin{description} 
\item[\qquad {\sf R1}:] $G^{\ \,\alpha}_{{\rm R}\,\beta'}(x,x')$
satisfies the homogeneous wave equation, 
\begin{equation}
\Box G^{\ \,\alpha}_{{\rm R}\,\beta'}(x,x') - R^\alpha_{\ \beta}(x)  
G^{\ \,\beta}_{{\rm R}\,\beta'}(x,x') = 0;  
\label{14.4.10} 
\end{equation} 
\item[\qquad {\sf R2}:] $G^{\ \,\alpha}_{{\rm R}\,\beta'}(x,x')$
agrees with the retarded Green's function if $x$ is in the
chronological future of $x'$,  
\begin{equation}
G^{\ \,\alpha}_{{\rm R}\,\beta'}(x,x') = 
G^{\ \alpha}_{+\beta'}(x,x') 
\qquad \mbox{when $x \in I^+(x')$}; 
\label{14.4.11} 
\end{equation} 
\item[\qquad {\sf R3}:] $G^{\ \,\alpha}_{{\rm R}\,\beta'}(x,x')$
vanishes if $x$ is in the chronological past of $x'$,  
\begin{equation}
G^{\ \,\alpha}_{{\rm R}\,\beta'}(x,x') = 0 
\qquad \mbox{when $x \in I^-(x')$}.  
\label{14.4.12} 
\end{equation} 
\end{description}
Those follow immediately from {\sf S1}--{\sf S3} and the properties of
the retarded Green's function. 

When $x$ is restricted to the normal convex neighbourhood of $x'$, we
have the explicit relations 
\begin{eqnarray} 
H^\alpha_{\ \beta'}(x,x') &=& V^\alpha_{\ \beta'}(x,x'), 
\label{14.4.13} \\ 
G^{\ \alpha}_{{\rm S}\,\beta'}(x,x') &=& 
\frac{1}{2} U^\alpha_{\ \beta'}(x,x') \delta(\sigma) - \frac{1}{2} 
V^\alpha_{\ \beta'}(x,x') \theta(\sigma), 
\label{14.4.14} \\ 
G^{\ \,\alpha}_{{\rm R}\,\beta'}(x,x') &=& 
\frac{1}{2} U^\alpha_{\ \beta'}(x,x') \Bigl[ \delta_+(\sigma) -
\delta_-(\sigma) \Bigr] + V^\alpha_{\ \beta'}(x,x') \Bigl[
\theta_+(-\sigma) + \frac{1}{2} \theta(\sigma) \Bigr]. 
\label{14.4.15}
\end{eqnarray} 
From these we see clearly that the singular Green's function does not
distinguish between past and future (property {\sf S2}), and that its 
support excludes $I^\pm(x')$ (property {\sf S3}). We see also that
the regular two-point function coincides with 
$G^{\ \alpha}_{+\beta'}(x,x')$ in $I^+(x')$ (property {\sf R2}), and  
that its support does not include $I^-(x')$ (property {\sf R3}). 
 
\section{Gravitational Green's functions}
\label{15} 

\subsection{Equations of linearized gravity}  
\label{15.1} 

We are given a background spacetime for which the metric
$g_{\alpha\beta}$ satisfies the Einstein field equations 
{\it in vacuum}. We then perturb the metric from $g_{\alpha\beta}$ to 
\begin{equation} 
{\sf g}_{\alpha\beta} = g_{\alpha\beta} + h_{\alpha\beta}.   
\label{15.1.1}
\end{equation} 
The metric perturbation $h_{\alpha\beta}$ is assumed to be small, and
when working out the Einstein field equations to be satisfied by the
new metric ${\sf g}_{\alpha\beta}$, we work consistently to
first order in $h_{\alpha\beta}$. To simplify the expressions we use
the trace-reversed potentials $\gamma_{\alpha\beta}$ defined by      
\begin{equation} 
\gamma_{\alpha\beta} = h_{\alpha\beta} - \frac{1}{2} \bigl(
g^{\gamma\delta} h_{\gamma\delta} \bigr) g_{\alpha\beta},  
\label{15.1.2}
\end{equation} 
and we impose the Lorenz gauge condition,  
\begin{equation}
\gamma^{\alpha\beta}_{\ \ \ ;\beta} = 0. 
\label{15.1.3} 
\end{equation} 
In this equation, and in all others below, indices are raised and
lowered with the background metric $g_{\alpha\beta}$. Similarly, the
connection involved in Eq.~(\ref{15.1.3}), and in all other equations
below, is the one that is compatible with the background metric. If 
$T^{\alpha\beta}$ is the perturbing energy-momentum tensor, then by 
virtue of the linearized Einstein field equations the perturbation
field obeys the wave equation   
\begin{equation} 
\Box \gamma^{\alpha\beta} + 2 R_{\gamma\ \delta}^{\ \alpha\ \beta}
\gamma^{\gamma\delta} = -16\pi T^{\alpha\beta}, 
\label{15.1.4}
\end{equation} 
in which $\Box = g^{\alpha\beta} \nabla_\alpha \nabla_\beta$ is the
wave operator and $R_{\gamma\alpha\delta\beta}$ the Riemann tensor. In   
first-order perturbation theory, the energy-momentum tensor must be
conserved in the background spacetime: 
$T^{\alpha\beta}_{\ \ \ ;\beta} = 0$. 

The solution to the wave equation is written as 
\begin{equation}
\gamma^{\alpha\beta}(x) = 4\int 
G^{\alpha\beta}_{\ \ \gamma'\delta'}(x,x') 
T^{\gamma'\delta'}(x') \sqrt{-g'}\, d^4 x', 
\label{15.1.5} 
\end{equation}
in terms of a Green's function 
$G^{\alpha\beta}_{\ \ \gamma'\delta'}(x,x')$ that satisfies 
\cite{sciama-etal:69}  
\begin{equation} 
\Box G^{\alpha\beta}_{\ \ \gamma'\delta'}(x,x') + 
2 R_{\gamma\ \delta}^{\ \alpha\ \beta} (x)  
G^{\gamma\delta}_{\ \ \gamma'\delta'}(x,x')
= -4\pi g^{(\alpha}_{\ \gamma'}(x,x')
g^{\beta)}_{\ \delta'}(x,x') \delta_4(x,x'),  
\label{15.1.6} 
\end{equation}
where $g^\alpha_{\ \gamma'}(x,x')$ is a parallel propagator and
$\delta_4(x,x')$ an invariant Dirac functional. The parallel
propagators are inserted on the right-hand side of Eq.~(\ref{15.1.6})
to keep the index structure of the equation consistent from side to
side; in particular, both sides of the equation are symmetric in
$\alpha$ and $\beta$, and in $\gamma'$ and $\delta'$. It is easy to
check that by virtue of Eq.~(\ref{15.1.6}), the perturbation field of  
Eq.~(\ref{15.1.5}) satisfies the wave equation of Eq.~(\ref{15.1.4}).      
Once $\gamma_{\alpha\beta}$ is known, the metric perturbation can  
be reconstructed from the relation $h_{\alpha\beta} =
\gamma_{\alpha\beta} - \frac{1}{2} (g^{\gamma\delta}
\gamma_{\gamma\delta}) g_{\alpha\beta}$. 

We will assume that the retarded Green's function 
$G^{\ \alpha\beta}_{+\ \gamma'\delta'}(x,x')$, which is nonzero if $x$
is in the causal future of $x'$, and the advanced Green's function 
$G^{\ \alpha\beta}_{-\ \gamma'\delta'}(x,x')$, which is nonzero if $x$
is in the causal past of $x'$, exist as distributions and can be
defined globally in the entire background spacetime.     

\subsection{Hadamard construction of the Green's functions} 
\label{15.2}  

Assuming throughout this subsection that $x$ is in the normal convex
neighbourhood of $x'$, we make the ansatz 
\begin{equation}
G^{\ \alpha\beta}_{\pm\ \gamma'\delta'}(x,x') = 
U^{\alpha\beta}_{\ \ \gamma'\delta'}(x,x') 
\delta_\pm(\sigma) 
+ V^{\alpha\beta}_{\ \ \gamma'\delta'}(x,x') \theta_\pm(-\sigma),  
\label{15.2.1}  
\end{equation}
where $\theta_\pm(-\sigma)$, $\delta_\pm(\sigma)$ are the
light-cone distributions introduced in Sec.~\ref{12.2}, and where   
$U^{\alpha\beta}_{\ \ \gamma'\delta'}(x,x')$, 
$V^{\alpha\beta}_{\ \ \gamma'\delta'}(x,x')$ are smooth bitensors. 

To conveniently manipulate the Green's functions we shift $\sigma$ by  
a small positive quantity $\epsilon$. The Green's functions are then 
recovered by the taking the limit of 
\[
G^{\epsilon\ \alpha\beta}_{\pm\ \ \gamma'\delta'}(x,x') = 
U^{\alpha\beta}_{\ \ \gamma'\delta'}(x,x') 
\delta_\pm(\sigma+\epsilon) 
+ V^{\alpha\beta}_{\ \ \gamma'\delta'}(x,x')
\theta_\pm(-\sigma-\epsilon)
\]
as $\epsilon \to 0^+$. When we substitute this into the left-hand side 
of Eq.~(\ref{15.1.6}) and then take the limit, we obtain  
\begin{eqnarray*}
\Box G^{\ \alpha\beta}_{\pm \ \gamma'\delta'}  
+ 2 R_{\gamma\ \delta}^{\ \alpha\ \beta} 
G^{\ \gamma\delta}_{\pm \ \gamma'\delta'}   
&=& 
- 4\pi \delta_4(x,x') U^{\alpha\beta}_{\ \ \gamma'\delta'}   
+ \delta'_\pm(\sigma) \Bigl\{ 
   2 U^{\alpha\beta}_{\ \ \gamma'\delta';\gamma} \sigma^\gamma  
  + (\sigma^\gamma_{\ \gamma} - 4) U^{\alpha\beta}_{\ \ \gamma'\delta'}
\Bigr\}  
\\ & & \mbox{} 
+ \delta_\pm(\sigma) \Bigl\{ 
  -2 V^{\alpha\beta}_{\ \ \gamma'\delta';\gamma} \sigma^\gamma 
  + (2 - \sigma^\gamma_{\ \gamma}) 
    V^{\alpha\beta}_{\ \ \gamma'\delta'}  
  + \Box U^{\alpha\beta}_{\ \ \gamma'\delta'} 
  + 2 R_{\gamma\ \delta}^{\ \alpha\ \beta} 
    U^{\gamma\delta}_{\ \ \gamma'\delta'} 
  \Bigr\}
\\ & & \mbox{} 
+ \theta_\pm(-\sigma) \Bigl\{ 
\Box V^{\alpha\beta}_{\ \ \gamma'\delta'} 
  + 2 R_{\gamma\ \delta}^{\ \alpha\ \beta} 
    V^{\gamma\delta}_{\ \ \gamma'\delta'}
\Bigr\}
\end{eqnarray*}
after a routine computation similar to the one presented at the
beginning of Sec.~\ref{13.2}. Comparison with Eq.~(\ref{15.1.6})
returns: (i) the equations  
\begin{equation}
\bigl[ U^{\alpha\beta}_{\ \ \gamma'\delta'} \bigr] = 
\Bigl[ g^{(\alpha}_{\ \gamma'} g^{\beta)}_{\ \delta'} \Bigr]
= \delta^{(\alpha'}_{\ \ \gamma'} \delta^{\beta')}_{\ \delta'}
\label{15.2.2} 
\end{equation}
and 
\begin{equation} 
2 U^{\alpha\beta}_{\ \ \gamma'\delta';\gamma} \sigma^\gamma 
  + (\sigma^\gamma_{\ \gamma} - 4) 
  U^{\alpha\beta}_{\ \ \gamma'\delta'} = 0 
\label{15.2.3} 
\end{equation}
that determine $U^{\alpha\beta}_{\ \ \gamma'\delta'}(x,x')$; (ii) the 
equation  
\begin{equation} 
\check{V}^{\alpha\beta}_{\ \ \gamma'\delta';\gamma} \sigma^\gamma   
+ \frac{1}{2} (\sigma^\gamma_{\ \gamma} - 2) 
\check{V}^{\alpha\beta}_{\ \ \gamma'\delta'}  
= \frac{1}{2} \bigl( \Box U^{\alpha\beta}_{\ \ \gamma'\delta'}  
+ 2 R_{\gamma\ \delta}^{\ \alpha\ \beta} 
    U^{\gamma\delta}_{\ \ \gamma'\delta'} \bigr) \Bigr|_{\sigma=0}  
\label{15.2.4}  
\end{equation} 
that determine $\check{V}^{\alpha\beta}_{\ \ \gamma'\delta'}(x,x')$,
the restriction of $V^{\alpha\beta}_{\ \ \gamma'\delta'}(x,x')$ on the
light cone $\sigma(x,x') = 0$; and (iii) the wave equation 
\begin{equation} 
\Box V^{\alpha\beta}_{\ \ \gamma'\delta'} 
  + 2 R_{\gamma\ \delta}^{\ \alpha\ \beta} 
    V^{\gamma\delta}_{\ \ \gamma'\delta'} = 0 
\label{15.2.5}
\end{equation}
that determines $V^{\alpha\beta}_{\ \ \gamma'\delta'}(x,x')$ inside
the light cone.  

Equation (\ref{15.2.3}) can be integrated along the unique geodesic  
$\beta$ that links $x'$ to $x$. The initial conditions are provided 
by Eq.~(\ref{15.2.2}), and if we set 
$U^{\alpha\beta}_{\ \ \gamma'\delta'}(x,x') = g^{(\alpha}_{\ \gamma'} 
g^{\beta)}_{\ \delta'} U(x,x')$, we find that these equations reduce
to Eqs.~(\ref{13.2.4}) and (\ref{13.2.3}), respectively. According to 
Eq.~(\ref{13.2.5}), then, we have   
\begin{equation}
U^{\alpha\beta}_{\ \ \gamma'\delta'}(x,x') = 
g^{(\alpha}_{\ \gamma'}(x,x') g^{\beta)}_{\ \delta'}(x,x') 
\Delta^{1/2}(x,x'), 
\label{15.2.6} 
\end{equation} 
which reduces to 
\begin{equation} 
U^{\alpha\beta}_{\ \ \gamma'\delta'} = 
g^{(\alpha}_{\ \gamma'} g^{\beta)}_{\ \delta'} 
\Bigl( 1 + O(\lambda^3) \Bigr) 
\label{15.2.7} 
\end{equation}
near coincidence, with $\lambda$ denoting the affine-parameter
distance between $x'$ and $x$; there is no term of order $\lambda^2$
because by assumption, the background Ricci tensor vanishes at $x'$
(as it does in the entire spacetime). Differentiation of this relation
gives     
\begin{eqnarray}
U^{\alpha\beta}_{\ \ \gamma'\delta';\epsilon} &=& \frac{1}{2}  
g^{(\alpha}_{\ \alpha'} g^{\beta)}_{\ \beta'} 
g^{\epsilon'}_{\ \epsilon}  
\Bigl( R^{\alpha'}_{\ \gamma'\epsilon'\iota'} 
\delta^{\beta'}_{\ \delta'} 
+ R^{\alpha'}_{\ \delta'\epsilon'\iota'} \delta^{\beta'}_{\ \gamma'}
\Bigr) \sigma^{\iota'} + O(\lambda^2),
\label{15.2.8} \\ 
U^{\alpha\beta}_{\ \ \gamma'\delta';\epsilon'} &=& \frac{1}{2}  
g^{(\alpha}_{\ \alpha'} g^{\beta)}_{\ \beta'} 
\Bigl( R^{\alpha'}_{\ \gamma'\epsilon'\iota'} 
\delta^{\beta'}_{\ \delta'} 
+ R^{\alpha'}_{\ \delta'\epsilon'\iota'} 
\delta^{\beta'}_{\ \gamma'} \Bigr)   
\sigma^{\iota'} + O(\lambda^2), 
\label{15.2.9} 
\end{eqnarray}
and eventually, 
\begin{equation} 
\bigl[ \Box U^{\alpha\beta}_{\ \ \gamma'\delta'} ] = 0;
\label{15.2.10}
\end{equation} 
this last result follows from the fact that 
$[U^{\alpha\beta}_{\ \ \gamma'\delta';\epsilon\iota}]$ is
antisymmetric in the last pair of indices.    

Similarly, Eq.~(\ref{15.2.4}) can be integrated along each null
geodesic that generates the null cone $\sigma(x,x')=0$. The initial
values are obtained by taking the coincidence limit of this equation,
using Eqs.~(\ref{15.2.2}), (\ref{15.2.10}), and the additional
relation $[\sigma^\gamma_{\ \gamma}] = 4$. We arrive at  
\begin{equation} 
\bigl[ V^{\alpha\beta}_{\ \ \gamma'\delta'} \bigr] = \frac{1}{2}
\Bigl( R^{\alpha'\ \beta'}_{\ \gamma'\ \delta'} + 
R^{\beta'\ \alpha'}_{\ \gamma'\ \delta'} \Bigr).    
\label{15.2.11} 
\end{equation}
With the characteristic data obtained by integrating
Eq.~(\ref{15.2.4}), the wave equation of Eq.~(\ref{15.2.5}) admits a
unique solution.   

To summarize, the retarded and advanced gravitational Green's
functions are given by Eq.~(\ref{15.2.1}) with  
$U^{\alpha\beta}_{\ \ \gamma'\delta'}(x,x')$ given by
Eq.~(\ref{15.2.6}) and $V^{\alpha\beta}_{\ \ \gamma'\delta'}(x,x')$
determined by Eq.~(\ref{15.2.5}) and the characteristic data
constructed with Eqs.~(\ref{15.2.4}) and (\ref{15.2.11}). It should be
emphasized that the construction provided in this subsection is
restricted to ${\cal N}(x')$, the normal convex neighbourhood of the
reference point $x'$.    

\subsection{Reciprocity and Kirchhoff representation}
\label{15.3}   

The (globally defined) gravitational Green's functions satisfy the
reciprocity relation   
\begin{equation} 
G^-_{\gamma'\delta'\alpha\beta}(x',x) =
G^+_{\alpha\beta\gamma'\delta'}(x,x').  
\label{15.3.1} 
\end{equation} 
The derivation of this result is virtually identical to what was
presented in Secs.~\ref{13.3} and \ref{14.3}. A direct consequence of
the reciprocity relation is the statement
\begin{equation}
V_{\gamma'\delta'\alpha\beta}(x',x) =
V_{\alpha\beta\gamma'\delta'}(x,x').  
\label{15.3.2}
\end{equation} 

The Kirchhoff representation for the trace-reversed gravitational
perturbation $\gamma_{\alpha\beta}$ is formulated as follows. Suppose
that $\gamma^{\alpha\beta}(x)$ satisfies the homogeneous version of
Eq.~(\ref{15.1.4}) and that initial values 
$\gamma^{\alpha'\beta'}(x')$, $n^{\gamma'} \nabla_{\gamma'}
\gamma^{\alpha'\beta'}(x')$ are specified on a spacelike hypersurface
$\Sigma$. Then the value of the perturbation field at a point $x$ in
the future of $\Sigma$ is given by   
\begin{equation}
\gamma^{\alpha\beta}(x) = -\frac{1}{4\pi} \int_{\Sigma} \biggl( 
G^{\ \alpha\beta}_{+\ \gamma'\delta'}(x,x') \nabla^{\epsilon'}
\gamma^{\gamma'\delta'}(x') - \gamma^{\gamma'\delta'}(x') 
\nabla^{\epsilon'} G^{\ \alpha\beta}_{+\ \gamma'\delta'}(x,x')
\biggr)\, d\Sigma_{\epsilon'},  
\label{15.3.3}  
\end{equation}
where $d\Sigma_{\epsilon'} = -n_{\epsilon'} dV$ is a surface element
on $\Sigma$; $n_{\epsilon'}$ is the future-directed unit normal and
$dV$ is the invariant volume element on the hypersurface. The
derivation of Eq.~(\ref{15.3.3}) is virtually identical to what was
presented in Secs.~\ref{13.4} and \ref{14.3}.     

\subsection{Relation with electromagnetic and scalar Green's
  functions} 
\label{15.3b} 

In a spacetime that satisfies the Einstein field equations in vacuum,
so that $R_{\alpha\beta} = 0$ everywhere in the spacetime, the
(retarded and advanced) gravitational Green's functions satisfy the 
identities \cite{pound:10a} 
\begin{equation} 
G^{\ \alpha\beta}_{\pm\ \gamma'\delta'; \beta}= 
-G^{\ \alpha}_{\pm(\gamma';\delta')} 
\label{15.3b.1} 
\end{equation}
and 
\begin{equation} 
g^{\gamma'\delta'}  G^{\ \alpha\beta}_{\pm\ \gamma'\delta'}
= g^{\alpha\beta} G_\pm, 
\label{15.3b.2} 
\end{equation} 
where $G^{\ \alpha}_{\pm\beta'}$ are the corresponding electromagnetic 
Green's functions, and $G_{\pm}$ the corresponding scalar Green's
functions.   

To prove Eq.~(\ref{15.3b.1}) we differentiate Eq.~(\ref{15.1.6})
covariantly with respect to $x^\beta$, use Eq.~(\ref{12.1.3}) to
work on the right-hand side, and invoke Ricci's identity to permute
the ordering of the covariant derivatives on the left-hand side. After
simplification and involvement of the Ricci-flat condition (which,
together with the Bianchi identities, implies that 
$R^{\ \ \ \ \ \: ;\beta}_{\alpha\gamma\beta\delta} = 0$),  
we arrive at the equation 
\begin{equation} 
\Box \bigl( -G^{\alpha\beta}_{\ \ \gamma'\delta';\beta} \bigr) = -4\pi    
g^\alpha_{\ (\gamma'} \partial_{\delta')} \delta_4(x,x').
\label{15.3b.3}
\end{equation} 
Because this is also the differential equation satisfied by
$G^\alpha_{\ (\beta';\gamma')}$, and because the solutions are chosen
to satisfy the same boundary conditions, we have established the
validity of Eq.~(\ref{15.3b.1}). 

The identity of Eq.~(\ref{15.3b.2}) follows simply from the fact that     
$g^{\gamma'\delta'}  G^{\alpha\beta}_{\ \ \gamma'\delta'}$ and 
$g^{\alpha\beta} G$ satisfy the same tensorial wave equation in a
Ricci-flat spacetime. 

\subsection{Singular and regular Green's functions} 
\label{15.4} 

We shall now construct singular and regular Green's functions for
the linearized gravitational field. The treatment here
parallels closely what was presented in Secs.~\ref{13.5} and
\ref{14.4}.  

We begin by introducing the bitensor 
$H^{\alpha\beta}_{\ \ \gamma'\delta'}(x,x')$ with properties  
\begin{description} 
\item[\qquad {\sf H1}:] $H^{\alpha\beta}_{\ \ \gamma'\delta'}(x,x')$
satisfies the homogeneous wave equation, 
\begin{equation}
\Box H^{\alpha\beta}_{\ \ \gamma'\delta'}(x,x') 
  + 2 R_{\gamma\ \delta}^{\ \alpha\ \beta}(x) 
    H^{\gamma\delta}_{\ \ \gamma'\delta'}(x,x')  
= 0; 
\label{15.4.1} 
\end{equation} 
\item[\qquad {\sf H2}:] $H^{\alpha\beta}_{\ \ \gamma'\delta'}(x,x')$
is symmetric in its indices and arguments,   
\begin{equation}
H_{\gamma'\delta'\alpha\beta}(x',x) 
= H_{\alpha\beta\gamma'\delta'}(x,x'); 
\label{15.4.2}
\end{equation} 
\item[\qquad {\sf H3}:] $H^{\alpha\beta}_{\ \ \gamma'\delta'}(x,x')$
agrees with the retarded Green's function if $x$ is in the
chronological future of $x'$,   
\begin{equation}
H^{\alpha\beta}_{\ \ \gamma'\delta'}(x,x') 
= G^{\ \alpha\beta}_{+\ \gamma'\delta'}(x,x') \qquad
\mbox{when $x \in I^+(x')$};   
\label{15.4.3} 
\end{equation} 
\item[\qquad {\sf H4}:] $H^{\alpha\beta}_{\ \ \gamma'\delta'}(x,x')$
agrees with the advanced Green's function if $x$ is in the
chronological past of $x'$,    
\begin{equation}
H^{\alpha\beta}_{\ \ \gamma'\delta'}(x,x') 
= G^{\ \alpha\beta}_{-\ \gamma'\delta'}(x,x') \qquad
\mbox{when $x \in I^-(x')$}.    
\label{15.4.4} 
\end{equation} 
\end{description}
It is easy to prove that property {\sf H4} follows from {\sf H2}, 
{\sf H3}, and the reciprocity relation (\ref{15.3.1}) satisfied by the  
retarded and advanced Green's functions. That such a bitensor exists
can be argued along the same lines as those presented in
Sec.~\ref{13.5}.  

Equipped with $H^{\alpha\beta}_{\ \ \gamma'\delta'}(x,x')$ we define
the singular Green's function to be  
\begin{equation}
G^{\ \alpha\beta}_{{\rm S}\ \,\gamma'\delta'}(x,x') = \frac{1}{2}
\Bigl[ G^{\ \alpha\beta}_{+\ \gamma'\delta'}(x,x') 
+ G^{\ \alpha\beta}_{-\ \gamma'\delta'}(x,x') 
- H^{\alpha\beta}_{\ \ \gamma'\delta'}(x,x') \Bigr]. 
\label{15.4.5}
\end{equation} 
This comes with the properties  
\begin{description} 
\item[\qquad {\sf S1}:] 
$G^{\ \alpha\beta}_{{\rm S}\ \,\gamma'\delta'}(x,x')$ satisfies the
inhomogeneous wave equation,      
\begin{equation} 
\Box G^{\ \alpha\beta}_{{\rm S}\ \,\gamma'\delta'}(x,x') +  
2 R_{\gamma\ \delta}^{\ \alpha\ \beta} (x)  
G^{\ \gamma\delta}_{{\rm S}\ \,\gamma'\delta'}(x,x') 
= -4\pi g^{(\alpha}_{\ \gamma'}(x,x')
g^{\beta)}_{\ \delta'}(x,x') \delta_4(x,x');   
\label{15.4.6}  
\end{equation} 
\item[\qquad {\sf S2}:] 
$G^{\ \alpha\beta}_{{\rm S}\ \,\gamma'\delta'}(x,x')$ is symmetric in
its indices and arguments,   
\begin{equation} 
G^{\rm S}_{\gamma'\delta'\alpha\beta}(x',x) 
= G^{\rm S}_{\alpha\beta\gamma'\delta'}(x,x');  
\label{15.4.7}
\end{equation} 
\item[\qquad {\sf S3}:] 
$G^{\ \alpha\beta}_{{\rm S}\ \,\gamma'\delta'}(x,x')$ vanishes if $x$ is 
in the chronological past or future of $x'$,  
\begin{equation}
G^{\ \alpha\beta}_{{\rm S}\ \,\gamma'\delta'}(x,x') = 0 \qquad  
\mbox{when $x \in I^\pm(x')$}.  
\label{15.4.8} 
\end{equation} 
\end{description}  
These can be established as consequences of {\sf H1}--{\sf H4} and the 
properties of the retarded and advanced Green's functions.  

The regular two-point function is then defined by 
\begin{equation}  
G^{\ \,\alpha\beta}_{{\rm R}\ \,\gamma'\delta'}(x,x') = 
G^{\ \alpha\beta}_{+\ \gamma'\delta'}(x,x') 
- G^{\ \alpha\beta}_{{\rm S}\ \,\gamma'\delta'}(x,x'), 
\label{15.4.9}
\end{equation}
and it comes with the properties 
\begin{description} 
\item[\qquad {\sf R1}:] 
$G^{\ \,\alpha\beta}_{{\rm R}\ \,\gamma'\delta'}(x,x')$ satisfies the 
homogeneous wave equation,  
\begin{equation}
\Box G^{\ \,\alpha\beta}_{{\rm R}\ \,\gamma'\delta'}(x,x') +  
2 R_{\gamma\ \delta}^{\ \alpha\ \beta} (x)  
G^{\ \,\gamma\delta}_{{\rm R}\ \,\gamma'\delta'}(x,x') = 0; 
\label{15.4.10} 
\end{equation} 
\item[\qquad {\sf R2}:] 
$G^{\ \,\alpha\beta}_{{\rm R}\ \,\gamma'\delta'}(x,x')$ agrees with
the retarded Green's function if $x$ is in the chronological future of 
$x'$,   
\begin{equation}
G^{\ \,\alpha\beta}_{{\rm R}\ \,\gamma'\delta'}(x,x') = 
G^{\ \alpha\beta}_{+\ \gamma'\delta'}(x,x') 
\qquad \mbox{when $x \in I^+(x')$}; 
\label{15.4.11} 
\end{equation} 
\item[\qquad {\sf R3}:] 
$G^{\ \,\alpha\beta}_{{\rm R}\ \,\gamma'\delta'}(x,x')$ vanishes if
$x$ is in the chronological past of $x'$,  
\begin{equation}
G^{\ \,\alpha\beta}_{{\rm R}\ \,\gamma'\delta'}(x,x') = 0 
\qquad \mbox{when $x \in I^-(x')$}.  
\label{15.4.12} 
\end{equation} 
\end{description}
Those follow immediately from {\sf S1}--{\sf S3} and the properties of 
the retarded Green's function. 

When $x$ is restricted to the normal convex neighbourhood of $x'$, we 
have the explicit relations 
\begin{eqnarray} 
H^{\alpha\beta}_{\ \ \gamma'\delta'}(x,x') &=& 
V^{\alpha\beta}_{\ \ \gamma'\delta'}(x,x'), 
\label{15.4.13} \\ 
G^{\ \alpha\beta}_{{\rm S}\ \,\gamma'\delta'}(x,x') &=& 
\frac{1}{2} U^{\alpha\beta}_{\ \ \gamma'\delta'}(x,x') \delta(\sigma)
- \frac{1}{2} V^{\alpha\beta}_{\ \ \gamma'\delta'}(x,x')
  \theta(\sigma),  
\label{15.4.14} \\ 
G^{\ \,\alpha\beta}_{{\rm R}\ \,\gamma'\delta'}(x,x') &=&
\frac{1}{2} U^{\alpha\beta}_{\ \ \gamma'\delta'}(x,x')
\Bigl[ \delta_+(\sigma) - \delta_-(\sigma) \Bigr] 
+ V^{\alpha\beta}_{\ \ \gamma'\delta'}(x,x') \Bigl[ 
\theta_+(-\sigma) + \frac{1}{2} \theta(\sigma) \Bigr]. 
\label{15.4.15}
\end{eqnarray} 
From these we see clearly that the singular Green's function does not
distinguish between past and future (property {\sf S2}), and that its 
support excludes $I^\pm(x')$ (property {\sf S3}). We see also that 
the regular two-point function coincides with 
$G^{\ \alpha\beta}_{+\ \gamma'\delta'}(x,x')$ in $I^+(x')$ (property
{\sf R2}), and that its support does not include $I^-(x')$ (property 
{\sf R3}).

%% file: part4.tex
%
\section{Motion of a scalar charge} 
\label{16} 

\subsection{Dynamics of a point scalar charge}   
\label{16.1} 

A point particle carries a scalar charge $q$ and moves on a world line 
$\gamma$ described by relations $z^\mu(\lambda)$, in which $\lambda$
is an arbitrary parameter. The particle generates a scalar potential 
$\Phi(x)$ and a field $\Phi_\alpha(x) := \nabla_\alpha
\Phi(x)$. The dynamics of the entire system is governed by the action 
\begin{equation}
S = S_{\rm field} + S_{\rm particle} + S_{\rm interaction}, 
\label{16.1.1}
\end{equation} 
where $S_{\rm field}$ is an action functional for a free scalar field 
in a spacetime with metric $g_{\alpha\beta}$, $S_{\rm particle}$ is
the action of a free particle moving on a world line $\gamma$ in this  
spacetime, and $S_{\rm interaction}$ is an interaction term that
couples the field to the particle. 

The field action is given by  
\begin{equation}
S_{\rm field} = -\frac{1}{8\pi} \int \bigl( g^{\alpha\beta}
\Phi_{\alpha} \Phi_{\beta} + \xi R \Phi^2 \bigr) \sqrt{-g}\, d^4 x,  
\label{16.1.2}
\end{equation} 
where the integration is over all of spacetime; the field is coupled
to the Ricci scalar $R$ by an arbitrary constant $\xi$. The particle
action is   
\begin{equation} 
S_{\rm particle} = -m_0 \int_\gamma d\tau, 
\label{16.1.3}
\end{equation}
where $m_0$ is the bare mass of the particle and $d\tau =
\sqrt{-g_{\mu\nu}(z) \dot{z}^\mu \dot{z}^\nu}\, d\lambda$ is the
differential of proper time along the world line; we use an overdot on 
$z^\mu(\lambda)$ to indicate differentiation with respect to the
parameter $\lambda$. Finally, the interaction term is given by  
\begin{equation}  
S_{\rm interaction} = q \int_\gamma \Phi(z)\, d\tau = q \int
\Phi(x) \delta_4(x,z)\sqrt{-g}\, d^4x d\tau. 
\label{16.1.4}
\end{equation}
Notice that both $S_{\rm particle}$ and $S_{\rm interaction}$ are
invariant under a reparameterization $\lambda \to \lambda'(\lambda)$
of the world line.  

Demanding that the total action be stationary under a variation
$\delta\Phi(x)$ of the field configuration yields the wave equation  
\begin{equation} 
\bigl( \Box - \xi R \bigr) \Phi(x) = -4\pi \mu(x) 
\label{16.1.5}
\end{equation} 
for the scalar potential, with a charge density $\mu(x)$ defined by  
\begin{equation}
\mu(x) = q \int_\gamma \delta_4(x,z)\, d\tau. 
\label{16.1.6}
\end{equation} 
These equations determine the field $\Phi_\alpha(x)$ once the motion
of the scalar charge is specified. On the other hand, demanding that
the total action be stationary under a variation 
$\delta z^\mu(\lambda)$ of the world line yields the equations of
motion  
\begin{equation} 
m(\tau) \frac{D u^\mu}{d\tau} = q \bigl( g^{\mu\nu} + u^\mu u^\nu
\bigr) \Phi_{\nu}(z) 
\label{16.1.7}
\end{equation} 
for the scalar charge. We have here adopted $\tau$ as the parameter on
the world line, and introduced the four-velocity $u^\mu(\tau) := 
dz^\mu/d\tau$. The dynamical mass that appears in Eq.~(\ref{16.1.7})
is defined by $m(\tau) := m_0 - q \Phi(z)$, which can also be
expressed in differential form as 
\begin{equation} 
\frac{d m}{d\tau} = -q \Phi_{\mu}(z) u^\mu. 
\label{16.1.8}
\end{equation} 
It should be clear that Eqs.~(\ref{16.1.7}) and (\ref{16.1.8}) are
valid only in a formal sense, because the scalar potential obtained
from Eqs.~(\ref{16.1.5}) and (\ref{16.1.6}) diverges on the world
line. Before we can make sense of these equations we have to analyze
the field's singularity structure near the world line.   

\subsection{Retarded potential near the world line} 
\label{16.2}

The retarded solution to Eq.~(\ref{16.1.5}) is $\Phi(x) = \int
G_+(x,x') \mu(x') \sqrt{g'}\, d^4x'$, where $G_+(x,x')$ is the
retarded Green's function introduced in Sec.~\ref{13}. After
substitution of Eq.~(\ref{16.1.6}) we obtain
\begin{equation}
\Phi(x) = q \int_\gamma G_+(x,z)\, d\tau, 
\label{16.2.1}
\end{equation}
in which $z(\tau)$ gives the description of the world line
$\gamma$. Because the retarded Green's function is defined globally in
the entire spacetime, Eq.~(\ref{16.2.1}) applies to any field point
$x$. 

\begin{figure}
\begin{center}
\vspace*{-20pt} 
\includegraphics[width=0.5\linewidth]{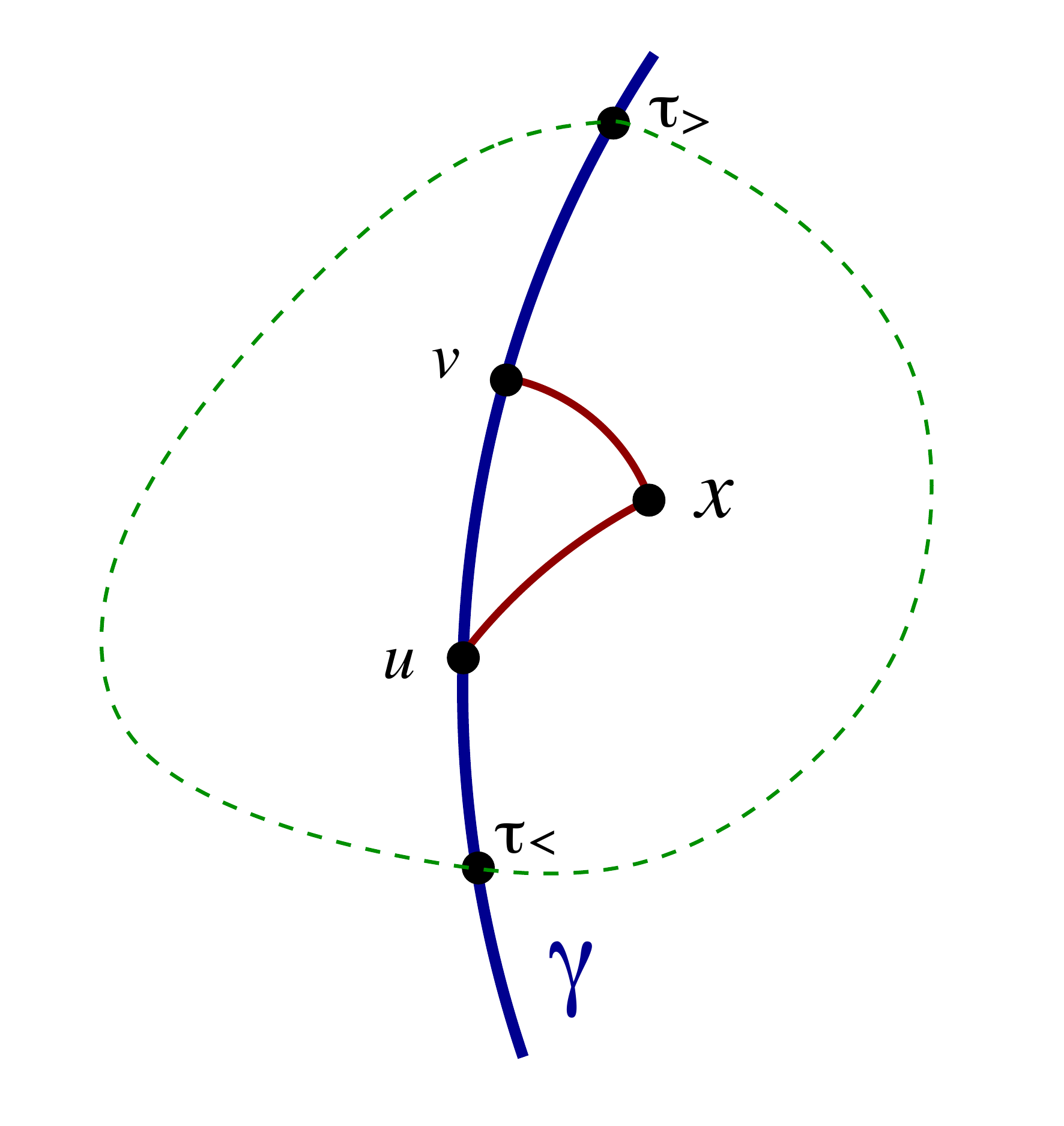}
\vspace*{-20pt}
\end{center} 
\caption{The region within the dashed boundary represents the normal 
convex neighbourhood of the point $x$. The world line $\gamma$ enters 
the neighbourhood at proper time $\tau_<$ and exits at proper time
$\tau_>$. Also shown are the retarded point $z(u)$ and the advanced
point $z(v)$.}   
\end{figure} 

We now specialize Eq.~(\ref{16.2.1}) to a point $x$ near the world
line; see Fig.~9. We let ${\cal N}(x)$ be the normal convex
neighbourhood of this point, and we assume that the world line
traverses ${\cal N}(x)$. Let $\tau_<$ be the value of the proper-time
parameter at which $\gamma$ enters ${\cal N}(x)$ from the past, and
let $\tau_>$ be its value when the world line leaves 
${\cal N}(x)$. Then Eq.~(\ref{16.2.1}) can be broken up into the
three integrals    
\[
\Phi(x) = q \int_{-\infty}^{\tau_<} G_+(x,z)\, d\tau 
+ q \int_{\tau_<}^{\tau_>} G_+(x,z)\, d\tau 
+ q \int_{\tau_>}^{\infty} G_+(x,z)\, d\tau. 
\]
The third integration vanishes because $x$ is then in the past of
$z(\tau)$, and $G_+(x,z) = 0$. For the second integration, $x$ is the 
normal convex neighbourhood of $z(\tau)$, and the retarded Green's 
function can be expressed in the Hadamard form produced in 
Sec.~\ref{13.2}. This gives   
\[ 
\int_{\tau_<}^{\tau_>} G_+(x,z)\, d\tau = \int_{\tau_<}^{\tau_>}
U(x,z) \delta_+(\sigma)\, d\tau + \int_{\tau_<}^{\tau_>} V(x,z)
\theta_+(-\sigma)\, d\tau, 
\] 
and to evaluate this we refer back to Sec.~\ref{9} and let $x' 
:= z(u)$ be the retarded point associated with $x$; these points
are related by $\sigma(x,x') = 0$ and $r := \sigma_{\alpha'}
u^{\alpha'}$ is the retarded distance between $x$ and the world
line. We resume the index convention of Sec.~\ref{9}: to tensors at
$x$ we assign indices $\alpha$, $\beta$, etc.; to tensors at $x'$
we assign indices $\alpha'$, $\beta'$, etc.; and to tensors at a
generic point $z(\tau)$ on the world line we assign indices $\mu$,
$\nu$, etc.    

To perform the first integration we change variables from $\tau$
to $\sigma$, noticing that $\sigma$ increases as $z(\tau)$ passes
through $x'$. The change of $\sigma$ on the world line is given
by $d\sigma := \sigma(x,z+dz) - \sigma(x,z) = \sigma_\mu u^\mu\,
d\tau$, and we find that the first integral evaluates to
$U(x,z)/(\sigma_\mu u^\mu)$ with $z$ identified with $x'$. The second
integration is cut off at $\tau = u$ by the step function, and we
obtain our final expression for the retarded potential of a point
scalar charge:  
\begin{equation}
\Phi(x) = \frac{q}{r} U(x,x') + q \int_{\tau_<}^u V(x,z)\, d\tau + q
\int_{-\infty}^{\tau_<} G_+(x,z)\, d\tau.   
\label{16.2.2}
\end{equation} 
This expression applies to a point $x$ sufficiently close to the world
line that there exists a nonempty intersection between 
${\cal N}(x)$ and $\gamma$.  

\subsection{Field of a scalar charge in retarded coordinates}  
\label{16.3} 

When we differentiate the potential of Eq.~(\ref{16.2.2}) we must keep
in mind that a variation in $x$ induces a variation in $x'$ because
the new points $x + \delta x$ and $x' + \delta x'$ must also be linked
by a null geodesic --- you may refer back to Sec.~\ref{9.2} for a
detailed discussion. This means, for example, that the total variation
of $U(x,x')$ is $\delta U = U(x+\delta x,x'+\delta x') - U(x,x') =
U_{;\alpha} \delta x^\alpha + U_{;\alpha'} u^{\alpha'}\, 
\delta u$. The gradient of the scalar potential is therefore given by    
\begin{equation}
\Phi_{\alpha}(x) = -\frac{q}{r^2} U(x,x') \partial_\alpha r + 
\frac{q}{r} U_{;\alpha}(x,x') + \frac{q}{r} U_{;\alpha'}(x,x')
u^{\alpha'} \partial_\alpha u + q V(x,x') \partial_\alpha u +  
\Phi^{\rm tail}_{\alpha}(x),   
\label{16.3.1}
\end{equation}
where the ``tail integral'' is defined by    
\begin{eqnarray}
\Phi^{\rm tail}_{\alpha}(x) &=& q \int_{\tau_<}^u \nabla_\alpha 
V(x,z)\, d\tau + q \int_{-\infty}^{\tau_<} \nabla_\alpha G_+(x,z)\, 
d\tau \nonumber \\  
&=& q \int_{-\infty}^{u^-} \nabla_\alpha G_+(x,z)\, d\tau.   
\label{16.3.2}
\end{eqnarray} 
In the second form of the definition we integrate $\nabla_\alpha 
G_+(x,z)$ from $\tau = -\infty$ to almost $\tau = u$, but we cut the  
integration short at $\tau = u^- := u - 0^+$ to avoid the singular  
behaviour of the retarded Green's function at $\sigma = 0$. This
limiting procedure gives rise to the first form of the definition,
with the advantage that the integral need not be broken up into
contributions that refer to ${\cal N}(x)$ and its complement,
respectively.  

We shall now expand $\Phi_{\alpha}(x)$ in powers of $r$, and express 
the results in terms of the retarded coordinates $(u,r,\Omega^a)$
introduced in Sec.~\ref{9}. It will be convenient to decompose
$\Phi_{\alpha}(x)$ in the tetrad
$(\base{\alpha}{0},\base{\alpha}{a})$ that is obtained by parallel
transport of $(u^{\alpha'},\base{\alpha'}{a})$ on the null geodesic
that links $x$ to $x' := z(u)$; this construction is detailed in 
Sec.~\ref{9}. The expansion relies on Eq.~(\ref{9.5.3}) for 
$\partial_\alpha u$, Eq.~(\ref{9.5.5}) for $\partial_\alpha r$, and we
shall need     
\begin{equation}
U(x,x') = 1 + \frac{1}{12} r^2 \bigl( R_{00} + 2 R_{0a} \Omega^a + 
R_{ab} \Omega^a \Omega^b \bigr) + O(r^3), 
\label{16.3.3}
\end{equation} 
which follows from Eq.~(\ref{13.2.7}) and the relation 
$\sigma^{\alpha'} = -r (u^{\alpha'} + \Omega^a \base{\alpha'}{a})$
first encountered in Eq.~(\ref{9.2.3}); recall that
\[
R_{00}(u) = R_{\alpha'\beta'} u^{\alpha'} u^{\beta'}, 
\qquad 
R_{0a}(u) = R_{\alpha'\beta'} u^{\alpha'} \base{\beta'}{a},
\qquad 
R_{ab}(u) = R_{\alpha'\beta'} \base{\alpha'}{a} \base{\beta'}{b}
\]
are frame components of the Ricci tensor evaluated at $x'$. We shall
also need the expansions    
\begin{equation} 
U_{;\alpha}(x,x') = \frac{1}{6} r g^{\alpha'}_{\ \alpha} 
\bigl( R_{\alpha'0} + R_{\alpha' b} \Omega^b \bigr) + O(r^2)   
\label{16.3.4} 
\end{equation}
and 
\begin{equation} 
U_{;\alpha'}(x,x') u^{\alpha'} = -\frac{1}{6} r \bigl( R_{00} 
+ R_{0a} \Omega^a \bigr) + O(r^2) 
\label{16.3.5}
\end{equation}
which follow from Eqs.~(\ref{13.2.8}); recall from Eq.~(\ref{9.1.4}) 
that the parallel propagator can be expressed as 
$g^{\alpha'}_{\ \alpha} = u^{\alpha'} \base{0}{\alpha} 
+ \base{\alpha'}{a} \base{a}{\alpha}$. And finally, we shall need  
\begin{equation}
V(x,x') = \frac{1}{12} \bigl( 1 - 6\xi \bigr) R + O(r), 
\label{16.3.6}
\end{equation} 
a relation that was first established in Eq.~(\ref{13.2.10}); here
$R := R(u)$ is the Ricci scalar evaluated at $x'$.  

Collecting all these results gives    
\begin{eqnarray} 
\Phi_0(u,r,\Omega^a) &:=& \Phi_{\alpha}(x) \base{\alpha}{0}(x)
\nonumber \\  
&=& \frac{q}{r} a_a \Omega^a + \frac{1}{2} q R_{a0b0}\Omega^a
\Omega^b + \frac{1}{12} \bigl( 1 - 6\xi \bigr) q R 
+ \Phi_0^{\rm tail} + O(r), 
\label{16.3.7} \\ 
\Phi_a(u,r,\Omega^a) &:=& \Phi_{\alpha}(x) \base{\alpha}{a}(x)
\nonumber \\  
&=& -\frac{q}{r^2} \Omega_a - \frac{q}{r} a_b \Omega^b \Omega_a -
\frac{1}{3} q R_{b0c0} \Omega^b \Omega^c \Omega_a - \frac{1}{6} q
\bigl( R_{a0b0} \Omega^b - R_{ab0c} \Omega^b \Omega^c \bigr) 
\nonumber \\ & & \mbox{}
+ \frac{1}{12} q \bigl[ R_{00} - R_{bc} \Omega^b\Omega^c -
(1-6\xi) R \bigr] \Omega_a + \frac{1}{6} q \bigl( R_{a0} + R_{ab} 
\Omega^b \bigr) + \Phi_{a}^{\rm tail} + O(r),  \qquad
\label{16.3.8}
\end{eqnarray} 
where $a_a = a_{\alpha'} \base{\alpha'}{a}$ are the frame  
components of the acceleration vector, 
\[
R_{a0b0}(u) = R_{\alpha'\gamma'\beta'\delta'} \base{\alpha'}{a}
u^{\gamma'}  \base{\beta'}{b} u^{\delta'}, \qquad 
R_{ab0c}(u) = R_{\alpha'\gamma'\beta'\delta'} \base{\alpha'}{a}
\base{\gamma'}{b} u^{\beta'} \base{\delta'}{c}
\]
are frame components of the Riemann tensor evaluated at $x'$, and
\begin{equation} 
\Phi_0^{\rm tail}(u) = \Phi_{\alpha'}^{\rm tail}(x')
u^{\alpha'},  
\qquad  
\Phi_a^{\rm tail}(u) = \Phi_{\alpha'}^{\rm tail}(x') 
\base{\alpha'}{a}  
\label{16.3.9}
\end{equation}
are the frame components of the tail integral evaluated at 
$x'$. Equations (\ref{16.3.7}) and (\ref{16.3.8}) show clearly 
that $\Phi_{\alpha}(x)$ is singular on the world line: the field
diverges as $r^{-2}$ when $r\to 0$, and many of the terms that stay 
bounded in the limit depend on $\Omega^a$ and therefore possess a
directional ambiguity at $r=0$.  

\subsection{Field of a scalar charge in Fermi normal coordinates}   
\label{16.4} 

The gradient of the scalar potential can also be expressed in the
Fermi normal coordinates of Sec.~\ref{8}. To effect this translation
we make $\bar{x} := z(t)$ the new
reference point on the world line. We resume here the notation of 
Sec.~\ref{10} and assign indices $\bar{\alpha}$, $\bar{\beta}$,
\ldots to tensors at $\bar{x}$. The Fermi normal coordinates are
denoted $(t,s,\omega^a)$, and we let $(\bar{e}^\alpha_0,
\bar{e}^\alpha_a)$ be the tetrad at $x$ that is obtained by parallel
transport of $(u^{\bar{\alpha}}, \base{\bar{\alpha}}{a})$ on the
spacelike geodesic that links $x$ to $\bar{x}$.  

Our first task is to decompose $\Phi_{\alpha}(x)$ in the tetrad
$(\bar{e}^\alpha_0, \bar{e}^\alpha_a)$, thereby defining $\bar{\Phi}_0 
:= \Phi_{\alpha} \bar{e}^\alpha_0$ and $\bar{\Phi}_a :=
\Phi_{\alpha} \bar{e}^\alpha_a$. For this purpose we use 
Eqs.~(\ref{10.3.1}), (\ref{10.3.2}), (\ref{16.3.7}), and
(\ref{16.3.8}) to obtain  
\begin{eqnarray*} 
\bar{\Phi}_0 &=& 
\Bigl[ 1 + O(r^2) \Bigr] \Phi_0 + \Bigl[ r \bigl( 1 - a_b \Omega^b 
\bigr) a^a + \frac{1}{2} r^2 \dot{a}^a + \frac{1}{2} r^2 
R^a_{\ 0b0} \Omega^b + O(r^3) \Bigr] \Phi_a \\ 
&=& -\frac{1}{2} q \dot{a}_a \Omega^a + \frac{1}{12} (1-6\xi) q R +
\bar{\Phi}_0^{\rm tail} + O(r) 
\end{eqnarray*} 
and 
\begin{eqnarray*} 
\bar{\Phi}_a &=& 
\Bigl[ \delta^b_{\ a} + \frac{1}{2} r^2 a^b a_a - \frac{1}{2} r^2
R^b_{\ a0c} \Omega^c + O(r^3) \Bigr] \Phi_b + \Bigl[ r a_a + O(r^2)
\Bigr] \Phi_0 \\ 
&=& -\frac{q}{r^2} \Omega_a - \frac{q}{r} a_b \Omega^b \Omega_a 
+ \frac{1}{2} q a_b \Omega^b a_a 
- \frac{1}{3} q R_{b0c0} \Omega^b \Omega^c \Omega_a 
- \frac{1}{6} q R_{a0b0} \Omega^b 
- \frac{1}{3} q R_{ab0c} \Omega^b \Omega^c  
\nonumber \\ & & \mbox{}
+ \frac{1}{12} q \bigl[ R_{00} - R_{bc} \Omega^b\Omega^c 
- (1-6\xi) R \bigr] \Omega_a 
+ \frac{1}{6} q \bigl( R_{a0} + R_{ab} \Omega^b \bigr) 
+ \bar{\Phi}_{a}^{\rm tail} + O(r), 
\end{eqnarray*} 
where all frame components are still evaluated at $x'$, except for 
$\bar{\Phi}_{0}^{\rm tail}$ and $\bar{\Phi}_{a}^{\rm tail}$ which are
evaluated at $\bar{x}$.   

We must still translate these results into the Fermi normal
coordinates $(t,s,\omega^a)$. For this we involve Eqs.~(\ref{10.2.1}),
(\ref{10.2.2}), and (\ref{10.2.3}), from which we deduce, for example,   
\begin{eqnarray*} 
\frac{1}{r^2}\Omega_a &=& \frac{1}{s^2}\omega_a + \frac{1}{2s} a_a  
- \frac{3}{2s} a_b \omega^b \omega_a - \frac{3}{4} a_b \omega^b a_a 
+ \frac{15}{8} \bigl( a_b \omega^b \bigr)^2 \omega_a 
+ \frac{3}{8} \dot{a}_0 \omega_a - \frac{1}{3} \dot{a}_a 
\\ & & \mbox{} 
+ \dot{a}_b \omega^b \omega_a 
+ \frac{1}{6} R_{a0b0} \omega^b    
- \frac{1}{2} R_{b0c0} \omega^b \omega^c \omega_a 
- \frac{1}{3} R_{ab0c} \omega^b \omega^c + O(s) 
\end{eqnarray*}
and 
\[
\frac{1}{r} a_b \Omega^b \Omega_a = \frac{1}{s} a_b \omega^b \omega_a 
+ \frac{1}{2} a_b \omega^b a_a 
- \frac{3}{2} \bigl( a_b \omega^b \bigr)^2 \omega_a   
- \frac{1}{2} \dot{a}_0 \omega_a - \dot{a}_b \omega^b \omega_a 
+ O(s), 
\]
in which all frame components (on the right-hand side of these
relations) are now evaluated at $\bar{x}$; to obtain the second
relation we expressed $a_a(u)$ as $a_a(t) - s \dot{a}_a(t) + O(s^2)$
since according to Eq.~(\ref{10.2.1}), $u = t - s + O(s^2)$.  

Collecting these results yields 
\begin{eqnarray} 
\bar{\Phi}_0(t,s,\omega^a) &:=& \Phi_{\alpha}(x)
\bar{e}^\alpha_0(x) 
\nonumber \\  
&=& -\frac{1}{2} q \dot{a}_a \omega^a + \frac{1}{12} (1-6\xi) q R + 
\bar{\Phi}_0^{\rm tail} + O(s), 
\label{16.4.1} \\  
\bar{\Phi}_a(t,s,\omega^a) &:=& \Phi_{\alpha}(x)
\bar{e}^\alpha_a(x) 
\nonumber \\  
&=& -\frac{q}{s^2} \omega_a 
- \frac{q}{2s} \bigl( a_a - a_b \omega^b \omega_a \bigr)  
+ \frac{3}{4} q a_b \omega^b a_a 
- \frac{3}{8} q \bigl( a_b \omega^b \bigr)^2 \omega_a 
+ \frac{1}{8} q \dot{a}_0 \omega_a 
+ \frac{1}{3} q \dot{a}_a 
\nonumber \\ & & \mbox{} 
- \frac{1}{3} q R_{a0b0} \omega^b 
+ \frac{1}{6} q R_{b0c0} \omega^b \omega^c \omega_a 
+ \frac{1}{12} q \bigl[ R_{00} - R_{bc} \omega^b\omega^c 
- (1-6\xi) R \bigr] \omega_a 
\nonumber \\ & & \mbox{} 
+ \frac{1}{6} q \bigl( R_{a0} + R_{ab} \omega^b \bigr) 
+ \bar{\Phi}_{a}^{\rm tail} + O(s). 
\label{16.4.2}
\end{eqnarray} 
In these expressions, $a_a(t) = a_{\bar{\alpha}}
\base{\bar{\alpha}}{a}$ are the frame components of the acceleration
vector evaluated at $\bar{x}$, $\dot{a}_0(t) =
\dot{a}_{\bar{\alpha}} u^{\bar{\alpha}}$ and $\dot{a}_a(t) =  
\dot{a}_{\bar{\alpha}} \base{\bar{\alpha}}{a}$ are frame components of
its covariant derivative, $R_{a0b0}(t) =
R_{\bar{\alpha}\bar{\gamma}\bar{\beta}\bar{\delta}}
\base{\bar{\alpha}}{a} u^{\bar{\gamma}} \base{\bar{\beta}}{b}
u^{\bar{\delta}}$ are frame components of the Riemann tensor evaluated
at $\bar{x}$, 
\[
R_{00}(t) = R_{\bar{\alpha}\bar{\beta}} u^{\bar{\alpha}}
u^{\bar{\beta}},  
\qquad 
R_{0a}(t) = R_{\bar{\alpha}\bar{\beta}} u^{\bar{\alpha}}
\base{\bar{\beta}}{a}, 
\qquad 
R_{ab}(t) = R_{\bar{\alpha}\bar{\beta}} \base{\bar{\alpha}}{a}
\base{\bar{\beta}}{b} 
\]
are frame components of the Ricci tensor, and $R(t)$ is the Ricci 
scalar evaluated at $\bar{x}$. Finally, we have that 
\begin{equation} 
\bar{\Phi}_0^{\rm tail}(t) = 
\Phi_{\bar{\alpha}}^{\rm tail}(\bar{x}) u^{\bar{\alpha}},   
\qquad  
\bar{\Phi}_a^{\rm tail}(t) =
\Phi_{\bar{\alpha}}^{\rm tail}(\bar{x}) \base{\bar{\alpha}}{a}   
\label{16.4.3}
\end{equation}
are the frame components of the tail integral --- see
Eq.~(\ref{16.3.2}) --- evaluated at $\bar{x} := z(t)$. 

We shall now compute the averages of $\bar{\Phi}_0$ and $\bar{\Phi}_a$ 
over $S(t,s)$, a two-surface of constant $t$ and $s$; these will
represent the mean value of the field at a fixed proper distance away
from the world line, as measured in a reference frame that is
momentarily comoving with the particle. The two-surface is
charted by angles $\theta^A$ ($A = 1, 2$) and it is described,
in the Fermi normal coordinates, by the parametric relations
$\hat{x}^a = s \omega^a(\theta^A)$; a canonical choice of
parameterization is $\omega^a = (\sin\theta\cos\phi,
\sin\theta\sin\phi, \cos\theta)$. Introducing the transformation
matrices $\omega^a_A := \partial \omega^a/\partial \theta^A$, we
find from Eq.~(\ref{8.5.5}) that the induced metric on $S(t,s)$ is
given by    
\begin{equation} 
ds^2 = s^2 \Bigl[ \omega_{AB} - \frac{1}{3} s^2 R_{AB} + O(s^3) 
\Bigr]\, d\theta^A d\theta^B, 
\label{16.4.4}
\end{equation} 
where $\omega_{AB} := \delta_{ab} \omega^a_A \omega^b_B$ is the 
metric of the unit two-sphere, and where $R_{AB} := R_{acbd}
\omega^a_A \omega^c \omega^b_B \omega^d$ depends on $t$ and the angles
$\theta^A$. From this we infer that the element of surface area is
given by 
\begin{equation} 
d{\cal A} = s^2 \Bigl[ 1 - \frac{1}{6} s^2 R^c_{\ acb}(t) \omega^a 
\omega^b + O(s^3) \Bigr]\, d\omega, 
\label{16.4.5}
\end{equation}
where $d\omega = \sqrt{\mbox{det}[\omega_{AB}]}\, d^2\theta$ is an
element of solid angle --- in the canonical parameterization, $d\omega
= \sin\theta\, d\theta d\phi$. Integration of Eq.~(\ref{16.4.5})
produces the total surface area of $S(t,s)$, and ${\cal A} 
= 4\pi s^2 [1 - \frac{1}{18} s^2 R^{ab}_{\ \ ab} + O(s^3)]$.  

The averaged fields are defined by 
\begin{equation} 
\bigl\langle \bar{\Phi}_0 \bigr\rangle(t,s) = \frac{1}{\cal A} 
\oint_{S(t,s)} \bar{\Phi}_0(t,s,\theta^A)\, d{\cal A}, \qquad  
\bigl\langle \bar{\Phi}_a \bigr\rangle(t,s) = \frac{1}{\cal A} 
\oint_{S(t,s)} \bar{\Phi}_a(t,s,\theta^A)\, d{\cal A}, 
\label{16.4.6}
\end{equation} 
where the quantities to be integrated are scalar functions of the
Fermi normal coordinates. The results 
\begin{equation}
\frac{1}{4\pi} \oint \omega^a\, d\omega = 0, \qquad 
\frac{1}{4\pi} \oint \omega^a \omega^b\, d\omega = 
\frac{1}{3} \delta^{ab}, \qquad  
\frac{1}{4\pi} \oint \omega^a \omega^b \omega^c\, d\omega = 0, 
\label{16.4.7}
\end{equation} 
are easy to establish, and we obtain 
\begin{eqnarray} 
\bigl\langle \bar{\Phi}_0 \bigr\rangle &=& \frac{1}{12} (1-6\xi)
q R + \bar{\Phi}_0^{\rm tail} + O(s), 
\label{16.4.8} \\ 
\bigl\langle \bar{\Phi}_a \bigr\rangle &=& -\frac{q}{3s} a_a +
\frac{1}{3} q \dot{a}_a + \frac{1}{6} q R_{a0} + 
\bar{\Phi}_a^{\rm tail} + O(s).  
\label{16.4.9}
\end{eqnarray} 
The averaged field is still singular on the world line. Regardless, we
shall take the formal limit $s \to 0$ of the expressions displayed in 
Eqs.~(\ref{16.4.8}) and (\ref{16.4.9}). In the limit the tetrad
$(\bar{e}^\alpha_0, \bar{e}^\alpha_a)$ reduces to 
$(u^{\bar{\alpha}}, \base{\bar{\alpha}}{a})$, and we can
reconstruct the field at $\bar{x}$ by invoking the completeness
relations $\delta^{\bar{\alpha}}_{\ \bar{\beta}} =
-u^{\bar{\alpha}} u_{\bar{\beta}} + \base{\bar{\alpha}}{a}
\base{a}{\bar{\beta}}$. We thus obtain 
\begin{equation} 
\bigl\langle \Phi_{\bar{\alpha}} \bigr\rangle = \lim_{s\to 0} \biggl(
- \frac{q}{3s} \biggr) a_{\bar{\alpha}}  
- \frac{1}{12} (1-6\xi) q R u_{\bar{\alpha}} 
+ q \bigl( g_{\bar{\alpha}\bar{\beta}} 
+ u_{\bar{\alpha}} u_{\bar{\beta}} \bigr) 
\biggl( \frac{1}{3} \dot{a}^{\bar{\beta}} 
+ \frac{1}{6} R^{\bar{\beta}}_{\ \bar{\gamma}} u^{\bar{\gamma}}
\biggr) + \Phi_{\bar{\alpha}}^{\rm tail}, 
\label{16.4.10}
\end{equation}
where the tail integral can be copied from Eq.~(\ref{16.3.2}), 
\begin{equation} 
\Phi_{\bar{\alpha}}^{\rm tail}(\bar{x}) = q \int_{-\infty}^{t^-} 
\nabla_{\bar{\alpha}} G_+(\bar{x},z)\, d\tau. 
\label{16.4.11}
\end{equation} 
The tensors appearing in Eq.~(\ref{16.4.10}) all refer to $\bar{x}
:= z(t)$, which now stands for an arbitrary point on the world
line $\gamma$.   

\subsection{Singular and regular fields} 
\label{16.5} 

The singular potential 
\begin{equation} 
\Phi^{\rm S}(x) = q \int_\gamma G_{\rm S}(x,z)\, d\tau
\label{16.5.1}
\end{equation} 
is the (unphysical) solution to Eqs.~(\ref{16.1.5}) and (\ref{16.1.6})
that is obtained by adopting the singular Green's function of
Eq.~(\ref{13.5.12}) instead of the retarded Green's function. As we
shall see, the resulting singular field $\Phi^{\rm S}_\alpha(x)$
reproduces the singular behaviour of the retarded solution; the
difference, $\Phi^{\rm R}_\alpha(x) = \Phi_\alpha(x) 
- \Phi^{\rm S}_\alpha(x)$, is regular on the world line.     

To evaluate the integral of Eq.~(\ref{16.5.1}) we assume once more
that $x$ is sufficiently close to $\gamma$ that the world line
traverses ${\cal N}(x)$; refer back to Fig.~9. As before we let
$\tau_<$ and $\tau_>$ be the values of the proper-time parameter at
which $\gamma$ enters and leaves ${\cal N}(x)$, respectively. Then
Eq.~(\ref{16.5.1}) can be broken up into the three integrals   
\[
\Phi^{\rm S}(x) = q \int_{-\infty}^{\tau_<} G_{\rm S}(x,z)\, d\tau  
+ q \int_{\tau_<}^{\tau_>} G_{\rm S}(x,z)\, d\tau 
+ q \int_{\tau_>}^{\infty} G_{\rm S}(x,z)\, d\tau.  
\]
The first integration vanishes because $x$ is then in the
chronological future of $z(\tau)$, and $G_{\rm S}(x,z) = 0$ by
Eq.~(\ref{13.5.3}). Similarly, the third integration vanishes because
$x$ is then in the chronological past of $z(\tau)$. For the second
integration, $x$ is the normal convex neighbourhood of $z(\tau)$, the
singular Green's function can be expressed in the Hadamard form of
Eq.~(\ref{13.5.14}), and we have      
\[ 
\int_{\tau_<}^{\tau_>} G_{\rm S}(x,z)\, d\tau = 
\frac{1}{2} \int_{\tau_<}^{\tau_>} U(x,z) \delta_+(\sigma)\, d\tau 
+ \frac{1}{2} \int_{\tau_<}^{\tau_>} U(x,z) \delta_-(\sigma)\, d\tau 
- \frac{1}{2} \int_{\tau_<}^{\tau_>} V(x,z) \theta(\sigma)\, d\tau.  
\] 
To evaluate these we re-introduce the retarded point $x' := z(u)$
and let $x'' := z(v)$ be the {\it advanced point} associated with
$x$; we recall from Sec.~\ref{10.4} that these points are related by
$\sigma(x,x'') = 0$ and that $r_{\rm adv} := - \sigma_{\alpha''}
u^{\alpha''}$ is the advanced distance between $x$ and the world
line. 

To perform the first integration we change variables from $\tau$
to $\sigma$, noticing that $\sigma$ increases as $z(\tau)$ passes
through $x'$; the integral evaluates to $U(x,x')/r$. We do the same
for the second integration, but we notice now that $\sigma$ decreases
as $z(\tau)$ passes through $x''$; the integral evaluates to
$U(x,x'')/r_{\rm adv}$. The third integration is restricted to the
interval $u \leq \tau \leq v$ by the step function, and we obtain our
final expression for the singular potential of a point scalar charge:   
\begin{equation}
\Phi^{\rm S}(x) = \frac{q}{2r} U(x,x') 
+ \frac{q}{2r_{\rm adv}} U(x,x'') 
- \frac{1}{2} q \int_u^v V(x,z)\, d\tau.   
\label{16.5.2} 
\end{equation} 
We observe that $\Phi^{\rm S}(x)$ depends on the state of motion of 
the scalar charge between the retarded time $u$ and the advanced time
$v$; contrary to what was found in Sec.~\ref{16.2} for the retarded
potential, there is no dependence on the particle's remote past.          

We use the techniques of Sec.~\ref{16.3} to differentiate the
potential of Eq.~(\ref{16.5.2}). We find 
\begin{eqnarray}
\Phi^{\rm S}_{\alpha}(x) &=& 
-\frac{q}{2r^2} U(x,x') \partial_\alpha r  
- \frac{q}{2 {r_{\rm adv}}^2} U(x,x'') \partial_\alpha r_{\rm adv} 
+ \frac{q}{2r} U_{;\alpha}(x,x') 
+ \frac{q}{2r} U_{;\alpha'}(x,x') u^{\alpha'} \partial_\alpha u 
\nonumber \\  & & \mbox{} 
+ \frac{q}{2r_{\rm adv}} U_{;\alpha}(x,x'') 
+ \frac{q}{2r_{\rm adv}} U_{;\alpha''}(x,x'') u^{\alpha''}
\partial_\alpha v + \frac{1}{2} q V(x,x') \partial_\alpha u 
- \frac{1}{2} q V(x,x'') \partial_\alpha v 
\nonumber \\  & & \mbox{} 
- \frac{1}{2} q \int_u^v \nabla_\alpha V(x,z)\, d\tau, 
\label{16.5.3}
\end{eqnarray}
and we would like to express this as an expansion in powers of
$r$. For this we shall rely on results already established in 
Sec.~\ref{16.3}, as well as additional expansions that will involve
the advanced point $x''$. Those we develop now.  

We recall first that a relation between retarded and advanced times
was worked out in Eq.~(\ref{10.4.2}), that an expression for the
advanced distance was displayed in Eq.~(\ref{10.4.3}), and that
Eqs.~(\ref{10.4.4}) and (\ref{10.4.5}) give expansions for
$\partial_\alpha v$ and $\partial_\alpha r_{\rm adv}$,
respectively. 

To derive an expansion for $U(x,x'')$ we follow the general method of 
Sec.~\ref{10.4} and define a function $U(\tau) := U(x,z(\tau))$
of the proper-time parameter on $\gamma$. We have that 
\[
U(x,x'') := U(v) = U(u+\Delta^{\!\prime}) = U(u) + \dot{U}(u) 
\Delta^{\!\prime} + \frac{1}{2} \ddot{U}(u) \Delta^{\!\prime 2} 
+ O\bigl( \Delta^{\!\prime 3} \bigr), 
\]
where overdots indicate differentiation with respect to $\tau$, and
where $\Delta^{\!\prime} := v-u$. The leading term $U(u) :=
U(x,x')$ was worked out in Eq.~(\ref{16.3.3}), and the derivatives of 
$U(\tau)$ are given by 
\[
\dot{U}(u) = U_{;\alpha'} u^{\alpha'} = -\frac{1}{6} r \bigl( R_{00} 
+ R_{0a} \Omega^a \bigr) + O(r^2)
\]
and 
\[
\ddot{U}(u) = U_{;\alpha'\beta'} u^{\alpha'} u^{\beta'} + U_{;\alpha'}
a^{\alpha'} = \frac{1}{6} R_{00} + O(r),
\]
according to Eqs.~(\ref{16.3.5}) and (\ref{13.2.8}). Combining these
results together with Eq.~(\ref{10.4.2}) for $\Delta^{\!\prime}$ gives   
\begin{equation} 
U(x,x'') = 1 + \frac{1}{12} r^2 \bigl( R_{00} - 2 R_{0a} \Omega^a +
R_{ab} \Omega^a \Omega^b \bigr) + O(r^3), 
\label{16.5.4}
\end{equation}
which should be compared with Eq.~(\ref{16.3.3}). It should be
emphasized that in Eq.~(\ref{16.5.4}) and all equations below, the
frame components of the Ricci tensor are evaluated at the retarded
point $x' := z(u)$, and not at the advanced point. The preceding
computation gives us also an expansion for 
$U_{;\alpha''} u^{\alpha''} := \dot{U}(v) =
\dot{U}(u) + \ddot{U}(u) \Delta^{\!\prime} + O(\Delta^{\!\prime 2})$. 
This becomes     
\begin{equation}
U_{;\alpha''}(x,x'') u^{\alpha''} = \frac{1}{6} r \bigl( R_{00} 
- R_{0a} 
\Omega^a \bigr) + O(r^2), 
\label{16.5.5}
\end{equation}
which should be compared with Eq.~(\ref{16.3.5}).  

We proceed similarly to derive an expansion for
$U_{;\alpha}(x,x'')$. Here we introduce the functions
$U_{\alpha}(\tau) := U_{;\alpha}(x,z(\tau))$ and express
$U_{;\alpha}(x,x'')$ as $U_\alpha(v) = U_\alpha(u) + \dot{U}_\alpha(u)
\Delta^{\!\prime} + O(\Delta^{\!\prime 2})$. The leading term
$U_\alpha(u) := U_{;\alpha}(x,x')$ was computed in
Eq.~(\ref{16.3.4}), and  
\[
\dot{U}_\alpha(u) = U_{;\alpha\beta'} u^{\beta'} = -\frac{1}{6}
g^{\alpha'}_{\ \alpha} R_{\alpha'0} + O(r) 
\]
follows from Eq.~(\ref{13.2.8}). Combining these results together with
Eq.~(\ref{10.4.2}) for $\Delta^{\!\prime}$ gives
\begin{equation}
U_{;\alpha}(x,x'') = -\frac{1}{6} r g^{\alpha'}_{\ \alpha} 
\bigr( R_{\alpha'0} - R_{\alpha'b} \Omega^b \Bigr) + O(r^2), 
\label{16.5.6}
\end{equation}
and this should be compared with Eq.~(\ref{16.3.4}).  

The last expansion we shall need is 
\begin{equation}
V(x,x'') = \frac{1}{12} \bigl( 1 - 6\xi \bigr) R + O(r), 
\label{16.5.7} 
\end{equation} 
which follows at once from Eq.~(\ref{16.3.6}) and the fact that
$V(x,x'') - V(x,x') = O(r)$; the Ricci scalar is evaluated at the
retarded point $x'$. 

It is now a straightforward (but tedious) matter to substitute these
expansions (all of them!)\ into Eq.~(\ref{16.5.3}) and obtain the
projections of the singular field $\Phi^{\rm S}_\alpha(x)$ in the
same tetrad $(\base{\alpha}{0}, \base{\alpha}{a})$ that was employed
in Sec.~\ref{16.3}. This gives 
\begin{eqnarray} 
\Phi^{\rm S}_0(u,r,\Omega^a) &:=& \Phi^{\rm S}_{\alpha}(x)
\base{\alpha}{0}(x) 
\nonumber \\  
&=& \frac{q}{r} a_a \Omega^a 
+ \frac{1}{2} q R_{a0b0}\Omega^a \Omega^b + O(r), 
\label{16.5.8} \\ 
\Phi^{\rm S}_a(u,r,\Omega^a) &:=& \Phi^{\rm S}_{\alpha}(x)
\base{\alpha}{a}(x) 
\nonumber \\  
&=& -\frac{q}{r^2} \Omega_a - \frac{q}{r} a_b \Omega^b \Omega_a 
- \frac{1}{3} q \dot{a}_a 
- \frac{1}{3} q R_{b0c0} \Omega^b \Omega^c \Omega_a 
- \frac{1}{6} q \bigl( R_{a0b0} \Omega^b 
- R_{ab0c} \Omega^b \Omega^c \bigr)  
\nonumber \\ & & \mbox{}
+ \frac{1}{12} q \bigl[ R_{00} - R_{bc} \Omega^b\Omega^c 
- (1-6\xi) R \bigr] \Omega_a 
+ \frac{1}{6} q R_{ab} \Omega^b, 
\label{16.5.9}
\end{eqnarray} 
in which all frame components are evaluated at the retarded point 
$x' := z(u)$. Comparison of these expressions with 
Eqs.~(\ref{16.3.7}) and (\ref{16.3.8}) reveals that the retarded and
singular fields share the same singularity structure.    

The difference between the retarded field of Eqs.~(\ref{16.3.7}),
(\ref{16.3.8}) and the singular field of Eqs.~(\ref{16.5.8}),
(\ref{16.5.9}) defines the regular field 
$\Phi^{\rm R}_\alpha(x)$. Its frame components are 
\begin{eqnarray} 
\Phi^{\rm R}_0 &=& \frac{1}{12} (1-6\xi) q R + \Phi_0^{\rm tail} 
+ O(r),  
\label{16.5.10} \\ 
\Phi^{\rm R}_a &=& \frac{1}{3} q \dot{a}_a + \frac{1}{6} q R_{a0} 
+ \Phi_a^{\rm tail} + O(r), 
\label{16.5.11}
\end{eqnarray} 
and we see that $\Phi^{\rm R}_\alpha(x)$ is a regular vector field on
the world line. There is therefore no obstacle in evaluating the
regular field directly at $x=x'$, where the tetrad
$(\base{\alpha}{0},\base{\alpha}{a})$ becomes $(u^{\alpha'}, 
\base{\alpha'}{a})$. Reconstructing the field at $x'$ from its
frame components, we obtain 
\begin{equation} 
\Phi^{\rm R}_{\alpha'}(x') = 
- \frac{1}{12} (1-6\xi) q R u_{\alpha'} 
+ q \bigl( g_{\alpha'\beta'} + u_{\alpha'} u_{\beta'} \bigr) 
\biggl( \frac{1}{3} \dot{a}^{\beta'} 
+ \frac{1}{6} R^{\beta'}_{\ \gamma'} u^{\gamma'}
\biggr) + \Phi_{\alpha'}^{\rm tail}, 
\label{16.5.12}
\end{equation}
where the tail term can be copied from Eq.~(\ref{16.3.2}), 
\begin{equation} 
\Phi_{\alpha'}^{\rm tail}(x') = q \int_{-\infty}^{u^-} 
\nabla_{\alpha'} G_+(x',z)\, d\tau. 
\label{16.5.13} 
\end{equation} 
The tensors appearing in Eq.~(\ref{16.5.12}) all refer to the 
retarded point $x' := z(u)$, which now stands for an  
arbitrary point on the world line $\gamma$.  
  
\subsection{Equations of motion} 
\label{16.6}

The retarded field $\Phi_\alpha(x)$ of a point scalar charge is
singular on the world line, and this behaviour makes it difficult to
understand how the field is supposed to act on the particle and affect
its motion. The field's singularity structure was analyzed in
Secs.~\ref{16.3} and \ref{16.4}, and in Sec.~\ref{16.5} it was shown
to originate from the singular field $\Phi^{\rm S}_\alpha(x)$; the
regular field $\Phi^{\rm R}_\alpha(x) = \Phi_\alpha(x) 
- \Phi^{\rm S}_\alpha(x)$ was then shown to be regular on the world
line.  

To make sense of the retarded field's action on the particle we
temporarily model the scalar charge not as a point particle, but as a  
small hollow shell that appears spherical when observed in a reference 
frame that is momentarily comoving with the particle; the shell's
radius is $s_0$ in Fermi normal coordinates, and it is independent of
the angles contained in the unit vector $\omega^a$. The {\it net
force} acting at proper time $\tau$ on this hollow shell is the
average of $q\Phi_\alpha(\tau,s_0,\omega^a)$ over the surface of the 
shell. Assuming that the field on the shell is equal to the field of a
point particle evaluated at $s=s_0$, and ignoring terms that
disappear in the limit $s_0 \to 0$, we obtain from Eq.~(\ref{16.4.10})  
\begin{equation} 
q\bigl\langle \Phi_\mu \bigr\rangle = -(\delta m) a_\mu       
- \frac{1}{12} (1-6\xi) q^2 R u_{\mu} 
+ q^2 \bigl( g_{\mu\nu} + u_{\mu} u_{\nu} \bigr) 
\biggl( \frac{1}{3} \dot{a}^{\nu} 
+ \frac{1}{6} R^{\nu}_{\ \lambda} u^{\lambda}
\biggr) + q\Phi_{\mu}^{\rm tail},
\label{16.6.1}
\end{equation}
where 
\begin{equation} 
\delta m := \lim_{s_0 \to 0} \frac{q^2}{3 s_0} 
\label{16.6.2}
\end{equation}
is formally a divergent quantity and 
\begin{equation} 
q \Phi_\mu^{\rm tail} = q^2 \int_{-\infty}^{\tau^-}  
\nabla_{\mu} G_+\bigl(z(\tau),z(\tau')\bigr)\, d\tau' 
\label{16.6.3}
\end{equation}
is the tail part of the force; all tensors in Eq.~(\ref{16.6.1}) are
evaluated at an arbitrary point $z(\tau)$ on the world line.  

Substituting Eqs.~(\ref{16.6.1}) and (\ref{16.6.3}) into
Eq.~(\ref{16.1.7}) gives rise to the equations of motion  
\begin{equation} 
\bigl( m + \delta m) a^\mu = q^2 \bigl( \delta^\mu_{\ \nu} 
+ u^\mu u_\nu \bigr) \Biggl[ \frac{1}{3} \dot{a}^{\nu} 
+ \frac{1}{6} R^{\nu}_{\ \lambda} u^{\lambda} 
+ \int_{-\infty}^{\tau^-}   
\nabla^{\nu} G_+\bigl(z(\tau),z(\tau')\bigr)\, d\tau' \Biggr] 
\label{16.6.4} 
\end{equation} 
for the scalar charge, with $m := m_0 -q \Phi(z)$ denoting
the (also formally divergent) dynamical mass of the particle. We see
that $m$ and $\delta m$ combine in Eq.~(\ref{16.6.4}) to form the 
particle's observed mass $m_{\rm obs}$, which is taken to be finite
and to give a true measure of the particle's inertia. All diverging
quantities have thus disappeared into the process of mass
renormalization. Substituting Eqs.~(\ref{16.6.1}) and (\ref{16.6.3})
into Eq.~(\ref{16.1.8}), in which we replace $m$ by $m_{\rm obs} 
= m + \delta m$, returns an expression for the rate of change of the 
observed mass,  
\begin{equation}
\frac{d m_{\rm obs}}{d\tau} = - \frac{1}{12} (1-6\xi) q^2 R 
- q^2 u^\mu \int_{-\infty}^{\tau^-} \nabla_{\mu}
G_+\bigl(z(\tau),z(\tau')\bigr)\, d\tau'. 
\label{16.6.5}
\end{equation} 
That the observed mass is {\it not} conserved is a remarkable property
of the dynamics of a scalar charge in a curved spacetime. Physically, 
this corresponds to the fact that in a spacetime with a time-dependent  
metric, a scalar charge radiates monopole waves and the radiated
energy comes at the expense of the particle's inertial mass.  

We must confess that the derivation of the equations of motion
outlined above returns the {\it wrong expression} for the self-energy
of a spherical shell of scalar charge. We obtained 
$\delta m = q^2/(3s_0)$, while the correct expression is 
$\delta m = q^2/(2s_0)$; we are wrong by a factor of $2/3$.  We
believe that this discrepancy originates in a previously stated
assumption, that the field on the shell (as produced by the shell
itself) is equal to the field of a point particle evaluated at
$s=s_0$. We believe that this assumption is in fact wrong, and  
that a calculation of the field actually produced by a spherical shell
would return the correct expression for $\delta m$. We also believe,
however, that except for the diverging terms that determine 
$\delta m$, the difference between the shell's field and the
particle's field should vanish in the limit $s_0 \to 0$. Our
conclusion is therefore that while our expression for $\delta m$ is
admittedly incorrect, the statement of the equations of motion is
reliable.  

Apart from the term proportional to $\delta m$, the averaged field of
Eq.~(\ref{16.6.1}) has exactly the same form as the regular field of
Eq.~(\ref{16.5.12}), which we re-express as 
\begin{equation} 
q \Phi^{\rm R}_{\mu} = - \frac{1}{12} (1-6\xi) q^2 R u_{\mu}  
+ q^2 \bigl( g_{\mu\nu} + u_{\mu} u_{\nu} \bigr) 
\biggl( \frac{1}{3} \dot{a}^{\nu} 
+ \frac{1}{6} R^{\nu}_{\ \lambda} u^{\lambda}
\biggr) + q\Phi_{\mu}^{\rm tail}. 
\label{16.6.6}
\end{equation}
The force acting on the point particle can therefore be thought of as 
originating from the regular field, while the singular field simply
contributes to the particle's inertia. After mass renormalization,
Eqs.~(\ref{16.6.4}) and (\ref{16.6.5}) are equivalent to the
statements  
\begin{equation} 
m a^\mu = q \bigl( g^{\mu\nu} + u^\mu u^\nu \bigr) 
\Phi^{\rm R}_{\nu}(z), \qquad 
\frac{d m}{d\tau} = -q u^\mu \Phi^{\rm R}_{\mu}(z), 
\label{16.6.7}
\end{equation} 
where we have dropped the superfluous label ``obs'' on the
particle's observed mass. Another argument in support of the claim
that the motion of the particle should be affected by the regular
field only was presented in Sec.~\ref{13.5}. 

The equations of motion displayed in Eqs.~(\ref{16.6.4}) and
(\ref{16.6.5}) are third-order differential equations for the
functions $z^\mu(\tau)$. It is well known that such a system of
equations admits many unphysical solutions, such as runaway situations
in which the particle's acceleration increases exponentially with
$\tau$, even in the absence of any external force \cite{dirac:38, 
jackson:99}. And indeed, our equations of motion do not yet
incorporate an external force which presumably is mostly responsible
for the particle's acceleration. Both defects can be cured in one
stroke. We shall take the point of view,  
the only admissible one in a classical treatment, that a point
particle is merely an idealization for an extended object whose
internal structure --- the details of its charge distribution --- can
be considered to be irrelevant. This view automatically implies that
our equations are meant to provide only an {\it approximate}
description of the object's motion. It can then be shown
\cite{landau-lifshitz:b2, flanagan-wald:96} that within the context of
this approximation, it is consistent to replace, on the 
right-hand side of the equations of motion, any occurrence of the
acceleration vector by $f_{\rm ext}^\mu/m$, where $f_{\rm ext}^\mu$ is
the external force acting on the particle. Because $f_{\rm ext}^\mu$
is a prescribed quantity, differentiation of the external force does
not produce higher derivatives of the functions $z^\mu(\tau)$, and the 
equations of motion are properly of the second order. 

We shall strengthen this conclusion in part~\ref{part5} of the
review, when we consider the motion of an extended body in a curved   
external spacetime. While the discussion there will concern
the gravitational self-force, many of the lessons learned in 
part~\ref{part5} apply just as well to the case of a scalar (or
electric) charge. And the main lesson is this: It is natural ---
indeed it is an imperative --- to view an equation of motion such as 
Eq.~(\ref{16.6.4}) as an expansion of the acceleration in powers of
$q^2$, and it is therefore appropriate --- indeed imperative --- to
insert the zeroth-order expression for $\dot{a}^\nu$ within the term
of order $q^2$. The resulting expression for the acceleration is then
valid up to correction terms of order $q^4$. Omitting these error
terms, we shall write, in final analysis, the equations of motion in
the form   
\begin{equation} 
m \frac{D u^\mu}{d\tau} = f_{\rm ext}^\mu 
+ q^2 \bigl( \delta^\mu_{\ \nu} + u^\mu u_\nu \bigr) 
\Biggl[ \frac{1}{3m} \frac{D f_{\rm ext}^\nu}{d \tau}   
+ \frac{1}{6} R^{\nu}_{\ \lambda} u^{\lambda} 
+ \int_{-\infty}^{\tau^-}   
\nabla^{\nu} G_+\bigl(z(\tau),z(\tau')\bigr)\, d\tau' \Biggr] 
\label{16.6.8} 
\end{equation}   
and 
\begin{equation}
\frac{d m}{d\tau} = - \frac{1}{12} (1-6\xi) q^2 R 
- q^2 u^\mu \int_{-\infty}^{\tau^-} \nabla_{\mu}
G_+\bigl(z(\tau),z(\tau')\bigr)\, d\tau',  
\label{16.6.9}
\end{equation} 
where $m$ denotes the observed inertial mass of the scalar charge, and
where all tensors are evaluated at $z(\tau)$. We recall that the tail
integration must be cut short at $\tau' = \tau^- := \tau - 0^+$ to
avoid the singular behaviour of the retarded Green's function at
coincidence; this procedure was justified at the beginning of
Sec.~\ref{16.3}. Equations (\ref{16.6.8}) and (\ref{16.6.9}) were
first derived by Theodore C.\ Quinn in 2000 \cite{quinn:00}. In his
paper Quinn also establishes that the total work done by the scalar
self-force matches the amount of energy radiated away by the particle.      

\section{Motion of an electric charge} 
\label{17} 

\subsection{Dynamics of a point electric charge}   
\label{17.1} 

A point particle carries an electric charge $e$ and moves on a world
line $\gamma$ described by relations $z^\mu(\lambda)$, in which
$\lambda$ is an arbitrary parameter. The particle generates a vector
potential $A^\alpha(x)$ and an electromagnetic field
$F_{\alpha\beta}(x) = \nabla_\alpha A_\beta - \nabla_\beta
A_\alpha$. The dynamics of the entire system is governed by the action  
\begin{equation}
S = S_{\rm field} + S_{\rm particle} + S_{\rm interaction}, 
\label{17.1.1}
\end{equation} 
where $S_{\rm field}$ is an action functional for a free
electromagnetic field in a spacetime with metric $g_{\alpha\beta}$,
$S_{\rm particle}$ is the action of a free particle moving on a world
line $\gamma$ in this spacetime, and $S_{\rm interaction}$ is an
interaction term that couples the field to the particle. 

The field action is given by  
\begin{equation}
S_{\rm field} = -\frac{1}{16\pi} \int F_{\alpha\beta} F^{\alpha\beta} 
\sqrt{-g}\, d^4 x,  
\label{17.1.2}
\end{equation} 
where the integration is over all of spacetime. The particle action is    
\begin{equation} 
S_{\rm particle} = -m \int_\gamma d\tau, 
\label{17.1.3}
\end{equation}
where $m$ is the bare mass of the particle and $d\tau =
\sqrt{-g_{\mu\nu}(z) \dot{z}^\mu \dot{z}^\nu}\, d\lambda$ is the
differential of proper time along the world line; we use an overdot to
indicate differentiation with respect to the parameter
$\lambda$. Finally, the interaction term is given by  
\begin{equation}  
S_{\rm interaction} = e \int_\gamma A_\mu(z) \dot{z}^\mu\, d\lambda =
e \int A_\alpha(x) g^\alpha_{\ \mu}(x,z) \dot{z}^\mu \delta_4(x,z)
\sqrt{-g}\, d^4x d\lambda.   
\label{17.1.4}
\end{equation}
Notice that both $S_{\rm particle}$ and $S_{\rm interaction}$ are
invariant under a reparameterization $\lambda \to \lambda'(\lambda)$
of the world line.  

Demanding that the total action be stationary under a variation
$\delta A^\alpha(x)$ of the vector potential yields Maxwell's
equations 
\begin{equation} 
F^{\alpha\beta}_{\ \ \ ;\beta} = 4\pi j^\alpha 
\label{17.1.5}
\end{equation} 
with a current density $j^\alpha(x)$ defined by  
\begin{equation}
j^\alpha(x) = e \int_\gamma g^{\alpha}_{\ \mu}(x,z) \dot{z}^\mu 
\delta_4(x,z)\, d\lambda.  
\label{17.1.6}
\end{equation} 
These equations determine the electromagnetic field $F_{\alpha\beta}$
once the motion of the electric charge is specified. On the other
hand, demanding that the total action be stationary under a variation  
$\delta z^\mu(\lambda)$ of the world line yields the equations of
motion  
\begin{equation} 
m \frac{D u^\mu}{d\tau} = e F^\mu_{\ \nu}(z) u^\nu  
\label{17.1.7}
\end{equation} 
for the electric charge. We have adopted $\tau$ as the parameter on
the world line, and introduced the four-velocity $u^\mu(\tau)
:= dz^\mu/d\tau$. 

The electromagnetic field $F_{\alpha\beta}$ is invariant under a gauge
transformation of the form $A_\alpha \to A_\alpha + \nabla_\alpha
\Lambda$, in which $\Lambda(x)$ is an arbitrary scalar function. This 
function can always be chosen so that the vector potential satisfies
the Lorenz gauge condition,  
\begin{equation}
\nabla_\alpha A^\alpha = 0. 
\label{17.1.8}
\end{equation} 
Under this condition the Maxwell equations of Eq.~(\ref{17.1.5})
reduce to a wave equation for the vector potential, 
\begin{equation} 
\Box A^\alpha - R^\alpha_{\ \beta} A^\beta = -4\pi j^\alpha, 
\label{17.1.9}
\end{equation}
where $\Box = g^{\alpha\beta} \nabla_\alpha \nabla_\beta$ is the wave 
operator and $R^\alpha_{\ \beta}$ is the Ricci tensor. Having adopted
$\tau$ as the parameter on the world line, we can re-express the
current density of Eq.~(\ref{17.1.6}) as 
\begin{equation}
j^\alpha(x) = e \int_\gamma g^{\alpha}_{\ \mu}(x,z) u^\mu 
\delta_4(x,z)\, d\tau, 
\label{17.1.10}
\end{equation} 
and we shall use Eqs.~(\ref{17.1.9}) and (\ref{17.1.10}) to determine 
the electromagnetic field of a point electric charge. The motion of
the particle is in principle determined by Eq.~(\ref{17.1.7}), but 
because the vector potential obtained from Eq.~(\ref{17.1.9}) is
singular on the world line, these equations have only formal validity.   
Before we can make sense of them we will have to analyze the field's
singularity structure near the world line. The calculations to be
carried out parallel closely those presented in Sec.~\ref{16} for the
case of a scalar charge; the details will therefore be kept to a 
minimum and the reader is referred to Sec.~\ref{16} for
additional information.     

\subsection{Retarded potential near the world line} 
\label{17.2}

The retarded solution to Eq.~(\ref{17.1.9}) is $A^\alpha(x) = \int 
G^{\ \alpha}_{+\beta'}(x,x') j^{\beta'}(x') \sqrt{g'}\, d^4x'$, where
$G^{\ \alpha}_{+\beta'}(x,x')$ is the retarded Green's function
introduced in Sec.~\ref{14}. After substitution of Eq.~(\ref{17.1.10})
we obtain 
\begin{equation}
A^\alpha(x) = e \int_\gamma G^{\ \alpha}_{+\mu}(x,z) u^\mu\, d\tau, 
\label{17.2.1}
\end{equation}
in which $z^\mu(\tau)$ gives the description of the world line
$\gamma$ and $u^\mu(\tau) = dz^\mu/d\tau$. Because the retarded
Green's function is defined globally in the entire spacetime,
Eq.~(\ref{17.2.1}) applies to any field point $x$. 

We now specialize Eq.~(\ref{17.2.1}) to a point $x$ close to the world 
line. We let ${\cal N}(x)$ be the normal convex neighbourhood
of this point, and we assume that the world line traverses 
${\cal N}(x)$; refer back to Fig.~9. As in Sec.~\ref{16.2} we let
$\tau_<$ and $\tau_>$ be the values of the proper-time parameter at
which $\gamma$ enters and leaves ${\cal N}(x)$, respectively. Then
Eq.~(\ref{17.2.1}) can be expressed as 
\[
A^\alpha(x) = e \int_{-\infty}^{\tau_<} G^{\ \alpha}_{+\mu}(x,z)
u^\mu\, d\tau + e \int_{\tau_<}^{\tau_>} G^{\ \alpha}_{+\mu}(x,z)
u^\mu\, d\tau + e \int_{\tau_>}^{\infty} G^{\ \alpha}_{+\mu}(x,z)
u^\mu\, d\tau.  
\]
The third integration vanishes because $x$ is then in the past of
$z(\tau)$, and $G^{\ \alpha}_{+\mu}(x,z) = 0$. For the second
integration, $x$ is the normal convex neighbourhood of $z(\tau)$, and
the retarded Green's function can be expressed in the Hadamard form
produced in Sec.~\ref{14.2}. This gives   
\[ 
\int_{\tau_<}^{\tau_>} G^{\ \alpha}_{+\mu}(x,z) u^\mu\, d\tau 
= \int_{\tau_<}^{\tau_>} U^\alpha_{\ \mu}(x,z) u^\mu
\delta_+(\sigma)\, d\tau 
+ \int_{\tau_<}^{\tau_>} V^\alpha_{\ \mu}(x,z) u^\mu 
\theta_+(-\sigma)\, d\tau, 
\] 
and to evaluate this we let $x' := z(u)$ be the retarded point
associated with $x$; these points are related by $\sigma(x,x') = 0$
and $r := \sigma_{\alpha'} u^{\alpha'}$ is the retarded distance
between $x$ and the world line. To perform the first integration we
change variables from $\tau$ to $\sigma$, noticing that $\sigma$
increases as $z(\tau)$ passes through $x'$; the integral evaluates to  
$U^\alpha_{\ \beta'} u^{\beta'}/r$. The second integration is cut off
at $\tau = u$ by the step function, and we obtain our final expression
for the vector potential of a point electric charge:  
\begin{equation}
A^\alpha(x) = \frac{e}{r} U^\alpha_{\ \beta'}(x,x') u^{\beta'} 
+ e \int_{\tau_<}^u V^\alpha_{\ \mu}(x,z) u^\mu\, d\tau 
+ e \int_{-\infty}^{\tau_<} G^{\ \alpha}_{+\mu}(x,z) u^\mu\, d\tau.    
\label{17.2.2}
\end{equation} 
This expression applies to a point $x$ sufficiently close to the world 
line that there exists a nonempty intersection between 
${\cal N}(x)$ and $\gamma$.  

\subsection{Electromagnetic field in retarded coordinates}  
\label{17.3} 

When we differentiate the vector potential of Eq.~(\ref{17.2.2}) we
must keep in mind that a variation in $x$ induces a variation in $x'$,
because the new points $x + \delta x$ and
$x' + \delta x'$ must also be linked by a null geodesic. Taking this
into account, we find that the gradient of the vector potential is
given by     
\begin{equation}
\nabla_\beta A_\alpha(x) = 
-\frac{e}{r^2} U_{\alpha \beta'} u^{\beta'} \partial_\beta r  
+ \frac{e}{r} U_{\alpha \beta';\beta} u^{\beta'} 
+ \frac{e}{r} \Bigl( U_{\alpha \beta';\gamma'} u^{\beta'}
   u^{\gamma'} 
+ U_{\alpha \beta'} a^{\beta'} \Bigr) \partial_\beta u   
+ e V_{\alpha \beta'} u^{\beta'} \partial_\beta u 
+ A^{\rm tail}_{\alpha\beta}(x),  
\label{17.3.1}
\end{equation}
where the ``tail integral'' is defined by    
\begin{eqnarray}
A^{\rm tail}_{\alpha\beta}(x) &=& e \int_{\tau_<}^u 
\nabla_\beta V_{\alpha\mu}(x,z) u^\mu\, d\tau 
+ e \int_{-\infty}^{\tau_<} 
\nabla_\beta G_{+\alpha\mu}(x,z) u^\mu\, d\tau \nonumber \\    
&=& e \int_{-\infty}^{u^-} 
\nabla_\beta G_{+\alpha\mu}(x,z) u^\mu\, d\tau.  
\label{17.3.2}
\end{eqnarray} 
The second form of the definition, in which we integrate the gradient
of the retarded Green's function from $\tau = -\infty$ to $\tau = u^-
:= u - 0^+$ to avoid the singular behaviour of the retarded
Green's function at $\sigma = 0$, is equivalent to the first form.      

We shall now expand $F_{\alpha\beta} = \nabla_\alpha A_\beta 
- \nabla_\beta A_\alpha$ in powers of $r$, and express the result in
terms of the retarded coordinates $(u,r,\Omega^a)$ introduced in
Sec.~\ref{9}. It will be convenient to decompose the electromagnetic
field in the tetrad $(\base{\alpha}{0},\base{\alpha}{a})$ that is
obtained by parallel transport of $(u^{\alpha'},\base{\alpha'}{a})$ on
the null geodesic that links $x$ to $x' := z(u)$; this
construction is detailed in Sec.~\ref{9}. We recall from
Eq.~(\ref{9.1.4}) that the parallel propagator can be expressed as  
$g^{\alpha'}_{\ \alpha} = u^{\alpha'} \base{0}{\alpha} 
+ \base{\alpha'}{a} \base{a}{\alpha}$. The expansion relies on
Eq.~(\ref{9.5.3}) for $\partial_\alpha u$, Eq.~(\ref{9.5.5}) for
$\partial_\alpha r$, and we shall need    
\begin{equation}
U_{\alpha \beta'} u^{\beta'} = g^{\alpha'}_{\ \alpha} \biggl[
u_{\alpha'} + \frac{1}{12} r^2 \bigl( R_{00} + 2 R_{0a} \Omega^a 
+ R_{ab} \Omega^a \Omega^b \bigr)u_{\alpha'} + O(r^3) \biggr],  
\label{17.3.3}
\end{equation} 
which follows from Eq.~(\ref{14.2.7}) and the relation 
$\sigma^{\alpha'} = -r (u^{\alpha'} + \Omega^a \base{\alpha'}{a})$ 
first encountered in Eq.~(\ref{9.2.3}). We shall also need the
expansions     
\begin{equation} 
U_{\alpha \beta';\beta} u^{\beta'} = -\frac{1}{2} r  
g^{\alpha'}_{\ \alpha} g^{\beta'}_{\ \beta} \biggl[ 
R_{\alpha'0\beta'0} + R_{\alpha' 0\beta'c} \Omega^c 
- \frac{1}{3} \bigl( R_{\beta'0} + R_{\beta' c} \Omega^c 
\bigr) u_{\alpha'} + O(r) \biggr]    
\label{17.3.4}  
\end{equation}
and 
\begin{equation} 
U_{\alpha\beta';\gamma'} u^{\beta'}u^{\gamma'} +
U_{\alpha\beta'} a^{\beta'} = g^{\alpha'}_{\ \alpha} \biggl[ 
a_{\alpha'} + \frac{1}{2} r R_{\alpha'0b0}\Omega^b 
-\frac{1}{6} r \bigl( R_{00} + R_{0b} \Omega^b \bigr) u_{\alpha'} 
+ O(r^2) \biggr]   
\label{17.3.5}
\end{equation}
that follow from Eqs.~(\ref{14.2.7})--(\ref{14.2.9}). And finally, 
we shall need    
\begin{equation}
V_{\alpha\beta'} u^{\beta'} = 
-\frac{1}{2} g^{\alpha'}_{\ \alpha} \biggl[
R_{\alpha'0} - \frac{1}{6} R u_{\alpha'} + O(r) \biggr], 
\label{17.3.6}
\end{equation} 
a relation that was first established in Eq.~(\ref{14.2.11}).  

Collecting all these results gives    
\begin{eqnarray} 
F_{a0}(u,r,\Omega^a) &:=& F_{\alpha\beta}(x) 
\base{\alpha}{a}(x) \base{\beta}{0}(x) 
\nonumber \\  
&=& \frac{e}{r^2} \Omega_a 
- \frac{e}{r} \bigl( a_a - a_b \Omega^b \Omega_a \bigr) 
+ \frac{1}{3} e R_{b0c0} \Omega^b \Omega^c \Omega_a 
- \frac{1}{6} e \bigl( 5R_{a0b0} \Omega^b + R_{ab0c} \Omega^b \Omega^c
\bigr) 
\nonumber \\ & & \mbox{}
+ \frac{1}{12} e \bigl( 5 R_{00} + R_{bc} \Omega^b\Omega^c + R \bigr)
\Omega_a 
+ \frac{1}{3} e R_{a0} - \frac{1}{6} e R_{ab} \Omega^b 
+ F_{a0}^{\rm tail} + O(r), 
\label{17.3.7} \\ 
F_{ab}(u,r,\Omega^a) &:=& F_{\alpha\beta}(x) 
\base{\alpha}{a}(x) \base{\beta}{b}(x) 
\nonumber \\  
&=& \frac{e}{r} \bigl( a_a \Omega_b - \Omega_a a_b \bigr) 
+ \frac{1}{2} e \bigl( R_{a0bc} - R_{b0ac} + R_{a0c0} \Omega_b 
- \Omega_a R_{b0c0} \bigr) \Omega^c 
\nonumber \\ & & \mbox{}
- \frac{1}{2} e \bigl( R_{a0} \Omega_b - \Omega_a R_{b0} \bigr)
+ F_{ab}^{\rm tail} + O(r),  
\label{17.3.8}
\end{eqnarray} 
where 
\begin{equation} 
F_{a0}^{\rm tail} = F_{\alpha'\beta'}^{\rm tail}(x') \base{\alpha'}{a}
u^{\beta'}, \qquad 
F_{ab}^{\rm tail} = F_{\alpha'\beta'}^{\rm tail}(x') \base{\alpha'}{a}
\base{\beta'}{b} 
\label{17.3.9}
\end{equation}
are the frame components of the tail integral; this is
obtained from Eq.~(\ref{17.3.2}) evaluated at $x'$:   
\begin{equation}
F_{\alpha'\beta'}^{\rm tail}(x') = 2 e \int_{-\infty}^{u^-}
\nabla_{[\alpha'} G_{+\beta']\mu}(x',z) u^\mu\, d\tau.  
\label{17.3.10}
\end{equation} 
It should be emphasized that in Eqs.~(\ref{17.3.7}) and
(\ref{17.3.8}), all frame components are evaluated at the retarded
point $x' := z(u)$ associated with $x$; for example, $a_a :=
a_a(u) := a_{\alpha'} \base{\alpha'}{a}$. It is clear from these
equations that the electromagnetic field $F_{\alpha\beta}(x)$ is
singular on the world line. 

\subsection{Electromagnetic field in Fermi normal coordinates}   
\label{17.4} 

We now wish to express the electromagnetic field in the Fermi normal
coordinates of Sec.~\ref{8}; as before those will be denoted
$(t,s,\omega^a)$. The translation will be carried out as in
Sec.~\ref{16.4}, and we will decompose the field in the tetrad 
$(\bar{e}^\alpha_0, \bar{e}^\alpha_a)$ that is obtained by parallel
transport of $(u^{\bar{\alpha}}, \base{\bar{\alpha}}{a})$ on the
spacelike geodesic that links $x$ to the simultaneous point
$\bar{x} := z(t)$.    

Our first task is to decompose $F_{\alpha\beta}(x)$ in the tetrad 
$(\bar{e}^\alpha_0, \bar{e}^\alpha_a)$, thereby defining $\bar{F}_{a0}  
:= F_{\alpha\beta} \bar{e}^\alpha_a \bar{e}^\beta_0$ and
$\bar{F}_{ab} := F_{\alpha\beta} \bar{e}^\alpha_a
\bar{e}^\beta_b$. For this purpose we use Eqs.~(\ref{10.3.1}),
(\ref{10.3.2}), (\ref{17.3.7}), and (\ref{17.3.8}) to obtain  
\begin{eqnarray*} 
\bar{F}_{a0} &=& 
\frac{e}{r^2} \Omega_a 
- \frac{e}{r} \bigl( a_a - a_b \Omega^b \Omega_a \bigr) 
+ \frac{1}{2} e a_b \Omega^b a_a 
+ \frac{1}{2} e \dot{a}_0 \Omega_a 
- \frac{5}{6} e R_{a0b0} \Omega^b 
+ \frac{1}{3} e R_{b0c0} \Omega^b \Omega^c \Omega_a 
\\ & & \mbox{}
+ \frac{1}{3} e R_{ab0c} \Omega^b \Omega^c 
+ \frac{1}{12} e \bigl( 5 R_{00} + R_{bc} \Omega^b\Omega^c + R \bigr)
\Omega_a 
+ \frac{1}{3} e R_{a0} - \frac{1}{6} e R_{ab} \Omega^b 
+ \bar{F}_{a0}^{\rm tail} + O(r) 
\end{eqnarray*} 
and 
\[
\bar{F}_{ab} = 
\frac{1}{2} e \bigl( \Omega_a \dot{a}_b - \dot{a}_a \Omega_b \bigr) 
+ \frac{1}{2} e \bigl( R_{a0bc} - R_{b0ac} \bigr) \Omega^c 
- \frac{1}{2} e \bigl( R_{a0} \Omega_b - \Omega_a R_{b0} \bigr)
+ \bar{F}_{ab}^{\rm tail} + O(r), 
\]
where all frame components are still evaluated at $x'$, except for 
\[
\bar{F}_{a0}^{\rm tail} := 
F^{\rm tail}_{\bar{\alpha}\bar{\beta}}(\bar{x})
\base{\bar{\alpha}}{a} u^{\bar{\beta}}, \qquad 
\bar{F}_{ab}^{\rm tail} := 
F^{\rm tail}_{\bar{\alpha}\bar{\beta}}(\bar{x})
\base{\bar{\alpha}}{a} \base{\bar{\beta}}{b}, 
\] 
which are evaluated at $\bar{x}$.   

We must still translate these results into the Fermi normal
coordinates $(t,s,\omega^a)$. For this we involve Eqs.~(\ref{10.2.1}),
(\ref{10.2.2}), and (\ref{10.2.3}), and we recycle some computations
that were first carried out in Sec.~\ref{16.4}. After some algebra, we
arrive at       
\begin{eqnarray} 
\bar{F}_{a0}(t,s,\omega^a) &:=& F_{\alpha\beta}(x) 
\bar{e}^\alpha_a(x) \bar{e}^\beta_0(x) 
\nonumber \\ 
&=& \frac{e}{s^2} \omega_a 
- \frac{e}{2s} \bigl( a_a + a_b \omega^b \omega_a \bigr) 
+ \frac{3}{4} e a_b \omega^b a_a 
+ \frac{3}{8} e \bigl( a_b \omega^b \bigr)^2 \omega_a 
+ \frac{3}{8} e \dot{a}_0 \omega_a
+ \frac{2}{3} e \dot{a}_a  
\nonumber \\ & & \mbox{}
- \frac{2}{3} e R_{a0b0} \omega^b 
- \frac{1}{6} e R_{b0c0} \omega^b \omega^c \omega_a 
+ \frac{1}{12} e \bigl( 5 R_{00} + R_{bc} \omega^b\omega^c + R \bigr) 
\omega_a 
\nonumber \\ & & \mbox{}
+ \frac{1}{3} e R_{a0} - \frac{1}{6} e R_{ab} \omega^b 
+ \bar{F}_{a0}^{\rm tail} + O(s), 
\label{17.4.1} \\ 
\bar{F}_{ab}(t,s,\omega^a) &:=& F_{\alpha\beta}(x) 
\bar{e}^\alpha_a(x) \bar{e}^\beta_b(x) 
\nonumber \\
&=& \frac{1}{2} e \bigl( \omega_a \dot{a}_b - \dot{a}_a \omega_b
\bigr)   
+ \frac{1}{2} e \bigl( R_{a0bc} - R_{b0ac} \bigr) \omega^c 
- \frac{1}{2} e \bigl( R_{a0} \omega_b - \omega_a R_{b0} \bigr)
\nonumber \\ & & \mbox{}
+ \bar{F}_{ab}^{\rm tail} + O(s),
\label{17.4.2}
\end{eqnarray} 
where all frame components are now evaluated at $\bar{x} := z(t)$;
for example, $a_a := a_a(t) :=
a_{\bar{\alpha}} \base{\bar{\alpha}}{a}$. 

Our next task is to compute the averages of $\bar{F}_{a0}$ and
$\bar{F}_{ab}$ over $S(t,s)$, a two-surface of constant $t$ and $s$.  
These are defined by 
\begin{equation} 
\bigl\langle \bar{F}_{a0} \bigr\rangle(t,s) = \frac{1}{\cal A}  
\oint_{S(t,s)} \bar{F}_{a0}(t,s,\omega^a)\, d{\cal A}, \qquad   
\bigl\langle \bar{F}_{ab} \bigr\rangle(t,s) = \frac{1}{\cal A} 
\oint_{S(t,s)} \bar{F}_{ab}(t,s,\omega^a)\, d{\cal A},  
\label{17.4.3}
\end{equation} 
where $d{\cal A}$ is the element of surface area on $S(t,s)$, and 
${\cal A} = \oint d {\cal A}$. Using the methods developed in
Sec.~\ref{16.4}, we find 
\begin{eqnarray} 
\bigl\langle \bar{F}_{a0} \bigr\rangle &=& -\frac{2e}{3s} a_a 
+ \frac{2}{3} e \dot{a}_a + \frac{1}{3} e R_{a0} 
+ \bar{F}_{a0}^{\rm tail} + O(s), 
\label{17.4.4} \\ 
\bigl\langle \bar{F}_{ab} \bigr\rangle &=& 
\bar{F}_{ab}^{\rm tail} + O(s). 
\label{17.4.5}
\end{eqnarray} 
The averaged field is singular on the world line, but we nevertheless
take the formal limit $s \to 0$ of the expressions displayed in 
Eqs.~(\ref{17.4.4}) and (\ref{17.4.5}). In the limit the tetrad 
$(\bar{e}^\alpha_0, \bar{e}^\alpha_a)$ becomes 
$(u^{\bar{\alpha}}, \base{\bar{\alpha}}{a})$, and we can easily 
reconstruct the field at $\bar{x}$ from its frame components. 
We thus obtain  
\begin{equation} 
\bigl\langle F_{\bar{\alpha}\bar{\beta}} \bigr\rangle = 
\lim_{s\to 0} \biggl( - \frac{4e}{3s} \biggr) 
u_{[\bar{\alpha}} a_{\bar{\beta}]}   
+ 2e u_{[\bar{\alpha}} \bigl( g_{\bar{\beta}]\bar{\gamma}}  
+ u_{\bar{\beta}]} u_{\bar{\gamma}} \bigr) 
\biggl( \frac{2}{3} \dot{a}^{\bar{\gamma}} 
+ \frac{1}{3} R^{\bar{\gamma}}_{\ \bar{\delta}} u^{\bar{\delta}}
\biggr) + F_{\bar{\alpha}\bar{\beta}}^{\rm tail}, 
\label{17.4.6}
\end{equation}
where the tail term can be copied from Eq.~(\ref{17.3.10}), 
\begin{equation} 
F_{\bar{\alpha}\bar{\beta}}^{\rm tail}(\bar{x}) = 
2 e \int_{-\infty}^{t^-}
\nabla_{[\bar{\alpha}} G_{+\bar{\beta}]\mu}(\bar{x},z) u^\mu\, d\tau. 
\label{17.4.7}
\end{equation} 
The tensors appearing in Eq.~(\ref{17.4.6}) all refer to $\bar{x}
:= z(t)$, which now stands for an    
arbitrary point on the world line $\gamma$.  

\subsection{Singular and regular fields} 
\label{17.5} 

The singular vector potential 
\begin{equation} 
A^\alpha_{\rm S}(x) = e \int_\gamma 
G^{\ \alpha}_{{\rm S}\,\mu}(x,z) u^\mu\, d\tau 
\label{17.5.1}
\end{equation} 
is the (unphysical) solution to Eqs.~(\ref{17.1.9}) and
(\ref{17.1.10}) that is obtained by adopting the singular Green's
function of Eq.~(\ref{14.4.5}) instead of the retarded Green's
function. We will see that the singular field
$F^{\rm S}_{\alpha\beta}$ reproduces the singular behaviour of the
retarded solution, and that the difference, $F^{\rm R}_{\alpha\beta} 
= F_{\alpha\beta} - F^{\rm S}_{\alpha\beta}$, is regular on the world
line.      

To evaluate the integral of Eq.~(\ref{17.5.1}) we assume once more 
that $x$ is sufficiently close to $\gamma$ that the world line
traverses ${\cal N}(x)$; refer back to Fig.~9. As before we let
$\tau_<$ and $\tau_>$ be the values of the proper-time parameter at
which $\gamma$ enters and leaves ${\cal N}(x)$, respectively. Then
Eq.~(\ref{17.5.1}) becomes   
\[
A^\alpha_{\rm S}(x) = e \int_{-\infty}^{\tau_<} 
G^{\ \alpha}_{{\rm S}\,\mu}(x,z) u^\mu \, d\tau  
+ e \int_{\tau_<}^{\tau_>} G^{\ \alpha}_{{\rm S}\,\mu}(x,z) u^\mu\,
d\tau + e \int_{\tau_>}^{\infty} G^{\ \alpha}_{{\rm S}\,\mu}(x,z)
u^\mu\, d\tau.   
\]
The first integration vanishes because $x$ is then in the
chronological future of $z(\tau)$, and 
$G^{\ \alpha}_{{\rm S}\,\mu}(x,z) = 0$ by
Eq.~(\ref{14.4.8}). Similarly, the third integration vanishes because 
$x$ is then in the chronological past of $z(\tau)$. For the second
integration, $x$ is the normal convex neighbourhood of $z(\tau)$, the
singular Green's function can be expressed in the Hadamard form of
Eq.~(\ref{14.4.14}), and we have      
\begin{eqnarray*} 
\int_{\tau_<}^{\tau_>} G^{\ \alpha}_{{\rm S}\,\mu}(x,z) u^\mu\, 
d\tau &=& \frac{1}{2} \int_{\tau_<}^{\tau_>} U^\alpha_{\ \mu}(x,z)
u^\mu \delta_+(\sigma)\, d\tau + \frac{1}{2} \int_{\tau_<}^{\tau_>} 
U^\alpha_{\ \mu}(x,z) u^\mu \delta_-(\sigma)\, d\tau \\ 
& & \mbox{} 
- \frac{1}{2} \int_{\tau_<}^{\tau_>} V^\alpha_{\ \mu}(x,z) u^\mu 
\theta(\sigma)\, d\tau.   
\end{eqnarray*} 
To evaluate these we let $x' := z(u)$ and $x'' := z(v)$ be the
retarded and advanced points associated with $x$, respectively. To
perform the first integration we change variables from $\tau$ to
$\sigma$, noticing that $\sigma$ increases as $z(\tau)$ passes
through $x'$; the integral evaluates to $U^\alpha_{\ \beta'}
u^{\beta'}/r$. We do the same for the second integration, but we
notice now that $\sigma$ decreases as $z(\tau)$ passes through $x''$;
the integral evaluates to 
$U^\alpha_{\ \beta''} u^{\beta''}/r_{\rm adv}$, where $r_{\rm adv}
:= - \sigma_{\alpha''} u^{\alpha''}$ is the advanced distance
between $x$ and the world line. The third integration is restricted to
the interval $u \leq \tau \leq v$ by the step function, and we obtain
the expression
\begin{equation}
A^\alpha_{\rm S}(x) = \frac{e}{2r} U^\alpha_{\ \beta'} u^{\beta'}  
+ \frac{e}{2r_{\rm adv}} U^\alpha_{\ \beta''} u^{\beta''} 
- \frac{1}{2} e \int_u^v V^\alpha_{\ \mu}(x,z) u^\mu\, d\tau   
\label{17.5.2} 
\end{equation} 
for the singular vector potential. 

Differentiation of Eq.~(\ref{17.5.2}) yields 
\begin{eqnarray}
\nabla_\beta A_\alpha^{\rm S}(x) &=& 
-\frac{e}{2r^2} U_{\alpha\beta'} u^{\beta'} \partial_\beta r   
- \frac{e}{2 {r_{\rm adv}}^2} U_{\alpha\beta''} u^{\beta''}
  \partial_\beta r_{\rm adv}  
+ \frac{e}{2r} U_{\alpha\beta';\beta} u^{\beta'}  
\nonumber \\  & & \mbox{} 
+ \frac{e}{2r} \Bigl( U_{\alpha \beta';\gamma'} u^{\beta'}
   u^{\gamma'} 
+ U_{\alpha\beta'} a^{\beta'} \Bigr) \partial_\beta u   
+ \frac{e}{2r_{\rm adv}} U_{\alpha\beta'';\beta} u^{\beta''}   
\nonumber \\  & & \mbox{} 
+ \frac{e}{2r_{\rm adv}} \Bigl( U_{\alpha \beta'';\gamma''}
   u^{\beta''} u^{\gamma''} 
+ U_{\alpha\beta''} a^{\beta''} \Bigr) \partial_\beta v   
+ \frac{1}{2} e V_{\alpha \beta'} u^{\beta'} \partial_\beta u   
\nonumber \\  & & \mbox{} 
- \frac{1}{2} e V_{\alpha \beta''} u^{\beta''} \partial_\beta v    
- \frac{1}{2} e \int_u^v \nabla_\beta V_{\alpha\mu}(x,z) u^\mu\,
d\tau, 
\label{17.5.3}
\end{eqnarray}
and we would like to express this as an expansion in powers of
$r$. For this we will rely on results already established in 
Sec.~\ref{17.3}, as well as additional expansions that will involve
the advanced point $x''$. We recall that a relation between retarded
and advanced times was worked out in Eq.~(\ref{10.4.2}), that an
expression for the advanced distance was displayed in
Eq.~(\ref{10.4.3}), and that Eqs.~(\ref{10.4.4}) and (\ref{10.4.5})
give expansions for $\partial_\alpha v$ and 
$\partial_\alpha r_{\rm adv}$, respectively.  

To derive an expansion for $U_{\alpha\beta''} u^{\beta''}$ we follow
the general method of Sec.~\ref{10.4} and introduce the functions
$U_\alpha(\tau) := U_{\alpha\mu}(x,z) u^\mu$. We have 
that  
\[
U_{\alpha\beta''} u^{\beta''} := U_\alpha(v) = U_\alpha(u) 
+ \dot{U}_\alpha(u) \Delta^{\!\prime} + \frac{1}{2} \ddot{U}_\alpha(u)
\Delta^{\!\prime 2} + O\bigl( \Delta^{\!\prime 3} \bigr),  
\]
where overdots indicate differentiation with respect to $\tau$, and
$\Delta^{\!\prime} := v-u$. The leading term $U_\alpha(u)
:= U_{\alpha\beta'} u^{\beta'}$ was worked out in
Eq.~(\ref{17.3.3}), and the derivatives of $U_\alpha(\tau)$ are given
by  
\[
\dot{U}_\alpha(u) = U_{\alpha\beta';\gamma'} u^{\beta'} u^{\gamma'} +
U_{\alpha\beta'} a^{\beta'} = g^{\alpha'}_{\ \alpha} \biggl[ 
a_{\alpha'} + \frac{1}{2} r R_{\alpha'0b0}\Omega^b 
-\frac{1}{6} r \bigl( R_{00} + R_{0b} \Omega^b \bigr) u_{\alpha'} 
+ O(r^2) \biggr]   
\] 
and 
\[
\ddot{U}_\alpha(u) = U_{\alpha\beta';\gamma'\delta'} u^{\beta'}
u^{\gamma'} u^{\delta'} + U_{\alpha\beta';\gamma'} \bigl( 2 a^{\beta'}
u^{\gamma'} + u^{\beta'} a^{\gamma'} \bigr) + U_{\alpha\beta'}
\dot{a}^{\beta'} = g^{\alpha'}_{\ \alpha} \biggl[ \dot{a}_{\alpha'} 
+ \frac{1}{6} R_{00} u_{\alpha'} + O(r) \biggr], 
\]
according to Eqs.~(\ref{17.3.5}) and (\ref{14.2.9}). Combining these
results together with Eq.~(\ref{10.4.2}) for $\Delta^{\!\prime}$ gives    
\begin{eqnarray} 
U_{\alpha\beta''} u^{\beta''} &=& g^{\alpha'}_{\ \alpha} \biggl[ 
u_{\alpha'} + 2r \bigl(1 - r a_b \Omega^b \bigr) a_{\alpha'} 
+ 2r^2 \dot{a}_{\alpha'} + r^2 R_{\alpha'0b0}\Omega^b \qquad \qquad
\nonumber \\ & & \qquad \mbox{} 
+ \frac{1}{12} r^2 \bigl( R_{00} - 2 R_{0a} \Omega^a 
+ R_{ab} \Omega^a \Omega^b \bigr)u_{\alpha'} 
+ O(r^3) \biggr],  
\label{17.5.4}
\end{eqnarray} 
which should be compared with Eq.~(\ref{17.3.3}). It should be
emphasized that in Eq.~(\ref{17.5.4}) and all equations below, all
frame components are evaluated at the retarded point $x'$, and not at
the advanced point. The preceding computation gives us also
an expansion for  
\[
U_{\alpha\beta'';\gamma''} u^{\beta''} u^{\gamma''} +
U_{\alpha\beta''} a^{\beta''} := \dot{U}_\alpha(v) =
\dot{U}_\alpha(u) + \ddot{U}_\alpha(u) \Delta^{\!\prime} +
O(\Delta^{\!\prime 2}), 
\]
which becomes     
\begin{equation}
U_{\alpha\beta'';\gamma''} u^{\beta''} u^{\gamma''} +
U_{\alpha\beta''} a^{\beta''} = g^{\alpha'}_{\ \alpha} \biggl[  
a_{\alpha'} + 2 r \dot{a}_{\alpha'} 
+ \frac{1}{2} r R_{\alpha'0b0}\Omega^b  
+ \frac{1}{6} r \bigl( R_{00} - R_{0b} \Omega^b \bigr) u_{\alpha'} 
+ O(r^2) \biggr],   
\label{17.5.5}
\end{equation}
and which should be compared with Eq.~(\ref{17.3.5}).   

We proceed similarly to derive an expansion for
$U_{\alpha\beta'';\beta} u^{\beta''}$. Here we introduce the functions 
$U_{\alpha\beta}(\tau) := U_{\alpha\mu;\beta}u^\mu$ and express
$U_{\alpha\beta'';\beta} u^{\beta''}$ as $U_{\alpha\beta}(v) =
U_{\alpha\beta}(u) + \dot{U}_{\alpha\beta}(u) \Delta^{\!\prime} 
+ O(\Delta^{\!\prime 2})$. The leading term $U_{\alpha\beta}(u) :=
U_{\alpha\beta';\beta}u^{\beta'}$ was computed in Eq.~(\ref{17.3.4}),
and   
\[
\dot{U}_{\alpha\beta}(u) = U_{\alpha\beta';\beta\gamma'} u^{\beta'}
u^{\gamma'} + U_{\alpha\beta';\beta} a^{\beta'} = \frac{1}{2}
g^{\alpha'}_{\ \alpha} g^{\beta'}_{\ \beta} \biggl[
R_{\alpha'0\beta'0} - \frac{1}{3} u_{\alpha'} R_{\beta'0} + O(r)
\biggr] 
\]
follows from Eq.~(\ref{14.2.8}). Combining these results together with 
Eq.~(\ref{10.4.2}) for $\Delta^{\!\prime}$ gives
\begin{equation}
U_{\alpha \beta'';\beta} u^{\beta''} = \frac{1}{2} r  
g^{\alpha'}_{\ \alpha} g^{\beta'}_{\ \beta} \biggl[ 
R_{\alpha'0\beta'0} - R_{\alpha'0\beta'c} \Omega^c 
- \frac{1}{3} \bigl( R_{\beta'0} - R_{\beta' c} \Omega^c 
\bigr) u_{\alpha'} + O(r) \biggr],     
\label{17.5.6}
\end{equation}
and this should be compared with Eq.~(\ref{17.3.4}). The last
expansion we shall need is  
\begin{equation}
V_{\alpha\beta''} u^{\beta''} = 
-\frac{1}{2} g^{\alpha'}_{\ \alpha} \biggl[
R_{\alpha'0} - \frac{1}{6} R u_{\alpha'} + O(r) \biggr],
\label{17.5.7} 
\end{equation} 
which follows at once from Eq.~(\ref{17.3.6}).

It is now a straightforward (but still tedious) matter to substitute
these expansions into Eq.~(\ref{17.5.3}) to obtain the projections of
the singular electromagnetic field $F^{\rm S}_{\alpha\beta} =
\nabla_\alpha A^{\rm S}_\beta - \nabla_\beta A^{\rm S}_\alpha$ in the  
same tetrad $(\base{\alpha}{0}, \base{\alpha}{a})$ that was employed
in Sec.~\ref{17.3}. This gives 
\begin{eqnarray} 
F^{\rm S}_{a0}(u,r,\Omega^a) &:=& F^{\rm S}_{\alpha\beta}(x)  
\base{\alpha}{a}(x) \base{\beta}{0}(x) 
\nonumber \\  
&=& \frac{e}{r^2} \Omega_a 
- \frac{e}{r} \bigl( a_a - a_b \Omega^b \Omega_a \bigr) 
- \frac{2}{3} e \dot{a}_a 
+ \frac{1}{3} e R_{b0c0} \Omega^b \Omega^c \Omega_a 
- \frac{1}{6} e \bigl( 5R_{a0b0} \Omega^b + R_{ab0c} \Omega^b \Omega^c
\bigr) 
\nonumber \\ & & \mbox{}
+ \frac{1}{12} e \bigl( 5 R_{00} + R_{bc} \Omega^b\Omega^c + R \bigr)
\Omega_a 
- \frac{1}{6} e R_{ab} \Omega^b 
+ O(r), 
\label{17.5.8} \\ 
F^{\rm S}_{ab}(u,r,\Omega^a) &:=& F^{\rm S}_{\alpha\beta}(x)  
\base{\alpha}{a}(x) \base{\beta}{b}(x) 
\nonumber \\  
&=& \frac{e}{r} \bigl( a_a \Omega_b - \Omega_a a_b \bigr) 
+ \frac{1}{2} e \bigl( R_{a0bc} - R_{b0ac} + R_{a0c0} \Omega_b 
- \Omega_a R_{b0c0} \bigr) \Omega^c 
\nonumber \\ & & \mbox{}
- \frac{1}{2} e \bigl( R_{a0} \Omega_b - \Omega_a R_{b0} \bigr)
+ O(r),  
\label{17.5.9}
\end{eqnarray} 
in which all frame components are evaluated at the retarded point 
$x'$. Comparison of these expressions with Eqs.~(\ref{17.3.7}) and
(\ref{17.3.8}) reveals that the retarded and singular fields share the
same singularity structure.    

The difference between the retarded field of Eqs.~(\ref{17.3.7}),
(\ref{17.3.8}) and the singular field of Eqs.~(\ref{17.5.8}),
(\ref{17.5.9}) defines the regular field 
$F^{\rm R}_{\alpha\beta}(x)$. Its tetrad components are 
\begin{eqnarray} 
F^{\rm R}_{a0} &=& \frac{2}{3} e \dot{a}_a + \frac{1}{3} e R_{a0} 
+ F_{a0}^{\rm tail} + O(r),  
\label{17.5.10} \\ 
F^{\rm R}_{ab} &=& F_{ab}^{\rm tail} + O(r), 
\label{17.5.11}
\end{eqnarray} 
and we see that $F^{\rm R}_{\alpha\beta}$ is a regular tensor field
on the world line. There is therefore no obstacle in evaluating the
regular field directly at $x=x'$, where the tetrad
$(\base{\alpha}{0},\base{\alpha}{a})$ becomes $(u^{\alpha'}, 
\base{\alpha'}{a})$. Reconstructing the field at $x'$ from its 
frame components, we obtain 
\begin{equation} 
F^{\rm R}_{\alpha'\beta'}(x') = 
2e u_{[\alpha'} \bigl( g_{\beta']\gamma'}  
+ u_{\beta']} u_{\gamma'} \bigr) 
\biggl( \frac{2}{3} \dot{a}^{\gamma'} 
+ \frac{1}{3} R^{\gamma'}_{\ \delta'} u^{\delta'}
\biggr) + F_{\alpha'\beta'}^{\rm tail}, 
\label{17.5.12}
\end{equation}
where the tail term can be copied from Eq.~(\ref{17.3.10}), 
\begin{equation} 
F_{\alpha'\beta'}^{\rm tail}(x') = 2 e \int_{-\infty}^{u^-}
\nabla_{[\alpha'} G_{+\beta']\mu}(x',z) u^\mu\, d\tau.  
\label{17.5.13} 
\end{equation} 
The tensors appearing in Eq.~(\ref{17.5.12}) all refer to the  
retarded point $x' := z(u)$, which now stands for an  
arbitrary point on the world line $\gamma$.  
  
\subsection{Equations of motion} 
\label{17.6}

The retarded field $F_{\alpha\beta}$ of a point electric charge is 
singular on the world line, and this behaviour makes it difficult to
understand how the field is supposed to act on the particle and exert
a force. The field's singularity structure was analyzed in
Secs.~\ref{17.3} and \ref{17.4}, and in Sec.~\ref{17.5} it was shown 
to originate from the singular field $F^{\rm S}_{\alpha\beta}$; the 
regular field $F^{\rm R}_{\alpha\beta} = F_{\alpha\beta} - 
F^{\rm S}_{\alpha\beta}$ was then shown to regular on the world
line.  

To make sense of the retarded field's action on the particle we follow
the discussion of Sec.~\ref{16.6} and temporarily picture the electric
charge as a spherical hollow shell; the shell's radius is $s_0$ in
Fermi normal coordinates, and it is independent of the angles
contained in the unit vector $\omega^a$. The {\it net force} acting at
proper time $\tau$ on this shell is proportional to the average of 
$F_{\alpha\beta}(\tau,s_0,\omega^a)$ over the shell's surface. 
Assuming that the field on the shell is equal to the field of a
point particle evaluated at $s=s_0$, and ignoring terms that
disappear in the limit $s_0 \to 0$, we obtain from Eq.~(\ref{17.4.6})
\begin{equation} 
e\bigl\langle F_{\mu\nu} \bigr\rangle u^\nu = -(\delta m) a_\mu        
+ e^2 \bigl( g_{\mu\nu} + u_{\mu} u_{\nu} \bigr) 
\biggl( \frac{2}{3} \dot{a}^{\nu} 
+ \frac{1}{3} R^{\nu}_{\ \lambda} u^{\lambda}
\biggr) + eF_{\mu\nu}^{\rm tail} u^\nu, 
\label{17.6.1}
\end{equation}
where 
\begin{equation} 
\delta m := \lim_{s_0 \to 0} \frac{2 e^2}{3 s_0} 
\label{17.6.2}
\end{equation}
is formally a divergent quantity and 
\begin{equation} 
eF_{\mu\nu}^{\rm tail} u^\nu = 2 e^2 u^\nu \int_{-\infty}^{\tau^-}     
\nabla_{[\mu} G_{+\nu]\lambda'}\bigl(z(\tau),z(\tau') \bigr)
u^{\lambda'}\, d\tau' 
\label{17.6.3}
\end{equation}
is the tail part of the force; all tensors in Eq.~(\ref{17.6.1}) are 
evaluated at an arbitrary point $z(\tau)$ on the world line.  

Substituting Eqs.~(\ref{17.6.1}) and (\ref{17.6.3}) into
Eq.~(\ref{17.1.7}) gives rise to the equations of motion  
\begin{equation} 
\bigl( m + \delta m) a^\mu = e^2 \bigl( \delta^\mu_{\ \nu} 
+ u^\mu u_\nu \bigr) \biggl( \frac{2}{3} \dot{a}^{\nu} 
+ \frac{1}{3} R^{\nu}_{\ \lambda} u^{\lambda} \biggr)  
+ 2 e^2 u_\nu \int_{-\infty}^{\tau^-}     
\nabla^{[\mu} G^{\ \nu]}_{+\,\lambda'}\bigl(z(\tau),z(\tau')\bigr)   
u^{\lambda'}\, d\tau' 
\label{17.6.4} 
\end{equation} 
for the electric charge, with $m$ denoting the (also formally
divergent) bare mass of the particle. We see that $m$ and $\delta m$ 
combine in Eq.~(\ref{17.6.4}) to form the particle's observed mass
$m_{\rm obs}$, which is finite and gives a true measure of the
particle's inertia. All diverging quantities have thus disappeared
into the procedure of mass renormalization.  

We must confess, as we did in the case of the scalar self-force, that
the derivation of the equations of motion outlined above returns the 
{\it wrong expression} for the self-energy of a spherical shell of
electric charge. We obtained $\delta m = 2e^2/(3s_0)$, while the
correct expression is $\delta m = e^2/(2s_0)$; we are wrong by a
factor of $4/3$.  As before we believe that this discrepancy
originates in a previously stated assumption, that the field on the
shell (as produced by the shell itself) is equal to the field of a
point particle evaluated at $s=s_0$. We believe that this assumption
is in fact wrong, and that a calculation of the field actually
produced by a spherical shell would return the correct expression for
$\delta m$. We also believe, however, that except for the diverging
terms that determine $\delta m$, the difference between the shell's
field and the particle's field should vanish in the limit $s_0 \to
0$. Our conclusion is therefore that while our expression for 
$\delta m$ is admittedly incorrect, the statement of the equations of
motion is reliable.  

Apart from the term proportional to $\delta m$, the averaged force of 
Eq.~(\ref{17.6.1}) has exactly the same form as the force that arises
from the regular field of Eq.~(\ref{17.5.12}), which we express as 
\begin{equation} 
e F^{\rm R}_{\mu\nu} u^\nu = e^2 \bigl( g_{\mu\nu} 
+ u_{\mu} u_{\nu} \bigr) \biggl( \frac{2}{3} \dot{a}^{\nu} 
+ \frac{1}{3} R^{\nu}_{\ \lambda} u^{\lambda} \biggr) 
+ eF_{\mu\nu}^{\rm tail} u^\nu. 
\label{17.6.5}
\end{equation}
The force acting on the point particle can therefore be thought of as 
originating from the regular field, while the singular
field simply contributes to the particle's inertia. After mass
renormalization, Eq.~(\ref{17.6.4}) is equivalent to the statement  
\begin{equation} 
m a_\mu = e F^{\rm R}_{\mu\nu}(z) u^\nu,
\label{17.6.6}
\end{equation} 
where we have dropped the superfluous label ``obs'' on the 
particle's observed mass. 

For the final expression of the equations of motion we follow the 
discussion of Sec.~\ref{16.6} and allow an external force 
$f^\mu_{\rm ext}$ to act on the particle, and we replace, on the 
right-hand side of the equations, the acceleration vector by
$f^\mu_{\rm ext}/m$. This produces  
\begin{equation} 
m \frac{D u^\mu}{d\tau} = f_{\rm ext}^\mu 
+ e^2 \bigl( \delta^\mu_{\ \nu} + u^\mu u_\nu \bigr) 
\biggl( \frac{2}{3m} \frac{D f_{\rm ext}^\nu}{d \tau}   
+ \frac{1}{3} R^{\nu}_{\ \lambda} u^{\lambda} \biggr)  
+ 2 e^2 u_\nu \int_{-\infty}^{\tau^-}     
\nabla^{[\mu} G^{\ \nu]}_{+\,\lambda'}\bigl(z(\tau),z(\tau')\bigr)   
u^{\lambda'}\, d\tau',  
\label{17.6.7} 
\end{equation}   
in which $m$ denotes the observed inertial mass of the electric charge 
and all tensors are evaluated at $z(\tau)$, the current position of
the particle on the world line; the primed indices in the tail
integral refer to the point $z(\tau')$, which represents a prior
position. We recall that the integration must be cut short at $\tau' =
\tau^- := \tau - 0^+$ to avoid the singular behaviour of the
retarded Green's function at coincidence; this procedure was justified
at the beginning of Sec.~\ref{17.3}. Equation (\ref{17.6.7}) was first
derived (without the Ricci-tensor term) by Bryce S.\ DeWitt and Robert
W.\ Brehme in 1960 \cite{dewitt-brehme:60}, and then corrected by
J.M.\ Hobbs in 1968 \cite{hobbs:68}. An alternative derivation was
produced by Theodore C.\ Quinn and Robert M.\ Wald in 1997
\cite{quinn-wald:97}. In a subsequent publication
\cite{quinn-wald:99}, Quinn and Wald proved that the total work done
by the electromagnetic self-force matches the energy radiated away by
the particle.   

\section{Motion of a point mass} 
\label{18} 

\subsection{Dynamics of a point mass} 
\label{18.1}

\subsubsection*{Introduction} 

In this section we consider the motion of a point particle of mass $m$
subjected to its own gravitational field in addition to an external
field. The particle moves on a world line $\gamma$ in a curved
spacetime whose background metric $g_{\alpha\beta}$ is assumed to be a
{\it vacuum} solution to the Einstein field equations. We shall
suppose that $m$ is small, so that the perturbation $h_{\alpha\beta}$
created by the particle can also be considered to be small. In the
final analysis we shall find that $h_{\alpha\beta}$ obeys a linear
wave equation in the background spacetime, and this linearization of
the field equations will allow us to fit the problem of determining
the motion of a point mass within the general framework developed in
Secs.~\ref{16} and \ref{17}. We shall find that $\gamma$ is not a
geodesic of the background spacetime because $h_{\alpha\beta}$ acts on
the particle and produces an acceleration proportional to $m$; the
motion is geodesic in the test-mass limit only.     

While we can make the problem fit within the general framework, it is
important to understand that the problem of motion in gravitation is
conceptually very different from the versions encountered previously
in the case of a scalar or electromagnetic field. In these cases, the
field equations satisfied by the scalar potential $\Phi$ or the vector
potential $A^\alpha$ are fundamentally linear; in general relativity
the field equations satisfied by $h_{\alpha\beta}$ are fundamentally
nonlinear, and this makes a major impact on the formulation of the
problem. (In all cases the coupled problem of determining the field 
{\it and} the motion of the particle is nonlinear.) 
Another difference resides with the fact that in the previous
cases, the field equations and the law of motion could be formulated
independently of each other (because the action functional could be
varied independently with respect to the field and the world line); in
general relativity the law of motion follows from energy-momentum
conservation, which is itself a consequence of the field equations.  

The dynamics of a point mass in general relativity must therefore be
formulated with care. We shall describe a formal approach to this 
problem, based on the fiction that the spacetime of a
point particle can be constructed exactly in general relativity. (This
is indeed a fiction, because it is known \cite{geroch-traschen:87}
that the metric of a point particle, as described by a Dirac
distribution on a world line, is much too singular to be defined as a
distribution in spacetime. The construction, however, makes
distributional sense at the level of the linearized theory.) The
outcome of this approach will be an approximate formulation of the
equations of motion that relies on a linearization of the field
equations, and which turns out to be closely analogous to the scalar   
and electromagnetic cases encountered previously. We shall put the 
motion of a small mass on a much sounder foundation in
Part~\ref{part5}, where we take $m$ to be a (small) extended body
instead of a point particle.     

\subsubsection*{Exact formulation} 

Let a point particle of mass $m$ move on a world line $\gamma$ in a
curved spacetime with metric ${\sf g}_{\alpha\beta}$. This is the
{\it exact metric} of the {\it perturbed spacetime}, and it depends on 
$m$ as well as all other relevant parameters. At a later stage of the
discussion ${\sf g}_{\alpha\beta}$ will be expressed as sum of a 
``background'' part $g_{\alpha\beta}$ that is independent of $m$, and 
a ``perturbation'' part $h_{\alpha\beta}$ that contains the dependence
on $m$. The world line is described by relations $z^\mu(\lambda)$ in 
which $\lambda$ is an arbitrary parameter --- this will later be
identified with proper time $\tau$ in the {\it background}
spacetime. We use {\sf sans-serif symbols} to denote tensors that
refer to the perturbed spacetime; tensors in the background spacetime
will be denoted, as usual, by italic symbols.         

The particle's action functional is 
\begin{equation} 
S_{\rm particle} = - m \int_\gamma \sqrt{ -{\sf g}_{\mu\nu}
\dot{z}^\mu \dot{z}^\nu }\, d\lambda 
\label{18.1.1}
\end{equation}
where $\dot{z}^\mu = dz^\mu/d\lambda$ is tangent to the world line and
the metric is evaluated at $z$. We assume that the particle provides
the {\it only} source of matter in the spacetime --- an explanation
will be provided below --- so that the Einstein field equations take
the form of    
\begin{equation}
{\sf G}^{\alpha\beta} = 8\pi {\sf T}^{\alpha\beta}, 
\label{18.1.2}
\end{equation} 
where ${\sf G}^{\alpha\beta}$ is the Einstein tensor constructed from 
${\sf g}_{\alpha\beta}$ and 
\begin{equation} 
{\sf T}^{\alpha\beta}(x) = m \int_\gamma 
\frac{ {\sf g}^\alpha_{\ \mu}(x,z) {\sf g}^\beta_{\ \nu}(x,z)  
\dot{z}^\mu \dot{z}^\nu}{ \sqrt{ -{\sf g}_{\mu\nu}
\dot{z}^\mu \dot{z}^\nu }}\, \delta_4(x,z)\, d\lambda 
\label{18.1.3}
\end{equation} 
is the particle's energy-momentum tensor, obtained by functional  
differentiation of $S_{\rm particle}$ with respect to 
${\sf g}_{\alpha\beta}(x)$; the parallel propagators appear naturally 
by expressing ${\sf g}_{\mu\nu}$ as ${\sf g}^\alpha_{\ \mu}   
{\sf g}^\beta_{\ \nu} {\sf g}_{\alpha\beta}$. 

On a formal level the metric ${\sf g}_{\alpha\beta}$ is obtained by 
solving the Einstein field equations, and the world line is determined
by the equations of energy-momentum conservation, which follow 
from the field equations. From Eqs.~(\ref{4.3.2}),
(\ref{12.1.3}), and (\ref{18.1.3}) we obtain  
\[
\nabla_\beta {\sf T}^{\alpha\beta} = m \int_\gamma \frac{d}{d\lambda} 
\biggl( \frac{ {\sf g}^\alpha_{\ \mu} \dot{z}^\mu }{     
\sqrt{ -{\sf g}_{\mu\nu} \dot{z}^\mu \dot{z}^\nu }} \biggr)
\delta_4(x,z)\, d\lambda,    
\]
and additional manipulations reduce this to 
\begin{equation} 
\nabla_\beta {\sf T}^{\alpha\beta} = m \int_\gamma  
\frac{ {\sf g}^\alpha_{\ \mu} }{     
\sqrt{ -{\sf g}_{\mu\nu} \dot{z}^\mu \dot{z}^\nu }}
\biggl( \frac{ {\sf D} \dot{z}^\mu}{d\lambda} - {\sf k}
\dot{z}^\mu \biggr) \delta_4(x,z)\, d\lambda,  
\label{18.1.4} 
\end{equation} 
where ${\sf D} \dot{z}^\mu/d\lambda$ is the covariant acceleration and
${\sf k}$ is a scalar field on the world line. Energy-momentum 
conservation therefore produces the geodesic equation   
\begin{equation} 
\frac{ {\sf D} \dot{z}^\mu}{d\lambda} = {\sf k} \dot{z}^\mu, 
\label{18.1.5} 
\end{equation} 
and 
\begin{equation} 
{\sf k} := \frac{1}{\sqrt{ -{\sf g}_{\mu\nu}
\dot{z}^\mu \dot{z}^\nu }} \frac{d}{d\lambda} 
\sqrt{ -{\sf g}_{\mu\nu} \dot{z}^\mu \dot{z}^\nu}  
\label{18.1.6}
\end{equation} 
measures the failure of $\lambda$ to be an affine parameter on the 
geodesic $\gamma$.  

\subsubsection*{Decomposition into background and perturbation}   

At this stage we begin treating $m$ as a small quantity, and we write   
\begin{equation}
{\sf g}_{\alpha\beta} = g_{\alpha\beta} + h_{\alpha\beta}, 
\label{18.1.7}
\end{equation}
with $g_{\alpha\beta}$ denoting the $m \to 0$ limit of the 
metric ${\sf g}_{\alpha\beta}$, and $h_{\alpha\beta}$ containing the
dependence on $m$. We shall refer to $g_{\alpha\beta}$ as the
``metric of the background spacetime'' and to $h_{\alpha\beta}$ as the
``perturbation'' produced by the particle. We insist, however, that no
approximation is introduced at this stage; the perturbation
$h_{\alpha\beta}$ is the {\it exact difference} between the exact
metric ${\sf g}_{\alpha\beta}$ and the background metric
$g_{\alpha\beta}$. Below we shall use the background metric to lower
and raise indices. 

We introduce the tensor field 
\begin{equation} 
C^\alpha_{\beta\gamma} := {\sf \Gamma}^\alpha_{\beta\gamma} 
- \Gamma^\alpha_{\beta\gamma}
\label{18.1.8} 
\end{equation} 
as the {\it exact difference} between 
${\sf \Gamma}^\alpha_{\beta\gamma}$, the connection compatible with
the exact metric ${\sf g}_{\alpha\beta}$, and 
$\Gamma^\alpha_{\beta\gamma}$, the connection compatible with the
background metric $g_{\alpha\beta}$. A covariant differentiation
indicated by $;\alpha$ will refer to $\Gamma^\alpha_{\beta\gamma}$,
while a covariant differentiation indicated by $\nabla_\alpha$ will
continue to refer to ${\sf \Gamma}^\alpha_{\beta\gamma}$.  

We express the exact Einstein tensor as 
\begin{equation} 
{\sf G}^{\alpha\beta} = G^{\alpha\beta}[g] 
+ \delta G^{\alpha\beta}[g,h] 
+ \Delta G^{\alpha\beta}[g,h], 
\label{18.1.9}
\end{equation} 
where $G^{\alpha\beta}$ is the Einstein tensor of the background
spacetime, which is assumed to vanish. The second term $\delta
G^{\alpha\beta}$ is the {\it linearized Einstein operator} defined by 
\begin{equation} 
\delta G^{\alpha\beta} := -\frac{1}{2} \bigl( \Box
\gamma^{\alpha\beta} + 2 R_{\gamma\ \delta}^{\ \alpha\ \beta} 
\gamma^{\gamma\delta} \bigr) 
+ \frac{1}{2} \bigl( \gamma^{\alpha\gamma\ \ \beta}_{\ \ \ ;\gamma} 
+ \gamma^{\beta\gamma\ \ \alpha}_{\ \ \ ;\gamma}
- g^{\alpha\beta} \gamma^{\gamma\delta}_{\ \ \, ;\gamma\delta} \bigr),  
\label{18.1.10} 
\end{equation} 
where $\Box \gamma^{\alpha\beta} := g^{\gamma\delta}
\gamma^{\alpha\beta}_{\ \ \ ;\gamma\delta}$ is the wave operator in
the background spacetime, and  
\begin{equation} 
\gamma^{\alpha\beta} := h^{\alpha\beta} - \frac{1}{2} g^{\alpha\beta}
\bigl(g_{\gamma\delta} h^{\gamma\delta}\bigr) 
\label{18.1.11}
\end{equation} 
is the ``trace-reversed'' metric perturbation (with all indices
raised with the background metric). The third term 
$\Delta G^{\alpha\beta}$ contains the remaining nonlinear pieces
that are excluded from $\delta G^{\alpha\beta}$.  

\subsubsection*{Field equations and conservation statement} 

The exact Einstein field equations can be expressed as 
\begin{equation} 
\delta G^{\alpha\beta} = 8\pi T^{\alpha\beta}_{\rm eff}, 
\label{18.1.12} 
\end{equation} 
where the effective energy-momentum tensor is defined by 
\begin{equation} 
T^{\alpha\beta}_{\rm eff}:= {\sf T}^{\alpha\beta} 
- \frac{1}{8\pi} \Delta G^{\alpha\beta}. 
\label{18.1.13} 
\end{equation} 
Because $\delta G^{\alpha\beta}$ satisfies the Bianchi-like identities
$\delta G^{\alpha\beta}_{\ \ \, ;\beta} = 0$, the effective
energy-momentum tensor is conserved in the background spacetime: 
\begin{equation} 
T^{\alpha\beta}_{{\rm eff}\, ;\beta} = 0. 
\label{18.1.14} 
\end{equation} 
This statement is equivalent to $\nabla_\beta {\sf T}^{\alpha\beta} =
0$, as can be inferred from the equations 
$\nabla_\beta {\sf G}^{\alpha\beta} 
= {\sf G}^{\alpha\beta}_{\ \ \, ;\beta} + C^\alpha_{\gamma\beta} 
{\sf G}^{\gamma\beta} + C^\beta_{\gamma\beta} {\sf G}^{\alpha\gamma}$, 
$\nabla_\beta {\sf T}^{\alpha\beta} 
= {\sf T}^{\alpha\beta}_{\ \ \, ;\beta} + C^\alpha_{\gamma\beta} 
{\sf T}^{\gamma\beta} + C^\beta_{\gamma\beta} {\sf T}^{\alpha\gamma}$,
and the definition of $T^{\alpha\beta}_{\rm eff}$. Equation
(\ref{18.1.14}), in turn, is equivalent to Eq.~(\ref{18.1.5}), which
states that the motion of the point particle is geodesic in the
perturbed spacetime.  

\subsubsection*{Integration of the field equations} 

Equation (\ref{18.1.12}) expresses the full and exact content of
Einstein's field equations. It is written in such a way that the
left-hand side is linear in the perturbation $h_{\alpha\beta}$, while
the right-hand side contains all nonlinear terms. It may be
viewed formally as a set of linear differential equations for
$h_{\alpha\beta}$ with a specified source term 
$T^{\alpha\beta}_{\rm eff}$. This equation is of mixed
hyperbolic-elliptic type, and as such it is a poor starting point for
the selection of retarded solutions that enforce a strict causal link
between the source and the field. This inadequacy, however, can be
remedied by imposing the {\it Lorenz gauge condition} 
\begin{equation} 
\gamma^{\alpha\beta}_{\ \ \ ;\beta} = 0,  
\label{18.1.15} 
\end{equation}   
which converts $\delta G^{\alpha\beta}$ into a strictly hyperbolic
differential operator. In this gauge the field equations become  
\begin{equation} 
\Box \gamma^{\alpha\beta} + 2 R_{\gamma\ \delta}^{\ \alpha\ \beta}
\gamma^{\gamma\delta} = -16\pi T_{\rm eff}^{\alpha\beta}.  
\label{18.1.16}
\end{equation} 
This is a tensorial wave equation formulated in the background
spacetime, and while the left-hand side is manifestly linear in
$h_{\alpha\beta}$, the right-hand side continues to incorporate all 
nonlinear terms. Equations (\ref{18.1.15}) and (\ref{18.1.16}) still
express the full content of the exact field equations.  

A formal solution to Eq.~(\ref{18.1.16}) is 
\begin{equation}
\gamma^{\alpha\beta}(x) = 4\int 
G^{\ \alpha\beta}_{+\ \gamma'\delta'}(x,x') 
T_{\rm eff}^{\gamma'\delta'}(x') \sqrt{-g'}\, d^4 x', 
\label{18.1.17} 
\end{equation}
where $G^{\ \alpha\beta}_{+\ \gamma'\delta'}(x,x')$ is the retarded     
Green's function introduced in Sec.~\ref{15}. With the
help of Eq.~(\ref{15.3b.1}), it is easy to show that  
\begin{equation} 
\gamma^{\alpha\beta}_{\ \ \ ;\beta}= 4\int 
G^{\ \alpha}_{+ \gamma'} 
T^{\gamma'\delta'}_{{\rm eff}\ ;\delta'}\sqrt{-g'}\, d^4 x' 
\label{18.1.18} 
\end{equation}
follows directly from Eq.~(\ref{18.1.17}); 
$G^{\ \alpha}_{+\ \gamma'}(x,x')$ is the electromagnetic
Green's function introduced in Sec.~\ref{14}. This equation indicates
that the Lorenz gauge condition is automatically enforced when the
conservation equation $T^{\alpha\beta}_{{\rm eff}\, ;\beta} = 0$ is
imposed. Conversely, Eq.~(\ref{18.1.18}) implies that 
$\Box(\gamma^{\alpha\beta}_{\ \ \ ;\beta}) = -16\pi 
T^{\alpha\beta}_{{\rm eff}\, ;\beta}$, which indicates that imposition
of $\gamma^{\alpha\beta}_{\ \ ;\beta} = 0$ automatically enforces the 
conservation equation. There is a one-to-one correspondence  
between the conservation equation and the Lorenz gauge condition.   

The split of the Einstein field equations into a wave equation and a
gauge condition directly tied to the conservation of the effective 
energy-momentum tensor is a most powerful tool, because it allows us 
to disentangle the problems of obtaining $h_{\alpha\beta}$ and
determining the motion of the particle. This comes about because the
wave equation can be solved first, independently of the gauge
condition, for a particle moving on an arbitrary world line $\gamma$;
the world line is determined next, by imposing the Lorenz gauge
condition on the solution to the wave equation. More precisely stated,
the source term $T^{\alpha\beta}_{\rm eff}$ for the wave equation can
be evaluated for any world line $\gamma$, without demanding that the
effective energy-momentum tensor be conserved, and without demanding
that $\gamma$ be a geodesic of the perturbed spacetime. Solving the
wave equation then returns $h_{\alpha\beta}[\gamma]$ as a functional
of the arbitrary world line, and the metric is not yet fully
specified. Because imposing the Lorenz gauge condition is equivalent
to imposing conservation of the effective energy-momentum tensor,
inserting $h_{\alpha\beta}[\gamma]$ within Eq.~(\ref{18.1.15}) finally 
determines $\gamma$, and forces it to be a geodesic of the perturbed
spacetime. At this stage the full set of Einstein field equations is
accounted for, and the metric is fully specified as a tensor field in 
spacetime. The split of the field equations into a wave equation and a
gauge condition is key to the formulation of the gravitational
self-force; in this specific context the Lorenz gauge is conferred a
preferred status among all choices of gauge.   

An important question to be addressed is how the wave equation is to
be integrated. A method of principle, based on the assumed smallness
of $m$ and $h_{\alpha\beta}$, is suggested by post-Minkowskian theory 
\cite{walker-will:80, blanchet:06}. One proceeds by iterations. In the
first iterative stage, one fixes $\gamma$ and substitutes 
$h^{\alpha\beta}_0 = 0$ within $T^{\alpha\beta}_{\rm eff}$; evaluation
of the integral in Eq.~(\ref{18.1.17}) returns the first-order
approximation $h^{\alpha\beta}_1[\gamma] = O(m)$ for the
perturbation. In the second stage $h^{\alpha\beta}_1$ is inserted
within $T^{\alpha\beta}_{\rm eff}$ and Eq.~(\ref{18.1.17}) returns
the second-order approximation $h^{\alpha\beta}_2[\gamma] = O(m,m^2)$
for the perturbation. Assuming that this procedure can be repeated at 
will and produces an adequate asymptotic series for the exact 
perturbation, the iterations are stopped when the $n^{\rm th}$-order 
approximation $h_n^{\alpha\beta}[\gamma] = O(m, m^2, \cdots, m^n)$ is 
deemed to be sufficiently accurate. The world line is then determined,
to order $m^n$, by subjecting the approximated field to the Lorenz
gauge condition. It is to be noted that the procedure necessarily
produces an approximation of the field, and an approximation of the 
motion, because the number of iterations is necessarily finite. This
is the {\it only source of approximation} in our formulation of the
dynamics of a point mass.        

\subsubsection*{Equations of motion} 

Conservation of energy-momentum implies Eq.~(\ref{18.1.5}), which
states that the motion of the point mass is geodesic in the perturbed
spacetime. The equation is expressed in terms of the exact connection  
${\sf \Gamma}^\alpha_{\beta\gamma}$, and with the help of
Eq.~(\ref{18.1.8}) it can be re-written in terms of the background
connection $\Gamma^\alpha_{\beta\gamma}$. We get 
$D\dot{z}^\mu/d\lambda = 
- C^\mu_{\nu\lambda} \dot{z}^\nu \dot{z}^\lambda 
+ {\sf k} \dot{z}^\mu$, 
where the left-hand side is the covariant acceleration in the  
background spacetime, and $\sf k$ is given by Eq.~(\ref{18.1.6}). At
this stage the arbitrary parameter $\lambda$ can be identified with
proper time $\tau$ in the background spacetime. With this choice
the equations of motion become 
\begin{equation} 
a^\mu = - C^\mu_{\nu\lambda} u^\nu u^\lambda 
+ {\sf k} u^\mu,  
\label{18.1.19}
\end{equation} 
where $u^\mu := dz^\mu/d\tau$ is the velocity vector in the background 
spacetime, $a^\mu := Du^\mu/d\tau$ the covariant acceleration, and 
\begin{equation} 
{\sf k} = \frac{1}{\sqrt{1-h_{\mu\nu} u^\mu u^\nu}} \frac{d}{d\tau} 
\sqrt{1-h_{\mu\nu} u^\mu u^\nu}. 
\label{18.1.20}
\end{equation} 
Equation (\ref{18.1.19}) is an exact statement of the equations of
motion. It expresses the fact that while the motion is geodesic in the
perturbed spacetime, it may be viewed as accelerated motion in the
background spacetime. Because $h_{\alpha\beta}$ is calculated as an
expansion in powers of $m$, the acceleration also is eventually
obtained as an expansion in powers of $m$. Here, in keeping with the
preceding sections, we will use order-reduction to make that expansion
well-behaved; in Part V of the review, we will formulate the expansion
more clearly as part of more systematic approach. 

\subsubsection*{Implementation to first order in $m$} 

While our formulation of the dynamics of a point mass is
in principle exact, any practical implementation will rely on an
approximation method. As we saw previously, the most immediate source
of approximation concerns the number of iterations involved in the 
integration of the wave equation. Here we perform a single iteration
and obtain the perturbation $h_{\alpha\beta}$ and the equations of
motion to first order in the mass $m$.  

In a first iteration of the wave equation we fix $\gamma$ and set
$\Delta G^{\alpha\beta} = 0$, ${\sf T}^{\alpha\beta} =
T^{\alpha\beta}$, where 
\begin{equation} 
T^{\alpha\beta} =  
m \int_\gamma  g^\alpha_{\ \mu}(x,z) g^\beta_{\ \nu}(x,z)   
u^\mu u^\nu\, \delta_4(x,z)\, d\tau
\label{18.1.21}
\end{equation}
is the particle's energy-momentum tensor in the background
spacetime. This implies that $T^{\alpha\beta}_{\rm eff} =
T^{\alpha\beta}$, and Eq.~(\ref{18.1.16}) becomes 
\begin{equation} 
\Box \gamma^{\alpha\beta} + 2 R_{\gamma\ \delta}^{\ \alpha\ \beta}
\gamma^{\gamma\delta} = -16\pi T^{\alpha\beta} + O(m^2).   
\label{18.1.22}
\end{equation} 
Its solution is 
\begin{equation} 
\gamma^{\alpha\beta}(x) = 4 m \int_\gamma 
G^{\ \alpha\beta}_{+\ \mu\nu}(x,z) u^\mu u^\nu\, d\tau 
+ O(m^2),  
\label{18.1.23}
\end{equation}
and from this we obtain $h^{\alpha\beta}$. Equation (\ref{18.1.8})
gives rise to $C^\alpha_{\beta\gamma} = \frac{1}{2}(
h^\alpha_{\ \beta;\gamma} + h^\alpha_{\ \gamma;\beta} -
h_{\beta\gamma}^{\ \ \ ;\alpha}) + O(m^2)$, and from
Eq.~(\ref{18.1.20}) we obtain ${\sf k} 
= -\frac{1}{2} h_{\nu\lambda;\rho} u^\nu u^\lambda
u^\rho - h_{\nu\lambda} u^\nu a^\lambda + O(m^2)$; we can  
discard the second term because it is clear that the acceleration will 
be of order $m$. Inserting these results within Eq.~(\ref{18.1.19}), we
obtain  
\begin{equation} 
a^\mu = -\frac{1}{2} \Bigl( h^\mu_{\ \nu;\lambda} +
h^\mu_{\ \lambda;\nu} - h_{\nu\lambda}^{\ \ ;\mu} + u^\mu
h_{\nu\lambda;\rho} u^\rho \Bigr) u^\nu u^\lambda + O(m^2).
\end{equation}
We express this in the equivalent form  
\begin{equation}
a^\mu = -\frac{1}{2} \bigl( g^{\mu\nu} + u^\mu
u^\nu \bigr) \bigl( 2 h_{\nu\lambda;\rho} - h_{\lambda\rho;\nu} \bigr)
u^\lambda u^\rho + O(m^2) 
\label{18.1.24} 
\end{equation}   
to emphasize the fact that the acceleration is orthogonal to the
velocity vector.  

It should be clear that Eq.~(\ref{18.1.24}) is valid only in a formal
sense, because the potentials obtained from Eqs.~(\ref{18.1.23})
diverge on the world line. To make sense of these equations we will
proceed as in Secs.~\ref{16} and \ref{17} with a careful analysis of
the field's singularity structure; regularization will produce a
well-defined version of Eq.~(\ref{18.1.24}). Our formulation of the
dynamics of a point mass makes it clear that a proper implementation  
requires that the wave equation of Eq.~(\ref{18.1.22}) and the equations of
motion of Eq.~(\ref{18.1.24}) be integrated simultaneously, in a 
self-consistent manner. 

\subsubsection*{Failure of a strictly linearized formulation} 

In the preceding discussion we started off with an exact formulation
of the problem of motion for a small mass $m$ in a background
spacetime with metric $g_{\alpha\beta}$, but eventually boiled  
it down to an implementation accurate to first order in $m$. Would it 
not be simpler and more expedient to formulate the problem directly to
first order? The answer is a resounding no: By doing so we would be
driven toward a grave inconsistency; the nonlinear formulation is
absolutely necessary if one wishes to contemplate a self-consistent
integration of Eqs.~(\ref{18.1.22}) and (\ref{18.1.24}). 

A strictly linearized formulation of the problem would be based on the
field equations $\delta G^{\alpha\beta} = 8\pi T^{\alpha\beta}$, where
$T^{\alpha\beta}$ is the energy-momentum tensor of
Eq.~(\ref{18.1.21}). The Bianchi-like identities $\delta
G^{\alpha\beta}_{\ \ \, ;\beta} = 0$ dictate that $T^{\alpha\beta}$ must
be {\it conserved in the background spacetime}, and a calculation
identical to the one leading to Eq.~(\ref{18.1.5}) would reveal that
the particle's motion must be {\it geodesic in the background
spacetime}. In the strictly linearized formulation, therefore, the
gravitational potentials of Eq.~(\ref{18.1.23}) must be sourced by a
particle moving on a geodesic, and there is no opportunity for these
potentials to exert a self-force. To get the self-force, one must
provide a formulation that extends beyond linear order. To be sure,
one could persist in adopting the linearized formulation and ``save
the phenomenon'' by relaxing the conservation equation. In practice
this could be done by adopting the solutions of Eq.~(\ref{18.1.23}),
demanding that the motion be geodesic in the perturbed spacetime, 
and relaxing the linearized gauge condition to 
$\gamma^{\alpha\beta}_{\ \ \ ;\beta} = O(m^2)$. While this
prescription would produce the correct answer, it is largely 
{\it ad hoc} and does not come with a clear justification. Our exact
formulation provides much more control, at least in a formal sense. We
shall do even better in Part~\ref{part5}.   

An alternative formulation of the problem provided by Gralla and Wald
\cite{gralla-wald:08} avoids the inconsistency by refraining from
performing a self-consistent integration of Eqs.~(\ref{18.1.22}) and
(\ref{18.1.24}). Instead of an expansion of the acceleration in powers
of $m$, their approach is based on an expansion of the world line
itself, and it returns the equations of motion for a deviation vector
which describes the offset of the true world line relative to a reference
geodesic. While this approach is mathematically sound, it eventually
breaks down as the deviation vector becomes large, and it does not
provide a justification of the self-consistent treatment of the
equations. 

The difference between the Gralla-Wald approach and a self-consistent 
one is the difference between a regular expansion and a general
one. In a regular expansion, all dependence on a small quantity $m$ is  
expanded in powers: 
\begin{equation}
h_{\alpha\beta}(x,m)=\sum_{n=0}^\infty m^nh^{(n)}_{\alpha\beta}(x).
\end{equation}
In a general expansion, on the other hand, the functions
$h^{(n)}_{\alpha\beta}$ are allowed to retain some dependence on the
small quantity:
\begin{equation}\label{singular_expansion}
h_{\alpha\beta}(x,m)=\sum_{n=0}^\infty m^nh^{(n)}_{\alpha\beta}(x,m).
\end{equation}
Put simply, the goal of a general expansion is to expand only
\emph{part} of a function's dependence on a small quantity, while
holding fixed some specific dependence that captures one or more of
the function's essential features. In the self-consistent expansion
that we advocate here, our iterative solution returns
\begin{equation} 
h^N_{\alpha\beta}(x,m)=\sum_{n=0}^N
m^nh^{(n)}_{\alpha\beta}(x;\gamma(m)),
\end{equation} 
 in which the functional dependence on the
world line $\gamma$ incorporates a dependence on the expansion
parameter $m$.  We deliberately introduce this functional dependence
on a mass-dependent world line in order to maintain a meaningful and  
accurate description of the particle's motion. Although the regular
expansion can be retrieved by further expanding the dependence within 
$\gamma(m)$, the reverse statement does not hold: the general
expansion cannot be justified on the basis of the regular one. The
notion of a general expansion is at the core of singular perturbation
theory \cite{eckhaus:79, holmes:95, kevorkian-cole:96, lagerstrom:88, 
  verhulst:05, pound:10b}. We shall return to these issues in our
treatment of asymptotically small bodies, and in particular, in 
Sec.~\ref{comments on force} below.  

\subsubsection*{Vacuum background spacetime} 

To conclude this subsection we should explain why it is desirable to  
restrict our discussion to spacetimes that contain no matter except
for the point particle. Suppose, in contradiction with this
assumption, that the background spacetime contains a distribution of
matter around which the particle is moving. (The corresponding vacuum
situation has the particle moving around a black hole. Notice that we
are still assuming that the particle moves in a region of spacetime in
which there is no matter; the issue is whether we can allow for a
distribution of matter {\it somewhere else}.) Suppose also that the
matter distribution is described by a collection of matter fields
$\Psi$. Then the field equations satisfied by the matter have the
schematic form $E[\Psi;g] = 0$, and the background metric is
determined by the Einstein field equations $G[g] = 8\pi M[\Psi;g]$, in
which $M[\Psi;g]$ stands for the matter's energy-momentum tensor. We
now insert the point particle in the spacetime, and recognize that
this displaces the background solution $(\Psi,g)$ to a new solution
($\Psi + \delta \Psi, g + \delta g)$. The perturbations are determined
by the coupled set of equations $E[\Psi+\delta \Psi;g+\delta g] = 0$
and $G[g + \delta g] = 8\pi M[\Psi+\delta \Psi;g + \delta g] 
+ 8\pi T[g]$. After linearization these take the form of 
\[
E_\Psi \cdot \delta \Psi + E_g \cdot \delta g = 0, \qquad 
G_g \cdot \delta g = 8\pi \bigl( M_\Psi \cdot \delta \Psi + M_g \cdot
\delta g + T \bigr), 
\]
where $E_\Psi$, $E_g$, $M_\Phi$, and $M_g$ are suitable differential
operators acting on the perturbations. This is a {\it coupled set} of
partial differential equations for the perturbations $\delta \Psi$ and
$\delta g$. These equations are linear, but they are much more
difficult to deal with than the single equation for $\delta g$ that
was obtained in the vacuum case. And although it is still possible to
solve the coupled set of equations via a Green's function technique,
the degree of difficulty is such that we will not attempt this
here. We shall, therefore, continue to restrict our attention to the
case of a point particle moving in a vacuum (globally Ricci-flat)
background spacetime. 

\subsection{Retarded potentials near the world line} 
\label{18.2}

Going back to Eq.~(\ref{18.1.23}), we have that the gravitational
potentials associated with a point particle of mass $m$ moving on
world line $\gamma$ are given by 
\begin{equation} 
\gamma^{\alpha\beta}(x) = 4 m \int_\gamma 
G^{\ \alpha\beta}_{+\ \mu\nu}(x,z) u^\mu u^\nu\, d\tau, 
\label{18.2.1}
\end{equation}
up to corrections of order $m^2$; here $z^\mu(\tau)$ gives the
description of the world line, $u^\mu = d z^\mu/d\tau$ is the velocity
vector, and $G^{\ \alpha\beta}_{+\ \gamma'\delta'}(x,x')$ is the
retarded Green's function introduced in Sec.~\ref{15}. Because the
retarded Green's function is defined globally in the entire background 
spacetime, Eq.~(\ref{18.2.1}) describes the gravitational perturbation
created by the particle at any point $x$ in that spacetime.  

For a more concrete expression we take $x$ to be in a
neighbourhood of the world line. The manipulations that follow are
very close to those performed in Sec.~\ref{16.2} for the case of a
scalar charge, and in Sec.~\ref{17.2} for the case of an electric
charge. Because these manipulations are by now familiar, it will be
sufficient here to present only the main steps. There are two 
important simplifications that occur in the case of a massive
particle. First, it is clear that
\begin{equation}
a^\mu = O(m) = \dot a^\mu,
\label{18.2.2} 
\end{equation}
and we will take the liberty of performing a pre-emptive
order-reduction by dropping all terms involving the acceleration
vector when computing $\gamma^{\alpha\beta}$ and 
$\gamma_{\alpha\beta;\gamma}$ to first order in $m$; otherwise we 
would arrive at an equation for the acceleration that would include an
antidamping term $-\frac{11}{3}m\dot a^\mu$ \cite{havas:57,
  havas-goldberg:62, quinn-wald:97}. Second, because we take
$g_{\alpha\beta}$ to be a solution to the vacuum field equations, we
are also allowed to set    
\begin{equation}
R_{\mu\nu}(z) = 0 
\label{18.2.3}
\end{equation} 
in our computations. 

With the understanding that $x$ is close to the world line (refer back
to Fig.~9), we substitute the Hadamard construction of
Eq.~(\ref{15.2.1}) into Eq.~(\ref{18.2.1}) and integrate over the
portion of $\gamma$ that is contained in ${\cal N}(x)$. The result is  
\begin{equation}
\gamma^{\alpha\beta}(x) = \frac{4m}{r} 
U^{\alpha\beta}_{\ \ \gamma'\delta'}(x,x') u^{\gamma'} u^{\delta'} 
+ 4 m \int_{\tau_<}^u V^{\alpha\beta}_{\ \ \mu\nu}(x,z) u^\mu u^\nu\,
d\tau + 4 m \int_{-\infty}^{\tau_<} 
G^{\ \alpha\beta}_{+\ \mu\nu}(x,z) u^\mu u^\nu\, d\tau, 
\label{18.2.4}
\end{equation}
in which primed indices refer to the retarded point $x' := z(u)$ 
associated with $x$, $r := \sigma_{\alpha'} u^{\alpha'}$ is the
retarded distance from $x'$ to $x$, and $\tau_<$ is the proper time at
which $\gamma$ enters ${\cal N}(x)$ from the past. 

In the following subsections we shall refer to
$\gamma_{\alpha\beta}(x)$ as the {\it gravitational potentials} at $x$
produced by a particle of mass $m$ moving on the world line $\gamma$, 
and to $\gamma_{\alpha\beta;\gamma}(x)$ as the {\it gravitational
field} at $x$. To compute this is our next task.   

\subsection{Gravitational field in retarded coordinates} 
\label{18.3}
    
Keeping in mind that $x'$ and $x$ are related by $\sigma(x,x') = 0$, a
straightforward computation reveals that the covariant derivatives of
the gravitational potentials are given by  
\begin{eqnarray}
\gamma_{\alpha\beta;\gamma}(x) &=& -\frac{4m}{r^2}
U_{\alpha\beta\alpha'\beta'} u^{\alpha'} u^{\beta'} \partial_\gamma r 
+ \frac{4m}{r} U_{\alpha\beta\alpha'\beta';\gamma} u^{\alpha'}
u^{\beta'} 
+ \frac{4m}{r} U_{\alpha\beta\alpha'\beta';\gamma'} u^{\alpha'}
u^{\beta'} u^{\gamma'} \partial_\gamma u
\nonumber \\ & & \mbox{} 
+ 4m V_{\alpha\beta\alpha'\beta'} u^{\alpha'} u^{\beta'}
\partial_\gamma u 
+ \gamma_{\alpha\beta\gamma}^{\rm tail}(x), 
\label{18.3.1}
\end{eqnarray} 
where the ``tail integral'' is defined by 
\begin{eqnarray} 
\gamma_{\alpha\beta\gamma}^{\rm tail}(x) &=& 4 m \int_{\tau_<}^u
\nabla_\gamma V_{\alpha\beta\mu\nu}(x,z) u^\mu u^\nu\, d\tau
+ 4 m \int_{-\infty}^{\tau_<} \nabla_\gamma
G_{+\alpha\beta\mu\nu}(x,z) u^\mu u^\nu\, d\tau 
\nonumber \\ 
&=& 4m \int_{-\infty}^{u^-} \nabla_\gamma
G_{+\alpha\beta\mu\nu}(x,z) u^\mu u^\nu\, d\tau.  
\label{18.3.2}
\end{eqnarray} 
The second form of the definition, in which the integration is cut 
short at $\tau = u^- := u - 0^+$ to avoid the singular behaviour
of the retarded Green's function at $\sigma = 0$, is equivalent to the
first form. 

We wish to express $\gamma_{\alpha\beta;\gamma}(x)$ in the retarded
coordinates of Sec.~\ref{9}, as an expansion in powers of $r$. For
this purpose we decompose the field in the tetrad
$(\base{\alpha}{0},\base{\alpha}{a})$ that is obtained by parallel
transport of $(u^{\alpha'},\base{\alpha'}{a})$ on the null geodesic
that links $x$ to $x'$; this construction is detailed in
Sec.~\ref{9}. We recall from
Eq.~(\ref{9.1.4}) that the parallel propagator can be expressed as  
$g^{\alpha'}_{\ \alpha} = u^{\alpha'} \base{0}{\alpha}  
+ \base{\alpha'}{a} \base{a}{\alpha}$. The expansion relies on
Eq.~(\ref{9.5.3}) for $\partial_\gamma u$ and Eq.~(\ref{9.5.5}) for 
$\partial_\gamma r$, both simplified by setting $a_a = 0$. We shall
also need  
\begin{equation} 
U_{\alpha\beta\alpha'\beta'} u^{\alpha'} u^{\beta'} = 
g^{\alpha'}_{\ (\alpha} g^{\beta'}_{\ \beta)} \Bigl[ u_{\alpha'}
u_{\beta'} + O(r^3) \Bigr], 
\label{18.3.3}
\end{equation} 
which follows from Eq.~(\ref{15.2.7}), 
\begin{eqnarray} 
U_{\alpha\beta\alpha'\beta';\gamma} u^{\alpha'} u^{\beta'} &=&  
g^{\alpha'}_{\ (\alpha} g^{\beta'}_{\ \beta)} g^{\gamma'}_{\ \gamma}
\Bigl[ -r\bigl( R_{\alpha'0\gamma'0} + R_{\alpha'0\gamma'd} \Omega^d
\bigr) u_{\beta'} + O(r^2) \Bigr], 
\label{18.3.4} \\ 
U_{\alpha\beta\alpha'\beta';\gamma'} u^{\alpha'} u^{\beta'}
u^{\gamma'} &=& g^{\alpha'}_{\ (\alpha} g^{\beta'}_{\ \beta)} \Bigl[ 
r R_{\alpha'0d0} \Omega^d u_{\beta'} + O(r^2) \Bigr], 
\label{18.3.5}
\end{eqnarray} 
which follow from Eqs.~(\ref{15.2.8}) and (\ref{15.2.9}),
respectively, as well as the relation $\sigma^{\alpha'} = -r
(u^{\alpha'} + \Omega^a \base{\alpha'}{a})$ first encountered
in Eq.~(\ref{9.2.3}). And finally, we shall need  
\begin{equation}
V_{\alpha\beta\alpha'\beta'} u^{\alpha'} u^{\beta'} = 
g^{\alpha'}_{\ (\alpha} g^{\beta'}_{\ \beta)} \Bigl[
R_{\alpha'0\beta'0} + O(r) \Bigr], 
\label{18.3.6}
\end{equation} 
which follows from Eq.~(\ref{15.2.11}).  

Making these substitutions in Eq.~(\ref{18.1.3}) and projecting
against various members of the tetrad gives 
\begin{eqnarray} 
\gamma_{000}(u,r,\Omega^a) &:=& \gamma_{\alpha\beta;\gamma}(x) 
\base{\alpha}{0}(x) \base{\beta}{0}(x) \base{\gamma}{0}(x) 
= 2m R_{a0b0} \Omega^a \Omega^b + \gamma^{\rm tail}_{000} + O(r), 
\label{18.3.7} \\ 
\gamma_{0b0}(u,r,\Omega^a) &:=& \gamma_{\alpha\beta;\gamma}(x) 
\base{\alpha}{0}(x) \base{\beta}{b}(x) \base{\gamma}{0}(x) 
= -4m R_{b0c0} \Omega^c + \gamma^{\rm tail}_{0b0} + O(r),
\label{18.3.8} \\ 
\gamma_{ab0}(u,r,\Omega^a) &:=& \gamma_{\alpha\beta;\gamma}(x) 
\base{\alpha}{a}(x) \base{\beta}{b}(x) \base{\gamma}{0}(x) 
= 4m R_{a0b0} + \gamma^{\rm tail}_{ab0} + O(r), 
\label{18.3.9} \\ 
\gamma_{00c}(u,r,\Omega^a) &:=& \gamma_{\alpha\beta;\gamma}(x) 
\base{\alpha}{0}(x) \base{\beta}{0}(x) \base{\gamma}{c}(x) 
\nonumber \\ 
&=& -4m \biggl[ \Bigl(\frac{1}{r^2} + \frac{1}{3} R_{a0b0} \Omega^a
\Omega^b \Bigr) \Omega_c + \frac{1}{6} R_{c0b0} \Omega^b - \frac{1}{6}
R_{ca0b} \Omega^a \Omega^b \biggr] + \gamma^{\rm tail}_{00c} + O(r),\qquad 
\label{18.3.10} \\ 
\gamma_{0bc}(u,r,\Omega^a) &:=& \gamma_{\alpha\beta;\gamma}(x) 
\base{\alpha}{0}(x) \base{\beta}{b}(x) \base{\gamma}{c}(x) 
\nonumber \\ 
&=& 2m \bigl( R_{b0c0} + R_{b0cd} \Omega^d + R_{b0d0} \Omega^d
\Omega_c \bigr) + \gamma^{\rm tail}_{0bc} + O(r), 
\label{18.3.11} \\ 
\gamma_{abc}(u,r,\Omega^a) &:=& \gamma_{\alpha\beta;\gamma}(x) 
\base{\alpha}{a}(x) \base{\beta}{b}(x) \base{\gamma}{c}(x) 
= -4m R_{a0b0} \Omega_c + \gamma^{\rm tail}_{abc} + O(r), 
\label{18.3.12}
\end{eqnarray} 
where, for example, $R_{a0b0}(u) :=
R_{\alpha'\gamma'\beta'\delta'} \base{\alpha'}{a} u^{\gamma'}
\base{\beta'}{b} u^{\delta'}$ are frame components of the Riemann
tensor evaluated at $x' := z(u)$. We have also introduced the
frame components of the tail part of the gravitational field, which
are obtained from Eq.~(\ref{18.3.2}) evaluated at $x'$ instead of $x$;
for example, $\gamma^{\rm tail}_{000} = u^{\alpha'} u^{\beta'}
u^{\gamma'} \gamma^{\rm tail}_{\alpha'\beta'\gamma'}(x')$. We 
may note here that while $\gamma_{00c}$ is the only component of the
gravitational field that diverges when $r \to 0$, the other components
are nevertheless singular because of their dependence on
the unit vector $\Omega^a$; the only exception is $\gamma_{ab0}$,
which is regular.  

\subsection{Gravitational field in Fermi normal coordinates}   
\label{18.4} 

The translation of the results contained in
Eqs.~(\ref{18.3.7})--(\ref{18.3.12}) into the Fermi normal coordinates
of Sec.~\ref{8} proceeds as in Secs.~\ref{16.4} and \ref{17.4}, but 
is simplified by setting $a_a = \dot{a}_0 = \dot{a}_a = 0$ in
Eqs.~(\ref{10.3.1}), (\ref{10.3.2}), (\ref{10.2.1}), (\ref{10.2.2}),
and (\ref{10.2.3}) that relate the Fermi normal coordinates
$(t,s,\omega^a)$ to the retarded coordinates. We recall
that the Fermi normal coordinates refer to a point $\bar{x} :=
z(t)$ on the world line that is linked to $x$ by a spacelike geodesic
that intersects $\gamma$ orthogonally. 

The translated results are     
\begin{eqnarray} 
\bar{\gamma}_{000}(t,s,\omega^a) &:=&
\gamma_{\alpha\beta;\gamma}(x)  
\bar{e}^\alpha_0(x) \bar{e}^\beta_0(x) \bar{e}^\gamma_0(x) 
= \bar{\gamma}^{\rm tail}_{000} + O(s),  
\label{18.4.1} \\ 
\bar{\gamma}_{0b0}(t,s,\omega^a) &:=&
\gamma_{\alpha\beta;\gamma}(x)  
\bar{e}^\alpha_0(x) \bar{e}^\beta_b(x) \bar{e}^\gamma_0(x) 
= -4m R_{b0c0} \omega^c + \bar{\gamma}^{\rm tail}_{0b0} + O(s),
\label{18.4.2} \\ 
\bar{\gamma}_{ab0}(t,s,\omega^a) &:=&
\gamma_{\alpha\beta;\gamma}(x)  
\bar{e}^\alpha_a(x) \bar{e}^\beta_b(x) \bar{e}^\gamma_0(x) 
= 4m R_{a0b0} + \bar{\gamma}^{\rm tail}_{ab0} + O(s), 
\label{18.4.3} \\ 
\bar{\gamma}_{00c}(t,s,\omega^a) &:=&
\gamma_{\alpha\beta;\gamma}(x)  
\bar{e}^\alpha_0(x) \bar{e}^\beta_0(x) \bar{e}^\gamma_c(x) 
\nonumber \\ 
&=& -4m \biggl[ \Bigl(\frac{1}{s^2} - \frac{1}{6} R_{a0b0} \omega^a
\omega^b \Bigr) \omega_c + \frac{1}{3} R_{c0b0} \omega^b \biggr] 
+ \bar{\gamma}^{\rm tail}_{00c} + O(s), 
\label{18.4.4} \\ 
\bar{\gamma}_{0bc}(t,s,\omega^a) &:=&
\gamma_{\alpha\beta;\gamma}(x)  
\bar{e}^\alpha_0(x) \bar{e}^\beta_b(x) \bar{e}^\gamma_c(x) 
= 2m \bigl( R_{b0c0} + R_{b0cd} \omega^d \bigr) 
+ \bar{\gamma}^{\rm tail}_{0bc} + O(s), 
\label{18.4.5} \\ 
\bar{\gamma}_{abc}(t,s,\omega^a) &:=&
\gamma_{\alpha\beta;\gamma}(x)  
\bar{e}^\alpha_a(x) \bar{e}^\beta_b(x) \bar{e}^\gamma_c(x) 
= -4m R_{a0b0} \omega_c + \bar{\gamma}^{\rm tail}_{abc} + O(s), 
\label{18.4.6}
\end{eqnarray} 
where all frame components are now evaluated at $\bar{x}$ instead of
$x'$.  

It is then a simple matter to average these results over a two-surface
of constant $t$ and $s$. Using the area element of Eq.~(\ref{16.4.5})
and definitions analogous to those of Eq.~(\ref{16.4.6}), we obtain 
\begin{eqnarray}
\langle \bar{\gamma}_{000} \rangle &=& \bar{\gamma}^{\rm tail}_{000}  
+ O(s), 
\label{18.4.7} \\ 
\langle \bar{\gamma}_{0b0} \rangle &=& \bar{\gamma}^{\rm tail}_{0b0}  
+ O(s), 
\label{18.4.8} \\ 
\langle \bar{\gamma}_{ab0} \rangle &=& 4 m R_{a0b0} 
+ \bar{\gamma}^{\rm tail}_{ab0} + O(s), 
\label{18.4.9} \\ 
\langle \bar{\gamma}_{00c} \rangle &=& \bar{\gamma}^{\rm tail}_{00c}  
+ O(s), 
\label{18.4.10} \\ 
\langle \bar{\gamma}_{0bc} \rangle &=& 2 m R_{b0c0} 
+ \bar{\gamma}^{\rm tail}_{0bc} + O(s), 
\label{18.4.11} \\ 
\langle \bar{\gamma}_{abc} \rangle &=& \bar{\gamma}^{\rm tail}_{abc}  
+ O(s).   
\label{18.4.12} 
\end{eqnarray} 
The averaged gravitational field is regular in the limit $s \to 0$, in
which the tetrad $(\bar{e}^\alpha_0,\bar{e}^\alpha_a)$ coincides with 
$(u^{\bar{\alpha}},\base{\bar{\alpha}}{a})$. Reconstructing the field 
at $\bar{x}$ from its frame components gives 
\begin{equation}
\langle \gamma_{\bar{\alpha}\bar{\beta};\bar{\gamma}} \rangle = 
-4 m \Bigl( u_{(\bar{\alpha}}
R_{\bar{\beta})\bar{\delta}\bar{\gamma}\bar{\epsilon}}
+ R_{\bar{\alpha}\bar{\delta}\bar{\beta}\bar{\epsilon}}
u_{\bar{\gamma}} \Bigr) u^{\bar{\delta}} u^{\bar{\epsilon}} 
+ \gamma^{\rm tail}_{\bar{\alpha}\bar{\beta}\bar{\gamma}}, 
\label{18.4.13}
\end{equation}
where the tail term can be copied from Eq.~(\ref{18.3.2}), 
\begin{equation} 
\gamma^{\rm tail}_{\bar{\alpha}\bar{\beta}\bar{\gamma}}(\bar{x}) 
= 4m \int_{-\infty}^{t^-} \nabla_{\bar{\gamma}}
G_{+\bar{\alpha}\bar{\beta}\mu\nu}(\bar{x},z) u^\mu u^\nu\, d\tau. 
\label{18.4.14}
\end{equation}
The tensors that appear in Eq.~(\ref{18.4.13}) all refer to the  
simultaneous point $\bar{x} := z(t)$, which can now be treated as
an arbitrary point on the world line $\gamma$. 

\subsection{Singular and regular fields}
\label{18.5} 

The singular gravitational potentials
\begin{equation} 
\gamma^{\alpha\beta}_{\rm S}(x) = 4 m \int_\gamma 
G^{\ \alpha\beta}_{{\rm S}\ \mu\nu}(x,z) u^\mu u^\nu\, d\tau 
\label{18.5.1}
\end{equation}
are solutions to the wave equation of Eq.~(\ref{18.1.22}); the
singular Green's function was introduced in Sec.~\ref{15.4}. We will
see that the singular field $\gamma^{\rm S}_{\alpha\beta;\gamma}$
reproduces the singular behaviour of the retarded solution near the
world line, and that the difference, 
$\gamma^{\rm R}_{\alpha\beta;\gamma} = 
\gamma_{\alpha\beta;\gamma} - \gamma^{\rm S}_{\alpha\beta;\gamma}$, 
is regular on the world line. 

To evaluate the integral of Eq.~(\ref{18.5.1}) we take $x$ to be close
to the world line (see Fig.~9), and we invoke Eq.~(\ref{15.4.8}) as
well as the Hadamard construction of Eq.~(\ref{15.4.14}). This gives 
\begin{equation} 
\gamma^{\alpha\beta}_{\rm S}(x) = \frac{2m}{r} 
U^{\alpha\beta}_{\ \ \gamma'\delta'} u^{\gamma'} u^{\delta'} 
+ \frac{2m}{r_{\rm adv}} U^{\alpha\beta}_{\ \ \gamma''\delta''}
u^{\gamma''} u^{\delta''} 
- 2 m \int_u^v V^{\alpha\beta}_{\ \ \mu\nu}(x,z) u^\mu u^\nu\, d\tau, 
\label{18.5.2}
\end{equation} 
where primed indices refer to the retarded point $x' := z(u)$,
double-primed indices refer to the advanced point $x'' := z(v)$,
and where $r_{\rm adv} := -\sigma_{\alpha''} u^{\alpha''}$ is the
advanced distance between $x$ and the world line. 

Differentiation of Eq.~(\ref{18.5.2}) yields 
\begin{eqnarray} 
\gamma^{\rm S}_{\alpha\beta;\gamma}(x) &=&    
-\frac{2m}{r^2}
U_{\alpha\beta\alpha'\beta'} u^{\alpha'} u^{\beta'} \partial_\gamma r 
-\frac{2m}{{r_{\rm adv}}^2}
U_{\alpha\beta\alpha''\beta''} u^{\alpha''} u^{\beta''}
\partial_\gamma r_{\rm adv} 
+ \frac{2m}{r} U_{\alpha\beta\alpha'\beta';\gamma} u^{\alpha'}
u^{\beta'} 
\nonumber \\ & & \mbox{} 
+ \frac{2m}{r} U_{\alpha\beta\alpha'\beta';\gamma'} u^{\alpha'}
u^{\beta'} u^{\gamma'} \partial_\gamma u
+ \frac{2m}{r_{\rm adv}} U_{\alpha\beta\alpha''\beta'';\gamma}
u^{\alpha''} u^{\beta''} 
+ \frac{2m}{r_{\rm adv}} U_{\alpha\beta\alpha''\beta'';\gamma''}
u^{\alpha''} u^{\beta''} u^{\gamma''} \partial_\gamma v
\nonumber \\ & & \mbox{} 
+ 2m V_{\alpha\beta\alpha'\beta'} u^{\alpha'} u^{\beta'}
\partial_\gamma u    
- 2m V_{\alpha\beta\alpha''\beta''} u^{\alpha''} u^{\beta''}
\partial_\gamma v
- 2m \int_u^v \nabla_\gamma V_{\alpha\beta\mu\nu}(x,z) 
u^\mu u^\nu\, d\tau, 
\label{18.5.3}
\end{eqnarray}
and we would like to express this as an expansion in powers of
$r$. For this we will rely on results already established in 
Sec.~\ref{18.3}, as well as additional expansions that will involve 
the advanced point $x''$. We recall that a relation between retarded
and advanced times was worked out in Eq.~(\ref{10.4.2}), that an
expression for the advanced distance was displayed in
Eq.~(\ref{10.4.3}), and that Eqs.~(\ref{10.4.4}) and (\ref{10.4.5})
give expansions for $\partial_\gamma v$ and 
$\partial_\gamma r_{\rm adv}$, respectively; these results can be
simplified by setting $a_a = \dot{a}_0 = \dot{a}_a = 0$, which is
appropriate in this computation.  

To derive an expansion for $U_{\alpha\beta\alpha''\beta''}
u^{\alpha''} u^{\beta''}$ we follow the general method of
Sec.~\ref{10.4} and introduce the functions $U_{\alpha\beta}(\tau)
:= U_{\alpha\beta\mu\nu}(x,z) u^\mu u^\nu$. We have that  
\[
U_{\alpha\beta\alpha''\beta''} u^{\alpha''} u^{\beta''} :=
U_{\alpha\beta}(v) = U_{\alpha\beta}(u)  
+ \dot{U}_{\alpha\beta}(u) \Delta^{\!\prime} 
+ \frac{1}{2} \ddot{U}_{\alpha\beta}(u) \Delta^{\!\prime 2}  
+ O\bigl( \Delta^{\!\prime 3} \bigr),  
\]
where overdots indicate differentiation with respect to $\tau$ and
$\Delta^{\!\prime} := v-u$. The leading term $U_{\alpha\beta}(u) 
:= U_{\alpha\beta\alpha'\beta'} u^{\alpha'} u^{\beta'}$ was worked
out in Eq.~(\ref{18.3.3}), and the derivatives of
$U_{\alpha\beta}(\tau)$ are given by  
\[
\dot{U}_{\alpha\beta}(u) = U_{\alpha\beta\alpha'\beta';\gamma'}
u^{\alpha'} u^{\beta'} u^{\gamma'} = g^{\alpha'}_{\ (\alpha} 
g^{\beta'}_{\ \beta)} \Bigl[ r R_{\alpha'0d0}\Omega^d u_{\beta'}  
+ O(r^2) \Bigr]   
\] 
and 
\[
\ddot{U}_{\alpha\beta}(u) =
U_{\alpha\beta\alpha'\beta';\gamma'\delta'} u^{\alpha'} u^{\beta'}
u^{\gamma'} u^{\delta'} = O(r),  
\] 
according to Eqs.~(\ref{18.3.5}) and (\ref{15.2.9}). Combining these
results together with Eq.~(\ref{10.4.2}) for $\Delta^{\!\prime}$ gives    
\begin{equation} 
U_{\alpha\beta\alpha''\beta''} u^{\alpha''} u^{\beta''} = 
g^{\alpha'}_{\ (\alpha} g^{\beta'}_{\ \beta)} \Bigl[ u_{\alpha'}
u_{\beta'} + 2 r^2 R_{\alpha'0d0} \Omega^d u_{\beta'} + O(r^3) \Bigr], 
\label{18.5.4}
\end{equation} 
which should be compared with Eq.~(\ref{18.3.3}). It should be
emphasized that in Eq.~(\ref{18.5.4}) and all equations below, all
frame components are evaluated at the retarded point $x'$,
and not at the advanced point. The preceding computation gives us also
an expansion for  
\[
U_{\alpha\beta\alpha''\beta'';\gamma''} u^{\alpha'} u^{\beta''}
u^{\gamma''} = \dot{U}_{\alpha\beta}(u) + \ddot{U}_{\alpha\beta}(u)
\Delta^{\!\prime} + O(\Delta^{\!\prime 2}), 
\]
which becomes     
\begin{equation}
U_{\alpha\beta\alpha''\beta'';\gamma''} u^{\alpha''} u^{\beta''}
u^{\gamma''} = g^{\alpha'}_{\ (\alpha} g^{\beta'}_{\ \beta)} \Bigl[ 
r R_{\alpha'0d0} \Omega^d u_{\beta'} + O(r^2) \Bigr], 
\label{18.5.5}
\end{equation}
and which is identical to Eq.~(\ref{18.3.5}).    

We proceed similarly to obtain an expansion for
$U_{\alpha\beta\alpha''\beta'';\gamma} u^{\alpha''} u^{\beta''}$. Here
we introduce the functions $U_{\alpha\beta\gamma}(\tau) :=
U_{\alpha\beta\mu\nu;\gamma} u^\mu u^\nu$ and express
$U_{\alpha\beta\alpha''\beta'';\gamma} u^{\alpha''} u^{\beta''}$ as
$U_{\alpha\beta\gamma}(v) = U_{\alpha\beta\gamma}(u) +
\dot{U}_{\alpha\beta\gamma}(u) \Delta^{\!\prime} 
+ O(\Delta^{\!\prime 2})$. The leading term $U_{\alpha\beta\gamma}(u)
:= U_{\alpha\beta\alpha'\beta';\gamma} u^{\alpha'} u^{\beta'}$ was
computed in Eq.~(\ref{18.3.4}), and   
\[
\dot{U}_{\alpha\beta\gamma}(u) =
U_{\alpha\beta\alpha'\beta';\gamma\gamma'} u^{\alpha'} u^{\beta'}
u^{\gamma'} = g^{\alpha'}_{\ (\alpha} g^{\beta'}_{\ \beta)}
g^{\gamma'}_{\ \gamma} \Bigl[ R_{\alpha'0\gamma'0} u_{\beta'} + O(r)
\Bigr] 
\]
follows from Eq.~(\ref{15.2.8}). Combining these results together with  
Eq.~(\ref{10.4.2}) for $\Delta^{\!\prime}$ gives
\begin{equation}
U_{\alpha\beta\alpha''\beta'';\gamma} u^{\alpha''} u^{\beta''} =    
g^{\alpha'}_{\ (\alpha} g^{\beta'}_{\ \beta)} g^{\gamma'}_{\ \gamma} 
\Bigl[ r\bigl( R_{\alpha'0\gamma'0} 
- R_{\alpha'0\gamma'd} \Omega^d \bigr) u_{\beta'} + O(r^2) \Bigr],     
\label{18.5.6}
\end{equation}
and this should be compared with Eq.~(\ref{18.3.4}). The last
expansion we shall need is  
\begin{equation}
V_{\alpha\beta\alpha''\beta''} u^{\alpha''} u^{\beta''} = 
g^{\alpha'}_{\ (\alpha} g^{\beta'}_{\ \beta)} \Bigl[
R_{\alpha'0\beta'0} + O(r) \Bigr], 
\label{18.5.7}
\end{equation} 
which is identical to Eq.~(\ref{18.3.6}). 

We obtain the frame components of the singular gravitational field by
substituting these expansions into Eq.~(\ref{18.5.3}) and projecting
against the tetrad  $(\base{\alpha}{0},\base{\alpha}{a})$. 
After some algebra we arrive at    
\begin{eqnarray} 
\gamma^{\rm S}_{000}(u,r,\Omega^a) &:=& 
\gamma^{\rm S}_{\alpha\beta;\gamma}(x) 
\base{\alpha}{0}(x) \base{\beta}{0}(x) \base{\gamma}{0}(x) 
= 2m R_{a0b0} \Omega^a \Omega^b + O(r), 
\label{18.5.8} \\ 
\gamma^{\rm S}_{0b0}(u,r,\Omega^a) &:=& 
\gamma^{\rm S}_{\alpha\beta;\gamma}(x) 
\base{\alpha}{0}(x) \base{\beta}{b}(x) \base{\gamma}{0}(x) 
= -4m R_{b0c0} \Omega^c + O(r),
\label{18.5.9} \\ 
\gamma^{\rm S}_{ab0}(u,r,\Omega^a) &:=& 
\gamma^{\rm S}_{\alpha\beta;\gamma}(x) 
\base{\alpha}{a}(x) \base{\beta}{b}(x) \base{\gamma}{0}(x) 
= O(r), 
\label{18.5.10} \\ 
\gamma^{\rm S}_{00c}(u,r,\Omega^a) &:=& 
\gamma^{\rm S}_{\alpha\beta;\gamma}(x) 
\base{\alpha}{0}(x) \base{\beta}{0}(x) \base{\gamma}{c}(x) 
\nonumber \\ 
&=& -4m \biggl[ \Bigl(\frac{1}{r^2} + \frac{1}{3} R_{a0b0} \Omega^a
\Omega^b \Bigr) \Omega_c + \frac{1}{6} R_{c0b0} \Omega^b - \frac{1}{6}
R_{ca0b} \Omega^a \Omega^b \biggr] + O(r),
\label{18.5.11} \\ 
\gamma^{\rm S}_{0bc}(u,r,\Omega^a) &:=& 
\gamma^{\rm S}_{\alpha\beta;\gamma}(x) 
\base{\alpha}{0}(x) \base{\beta}{b}(x) \base{\gamma}{c}(x) 
= 2m \bigl( R_{b0cd} \Omega^d + R_{b0d0} \Omega^d
\Omega_c \bigr) + O(r), 
\label{18.5.12} \\ 
\gamma^{\rm S}_{abc}(u,r,\Omega^a) &:=& 
\gamma^{\rm S}_{\alpha\beta;\gamma}(x) 
\base{\alpha}{a}(x) \base{\beta}{b}(x) \base{\gamma}{c}(x) 
= -4m R_{a0b0} \Omega_c + O(r), 
\label{18.5.13}
\end{eqnarray} 
in which all frame components are evaluated at the retarded point
$x'$. Comparison of these expressions with 
Eqs.~(\ref{18.3.7})--(\ref{18.3.12}) reveals identical singularity 
structures for the retarded and singular gravitational fields.  

The difference between the retarded field of
Eqs.~(\ref{18.3.7})--(\ref{18.3.12}) and the singular field of
Eqs.~(\ref{18.5.8})--(\ref{18.5.13}) defines the regular
gravitational field $\gamma^{\rm R}_{\alpha\beta;\gamma}$. Its 
frame components are 
\begin{eqnarray} 
\gamma^{\rm R}_{000} &=& \gamma^{\rm tail}_{000} + O(r), 
\label{18.5.14} \\ 
\gamma^{\rm R}_{0b0} &=& \gamma^{\rm tail}_{0b0} + O(r),
\label{18.5.15} \\ 
\gamma^{\rm R}_{ab0} &=& 4m R_{a0b0} + \gamma^{\rm tail}_{ab0} + O(r), 
\label{18.5.16} \\ 
\gamma^{\rm R}_{00c} &=& \gamma^{\rm tail}_{00c} + O(r),
\label{18.5.17} \\ 
\gamma^{\rm R}_{0bc} &=& 2m R_{b0c0} + \gamma^{\rm tail}_{0bc} + O(r),  
\label{18.5.18} \\ 
\gamma^{\rm R}_{abc} &=& \gamma^{\rm tail}_{abc} + O(r), 
\label{18.5.19}
\end{eqnarray} 
and we see that $\gamma^{\rm R}_{\alpha\beta;\gamma}$ is regular in the
limit $r \to 0$. We may therefore evaluate the regular field
directly at $x=x'$, where the tetrad $(\base{\alpha}{0},
\base{\alpha}{a})$ coincides with $(u^{\alpha'},
\base{\alpha'}{a})$. After reconstructing the field at $x'$ from its
frame components, we obtain    
\begin{equation}
\gamma^{\rm R}_{\alpha'\beta';\gamma'}(x') =  
-4 m \Bigl( u_{(\alpha'} R_{\beta')\delta'\gamma'\epsilon'} 
+ R_{\alpha'\delta'\beta'\epsilon'} u_{\gamma'} \Bigr) u^{\delta'}
u^{\epsilon'} + \gamma^{\rm tail}_{\alpha'\beta'\gamma'},  
\label{18.5.20}
\end{equation}
where the tail term can be copied from Eq.~(\ref{18.3.2}), 
\begin{equation} 
\gamma^{\rm tail}_{\alpha'\beta'\gamma'}(x') 
= 4m \int_{-\infty}^{u^-} \nabla_{\gamma'}
G_{+\alpha'\beta'\mu\nu}(x',z) u^\mu u^\nu\, d\tau. 
\label{18.5.21}
\end{equation}
The tensors that appear in Eq.~(\ref{18.5.21}) all refer to the  
retarded point $x' := z(u)$, which can now be treated as an
arbitrary point on the world line $\gamma$.  

\subsection{Equations of motion}
\label{18.6}

The retarded gravitational field $\gamma_{\alpha\beta;\gamma}$ of a
point particle is singular on the world line, and this behaviour makes
it difficult to understand how the field is supposed to act on the
particle and influence its motion. The field's singularity structure
was analyzed in Secs.~\ref{18.3} and \ref{18.4}, and in
Sec.~\ref{18.5} it was shown to originate from the singular field
$\gamma^{\rm S}_{\alpha\beta;\gamma}$; the regular field 
$\gamma^{\rm R}_{\alpha\beta;\gamma}$ was then shown to be regular on 
the world line. 

To make sense of the retarded field's action on the particle we can
follow the discussions of Sec.~\ref{16.6} and \ref{17.6} and postulate
that the self gravitational field of the point particle is either
$\langle \gamma_{\mu\nu;\lambda} \rangle$, as worked out in
Eq.~(\ref{18.4.13}), or $\gamma^{\rm R}_{\mu\nu;\lambda}$, as worked
out in Eq.~(\ref{18.5.20}). These regularized fields are both given by  
\begin{equation}
\gamma^{\rm reg}_{\mu\nu;\lambda} 
= -4 m \Bigl( u_{(\mu} R_{\nu)\rho\lambda\xi}
+ R_{\mu\rho\nu\xi} u_\lambda \Bigr) u^\rho u^\xi 
+ \gamma^{\rm tail}_{\mu\nu\lambda}
\label{18.6.1}
\end{equation}
and 
\begin{equation}
\gamma^{\rm tail}_{\mu\nu\lambda} = 4 m \int_{-\infty}^{\tau^-}
\nabla_\lambda G_{+\mu\nu\mu'\nu'}\bigl( z(\tau), z(\tau') \bigr)
u^{\mu'} u^{\nu'}\, d\tau', 
\label{18.6.2} 
\end{equation}     
in which all tensors are now evaluated at an arbitrary point $z(\tau)$
on the world line $\gamma$.  

The actual gravitational perturbation $h_{\alpha\beta}$ is obtained by
inverting Eq.~(\ref{18.1.10}), which leads to $h_{\mu\nu;\lambda} =
\gamma_{\mu\nu;\gamma} - \frac{1}{2} g_{\mu\nu} 
\gamma^\rho_{\ \rho;\lambda}$. Substituting Eq.~(\ref{18.6.1}) yields  
\begin{equation} 
h^{\rm reg}_{\mu\nu;\lambda} 
= -4 m \Bigl( u_{(\mu} R_{\nu)\rho\lambda\xi}
+ R_{\mu\rho\nu\xi} u_\lambda \Bigr) u^\rho u^\xi 
+ h^{\rm tail}_{\mu\nu\lambda}, 
\label{18.6.3}
\end{equation}
where the tail term is given by the trace-reversed counterpart to 
Eq.~(\ref{18.6.2}): 
\begin{equation}
h^{\rm tail}_{\mu\nu\lambda} = 4 m \int_{-\infty}^{\tau^-}
\nabla_\lambda \biggl( G_{+\mu\nu\mu'\nu'}
- \frac{1}{2} g_{\mu\nu} G^{\ \ \rho}_{+\ \rho\mu'\nu'}
\biggr) \bigl( z(\tau), z(\tau')\bigr) u^{\mu'} u^{\nu'}\, d\tau'. 
\label{18.6.4} 
\end{equation}     
When this regularized field is substituted into Eq.~(\ref{18.1.13}),
we find that the terms that depend on the Riemann tensor cancel out,
and we are left with 
\begin{equation}
\frac{D u^\mu}{d\tau} = -\frac{1}{2} \bigl( g^{\mu\nu} + u^\mu
u^\nu \bigr) \bigl( 2 h^{\rm tail}_{\nu\lambda\rho} 
- h^{\rm tail}_{\lambda\rho\nu} \bigr) u^\lambda u^\rho. 
\label{18.6.5} 
\end{equation}   
We see that only the tail term is involved
in the final form of the equations of motion. The tail integral of
Eq.~(\ref{18.6.4}) involves the current position $z(\tau)$ of the
particle, at which all tensors with unprimed indices are evaluated,
as well as all prior positions $z(\tau')$, at which all tensors with
primed indices are evaluated. The tail integral is cut short at
$\tau' = \tau^- := \tau - 0^+$ to avoid the singular behaviour of
the retarded Green's function at coincidence; this limiting procedure
was justified at the beginning of Sec.~\ref{18.3}.    
 
Equation (\ref{18.6.5}) was first derived by Yasushi Mino, Misao
Sasaki, and Takahiro Tanaka in 1997 \cite{mino-etal:97a}. (An
incomplete treatment had been given previously by Morette-DeWitt and
Ging \cite{morette-ging:60}.) An alternative derivation was then
produced, also in 1997, by Theodore C.\ Quinn and Robert M.\ Wald
\cite{quinn-wald:97}. These equations are now known as the MiSaTaQuWa
equations of motion, and other derivations \cite{gralla-wald:08,
  pound:10a}, based on an extended-body approach, will be reviewed
below in Part~\ref{part5}. It should be noted that Eq.~(\ref{18.6.5})
is formally equivalent to the statement that the point particle moves
on a geodesic in a spacetime with metric 
$g_{\alpha\beta} + h^{\rm R}_{\alpha\beta}$, where 
$h^{\rm R}_{\alpha\beta}$ is the regular metric perturbation
obtained by trace-reversal of the potentials 
$\gamma^{\rm R}_{\alpha\beta} := \gamma_{\alpha\beta} 
- \gamma^{\rm S}_{\alpha\beta}$; this perturbed metric is regular on
the world line, and it is a solution to the vacuum field
equations. This elegant interpretation of the MiSaTaQuWa equations was  
proposed in 2003 by Steven Detweiler and Bernard F.\ Whiting
\cite{detweiler-whiting:03}. Quinn and Wald \cite{quinn-wald:99} have
shown that under some conditions, the total work done by the
gravitational self-force is equal to the energy radiated (in
gravitational waves) by the particle.

%% file: part5.tex
%
\section{Point-particle limits and matched asymptotic expansions}  
\label{20}

The expansion presented in the previous section is based on an exact
point-particle source. But in the full, nonlinear theory, no
distributional solution would exist for such a source
\cite{geroch-traschen:87}. Although the expansion nevertheless yields
a well-behaved linear approximation, it is ill-behaved beyond that
order, since the second- and higher-order Einstein tensors will
contain products of delta functions, which have no meaning as
distributions. It may be possible to overcome this limitation using
more advanced methods such as Colombeau algebras
\cite{steinbauer-vickers:06}, which allow for the multiplication of
distributions, but little work has been done to that end. Instead, the
common approach, and the one we shall pursue here, has been to abandon
the fiction of a point particle in favor of considering an
asymptotically small body. As we shall see, we can readily generalize
the self-consistent expansion scheme to this case. Furthermore, we
shall find that the results of the previous section are justified by
this approach: at linear order, the metric perturbation due to an
asymptotically small body is precisely that of a point particle moving
on a world line with an acceleration given by the MiSaTaQuWa equation
(plus higher-order corrections). 

In order for the body to be considered ``small,'' its mass and size
must be much smaller than all external lengthscales. We denote these
external scales collectively as $\mathscr{R}$, which we may define to
be the radius of curvature of the spacetime (were the small body
removed from it) in the region in which we seek an
approximation. Given this definition, a typical component of the
spacetime's Riemann tensor is equal to $1/\mathscr{R}^2$ up to a
numerical factor of order unity. Now, we consider a family of metrics
$g_{\alpha\beta}(\e)$ containing a body whose mass scales as $\e$ in
the limit $\e\to0$; that is, $\e\sim m/\mathscr{R}$. If each member of
the family is to contain a body of the same type, then the size of the
body must also approach zero with $\e$. The precise scaling of size
with $\e$ is determined by the type of body, but it is not generally
relevant. What \emph{is} relevant is the ``gravitational size" --- the
length scale determining the metric outside the body --- and this size 
always scales linearly with the mass. If the body is compact, as is a
neutron star or a black hole, then its gravitational size is also its
actual linear size. In what remains, we assume that all lengths have
been scaled by $\mathscr{R}$, such that we can write, for example,
$m\ll1$. Our goal is to determine the metric perturbation and the
equation of motion produced by the body in this limit. 

Point-particle limits such as this have been used to derive equations
of motion many times in the past, including in derivations of geodesic
motion at leading order
\cite{infeld-schild:49,geroch-jang:75,ehlers-geroch:03} and in
constructing post-Newtonian limits \cite{futamase-itoh:07}. Perhaps
the most obvious means of approaching the problem is to first work
nonperturbatively, with a body of arbitrary size, and then take the
limit. Using this approach (though with some restrictions on the
body's size and compactness) and generalized definitions of momenta,
Harte has calculated the self-force in the case of scalar
\cite{harte:08} and electromagnetic~\cite{harte:09a} charge
distributions in fixed backgrounds, following the earlier work of
Dixon \cite{dixon:70a, dixon:70b, dixon:74}. However, while this
approach is conceptually compelling, at this stage it applies only to
material bodies, not black holes, and has not yet been presented as
part of a systematic expansion of the Einstein equation. Here, we
focus instead on a more general method. 

Alternatively, one could take the opposite approach, essentially
taking the limit first and then trying to recover the higher-order
effects, by treating the body as an \emph{effective} point particle at
leading order, with finite size effects introduced as higher-order
effective fields, as done by Galley and Hu
\cite{galley-etal:06, galley-hu:09}. However, while this approach is
computationally efficient, allowing one to perform high-order
calculations with (relative) ease, it requires methods such as
dimensional regularization and mass renormalization in order to arrive
at meaningful results. Because of these undesirable requirements, we
will not consider it here. 

In the approach we review, we make use of the method of matched
asymptotic expansions \cite{death:75, death:96,
  kates:80a,kates:80b,thorne-hartle:85, mino-etal:97a,
  mino-etal:97b,alvi:00,poisson:04a,detweiler:05,
  futamase-itoh:07,taylor-poisson:08,gralla-wald:08, gralla-etal:09,
  pound:10a,pound:10b}. Broadly speaking, this method consists of
constructing two different asymptotic expansions, each valid in a
specific region, and combining them to form a global expansion. In the
present context, the method begins with two types of point-particle
limits: an \emph{outer limit}, in which $\e\to0$ at fixed coordinate
values (we will slightly modify this in a moment); and an 
\emph{inner limit}, in which $\e\to0$ at fixed values of 
$\tilde R:= R/\e$, where $R$ is 
a measure of radial distance from the body. In the outer limit, the
body shrinks toward zero size as all other distances remain roughly
constant; in the inner limit, the small body keeps a constant size
while all other distances blow up toward infinity. Thus, the inner
limit serves to ``zoom in" on a small region around the body. The
outer limit can be expected to be valid in regions where $R\sim 1$,
while the inner limit can be expected to be valid in regions where
$\tilde R\sim 1$ (or $R\sim\e$), though both of these regions can be
extended into larger domains. 

More precisely, consider an exact solution $\exact{g}_{\alpha\beta}$
on a manifold $\man_\e$ with two coordinate systems: a local
coordinate system $X^\alpha=(T,R,\Theta^A)$ that is centered (in some
approximate sense) on the small body, and a global coordinate system
$x^\alpha$. For example, in an extreme-mass-ratio inspiral, the local
coordinates might be the Schwarzschild-type coordinates of the small
body, and the global coordinates might be the Boyer-Lindquist
coordinates of the supermassive Kerr black hole. In the outer limit,
we expand $\exact{g}_{\alpha\beta}$ for small $\e$ while holding
$x^\alpha$ fixed. The leading-order solution in this case is the
background metric $g_{\alpha\beta}$ on a manifold $\man_E$; this is
the external spacetime, which contains no small body. It might, for
example, be the spacetime of the supermassive black hole. In the inner
limit, we expand $\exact{g}_{\alpha\beta}$ for small $\e$ while
holding $(T,\tilde R,\Theta^A)$ fixed. The leading-order solution in
this case is the metric $g^{\rm body}_{\alpha\beta}$ on a manifold
$\man_I$; this is the spacetime of the small body if it were isolated
(though it may include slow evolution due to its interaction with the
external spacetime --- this will be discussed below). Note that
$\man_E$ and $\man_I$ generically differ: in an extreme-mass-ratio
inspiral, for example, if the small body is a black hole, then
$\man_I$ will contain a spacelike singularity in
the black hole's interior, while $\man_E$ will be smooth at the
``position'' where the small black hole would be. What we are
interested in is that ``position'' --- the world line in the smooth
external spacetime $\man_E$ that represents the motion of the small
body. Note that this world line generically appears \emph{only} in the
external spacetime, rather than as a curve in the exact spacetime
$(\exact{g}_{\alpha\beta},\man_\e)$; in fact, if the small body is a
black hole, then obviously no such curve exists. 

Determining this world line presents a fundamental problem. In the
outer limit, the body vanishes at $\e=0$, leaving only a remnant,
$\e$-independent curve in $\man_E$. (Outside any small body, the
metric will contain terms such as $m/R$, such that in the limit
$m\to0$, the limit exists everywhere except at $R=0$, which leaves a
removable discontinuity in the external spacetime; the removal of this
discontinuity defines the remnant world line of the small body.) But
the true motion of the body will generically be $\e$-dependent. If we
begin with the remnant world line and correct it with the effects of the
self-force, for example, then the corrections must be small: they are
small deviation vectors defined on the remnant world line. Put another
way, if we expand $\exact{g}_{\alpha\beta}$ in powers of $\e$, then
all functions in it must similarly be expanded, including any
representation of the motion, and in particular, any representative
world line. We would then have a representation of the form
$z^\alpha(t,\e)=z_{(0)}^\alpha(t)+\e z_{(1)}^\alpha(t)+\ldots$, where
$z_{(1)}^\alpha(t)$ is a vector defined on the remnant curve described
by $z_{(0)}^\alpha(t)$. The remnant curve would be a geodesic, and the
small corrections would incorporate the self-force and finite-size
effects \cite{gralla-wald:08} (see also \cite{kates:80a}). However,
because the body will generically drift away from any such geodesic,
the small corrections will generically grow large with time, leading
to the failure of the regular expansion. So we will modify this
approach by performing a self-consistent expansion in the outer limit,
following the same scheme as presented in the point-particle
case. Refs.~\cite{pound:10a,pound:10b,pound:10c} contain far more
detailed discussions of these points. 

Regardless of whether the self-consistent expansion is used, the
success of matched asymptotic expansions relies on the buffer region
defined by $\e\ll R\ll 1$ (see Fig.~10). In this region, both the
inner and outer expansions are valid. From the perspective of the
outer expansion, this corresponds to an asymptotically small region
around the world line: $R\ll 1$. From the perspective of the inner
expansion, it corresponds to asymptotic spatial infinity: 
$1/\tilde R=\e/R\ll 1$. Because both expansions are valid in this
region, and because both are expansions of the same exact metric 
$\exact{g}_{\alpha\beta}$ and hence must ``match,'' by working in the
buffer region we can use information from the inner expansion to
determine information about the outer expansion (or vice versa). We
shall begin by solving the Einstein equation in the buffer region,
using information from the inner expansion to determine the form of
the external metric perturbation therein. In so doing, we shall
determine the acceleration of the small body's world line. Finally,
using the field values in the buffer region, we shall construct a
global solution for the metric perturbation.  

In this calculation, the structure of the body is left
unspecified. Our only condition is that part of the buffer region must  
lie outside the body, because we wish to solve the Einstein field
equations in vacuum.
This requires the body to be sufficiently compact. For example, our
calculation would fail for a diffuse body such as our Sun; likewise,
it would fail if a body became tidally disrupted. Although we will
detail only the case of an uncharged body, the same techniques would
apply to charged bodies; Gralla {\it et al.}~\cite{gralla-etal:09}
have recently performed a similar calculation for the electromagnetic
self-force on an asymptotically small body in a flat background
spacetime. Using very different methods, 
Futamase {\it et al.}~\cite{futamase-etal:08} have calculated
equations of motion for an asymptotically small charged black hole.  

The structure of our discussion is as follows: In
Sec.~\ref{self_consistent_expansion}, we present the self-consistent
expansion of the Einstein equation. Next, in Sec.~\ref{buffer_region},
we solve the equations in the buffer region up to second order in the
outer expansion. Last, in Sec.~\ref{global_solution}, we discuss the
global solution in the outer expansion and show that it is that of a
point particle at first order. Over the course of this calculation, we
will take the opportunity to incorporate several details that we could
have accounted for in the point-particle case but opted to neglect for
simplicity: an explicit expansion of the acceleration vector that
makes the self-consistent expansion properly systematic, and a finite
time domain that accounts for the fact that large errors eventually 
accumulate if the approximation is truncated at any finite order.
For more formal discussions of matched asymptotic expansions in
general relativity, see Refs.~\cite{kates:81, pound:10b}; the latter
reference, in particular, discusses the method as it pertains to the
motion of small bodies. For background on the use of matched
asymptotic expansions in applied mathematics, see 
Refs.~\cite{eckhaus:79, holmes:95, kevorkian-cole:96,   
lagerstrom:88,verhulst:05}; the text by Eckhaus~\cite{eckhaus:79}
provides the most rigorous treatment. 

\begin{figure}
\begin{center}
\vspace*{-5pt} 
\includegraphics[width=0.5\linewidth]{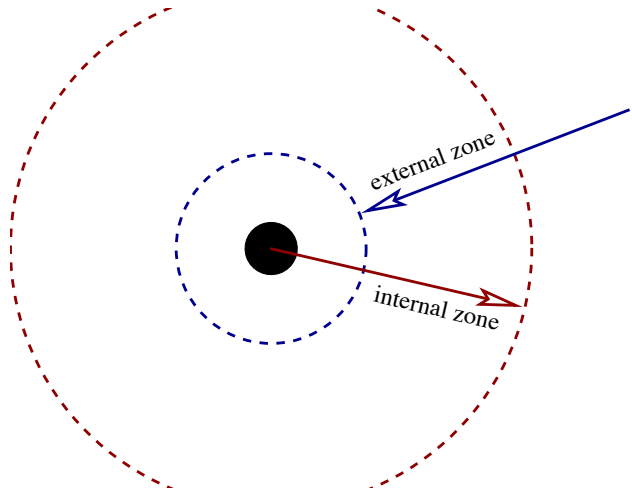}
\vspace*{-5pt}
\end{center} 
\caption{A small body, represented by the black disk, is immersed in 
a background spacetime. The internal zone is defined by $R\ll 1$,
while the external zone is defined by $R\gg \e$. Since $\e \ll 1$,
there exists a buffer region defined by $\e\ll R\ll 1$. In the buffer
region $\e/R$ and $R$ are both small.}   
\end{figure} 

\section{Self-consistent expansion}
\label{self_consistent_expansion}

\subsection{Introduction} 

We wish to represent the motion of the body through the external
background spacetime $(g_{\alpha\beta},\man_E)$, rather than through
the exact spacetime $(\exact{g}_{\alpha\beta},\man_\e)$. In order to
achieve this, we begin by surrounding the body with a (hollow,
three-dimensional) world tube $\Gamma$ embedded in the buffer
region. We define the tube to be a surface of constant radius
$s=\rad(\e)$ in Fermi normal coordinates centered on a world line
$\gamma\subset\man_E$, though the exact definition of the tube is
immaterial. Since there exists a diffeomorphism between $\man_E$ and
$\man_I$ in the buffer region, this defines a tube
$\Gamma_I\subset\man_I$. Now, the problem is the following: what
equation of motion must $\gamma$ satisfy in order for $\Gamma_I$ to be
``centered'' about the body? 

How shall we determine if the body lies at the centre of the tube's
interior? Since the tube is close to the small body (relative to all
external length scales), the metric on the tube is primarily
determined by the small body's structure. Recall that the buffer
region corresponds to an asymptotically large spatial distance in the
inner expansion. Hence, on the tube, we can construct a multipole
expansion of the body's field, with the form $\sum R^{-n}$ (or 
$\sum s^{-n}$---we will assume $s\sim R$ in the buffer
region). Although alternative definitions could be used, we define the
tube to be centered about the body if the mass dipole moment vanishes
in this expansion. Note that this is the typical approach in general
relativity: Whereas in Newtonian mechanics one directly finds the
equation of motion for the centre of mass of a body, in general
relativity one typically seeks a world line about which the mass
dipole of the body vanishes (or an equation of motion for the mass
dipole relative to a given nearby world line) 
\cite{einstein-infeld:49, racine-flanagan:05,
  gralla-wald:08, pound:10a}. This definition of the world line is
sufficiently general to apply to a black hole. If the body is
material, one could instead imagine a centre-of-mass world line that
lies in the interior of the body in the exact spacetime. This world
line would then be the basis of our self-consistent expansion. We use
our more general definition to cover both cases. 
See Ref.~\cite{thorne:80} and references 
therein for discussion of multipole expansions in general relativity,
see Refs.~\cite{thorne:80, thorne-hartle:85} for discussions of
mass-centered coordinates in the buffer region, and see, e.g.,
Refs.~\cite{schattner:79, ehlers-rudolph:77} for alternative
definitions of centre of mass in general relativity.  

As in the point-particle case, in order to determine the equation of
motion of the world line, we consider a family of metrics, now denoted 
$g_E(x,\e;\gamma)$, parametrized by $\gamma$, such that when $\gamma$ 
is given by the correct equation of motion for a given value of $\e$,
we have $g_E(x,\e;\gamma(\e))=\exact{g}(x)$. The metric in the outer
limit is thus taken to be the general expansion 
\begin{equation}\label{external ansatz}
\exact{g}_{\alpha\beta}(x,\e) = g_{E\alpha\beta}(x,\e;\gamma)
= g_{\alpha\beta}(x)+h_{\alpha\beta}(x,\e;\gamma),
\end{equation}
where
\begin{equation}
h_{\alpha\beta}(x,\e;\gamma) = 
\sum_{n=1}^{\infty}\e^n\hmn{\alpha\beta}{\emph{n}}(x;\gamma).
\end{equation}
In the point-particle case, solving Einstein's equations determined
the equation of motion of the particle's world line; in this case, it
will determine the world line $\gamma$ for which the inner expansion is
mass-centered. In this self-consistent expansion, the perturbations
produced by the body are constructed about a fixed world line
determined by the particular value of $\e$ at which one seeks an
approximation. 

In the remainder of this section, we present a sequence of
perturbation equations that arise in this expansion scheme, along with
a complementary sequence for the inner expansion. 
 
\subsection{Field equations in outer expansion}

\begin{figure}
\begin{center}
\vspace*{-5pt} 
\includegraphics[width=0.5\linewidth]{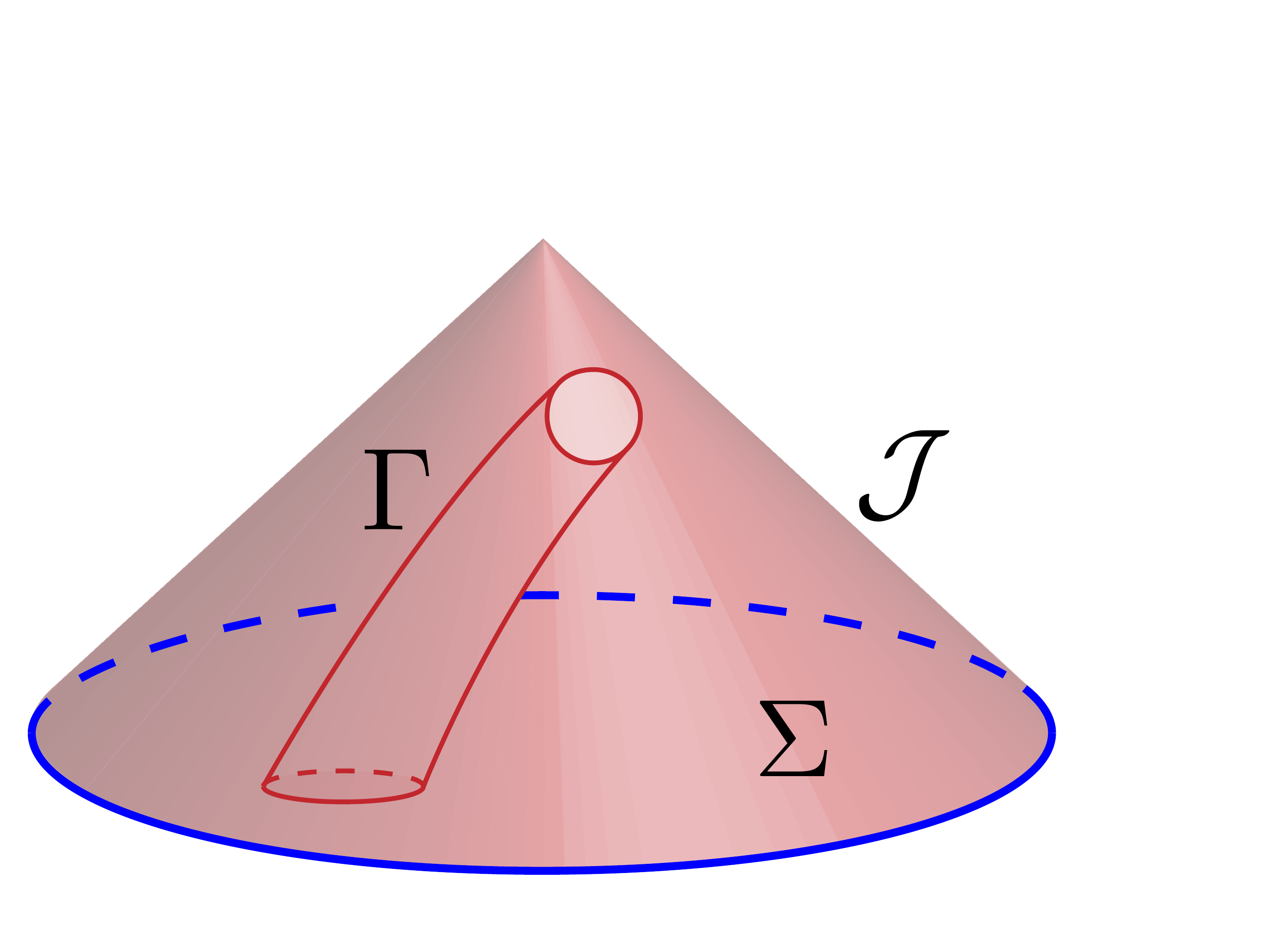}
\vspace*{-5pt}
\end{center} 
\caption[The vacuum region $\Omega$]{The spacetime region $\Omega$ is 
  bounded by the union of the spacelike surface $\Sigma$, the timelike
  tube $\Gamma$, and the null surface $\mathcal{J}$.}   
\label{volume}
\end{figure}

In the outer expansion, we seek a solution in a vacuum region $\Omega$
outside of $\Gamma$. We specify $\Omega\subset\man_E$ to be an open
set consisting of the future domain of dependence of the spacelike
initial-data surface $\Sigma$, excluding the interior of the world
tube $\Gamma$. This implies that the future boundary of $\Omega$ is a
null surface $\mathcal{J}$. Refer to Fig.~\ref{volume} for an
illustration. The boundary of the domain is $\partial\Omega
:= \Gamma\cup\mathcal{J}\cup\Sigma$. The spatial surface $\Sigma$
is chosen to intersect $\Gamma$ at the initial time $t=0$. 

Historically, in derivations of the self-force, solutions to the perturbative 
field equations were taken to be global in time, with tail integrals
extended to negative infinity, as we wrote them in the preceding
sections. But as was first noted in Ref.~\cite{pound:10a}, because the
self-force drives long-term, cumulative changes, any approximation
truncated at a given order will be accurate to that order only for a
finite time; and this necessites working in a finite region such as
$\Omega$. This is also true in the case of point charges and masses.   
For simplicity, we neglected this detail in the preceding sections,
but for completeness, we account for it here.

\subsubsection*{Field equations}  

Within this region, we follow the methods presented in the case of a
point mass. We begin by reformulating the Einstein equation such that
it can be solved for an arbitrary world line. To accomplish this, we
assume that the Lorenz gauge can be imposed on the whole of
$h_{\alpha\beta}$, everywhere in $\Omega$, such that
$L_\mu[h]=0$. Here 
\begin{equation} 
L_\mu[h] := \biggl( g^\alpha_{\ \mu} g^{\beta\nu} 
- \frac{1}{2} g^{\alpha\beta} g_{\mu}^{\ \nu} \biggr) 
\nabla_\nu h_{\alpha\beta} 
\end{equation} 
is the Lorenz-gauge operator that was first introduced in
Secs.~\ref{15.1} and \ref{18.1}; the condition $L_\mu[h] = 0$ is the
same statement as $\nabla_\nu \gamma^{\mu\nu} = 0$, where 
$\gamma_{\mu\nu} := h_{\mu\nu} - \frac{1}{2} g_{\mu\nu}
g^{\alpha\beta} h_{\alpha\beta}$ is the  ``trace-reversed'' metric
perturbation. We discuss the validity of this assumption below. 

Just as in the case of a point mass, this choice of gauge reduces
the vacuum Einstein equation $\exact{R}_{\mu\nu}=0$ to a weakly
nonlinear wave equation that can be expanded and solved at fixed
$\gamma$. However, we now seek a solution only in the region $\Omega$,
where the energy-momentum tensor vanishes, so the resulting sequence
of wave equations reads 
\begin{align}
E_{\mu\nu}[\hmn{}{1}] &=0, \label{h1 eqn}\\
E_{\mu\nu}[\hmn{}{2}] &=2\delta^2 R_{\mu\nu}[\hmn{}{1}], 
\label{h2 eqn}\\ 
&\vdots\nonumber
\end{align}
where 
\begin{equation} 
E_{\mu\nu}[h] := \bigl( g^{\alpha}_{\ \mu} g^\beta_{\ \nu} 
\nabla^\gamma \nabla_\gamma 
+ 2 R^{\ \alpha\ \beta}_{\mu\ \nu} \bigr) h_{\alpha\beta} 
\end{equation} 
is the tensorial wave operator that was first introduced in
Secs.~\ref{15.1} and \ref{18.1}, and the second-order Ricci tensor
$\delta^2 R_{\mu\nu}$, which is quadratic in its argument, is shown
explicitly in Eq~\eqref{second-order Ricci}. More generally, we can
write the $n$th-order equation as 
\begin{equation}
E_{\mu\nu}\big[\hmn{}{\emph{n}}\big] = 
S^{(n)}_{\mu\nu}\big[\hmn{}{1},\ldots,\hmn{}{n-1},\gamma\big],
\label{21.7} 
\end{equation}
where the source term $S^{(n)}_{\mu\nu}$ consists of nonlinear terms
in the expansion of the Ricci tensor. 

Again as in the case of a point particle, we can easily write down
formal solutions to the wave equations, for arbitrary $\gamma$. Using
the same methods as were used to derive the Kirchoff representation in
Sec.~\ref{15.3}, we find 
\begin{align}
\label{formal_solution}
\hmn{\alpha\beta}{n} = 
\frac{1}{4\pi}\!\oint\limits_{\partial\Omega}\!\!
\Big(G^+_{\alpha\beta}{}^{\gamma'\delta'} 
\del{\mu'}\hmn{\gamma'\delta'}{n} 
- \hmn{\gamma'\delta'}{n}\del{\mu'} 
G^+_{\alpha\beta}{}^{\gamma'\delta'}\Big) dS^{\mu'} 
+\frac{1}{4\pi}\int_\Omega G^+_{\alpha\beta}{}^{\gamma'\delta'} 
S^{(n)}_{\gamma'\delta'}dV'.
\end{align}
Because $\mathcal{J}$ is a future null surface, the integral over it
vanishes. Hence, this formal solution requires only initial data on
$\Sigma$ and boundary data on $\Gamma$. Since $\Gamma$ lies in the
buffer region, the boundary data on it is determined by information
from the inner expansion. 

One should note several important properties of these integral
representations: First, $x$ must lie in the interior of $\Omega$; an
alternative expression must be derived if $x$ lies on the boundary
\cite{roach:82}. Second, the integral over the boundary is, in each
case, a homogeneous solution to the wave equation, while the integral
over the volume is an inhomogeneous solution. Third, if the field
at the boundary satisfies the Lorenz gauge condition, then by virtue
of the wave equation, it satisfies the gauge condition everywhere;
hence, imposing the gauge condition to some order in the buffer region
ensures that it is imposed to the same order everywhere. 

While the integral representation is satisfied by any
solution to the associated wave equation, it does not \emph{provide} a
solution. That is, one cannot prescribe arbitrary boundary values on
$\Gamma$ and then arrive at a solution. The reason is that the tube is
a timelike boundary, which means that field data on it can propagate
forward in time and interfere with the data at a later time. However,
by applying the wave operator $E_{\alpha\beta}$ onto equation
\eqref{formal_solution}, we see that the integral representation of
$\hmn{\alpha\beta}{n}$ is guaranteed to satisfy the wave equation at
each point $x\in\Omega$. In other words, the problem arises not in
satisfying the wave equation in a pointwise sense, but in
simultaneously satisfying the boundary conditions. But since the tube
is chosen to lie in the buffer region, these boundary conditions can
be supplied by the buffer-region expansion. And as we will discuss in
Sec.~\ref{global_solution}, because of the asymptotic smallness of the
tube, the pieces of the buffer-region expansion diverging as $s^{-n}$
are sufficient boundary data to fully determine the global solution.  

Finally, just as in the point-particle case, in order to split the
gauge condition into a set of exactly solvable equations, we assume
that the acceleration of $\gamma$ possesses an expansion 
\begin{equation}
a^\mu(t,\e)=a_{(0)}^\mu(t)+\e a_{(1)}^\mu(t;\gamma)
+\ldots
\label{a expansion}.
\end{equation}
This leads to the set of equations 
\begin{align}
L^{(0)}_\mu\big[\hmn{}{1}\big] &=0, 
\label{gauge_expansion 1}\\
L^{(1)}_\mu\big[\hmn{}{1}\big] &= -L^{(0)}_\mu\big[\hmn{}{2}\big],
 \label{gauge_expansion 2}\\
&\ \ \vdots\nonumber
\end{align}
or, more generally, for $n>0$,
\begin{equation}
\label{compact_gauge}
L^{(n)}_\mu\big[\hmn{}{1}\big] = 
-\sum_{m=1}^n L_\mu^{(n-m)}\big[\hmn{}{m+1}\big]. 
\end{equation}
In these expressions, $L^{(0)}_\mu[f]$ is the Lorenz-gauge operator
acting on the tensor field $f_{\alpha\beta}$ evaluated with $a^\mu =
a^\mu_{(0)}$, $L^{(1)}_\mu[f]$ consists of the terms in $L_\mu[f]$
that are linear in $a^\mu_{(1)}$, and $L^{(n)}_\mu[f]$ contains the
terms linear in $a^{(n)\mu}$, the combinations
$a^{(n-1)\mu}a^{(1)\nu}$, and so on. Imposing these gauge
conditions on the solutions to the wave equations will determine the
acceleration of the world line.  Although we introduce an expansion of
the acceleration vector in order to obtain a systematic sequence of
equations that can be solved exactly, such an expansion also trivially
eliminates the need for order-reduction of the resulting equations of
motion, since it automatically leads to equations for $a^{(1)\mu}$ in
terms of $a^{(0)\mu}$, $a^{(2)\mu}$ in terms of $a^{(0)\mu}$ and
$a^{(1)\mu}$, and so on.

\subsubsection*{Gauge transformations and the Lorenz condition}
\label{Lorenz_gauge}

The outer expansion is defined not only by holding $x^\alpha$ fixed,
but also by demanding that the mass dipole of the body vanishes when
calculated in coordinates centered on $\gamma$. If we perform a gauge
transformation generated by a vector $\xi^{(1)\alpha}(x;\gamma)$, then
the mass dipole will no longer vanish in those coordinates. Hence, a
new world line $\gamma'$ must be constructed, such that the mass dipole
vanishes when calculated in coordinates centered on that new
world line. In other words, in the outer expansion we have the usual
gauge freedom of regular perturbation theory, so long as the world line
is appropriately transformed as well:
$(h_{\alpha\beta},\gamma)\to(h'_{\alpha\beta},\gamma')$. The
transformation law for the world line was first derived by Barack and
Ori~\cite{barack-ori:01}; it was displayed in Eq.~(\ref{1.9.8}), and
it will be worked out again in Sec.~\ref{gauge_and_force}. 

Using this gauge freedom, we now justify, to some extent, the
assumption that the Lorenz gauge condition can be imposed on the
entirety of $h_{\alpha\beta}$. If we begin with the metric in an
arbitrary gauge, then the gauge vectors $\e\xi^\alpha_{(1)}[\gamma]$,
$\e^2\xi^\alpha_{(2)}[\gamma]$, etc., induce the transformation 
\begin{align}
h_{\alpha\beta}\to h'_{\alpha\beta} &= h_{\alpha\beta} 
+\Delta h_{\alpha\beta} \nonumber\\ 
&=h_{\alpha\beta}+\e\Lie{\xi_{(1)}}g_{\alpha\beta}
+ \tfrac{1}{2}\e^2(\Lie{\xi_{(2)}}+\Lie{\xi_{(1)}}^2)g_{\alpha\beta}
+\e^2\Lie{\xi_{(1)}}\hmn{\alpha\beta}{1}+O(\e^3).
\end{align}
If $h'_{\alpha\beta}$ is to satisfy the gauge condition $L_\mu[h']$,
then $\xi$ must satisfy $L_\mu[\Delta h]=-L_\mu[h]$. After a trivial
calculation, this equation becomes 
\begin{align}
\sum_{n>0}\frac{\e^n}{n!}\Box\xi_{(n)}^\alpha &= 
-\e L^{\alpha}\big[\hmn{}{1}\big]-\e^2 L^{\alpha}\big[\hmn{}{2}\big]
-\e^2 L^\alpha\big[\tfrac{1}{2}\Lie{\xi_{(1)}}^2g
+\Lie{\xi_{(1)}}\hmn{}{1}\big]+O(\e^3).
\end{align}
Solving this equation for arbitrary $\gamma$, we equate coefficients
of powers of $\e$, leading to a sequence of wave equations of the form 
\begin{equation}
\Box\xi_{(n)}^\alpha = W^{\alpha}_{(n)},
\end{equation}
where $W^{\alpha}_{(n)}$ is a functional of
$\xi_{(1)}^\alpha,\ldots,\xi_{(n-1)}^\alpha$ and
$\hmn{\alpha\beta}{1},\ldots,\hmn{\alpha\beta}{\emph{n}}$. We seek a
solution in the region $\Omega$ described in the preceding
section. The formal solution reads 
\begin{align}
\xi_{(n)}^\alpha &=
-\frac{1}{4\pi}\int_\Omega G_+^\alpha{}_{\alpha'}
W^{\alpha'}_{(n)}dV'
+\frac{1}{4\pi}\!\oint\limits_{\partial\Omega}\!\!
\Big(G_+^{\alpha}{}_{\gamma'}\del{\mu'} \xi_{(n)}^{\gamma'}
-\xi_{(n)}^{\gamma'}\del{\mu'} G_+^{\alpha}{}_{\gamma'}\Big)
dS^{\mu'}. 
\end{align}
From this we see that the Lorenz gauge condition can be adopted to any
desired order of accuracy, given the existence of self-consistent data
on a tube $\Gamma$ of asymptotically small radius. We leave the
question of the data's existence to future work. This argument was
first presented in Ref.~\cite{pound:10b}. 

\subsection{Field equations in inner expansion}

For the inner expansion, we assume the existence of some local polar
coordinates $X^\alpha=(T,R,\Theta^A)$, such that the metric can be
expanded for $\e\to 0$ while holding fixed $\tilde R:= R/\e$,
$\Theta^A$, and $T$; to relate the inner and outer expansions, we
assume $R\sim s$, but otherwise leave the inner expansion completely
general. 

This leads to the ansatz
\begin{equation}
{\sf g}_{\alpha\beta}(T,\tilde R,\Theta^A,\e) = 
g^{\rm body}_{\alpha\beta}(T,\tilde R,\Theta^A)
+ H_{\alpha\beta}(T,\tilde R,\Theta^A,\e),
\label{21.17} 
\end{equation}
where $H_{\alpha\beta}$ at fixed values of $(T,\tilde R,\Theta^A)$ is
a perturbation beginning at order $\e$. This equation represents an
asymptotic expansion along flow lines of constant $R/\e$ as
$\e\to0$. It is tensorial in the usual sense of perturbation theory:
the decomposition into $g^{\rm body}_{\alpha\beta}$ and
$H_{\alpha\beta}$ is valid in any coordinates that can be decomposed
into $\e$-independent functions of the scaled coordinates plus $O(\e)$
functions of them. As written, with $g^{\rm body}_{\alpha\beta}$
depending only on the scaled coordinates and independent of $\e$, the
indices in Eq.~(\ref{21.17}) can be taken to refer to the unscaled
coordinates $(T,R,\Theta^A)$. However, writing the components in the
scaled coordinates will not alter the form of the expansion, but only
introduce an overall rescaling of spatial components due to the
spatial forms transforming as, e.g., $dR\to\e d\tilde R$. For example,
if the body is a small Schwarzschild black hole of ADM mass $\e m(T)$,
then in scaled Schwarzschild coordinates $(T,\tilde R,\Theta^A)$, 
$g^{\rm body}_{\alpha\beta}(T,\tilde R,\Theta^A)$ is given by
\begin{equation}
ds^2 = -(1 - 2m(T)/\tilde R)dT^2 +(1 - 2m(T)/\tilde R)^{-1}\e^2 d\tilde R^2
 + \e^2 \tilde R^2 (d\Theta^2 + \sin \Theta d\Phi^2).
\end{equation}
As we would expect from the fact that the inner limit follows the body
down as it shrinks, all points are mapped to the curve $R=0$ at
$\e=0$, such that the metric in the scaled coordinates naturally
becomes one-dimensional at $\e = 0$. This singular limit can be made
regular by rescaling time as well, such that $\tilde T=(T-T_0)/\e$,
and then rescaling the entire metric by a conformal factor
$1/\e^2$. In order to arrive at a global-in-time inner expansion,
rather than a different expansion at each time $T_0$, 
we forgo this extra step. We do, however, make an equivalent
assumption, which is that the metric $g^{\rm body}_{\alpha\beta}$ and
its perturbations are quasistatic (evolving only on timescales 
$\sim 1$). Both approaches are equivalent to assuming that the exact
metric contains no high-frequency oscillations occurring on the body's 
natural timescale $\sim\e$. In other words, the body is assumed to be
in equilibrium. If we did not make this assumption, high-frequency
oscillations could propagate throughout the external spacetime,
invalidating our external expansion. 

Since we are interested in the inner expansion only insofar as it
informs the outer expansion, we shall not seek to explicitly solve the
perturbative Einstein equation in the inner expansion. See
Ref.~\cite{pound:10a} for the forms of the equations and an example of
an explicit solution in the case of a perturbed black hole. 

\section{General expansion in the buffer region}
\label{buffer_region}

We now seek the general solution to the equations of the outer 
expansion in the buffer region. To perform the expansion, we adopt
Fermi coordinates centered about $\gamma$ and expand for small $s$. In 
solving the first-order equations, we will determine $a^{(0)\mu}$; in
solving the second-order equations, we will determine
$a^{(1)\mu}$, including the self-force on the body. Although we
perform this calculation in the Lorenz gauge, the choice of gauge is
not essential for our purposes here --- the essential aspect is our
assumed expansion of the acceleration of the world line $\gamma$.  

\subsection{Metric expansions} 

The method of matched asymptotic expansions relies on the fact that
the inner and outer expansion agree term by term when re-expanded
in the buffer region, where $\e\ll s\ll 1$. To illustrate this idea of
matching, consider the forms of the two expansions in the buffer
region. The inner expansion holds $\tilde s$ constant (since 
$R\sim s$) while expanding for small $\e$. But if $\tilde s$ is
replaced with its value $s/\e$, the inner expansion takes the form
${\sf g}_{\alpha\beta} = g^{\rm body}_{\alpha\beta}(s/\e) 
+ \e H^{(1)}_{\alpha\beta}(s/\e) + \cdots$, where each term has a
dependence on $\e$ that can be expanded in the limit $\e\to0$ 
to arrive at the schematic forms $g^{\rm body}(s/\e) 
= 1 \oplus \e/s \oplus \e^2/s^2\oplus\ldots$ and 
$\e H^{(1)}(s/\e) = s\oplus\e\oplus\e^2/s\oplus\cdots$, where $\oplus$ 
signifies  ``plus terms of the form'' and the expanded quantities can
be taken to be components in Fermi coordinates. Here we have
preemptively restricted the form of the expansions, since terms such
as $s^2/\e$ must vanish because they would have no corresponding  
terms in the outer expansion. Putting these two expansions together,
we arrive at 
\begin{equation}
{\sf g}({\rm buffer}) = 1 \oplus \frac{\e}{s}\oplus
s\oplus\e\oplus\cdots.
\end{equation}
Since this expansion relies on both an expansion at fixed $\tilde s$
and an expansion at fixed $s$, it can be expected to be accurate if
$s\ll 1$ and $\e\ll s$ --- that  is, in the buffer region 
$\e\ll s\ll1$.  

On the other hand, the outer expansion holds $s$ constant (since $s$
is formally of the order of the global external coordinates) while
expanding for small $\e$, leading to the form 
${\sf g}_{\alpha\beta} = g_{\alpha\beta}(s)
+ \e h^{(1)}_{\alpha\beta}(s)+\cdots$. But very near the world line,
each term in this expansion can be expanded for small $s$, leading to 
$g = 1 \oplus s \oplus s^2 \oplus \cdots$ and 
$\e h^{(1)} = \e/s\oplus\e\oplus\e s\oplus\cdots$. (Again, we have 
restricted this form because terms such as $\e/s^2$ cannot arise in
the inner expansion.) Putting these two expansions together, we arrive
at 
\begin{equation}
{\sf g}({\rm buffer}) = 1\oplus s\oplus \frac{\e}{s}\oplus \e\cdots. 
\end{equation}
Since this expansion relies on both an expansion at fixed $s$ and an
expansion for small $s$, it can be expected to be accurate in the
buffer region $\e\ll s\ll1$. As we can see, the two buffer-region
expansions have an identical form; and because they are expansions  
of the same exact metric ${\sf g}$, they must agree term by term. 

One can make use of this fact by first determining the inner and outer
expansions as fully as possible, then fixing any unknown functions in
them by matching them term by term in the buffer region; this was the
route taken in, e.g., Refs.~\cite{mino-etal:97a, poisson:04a,
  detweiler:05, taylor-poisson:08}. However, such an approach is
complicated by the subtleties of matching in a
diffeomorphism-invariant theory, where the inner and outer expansions
are generically in different coordinate systems. See
Ref.~\cite{pound:10b} for an analysis of the limitations of this
approach as it has typically been implemented. Alternatively, one can
take the opposite approach, working in the buffer region first,
constraining the forms of the two expansions by making use of their
matching, then using the buffer-region information to construct a
global solution; this was the route taken in, e.g.,
Refs.~\cite{kates:80a,gralla-wald:08,pound:10a}. In general, some
mixture of these two approaches can be taken. Our calculation follows 
Ref.~\cite{pound:10a}. The only information we take from the inner
expansion is its general form, which is characterized by the multipole
moments of the body. From this information, we determine the external
expansion, and thence the equation of motion of the world line. 

Over the course of our calculation, we will find that the external
metric perturbation in the buffer region is expressed as the sum of
two solutions: one that formally diverges at $s=0$ and is entirely
determined from a combination of (i) the multipole moments of the
internal background metric $g^{\rm body}_{\alpha\beta}$, (ii) the
Riemann tensor of the external background $g_{\alpha\beta}$, and (iii)
the acceleration of the world line $\gamma$; and a second solution
that is formally regular at $s=0$ and depends on the past history of
the body and the initial conditions of the field. At leading order,
these two solutions are identified as the Detweiler-Whiting singular
and regular fields $h^{\rm S}_{\alpha\beta}$ and 
$h^{\rm R}_{\alpha\beta}$, respectively, and the
self-force is determined entirely by $h^{\rm R}_{\alpha\beta}$. Along 
with the self-force, the acceleration of the world line includes the
Papapetrou spin-force \cite{papapetrou:51}. This calculation leaves us
with the self-force in terms of the the metric perturbation in the 
neighbourhood of the body. In Sec.~\ref{global_solution}, we use the
local information from the buffer region to construct a global
solution for the metric perturbation, completing the solution of the
problem.  

\subsection{The form of the expansion} 
\label{buffer_expansion}

Before proceeding, we define some notation. We use the multi-index
notation $\omega^L:= \omega^{i_1}\cdots\omega^{i_\ell}:=
\omega^{i_1\cdots  i_\ell}$. Angular brackets denote the STF
combination of the enclosed indices, and a tensor bearing a hat is an
STF tensor. To accomodate this, we now write the Fermi spatial
coordinates as $x^a$, instead of $\hat{x}^a$ as they were written in
previous sections. Finally, we define the one-forms
$t_\alpha:=\partial_\alpha t$ and
$x^a_\alpha:= \partial_\alpha x^a$. 

One should note that the coordinate transformation $x^\alpha(t,x^a)$
between Fermi coordinates and the global coordinates is
$\e$-dependent, since Fermi coordinates are tethered to an
$\e$-dependent world line. If one were using a regular expansion, then
this coordinate transformation would devolve into a background
coordinate transformation to a Fermi coordinate system centered on a
geodesic world line, combined with a gauge transformation to account
for the $\e$-dependence. But in the self-consistent expansion, the
transformation is purely a background transformation, because the
$\e$-dependence in it is reducible to that of the fixed world line. 

Because the dependence on $\e$ in the coordinate transformation cannot
be reduced to a gauge transformation, in Fermi coordinates the
components $g_{\alpha\beta}$ of the background metric become
$\e$-dependent. This dependence takes the explicit form of factors of
the acceleration $a^\mu(t,\e)$ and its derivatives, for which we have
assumed the expansion
$a^i(t,\e)=a^{(0)i}(t)+a^{(1)i}(t;\gamma)+O(\e^2)$. 
There is also an implicit dependence on $\e$ in that the proper time
$t$ on the world line depends on $\e$ if written as a function of the
global coordinates; but this dependence can be ignored so long as we
work consistently with Fermi coordinates. 

Of course, even in these $\e$-dependent coordinates, $g_{\mu\nu}$
remains the background metric of the outer expansion, and
$h^{(n)}_{\mu\nu}$ is an exact solution to the wave equation  
(\ref{21.7}). At first order we will therefore obtain
$h^{(1)}_{\mu\nu}$ exactly in Fermi coordinates, for arbitrary
$a^\mu$. However, for some purposes an approximate solution 
of the wave equation may suffice, in which case we may utilize the 
expansion of $a^\mu$. Substituting that expansion into $g_{\mu\nu}$
and $h^{(n)}_{\mu\nu}$ yields the \emph{buffer-region expansions} 
\begin{align}
g_{\mu\nu}(t,x^a;a^i) &= g_{\mu\nu}(t, x^a;a^{(0)i}) 
+ \e g^{(1)}_{\mu\nu}(t, x^a;a^{(1)i}) + O(\e^2) 
\label{buffer_expansion g} 
\\
h^{(n)}_{\mu\nu}(t,x^a;a^i) &= h^{(n)}_{\mu\nu}(t, x^a; a^{(0)i}) 
+ O(\e), 
\label{buffer_expansion h} 
\end{align}
where indices refer to Fermi coordinates, $g^{(1)}_{\mu\nu}$ is linear
in $a^{(1)i}$ and its derivatives, and for future compactness of
notation we define $h^{(n)}_{B\mu\nu}(t,x^a) 
:= h^{(n)}_{\mu\nu}(t,x^a;a^{(0)i})$, where the subscript
`B' stands for `buffer'. In the case that $a^{(0)i}=0$, these
expansions will significantly reduce the complexity of calculations in
the buffer region. For that reason, we shall use them in solving the
second-order wave equation, but we stress that they are simply a means
of economizing calculations in Fermi coordinates; they do not play  
a fundamental role in the formalism, and one could readily do without
them. 

Now, we merely assume that in the buffer region there exists a smooth
coordinate transformation between the local coordinates
$(T,R,\Theta^A)$ and the Fermi coordinates $(t,x^a)$ such that $T\sim
t$, $R\sim s$, and $\Theta^A\sim\theta^A$. The buffer region
corresponds to asymptotic infinity $s\gg\e$ (or $\tilde s\gg1$) in the
internal spacetime. So after re-expressing $\tilde s$ as $s/\e$, the
internal background metric can be expanded as 
\begin{align}
g^{\rm body}_{\alpha\beta}(t,\tilde s,\theta^A) &= 
\sum_{n\geq0}\left(\frac{\e}{s}\right)^n 
g^{{\rm body}(n)}_{\alpha\beta}(t,\theta^A).
\end{align}
As mentioned above, since the outer expansion has no negative powers
of $\e$, we exclude them from the inner expansion. Furthermore, since
$g_{\alpha\beta}+h_{\alpha\beta}=g^{\rm body}_{\alpha\beta}
+H_{\alpha\beta}$, we must have $g^{{\rm body}(0)}_{\alpha\beta}
=g_{\alpha\beta}(x^a=0)$, since these are
the only terms independent of both $\e$ and $s$. Thus, noting that
$g_{\alpha\beta}(x^a=0)=\eta_{\alpha\beta}:=\text{diag}(-1,1,1,1)$, 
we can write 
\begin{align}
g^{\rm body}_{\alpha\beta}(t,\tilde s,\theta^A) &= 
\eta_{\alpha\beta}
+\frac{\e}{s} g^{{\rm body}(1)}_{\alpha\beta}(t,\theta^A)  
+\left(\frac{\e}{s} \right)^2 
g^{{\rm body}(2)}_{B\alpha\beta}(t,\theta^A)+O(\e^3/s^3), 
\end{align}
implying that the internal background spacetime is asymptotically
flat. 

We assume that the perturbation $H_{\alpha\beta}$ can be similarly
expanded in powers of $\e$ at fixed $\tilde s$, 
\begin{align}
H_{\alpha\beta}(t,\tilde s,\theta^A,\e) &= 
\e H^{(1)}_{\alpha\beta}(t,\tilde s,\theta^A;\gamma) 
+\e^2 H^{(2)}_{\alpha\beta}(t,\tilde s,\theta^A;\gamma)+O(\e^3), 
\end{align}
and that each coefficient can be expanded in powers of 
$1/\tilde s=\e/s$ to yield 
\begin{align}\label{buffer ansatz}
\e H^{(1)}_{\alpha\beta}(\tilde s) &= sH^{(0,1)}_{\alpha\beta}
+\e H^{(1,0)}_{\alpha\beta}+ \frac{\e^2}{s}H^{(2,-1)}_{\alpha\beta}
+O(\e^3/s^2),\\
\e^2 H^{(2)}_{\alpha\beta}(\tilde s) & = s^2 H^{(0,2)}_{\alpha\beta} 
+\e sH^{(1,1)}_{\alpha\beta}+\e^2 H^{(2,0)}_{\alpha\beta}
+\e^2\ln s\, H^{(2,0,\ln)}_{\alpha\beta}+O(\e^3/s),\\
\e^3 H^{(3)}_{\alpha\beta}(\tilde s) & = O(\e^3,\e^2s, \e s^2,s^3), 
\end{align}
where $H^{(n,m)}_{\alpha\beta}$, the coefficient of $\e^n$ and $s^m$,
is a function of $t$ and $\theta^A$ (and potentially a functional of
$\gamma$). Again, the form of this expansion is constrained by the
fact that no negative powers of $\e$ can appear in the buffer
region. (One might think that terms with negative powers of $\e$ could
be allowed in the expansion of $g^{\rm body}_{\alpha\beta}$ if they
are exactly canceled by terms in the expansion of $H_{\alpha\beta}$,
but the differing powers of $s$ in the two expansions makes this
impossible.) Note that explicit powers of $s$ appear because 
$\e\tilde{s}=s$. Also note that we allow for a logarithmic term at
second order in $\e$; this term arises because the retarded time in
the internal background includes a logarithmic correction of the form
$\e\ln s$ (e.g., $t-r\to t-r^*$ in Schwarzschild coordinates). Since
we seek solutions to a wave equation, this correction to the
characteristic curves induces a corresponding correction to the
first-order perturbations.  

The expansion of $H_{\alpha\beta}$ may or may not hold the
acceleration fixed. Regardless of this choice, the general form of the
expansion remains valid: incorporating the expansion of the
acceleration would merely shuffle terms from one coefficient to
another. And since the internal metric 
$g^{\rm body}_{\alpha\beta}+H_{\alpha\beta}$ must equal the external
metric $g_{\alpha\beta}+h_{\alpha\beta}$, the general form of the
above expansions of $g^{\rm body}_{\alpha\beta}$ and 
$H_{\alpha\beta}$ completely determines the general form of the
external perturbations:  
\begin{align}
\hmn{\alpha\beta}{1} & = \frac{1}{s}\hmn{\alpha\beta}{1,-1} 
+\hmn{\alpha\beta}{1,0} +s\hmn{\alpha\beta}{1,1}+O(s^2), 
\label{h1 expansion}\\
\hmn{\alpha\beta}{2} & = \frac{1}{s^2}\hmn{\alpha\beta}{2,-2} 
+\frac{1}{s}\hmn{\alpha\beta}{2,-1} +\hmn{\alpha\beta}{2,0}
+\ln s\,\hmn{\alpha\beta}{2,0,\ln}+O(s),
\label{h2 expansion}
\end{align}
where each $\hmn{\alpha\beta}{\emph{n,m}}$ depends only on $t$ and 
$\theta^A$, along with an implicit functional dependence on
$\gamma$. If the internal expansion is performed with $a^\mu$ held
fixed, then the internal and external quantities are related order by
order: e.g., $\sum_m H_{\alpha\beta}^{(0,m)}=g_{\alpha\beta}$,
$\hmn{\alpha\beta}{1,-1}=g^{{\rm body}(1)}_{\alpha\beta}$, and
$\hmn{\alpha\beta}{1,0}=H_{\alpha\beta}^{(1,0)}$. Since we are not
concerned with determining the internal perturbations, the only such
relationship of interest is $\hmn{\alpha\beta}{\emph{n,-n}}=
g^{{\rm body}(n)}_{\alpha\beta}$. This equality tells us that the most
divergent, $s^{-n}$ piece of the $n$th-order perturbation
$\hmn{\alpha\beta}{\emph{n}}$ is defined entirely by the $n$th-order
piece of the internal background metric $g^{\rm body}_{\alpha\beta}$,
which is the metric of the body if it were isolated. 

To obtain a general solution to the Einstein equation, we write each
$\hmn{\alpha\beta}{\emph{n,m}}$ as an expansion in terms of
irreducible symmetric trace-free pieces: 
\begin{align}
\hmn{tt}{{\it n,m}} &= 
\sum_{\ell\ge0}\A{L}{{\it n,m}}\hat{\omega}^L, \\
\hmn{ta}{{\it n,m}} &= 
\sum_{\ell\ge0}\B{L}{{\it n,m}}\hat{\omega}_a{}^L 
+\sum_{\ell\ge1}\left[\C{aL-1}{{\it n,m}}\hat{\omega}^{L-1} 
+\epsilon_{ab}{}^c\D{cL-1}{{\it n,m}}\hat{\omega}^{bL-1}\right], \\ 
\hmn{ab}{{\it n,m}} &= 
\delta_{ab}\sum_{\ell\ge0}\K{L}{{\it n,m}}\hat{\omega}^L
+\sum_{\ell\ge0}\E{L}{{\it n,m}}\hat{\omega}_{ab}{}^L 
+\sum_{\ell\ge1}\! 
\left[\F{L-1\langle a}{{\it n,m}}\hat{\omega}^{}_{b\rangle}{}^{L-1} 
+\epsilon^{cd}{}_{(a}\hat{\omega}_{b)c}{}^{L-1}\G{dL-1}
{{\it n,m}}\right] \nonumber\\ 
&\quad+\sum_{\ell\ge2}\!\left[\H{abL-2}{{\it n,m}}
\hat{\omega}^{L-2}+\epsilon^{cd}{}_{(a}\I{b)dL-2}{{\it n,m}} 
\hat{\omega}_c{}^{L-2}\right].
\end{align}
Here a hat indicates that a tensor is STF with respect to 
$\delta_{ab}$, angular brackets $\langle\rangle$ indicate the STF 
combination of enclosed indices, parentheses indicate the symmetric
combination of enclosed indices, and symbols such as
$\A{L}{{\it n,m}}$ are functions of time (and potentially functionals
of $\gamma$) and are STF in all their indices. Each term in this
expansion is linearly independent of all the other terms. All the
quantities on the right-hand side are flat-space Cartesian tensors;
their indices can be raised or lowered with $\delta_{ab}$. Refer to 
Appendix~\ref{STF tensors} for more details about this expansion.  

Now, since the wave equations \eqref{h1 eqn} and \eqref{h2 eqn} are
covariant, they must still hold in the new coordinate system, despite
the additional $\e$-dependence. Thus, both equations could be solved
for arbitrary acceleration in the buffer region. However, due to the
length of the calculations involved, we will instead solve the
equations 
\begin{align}
E_{\alpha\beta}[\hmn{}{1}] & = 0, 
\label{hB1 eqn}\\
E^{(0)}_{\alpha\beta}[\hmn{B}{2}] & = 
2\delta^2 R^{(0)}_{\alpha\beta}[\hmn{}{1}]+O(\e),
\label{hB2 eqn}
\end{align}
where $E^{(0)}[f]:= E[f]\big|_{a=a_{(0)}}$ and $\delta^2
R^{(0)}[f]:= \delta^2R[f]\big|_{a=a_{(0)}}$. In analogy with the
notation used for $L_\mu^{(n)}$, $E^{(1)}_{\mu\nu}[f]$ and 
$\delta^2 R_{\mu\nu}^{(1)}[f]$ would be linear in $a_{(1)}^\mu$, 
$E^{(2)}_{\mu\nu}[f]$ and $\delta^2 R_{\mu\nu}^{(2)}[f]$ would be
linear in $a_{(2)}^\mu$ and quadratic in $a_{(1)}^\mu$, and so on. For
a function $f\sim 1$, $L^{(n)}_\mu[f]$, $E^{(n)}_{\mu\nu}[f]$, and
$\delta^2 R_{\mu\nu}^{(n)}[f]$ correspond to the coefficients of
$\e^n$ in expansions in powers of $\e$. Equation~\eqref{hB1 eqn} is
identical to Eq.~\eqref{h1 eqn}. Equation~\eqref{hB2 eqn} follows
directly from substituting Eqs.~\eqref{buffer_expansion g} and
\eqref{buffer_expansion h} into Eq.~\eqref{h2 eqn}; in the buffer
region, it captures the dominant behaviour of $\hmn{\alpha\beta}{2}$,
represented by the approximation $\hmn{B\alpha\beta}{2}$, but it does 
not capture its full dependence on acceleration. If one desired a
global second-order solution, one might need to solve 
Eq.~\eqref{h2 eqn}, but for our purpose, which is to determine the
first-order acceleration $a_{(1)}^\mu$, Eq.~\eqref{hB2 eqn} will
suffice.  

Unlike the wave equations, the gauge conditions 
\eqref{gauge_expansion 1} and \eqref{gauge_expansion 2} already
incorporate the expansion of the acceleration. As such, they are
unmodified by the replacement of the second-order wave equation
\eqref{h2 eqn} with its approximation \eqref{hB2 eqn}. So we can write 
\begin{align}
L^{(0)}_\mu\big[\hmn{}{1}\big] &=0, \label{hB1 gauge}\\
L^{(1)}_\mu\big[\hmn{}{1}\big] &= 
-L^{(0)}_\mu\big[\hmn{B}{2}\big], \label{hB2 gauge}
\end{align}
where the first equation is identical to 
Eq.~\eqref{gauge_expansion 1}, and the second to 
Eq.~\eqref{gauge_expansion 2}. (The second identity holds because 
$L^{(0)}_\mu\big[\hmn{B}{2}\big]=L^{(0)}_\mu\big[\hmn{}{2}\big]$,
since $\hmn{B\alpha\beta}{2}$ differs from $\hmn{\alpha\beta}{2}$ by
$a_{(1)}^\alpha$ and higher acceleration terms, which are set to zero
in $L_\mu^{(0)}$.) We remind the reader that while this gauge choice
is important for finding the external perturbations globally, any
other choice would suffice in the buffer region calculation. 

In what follows, the reader may safely assume that all calculations
are lengthy unless noted otherwise. 

\subsection{First-order solution in the buffer region}
\label{buffer_expansion1}

In principle, solving the first-order Einstein equation in the buffer
region is straightforward. One need simply substitute the expansion of
$\hmn{\alpha\beta}{1}$, given in Eq.~\eqref{h1 expansion}, into the
linearized wave equation \eqref{hB1 eqn} and the gauge condition
\eqref{hB1 gauge}. Equating powers of $s$ in the resulting expansions
then yields a sequence of equations that can be solved for
successively higher-order terms in $\hmn{\alpha\beta}{1}$. Solving
these equations consists primarily of expressing each quantity in its
irreducible STF form, using the decompositions~\eqref{decomposition_1}
and \eqref{decomposition_2}; since the terms in this STF decomposition
are linearly independent, we can solve each equation term by
term. This calculation is aided by the fact that
$\del{\alpha}=x^a_\alpha\partial_a+O(s^0)$, so that, for example, the
wave operator $E_{\alpha\beta}$ consists of a flat-space Laplacian
$\partial^a\partial_a$ plus corrections of order
$1/s$. Appendix~\ref{STF tensors} also lists many useful
identities, particularly $\partial^a s =\omega^a := x^a/s$,
$\omega^a\partial_a\hat{\omega}^L=0$, and the fact that
$\hat{\omega}^L$ is an eigenvector of the flat-space Laplacian:
$s^2 \partial^a\partial_a\hat{\omega}^L
=-\ell(\ell+1) \hat{\omega}^L$. 

\subsubsection*{Summary of results} 

Before proceeding with the calculation, which consists mostly of
tedious and lengthy algebra, we summarize the results. The
first-order perturbation $\hmn{\alpha\beta}{1}$ consists of two
pieces, which we will eventually identify with the Detweiler-Whiting
regular and singular fields. In the buffer-region expansion, the
regular field consists entirely of unknowns, which is to be expected
since as a free radiation field, it must be provided by boundary
data. Only when we consider the global solution, in
Sec.~\ref{global_solution}, will we express it in terms of a tail
integral. On the other hand, the singular field is locally determined,
and it is characterized by the body's monopole moment $m$. More
precisely, it is fully determined by the tidal fields of the external
background spacetime and the Arnowitt-Deser-Misner mass of the
internal background spacetime $g^{\rm body}_{\alpha\beta}$. By itself
the wave equation does not restrict the behaviour of this monopole
moment, but imposing the gauge condition produces the evolution
equations  
\begin{align}
\partial_t m & = 0,\qquad a_{(0)}^i = 0.
\end{align}
Hence, at leading order, the body behaves as a test particle, with
constant mass and vanishing acceleration. 

\subsubsection*{Order $(1,-1)$} 

We now proceed to the details of the calculation. We begin with the
most divergent term in the wave equation: the order $1/s^3$,
flat-space Laplacian term 
\begin{equation}
\frac{1}{s}\partial^c\partial_c\hmn{\alpha\beta}{1,-1} = 0.
\end{equation}
The $tt$-component of this equation is
\begin{equation}
0=-\sum_{\ell\ge0}\ell(\ell+1)\A{L}{1,-1}\hat{\omega}^L,
\end{equation}
from which we read off that $\A{}{1,-1}$ is arbitrary and
$\A{L}{1,-1}$ must vanish for all $\ell\ge1$. The $ta$-component is 
\begin{align}
0&=-\sum_{\ell\ge0}(\ell+1)(\ell+2)\B{L}{1,-1}\hat{\omega}_a{}^L 
-\sum_{\ell\ge1}\ell(\ell-1)\C{aL-1}{1,-1}\hat{\omega}^{L-1}
\nonumber\\
&\quad-\sum_{\ell\ge1}\ell(\ell+1)\epsilon_{abc}
\D{cL-1}{1,-1}\hat{\omega}_b{}^{L-1},
\end{align}
from which we read off that $\C{a}{1,-1}$ is arbitrary and all other
coefficients must vanish. Lastly, the $ab$-component is 
\begin{align}
0&=-\delta_{ab}\sum_{\ell\ge0}\ell(\ell+1)\K{L}{1,-1}\hat{\omega}^L 
-\sum_{\ell\ge0}(\ell+2)(\ell+3)\E{L}{1,-1}\hat{\omega}_{ab}{}^L
\nonumber\\
&\quad-\sum_{\ell\ge1}\ell(\ell+1)\F{L-1\langle a}{1,-1}
\hat{\omega}^{}_{b\rangle}{}^{L-1} 
-\sum_{\ell\ge1}(\ell+1)(\ell+2)\epsilon_{cd(a}
\hat{\omega}_{b)}{}^{cL-1}\G{dL-1}{1,-1} 
\nonumber\\
&\quad-\sum_{\ell\ge2}(\ell-2)(\ell-1)\H{abL-2}{1,-1}\hat{\omega}^{L-2} 
-\sum_{\ell\ge2}\ell(\ell-1)\epsilon^{}_{cd(a}\I{b)dL-2}{1,-1}
\hat{\omega}_c{}^{L-2},
\end{align}
from which we read off that $\K{}{1,-1}$ and $\H{ab}{1,-1}$ are
arbitrary and all other coefficients must vanish. Thus, we find that
the wave equation constrains $\hmn{\alpha\beta}{1,-1}$ to be 
\begin{align}
\hmn{\alpha\beta}{1,-1}& =\A{}{1,-1}t_\alpha t_\beta
+ 2\C{a}{1,-1}t_{(\beta}x^a_{\alpha)} 
+(\delta_{ab}\K{}{1,-1}+\H{ab}{1,-1})x^a_\alpha x^b_\beta.
\end{align}
This is further constrained by the most divergent, $1/s^2$ term in the
gauge condition, which reads 
\begin{equation}
-\frac{1}{s^2}\hmn{\alpha c}{1,-1}\omega^c
+\frac{1}{2s^2}\omega_\alpha\eta^{\mu\nu}\hmn{\mu\nu}{1,-1}=0.
\end{equation}
From the $t$-component of this equation, we read off $\C{a}{1,-1}=0$;
from the $a$-component, $\K{}{1,-1}=\A{}{1,-1}$ and
$\H{ab}{1,-1}=0$. Thus, $\hmn{\alpha\beta}{1,-1}$ depends only on a
single function of time, $\A{}{1,-1}$. By the definition of the ADM
mass, this function (times $\e$) must be twice the mass of the
internal background spacetime. Thus, $\hmn{\alpha\beta}{1,-1}$ is
fully determined to be 
\begin{equation}\label{h1n1}
\hmn{\alpha\beta}{1,-1}=2m(t)(t_\alpha t_\beta 
+\delta_{ab}x^a_\alpha x^b_\beta),
\end{equation}
where $m(t)$ is defined to be the mass at time $t$ divided by the
initial mass $\e:= m_0$. (Because the mass will be found to be a
constant, $m(t)$ is merely a placeholder; it is identically unity. We
could instead set $\e$ equal to unity at the end of the calculation,
in which case $m$ would simply be the mass at time $t$. Obviously, the
difference between the two approaches is immaterial.) 

\subsubsection*{Order $(1,0)$} 
 
At the next order, $\hmn{\alpha\beta}{1,0}$, along with the
acceleration of the world line and the time-derivative of the mass,
first appears in the Einstein equation. The order $1/s^2$ term in the
wave equation is 
\begin{equation}
\partial^c\partial_c\hmn{\alpha\beta}{1,0}=
-\frac{2m}{s^2}a_c\omega^c(3t_\alpha t_\beta 
-\delta_{ab}x^a_\alpha x^b_\beta),
\end{equation}
where the terms on the right arise from the wave operator acting on 
$s^{-1} \hmn{\alpha\beta}{1,-1}$. This equation constrains
$\hmn{\alpha\beta}{1,0}$ to be 
\begin{equation}\label{h10}
\begin{split}
\hmn{tt}{1,0}&=\A{}{1,0}+3ma_c\omega^c, \\
\hmn{ta}{1,0}&=\C{a}{1,0}, \\
\hmn{ab}{1,0}&=\delta_{ab}\left(\K{}{1,0}-ma_c\omega^c\right)
+\H{ab}{1,0}.
\end{split}
\end{equation}
Substituting this result into the order $1/s$ term in the gauge
condition, we find 
\begin{equation}
-\frac{4}{s} t_\alpha\partial_t m +\frac{4m}{s}a^{(0)}_ax^a_\alpha=0.
\end{equation}
Thus, both the leading-order part of the acceleration and the rate of
change of the mass of the body vanish: 
\begin{equation}
\begin{array}{lcr}
\displaystyle\frac{\partial m}{\partial t}=0\,, && a_{(0)}^i =0.
\end{array}
\end{equation}

\subsubsection*{Order $(1,1)$} 
 
At the next order, $\hmn{\alpha\beta}{1,1}$, along with squares and
derivatives of the acceleration, first appears in the Einstein
equation, and the tidal fields of the external background couple to
$s^{-1} \hmn{\alpha\beta}{1,-1}$. The order $1/s$ term in the wave
equation becomes 
\begin{align}
\left(s\partial^c\partial_c+\frac{2}{s}\right)\hmn{tt}{1,1} &= 
-\frac{20m}{3s}\etide_{ij}\hat{\omega}^{ij}
-\frac{3m}{s}a_{\langle i}a_{j\rangle}\hat{\omega}^{ij} 
+\frac{8m}{s}a_ia^i, \\
\left(s\partial^c\partial_c+\frac{2}{s}\right)\hmn{ta}{1,1} & = 
-\frac{8m}{3s}\epsilon_{aij}\btide^j_k\hat{\omega}^{ik}
-\frac{4m}{s}\dot a_a,\\
\left(s\partial^c\partial_c+\frac{2}{s}\right)\hmn{ab}{1,1} & = 
\frac{20m}{9s}\delta_{ab}\etide_{ij}\hat{\omega}^{ij}  
-\frac{76m}{9s}\etide_{ab} 
-\frac{16m}{3s}\etide^i_{\langle a}\hat{\omega}_{b\rangle i}
+\frac{8m}{s}a_{\langle a}a_{b\rangle}\nonumber\\
&\quad+\frac{m}{s}\delta_{ab}\!\left(\tfrac{8}{3}a_ia^i\!
-3a_{\langle i}a_{j\rangle}\hat{\omega}^{ij}\right). 
\end{align}
From the $tt$-component, we read off that $\A{i}{1,1}$ is arbitrary,
$\A{}{1,1}=4ma_ia^i$, and $\A{ij}{1,1}=\tfrac{5}{3}m\etide_{ij}
+\tfrac{3}{4}ma_{\langle i}a_{j\rangle}$; from the $ta$-component,
$\B{}{1,1}$, $\C{ij}{1,1}$, and $\D{i}{1,1}$ are arbitrary,
$\C{i}{1,1}=-2m\dot a_i$, and $\D{ij}{1,1}=\tfrac{2}{3}m\btide_{ij}$;
from the $ab$ component, $\K{i}{1,1}$, $\F{i}{1,1}$, $\H{ijk}{1,1}$,
and $\I{ij}{1,1}$ are arbitrary, and $\K{}{1,1}=\tfrac{4}{3}ma_ia^i$,
$\K{ij}{1,1}=-\tfrac{5}{9}m\etide_{ij}+\tfrac{3}{4}ma_{\langle
  i}a_{j\rangle}$, $\F{ij}{1,1}=\tfrac{4}{3}m\etide_{ij}$, and
$\H{ij}{1,1}=-\tfrac{38}{9}m\etide_{ij}+4ma_{\langle i}a_{j\rangle}$. 

Substituting this into the order $s^0$ terms in the gauge condition,
we find 
\begin{align}
0&=(\omega^i+s\partial^i)\hmn{\alpha i}{1,1} 
-\tfrac{1}{2}\eta^{\mu\nu}(\omega_a
-s\partial_a)\hmn{\mu\nu}{1,1}x^a_\alpha 
-\partial_t\hmn{\alpha t}{1,0} 
-\tfrac{1}{2}\eta^{\mu\nu}\partial_t\hmn{\mu\nu}{1,0}t_\alpha
\nonumber\\
&\quad +\tfrac{4}{3}m\etide_{ij}\hat{\omega}^{ij}\omega_\alpha 
+\tfrac{2}{3}m\etide_{ai}\omega^ix^a_\alpha,
\end{align}
where the equation is to be evaluated at $a^i=a_{(0)}^i=0$. From the
$t$-component, we read off 
\begin{equation}\label{B11}
\B{}{1,1}=\tfrac{1}{6}\partial_t\left(\A{}{1,0}+3\K{}{1,0}\right).
\end{equation}
From the $a$-component,
\begin{equation}\label{F11}
\F{a}{1,1}=\tfrac{3}{10}\left(\K{a}{1,1}-\A{a}{1,1}
+\partial_t\C{a}{1,0}\right).
\end{equation}
It is understood that both these equations hold only when evaluated at 
$a^i=0$. 

Thus, the order $s$ component of $\hmn{\alpha\beta}{1}$ is
\begin{equation}\label{h11}
\begin{split}
\hmn{tt}{1,1}&=4ma_ia^i+\A{i}{1,1}\omega^i
+\tfrac{5}{3}m\etide_{ij}\hat{\omega}^{ij} 
+\tfrac{3}{4}ma_{\langle i}a_{j\rangle}\hat{\omega}^{ij}, \\ 
\hmn{ta}{1,1}&=\B{}{1,1}\omega_a-2m\dot a_a+\C{ai}{1,1}\omega^i
+\epsilon_{ai}{}^j\D{j}{1,1}\omega^i 
+\tfrac{2}{3}m\epsilon_{aij}\btide^j_k\hat{\omega}^{ik},\\
\hmn{ab}{1,1}&=\delta_{ab}\big(\tfrac{4}{3}ma_ia^i+\K{i}{1,1}\omega^i 
-\tfrac{5}{9}m\etide_{ij}\hat{\omega}^{ij} 
+\tfrac{3}{4}ma_{\langle i}a_{j\rangle}\hat{\omega}^{ij}\big) 
+\tfrac{4}{3}m\etide^i_{\langle a}\hat{\omega}_{b\rangle i}
\\
&\quad-\tfrac{38}{9}m\etide_{ab}+4ma_{\langle a}a_{b\rangle}
+\H{abi}{1,1}\omega^i 
+\epsilon\indices{_i^j_{(a}}\I{b)j}{1,1}\omega^i
+\F{\langle a}{1,1}\omega^{}_{b\rangle}.
\end{split}
\end{equation}
where $\B{}{1,1}$ and $\F{a}{1,1}$ are constrained to satisfy
Eqs.~\eqref{B11} and \eqref{F11}. 

\subsubsection*{First-order solution} 

To summarize the results of this section, we have
$\hmn{\alpha\beta}{1}=s^{-1} \hmn{\alpha\beta}{1,-1}
+\hmn{\alpha\beta}{1,0} +s\hmn{\alpha\beta}{1,1}+O(s^2)$, where
$\hmn{\alpha\beta}{1,-1}$ is given in Eq.~\eqref{h1n1},
$\hmn{\alpha\beta}{1,0}$ is given in Eq.~\eqref{h10}, and
$\hmn{\alpha\beta}{1,1}$ is given in Eq.~\eqref{h11}. In addition, we
have determined that the ADM mass of the internal background spacetime
is time-independent, and that the acceleration of the body's world line
vanishes at leading order. 

\subsection{Second-order solution in the buffer region}
\label{buffer_expansion2}

Though the calculations are much lengthier, solving the second-order
Einstein equation in the buffer region is essentially no different
from solving the first. We seek to solve the approximate wave equation
\eqref{hB2 eqn}, along with the gauge condition \eqref{hB2 gauge}, for
the second-order perturbation
$\hmn{B\alpha\beta}{2}:=\hmn{\alpha\beta}{2}\big|_{a=a_{0}}$; doing 
so will also, more importantly, determine the acceleration
$a_{(1)}^\mu$. In this calculation, the acceleration is set to
$a^i=a_{(0)}^i=0$ everywhere except in the left-hand side of the gauge
condition, $L^{(1)}_\mu[\hmn{}{1}]$, which is linear in $a_{(1)}^\mu$. 

\subsubsection*{Summary of results} 

We first summarize the results. As at first order, the metric
perturbation contains a regular, free radiation field and a singular,
bound field; but in addition to these pieces, it also contains terms
sourced by the first-order perturbation. Again, the regular field
requires boundary data to be fully determined. And again, the singular
field is characterized by the multipole moments of the body: the mass
dipole $M_i$ of the internal background metric 
$g^{\rm body}_{\alpha\beta}$, which measures the shift of the body's
centre of mass relative to the world line; the spin dipole $S_i$ of 
$g^{\rm body}_{\alpha\beta}$, which measures the spin of the body
about the world line; and an effective correction $\delta m$ to the
body's mass. The wave equation by itself imposes no restriction on
these quantities, but by imposing the gauge condition we find the
evolution equations 
\begin{align}
\partial_t\delta m &= \frac{m}{3}\partial_t\A{}{1,0} 
+\frac{5m}{6}\partial_t\K{}{1,0},\label{mdot}\\
\partial_t S_a &= 0,\\
\partial^2_tM_a+\etide_{ab}M^b &= -a^{(1)}_a+\tfrac{1}{2}\A{a}{1,1}
-\partial_t\C{a}{1,0}-\frac{1}{m} S_i\btide^i_a. 
\label{master_eqn_of_motion}
\end{align}
The first of these tells us that the free radiation field created by 
the body creates a time-varying shift in the body's mass. We can
immediately integrate it to find 
\begin{align}
\delta m(t)&=\delta m(0) +\tfrac{1}{6}m\left[2\A{}{1,0}(t)
+5\K{}{1,0}(t)\right]-\tfrac{1}{6}m\left[2\A{}{1,0}(0)
+5\K{}{1,0}(0)\right]. 
\end{align}
We note that this mass correction is entirely gauge dependent; it 
could be removed by redefining the time coordinate on the
world line. In addition, one could choose to incorporate $\delta m(0)$ 
into the leading-order mass $m$. The second of the equations tells us
that the body's spin is constant at this order; at higher orders,
time-dependent corrections to the spin dipole would arise. The last of
the equations is the principal result of this section. It tells us
that the relationship between the acceleration of the world line and
the drift of the body away from it is governed by (i) the local
curvature of the background spacetime, as characterized by
$\etide_{ab}$ --- this is the same term that appears in the geodesic
deviation equation --- (ii) the coupling of the body's spin to the local
curvature --- this is the Papapetrou spin force
\cite{papapetrou:51} --- and (iii) the free radiation field created by
the body --- this is the self-force. We identify the world line as the
body's by the condition $M_i=0$. If we start with initial conditions
$M_i(0)=0=\partial_t M_i(0)$, then the mass dipole remains zero for
all times if and only if the world line satisfies the equation 
\begin{equation}
\label{a1a}
a^{(1)}_a=\tfrac{1}{2}\A{a}{1,1}-\partial_t\C{a}{1,0}
-\frac{1}{m} S_i\btide^i_a.
\end{equation}
This is the equation of motion we sought. It, along with the more
general equation containing $M_i$, will be discussed further in the
following section. 

\subsubsection*{Order $(2,-2)$} 

We now proceed to the details of the calculation. Substituting the 
expansion 
\begin{align}
\hmn{\alpha\beta}{2}&=\frac{1}{s^2}\hmn{B\alpha\beta}{2,-2} 
+\frac{1}{s}\hmn{B\alpha\beta}{2,-1}
+\hmn{B\alpha\beta}{2,0} +\ln(s)\hmn{B\alpha\beta}{2,0,\ln} 
+O(\e,s)
\end{align}
and the results for $\hmn{\alpha\beta}{1}$ from the previous section 
into the wave equation and the gauge condition again yields a sequence
of equations that can be solved for coefficients of successively
higher-order powers (and logarithms) of $s$. Due to its length, the
expansion of the second-order Ricci tensor is given in
Appendix~\ref{second-order expansions}. Note that since the
approximate wave equation \eqref{hB2 eqn} contains an explicit $O(\e)$ 
correction, $\hmn{\alpha\beta}{2}$ will be determined only up to
$O(\e)$ corrections. For simplicity, we omit these $O(\e)$ symbols
from the equations in this section; note, however, that these
corrections do not effect the gauge condition, as discussed above.  

To begin, the most divergent, order $1/s^4$ term in the wave equation 
reads 
\begin{align}
\frac{1}{s^4}\left(2+s^2\partial^c\partial_c\right)
\hmn{B\alpha\beta}{2,-2} & = \frac{4m^2}{s^4}\left(7\hat{\omega}_{ab} 
+ \tfrac{4}{3}\delta_{ab}\right)x^a_\alpha x^b_\beta 
-\frac{4m^2}{s^4}t_\alpha t_\beta,
\end{align}
where the right-hand side is the most divergent part of the
second-order Ricci tensor, as given in Eq.~\eqref{ddR0n4}. From the
$tt$-component of this equation, we read off $\A{}{2,-2}=-2m^2$ and
that $\A{a}{2,-2}$ is arbitrary. From the $ta$-component,
$\B{}{2,-2}$, $\C{ab}{2,-2}$, and $\D{c}{2,-2}$ are arbitrary. From
the $ab$-component, $\K{}{2,-2}=\tfrac{8}{3}m^2$, $\E{}{2,-2}=-7m^2$,
and $\K{a}{2,-2}$, $\F{a}{2,-2}$, $\H{abc}{2,-2}$, and $\I{ab}{2,-2}$
are arbitrary. 

The most divergent, order $1/s^3$ terms in the gauge condition
similarly involve only $\hmn{\alpha\beta}{2,-2}$; they read 
\begin{equation}
\frac{1}{s^3}\left(s\partial^b-2\omega^b\right)
\hmn{B\alpha b}{2,-2}
-\frac{1}{2s^3}\eta^{\mu\nu}x^a_\alpha\left(s\partial_a 
-2\omega_a\right)\!\hmn{B\mu\nu}{2,-2}=0.
\end{equation}
After substituting the results from the wave equation, the
$t$-component of this equation determines that $\C{ab}{2,-2}=0$. The 
$a$-component determines that $\H{abc}{2,-2}=0$, $\I{ab}{2,-2}=0$, and  
\begin{equation}\label{F2n2}
\F{a}{2,-2}=3\K{a}{2,-2}-3\A{a}{2,-2}.
\end{equation}
Thus, the order $1/s^2$ part of $\hmn{\alpha\beta}{2}$ is given by 
\begin{equation}
\begin{split}
\hmn{Btt}{2,-2} & = -2m^2+\A{i}{2,-2}\omega^i, \\
\hmn{Bta}{2,-2} & = \B{}{2,-2}\omega_a
+\epsilon_{a}{}^{ij}\omega_i\D{j}{2,-2},\\
\hmn{Bab}{2,-2} & = \delta_{ab}\left(\tfrac{8}{3}m^2
+\K{i}{2,-2}\omega^i\right) -7m^2\hat{\omega}_{ab}
+\F{\langle a}{2,-2}\omega^{}_{b\rangle},
\end{split}
\end{equation}
where $\F{a}{2,-2}$ is given by Eq.~\eqref{F2n2}.

The metric perturbation in this form depends on five free functions of
time. However, from calculations in flat spacetime, we know that
order $\e^2/s^2$ terms in the metric perturbation can be written in
terms of two free functions: a mass dipole and a spin dipole. We
transform the perturbation into this ``canonical" form by performing a
gauge transformation (c.f. Ref.~\cite{damour-iyer:91}). The
transformation is generated by
$\xi_\alpha=-\frac{1}{s}\B{}{2,-2}t_\alpha
-\frac{1}{2s}\F{a}{2,-2}x^a_\alpha$,
the effect of which is to remove $\B{}{2,-2}$ and $\F{a}{2,-2}$ from
the metric.  This transformation is a refinement of the Lorenz
gauge. (Effects at higher order in $\e$ and $s$ will be automatically
incorporated into the higher-order perturbations.) The condition
$\F{a}{2,-2}-3\K{a}{2,-2}+3\A{a}{2,-2}=0$ then becomes
$\K{a}{2,-2}=\A{a}{2,-2}$. The remaining two functions are related to
the ADM momenta of the internal spacetime:  
\begin{equation}
\begin{array}{lcr}
\A{i}{2,-2} =2M_i\,, && \D{i}{2,-2}=2S_i,
\end{array}
\end{equation}
where $M_i$ is such that $\partial_t M_i$ is proportional to the ADM
linear momentum of the internal spacetime, and $S_i$ is the ADM
angular momentum. $M_i$ is a mass dipole term; it is what would result
from a transformation $x^a\to x^a+M^a/m$ applied to the $1/s$ term in
$\hmn{\alpha\beta}{1}$. $S_i$ is a spin dipole term. Thus, the
order $1/s^2$ part of $\hmn{B\alpha\beta}{2}$ reads 
\begin{equation}\label{h2n2}
\begin{split}
\hmn{Btt}{2,-2} & = -2m^2+2M_i\omega^i, \\
\hmn{Bta}{2,-2} & = 2\epsilon_{aij}\omega^iS^j,\\
\hmn{Bab}{2,-2} & = \delta_{ab}\left(\tfrac{8}{3}m^2
+2M_i\omega^i\right)-7m^2\hat{\omega}_{ab}.
\end{split}
\end{equation}

\subsubsection*{Order $(2,-1)$} 

At the next order, $1/s^3$, because the acceleration is set to zero,
$\hmn{B\alpha\beta}{2,-2}$ does not contribute to
$E^{(0)}_{\mu\nu}[\hmn{}{2}]$, and $\hmn{B\alpha\beta}{1,-1}$ does not
contribute to $\delta^2R^{(0)}_{\mu\nu}[\hmn{}{1}]$. The wave equation
hence reads 
\begin{equation}
\frac{1}{s}\partial^c\partial_c\hmn{B\alpha\beta}{2,-1}= 
\frac{2}{s^3}\ddR{\alpha\beta}{0,-3}{\hmn{}{1}},
\end{equation}
where $\ddR{\alpha\beta}{0,-3}{\hmn{}{1}}$ is given in
Eqs.~\eqref{ddR0n3_tt}--\eqref{ddR0n3_ab}. The $tt$-component of this 
equation implies
$s^2\partial^c\partial_c\hmn{Btt}{2,-1}=6m\H{ij}{1,0}\hat{\omega}^{ij}$,
from which we read off that $\A{}{2,-1}$ is arbitrary and
$\A{ij}{2,-1}=-m\H{ij}{1,0}$. The $ta$-component implies
$s^2\partial^c\partial_c\hmn{Bta}{2,-1}=6m\C{i}{1,0}\hat{\omega}_a^i$,
from which we read off $\B{i}{2,-1}=-m\C{i}{1,0}$ and that
$\C{a}{2,-1}$ is arbitrary. The $ab$-component implies 
\begin{align}
s^2\partial^c\partial_c\hmn{Bab}{2,-1}&=6m\left(\A{}{1,0} 
+\K{}{1,0}\right)\hat{\omega}_{ab}
-12m\H{i\langle a}{1,0}\hat{\omega}_{b\rangle}{}^{i}
+2m\delta_{ab}\H{ij}{1,0}\hat{\omega}^{ij},
\end{align}
from which we read off that $\K{}{2,-1}$ is arbitrary,
$\K{ij}{2,-1}=-\tfrac{1}{3}m\H{ij}{1,0}$,
$\E{}{2,-1}=-m\A{}{1,0}-m\K{}{1,0}$, $\F{ab}{2,-1}=2m\H{ab}{1,0}$, and
$\H{ab}{2,-1}$ is arbitrary. This restricts $\hmn{\alpha\beta}{2,-1}$
to the form 
\begin{equation}
\begin{split}
\hmn{Btt}{2,-1}&=\A{}{2,-1}-m\H{ij}{1,0}\hat{\omega}^{ij}, \\
\hmn{Bta}{2,-1}&=-m\C{i}{1,0}\hat{\omega}_a^i+\C{a}{2,-1}, \\
\hmn{Bab}{2,-1}&=\delta_{ab}\left(\K{}{2,-1} 
-\tfrac{1}{3}m\H{ij}{1,0}\hat{\omega}^{ij}\right) 
-m\left(\A{}{1,0}+\K{}{1,0}\right)\hat{\omega}_{ab} 
+2m\H{i\langle a}{1,0}\hat{\omega}^{}_{b\rangle}{}^i+\H{ab}{2,-1}.
\end{split}
\end{equation}

We next substitute $\hmn{B\alpha\beta}{2,-2}$ and
$\hmn{B\alpha\beta}{2,-1}$ into the order $1/s^2$ terms in the gauge
condition. The $t$-component becomes  
\begin{equation}
\frac{1}{s^2}\left(4m\C{i}{1,0}+12\partial_tM_i
+3\C{i}{2,-1}\right)\omega^i=0,
\end{equation}
from which we read off
\begin{equation}
\C{i}{2,-1}=-4\partial_tM_i-\tfrac{4}{3}m\C{i}{1,0}.
\end{equation}
And the $a$-component becomes
\begin{align}
0&=\left(-\tfrac{4}{3}m\A{}{1,0}-\tfrac{4}{3}m\K{}{1,0} 
-\tfrac{1}{2}\A{}{2,-1}+\tfrac{1}{2}\K{}{2,-1}\right)\omega_a  
+\left(\tfrac{2}{3}m\H{ai}{1,0}-\H{ai}{2,-1}\right)\omega^i 
\nonumber\\
&\quad -2\epsilon_{ija}\omega^i\partial_tS^j,
\end{align}
from which we read off
\begin{align}
\A{}{2,-1}&=\K{}{2,-1}-\tfrac{8}{3}m\left(\A{}{1,0}+\K{}{1,0}\right),\\
\H{ij}{2,-1}&=\tfrac{2}{3}m\H{ij}{1,0},
\end{align}
and that the angular momentum of the internal background is constant
at leading order: 
\begin{equation}
\partial_tS^i=0.\label{Si}
\end{equation}

Thus, the order $1/s$ term in $\hmn{B\alpha\beta}{2}$ is given by 
\begin{equation}\label{h2n1}
\begin{split}
\hmn{Btt}{2,-1}&=\K{}{2,-1}-\tfrac{8}{3}m\left(\A{}{1,0}+\K{}{1,0}\right) 
-m\H{ij}{1,0}\hat{\omega}^{ij}, \\
\hmn{Bta}{2,-1}&=-m\C{i}{1,0}\hat{\omega}_a^i-4\partial_tM_i
-\tfrac{4}{3}m\C{i}{2,-1},\\
\hmn{Bab}{2,-1}&=\delta_{ab}\left(\K{}{2,-1} 
-\tfrac{1}{3}m\H{ij}{1,0}\hat{\omega}^{ij}\right) -m\left(\A{}{1,0}
+\K{}{1,0}\right)\hat{\omega}_{ab} 
+2m\H{i\langle a}{1,0}\hat{\omega}^{}_{b\rangle}{}^i
+\tfrac{2}{3}m\H{ab}{1,0}.
\end{split}
\end{equation}
Note that the undetermined function $\K{}{2,-1}$ appears in precisely 
the form of a mass monopole. The value of this function will never be
determined (though its time-dependence will be). This ambiguity arises
because the mass $m$ that we have defined is the mass of the internal
\emph{background} spacetime, which is based on the inner limit that
holds $\e/R$ fixed. A term of the form $\e^2/R$ appears as a
perturbation of this background, even when, as in this case, it is
part of the mass monopole of the body. This is equivalent to the
ambiguity in any expansion in one's choice of small parameter: one
could expand in powers of $\e$, or one could expand in powers of
$\e+\e^2$, and so on. It is also equivalent to the ambiguity in
defining the mass of a non-isolated body; whether the ``mass" of the
body is taken to be $m$ or $m+\tfrac{1}{2}\K{}{2,-1}$ is a matter of
taste. As we shall discover, the time-dependent part of $\K{}{2,-1}$
is constructed from the tail terms in the first-order metric
perturbation. Hence, the ambiguity in the definition of the mass is,
at least in part, equivalent to whether or not one chooses to include
the free gravitational field induced by the body in what one calls its
mass. (In fact, any order $\e$ incoming radiation, not just that
originally produced by the body, will contribute to this effective
mass.) In any case, we will define the ``correction" to the mass as
$\delta m:=\tfrac{1}{2}\K{}{2,-1}$. 

\subsubsection*{Order $(2,0,\ln)$} 

We next move to the order $\ln(s)/s^2$ terms in the wave equation, and
the order $\ln(s)/s$ terms in the gauge condition, which read 
\begin{align}
\ln s\partial^c\partial_c\hmn{B\alpha\beta}{2,0,\ln} &= 0,\\
\ln s \left(\partial^b\hmn{B\alpha b}{2,0,\ln}
-\tfrac{1}{2}\eta^{\mu\nu}x^a_\alpha\partial_a 
\hmn{B\mu\nu}{2,0,\ln}\right) & = 0.
\end{align}
From this we determine
\begin{align}
\hmn{B\alpha\beta}{2,0,\ln} =\A{}{2,0,\ln}t_\alpha t_\beta
+ 2\C{a}{2,0,\ln}t_{(\beta}x^a_{\alpha)}+(\delta_{ab}\K{}{2,0,\ln}
+\H{ab}{2,0,\ln})x^a_\alpha x^b_\beta.
\end{align}

Finally, we arrive at the order $1/s^2$ terms in the wave equation. At
this order, the body's tidal moments become coupled to those of the
external background. The equation reads 
\begin{equation}\label{2n2 wave equation}
\partial^c\partial_c\hmn{B\alpha\beta}{2,0} 
+\frac{1}{s^2}\!\!\left(\hmn{B\alpha\beta}{2,0,\ln}\!\! 
+\tilde{E}_{\alpha\beta}\right)=
\frac{2}{s^2}\ddR{\alpha\beta}{0,-2}{\hmn{B}{1}},
\end{equation}
where $\tilde{E}_{\alpha\beta}$ comprises the contributions from
$\hmn{B\alpha\beta}{2,-2}$ and $\hmn{B\alpha\beta}{2,-1}$, given in 
Eqs.~\eqref{E_tt}, \eqref{E_ta}, and \eqref{E_ab}. The contribution
from the second-order Ricci tensor is given in
Eqs.~\eqref{ddR0n2_tt}--\eqref{ddR0n2_ab}. 

Foregoing the details, after some algebra we can read off the solution 
\begin{align}\label{h20}
\hmn{Btt}{2,0} & = \A{}{2,0}+\A{i}{2,0}\omega^i
+\A{ij}{2,0}\hat{\omega}^{ij}+\A{ijk}{2,0}\hat{\omega}^{ijk} \\
\hmn{Bta}{2,0} & = \B{}{2,0}\omega_a+\B{ij}{2,0}\hat{\omega}_a{}^{ij} 
+\C{a}{2,0}+\C{ai}{2,0}\hat{\omega}_a{}^i 
+\epsilon_a{}^{bc}\left(\D{c}{2,0}\omega_b
+\D{ci}{2,0}\hat{\omega}_b{}^i 
+\D{cij}{2,0}\hat{\omega}_b{}^{ij}\right)\\
\hmn{Bab}{2,0} & = \delta_{ab}\left(\K{}{2,0}+\K{i}{2,0}\omega^i 
+\K{ijk}{2,0}\hat{\omega}^{ijk}\right) 
+\E{i}{2,0}\hat{\omega}_{ab}{}^i+\E{ij}{2,0}\hat{\omega}_{ab}{}^{ij} 
+\F{\langle a}{2,0}\hat{\omega}_{b\rangle} 
+\F{i\langle a}{2,0}\hat{\omega}_{b\rangle}{}^i\nonumber\\
&\quad +\F{ij\langle a}{2,0}\hat{\omega}_{b\rangle}{}^{ij} 
+\epsilon^{cd}{}_{(a}\hat{\omega}_{b)c}{}^i\G{di}{2,0}
+\H{ab}{2,0}+\H{abi}{2,0}\omega^i
+\epsilon^{cd}{}_{(a}\I{b)d}{2,0}\omega_c,
\end{align}
where each one of the STF tensors is listed in Table~\ref{h20_tensors}. 

In solving Eq.~\eqref{2n2 wave equation}, we also find that the
logarithmic term in the expansion becomes uniquely determined: 
\begin{equation}\label{h2ln}
\hmn{B\alpha\beta}{2,0,\ln} = 
-\tfrac{16}{15}m^2\etide_{ab}x^a_\alpha x^b_\beta.
\end{equation}
This term arises because the sources in the wave equation 
\eqref{2n2 wave equation} contain a term $\propto\etide_{ab}$, which
cannot be equated to any term in
$\partial^c\partial_c\hmn{Bab}{2,0}$. Thus, the wave equation cannot
be satisfied without including a logarithmic term. 

\begin{table}[tb]
\caption[STF tensors in the order $\e^2s^0$ part of the metric
perturbation]{Symmetric trace-free tensors appearing in the
  order $\e^2s^0$ part of the metric perturbation in the buffer region
  around the body. Each tensor is a function of the proper time $t$ on
  the world line $\gamma$, and each is STF with respect to the
  Euclidean metric $\delta_{ij}$.} 
\begin{tabular*}{\linewidth}{l}
\hline\hline
$\begin{array}{rcl}
\A{}{2,0} & \phantom{=}& \text{ is arbitrary} \\ 
\A{i}{2,0} &=& -\partial^2_tM_i-\tfrac{4}{5}S^j\btide_{ji}
+\tfrac{1}{3}M^j\etide_{ji}
-\tfrac{7}{5}m\A{i}{1,1}-\tfrac{3}{5}m\K{i}{1,1} 
+\tfrac{4}{5}m\partial_t\C{i}{1,0}\\
\A{ij}{2,0} &=& -\tfrac{7}{3}m^2\etide_{ij}\\
\A{ijk}{2,0} &=& -2S_{\langle i}\btide_{jk\rangle} 
+\tfrac{5}{3}M_{\langle i} \etide_{jk\rangle}
-\tfrac{1}{2}m\H{ijk}{1,1} \\ 
\B{}{2,0} &=& m\partial_t\K{}{1,0} \\
\B{ij}{2,0} &=& \tfrac{1}{9}\left(2M^l\btide^k_{(i} 
-5S^l\etide^k_{(i}\right)\epsilon_{j)kl}-\tfrac{1}{2}m\C{ij}{1,1} \\ 
\C{i}{2,0} & \phantom{=}& \text{ is arbitrary} \\
\C{ij}{2,0} &=& 2\left(S^l\etide^k_{(i}
-\tfrac{14}{15}M^l\btide^k_{(i}\right)\epsilon_{j)lk}
-m\left(\tfrac{6}{5}\C{ij}{1,1}-\partial_t\H{ij}{1,0}\right)\\
\D{i}{2,0} &=& \tfrac{1}{5}\left(6M^j\btide_{ij} 
-7S^j\etide_{ij}\right)+2m\D{i}{1,1} \\
\D{ij}{2,0} &=& \tfrac{10}{3}m^2\btide_{ij} \\
\D{ijk}{2,0} &=&\tfrac{1}{3}S_{\langle i}\etide_{jk\rangle} 
+\tfrac{2}{3}M_{\langle i}\btide_{jk\rangle} \\
\K{}{2,0} &=& 2\delta m \\
\K{i}{2,0} &=& -\partial^2_tM_i-\tfrac{4}{5}S^j\btide_{ij} 
-\tfrac{5}{9}M^j\etide_{ij}+\tfrac{13}{15}m\A{i}{1,1}\!
+\tfrac{9}{5}m\K{i}{1,1}\! -\tfrac{16}{15}m\partial_t\C{i}{1,0}\\
\K{ijk}{2,0} &=& -\tfrac{5}{9}M_{\langle i}\etide_{jk\rangle} 
+\tfrac{2}{9}S_{\langle i}\btide_{jk\rangle}-\tfrac{1}{6}m\H{ijk}{1,1}\\
\E{i}{2,0} &=& \tfrac{2}{15}M^i\etide_{ij}+\tfrac{1}{5}S^j\btide_{ij} 
+\tfrac{1}{10}m\partial_t\C{i}{1,0}
-\tfrac{9}{20}m\K{i}{1,1} -\tfrac{11}{20}m\A{i}{1,1}\\
\E{ij}{2,0} &=& \tfrac{7}{5}m^2\etide_{ij}\\
\F{i}{2,0} &=&
\tfrac{184}{75}M^j\etide_{ij}+\tfrac{72}{25}S^j\btide_{ij} 
+\tfrac{46}{25}m\partial_t\C{i}{1,0}
-\tfrac{28}{25}m\A{i}{1,1}+\tfrac{18}{25}m\K{i}{1,1} \\
\F{ij}{2,0} &=& 4m^2\etide_{ij}\\
\F{ijk}{2,0} &=& \tfrac{4}{3}M_{\langle i}\etide_{jk\rangle} 
-\tfrac{4}{3}S_{\langle i}\btide_{jk\rangle} +m\H{ijk}{1,1}\\
\G{ij}{2,0} &=& -\tfrac{4}{9}\epsilon^{}_{lk(i}\etide_{j)}^kM^l\! 
-\tfrac{2}{9}\epsilon^{}_{lk(i}\btide_{j)}^kS^l\!+\tfrac{1}{2}m\I{ij}{1,1}\\ 
\H{ij}{2,0} & \phantom{=}& \text{ is arbitrary} \\
\H{ijk}{2,0} &=& \tfrac{58}{15}M_{\langle i}\etide_{jk\rangle} 
-\tfrac{28}{15}S_{\langle i}\btide_{jk\rangle}+\tfrac{2}{5}m\H{ijk}{1,1}\\
\I{ij}{2,0} &=& -\tfrac{104}{45}\epsilon^{}_{lk(i}\etide_{j)}^kM^l 
-\tfrac{112}{45}\epsilon^{}_{lk(i}\btide_{j)}^kS^l+\tfrac{8}{5}m\I{ij}{1,1}
\end{array}$\\
\hline\hline
\end{tabular*}
\label{h20_tensors}
\end{table}

\subsubsection*{Gauge condition} 

We now move to the final equation in the buffer region: the
order $1/s$ gauge condition. This condition will determine the
acceleration $a_{(1)}^\alpha$. At this order, $\hmn{\alpha\beta}{1}$
first contributes to Eq.~\eqref{hB2 gauge}: 
\begin{equation}
L_\alpha^{(1,-1)}[\hmn{}{1}]=\frac{4m}{s}a^{(1)}_ax^a_\alpha .
\end{equation}
The contribution from $\hmn{B\alpha\beta}{2}$ is most easily
calculated by making use of Eqs.~\eqref{gauge_help1} and
\eqref{gauge_help2}. After some algebra, we find that the
$t$-component of the gauge condition reduces to 
\begin{align}
0&=-\frac{4}{s}\partial_t\delta m +\frac{4m}{3s}\partial_t\A{}{1,0}
+\frac{10m}{3s}\partial_t\K{}{1,0}, 
\end{align}
and the $a$-component reduces to 
\begin{align}\label{accelerations}
0 &=\frac{4}{s}\partial_t^2 M_a+\frac{4m}{s}a^{(1)}_a
+\frac{4}{s}\etide_{ai}M^i+\frac{4}{s}\btide_{ai}S^i 
-\frac{2m}{s}\A{a}{1,1}+\frac{4m}{s}\partial_t\C{a}{1,0}.
\end{align}
After removing common factors, these equations become
Eqs.~\eqref{mdot} and \eqref{master_eqn_of_motion}. We remind the
reader that these equations are valid only when evaluated at
$a^a(t)=a_{(0)}^a(t)=0$, except in the term $4m a^{(1)}_a/s$
that arose from $L_\alpha^{(1)}[\hmn{}{1}]$. 

\subsubsection*{Second-order solution} 

We have now completed our calculation in the buffer region. In
summary, the second-order perturbation in the buffer region is given
by $h^{(2)}_{\alpha\beta} =s^{-2} \hmn{B\alpha\beta}{2,-2}
+s^{-1} \hmn{B\alpha\beta}{2,-1} +\hmn{B\alpha\beta}{2,0}
+\ln(s)\hmn{B\alpha\beta}{2,0,\ln} +O(\e,s)$, where
$\hmn{B\alpha\beta}{2,-2}$ is given in Eq.~\eqref{h2n2},
$\hmn{B\alpha\beta}{2,-1}$ in Eq.~\eqref{h2n1},
$\hmn{B\alpha\beta}{2,0}$ in Eq.~\eqref{h20}, and
$\hmn{B\alpha\beta}{2,0,\ln}$ in Eq.~\eqref{h2ln}. In addition, we
have found evolution equations for an effective correction to the
body's mass, given by Eq.~\eqref{mdot}, and mass and spin dipoles,
given by Eqs.~\eqref{Si} and \eqref{accelerations}. 

\subsection{The equation of motion}
\label{comments on force}

\subsubsection*{Master equation of motion} 

Equation~\eqref{accelerations} is the principal result of our
calculation. After simplification, it reads 
\begin{equation}
\partial^2_tM_a+\etide_{ab}M^b = -a^{(1)}_a+\tfrac{1}{2}\A{a}{1,1}
-\partial_t\C{a}{1,0}-\frac{1}{m} S_i\btide^i_a.
\label{accs} 
\end{equation}
We recall that $M_a$ is the body's mass dipole moment, $\etide_{ab}$
and $\btide_{ab}$ are components of the Riemann tensor of the
background spacetime evaluated on the world line, $a^{(1)}_a$ is the
first-order acceleration of the world line, $S_a$ is the body's spin
angular momentum, and $\A{a}{1,1}$, $\C{a}{1,0}$ are vector fields on
the world line that have yet to be determined. The equation is
formulated in Fermi normal coordinates.  

Equation (\ref{accs}) is a type of master equation of motion,
describing the position of the body relative to a world line of
unspecified (though small) acceleration, in terms of the metric
perturbation on the world line, the tidal fields of the spacetime it
lies in, and the spin of the body. It contains two types of
accelerations: $\partial_t^2 M_i$ and $a_{(1)}^i$. The first type is
the second time derivative of the body's mass dipole moment (or the
first derivative of its ADM linear momentum), as measured in a frame
centered on the world line $\gamma$. The second type is the covariant
acceleration of the world line through the external spacetime. In
other words, $\partial_t^2M_i$ measures the acceleration of the body's
centre of mass relative to the centre of the coordinate system, while
$a_i$ measures the acceleration of the coordinate system itself. As
discussed in Sec.~\ref{self_consistent_expansion}, our aim is to
identify the world line as that of the body, and we do so via the
condition that the mass dipole vanishes for all times, meaning that
the body is centered on the world line for all times. If we start with
initial conditions $M_i(0)=0=\partial_t M_i(0)$, then the mass dipole
remains zero for all times if and only if the world line satisfies the
equation 
\begin{equation}
\label{a1}
a^{(1)}_a=\tfrac{1}{2}\A{a}{1,1}-\partial_t\C{a}{1,0}
-\frac{1}{m} S_i\btide^i_a.
\end{equation}
This equation of motion contains two types of terms: a Papapetrou spin 
force, given by $-S_i\btide^i_a$, which arises due to the coupling of
the body's spin to the local magnetic-type tidal field of the external
spacetime; and a self-force, arising from homogenous terms in the wave
equation. Note that the right-hand side of this equation is to be
evaluated at $a^\mu = a^{(0)\mu} = 0$, and that it would contain an
antidamping term $-\frac{11}{3}m\dot a^\mu$ \cite{havas:57,
  havas-goldberg:62, quinn-wald:97} if we had not assumed that the
acceleration possesses an expansion of the form given in 
Eq.~(\ref{a expansion}).  

In our self-consistent approach, we began with the aim of identifying
$\gamma$ by the condition that the body must be centered about it for
all time. However, we could have begun with a regular expansion, in
which the world line is taken to be the remnant $\gamma^{(0)}$ of the
body in the outer limit of $\e\to0$ with only $x^\mu$ fixed. In that
case the acceleration of the world line would necessarily be
$\e$-independent, so $a_{(0)}^i$ would be the full acceleration of
$\gamma^{(0)}$. Hence, when we found $a_{(0)}^i=0$, we would have
identified the world line as a geodesic, and there would be no
corrections $a_{(n)}^i$ for $n>0$. We would then have arrived at the
equation of motion 
\begin{equation}\label{deviation_equation}
\partial^2_tM_a+\etide_{ab}M^b = \tfrac{1}{2}\A{a}{1,1}
-\partial_t\C{a}{1,0}-\frac{1}{m} S_i\btide^i_a.
\end{equation}
This equation of motion was first derived by Gralla and
Wald~\cite{gralla-wald:08} (although they phrased their expansion in
terms of an explicit expansion of the world line, with a deviation
vector on $\gamma^{(0)}$, rather than the mass dipole, measuring the
correction to the motion). It describes the drift of the body away
from the reference geodesic $\gamma^{(0)}$. This drift is driven
partially by the local curvature of the background, as seen in the
geodesic-deviation term $\etide_{ab}M^b$, and by the coupling between
the body's spin and the local curvature. It is also driven by the
self-force, as seen in the terms containing $\A{a}{1,1}$ and
$\C{a}{1,0}$, but unlike in the self-consistent equation, the fields
that produce the self-force are generated by a geodesic past history
(plus free propagation from initial data) rather than by the corrected
motion. 

Although perfectly valid, such an equation is of limited use. If the
external background is curved, then $M_i$ has meaning only if the body
is ``close" to the world line. Thus, $\partial_t^2M_i$ is a meaningful
acceleration only for a short time, since $M_i$ will generically grow
large as the body drifts away from the reference world line. On that
short timescale of validity, the deviation vector defined by $M^i$
accurately points from $\gamma^{(0)}$ to a ``corrected" world line
$\gamma$; that world line, the approximate equation of motion of which
is given in Eq.~\eqref{a1}, accurately tracks the motion of the
body. After a short time, when the mass dipole grows large and the
regular expansion scheme begins to break down, the deviation vector
will no longer correctly point to the corrected world line. Errors will
also accumulate in the field itself, because it is being sourced by
the geodesic, rather than corrected, motion. 

The self-consistent equation of motion appears to be more robust, and
offers a much wider range of validity. Furthermore, even beyond the
above step, where we had the option to choose to set either $M_i$ or
$a^i_{(0)}$ to zero, the self-consistent expansion continues to
contain within it the regular expansion. Starting from the solution in
the self-consistent expansion, one can recover the regular expansion,
and its equation of motion \eqref{deviation_equation}, simply by
assuming an expansion for the world line and following the usual steps
of deriving the geodesic deviation equation.  

\subsubsection*{Detweiler-Whiting decomposition} 
 
Regardless of which equation of motion we opt to use, we have now
completed the derivation of the gravitational self-force, in the sense
that, given the metric perturbation in the neighbourhood of the body,
the self-force is uniquely determined by irreducible pieces of that
perturbation. Explicitly, the terms that appear in the self-force are
given by 
\begin{align}
\A{a}{1,1} &= \frac{3}{4\pi}\int \omega_a\hmn{tt}{1,1}d\Omega,\\
\C{a}{1,0} &= \hmn{ta}{1,0}.
\end{align}
This is all that is needed to incorporate the motion of the body into
a dynamical system that can be numerically evolved; at each timestep,
one simply needs to calculate the field near the world line and
decompose it into irreducible pieces in order to determine the
acceleration of the body. The remaining difficulty is to actually
determine the field at each timestep. In the next section, we will use
the formal integral representation of the solution to determine the
metric perturbation at the location of the body in terms of a tail
integral. 

However, before doing so, we emphasize some important features of the
self-force and the field near the body. First, note that the
first-order external field $\hmn{\alpha\beta}{1}$ splits into two
distinct pieces. There is the singular piece $h^{\rm
  S}_{\alpha\beta}$, given by 
\begin{align}
h^{\rm S}_{tt} &= \frac{2m}{s}\Big\lbrace 1+\tfrac{3}{2}sa_i\omega^i
+2s^2a_ia^i +s^2\left(\tfrac{3}{8}a_{\langle i}a_{j\rangle} 
+\tfrac{5}{6}\etide_{ij}\right)\hat{\omega}^{ij}\Big\rbrace+ O(s^2) \\ 
h^{\rm S}_{ta} &= -2ms\dot a_a 
+\tfrac{2}{3}m s\epsilon_{aij}\btide^j_k\hat{\omega}^{ik}+O(s^2) \\
h^{\rm S}_{ab} &= \frac{2m}{s}\Big\lbrace\delta_{ab}\big[1
-\tfrac{1}{2}sa_i\omega^i +\tfrac{2}{3}s^2a_ia^i 
+s^2\left(\tfrac{3}{8}a_{\langle i}a_{j\rangle}
-\tfrac{5}{18}\etide_{ij}\right)\hat{\omega}^{ij}\big] 
+2s^2a_{\langle a}a_{b\rangle}\nonumber\\
&\quad -\tfrac{19}{9}s^2\etide_{ab}
+\tfrac{2}{3}s^2\etide^i_{\langle a}\hat{\omega}_{b\rangle i}
\Big\rbrace +O(s^2).
\end{align}
This field is a solution to the homogenous wave equation for $s>0$,
but it is divergent at $s=0$. It is the generalization of the $1/s$
Newtonian field of the body, as perturbed by the tidal fields of the
external spacetime $g_{\alpha\beta}$. Comparing with results to be
derived below in Sec.~\ref{first_order_perturbation}, we find that it
is precisely the Detweiler-Whiting singular field for a point mass.  

Next, there is the Detweiler-Whiting regular field 
$h^{\rm R}_{\alpha\beta}=\hmn{\alpha\beta}{1}
-h^{\rm S}_{\alpha\beta}$, given by
\begin{align}
h^{\rm R}_{tt} &= \A{}{1,0}+s\A{i}{1,1}\omega^i+O(s^2), \\
h^{\rm R}_{ta} &= \C{a}{1,0} +s\Big(\B{}{1,1}\omega_a
+\C{ai}{1,1}\omega^i +\epsilon_{ai}{}^j\D{j}{1,1}\omega^i\Big)
+O(s^2),\\
h^{\rm R}_{ab} &= \delta_{ab}\K{}{1,0}+\H{ab}{1,0}
+s\Big(\delta_{ab}\K{i}{1,1}\omega^i+\H{abi}{1,1}\omega^i 
+\epsilon\indices{_i^j_{(a}}\I{b)j}{1,1}\omega^i
+\F{\langle a}{1,1}\omega^{}_{b\rangle}\Big) +O(s^2).
\end{align}
This field is a solution to the homogeneous wave equation even at
$s=0$. It is a free radiation field in the neighbourhood of the
body. And it contains all the free functions in the buffer-region
expansion. 

Now, the acceleration of the body is given by
\begin{align}
a^{(1)}_a = \tfrac{1}{2}\partial_ah^{\rm R}_{tt} 
-\partial_t h^{\rm R}_{ta} 
-\frac{1}{m} S_i\btide^i_a,
\end{align}
which we can rewrite as 
\begin{align}\label{acceleration_hR}
a_{(1)}^\alpha &= -\tfrac{1}{2}\left(g^{\alpha\delta}
+u^{\alpha}u^{\delta}\right) \!\left(2h^{\rm R}_{\delta\beta;\gamma}
-h^{\rm R}_{\beta\gamma;\delta}\right)\!\!\big|_{a=0} 
u^{\beta}u^{\gamma} 
+\frac{1}{2m}R^{\alpha}{}_{\beta\gamma\delta}u^\beta 
S^{\gamma\delta}, 
\end{align}
where $S^{\gamma\delta}:= e_c^\gamma 
e_d^\delta\epsilon^{cdj}S_j$. In other words, up to order $\e^2$ 
errors, a body with order $\e$ or smaller spin (i.e., one for which
$S^{\gamma\delta}=0$), moves on a geodesic of a spacetime
$g_{\alpha\beta}+\e h^{\rm R}_{\alpha\beta}$, where 
$h^{\rm R}_{\alpha\beta}$ is a free radiation field in the
neighbourhood of the body; a local observer would measure the
``background spacetime,'' in which the body is in free fall, to have
the metric $g_{\alpha\beta}+\e h_{\alpha\beta}^R$ instead of  
$g_{\alpha\beta}$. If we performed a
transformation into Fermi coordinates in $g_{\alpha\beta}+\e
h_{\alpha\beta}^R$, the metric would contain no acceleration term, and
it would take the simple form of a smooth background plus a singular
perturbation. Hence, the Detweiler-Whiting axiom is a consequence,
rather than an assumption, of our derivation, and we have recovered
precisely the picture it provides in the point particle case. In the
electromagnetic and scalar cases, Harte has shown that this result is
quite general: even for a finite extended body, the field it produces
can be split into a homogeneous
field~\cite{harte:08,harte:09a,harte:09b} that exerts a direct force
on the body, and a nonhomogeneous field that exerts only an indirect
force by altering the body's multipole moments. His results should be
generalizable to the gravitational case as well. 

\subsection{The effect of a gauge transformation on the force}
\label{gauge_and_force}

We now turn to the question of how the world line transforms under a
gauge transformation. We begin with the equation of motion
\eqref{master_eqn_of_motion}, presented again here: 
\begin{align}
\partial_t^2 M_a+\etide_{ai}M^i &= -a^{(1)}_a 
- \frac{1}{m} \btide_{ai}S^i  
+ \left[ \tfrac{1}{2}\A{a}{1,1}-\partial_t\C{a}{1,0}\right]_{a^\mu=0}.
\label{master_copy} 
\end{align}
Setting $M_i=0$, we derive the first-order acceleration of $\gamma$,
given in Eq.~\eqref{a1}. If, for simplicity, we neglect the Papapetrou
spin term, then that acceleration is given by 
\begin{align}
a^{(1)}_a &= \lim_{s\to0}\left(\frac{3}{4\pi}\int\frac{\omega_a}{2s}
\hmn{tt}{1}d\Omega -\partial_t\hmn{ta}{1}\right)\nonumber\\
&=
\lim_{s\to0}\frac{3}{4\pi}\int\left(\tfrac{1}{2}\partial_i\hmn{tt}{1} 
-\partial_t\hmn{ti}{1}\right)\omega^i \omega_a d\Omega 
\label{reg force},
\end{align}
where it is understood that explicit appearances of the acceleration
are to be set to zero on the right-hand side. The first equality
follows directly from Eq.~\eqref{a1} and the definitions of
$\A{a}{1,1}$ and $\C{a}{1,0}$. The second equality follows from the
STF decomposition of $\hmn{\alpha\beta}{1}$ and the integral
identities \eqref{omega_integral}--\eqref{omegahat_integral}. We could 
also readily derive the form of the force given by the Quinn-Wald
method of regularization:
$\lim_{s\to0}\frac{1}{4\pi}\int\left(\tfrac{1}{2}\partial_a\hmn{tt}{1}
  -\partial_t\hmn{ta}{1}\right)d\Omega$. However, in order to derive a
gauge-invariant equation of motion, we shall use the form in
Eq.~\eqref{reg force}. 

Suppose that we had not chosen a world line for which the mass dipole
vanishes, but instead had chosen some ``nearby" world line. Then
Eq.~\eqref{master_copy} provides the relationship between the
acceleration of that world line, the mass dipole relative to it, and
the first-order metric perturbations (we again neglect spin for
simplicity). The mass dipole is given by $M_i= 
\frac{3}{8\pi}\lim_{s\to 0}\int s^2\hmn{tt}{2}\omega_i d\Omega$, which
has the covariant form  
\begin{equation}
M_{\alpha'} = \frac{3}{8\pi}\lim_{s\to0}\int\! 
g^\alpha_{\alpha'}\omega_\alpha s^2\hmn{\mu\nu}{2}
u^\mu u^\nu d\Omega,
\end{equation}
where a primed index corresponds to a point on the world line. Note
that the parallel propagator does not interfere with the
angle-averaging, because in Fermi coordinates,
$g^\alpha_{\beta'}=\delta^\alpha_\beta+O(\e,s^2)$. One can also
rewrite the first-order-metric-perturbation terms in
Eq.~\eqref{master_copy} using the form given in 
Eq.~\eqref{reg force}. We then have Eq.~\eqref{master_copy} in the
covariant form  
\begin{align}
\label{gauge-invariant form}
&\frac{3}{8\pi}\lim_{s\to0}\int\! g^\alpha_{\alpha'}\! 
\left(\!g_{\alpha\beta}\frac{D^2}{d\tau^2}
+\etide_{\alpha\beta}\!\right)\!\omega^\beta 
s^2\hmn{\mu\nu}{2}u^\mu u^\nu d\Omega\big|_{a=a^{(0)}} 
\nonumber\\
&=-\frac{3m}{8\pi}\lim_{s\to0}\int\! 
g^\alpha_{\alpha'}\left(2\hmn{\beta\mu;\nu}{1}
-\hmn{\mu\nu;\beta}{1}\right)u^\mu u^\nu 
\omega_\alpha^\beta d\Omega\big|_{a=a^{(0)}} -ma^{(1)}_{\alpha'}. 
\end{align}

Now consider a gauge transformation generated by
$\e\xi^{(1)\alpha}[\gamma]+\tfrac{1}{2}\e^2\xi^{(2)\alpha}[\gamma]+\cdots$,  
where $\xi^{(1)\alpha}$ is bounded as $s\to0$, and $\xi^{(2)\alpha}$
diverges as $1/s$. More specifically, we assume the expansions
$\xi^{(1)\alpha}=\xi^{(1,0)\alpha}(t,\theta^A)+O(s)$ and
$\xi^{(1)\alpha}=\frac{1}{s}\xi^{(2,-1)\alpha}(t,\theta^A)+O(1)$. (The
dependence on $\gamma$ appears in the form of dependence on proper
time $t$. Other dependences could appear, but it would not affect the
result.) This transformation preserves the presumed form of the outer
expansion, both in powers of $\e$ and in powers of $s$. The metric
perturbations transform as  
\begin{align}
\hmn{\mu\nu}{1} &\to \hmn{\mu\nu}{1}+2\xi^{(1)}_{(\mu;\nu)},\\
\hmn{\mu\nu}{2} &\to \hmn{\mu\nu}{2}+\xi^{(2)}_{(\mu;\nu)}
+\hmn{\mu\nu;\rho}{1}\xi_{(1)}^\rho 
+ 2\hmn{\rho(\mu}{1}\xi_{(1)}^\rho{}_{;\nu)} 
+\xi_{(1)}^\rho\xi^{(1)}_{(\mu;\nu)\rho}
+\xi_{(1);\mu}^\rho\xi^{(1)}_{\rho;\nu}
+\xi^\rho_{(1);(\mu}\xi^{(1)}_{\nu);\rho}.
\end{align}
Using the results for $\hmn{\alpha\beta}{1}$, the effect of this
transformation on $\hmn{tt}{2}$ is given by 
\begin{equation}
\hmn{tt}{2}\to\hmn{tt}{2}-\frac{2m}{s^2}\omega^i\xi^{(1)}_i+O(s^{-1}).
\end{equation}
The order $1/s^2$ term arises from
$\hmn{\mu\nu;\rho}{1}\xi_{(1)}^\rho$ in the gauge transformation. On
the right-hand side of Eq.~\eqref{gauge-invariant form}, the
metric-perturbation terms transform as 
\begin{align}
(2\hmn{\beta\mu;\nu}{1}-\hmn{\mu\nu;\beta}{1})
u^\mu u^\nu \omega^\beta &\to (2\hmn{\beta\mu;\nu}{1}
-\hmn{\mu\nu;\beta}{1})u^\mu u^\nu \omega^\beta 
+2\omega_\beta\left(g_\beta^\gamma\frac{D^2}{d\tau^2}
+\etide_\beta^\gamma\right)\xi^{(1)}_\gamma.
\end{align}
The only remaining term in the equation is $ma_{(1)}^\alpha$. If we
extend the acceleration off the world line in any smooth manner, then
it defines a vector field that transforms as $a_\alpha\to
a_\alpha+\e\Lie{\xi_{(1)}}a_\alpha+\cdots$. Since $a_{(0)}^\alpha=0$,
this means that $a_{(1)}^\alpha\to a_{(1)}^\alpha$ --- it is invariant
under a gauge transformation. It is important to note that this
statement applies to the acceleration on the original world line; it
does not imply that the acceleration of the body itself is
gauge-invariant. 

From these results, we find that the left- and right-hand sides of
Eq.~\eqref{gauge-invariant form} transform in the same way: 
\begin{align}\label{gauge_effect}
{\rm LHS/RHS} \to {\rm LHS/RHS} - \frac{3}{4\pi}\lim_{s\to0}\int 
g^\alpha_{\alpha'} \omega_\alpha^\beta
\left(g_\beta^\gamma\frac{D^2}{d\tau^2}
+\etide_\beta^\gamma\right)\xi^{(1)}_\gamma d\Omega.
\end{align}
Therefore, Eq.~\eqref{gauge-invariant form} provides a gauge-invariant
relationship between the acceleration of a chosen fixed world line, the
mass dipole of the body relative to that world line, and the
first-order metric perturbations. So suppose that we begin in the
Lorenz gauge, and we choose the fixed world line $\gamma$ such that the
mass dipole vanishes relative to it. Then in some other gauge, the
mass dipole will no longer vanish relative to $\gamma$, and we must
adopt a different, nearby fixed world line $\gamma'$. If the mass
dipole is to vanish relative to $\gamma'$, then the acceleration of
that new world line must be given by $a^\alpha=\e
a_{(1)}^\alpha+o(\e)$, where 
\begin{equation}
a^{(1)}_{\alpha'}=-\frac{3m}{8\pi}\lim_{s\to0}\int\! 
g^\alpha_{\alpha'}(2\hmn{\beta\mu;\nu}{1}
-\hmn{\mu\nu;\beta}{1})u^\mu u^\nu 
\omega_\alpha^\beta d\Omega.\big|_{a=a^{(0)}}.
\end{equation}
Hence, this is a covariant and gauge-invariant form of the first-order
acceleration. By that we mean the \emph{equation} is valid in any
gauge, not that the value of the acceleration is the same in every
gauge. Under a gauge transformation, a new fixed world line is adopted,
and the value of the acceleration on it is related to that on the old
world line according to Eq.~\eqref{gauge_effect}. In the particular
case that $\xi^{(1)}_\mu$ has no angle-dependence on the world line,
this relationship reduces to 
\begin{equation}
a^{(1)\alpha}_{\rm new} = a^{(1)\alpha}_{\rm old}
-\left(g^\alpha_\beta+u^\alpha u_\beta\right)
\left(\frac{D^2\xi^\beta_{(1)}}{d\tau^2}
+R^\beta{}_{\mu\nu\rho}u^\mu\xi^\nu_{(1)}u^\rho\right), 
\end{equation}
as first derived by Barack and Ori \cite{barack-ori:01}. (Here we've
replaced the tidal field with its expression in terms of the Riemann
tensor to more transparently agree with equations in the literature.)
An argument of this form was first presented by
Gralla~\cite{gralla:09} for the case of a regular expansion, and was
extended to the case of a self-consistent expansion in
Ref.~\cite{pound:10b}. 

\section{Global solution in the external spacetime}
\label{global_solution}

In the previous sections, we have determined the equation of motion of 
$\gamma$ in terms of the metric perturbation; we now complete the
first-order solution by determining the metric perturbation. In early
derivations of the gravitational self-force (excluding those in
Refs.~\cite{fukumoto-etal:06,gralla-wald:08}), the first-order
external perturbation was simply assumed to be that of a point
particle. This was first justified by Gralla and Wald
\cite{gralla-wald:08}. An earlier argument made by D'Eath
\cite{death:75, death:96} (and later used by Rosenthal
\cite{rosenthal:06a}) provided partial justification but was
incomplete \cite{pound:10a}. Here, we follow the derivation in
Ref.~\cite{pound:10a}, which makes use of the same essential elements
as D'Eath's: the integral formulation of the perturbative Einstein
equation and the asymptotically small radius of the tube $\Gamma$. 

\subsection{Integral representation} 

Suppose we take our buffer-region expansion of $\hmn{\alpha\beta}{1}$
to be valid everywhere in the interior of $\Gamma$ (in $\man_E$),
rather than just in the buffer region. This is a meaningful
supposition in a distributional sense, since the $1/s$ singularity in
$\hmn{\alpha\beta}{1}$ is locally integrable even at $\gamma$. Note
that the extension of the buffer-region expansion is not intended to
provide an accurate or meaningful approximation in the interior; it is
used only as a means of determining the field in the exterior. We can
do this because the field values in $\Omega$ are entirely determined
by the field values on $\Gamma$, so using the buffer-region expansion
in the interior of $\Gamma$ leaves the field values in $\Omega$
unaltered. Now, given the extension of the buffer-region expansion, it
follows from Stokes' law that the integral over $\Gamma$ in
Eq.~\eqref{formal_solution} can be replaced by a volume integral over
the interior of the tube, plus two surface integrals over the ``caps"
$\mathcal{J}_{\rm cap}$ and $\Sigma_{\rm cap}$, which fill the
``holes'' in $\mathcal{J}$ and $\Sigma$, respectively, where they
intersect $\Gamma$. Schematically, we can write Stokes' law as
$\int_{\text{Int}(\Gamma)}=\int_{\mathcal{J}_{\rm cap}}
+\int_{\Sigma_{\rm cap}}-\int_{\Gamma}$, where $\text{Int}(\Gamma)$ is
the interior of $\Gamma$. This is now valid as a distributional
identity. (Note that the ``interior" here means the region bounded by 
$\Gamma\cup\Sigma_{\rm cap}\cup\mathcal{J}_{\rm cap}$; 
$\text{Int}(\Gamma)$ does not refer to the set of interior points in
the point-set defined by $\Gamma$.) The minus sign in front of the
integral over $\Gamma$ accounts for the fact that the directed surface
element in Eq.~\eqref{formal_solution} points \emph{into} the
tube. Because $\mathcal{J}_{\rm cap}$ does not lie in the past of any
point in $\Omega$, it does not contribute to the perturbation at
$x\in\Omega$. Hence, we can rewrite Eq.~\eqref{formal_solution} as  
\begin{align}
\hmn{\alpha\beta}{1} &= -\frac{1}{4\pi}\!\!\!
\int\limits_{\text{Int}(\Gamma)}\!\!\! \del{\mu'}
\Big(G^+_{\alpha\beta}{}^{\alpha'\beta'}\nabla^{\mu'}
\hmn{\alpha'\beta'}{1} -\hmn{\alpha'\beta'}{1}\nabla^{\mu'}
G^+_{\alpha\beta}{}^{\alpha'\beta'}\Big)dV' 
+\hmn{\bar\Sigma\alpha\beta}{1}\nonumber\\
&=-\frac{1}{4\pi}\!\!\!\int\limits_{\text{Int}(\Gamma)}\!\!\! 
\Big(G_{\alpha\beta}{}^{\alpha'\beta'}E_{\alpha'\beta'}[\hmn{}{1}] 
-\hmn{\alpha'\beta'}{1}E^{\alpha'\beta'}[G^+_{\alpha\beta}]\Big)dV' 
+\hmn{\bar\Sigma\alpha\beta}{1},
\end{align}
where $\hmn{\bar\Sigma\alpha\beta}{1}$ is the contribution from the
spatial surface $\bar\Sigma:=\Sigma\cup\Sigma_{\rm cap}$, and
$E^{\alpha'\beta'}[G^+_{\alpha\beta}]$ denotes the action of the
wave-operator on $G^+_{\alpha\beta}{}^{\gamma'\delta'}$. Now note that 
$E^{\alpha'\beta'}[G^+_{\alpha\beta}]\propto\delta(x,x')$; since
$x\notin\text{Int}(\Gamma)$, this term integrates to zero. Next note
that $E_{\alpha'\beta'}[\hmn{\alpha\beta}{1}]$ vanishes everywhere
except at $\gamma$. This means that the field at $x$ can be written as 
\begin{align}
\hmn{\alpha\beta}{1} &= \frac{-1}{4\pi}\!\lim_{\rad\to 0}\!\!\!
\int\limits_{\text{Int}(\Gamma)} \!\!\!\!\!
G^+_{\alpha\beta}{}^{\alpha'\beta'}E_{\alpha'\beta'}[\hmn{}{1}]dV'
+\hmn{\bar\Sigma\alpha\beta}{1}.
\end{align}
Making use of the fact that $E_{\alpha\beta}[\hmn{}{1}]
= \partial^c\partial_c(1/s)\hmn{\alpha\beta}{1,-1}+O(s^{-2})$, along
with the identity $\partial^c\partial_c(1/s)=-4\pi\delta^3(x^a)$,
where $\delta^3$ is a coordinate delta function in Fermi coordinates,
we arrive at the desired result 
\begin{equation}
\hmn{\alpha\beta}{1} = 2m\int_\gamma 
G^+_{\alpha\beta\bar\alpha\bar\beta}\left(2u^{\bar\alpha}u^{\bar\beta} 
+g^{\bar\alpha\bar\beta}\right)d\bar t
+\hmn{\bar\Sigma\alpha\beta}{1}. 
\end{equation}
Therefore, in the region $\Omega$, the leading-order perturbation 
produced by the asymptotically small body is identical to the field
produced by a point particle. At second order, the same method can be 
used to simplify Eq.~\eqref{h2 eqn} by replacing at least part of the
integral over $\Gamma$ with an integral over $\gamma$. We will not
pursue this simplification here, however.  

Gralla and Wald \cite{gralla-wald:08} provided an alternative
derivation of the same result, using distributional methods to prove
that the distributional source for the linearized Einstein equation
must be that of a point particle in order for the solution to diverge
as $1/s$. One can understand this by considering that the most
divergent term in the linearized Einstein tensor is a Laplacian acting
on the perturbation, and the Laplacian of $1/s$ is a flat-space delta
function; the less divergent corrections are due to the curvature of
the background, which distorts the flat-space distribution into a
covariant curved-spacetime distribution. 

\subsection{Metric perturbation in Fermi coordinates}
\label{first_order_perturbation} 

\subsubsection*{Metric perturbation} 

We have just seen that the solution to the wave equation with a
point-mass source is given by     
\begin{equation}
\hmn{\alpha\beta}{1} = 2m\int_\gamma 
G_{\alpha\beta\alpha'\beta'}(2u^{\alpha'}u^{\beta'}
+g^{\alpha'\beta'})dt'+\hmn{\Sigma\alpha\beta}{1}.
\end{equation}
One can also obtain this result from Eq.~\eqref{18.2.1} by taking the 
trace-reversal and making use of the Green's function identity
\eqref{15.3b.2}. In this section, we seek an expansion of the
perturbation in Fermi coordinates. Following the same steps as in
Sec.~\ref{18.2}, we arrive at 
\begin{equation}
\hmn{\alpha\beta}{1} =\frac{2m}{r}U_{\alpha\beta\alpha'\beta'}
(2u^{\alpha'}u^{\beta'}+g^{\alpha'\beta'}) +\tail_{\alpha\beta}(u). 
\end{equation}
Here primed indices now refer to the retarded point $z^\alpha(u)$ on
the world line, $r$ is the retarded radial coordinate at $x$, and the
tail integral is given by 
\begin{align}
\tail_{\alpha\beta}(u) & = 
2m\int_{t^<}^{u} V_{\alpha\beta\alpha'\beta'}
(2u^{\alpha'}u^{\beta'}+g^{\alpha'\beta'})dt' 
+ 2m\int^{t^<}_0 G_{\alpha\beta\alpha'\beta'}
(2u^{\alpha'}u^{\beta'}+g^{\alpha'\beta'})dt'
+\hmn{\Sigma\alpha\beta}{1}\nonumber\\
&= 2m\int_0^{u^-} G_{\alpha\beta\alpha'\beta'}
(2u^{\alpha'}u^{\beta'}+g^{\alpha'\beta'})dt'
+\hmn{\Sigma\alpha\beta}{1},
\end{align}
where $t^<$ is the first time at which the world line enters
$\mathcal{N}(x)$, and $t=0$ denotes the time when it crosses the
initial-data surface $\Sigma$.   
 
We expand the direct term in $\hmn{\alpha\beta}{1}$ in powers of $s$
using the following: the near-coincidence expansion
$U_{\alpha\beta}{}^{\alpha'\beta'}=g^{\alpha'}_\alpha
g^{\beta'}_\beta(1+O(s^3))$; the relationship between $r$ and $s$,
given by Eq.~\eqref{10.2.2}; and the coordinate expansion of the
parallel-propagators, obtained from the formula
$g^{\alpha'}_\alpha=u^{\alpha'}e^0_\alpha+e^{\alpha'}_a e^a_{\alpha}$,
where the retarded tetrad $(u^\alpha,e_a^\alpha)$ can be expanded in
terms of $s$ using Eqs.~\eqref{10.3.3}, \eqref{10.3.4}, \eqref{8.5.1},
and \eqref{8.5.2}. We expand the tail integral similarly: noting that
$u=t-s+O(s^2)$, we expand $\tail_{\alpha\beta}(u)$ about $t$ as
$\tail_{\alpha\beta}(t)-s\partial_t \tail_{\alpha\beta}(t)+O(s^2)$;
each term is then expanded using the near-coincidence expansions
$V_{\alpha\beta}{}^{\alpha''\beta''}=g^{\gamma''}_{(\alpha}
g^{\delta''}_{\beta)}R^{\alpha''}{}_{\gamma''}{}^{\beta''}{}_{\delta''}+O(s)$
and $\tail_{\alpha\beta}(t)=g^{\bar\alpha}_\alpha
g^{\bar\beta}_\beta(\tail_{\bar\alpha\bar\beta}
+s\tail_{\bar\alpha\bar\beta i}\omega^i)+O(s^2)$, where barred indices
correspond to the point $\bar x=z(t)$, and 
$\tail_{\bar\alpha\bar\beta\bar\gamma}$ is given by 
\begin{equation}
\tail_{\bar\alpha\bar\beta\bar\gamma}=2m\int_0^{t^-} 
\del{\bar\gamma}G_{\bar\alpha\bar\beta\alpha'\beta'}
(2u^{\alpha'}u^{\beta'}+g^{\alpha'\beta'})dt'
+\hmn{\Sigma\bar\alpha\bar\beta\bar\gamma}{1}.
\end{equation}
This yields the expansion
\begin{equation}
\tail_{\alpha\beta}(u)=g^{\bar\alpha}_\alpha g^{\bar\beta}_\beta
(\tail_{\bar\alpha\bar\beta}+s\tail_{\bar\alpha\bar\beta i}\omega^i
-4ms\etide_{\bar\alpha\bar\beta}) +O(s^2).
\end{equation}
As with the direct part, the final coordinate expansion is found by
expressing $g^{\bar\alpha}_\alpha$ in terms of the Fermi tetrad. 

Combining the expansions of the direct and tail parts of the
perturbation, we arrive at the expansion in Fermi coordinates: 
\begin{align}
\hmn{tt}{1} & = \frac{2m}{s}\left(1+\tfrac{3}{2}sa_i\omega^i 
+\tfrac{3}{8}s^2a_ia_j\omega^{ij} 
-\tfrac{15}{8}s^2\dot a_{\bar\alpha}u^{\bar\alpha} 
+\tfrac{1}{3}s^2\dot a_i \omega^i 
+\tfrac{5}{6}s^2\etide_{ij}\omega^{ij}\right) 
+(1+2sa_i\omega^i)\tail_{00} \nonumber\\
&\quad +s\tail_{00i}\omega^i+O(s^2),\\
\hmn{ta}{1} & = 4ma_a-\tfrac{2}{3}msR_{0iaj}\omega^{ij}
+2ms\etide_{ai}\omega^i-2ms\dot a_a 
+(1+sa_i\omega^i)\tail_{0a}+s\tail_{0ai}\omega^i+O(s^2), \\
\hmn{ab}{1} & = \frac{2m}{s}\left(1-\tfrac{1}{2}sa_i\omega^i 
+\tfrac{3}{8}s^2a_ia_j\omega^{ij} 
+\tfrac{1}{8}s^2\dot a_{\bar\alpha}u^{\bar\alpha}  
+\tfrac{1}{3}s^2\dot a_i \omega^i 
-\tfrac{1}{6}s^2\etide_{ij}\omega^{ij}\right)\delta_{ab}
+4msa_aa_b \nonumber\\
&\quad -\tfrac{2}{3}msR_{aibj}\omega^{ij}-4ms\etide_{ab}
+\tail_{ab} +s\tail_{abi}\omega^i +O(s^2).
\end{align}
As the final step, each of these terms is decomposed into irreducible
STF pieces using the formulas \eqref{omegahat_expansion},
\eqref{decomposition_1}, and \eqref{decomposition_2}, yielding 
\begin{align}
\hmn{tt}{1} &= \frac{2m}{s}+\A{}{1,0}+3ma_i\omega^i
+s\big[4ma_ia^i+\A{i}{1,1}\omega^i 
+m\left(\tfrac{3}{4}a_{\langle i}a_{j\rangle} 
+\tfrac{5}{3}\etide_{ij}\right)\hat{\omega}^{ij}\big]+O(s^2),\\
\hmn{ta}{1} &= \C{a}{1,0}+s\big(\B{}{1,1}\omega_a-2m\dot a_a
+\C{ai}{1,1}\omega^i +\epsilon_{ai}{}^j\D{j}{1,1}\omega^i 
+\tfrac{2}{3}m\epsilon_{aij}\btide^j_k\hat{\omega}^{ik}\big)
+O(s^2), \\
\hmn{ab}{1} &= \frac{2m}{s}\delta_{ab}+(\K{}{1,0}
-ma_i\omega^i)\delta_{ab}+\H{ab}{1,0} 
+s\Big\lbrace\delta_{ab}\big[\tfrac{4}{3}ma_ia^i
+\K{i}{1,1}\omega^i 
+\tfrac{3}{4}ma_{\langle i}a_{j\rangle}\hat{\omega}^{ij} 
-\tfrac{5}{9}m\etide_{ij}\hat{\omega}^{ij}\big]\nonumber\\
&\quad +\tfrac{4}{3}m\etide^i_{\langle a}\hat{\omega}_{b\rangle i} 
+4ma_{\langle a}a_{b\rangle} 
-\tfrac{38}{9}m\etide_{ab}+\H{abi}{1,1}\omega^i 
+\epsilon_i{}^j{}_{(a}\I{b)j}{1,1}\omega^i 
+\F{\langle a}{1,1}\omega^{}_{b\rangle}\Big\rbrace+O(s^2),
\end{align}
where the uppercase hatted tensors are specified in 
Table~\ref{STF wrt tail}. Because the STF decomposition is unique,
these tensors must be identical to the free functions in
Eq.~(\ref{h11}); hence, those free functions, comprising a regular,
homogenous solution in the buffer region, have been uniquely
determined by boundary conditions and waves emitted by the particle in
the past.

\begin{table}[tb]
\caption[The regular field in terms of $\etide_{ab}$ and
$\tail_{\alpha\beta}$]{Symmetric trace-free tensors in the first-order
  metric perturbation in the buffer region, written in terms of the
  electric-type tidal field $\etide_{ab}$, the acceleration $a_i$, and
  the tail of the perturbation.}   
\begin{tabular*}{\textwidth}
{@{\hspace{0.05\textwidth}}
c@{\hspace{0.05\textwidth}}|@{\hspace{0.05\textwidth}}r}
\hline\hline
$\begin{array}{ll}
\A{}{1,0} &= \tail_{00}\\
\C{a}{1,0} &= \tail_{0a}+ma_a\\
\K{}{1,0} &= \tfrac{1}{3}\delta^{ab}\tail_{ab}\\
\H{ab}{1,0} &= \tail_{\langle ab\rangle}\\
\A{a}{1,1} &= \tail_{00a}+2\tail_{00}a_a+\tfrac{2}{3}m\dot a_a\\
\B{}{1,1} &= \tfrac{1}{3}\tail_{0ij}\delta^{ij}+\tfrac{1}{3}\tail_{0i}a^i
\end{array}$
& 
$\begin{array}{ll}
\C{ab}{1,1} &= \tail_{0\langle ab\rangle}+2m\etide_{ab}
+\tail_{0\langle a}a_{b\rangle} \\
\D{a}{1,1} &= \tfrac{1}{2}\epsilon_a{}^{bc}(\tail_{0bc}
+\tail_{0b}a_c)\\
\K{a}{1,1} &= \frac{1}{3}\delta^{bc}\tail_{bca}
+\tfrac{2}{3}m\dot a_a\\
\H{abc}{1,1} &= \tail_{\langle abc\rangle}\\
\F{a}{1,1} &= \tfrac{3}{5}\delta^{ij}\tail_{\langle ia\rangle j}\\
\I{ab}{1,1} &= \tfrac{2}{3}\displaystyle{\mathop{\STF}_{ab}} 
\left(\epsilon_b{}^{ij}\tail_{\langle ai\rangle j}\right)
\end{array}$\\
\hline\hline
\end{tabular*}
\label{STF wrt tail}
\end{table}

\subsubsection*{Singular and regular pieces}

The Detweiler-Whiting singular field is given by
\begin{equation}
h^{\rm S}_{\alpha\beta} = 2m\int G^S_{\alpha\beta\alpha'\beta'}
(2u^{\alpha'}u^{\beta'}+g^{\alpha'\beta'})dt'.
\end{equation}
Using the Hadamard decomposition
$G^S_{\alpha\beta\alpha'\beta'}=\tfrac{1}{2}
U_{\alpha\beta\alpha'\beta'}\delta(\sigma)
-\tfrac{1}{2}V_{\alpha\beta\alpha'\beta'}\theta(\sigma)$, we can write  
this as 
\begin{align}
h^{\rm S}_{\alpha\beta} &= \frac{m}{r}U_{\alpha\beta\alpha'\beta'}
(2u^{\alpha'}u^{\beta'}+g^{\alpha'\beta'})
+ \frac{m}{\advr}U_{\alpha\beta\alpha''\beta''}
(2u^{\alpha''}u^{\beta''}+g^{\alpha''\beta''}) \nonumber\\
&\quad-2m\int^v_u V_{\alpha\beta\bar\alpha\bar\beta}
(u^{\bar\alpha}u^{\bar\beta}
+\tfrac{1}{2} g^{\bar\alpha\bar\beta})d\bar t, \label{hS}
\end{align}
where primed indices now refer to the retarded point $x'=z(u)$;
double-primed indices refer to the advanced point $x''=z(v)$; barred
indices refer to points in the segment of the world line between $z(u)$
and $z(v)$. The first term in Eq.~\eqref{hS} can be read off from the
calculation of the retarded field. The other terms are expanded using
the identities $v=u+2s+O(s^2)$ and 
$\advr=r(1+\tfrac{2}{3}s^2\dot a_i\omega^i)$. The final result is 
\begin{align}
h^{\rm S}_{tt} &= \frac{2m}{s}+3ma_i\omega^i
+ms\big[4a_ia^i+\tfrac{3}{4}a_{\langle i}a_{j\rangle}\hat{\omega}^{ij} 
+\tfrac{5}{3}\etide_{ij}\hat{\omega}^{ij}\big]+O(s^2),\\
h^{\rm S}_{ta} &= s\big(-2m\dot a_a 
+\tfrac{2}{3}m\epsilon_{aij}\btide^j_k\hat{\omega}^{ik}\big)
+O(s^2), \\
h^{\rm S}_{ab} &= \frac{2m}{s}\delta_{ab}-ma_i\omega^i\delta_{ab} 
+s\Big\lbrace\delta_{ab}\big[\tfrac{4}{3}ma_ia^i
+\left(\tfrac{3}{4}ma_{\langle i}a_{j\rangle}
-\tfrac{5}{9}m\etide_{ij}\right)\hat{\omega}^{ij}\big] 
+\tfrac{4}{3}m\etide^i_{\langle a}\hat{\omega}_{b\rangle i} 
\nonumber\\
&\quad  +4ma_{\langle a}a_{b\rangle}
-\tfrac{38}{9}m\etide_{ab} \Big\rbrace+O(s^2).
\end{align}

The regular field could be calculated from the regular Green's
function. But it is more straightforwardly calculated using 
$h^{\rm R}_{\alpha\beta}=\hmn{\alpha\beta}{1}
-h^{\rm S}_{\alpha\beta}$. The result is 
\begin{align}
h^{\rm R}_{tt} &= \A{}{1,0}+s\A{i}{1,1}\omega^i+O(s^2), \\
h^{\rm R}_{ta} &= \C{a}{1,0} +s\Big(\B{}{1,1}\omega_a
+\C{ai}{1,1}\omega^i +\epsilon_{ai}{}^j\D{j}{1,1}\omega^i\Big)
+O(s^2),\\
h^{\rm R}_{ab} &= \delta_{ab}\K{}{1,0}+\H{ab}{1,0}
+s\Big(\delta_{ab}\K{i}{1,1}\omega^i+\H{abi}{1,1}\omega^i 
+\epsilon_i{}^j{}_{(a}\I{b)j}{1,1}\omega^i
+\F{\langle a}{1,1}\omega^{}_{b\rangle}\Big) +O(s^2). 
\end{align}

\subsection{Equation of motion} 

With the metric perturbation fully determined, we can now express the
self-force in terms of tail integrals. Reading off the components of
$h^{\rm R}_{\alpha\beta}$ from Table~\ref{STF wrt tail} and inserting
the results into Eq.~\eqref{acceleration_hR}, we arrive at 
\begin{equation}
a^\mu_{(1)} = -\frac{1}{2} \left( g^{\mu\nu} + u^\mu
u^\nu \right) \left( 2 h^{\rm tail}_{\nu\lambda\rho} 
- h^{\rm tail}_{\lambda\rho\nu} \right)\Big|_{a=0} u^\lambda u^\rho 
+\frac{1}{2m}R^\mu{}_{\nu\lambda\rho}u^\nu S^{\lambda\rho}. 
\end{equation}
We have now firmly established the results of the point-particle
analysis. 

%% file: conclusion.tex
%
\section{Concluding remarks}
\label{conclusion} 

We have presented a number of derivations of the equations that
determine the motion of a point scalar charge $q$, a point electric
charge $e$, and a point mass $m$ in a specified background
spacetime. In this concluding section we summarize these derivations
and their foundations. We conclude by describing the next step in the
gravitational case: obtaining an approximation scheme sufficiently
accurate to extract the parameters of an extreme-mass-ratio inspiral
from an observed gravitational waveform.    

Our derivations are of two types. The first is based on the notion of
an exact point particle. In this approach, we assume that the
self-force on the particle arises from a particular piece of its
field, either that which survives angle-averaging or the
Detweiler-Whiting regular field. The second type is based on the
notion of an asymptotically small body, and abandons the fiction of a
point particle. In this approach, we don't assume anything about the
body's equation of motion, but rather derive it directly from the field
equations. Although we have presented such a derivation only in the
gravitational case, analogous ones could be performed in the scalar
and electromagnetic cases, using conservation of energy-momentum
instead of the field equations alone. Such a calculation was
performed by Gralla {\it et al.}~in the restricted case of an electric
charge in a flat background \cite{gralla-etal:09}.  

Perhaps the essential result of our derivation based on an
asymptotically small body is that it confirms all of the results
derived using point particles: at linear order in the body's mass, the
field it creates is identical to that of a point particle, and its
equation of motion is precisely that derived from physically motivated
axioms for a point particle. In other words, at linear order, not only
can we get away with the fiction of a point particle, but our
assumptions about the physics governing its motion are also
essentially correct. 

\subsection{The motion of a point particle}

\subsubsection*{Spatial averaging}

Our first means of deriving equations of motion for point particles is
based on spatial averaging. In this approach, we assume the following
axiom: 
\begin{quote}
the force on the particle arises from the piece of the field that
survives angle averaging. 
\end{quote}
For convenience in our review, we consider the case of a point
electric charge and adopt the Detweiler-Whiting decomposition of the
Faraday tensor into singular and regular pieces,
$F_{\alpha\beta}=F^{\rm R}_{\alpha\beta}+F^{\rm S}_{\alpha\beta}$. We
average $F_{\alpha\beta}$ over a sphere of constant proper distance
from the particle. We then evaluate the averaged field at the
particle's position. Because the regular field is nonsingular on the
world line, this yields    
\[
e\langle F_{\mu\nu} \rangle u^\nu = e\langle F^{\rm S}_{\mu\nu}
\rangle u^\nu + e F^{\rm R}_{\mu\nu} u^\nu, 
\]
where 
\[
e\langle F^{\rm S}_{\mu\nu} \rangle u^\nu = -(\delta m) a_\mu, \qquad  
\delta m = \lim_{s\to 0} \biggl(\frac{2}{3} \frac{e^2}{s} \biggr),
\]
and 
\[ 
e F^{\rm R}_{\mu\nu} u^\nu = e^2 \bigl( g_{\mu\nu} 
+ u_\mu u_\nu \bigr) \biggl( \frac{2}{3} \dot{a}^{\nu} 
+ \frac{1}{3} R^{\nu}_{\ \lambda} u^{\lambda} \biggr)  
+ 2 e^2 u^\nu \int_{-\infty}^{\tau^-}     
\nabla_{[\mu} G^+_{\ \nu]\lambda'}\bigl(z(\tau),z(\tau')\bigr)   
u^{\lambda'}\, d\tau'.
\]
We now postulate that the equations of motion are $m a_\mu = 
e\langle F_{\mu\nu} \rangle u^\nu$, where $m$ is the particle's bare 
mass. With the preceding results we arrive at $m_{\rm obs} a_\mu = 
e F^{\rm R}_{\mu\nu} u^\nu$, where $m_{\rm obs} \equiv m + \delta m$ 
is the particle's observed (renormalized) inertial mass.   

In this approach, the fiction of a point particle manifests itself in
the need for mass renormalization. Such a requirement can be removed,
even within the point-particle picture, by adopting the ``comparison
axiom'' proposed by Quinn and Wald \cite{quinn-wald:97}. If we
consider extended (but small) bodies, no such renormalization is
required, and the equations of motion follow directly from the
conservation of energy-momentum. However, the essential assumption
about the nature of the force is valid: only the piece of the field
that survives angle-averaging exerts a force on the body. 

\subsubsection*{The Detweiler-Whiting Axiom}

Our second means of deriving equations of motion for point particles
is based on the Detweiler-Whiting axiom, which asserts that   
\begin{quote} 
the singular field exerts no force on the particle; the entire
self-force arises from the action of the regular field. 
\end{quote} 
This axiom, which is motivated by the symmetric nature of the singular
field, and also its causal structure, gives rise to the same equations of
motion as the 
averaging method. In this picture, the particle simply interacts with
a free field (whose origin can be traced to the particle's past),
and the procedure of mass renormalization is sidestepped. In the
scalar and electromagnetic cases, the picture of a particle
interacting with a free radiation field removes any tension between the
nongeodesic motion of the charge and the principle of equivalence. In
the gravitational case the Detweiler-Whiting axiom produces a
generalized equivalence principle (c.f. Ref.~\cite{futamase-itoh:07}):
up to order $\e^2$ errors, a point mass $m$ moves on a geodesic of the 
spacetime with metric $g_{\alpha\beta} + h^{\rm R}_{\alpha\beta}$, 
which is nonsingular and a solution to the vacuum field
equations. This is a conceptually powerful, and elegant, formulation
of the MiSaTaQuWa equations of motion. And it remains valid for
(non-spinning) small bodies.  

\subsubsection*{Resolving historical ambiguities}

Although they yield the correct physical description, the above axioms
are by themselves insufficient, and historically, two problems have
arisen in utilizing them: One, they led to ill-behaved equations of
motion, requiring a process of order reduction; and two, in the 
gravitational case they led to equations of motion that are
inconsistent with the field equations, requiring the procedure of
gauge-relaxation. Both of these problems arose because the expansions
were insufficiently systematic, in the sense that they did not
yield exactly solvable perturbation equations. In the approach taken
in our review, we have shown that these problems do not arise within
the context of a systematic expansion. Although we have done so only
in the case of an extended body, where we sought a higher degree of
rigor, one could do the same in the case of point particles by
expanding in the limit of small charge or mass (see, e.g., the
treatment of a point mass in Ref.~\cite{pound:10a}).  

Consider the Abraham-Lorentz-Dirac equation $ma^\mu 
= f^\mu_{\rm ext} + \frac{2}{3}e^2\dot a^\mu$. To be physically
meaningful and mathematically well-justified, it must be thought of as
an approximate equation of motion for a localized matter distribution
with small charge $e\ll1$. But it contains terms of differing orders
$e^0$ and $e^2$, and the acceleration itself is obviously a function
of $e$.  Hence, the equation has not been fully expanded. One might
think that it is somehow an exact equation, despite its ill
behaviour. Or one might replace it with the order-reduced equation  
$ma^\mu=f^\mu_{\rm ext} +\frac{2e^2}{3m}\dot f^\mu_{\rm ext}$ to
eliminate that ill behaviour. But one can instead assume that
$a^\mu(e)$, like other functions of $e$, possesses an expansion in
powers of $e$, leading to the two well-behaved equations 
$ma^{(0)\mu}=f^\mu_{\rm ext}$ and
$ma^{(1)\mu}=\frac{2}{3}e^2\dot a^{(0)\mu}$. However, the fact that 
such an equation can even arise indicates that one has not begun with
a systematic expansion of the governing field equations (in this case, 
the conservation equation and the Maxwell equations). If one began
with a systematic expansion, with equations exactly solvable at each
order, no such ambiguity would arise.  

The same can be said of the second problem. It is well known that in
general relativity, the motion of gravitating bodies is determined by
the Einstein field equations; the equations of motion cannot be
separately imposed. And specifically, if we deal with the linearized
Einstein equation $\delta G_{\alpha\beta}[\hmn{}{1}]=8\pi
T_{\alpha\beta}[\gamma]$, where $T_{\alpha\beta}$ is the
energy-momentum tensor of a point particle in the background
spacetime, then the linearized Bianchi identity requires the point
particle to move on a geodesic of the background spacetime. This seems
to contradict the MiSaTaQuWa equation and therefore the assumptions we
made in deriving it. In order to remove this inconsistency, the
earliest derivations \cite{mino-etal:97a,quinn-wald:97} invoked an a
posteriori gauge-relaxation: rather than solving a linearized Einstein
equation exactly, they solved the wave equation
$E_{\alpha\beta}[\hmn{}{1}]=-16\pi T_{\alpha\beta}$ in combination
with the relaxed gauge condition $L_\alpha[\e\hmn{}{1}]=O(\e^2)$. The
allowed errors in the gauge condition carry over into the linearized
Bianchi identity, such that it no longer restricts the motion to be
geodesic. In this approach, one is \emph{almost} solving the
linearized problem one set out to solve. But again, such an a
posteriori corrective measure is required only if one begins without a
systematic expansion. The first-order metric perturbation is a
functional of a world line; if we allow that world line to depend on
$\e$, then the metric has evidently not been fully expanded in powers
of $\e$. To resolve the problem, one needs to carefully deal with this
fact. 

In the approach we have adopted here, following Ref.~\cite{pound:10a},
we have resolved these problems via a self-consistent expansion in
which the world line is held fixed while expanding the metric. Rather
than beginning with the linear field equation, we began by
reformulating the exact equation in the form 
\begin{align}
E_{\alpha\beta}[h] & = T^{\rm eff}_{\alpha\beta},\\
L_\alpha[h] & = 0.
\end{align}
These equations we systematically expanded by (i) treating the metric
perturbation as a functional of a fixed world line, keeping the
dependence on the world line fixed while expanding, and (ii) expanding
the acceleration of the world line. Since we never sought a solution to
the equation $\delta G_{\alpha\beta}[\hmn{}{1}]=8\pi
T_{\alpha\beta}[\gamma]$, no tension arose between the equation of
motion and the field equation. In addition, we dealt only with exactly
solvable perturbative equations: rather then imposing the ad hoc a
posteriori gauge condition $L_\mu[\e\hmn{}{1}]=O(\e^2)$, our approach
systematically led to the conditions
$L^{(0)}_\mu\big[\hmn{}{1}\big]=0$ and $L^{(1)}_\mu\big[\hmn{}{1}\big]
= -L^{(0)}_\mu\big[\hmn{}{2}\big]$, which can be solved exactly. Like
in the issue of order reduction, the essential step in arriving at
exactly solvable equations is assuming an expansion of the particle's
acceleration.  

Historically, these issues were first resolved by Gralla and Wald
\cite{gralla-wald:08} using a different method. Rather than allowing
an $\e$-dependence in the first-order perturbation, they fully
expanded every function in the problem in a power series, including
the world line itself. In this approach, the world line of the body is
found to be a geodesic, but higher-order effects arise in deviation
vectors measuring the drift of the particle away from that
geodesic. Such a method has the drawback of being limited to short
timescales, since the deviation vectors will eventually grow large as
the body moves away from the initial reference geodesic. 

\subsection{The motion of a small body}

Although the above results for point particles require assumptions 
about the form of the force, their results have since been derived
from first principles, and the physical pictures they are based on
have proven to be justified: An asymptotically small body behaves as a
point particle moving on a geodesic of the smooth part of the
spacetime around it, or equivalently, it moves on a world line
accelerated by the asymmetric part of its own field. 

In addition to a derivation from first principles, we also seek a
useful approximation scheme. Any such scheme must deal with the
presence of multiple distinct scales. Most obviously, these are the
mass and size of the body itself and the lengthscales of the external
universe, but other scales also arise. For example, in order to
accurately represent the physics of an extreme-mass-ratio system, we
must consider changes over four scales: First, there is the large
body's mass, which is the characteristic lengthscale of the external
universe. For convenience, since all lengths are measured relative to
this one, we rescale them such that this global lengthscale is $\sim
1$. Second, there is the small body's mass $\sim \e$, which is the
scale over which the gravitational field changes near the body; since
the body is compact, this is also the scale of its linear size. Third,
there is the radiation-reaction time $\sim 1/\e$; this is the time
over which the small effects of the self-force accumulate to
produce significant changes, specifically the time required for
quantities such as the small body's energy and angular momentum to
accumulate order $1$ changes. Fourth, there is the large distance to
the wave zone. We will not discuss this last scale here, but dealing
with it analytically would likely require matching a wave solution at
null infinity to an expansion formally expected to be valid in a
region of size $\sim 1/\e$. 

\subsubsection*{Self-consistent and matched asymptotic expansions} 

In this review, we have focused on a self-consistent approximation
scheme first presented in
Refs.~\cite{pound:10a,pound:10b,pound:10c}. It deals with the small
size of the body using two expansions. Near the body, to capture
changes on the short distances $\sim\e$, we adopt an inner expansion
in which the body remains of constant size while all other distances
approach infinity. Outside of this small neibourhood around the body,
we adopt an outer expansion in which the body shrinks to zero mass and
size about an $\e$-dependent world line that accurately reflects its
long-term motion. The world line $\gamma$ is defined in the background
spacetime of the outer expansion. Its acceleration is found by solving
the Einstein equation in a buffer region surrounding the body, where
both expansions are valid; in this region, both expansions must agree,
and we can use the multipole moments of the inner expansion to
determine the outer one. In particular, we define $\gamma$ to be the
body's world line if and only if the body's mass dipole moment
vanishes when calculated in coordinates centered on it. As in the
point-particle calculation, the essential step in arriving at exactly
solvable equations and a well-behaved equation of motion is an assumed 
expansion of the acceleration on the world line. 
 
In order to construct a global solution in the outer expansion, we
first recast the Einstein equation in a form that can be expanded and
solved for an arbitrary world line. As in the point-particle case, we
accomplish this by adopting the Lorenz gauge for the total metric
perturbation. We can then write formal solutions to the perturbation
equations at each order as integrals over a small tube surrounding the
body (plus an initial data surface). By embedding the tube in the
buffer region, we can use the data from the buffer-region expansion to
determine the global solution. At first order, we find that the
metric perturbation is precisely that of a point particle moving on
the world line $\gamma$.  While the choice of gauge is not essential in
finding an expression for the force in terms of the field in the
buffer-region expansion, it \emph{is} essential in our method of
determining the global field. Without making use of the relaxed
Einstein equations, no clear method of globally solving the Einstein
equation presents itself, and no simple split between the perturbation
and the equation of motion arises. 

Because this expansion self-consistently incorporates the corrections
to the body's motion, it promises to be accurate on long
timescales. Specifically, when combined with the first-order metric
perturbation, the first-order equation of motion defines a solution to
the Einstein equation that we expect to be accurate up to order $\e^2$
errors over times $t\lesssim1/\e$. When combined with the second-order
perturbation, it defines a solution we expect to be accurate up to
order $\e^3$ errors on the shorter timescale $\sim 1$. 

This approach closely mirrors the extremely successful methods of
post-Newtonian theory~\cite{futamase-itoh:07}. In particular, both
schemes recast the Einstein equation in a relaxed form before
expanding it. Also, our use of an inner limit near the body is
analogous to the ``strong-field point particle limit'' exploited by
Futamase and his collaborators~\cite{futamase-itoh:07}. And our
calculation of the equation of motion is somewhat similar to the
methods used by Futamase and 
others~\cite{futamase-itoh:07, racine-flanagan:05}, in that it
utilizes a multipole expansion of the body's metric in the buffer
region. 

\subsubsection*{Alternative methods}

Various other approaches have been taken to deal with the multiple
scales in the problem. In particular, even the earliest paper on the
gravitational self-force \cite{mino-etal:97a} made use of inner and
outer limits, which were also used in different forms in later
derivations \cite{poisson:04a,detweiler:05,
  fukumoto-etal:06}. However, those early derivations are
problematic. Specifically, they never adequately define the world line
for which they derive equations of motion. 

In the method of matched asymptotic expansions used in
Refs.~\cite{mino-etal:97a, poisson:04a, detweiler:05}, the first-order
perturbation in the outer expansion is assumed to be that of a point
particle, with an attendant world line, and the inner expansion is
assumed to be that of a perturbed black hole; by matching the two
expansions in the buffer region, the acceleration of the world line is
determined. Since the forms of the inner and outer expansions are
already restricted, this approach's conclusions have somewhat limited
strength, but it has more fundamental problems as well. The matching
procedure begins by expanding the outer expansion in powers of
distance $s$ (or $r$) from the world line, and the inner expansion for
large spacial distances $R\gg \e$. But the two expansions begin in
different coordinate systems with an unknown relationship between
them. In particular, there is no given relationship between the
world line and the ``position" of the black hole. The two expansions
are matched by finding a coordinate transformation that makes them
agree in the buffer region. However, because there is no predetermined
relationship between the expansions, this transformation is not in
fact unique, and it does not yield a unique equation of motion. One
can increase the strength of the matching condition in order to arrive
at unique results, but that further weakens the strength of the
conclusions. Refer to Ref.~\cite{pound:10b} for a discussion of these
issues. 

Rather than finding the equation of motion from the field equations,
as in the above calculations, Fukumoto {\it et al.}~\cite{fukumoto-etal:06}
found the equation of motion by defining the body's linear momentum as
an integral over the body's interior and then taking the derivative of
that momentum. But they then required an assumed relationship between
the momentum and the four-velocity of a representative world line in
the body's interior. Hence, the problem is again an inadequate
defintion of the body's motion. Because it involves integrals over the
body's interior, and takes the world line to lie therein, this approach
is also limited to material bodies; it does not apply to black holes. 

The first reliable derivation of the first-order equation of motion
for an asymptotically small body was performed by Gralla and
Wald~\cite{gralla-wald:08}, who used a method of the same nature as
the one presented here, deriving equations of motion by solving the
field equations in the buffer region. As we have seen, however, their
derivation is based on an expansion of the world line in powers of
$\e$ instead of a self-consistent treatment that keeps it fixed.   

Using very different methods, Harte has also provided reliable
derivations of equations of motion for extended bodies interacting
with their own scalar and electromagnetic fields in fixed background
spacetimes~\cite{harte:08, harte:09a, harte:09b}. His approach is
based on generalized definitions of momenta, the evolution of which is
equivalent to energy-momentum conservation. The momenta are defined in 
terms of generalized Killing fields $\xi^\alpha$, the essential
property of which is that they satisfy
$\Lie{\xi}g_{\alpha\beta}|_{\gamma}
=\del{\gamma}\Lie{\xi}g_{\alpha\beta}|_{\gamma}=0$ --- that is, they
satisfy Killing's equation on the body's world line and approximately
satisfy it ``nearby.'' Here the world line can be defined in multiple
ways using, for example, center-of-mass conditions. This approach is 
nonperturbative, with no expansion in the limit of small mass and
size (though it does require an upper limit on the body's size and a
lower limit on its compactness). It has the advantage of very
naturally deriving a generalization of the Detweiler-Whiting axiom:
The field of an extended body can be split into (i) a solution to the
vacuum field equations which exerts a direct self-force, and (ii) a
solution to the equations sourced by the body, which shifts the body's
multipole moments. This approach has not yet been applied to gravity,
but such an application should be relatively straightforward. However,
as with the approach of Fukumoto {\it et al.}~\cite{fukumoto-etal:06},
this one does not apply to black holes.  

Other methods have also been developed (or suggested) to accomplish
the same goals as the self-consistent expansion. Most prominent among
these is the two-timescale expansion suggested by Hinderer and
Flanagan \cite{hinderer-flanagan:08}. As discussed in 
Sec.~\ref{lit_surv}, their method splits the orbital evolution into
slow and fast dynamics by introducing slow and fast time variables. In
the terminology of Sec.~\ref{buffer_region}, this method constructs a
general expansion by smoothly transitioning between regular expansions
constructed at each value of the slow time variable, with the
transition determined by the evolution with respect to the slow
time. On the scale of the fast time, the world line is a geodesic; but
when the slow time is allowed to vary, the world line transitions
between geodesics to form the true, accelerated world line. This
results in a global, uniform-in-time approximation. One should note
that simply patching together a sequence of regular expansions, by
shifting to a new geodesic every so often using the deviation vector,
would not accomplish this: Such a procedure would accumulate a secular
error in both the metric perturbation and the force, because the
perturbation would be sourced by a world line secularly deviating from
the position of the body, and the force would be calculated from this
erroneous perturbation. The error would be proportional to the number
of ``shifts" multiplied by a nonlinear factor depending on the time 
between them. And this error would, formally at least, be of the same
magnitude as the solution itself. 

The fundamental difference between the self-consistent expansion and
the two-timescale expansion is the following: In the two-timescale
method, the Einstein equation, coupled to the equation of motion of
the small body, is reduced to a dynamical system that can be evolved
in time. The true world line of the body then emerges from the
evolution of this system. In the method presented here, we have
instead sought global, formal solutions to the Einstein equation,
written in terms of global integrals; to accomplish this, we have
treated the world line of the body as a fixed structure in the external
spacetime. However, the two methods should agree. Note, though, that
the two-timescale expansion is limited to orbits in Kerr, and it
requires evolution equations for the slow evolution of the large black
hole's mass and spin parameters, which have not yet been
derived. Since the changes in mass and spin remain small on a
radiation-reaction timescale, in the self-consistent expansion
presented here they are automatically incorporated into the
perturbations $\hmn{\alpha\beta}{n}$. It is possible that this
incorporation leads to errors on long timescales, in which case a
different approach, naturally allowing slow changes in the background,
would be more advantageous. 

\subsection{Beyond first order}

The primary experimental motivation for researching the self-force is
to produce waveform templates for LISA. In order to extract the
parameters of an extreme-mass-ratio binary from a waveform, we require
a waveform that is accurate up to errors of order $\e$ after a
radiation-reaction time $\sim1/\e$. If we use the first-order equation
of motion, we will be neglecting an acceleration $\sim\e^2$, which
will lead to secular errors of order unity after a time
$\sim1/\e$. Thus, the second-order self-force is required in order to
obtain a sufficiently accurate waveform template. In order to achieve
the correct waveform, we must also obtain the second-order part of the
metric perturbation; this can be easily done, at least formally, using
the global integral representations outside a world tube. However, a
practical numerical calculation may prove difficult, since one would
not wish to excise the small tube from one's numerical domain, and the
second-order perturbation would diverge too rapidly on the world line
to be treated straightforwardly. 

A formal expression for the second-order force has already been
derived by Rosenthal~\cite{rosenthal:06a,rosenthal:06b}. However, he
expresses the second-order force in a very particular gauge in which
the first-order self-force vanishes. This is sensible on short
timescales, but not on long timescales, since it forces secular changes
into the first-order perturbation, presumably leading to the
first-order perturbation becoming large with time. Furthermore, it is
not a convenient gauge, since it does not provide what we wish it to:
a correction to the nonzero leading-order force in the Lorenz gauge. 

Thus, we wish to obtain an alternative to Rosenthal's
derivation. Based on the methods reviewed in this article, there is a
clear route to deriving the second-order force. One would construct a
buffer-region expansion accurate up to order $\e^3$. Since one would
require the order $\e^2 s$ terms in this expansion, in order to
determine the acceleration, one would need to increase the order of
the expansion in $r$ as well. Specifically, one would need terms up to
orders $\e^0 s^3$, $\e s^2$, $\e^2 s$, and $\e^3 s^0$. In such a
calculation, one would expect the following terms to appear: the
body's quadrupole moment $Q_{ab}$, corrections $\delta M_i$ and
$\delta S_i$ to its mass and spin dipoles, and a second-order
correction $\delta^2 m$ to its mass. Although some ambiguity may arise 
in defining the world line of the body at this order, a reasonable
definition appears to be to guarantee that $\delta M_i$
vanishes. However, at this order one may require some model of the
body's internal dynamics, since the equation of motion will involve
the body's quadrupole moment, for which the Einstein equation is not
expected to yield an evolution equation. But if one seeks only the
second-order self-force, one could simply neglect the quadrupole by
assuming that the body is spherically symmetric in isolation. In any
case, the force due to the body's quadrupole moment is already known
from various other methods: see, e.g., the work of 
Dixon~\cite{dixon:70a,dixon:70b,dixon:74}; more recent methods can be
found in Ref.~\cite{steinhoff-puetzfeld:10} and references therein. 

Because such a calculation could be egregiously lengthy, one may
consider simpler methods, perhaps requiring stronger assumptions. For
example, one could straightforwardly implement the method of matched
asymptotic expansions with the matching conditions discussed in
Ref.~\cite{pound:10b}, in which one makes strong assumptions about the
relationship between the inner and outer expansions. The effective
field theory method used by Galley and Hu \cite{galley-hu:09} offers
another possible route. 

Alternatively, one could calculate only part of the second-order
force. Specifically, as described in Sec.~\ref{subsec:adiabatic},
Hinderer and Flanagan~\cite{hinderer-flanagan:08,
  flanagan-hinderer:10} have shown that one requires in fact only the
averaged dissipative part of the second-order force. This piece of the
force can be calculated within an adiabatic approximation, in which
the rates of change of orbital parameters are calculated from the
radiative Green's function, asymptotic wave amplitudes, and
information about the orbit that sources them. Hence, we might be able
to forgo a complete calculation of the second-order force, and use
instead the complete first-order force in conjunction with an adiabatic
approximation for the second-order force.   

The self-force, however, is of interest beyond its relevance to
LISA. The full second-order force would be useful for more general
purposes, such as more accurate comparisons to post-Newtonian theory,
and analysis of other systems, such as intermediate mass ratio
binaries. Perhaps most importantly, it is of fundamental importance in
our understanding of the motion of small bodies. For these reasons,
proceeding to second order in a systematic expansion, and thereby
obtaining second-order expressions for the force on a small body,
remains an immediate goal.

%% file: appendices.tex
%
\section{Second-order expansions of the Ricci tensor}
\label{second-order expansions}

We present here various expansions used in solving the second-order
Einstein equation in Sec.~\ref{buffer_expansion2}. We require an
expansion of the second-order Ricci tensor $\delta^2R_{\alpha\beta}$,
defined by  
\begin{align}
\delta^2R_{\alpha\beta}[h] &=-\tfrac{1}{2}\gamma^{\mu\nu}{}_{;\nu}
\left(2h_{\mu(\alpha;\beta)}-h_{\alpha\beta;\mu}\right) 
+\tfrac{1}{4}h^{\mu\nu}{}_{;\alpha}h_{\mu\nu;\beta}
+\tfrac{1}{2}h^{\mu}{}_{\beta}{}^{;\nu}\left(h_{\mu\alpha;\nu} 
-h_{\nu\alpha;\mu}\right)
\nonumber\\
&\quad-\tfrac{1}{2}h^{\mu\nu}\left(2h_{\mu(\alpha;\beta)\nu}
-h_{\alpha\beta;\mu\nu}-h_{\mu\nu;\alpha\beta}\right),
\label{second-order Ricci}
\end{align}
where $\gamma^{\mu\nu}$ is the trace-reversed metric perturbation, 
and an expansion of a certain piece of
$E_{\mu\nu}[\hmn{}{2}]$. Specifically, we require an expansion of
$\delta^2 R^{(0)}_{\alpha\beta}[\hmn{}{1}]$ in powers of the Fermi
radial coordinate $s$, where for a function $f$, $\delta^2
R^{(0)}_{\alpha\beta}[f]$ consists of $\delta^2 R_{\alpha\beta}[f]$
with the acceleration $a^\mu$ set to zero. We write 
\begin{align}
\delta^2R^{(0)}_{\alpha\beta}[\hmn{}{1}] &= 
\frac{1}{s^4}\ddR{\alpha\beta}{0,-4}{\hmn{}{1}}\!\! 
+\frac{1}{s^3}\ddR{\alpha\beta}{0,-3}{\hmn{}{1}}
+\frac{1}{s^2}\ddR{\alpha\beta}{0,-2}{\hmn{}{1}} +O(1/s),
\end{align}
where the second superscript index in parentheses denotes the power of
$s$. Making use of the expansion of $\hmn{\alpha\beta}{1}$, obtained
by setting the acceleration to zero in the results for
$\hmn{\alpha\beta}{1}$ found in Sec.~\ref{buffer_expansion1}, one
finds 
\begin{align}
\ddR{\alpha\beta}{2,-4}{\hmn{}{1}} &= 2m^2\left(7\hat{\omega}_{ab}
+\tfrac{4}{3}\delta_{ab}\right)x^a_\alpha x^b_\beta 
-2m^2t_\alpha t_\beta,\label{ddR0n4}
\end{align}
and
\begin{align}
\ddR{tt}{2,-3}{\hmn{}{1}} &= 3m\H{ij}{1,0}\hat{\omega}^{ij},
\label{ddR0n3_tt}\\
\ddR{ta}{2,-3}{\hmn{}{1}} &= 3m\C{i}{1,0}\hat{\omega}_{a}^i,\\
\ddR{ab}{2,-3}{\hmn{}{1}} &= 3m\big(\A{}{1,0}
+\K{}{1,0}\big)\hat{\omega}_{ab}
-6m\H{i\langle a}{1,0}\hat{\omega}_{b\rangle}^i
+m\delta_{ab}\H{ij}{1,0}\hat{\omega}^{ij},
\label{ddR0n3_ab}
\end{align}
and
\begin{align}
\ddR{tt}{2,-2}{\hmn{}{1}} &=
-\tfrac{20}{3}m^2\etide_{ij}\hat{\omega}^{ij}
+3m\H{ijk}{1,1}\hat{\omega}^{ijk}
+\tfrac{7}{5}m\A{i}{1,1}\omega^i
+\tfrac{3}{5}m\K{i}{1,1}\omega^i\nonumber\\
&\quad-\tfrac{4}{5}m\partial_t\C{i}{1,0}\omega^i,
\label{ddR0n2_tt}\\
\ddR{ta}{2,-2}{\hmn{}{1}}&= -m\partial_t\K{}{1,0}\omega_a
+3m\C{ij}{1,1}\hat{\omega}_{a}{}^{ij}
+m\Big(\tfrac{6}{5}\C{ai}{1,1}-\partial_t\H{ai}{1,0}\Big)\omega^i
\nonumber\\
 &\quad+2m\epsilon_a{}^{ij}\D{i}{1,1}\omega_j
+\tfrac{4}{3}m^2\epsilon_{aik}\btide^k_j\hat{\omega}^{ij},\\
\ddR{ab}{2,-2}{\hmn{}{1}} &= 
\delta_{ab}m\left(\tfrac{16}{15}\partial_t\C{i}{1,0}
-\tfrac{13}{15}\A{i}{1,1} 
-\tfrac{9}{5}\K{i}{1,1}\right)\omega^i \nonumber\\
&\quad+\delta_{ab}\left(-\tfrac{50}{9}m^2\etide_{ij}\hat{\omega}^{ij}
+m\H{ijk}{1,1}\hat{\omega}^{ijk}\right) 
-\tfrac{14}{3}m^2\etide_{ij}\hat{\omega}_{ab}{}^{ij}\nonumber\\
&\quad+m\left(\tfrac{33}{10}\A{i}{1,1}+\tfrac{27}{10}\K{i}{1,1} 
-\tfrac{3}{5}\partial_t\C{i}{1,0}\right)\hat{\omega}_{ab}{}^i
\nonumber\\
&\quad+m\left(\tfrac{28}{25}\A{\langle a}{1,1}
-\tfrac{18}{25}\K{\langle a}{1,1}
-\tfrac{46}{25}\partial_t\C{\langle a}{1,0}\right)
\hat{\omega}^{}_{b\rangle}
\nonumber\\
&\quad-\tfrac{8}{3}m^2\etide_{i\langle a}\hat{\omega}_{b\rangle}{}^i
-6m\H{ij\langle a}{1,1}\hat{\omega}^{}_{b\rangle}{}^{ij}
+3m\epsilon_{ij(a}\hat{\omega}_{b)}{}^{jk}\I{ik}{1,1} \nonumber\\
&\quad+\tfrac{2}{45}m^2\etide_{ab}-\tfrac{2}{5}m\H{abi}{1,1}\omega^i 
+\tfrac{8}{5}m\epsilon^i{}_{j(a}^{}\I{b)i}{1,1}\omega^j.
\label{ddR0n2_ab}
\end{align}

Next, we require an analogous expansion of
$E^{(0)}_{\alpha\beta}\left[\frac{1}{r^2}\hmn{}{2,-2}
  +\frac{1}{r}\hmn{}{2,-1}\right]$, where $E^{(0)}_{\alpha\beta}[f]$
is defined for any $f$ by setting the acceleration to zero in
$E_{\alpha\beta}[f]$. The coefficients of the $1/r^4$ and $1/r^3$
terms in this expansion can be found in Sec.~\ref{buffer_expansion2};
the coefficient of $1/r^2$ will be given here. For compactness, we
define  this coefficient to be $\tilde E_{\alpha\beta}$. The
$tt$-component of this quantity is given by   
\begin{align}
\tilde E_{tt} &= 2\partial^2_tM_i\omega^i
+\tfrac{8}{5}S^j\btide_{ij}\omega^i
-\tfrac{2}{3}M^j\etide_{ij}\omega^i 
+\tfrac{82}{3}m^2\etide_{ij}\hat{\omega}^{ij} 
+ 24S_{\langle i}\btide_{jk\rangle}\hat{\omega}^{ijk}
-20M_{\langle i}\etide_{jk\rangle}\hat{\omega}^{ijk}.
\label{E_tt}
\end{align}
The $ta$-component is given by
\begin{align}
\tilde E_{ta} &= \tfrac{44}{15}\epsilon_{aij}M^k\btide^j_k\omega^i 
-\tfrac{2}{15}\left(11S^i\etide^j_k
+18M^i\btide^j\right)\epsilon_{ija}\omega^k 
+\tfrac{2}{15}\left(41S^j\etide^k_a 
-10M^j\btide^k_a\right)\epsilon_{ijk}\omega^i\nonumber\\ 
&\quad +4\epsilon_{aij}\left(S^j\etide_{kl} 
+2M_k\btide^j_l\right)\hat{\omega}^{ikl}
+4\epsilon^{}_{ij\langle k}\etide_{l\rangle}^jS^i\hat{\omega}_a{}^{kl} 
+\tfrac{68}{3}m^2\epsilon_{aij}\btide^j_k\hat{\omega}^{ik}.
\end{align}
This can be decomposed into irreducible STF pieces via the identities  
\begin{align}
\epsilon_{aij}S^i\etide^j_k & = S^i\etide^j_{(k}\epsilon_{a)ij}
+\tfrac{1}{2}\epsilon_{akj}S^i\etide_i^j\\
\epsilon_{aj\langle i}\etide_{kl\rangle}S^j&=
\mathop{\STF}_{ikl}\!\left[\epsilon^j{}_{al}
S_{\langle i}\etide_{jk\rangle} 
-\tfrac{2}{3}\delta_{al}S^p\etide^j_{(i}\epsilon^{}_{k)jp}\right]\\ 
\epsilon_{aj\langle i}M_l\btide_{k\rangle}{}^j&=
\mathop{\STF}_{ikl}\!\left[\epsilon^j{}_{al}M_{\langle i}
\btide_{jk\rangle}\! +\!\tfrac{1}{3}\delta_{al}M^p
\btide^j_{(i}\epsilon^{}_{k)jp}\right],
\end{align}
which follow from Eqs.~\eqref{decomposition_1} and
\eqref{decomposition_2}, and which lead to 
\begin{align}
\tilde E_{ta} &=\tfrac{2}{5}\epsilon_{aij}\Big(6M^k\btide^j_k
-7S^k\etide^j_k\Big)\omega^i
+\tfrac{4}{3}\left(2M^l\btide^k_{(i} 
-5S^l\etide^k_{(i}\right)\epsilon^{}_{j)kl}\hat{\omega}_a{}^{ij} 
\nonumber\\
&\quad+\left(4S^j\etide^k_{(a} 
-\tfrac{56}{15}M^j\btide^k_{(a}\right)\epsilon_{i)jk}\omega^i 
+ 4\epsilon_{ai}{}^l\left(S_{\langle j}\etide_{kl\rangle}
+2M_{\langle j}\btide_{kl\rangle}\right)\hat{\omega}^{ijk}
+\tfrac{68}{3}m^2\epsilon_{aij}\btide^j_k\hat{\omega}^{ik}.
\label{E_ta}
\end{align}
The $ab$-component is given by
\begin{align}
\tilde E_{ab} &= \tfrac{56}{3}m^2\etide_{ij}\hat{\omega}_{ab}{}^{ij} 
+\tfrac{52}{45}m^2\etide_{ab}
-\delta_{ab}\left[\left(2\partial^2_tM_i
+\tfrac{8}{5}S^j\btide_{ij} 
+\tfrac{10}{9}M^j\etide_{ij}\right)\omega^i
+\tfrac{100}{9}m^2\etide_{ij}\hat{\omega}^{ij}\right] \nonumber\\
&\quad -\delta_{ab}\left(\tfrac{20}{3}M_{\langle i}\etide_{jk\rangle} 
-\tfrac{8}{3}S_{\langle i}\btide_{jk\rangle}\right)\hat{\omega}^{ijk}
+\tfrac{8}{15}M_{\langle a}\etide_{b\rangle i}\omega^i
+\tfrac{8}{15}M^i\etide_{i\langle a}\omega_{b\rangle}
+\tfrac{56}{3}m^2\etide_{i\langle a}\hat{\omega}^{}_{b\rangle}{}^i
\nonumber\\
&\quad+16M_i\etide_{j\langle a}\hat{\omega}_{b\rangle}{}^{ij}
-\tfrac{32}{5}S_{\langle a}\btide_{b\rangle i}\omega^i
+\tfrac{4}{15}\left(10S_i\btide_{ab}
+27M_i\etide_{ab}\right)\omega^i\nonumber\\
&\quad+\tfrac{16}{3}S^i\btide_{i\langle a}\omega_{b\rangle}
-8\epsilon_{ij\langle a}\epsilon_{b\rangle kl}
S^j\btide^l_m\hat{\omega}^{ikm}
+\tfrac{16}{15}\epsilon_{ij\langle a}\epsilon_{b\rangle kl}
S^j\btide^{il}\omega^k.
\end{align}
Again, this can be decomposed, using the identities
\begin{align}
S_{\langle a}\btide_{b\rangle i} &=S_{\langle a}\btide_{bi\rangle} 
+\mathop{\STF}_{ab}\tfrac{1}{3}\epsilon_{ai}{}^j
\epsilon_{kl(b} \btide_{j)}{}^lS^k 
+\tfrac{1}{10}\delta_{i\langle a}\btide_{b\rangle j}S^j,\\
S_i\btide_{ab} &=S_{\langle a}\btide_{bi\rangle} 
-\mathop{\STF}_{ab}\tfrac{2}{3}\epsilon_{ai}{}^j\epsilon_{kl(b} 
\btide_{j)}{}^lS^k +\tfrac{3}{5}\delta_{i\langle a}
\btide_{b\rangle j}S^j,\\
\epsilon_{ij\langle a}\epsilon_{b\rangle kl}S^j\btide^{il} & = 
\mathop{\STF}_{ab}\epsilon_{akj}S^l\btide^i_{(j}\epsilon^{}_{b)il} 
-\tfrac{1}{2}\delta_{k\langle a}\btide_{b\rangle i}S^i,\\
\mathop{\STF}_{ikm}\epsilon_{ij\langle a}\epsilon_{b\rangle kl}
S^j\btide^l_m & = \mathop{\STF}_{ikm}\mathop{\STF}_{ab}
\Big(2\delta_{ai}S_{\langle b}\btide_{km\rangle}
+\tfrac{1}{3}\delta_{ai}\epsilon^l{}_{bk}S^j 
\btide^p_{(l}\epsilon^{}_{m)jp} 
-\tfrac{3}{10}\delta_{ai}\delta_{bk}\btide_{mj}S^j\Big),
\end{align}
which lead to
\begin{align}
\tilde E_{ab} & = -2\delta_{ab}\left[\left(\partial^2_tM_i
+\tfrac{4}{5}S^j\btide_{ij} 
+\tfrac{5}{9}M^j\etide_{ij}\right)\omega^i 
+\left(\tfrac{10}{3}M_{\langle i}\etide_{jk\rangle} 
-\tfrac{4}{3}S_{\langle i}\btide_{jk\rangle} 
\right)\hat{\omega}^{ijk}\right]\nonumber\\
 &\quad-\tfrac{100}{9}\delta_{ab}m^2\etide_{ij}\hat{\omega}^{ij}
+\tfrac{1}{5}\left(8M^j\etide_{ij} 
+12S^j\btide_{ij}\right)\hat{\omega}_{ab}{}^i 
+\tfrac{56}{3}m^2\etide_{ij}\hat{\omega}_{ab}{}^{ij}\nonumber\\
&\quad +\tfrac{4}{75}\left(92M^j\etide_{j\langle a} 
+108S^j\btide_{j\langle a}\right)\omega_{b\rangle}^{} 
+\tfrac{56}{3}m^2\etide_{i\langle a}\hat{\omega}_{b\rangle}{}^i 
\nonumber\\
&\quad+16\mathop{\STF}_{aij}\left(M_i\etide_{j\langle a}
-S_i\btide_{j\langle a}\right)\hat{\omega}^{}_{b\rangle}{}^{ij} 
-\tfrac{8}{3}\epsilon^{pq}{}_{\langle j}\left(2\etide_{k\rangle p}M_q
+\btide_{k\rangle p}S_q\right)
\epsilon^k{}_{i(a}\hat{\omega}_{b)}{}^{ij}
\nonumber\\
&\quad +\tfrac{16}{15}m^2\etide_{ab}
+\tfrac{4}{15} \left(29M_{\langle a}\etide_{bi\rangle}
-14S_{\langle a}\btide_{bi\rangle}\right)\omega^i \nonumber\\ 
&\quad-\tfrac{16}{45}\mathop{\STF}_{ab}
\epsilon_{ai}{}^j\omega^i\epsilon^{pq}{}_{(b} 
\left(13\etide_{j)q}M_p+14\btide_{j)q}S_p\right).
\label{E_ab}
\end{align}

\section{STF multipole decompositions}
\label{STF tensors}

All formulas in this Appendix are either taken directly from
Refs.~\cite{damour-blanchet:86} and \cite{damour-iyer:91} or are
easily derivable from formulas therein. 

Any Cartesian tensor field depending on two angles $\theta^A$ spanning
a sphere can be expanded in a unique decomposition in symmetric
trace-free tensors. Such a decomposition is equivalent to a 
decomposition in tensorial harmonics, but it is sometimes
more convenient. It begins with the fact that the angular dependence
of a Cartesian tensor $T_{S}(\theta^A)$ can be expanded in a series of
the form 
\begin{equation}\label{omegahat_expansion}
T_S(\theta^A)=\sum_{\ell\geq 0}T_{S\langle L\rangle}\hat{\omega}^L,
\end{equation}
where $S$ and $L$ denote multi-indices $S=i_1\ldots i_s$ and
$L=j_1\ldots j_\ell$, angular brackets denote an STF combination of
indices, $\omega^a$ is a Cartesian unit vector, $\omega^L:=
\omega^{j_1}\ldots \omega^{j_\ell}$, and $\hat{\omega}^L:=
\omega^{\langle L\rangle}$. This is entirely equivalent to an
expansion in spherical harmonics. Each coefficient 
$T_{S\langle L\rangle}$ can be found from the formula 
\begin{equation}
T_{S\langle L\rangle} = 
\frac{(2\ell+1)!!}{4\pi\ell!}\int T_S(\theta^A)\hat{\omega}_L d\Omega,
\end{equation}
where the double factorial is defined by
$x!!=x(x-2)\cdots1$. These coefficients can then be decomposed into
irreducible STF tensors. For example, for $s=1$, we have 
\begin{equation}\label{decomposition_1}
T_{a\langle L\rangle} = \hat T^{(+1)}_{aL}
+\epsilon^j{}_{a\langle i_\ell}\hat T^{(0)}_{L-1\rangle j}
+\delta_{a\langle i_\ell}\hat T^{(-1)}_{L-1\rangle},
\end{equation}
where the $\hat T^{(n)}$'s are STF tensors given by
\begin{align}
\hat T^{(+1)}_{L+1} & := T_{\langle L+1\rangle}, \\
\hat T^{(0)}_{L} & := \frac{\ell}{\ell+1}T_{pq\langle L-1}
\epsilon_{i_\ell\rangle}{}^{pq}, \\
\hat T^{(-1)}_{L-1} & := \frac{2\ell-1}{2\ell+1}T^j{}_{jL-1}.
\end{align} 
Similarly, for a symmetric tensor $T_S$ with $s=2$, we have
\begin{align}\label{decomposition_2}
T_{ab\langle L\rangle} & = \mathop{\STF}_L\mathop{\STF}_{ab}
\Big( \epsilon^p{}_{ai_\ell}\hat T^{(+1)}_{bpL-1} 
+ \delta_{ai_\ell}\hat T^{(0)}_{b L-1} 
+\delta_{a i_\ell}\epsilon^p{}_{bi_{\ell-1}}
\hat T^{(-1)}_{pL-2} +\delta_{ai_\ell}\delta_{bi_{\ell-1}}
\hat T^{(-2)}_{L-2}\Big) \nonumber\\
&\quad +\hat T^{(+2)}_{abL}+\delta_{ab}\hat K_L,
\end{align}
where
\begin{align}
\hat T^{(+2)}_{L+2} & := T_{\langle L+2\rangle}, \\
\hat T^{(+1)}_{L+1} & := \frac{2\ell}{\ell+2}
\mathop{\STF}_{L+1}(T_{\langle pi_\ell\rangle qL-1}
\epsilon_{i_{\ell+1}}{}^{pq}), \\
\hat T^{(0)}_L & := \frac{6\ell(2\ell-1)}{(\ell+1)(2\ell+3)}
\mathop{\STF}_L(T_{\langle ji_\ell\rangle}{}^j{}_{L-1}), \\
\hat T^{(-1)}_{L-1} & :=
\frac{2(\ell-1)(2\ell-1)}{(\ell+1)(2\ell+1)}
\mathop{\STF}_{L-1}(T_{\langle jp\rangle q}{}^j{}_{L-2}
\epsilon_{i_{\ell-1}}{}^{pq}), \\
\hat T^{(-2)}_{L-2} & := \frac{2\ell-3}{2\ell+1}
T_{\langle jk\rangle}{}^{jk}{}_{L-2} \\
\hat K_L & := \tfrac{1}{3}T^j{}_{jL}.
\end{align}
These decompositions are equivalent to the formulas for addition of
angular momenta, $J=S+L$, which results in terms with angular momentum 
$\ell-s\leq j\leq \ell+s$; the superscript labels $(\pm n)$ in these
formulas indicate by how much each term's angular momentum differs
from $\ell$. 

By substituting Eqs.~\eqref{decomposition_1} and
\eqref{decomposition_2} into Eq.~\eqref{omegahat_expansion}, we find
that a scalar, a Cartesian 3-vector, and the symmetric part of a
rank-2 Cartesian 3-tensor can be decomposed as, respectively, 
\begin{align}
T(\theta^A) &= \sum_{\ell\ge0}\hat A_L\hat{\omega}^L, 
\label{generic_STF tt}\\
T_a(\theta^A) &= \sum_{\ell\ge0}\hat B_L\hat{\omega}_{aL}
+\sum_{\ell\ge1}\left[\hat C_{aL-1}\hat{\omega}^{L-1} 
+ \epsilon^i{}_{aj}\hat D_{iL-1}\hat{\omega}^{jL-1}\right],
\label{generic_STF ta}\\
T_{(ab)}(\theta^A) &= \delta_{ab}\sum_{\ell\ge0}\hat K_L
\hat{\omega}^L+\sum_{\ell\ge0}\hat E_L\hat{\omega}_{ab}{}^L 
+\sum_{\ell\ge1}\left[\hat F_{L-1\langle a}
\hat{\omega}^{}_{b\rangle}{}^{L-1} 
+\epsilon^{ij}{}_{(a}\hat{\omega}_{b)i}{}^{L-1}\hat G_{jL-1}\right] 
\nonumber\\
&\quad+\sum_{\ell\ge2}\left[\hat H_{abL-2}\hat{\omega}^{L-2}
+\epsilon^{ij}{}_{(a}\hat I_{b)jL-2}\hat{\omega}_{i}{}^{L-2}\right].
\label{generic_STF ab}
\end{align}
Each term in these decompositions is algebraically independent of all
the other terms. 

We can also reverse a decomposition to ``peel off'' a fixed index from
an STF expression: 
\begin{align}
(\ell+1)\mathop{\STF}_{iL}T_{i\langle L\rangle} & = 
T_{i\langle L\rangle} 
+ \ell\mathop{\STF}_LT_{i_\ell \langle iL-1\rangle} 
-\frac{2\ell}{2\ell+1}\mathop{\STF}_L 
T^j{}_{\langle jL-1\rangle}\delta_{i_\ell i}.
\end{align}

In evaluating the action of the wave operator on a decomposed tensor,
the following formulas are useful: 
\begin{align}
\omega^c\hat{\omega}^L &= \hat{\omega}^{cL}
+\frac{\ell}{2\ell+1}\delta^{c\langle i_1}
\hat{\omega}^{i_2\ldots i_\ell\rangle},\\
\omega_c\hat{\omega}^{cL} & = \frac{\ell+1}{2\ell+1}\hat{\omega}^L, \\ 
r\partial_c\hat{\omega}_L &= -\ell\hat{\omega}_{cL}
+\frac{\ell(\ell+1)}{2\ell+1}\delta_{c\langle i_1}
\hat{\omega}_{i_2\ldots i_\ell\rangle}, \\
\partial^c\partial_c\hat{\omega}^L & = 
-\frac{\ell(\ell+1)}{r^2}\hat{\omega}^L, \\
\omega^c\partial_c\hat{\omega}^L &= 0, \\
r\partial_c\hat{\omega}^{cL} &= 
\frac{(\ell+1)(\ell+2)}{(2\ell+1)}\hat{\omega}^L.
\end{align}

In evaluating the $t$-component of the Lorenz gauge condition, the
following formula is useful for finding the most divergent term (in an
expansion in $r$): 
\begin{align}\label{gauge_help1}
r\partial^c\hmn{tc}{\emph{n,m}} &= 
\sum_{\ell\geq 0}\frac{(\ell+1)(\ell+2)}{2\ell+1}\B{L}{\emph{n,m}}
\hat{\omega}^L  -\sum_{\ell\geq 2}(\ell-1)\C{L}{\emph{n,m}}
\hat{\omega}^L.
\end{align}
And in evaluating the $a$-component, the following formula is useful
for the same purpose: 
\begin{align}\label{gauge_help2}
& r\partial^b\hmn{ab}{\emph{n,m}}-\tfrac{1}{2}r
\eta^{\beta\gamma}\partial_a\hmn{\beta\gamma}{\emph{n,m}}
\nonumber\\
&=
\sum_{\ell\geq0}\left[\tfrac{1}{2}\ell(\K{L}{\emph{n,m}}
-\A{L}{\emph{n,m}}) +\frac{(\ell+2)(\ell+3)}{2\ell+3}\E{L}{\emph{n,m}} 
-\tfrac{1}{6}\ell\F{L}{\emph{n,m}}\right]\hat{\omega}_a{}^L 
\nonumber\\
&\quad +\sum_{\ell\geq 1}\bigg[\frac{\ell(\ell+1)}{2(2\ell+1)}
(\A{aL-1}{\emph{n,m}} -\K{aL-1}{\emph{n,m}}) +
\frac{(\ell+1)^2(2\ell+3)}{6(2\ell+1)(2\ell-1)}\F{aL-1}{\emph{n,m}} 
\nonumber\\
&\quad-(\ell-2)\H{aL-1}{\emph{n,m}}\bigg]\hat{\omega}^{L-1} 
+\sum_{\ell\geq 1}\left[\frac{(\ell+2)^2}{2(2\ell+1)}
\G{dL-1}{\emph{n,m}} 
-\tfrac{1}{2}(\ell-1)\I{dL-1}{\emph{n,m}}\right]\epsilon_{ac}{}^d
\hat{\omega}^{cL-1}
\end{align}
where we have defined $\H{a}{\emph{n,m}}:= 0$ and
$\I{a}{\emph{n,m}}:= 0$. 

The unit vector $\omega_i$ satisfies the following integral
identities: 
\begin{align}
\int\hat{\omega}_Ld\Omega &=0 {\rm\ if\ } \ell>0, 
\label{omega_integral}\\
\int \omega_L d\Omega & = 0 {\rm\ if\ } \ell {\rm\ is\ odd}, \\ 
\int \omega_L d\Omega & = 4\pi\frac{\delta_{\lbrace i_1 i_2}\ldots
\delta_{i_{\ell-1}i_{\ell\rbrace}}}{(\ell+1)!!} {\rm\ if\ } \ell
{\rm\ is\ even}, \label{omegahat_integral}
\end{align}
where the curly braces indicate the smallest set  of permutations of
indices that make the result symmetric. For example, 
$\delta_{\lbrace ab}\omega_{c\rbrace}=
\delta_{ab}\omega_c+\delta_{bc}\omega_a+\delta_{ca}\omega_b$.